# Fundamentals In Quantum Algorithms: A Tutorial Series Using Qiskit Continued

Daniel Koch[1*], Saahil Patel[1], Laura Wessing[1], Paul M. Alsing[1]
[1] *Air Force Research Lab, Information Directorate, Rome, NY*

With the increasing rise of publicly available high level quantum computing languages, the field of Quantum Computing has reached an important milestone of separation of software from hardware. Consequently, the study of Quantum Algorithms is beginning to emerge as university courses and disciplines around the world, spanning physics, math, and computer science departments alike. As a continuation to its predecessor: "Introduction to Coding Quantum Algorithms: A Tutorial Series Using Qiskit", this tutorial series aims to help understand several of the most promising quantum algorithms to date, including Phase Estimation, Shor's, QAOA, VQE, and several others. Accompanying each algorithm's theoretical foundations are coding examples utilizing IBM's Qiskit, demonstrating the strengths and challenges of implementing each algorithm in gate-based quantum computing.

*Corresponding Author: dkoch.afrl@gmail.com

Code files available upon request.

# Table of Contents

The following lessons are adapted from their original Ipython Notebook formats. Citations and references can be found within each self-contained lesson. For more information / access to the original files, please contact the corresponding author * above.

# Lesson 6 - Quantum Fourier Transformation

In this tutorial, we will be switching gears from studying algorithms in order to cover a topic which is more akin to a subroutine: the Quantum Fourier Transformation (QFT). Much like how the Hadamard Transformation was the basis for all of the algorithms studied in lessons 5.1 - 5.4, the QFT will play a major role in algorithms studied in several of the coming lessons. At their core, the two transformations share a lot of similarities, both in their effect and usage in quantum algorithms.

In order to make sure that all cells of code run properly throughout this lesson, please run the following cell of code below:

In [ ]:
```
from qiskit import ClassicalRegister, QuantumRegister, QuantumCircuit, Aer, execute
import Our_Qiskit_Functions as oq
import math as m
import numpy as np
```

## Importance of Unitary Transformations

If we think back to lessons 5.1 - 5.4, we should ask ourselves: what was it about the Hadamard Transformation that allowed all of those algorithms to be successful. For the blackbox problems, we would say that it allowed us to work with all possible states at once, thus outperforming classical algorithms that are forced to check only one input at a time. Additionally for the Grover Algorithm, a second vital role of the Hadamard Transformation was that it allowed us to perform a 'reflection about the average' by transforming to a different basis.

The success of any transformation can always be traced to $\textit{the way}$ it maps states. In particular, by studying the way a certain transformation maps individual states, as well as how it maps combinations of states, we can learn about what types of advantages it can achieve. Or in other words, a transformation provides us with two 'domains' in which to work, where we can use the advantages of each to solve complex problems. Visually, moving to a transformed basis in order to achieve some desired effect looks like:

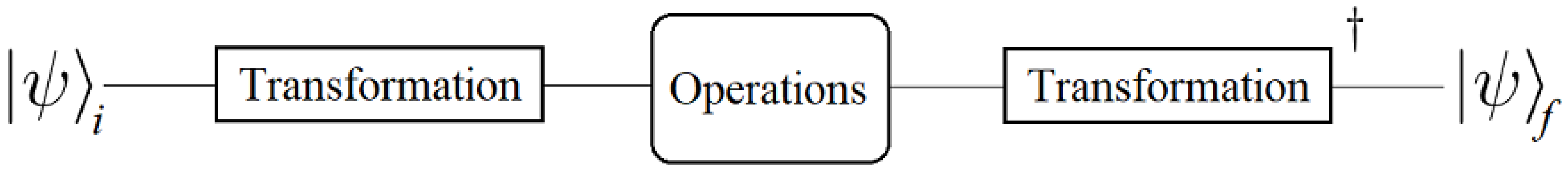

The operations we perform 'inside' the transformation are dependent on the algorithm, and what type of problem we are trying to solve. Sometimes, we need to perform transformations $\textit{within}$ transformations in order to get a certain effect. For example, the Grover Diffusion Operator from lesson 5.4 is essentially an X Transformation inside of a H Transformation, granting us the ability to flip the sign on the $|00...0\rangle$ state.

Another important property of transformations are the operators that map back and forth between the bases. In the figure above, this is represented by the $\text{Transformation}$ and $\text{Transformation}^{\dagger}$ operations. For certain algorithms the same operator is used for both transformations, like $H^N$ used in the Grover Diffusion Operator in lesson 5.4, but in general this is not always the case. For instance, several of the coming algorithms we shall see only require a single use of the QFT$^{\dagger}$, oftentimes paired with $H^N$ as the initial transformation. When using the same transformation multiple times however, it is important to consider the operator's inverse, as shown below for Hadamard Transformation:

$$\langle 01 | 10 \rangle = 0$$

$$H^2 |01\rangle \;=\; \frac{1}{2}\big( |00\rangle - |01\rangle + |10\rangle - |11\rangle \big) \qquad H^2 |10\rangle \;=\; \frac{1}{2}\big( |00\rangle + |01\rangle - |10\rangle - |11\rangle \big)$$

$$\frac{1}{4}\big( \langle 00| - \langle 01| + \langle 10| - \langle 11| \big)\big( |00\rangle + |01\rangle - |10\rangle - |11\rangle \big) \;=\; \frac{1}{4}\big(1 - 1 - 1 + 1\big) \;=\; 0$$

Or written in a more elegant way:

$$\begin{aligned} \langle 01 |\, H^{\dagger 2}\, H^2 \,| 10 \rangle &= \langle 01 |\, (H^{\dagger} H) \otimes (H^{\dagger} H) \,| 10 \rangle \\ &= \langle 01 |\, (H^{\dagger} H) \otimes (H^{\dagger} H) \,| 10 \rangle \\ &= \langle 01 |\, I \otimes I \,| 10 \rangle \\ &= \langle 01 | 10 \rangle \;=\; 0 \end{aligned}$$

What's important to note in the second example is the property $H^{\dagger} H = 1$. This is true of all unitary operators, $\textit{but}$ not all unitary operators are their own complex conjugate like $H^N$. That is to say, the Hadamard transformation is special in that $H = H^{\dagger}$, a property known as being Hermetian, which means that we can apply the same operation to transform back and forth between bases. And since an $H^N$ transformation is essentially $N$ individual 1-qubit Hadamard Transformations in parallel: $H \otimes H \otimes H \cdots$, the net result is that $H^{N^{\dagger}} = H^N$.

If we have an operation that acts on $N$ qubits, and can be decomposed into $N$ individual Hermetian operators: $O_0 \otimes O_1 \otimes O_2 \cdots$, then the total operator is Hermitian as well. For example:

In [ ]:

```
q  = QuantumRegister(4,name='q')
qc = QuantumCircuit(q,name='qc')
#------------------------------
qc.x( q[0] )
qc.x( q[2] )
print('__ Initial State __')
oq.Wavefunction(qc)

qc.h( q[0] )
qc.x( q[1] )
qc.y( q[2] )
qc.z( q[3] )
print('\n__ Opertor: H + X + Y + Z __')
oq.Wavefunction(qc)

qc.h( q[0] )
qc.x( q[1] )
qc.y( q[2] )
qc.z( q[3] )
print('\n__ Opertor: H + X + Y + Z __')
oq.Wavefunction(qc)
```

```
__ Initial State __
1.0 |1010>

__ Opertor: H + X + Y + Z __
-0.70711j |0100>    0.70711j |1100>

__ Opertor: H + X + Y + Z __
1.0 |1010>
```

In this example, we implement a 4-qubit operator which can be decomposed as: $H_0 \otimes X_1 \otimes Y_2 \otimes Z_3$. Each of the individual components is Hermtian, therefore the total operator is Hermitian as well. This is demonstrated by the fact that two applications of this operator return us back to our original state. However, as we pointed out earlier, not all multi-qubit operations are their own complex conjugate. For example, consider a single qubit operator that can be decomposed as: $U \equiv X_0 \otimes Z_0$

In [ ]:

```
q  = QuantumRegister(1,name='q')
qc = QuantumCircuit(q,name='qc')
#---------------------------------
print('__ Initial State __')
oq.Wavefunction(qc)

qc.x( q[0] )
qc.z( q[0] )
print('\n__ Opertor: XZ __')
oq.Wavefunction(qc)

qc.x( q[0] )
qc.z( q[0] )
print('\n__ Opertor: XZ __')
oq.Wavefunction(qc)
```

```
__ Initial State __
1.0 |0>

__ Opertor: XZ __
-1.0 |1>

__ Opertor: XZ __
-1.0 |0>
```

As we can see, applying this operator twice does not return us to our original state. Thus, $X_0 \otimes Z_0$ is not a Hermetian operation, *even though* it is made up of Hermetian components. If we define an operation composed of numerous gates that must act on the same qubit in a specific order, then chances are it won't be Hermetian. So then, if our algorithm requires us to use such an operator as a transformation, then we will need to find a *different* operator if we want to transform back, specifically the complex conjugate.

Luckily, if we know how to decompose an operation like the one in our example above, then finding the complex conjugate is as simple as reversing the order (with one caveat):

In [ ]:

```
q  = QuantumRegister(1,name='q')
qc = QuantumCircuit(q,name='qc')
#-------------------------------
print('__ Initial State __')
oq.Wavefunction(qc)

qc.x( q[0] )
qc.z( q[0] )
print('\n__ Opertor: XZ __')
oq.Wavefunction(qc)

qc.z( q[0] )
qc.x( q[0] )
print('\n__ Opertor: ZX __')
oq.Wavefunction(qc)
```

```
__ Initial State __
1.0 |0>

__ Opertor: XZ __
-1.0 |1>

__ Opertor: ZX __
1.0 |0>
```

As you may have guessed, the reason we've gone out of our way to discuss non-Hermitian operations is because the transformation we will be studying in this lesson is exactly that. The Quantum Fourier Transformation (QFT), which we will be using as a core element in several of the coming lessons, is an example where QFT and QFT$^\dagger$ are different operations. As we shall see, the relation between these two transformations is very straightforward.

## Discrete Fourier Transformation

Mathematically, it turns out that the QFT is equivilant to the Inverse Discrete Fourier Transformation (DFT$^\dagger$), but applied to the states of our quantum system. Thus, we will begin with a quick review of the DFT and its inverse. Formally written, the Discrete Fourier Transformation is as follows:

$$X = \{\, x_0,\ \ldots,\ x_k,\ \ldots,\ x_{N-1} \,\}$$
$$\tilde{X} = \{\, \tilde{x}_0,\ \ldots,\ \tilde{x}_k,\ \ldots,\ \tilde{x}_{N-1} \,\}$$

$$\tilde{x}_k \;=\; \sum_{j=0}^{N-1} x_j \cdot e^{-2\pi i \frac{k \cdot j}{N}}$$

Where the DFT maps all of the numbers in $X$ to $\tilde{X}$, and $e^{\pm i\theta} \;=\; \cos(\theta) \pm i \sin(\theta)$.

$$X \quad - DFT \rightarrow \quad \tilde{X}$$

The DFT is defined by the sum above, which shows that each output value $\tilde{x}_k$ receives a contribution from each input value $x_k$. Specifically, each input value is multiplied by a unique complex number of the form $e^{2\pi i\theta}$, which are then all summed together. The value of each $\theta$ is determined by the multiplication of $k \cdot j$. However, to properly compare this classical function with the quantum version to come, we need a positive phase in the exponential to implement DFT $^\dagger$:

$$X \;=\; \begin{bmatrix} 1 & -1 & -1 & 1 \end{bmatrix}$$

$$\begin{aligned} \tilde{x}_1 \;&=\; \sum_{j=0}^{3} x_j \cdot e^{2i\pi \frac{1 \cdot j}{4}} \\ &=\; 1 \cdot e^{0} - 1 \cdot e^{\frac{i\pi}{2}} - 1 \cdot e^{i\pi} + 1 \cdot e^{\frac{3i\pi}{2}} \\ &=\; 1 - i + 1 - i \\ &=\; 2 - 2i \end{aligned}$$

and the full transformation:

$$X = \begin{bmatrix} 1 & -1 & -1 & 1 \end{bmatrix} \qquad \longrightarrow \qquad \tilde{X} = \begin{bmatrix} 0 & 2-2i & 0 & 2+2i \end{bmatrix}$$

These $e^{i\theta}$ terms which multiply each $x_j$ input are derived from the concept of taking the roots of -1, which we will not cover here. I encourage you to work through all of the example above, as you will want to really develop a good feel for these transformations if you plan to continue onto the lesson 7 algorithms and beyond. For our goal of understanding the QFT here, we will only be taking from the DFT what we need. In particular, let's see what this DFT looks like in matrix representation:

$$\begin{bmatrix} 1 & 1 & 1 & 1 \\ 1 & i & -1 & -i \\ 1 & -1 & 1 & -1 \\ 1 & -i & -1 & i \end{bmatrix} \begin{bmatrix} 1 \\ -1 \\ -1 \\ 1 \end{bmatrix} \quad = \quad \begin{bmatrix} 0 \\ 2-2i \\ 0 \\ 2+2i \end{bmatrix}$$

where the values in the matrix above can all be expressed as:

$$\omega \equiv e^{\frac{2i\pi}{N}} \qquad\qquad \text{DFT matrix:} \quad F_4 = \begin{bmatrix} \omega^0 & \omega^0 & \omega^0 & \omega^0 \\ \omega^0 & \omega^1 & \omega^2 & \omega^3 \\ \omega^0 & \omega^2 & \omega^4 & \omega^6 \\ \omega^0 & \omega^3 & \omega^6 & \omega^9 \end{bmatrix}$$

The powers on all of the $\omega$'s come from the the products of $k \cdot j$, and $N$ refers to the total number of values being transformed ($N = 4$ for our example):

$$\begin{aligned} k \cdot j : \quad & 0 \cdot 1 \qquad 1 \cdot 2 \qquad 3 \cdot 1 \\ \omega^{k \cdot j} : \quad & \omega^0 \qquad \omega^2 \qquad \omega^3 \\ = \quad & e^0 \qquad e^{i\pi} \qquad e^{\frac{3i\pi}{2}} \\ = \quad & 1 \qquad -1 \qquad -i \end{aligned}$$

This covers everything we need from the Discrete Fourier Transformation, but I encourage you to check out other references for more information.

## Quantum Fourier Transformation

We now have a formal definition for the Inverse Discrete Fourier Transformation, so how do we make it quantum? Well, we've already shown how to represent the DFT$^\dagger$ as a matrix, so our task now is to implement it as an operator. Since we are dealing with quantum systems, we will naturally gravitate towards transformations of the size $2^N$. Let's use a 2-qubit example to illustrate how the DFT$^\dagger$ will look on a quantum system:

$$|\psi\rangle = \frac{1}{2}\Big( |00\rangle - |10\rangle - |01\rangle + |11\rangle \Big)$$

$$F_4 \, |\psi\rangle = \frac{1}{2}\Big( (1-i)\,|10\rangle + (1+i)\,|11\rangle \Big)$$

For clarity, the vector representing the state above is in the following order:

$$\begin{bmatrix} |00\rangle \\ |10\rangle \\ |01\rangle \\ |11\rangle \end{bmatrix}$$

This example is the quantum version of our $X \rightarrow \tilde{X}$ transformation from earlier. Our initial state corresponds to $X$, and our final state is $\tilde{X}$. Verifying that this operation is indeed unitary is simple enough (our initial and final amplitudes squared all sum to $1$), which means that $F_4$ is a legitimate quantum operator. And in general, any DFT matrix is guaranteed to be unitary with an accompanying normalization factor.

### Implementing a QFT

At this point, we have the structure for generating our QFT matrices, and the corresponding vector representations of our states. From the mathematical perspective, we have the full picture for the QFT. However, as we've already seen with past algorithms, simply writing it down doesn't do it justice. If we want to actually run a QFT in our quantum algorithms, then we need a way of translating the mathematical picture into gates.

The way in which we are going to achieve our QFT circuit is quite elegant, and by no means obvious at first. Proposed by Don Coppersmith **[1]**, it turns out that the only gates we need in order to construct a $2^N$ QFT are $H$ and control-$R_\phi$ ($CU1$), our trusty Hadmard gate along with some control-phase gates. Even better yet, we will not require any additional ancilla qubits. Below is the general template for how to construct a QFT circuit on $N$ qubits, acting on a $2^N$ space of states:

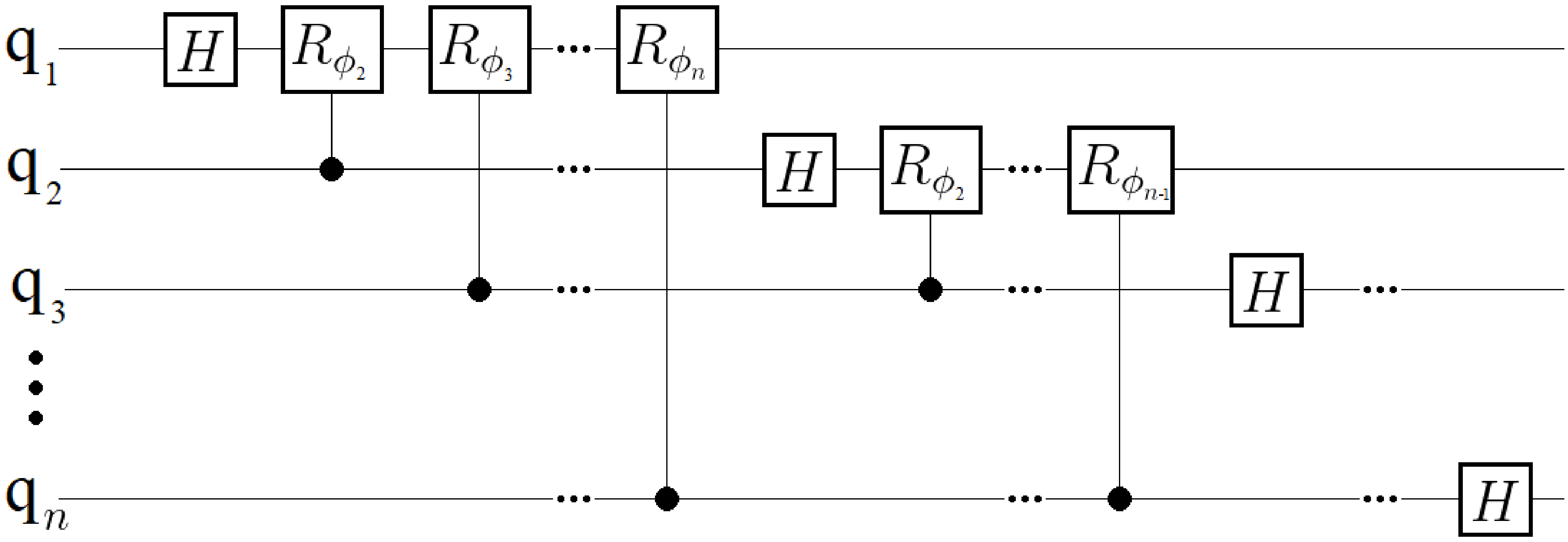

where

$$\phi_m = e^{\frac{2i\pi}{2^m}}$$

At first glance, this circuit may look a bit complex, but it's pattern is actually quite straightforward. Each qubit in the system undergoes the same process: a Hadamard gate followed by a series of control-phase gates. The number of $R_\phi$ gates that a qubit experiences is determined by its index, whereby the first qubit in the system receives $N-1$, and the last qubit doesn't receive any. In addition, the phases for each of the $R_\phi$ gates is a set pattern: $\phi = \frac{\pi}{2}, \frac{\pi}{4}, \frac{\pi}{8}$ ... It may not be immediately obvious why the circuit above works, but we're going to first test it out with a coding example (and then break it down afterwards):

In [ ]:

```
q  = QuantumRegister(2,name='q')
qc = QuantumCircuit(q,name='qc')
#-------------------------------
qc.x( q[0] )
qc.h( q[0] )
qc.x( q[1] )
qc.h( q[1] )
print('__ Initial State __')
oq.Wavefunction(qc)

qc.h( q[0] )
qc.cu1( m.pi/2,q[1],q[0] )
qc.h( q[1] )
print('\n__ After QFT __')
oq.Wavefunction(qc)
```

```
__ Initial State __
0.5 |00>    -0.5 |10>    -0.5 |01>    0.5 |11>

__ After QFT __
0.5-0.5j |10>    0.5+0.5j |11>
```

Try and match the pattern in the template above, with the steps we've implemented in this code example.:

$$1)\quad H_0 \qquad 2)\quad R_{\frac{\pi}{2}\,10} \qquad 2) H_1$$

where the $10$ subscript on the control-phase gate represents qubit $1$ is the control, and qubit $0$ is the target. Confirm for yourself that these are indeed the steps written into our coding example, and that they match the QFT template. Next, we will do one more example, this time with 3 qubits:

$$QFT\,|001\rangle = \frac{1}{4}\Big(\sqrt{2}\,|000\rangle - \sqrt{2}\,|001\rangle + \sqrt{2}i\,|010\rangle - \sqrt{2}i\,|011\rangle + (1+i)\,|100\rangle - (1+i)\,|101\rangle + (-1+i)\,|110\rangle + (1-i)\,|111\rangle\Big)$$

In [ ]:

```
q  = QuantumRegister(3,name='q')
qc = QuantumCircuit(q,name='qc')
#---------------------------------
qc.x( q[2] )
print('__ Initial State __')
oq.Wavefunction(qc)

#------------------------   qubit 0
qc.h( q[0] )
qc.cu1( m.pi/2,q[1],q[0] )
qc.cu1( m.pi/4,q[2],q[0] )
#------------------------   qubit 1
qc.h( q[1] )
qc.cu1( m.pi/2,q[2],q[1] )
#------------------------   qubit 2
qc.h( q[2] )
print('\n__ After QFT __')
oq.Wavefunction(qc)
```

```
__ Initial State __
1.0 |001>

__ After QFT __
0.35355 |000>    0.25+0.25j |100>    0.35355j |010>    -0.25+0.25j |110>    -0.35355 |001>    -0.25-0.25j |101>    -0.35355
j |011>    0.25-0.25j |111>
```

In this example, we've broken up the QFT instructions into three sections, where each section incorporates all of the operations being applied to one of the three qubits. Just like in the QFT template shown above, the number of operations decreases by 1 per qubit, where the last qubit only receives a single $H$.

Ultimately, writing out all the steps for a QFT is a tedious task, so just like the n_NOT function from lesson 4, we will use a custom function called **QFT()** from Our_Qiskit_Functions instead:

In [ ]:

```
q  = QuantumRegister(3,name='q')
qc = QuantumCircuit(q,name='qc')
#------------------------------
qc.x( q[2] )
print('__ Initial State __')
oq.Wavefunction(qc)

oq.QFT(qc,q,3)
print('\n__ After QFT __')
oq.Wavefunction(qc)
```

```
__ Initial State __
1.0 |001>

__ After QFT __
0.35355 |000>    0.25+0.25j |100>    0.35355j |010>    -0.25+0.25j |110>    -0.35355 |001>    -0.25-0.25j |101>    -0.35355
j |011>    0.25-0.25j |111>
```

The $\text{QFT}()$ function above handles all of the gates needed for the Quantum Fourier Transformation, only requiring that we pass the $\text{QuantumCircuit}$, $\text{QuantumRegister}$, and number of qubits.

### Why The QFT Circuit Works

Now that we have shown that we *can* implement a QFT, let's talk about why it works. If you followed along the derivation of the DFT$^{\dagger}$ matrix at the beginning of this lesson, then the way in which we are achieving these operations may seem surprisingly simple. For example, take a look at all of the complexity happening in the 2-qubit QFT matrix from earlier, and then note that we achieve all of this with only 2 $H$'s and one $R_{\phi}$. To make sense of how our pattern of gates is achieving all the desired phases, we will work through a 3-qubit example:

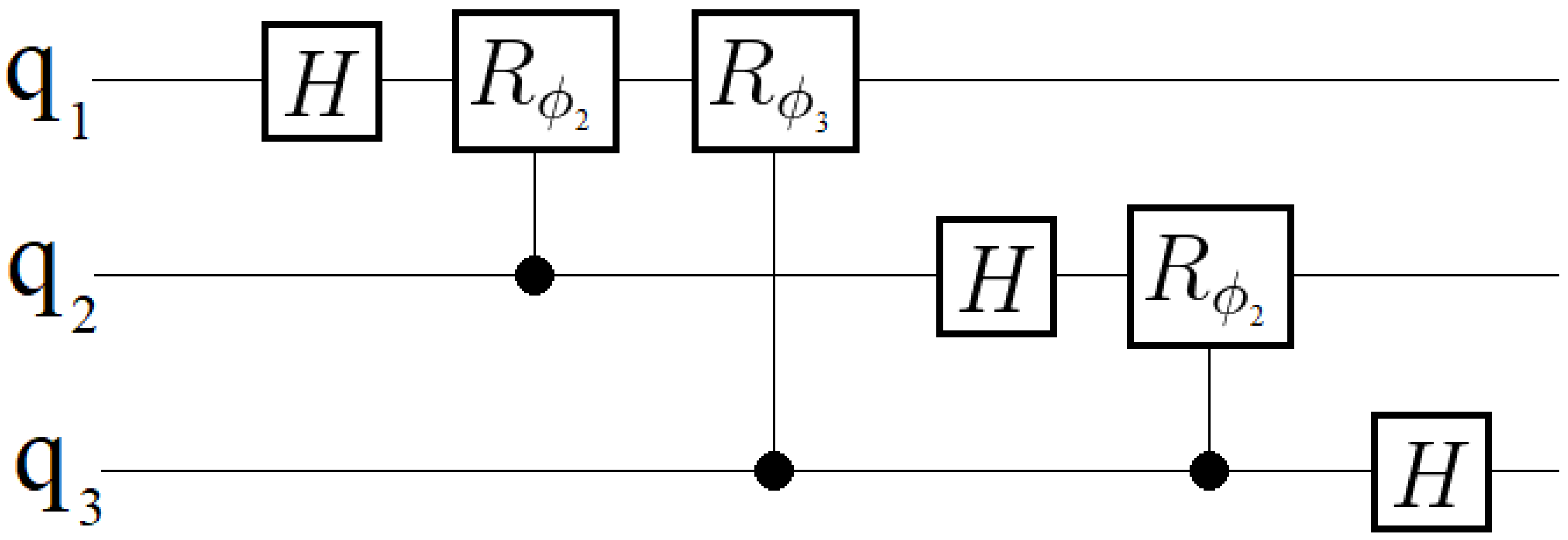

Let's start with $q_1$, and see what its final state will look like at the end of the circuit. We want to be general here, so we will say that our qubit starts off in the $|\,q_1\,\rangle$, where $q_1$ is either a $0$ or $1$. Following along with all of the operations that $q_1$ receives:

$$H: \quad \frac{1}{\sqrt{2}}\Big(\,|\,0\,\rangle \;+\; e^{q_1 \cdot i\pi}\,|\,1\,\rangle\,\Big)$$

$$R_{\phi_2}: \quad \frac{1}{\sqrt{2}}\Big(\,|\,0\,\rangle \;+\; e^{q_1 \cdot i\pi} \cdot e^{q_2 \cdot \frac{i\pi}{2}}\,|\,1\,\rangle\,\Big)$$

$$R_{\phi_3}: \quad \frac{1}{\sqrt{2}}\Big(\,|\,0\,\rangle \;+\; e^{q_1 \cdot i\pi} \cdot e^{q_2 \cdot \frac{i\pi}{2}} \cdot e^{q_3 \cdot \frac{i\pi}{4}}\,|\,1\,\rangle\,\Big)$$

First, take a look at how we've chosen to write the effect of our Hadamard gate on $q_1$: $\frac{1}{\sqrt{2}}\Big(\,|\,0\,\rangle \;+\; e^{q_1 \cdot i\pi}\,|\,1\,\rangle\,\Big)$. Typically we would write this with something like $(\text{-}1)^{q_1}$, where the state of $q_1$ determines whether or not the the Hadamard gate results in a positive or negative $|\,1\,\rangle$ state. Here however, we've chosen to express $\text{-}1$ as $e^{i\pi}$ in order to be consistent with the other gate effects.

Next are the control-phase gates, which produce a similar effect to that of the Hadamard gate at first glance, but have an important difference. Remember that control-phase gates only apply an effect when both the target and control qubits are in the $|\,1\,\rangle$ state. This is why a $H$ gate is necessary before any of the $R_{\phi}$'s, to ensure that $q_1$ is in a superposition state of both $|\,0\,\rangle$ and $|\,1\,\rangle$, and the effect of the $R_{\phi}$ gate applies an additional phase to the $|\,1\,\rangle$ component of $q_1$. However, because this is a control gate, and we must take into account that $q_2$ and $q_3$ may not be in the $|\,1\,\rangle$ state, there is an extra term multiplying each of the added phases, for example: $e^{q_2 \cdot \frac{i\pi}{2}}$. We can understand this extra term as the condition that $q_2$ is in the $|\,1\,\rangle$ state, otherwise no phase is applied.

This pattern continues for each qubit, all the way down to the last. Each qubit receives a number of phases added to their $|\,1\,\rangle$ component, which will then all be multiplied together in the final state:

$$\begin{aligned} |\,\psi\,\rangle_f &= |\,q_1\,\rangle_f \otimes |\,q_2\,\rangle_f \otimes |\,q_3\,\rangle_f \\ &= \frac{1}{2\sqrt{2}}\Big(\,|\,0\,\rangle + e^{q_1 \cdot i\pi} \cdot e^{q_2 \cdot \frac{i\pi}{2}} \cdot e^{q_3 \cdot \frac{i\pi}{4}}\,|\,1\,\rangle\,\Big) \otimes \Big(\,|\,0\,\rangle + e^{q_2 \cdot i\pi} \cdot e^{q_3 \cdot \frac{i\pi}{2}}\,|\,1\,\rangle\,\Big) \otimes \Big(\,|\,0\,\rangle + e^{q_3 \cdot i\pi}\,|\,1\,\rangle\,\Big) \end{aligned}$$

Hopefully now you are starting to see how we are able to achieve all of the various phases shown in the QFT matrices from earlier. If this is your first time seeing the math behind the QFT circuit, I highly recommend finishing the example above and multiplying out all of the states. In essence, all of the unique phase combinations are achieved through the $e^{q_j \cdot \frac{i\pi}{2^k}}\;|\,1\,\rangle$ states. The math is a little cumbersome, even for just three qubits, but hopefully this example illustrates the idea behind why we are able to achieve a QFT with this quantum circuit.

As a final optional exercise, I would encourage you to prove for yourself that mathematically our circuit is equal to the matrix representation from earlier, up to a normalization factor:

$$\text{show that} \quad H_1 R_\phi H_0 |\psi\rangle \quad = \quad \frac{1}{2}\begin{bmatrix} 1 & 1 & 1 & 1 \\ 1 & -1 & i & -i \\ 1 & 1 & -1 & -1 \\ 1 & -1 & -i & i \end{bmatrix}\begin{bmatrix} |00\rangle \\ |10\rangle \\ |01\rangle \\ |11\rangle \end{bmatrix} \qquad \phi = \frac{\pi}{2}$$

$$\text{hint: don't forget to represent } H_0 \text{ and } H_1 \text{ as 4x4 matrices!} \quad \longrightarrow \quad H_0 \equiv H_0 \otimes I_1$$

### Inverse QFT

Now that we have a way of transforming our system via a QFT, and hopefully a better intuition as to why it works, next we need to be able to transform back. Remember that the power of using transformations in quantum algorithms relies on being able to transform back and forth between bases. And as we mentioned earlier, our QFT transformation is not Hermetian, so the same construction of gates will not transform us back. To verify this, let's see what happens when we try to use our $\text{QFT}()$ function twice:

In [ ]:

```
q  = QuantumRegister(2,name='q')
qc = QuantumCircuit(q,name='qc')
#------------------------------
qc.x( q[0] )
qc.h( q[0] )
qc.x( q[1] )
qc.h( q[1] )
print('__ Initial State __')
oq.Wavefunction(qc)

oq.QFT(qc,q,2)
print('\n__ First QFT __')
oq.Wavefunction(qc)

oq.QFT(qc,q,2)
print('\n__ Second QFT __')
oq.Wavefunction(qc)
```

```
__ Initial State __
0.5 |00>    -0.5 |10>    -0.5 |01>    0.5 |11>

__ First QFT __
0.5-0.5j |10>    0.5+0.5j |11>

__ Second QFT __
0.5 |00>    -0.5j |01>    -0.5+0.5j |11>
```

Sure enough, we do not return to our original state. From our quantum computing perspective, we can understand why the QFT doesn't transform us back to our original state if we look at two QFTs in a row:

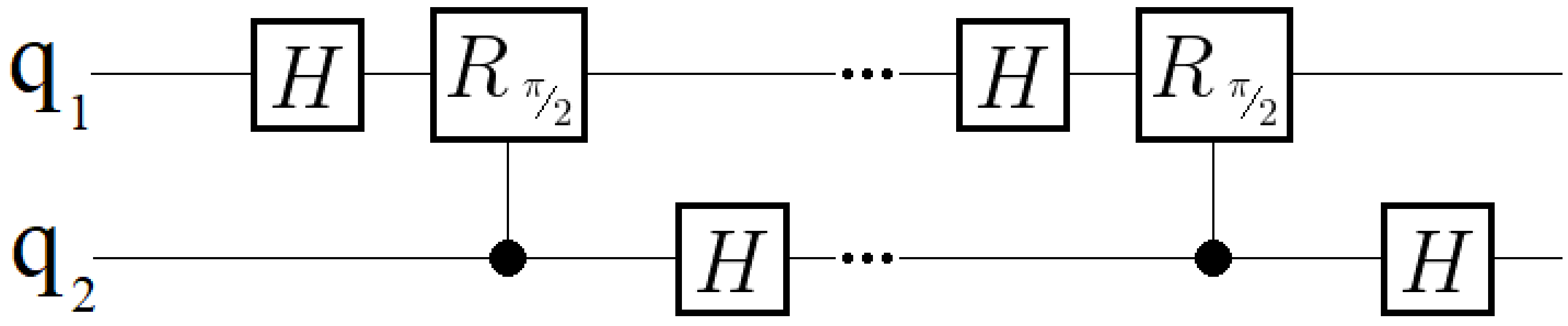

What should jump out at you is the apparent lack of symmetry here. Recall our example earlier of the gate $XZ$, and that the correct inverse transformation was to change the order: $ZX$. Here, if we want to implement the inverse of our QFT, we will need to invoke the same strategy of reversing the order of all the gates. In essence, imagine placing a mirror after our QFT, and the reflection will be our inverse QFT, with one slight change:

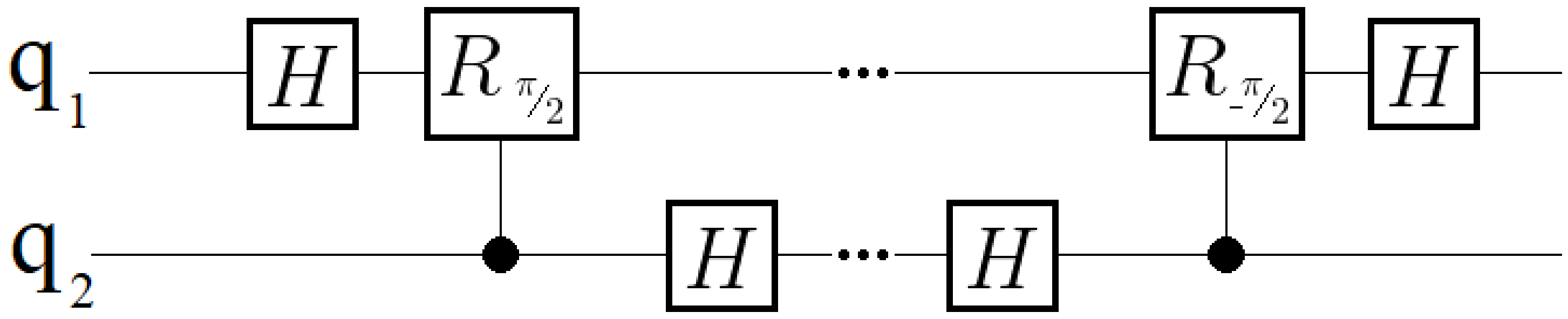

The slight change here is that our second $R_\phi$ has the opposite sign of our first. Conceptually, this should make sense: if our original transformation applies a phase $\theta$, then our inverse should apply the opposite phase, -$\theta$. As we pointed out earlier, the inverse of a transformation needs to be the complex conjugate of the original, which is why we need negative phases on all of the $\theta$'s. All together, our inverse QFT must be the *exact* reverse ordering our QFT, with all opposite phases on the $R_\phi$ gates:

In [ ]:
```
q  = QuantumRegister(2,name='q')
qc = QuantumCircuit(q,name='qc')
#------------------------------
qc.x( q[0] )
qc.h( q[0] )
qc.x( q[1] )
qc.h( q[1] )
print('__ Initial State __')
oq.Wavefunction(qc)

qc.h( q[0] )
qc.cu1( m.pi/2,q[1],q[0] )
qc.h( q[1] )
print('\n__ QFT __')
oq.Wavefunction(qc)

qc.h( q[1] )
qc.cu1( -m.pi/2,q[1],q[0] )
qc.h( q[0] )
print('\n__ Inverse QFT __')
oq.Wavefunction(qc)
```

```
__ Initial State __
0.5 |00>    -0.5 |10>    -0.5 |01>    0.5 |11>

__ QFT __
0.5-0.5j |10>    0.5+0.5j |11>

__ Inverse QFT __
0.5 |00>    -0.5 |10>    -0.5 |01>    0.5 |11>
```

Sure enough, we recover our original state, which means that we performed the correct inverse transformation. And like our $\mathrm{QFT()}$ function, we can use $\mathbf{QFT_dgr()}$ from $\mathrm{Our_Qiskit_Functions}$ to implement our QFT$^{\dagger}$:

In [ ]:
```
q  = QuantumRegister(2,name='q')
qc = QuantumCircuit(q,name='qc')
#--------------------------------
qc.x( q[0] )
qc.h( q[0] )
qc.x( q[1] )
qc.h( q[1] )
print('__ Initial State __')
oq.Wavefunction(qc)

oq.QFT(qc,q,2)
print('\n__ QFT __')
oq.Wavefunction(qc)

oq.QFT_dgr(qc,q,2)
print('\n__ Inverse QFT __')
oq.Wavefunction(qc)
```

```
__ Initial State __
0.5 |00>    -0.5 |10>    -0.5 |01>    0.5 |11>

__ QFT __
0.5-0.5j |10>    0.5+0.5j |11>

__ Inverse QFT __
0.5 |00>    -0.5 |10>    -0.5 |01>    0.5 |11>
```

Now that we have $\mathrm{QFT}$ and $\mathrm{QFT_dgr}$, we are nearly finished covering the basics of the Quantum Fourier Transformation.

### The Proper QFT: SWAP Gates

In the cell of code above, our 3-qubit QFT and QFT$^{\dagger}$ circuits successfully transform our state and then back. However, there is actually one final ingredient left to our Quantum Fourier Transformation, and it has to do with the way in which the phases are distributed amongst the states represented as binary numbers. More specifically, we constructed our QFT circuit with the intention that it be a parallel to the Inverse Discrete Fourier Transformation. But as we will see in the cell of code below, the two do not agree on what the final state should be after being applied to the state $|\,001\,\rangle$:

In [ ]:

```
X = [0,1/m.sqrt(8),0,0,0,0,0,0]
FX = oq.DFT( X, inverse=True)

print('_____ DFT\u2020 _____')
for i in np.arange(len(FX)):
    print('State:  ',oq.Binary(int(i),2**3,'R'),'    Amplitude: ',FX[i])

#=====================================================================

q  = QuantumRegister(3,name='q')
qc = QuantumCircuit(q,name='qc')
qc.x( q[2] )
oq.QFT(qc,q,3)
print('\n_____ QFT _____')
oq.Wavefunction(qc)
```

```
_____ DFT† _____
State:   [0, 0, 0]     Amplitude:  0.3536
State:   [0, 0, 1]     Amplitude:  (0.25+0.25j)
State:   [0, 1, 0]     Amplitude:  0.3536j
State:   [0, 1, 1]     Amplitude:  (-0.25+0.25j)
State:   [1, 0, 0]     Amplitude:  -0.3536
State:   [1, 0, 1]     Amplitude:  (-0.25-0.25j)
State:   [1, 1, 0]     Amplitude:  (-0-0.3536j)
State:   [1, 1, 1]     Amplitude:  (0.25-0.25j)

_____ QFT _____
0.35355 |000>    0.25+0.25j |100>    0.35355j |010>    -0.25+0.25j |110>    -0.35355 |001>    -0.25-0.25j |101>    -0.35355
j |011>    0.25-0.25j |111>
```

The cell of code above applies an Inverse Discrete Fourier Transformation via the function **DFT()** from Our_Qiskit_Functions. Passing the argument inverse as True performs the DFT$^\dagger$, which will match the corresponding phases for a regular QFT.

If you compare the amplitudes between the two examples above, the DFT$^\dagger$ and QFT, you will notice that they do not match. More specifically, they are almost in agreement, except for the ordering of the qubits. For example, our QFT circuit resulted in the amplitude $\frac{1+i}{4}$ on the state $|100\rangle$, whereas our inverse DFT gives us has placed this phase on the $|001\rangle$ state. To dissect this mismatch in final states, below is a breakdown of the individual qubit contributions resulting from a QFT operation on the $|001\rangle$ state:

$$\frac{1}{2\sqrt{2}}\Big(\,|0\rangle + e^{i\frac{\pi}{4}}|1\rangle\,\Big)\otimes\Big(\,|0\rangle + e^{i\frac{\pi}{2}}|1\rangle\,\Big)\otimes\Big(\,|0\rangle + e^{i\pi}|1\rangle\,\Big)$$

which when we distribute the phases gives us (ignoring the overall $\frac{1}{2\sqrt{2}}$ phase):

$$e^{0}|000\rangle + e^{i\pi}|001\rangle + e^{i\pi\frac{1}{2}}|010\rangle + e^{i\pi\frac{3}{2}}|011\rangle + e^{i\pi\frac{1}{4}}|100\rangle + e^{i\pi\frac{5}{4}}|101\rangle + e^{i\pi\frac{3}{4}}|110\rangle + e^{i\pi\frac{7}{4}}|111\rangle$$

Now, if we were to apply a DFT$^\dagger$ to the same state, the matrix multiplication would be as follows:

$$\begin{bmatrix} & \omega_8^{0\cdot 1} & & & & & & \\ \cdot & \omega_8^{1\cdot 1} & \cdot & \cdot & \cdot & \cdot & \cdot & \cdot \\ & \omega_8^{2\cdot 1} & & & & & & \\ & \omega_8^{3\cdot 1} & & & & & & \\ & \omega_8^{4\cdot 1} & & & & & & \\ & \omega_8^{5\cdot 1} & & & & & & \\ & \omega_8^{6\cdot 1} & & & & & & \\ & \omega_8^{7\cdot 1} & & & & & & \end{bmatrix}\begin{bmatrix} 0 \\ 1 \\ 0 \\ 0 \\ 0 \\ 0 \\ 0 \\ 0 \end{bmatrix} \qquad \omega_8 = e^{i\pi\frac{1}{4}}$$

where $\omega_8 = e^{i\pi\frac{1}{4}}$ and the vector ordering of all the states is: $\big[\,|000\rangle \quad |001\rangle \quad |010\rangle \quad |011\rangle \quad |100\rangle \quad |101\rangle \quad |110\rangle \quad |111\rangle\big]$

The result of the matrix multiplication above then gives us:

$$= \quad e^{0}|000\rangle + e^{i\pi\frac{1}{4}}|001\rangle + e^{i\pi\frac{1}{2}}|010\rangle + e^{i\pi\frac{3}{4}}|011\rangle + e^{i\pi}|100\rangle + e^{i\pi\frac{5}{4}}|101\rangle + e^{i\pi\frac{3}{2}}|110\rangle + e^{i\pi\frac{7}{4}}|111\rangle$$

Staring at the two results above, it is clear that they are not equal. But upon closer examination, you may notice that certain individual states are in agreement on phase. Specifically, the states $|000\rangle$, $|010\rangle$, $|101\rangle$, and $|111\rangle$ are correct. It is no coincidence that these exact states turned out to be equal, as they are the four states that are symmetric with respect to qubits 1 and 3. And if these states are any clue as to what's going on here, take a look at the remaining four states and notice what would happen if we were to switch the values of qubits 1 and 3:

$$QFT: \qquad e^{i\pi}|001\rangle \qquad e^{i\frac{\pi}{4}}|100\rangle \qquad \Big| \qquad e^{i\frac{3\pi}{2}}|011\rangle \qquad e^{i\frac{3\pi}{4}}|110\rangle$$

$$DFT^\dagger: \qquad e^{i\frac{\pi}{4}}|001\rangle \qquad e^{i\pi}|100\rangle \qquad \Big| \qquad e^{i\frac{3\pi}{4}}|011\rangle \qquad e^{i\frac{3\pi}{2}}|110\rangle$$

As you can see, all we need to do in order to have our QFT match the DFT$^\dagger$ is swap the first and last qubits. And in general, this pattern applies to QFTs for larger numbers of qubits, whereby we must switch qubits symmetric about the central qubit (for example, a 4-qubit QFT would require we switch qubits 1 & 4, and 2 & 3). In terms of our quantum circuit, this problem is a quick fix, simply requiring some SWAP gates at the end of our circuit:

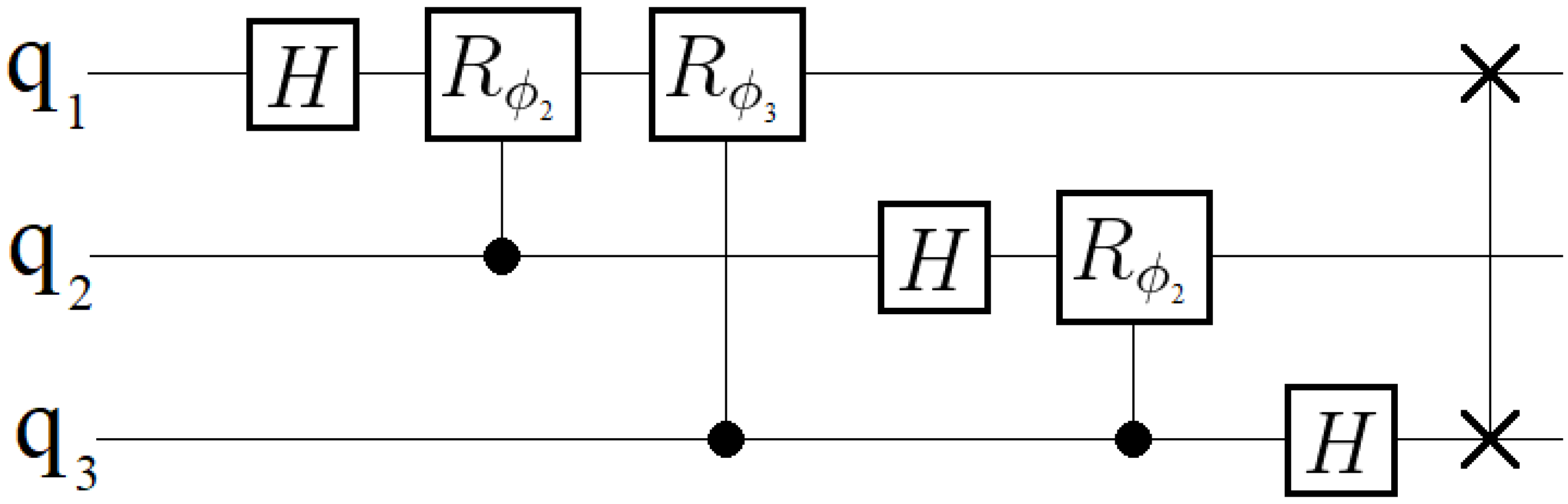

In [ ]:

```
X = [0,1/m.sqrt(8),0,0,0,0,0,0]
FX = oq.DFT( X, inverse=True)

print('_____ DFT\u2020 _____')
for i in np.arange(len(FX)):
    print('State:  ',oq.Binary(int(i),2**3,'R'),'    Amplitude: ',FX[i])

#======================================================================

q  = QuantumRegister(3,name='q')
qc = QuantumCircuit(q,name='qc')
qc.x( q[2] )
oq.QFT(qc,q,3)
qc.swap(q[0],q[2])
print('\n_____ QFT + SWAP _____')
oq.Wavefunction(qc)
```

```
_____ DFT† _____
State:   [0, 0, 0]     Amplitude:  0.3536
State:   [0, 0, 1]     Amplitude:  (0.25+0.25j)
State:   [0, 1, 0]     Amplitude:  0.3536j
State:   [0, 1, 1]     Amplitude:  (-0.25+0.25j)
State:   [1, 0, 0]     Amplitude:  -0.3536
State:   [1, 0, 1]     Amplitude:  (-0.25-0.25j)
State:   [1, 1, 0]     Amplitude:  (-0-0.3536j)
State:   [1, 1, 1]     Amplitude:  (0.25-0.25j)

_____ QFT + SWAP _____
0.35355 |000>    -0.35355 |100>    0.35355j |010>    -0.35355j |110>    0.25+0.25j |001>    -0.25-0.25j |101>    -0.25+0.25
j |011>    0.25-0.25j |111>
```

As demonstrated in the example above, the final SWAP gate at the end of our circuit remedies the state discrepancies between the QFT and DFT$^\dagger$. It's important to note however, that the reordering of our qubit states is only a cosmetic change, but one that can have very drastic implications if not taken into account. That is to say, many quantum algorithms rely critically on the QFT, and they may be designed with a certain ordering preference. Therefore, placing the final SWAP gates or not could ultimately determine the success or failure of a quantum algorithm.

In the next couple lessons, we will be using QFTs as the basis for some very important algorithms. If you would like to proceed to those lessons now, this is a sufficient concluding spot in the tutorial. The next and final section is an aside about the QFT, comparing some of its properties to the Hadamard transformation.

## Aside: Comparing QFT and H Transformations

Now that we have built up our full understanding of how to implement a QFT, let's discuss its similarities with the Hadamard Transformation, for which we've seen in several algorithms so far. First off, if you remove all of the $R_\phi$ gates from the QFT template, you're left with just a Hadamard Transformation. And in fact, our last qubit in the system only receives a single $H$. What this means, is that we can think of the QFT as a 'more complex' version of the Hadamard Transformation in some sense, where the extra bit of complexity is the additional phases. To see this, let's compare the 4X4 unitary matrices for the QFT and Hadamard Transformation on two qubits:

$$QFT \qquad\qquad\qquad\qquad H$$

$$\begin{bmatrix} 1 & 1 & 1 & 1 \\ 1 & i & -1 & -i \\ 1 & -1 & 1 & -1 \\ 1 & -i & -1 & i \end{bmatrix} \qquad \begin{bmatrix} 1 & 1 & 1 & 1 \\ 1 & 1 & -1 & -1 \\ 1 & -1 & 1 & -1 \\ 1 & -1 & -1 & 1 \end{bmatrix}$$

The two transformations are nearly identical, except for the extra presence of a couple $i$'s in the QFT. These $i$'s represent the extra complexity of the QFT for the 2-qubit case. And when we look at larger transformations, we will see more and more unique amplitudes accompanying states in the system. However, regardless of size, one property that both the Hadamard Transformation and QFT share is the way in which they map the state of all $0$'s:

$$|00...0\rangle \quad \longleftrightarrow \quad \frac{1}{\sqrt{2^N}}\big( |00...0\rangle + \cdots + |11...1\rangle \big)$$

Both transformations map the state of all $0$'s to an equal superposition, where all the states have the same positive phase. For $H^N$, we've shown that this result comes from the fact that all of the individual $H|0\rangle$ operations produce a state with positive $|0\rangle$ and $|1\rangle$ components. Similarly in the QFT, each qubit initially receives a $H$ gate followed by all control-gates. But since every qubit is initially in the $|0\rangle$ state when it acts as a control-qubit, none of the $R_\phi$ gates apply any phases. This mapping of $|00...0\rangle$ state was the core ingredient for the Grover Algorithm, allowing us to perform a reflection about the average. Thus, since our QFT also has this mapping property, we should be able to perform the Grover Algorithm using a QFT in place of the $H^N$ transformations:

In [ ]:

```
marked = [1,0]
print('marked state: ',marked)

q   = QuantumRegister(2,name='q')
anc = QuantumRegister(1,name='anc')
qc  = QuantumCircuit(q,anc,name='qc')
#-------------------------------------
qc.x( anc[0] )
oq.QFT(qc,q,2)
print('\n____ Initial State (QFT) ____')
oq.Wavefunction(qc, systems=[2,1], show_systems=[True,False])

oq.X_Transformation(qc, q, marked)
qc.h( anc[0] )
qc.ccx( q[0], q[1], anc[0] )
oq.X_Transformation(qc, q, marked)
qc.h( anc[0] )
print('\n____ Flip the Marked State ____')
oq.Wavefunction(qc, systems=[2,1], show_systems=[True,False])

oq.QFT(qc,q,2)
print('\n____ QFT ____')
oq.Wavefunction(qc, systems=[2,1], show_systems=[True,False])
qc.h( anc[0] )

oq.X_Transformation(qc, q, [0,0])
qc.ccx( q[0], q[1], anc[0] )
qc.h( anc[0] )
oq.X_Transformation(qc, q, [0,0])
print('\n____ Flip the |00> state ____')
oq.Wavefunction(qc, systems=[2,1], show_systems=[True,False])

oq.QFT_dgr(qc,q,2)
print('\n____ QFT\u2020  ____')
oq.Wavefunction(qc, systems=[2,1], show_systems=[True,False])
```

```
marked state:  [1, 0]

____ Initial State (QFT) ____
0.5 |00>    0.5 |10>    0.5 |01>    0.5 |11>

____ Flip the Marked State ____
0.5 |00>    -0.5 |10>    0.5 |01>    0.5 |11>

____ QFT ____
0.5 |00>    0.5 |10>    -0.5 |01>    0.5 |11>

____ Flip the |00> state ____
-0.5 |00>    0.5 |10>    -0.5 |01>    0.5 |11>

____ QFT†  ____
-1.0 |10>
```

Success! By using QFT and QFT$^\dagger$ we are able to perform a Grover Search for a marked state. For an explanation of the Grover Algorithm, please refer to lesson 5.4. Hopefully this example gives you an idea of just how similar the QFT and Hadamard transformation are at their core. But, the reason we will be using the QFT to accomplish the coming more complex algorithms comes from the fact that the states it maps to contain more phase differences. Or another way of saying that is, the QFT allows us to create 'more orthogonal' states (not literally), where the extra phases will prove very useful.

---

This concludes lesson 6, and our deep dive into the Quantum Fourier Transformation! Understanding the QFT is a bit tricky at first, so don't worry if everything in this tutorial doesn't feel second nature yet. Just like all of the practice we got with the Hadamard Transformation in lessons 5.1 - 5.4, we will be seeing quite a lot of the QFT in the next several tutorials, which is where we are really going to see the subtleties and advantages it has to offer.

I hope you enjoyed this lesson, and I encourage you to take a look at my other .ipynb tutorials!

### Citations

---

**[1]** D. Coppersmith, "An approximate Fourier transform useful in quantum factoring", arXiv:0201067 (1997)

# Lesson 6.1 - Quantum Adder

In this lesson, we will look at our first use of the Quantum Fourier Transformation in an algorithm: the quantum version of an adder. The way in which we will be achieving arithmetic addition in quantum via the QFT was first proposed by Draper in 2000 **[1]**. Although classical systems are already very proficient at adding numbers together, the quantum adder is a valuable algorithm worth studying, particularly for the insight into the capabilities of the QFT and QFT$^{\dagger}$ it provides. If you have not already, I recommend reading lesson 6 before proceeding:

Lesson 6 - Quantum Fourier Transformation

In order to make sure that all cells of code run properly throughout this lesson, please run the following cell of code below:

In [ ]:

```
from qiskit import ClassicalRegister, QuantumRegister, QuantumCircuit, Aer, execute
import Our_Qiskit_Functions as oq
import numpy as np
import math as m
```

## Classical Adder

To begin, let us briefly cover what it means to add classically. Although it may seem like a simple concept, it has been a very active field of research in computer science since the beginning of computers. This is because the ability to add, that is to add *quickly*, is a very powerful component for classical computing. In order for modern computers to carry out millions of computations per second, they rely on perfected hardware and software techniques to carry out basic arithmetic operations at lightning speeds.

So then, let's quickly review the basics of how a classical computer adds two integers together. As humans, we like to work in base-10, which is what we are taught from a young age. We first learn the ten basic integers: $0, 1, 2, 3, 4, 5, 6, 7, 8, 9$, and from there we advance to representing numbers of ten and greater by using combinations of these ten symbols. As you probably know however, classical computers only have the concept of 0 and 1, working exclusively in base-2. Thus, a classical adder must represent and compute all numbers in base-2, translating back to base-10 at the very end. Adding in base-2 is exactly like adding in base-10: when the sum of two numbers exceeds a power of $2^N$ we must carry over a $1$ into the $2^{N+1}$ digit. Let's take a look at how this process works with bits in a classical computer:

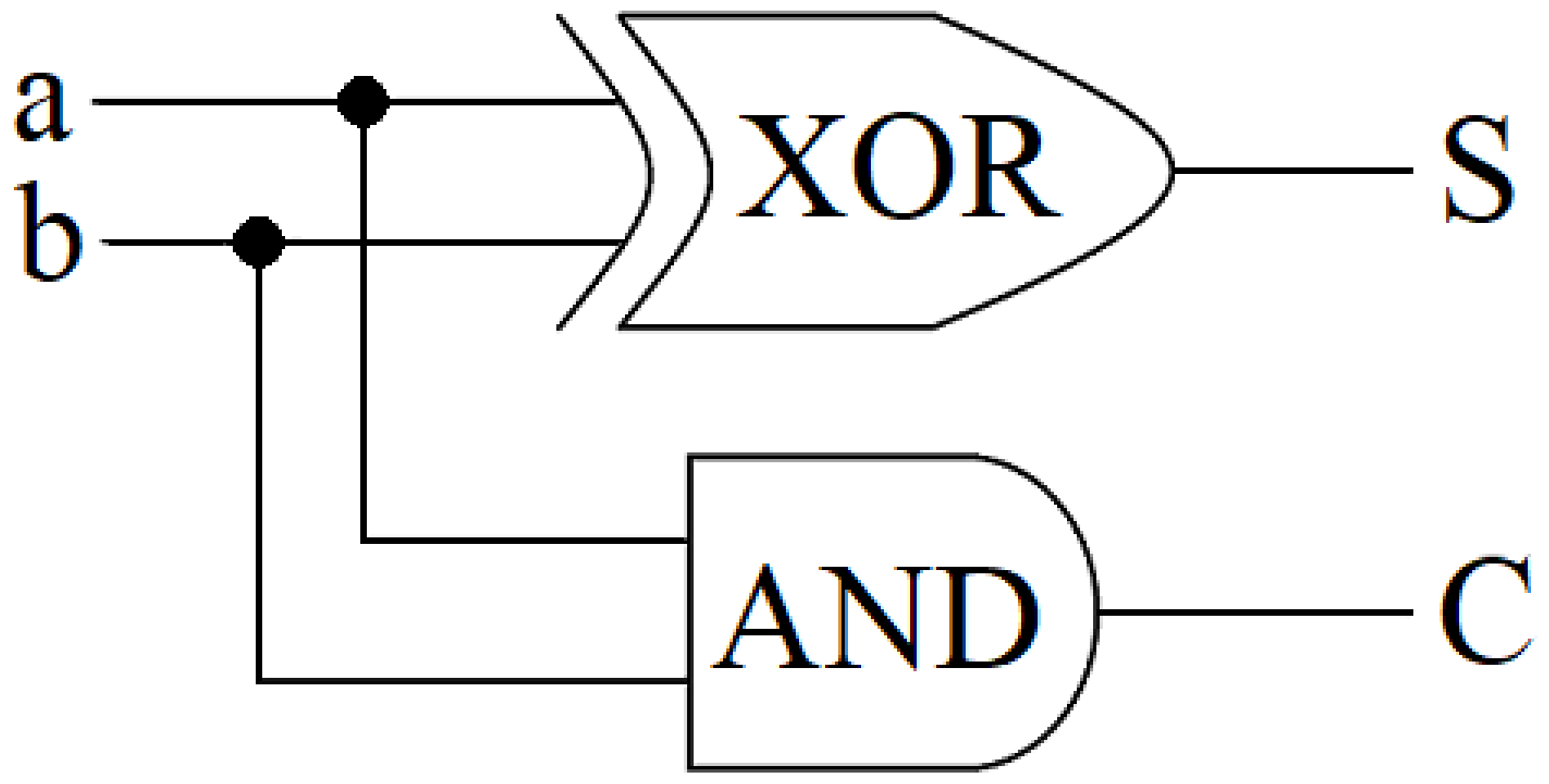

The figure above is an example of a classical adder, taking in two bits of information ($a$ and $b$), and outputting their sum in the form of two bits ($S$ and $C$). For completeness, the truth tables for the $\textbf{XOR}$ and $\textbf{AND}$ gates are provided below:

| a | b | XOR | AND |
|---|---|---|---|
| 0 | 0 | 0 | 0 |
| 1 | 0 | 1 | 0 |
| 0 | 1 | 1 | 0 |
| 1 | 1 | 0 | 1 |

Now, let's confirm that this circuit indeed adds two binary digits together as intended. As shown in the circuit, our outputs are the bits $S$ and $C$, which result from the $\textbf{XOR}$ and $\textbf{AND}$ gates respectively. $S$ refers to the sum of $a$ and $b$ (mod $2$), and $C$ represents the carry (for cases where the sum of $a$ and $b$ is greater than $1$). Working through the four possible combinations for $a$ and $b$, you should find the following results:

| $a$ | $b$ | $C$ | $S$ |
|---|---|---|---|
| 0 | 0 | 0 | 0 |
| 0 | 1 | 0 | 1 |
| 1 | 0 | 0 | 1 |
| 1 | 1 | 1 | 0 |

which are in fact the correct sums of $a$ and $b$ in binary. Now, in order to extend the circuit shown above into a full adder, we only need to incorporate one additional component: an incoming carry bit. If we want to add more than just the numbers $0$ and $1$ together, we need a circuit that can properly handle the addition of 3 bits. The circuit above can be thought of as the start of an adder, representing the addition of the least significant bit (LSB) of two large numbers $a$ and $b$. $S$ then is the correct LSB of the final answer, but $C$ must be added to the next set of bits. For example:

$$a = 3 = 11 \qquad b = 1 = 01$$

$$a_0 + b_0 = 1 + 1 \longrightarrow S_0 = 0 \quad C_0 = 1$$

$$a_1 + b_1 + C_0 = 1 + 0 + 1 \longrightarrow S_1 = 0 \quad C_1 = 1$$

$$\text{answer:} \quad C_1 S_1 S_0 = 100 = 4$$

Once we are able to incorporate carry bits, then we have a full methodology for adding together arbitrarily large numbers. An example circuit design that will allow for this is given below:

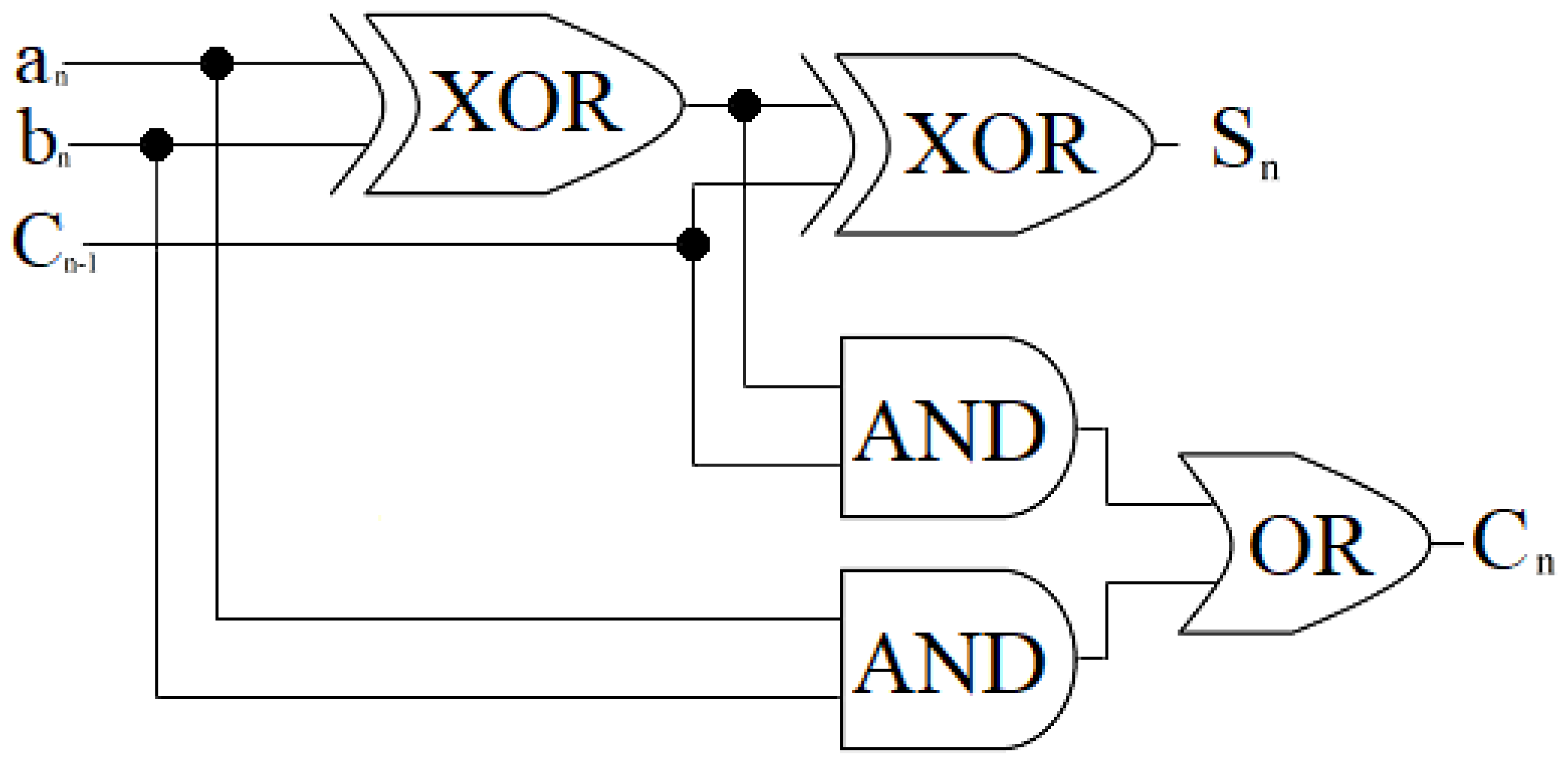

A bit more complex than the previous circuit, but still not too bad. In this circuit, we have our third input $C_{n-1}$, which is the carry bit from the previous addition of $a_{n-1}$ and $b_{n-1}$. The diagram provided above is modular, meaning that we can construct an $N$-bit adder by combining $N$ copies of the circuit above in series. Each module $n$ takes in a carry-bit from the $n-1$ component, computes the value $S_n$, and sends forward another carry-bit for the $n+1$ module. All together, the final output from the fuller adder is of the form: $C_n S_n S_{n-1} \cdots S_0$. To see such an adder in action, try out the cell of code below:

In [ ]:

```
def XOR(a,b):
    if( (a+b)==1 ):
        return 1
    else:
        return 0

def OR(a,b):
    if( (a+b)>=1 ):
        return 1
    else:
        return 0

def AND(a,b):
    if( (a+b)==2 ):
        return 1
    else:
        return 0

#====================================

a = [1,1,0,0]
b = [0,1,1,1]
C = 0
Sum = []

for i in np.arange( len(a) ):
    xor1 = XOR(a[0-i-1],b[0-i-1])
    S    = XOR(xor1,C)
    and1 = AND(a[0-i-1],b[0-i-1])
    and2 = AND(xor1,C)
    C    = OR(and1,and2)
    Sum.insert(0,S)
if( C == 1 ):
    Sum.insert(0,C)

print('Adding   a = ',a,'    +     b = ',b,'    ===>   ',Sum)
```

```
Adding    a =  [1, 1, 0, 0]     +     b =  [0, 1, 1, 1]     ===>    [1, 0, 0, 1, 1]
```

The important thing to note from the classical adder is the recursive process in which continuous pairs of bits from $a$ and $b$ are added together. The carry-bit from a single module within the adder influences the next operation, but nothing further. That is to say, the addition of $a_0$ and $b_0$ can indirectly affect the $a_n$ and $a_b$ sum, but only by passing through modules $1$ through $n-1$. By contrast, the quantum adder works quite differently. Because the foundation for the algorithm is based on using QFT transformations, each qubit representing a component of $a$ and $b$ can have a direct effect on every other qubit in the system.

## Defining QFT Notation

Before getting to the Quantum Adder Algorithm, we must first discuss the Quantum Fourier Transformation in a little more detail. To do this, we will work through an example that will serve to explain the inner workings of how the quantum adder is going to achieve the addition of two numbers. Recall from lesson 6 the way in which the QFT circuit achieves unique phases on each state. The resulting phases from the QFT are determined by the initial states of each qubit, where each combination of $|q_1\rangle|q_2\rangle|q_3\rangle$ will result in a unique state, orthogonal to any other combination:

$$|\psi\rangle = |q_1\rangle|q_2\rangle|q_3\rangle$$

$$|\Psi\rangle_{q_1q_2q_3} = QFT|\psi\rangle = \frac{1}{2\sqrt{2}}\Big(|0\rangle + e^{q_1\cdot i\pi}\cdot e^{q_2\cdot\frac{i\pi}{2}}\cdot e^{q_3\cdot\frac{i\pi}{4}}|1\rangle\Big)\otimes\Big(|0\rangle + e^{q_2\cdot i\pi}\cdot e^{q_3\cdot\frac{i\pi}{2}}|1\rangle\Big)\otimes\Big(|0\rangle + e^{q_3\cdot i\pi}|1\rangle\Big)$$

assuming each $q_n$ is in one of the computational basis states: $\Big\{|0\rangle, |1\rangle\Big\}$

$$\begin{aligned} {}_{q_1'q_2'q_3'}\langle\Psi|\Psi\rangle_{q_1q_2q_3} &= {}_{q_1'q_2'q_3'}\langle\psi|QFT^\dagger QFT|\psi\rangle_{q_1q_2q_3} \\ &= \langle q_1'q_2'q_3'|q_1q_2q_3\rangle \\ &= 0 \end{aligned}$$

Much like in lesson 5, where the trick to each algorithm was the steps implemented in between the $H^N$ transformations, our quantum adder will similarly implement some clever steps in between the QFT and QFT$^\dagger$ transformations. As shown above, if we don't do anything in between the transformations, the QFT and QFT$^\dagger$ steps will simply cancel each other out. Thus, to fully appreciate what will be taking place between them, let's begin by defining some new notation. Suppose we have some state $|a\rangle$, composed of $n$ qubits:

$$|a\rangle = |a_{n-1}a_{n-2}\cdots a_0\rangle = |a_{n_1}\rangle|a_{n_2}\rangle\cdots|a_0\rangle$$

where if $a$ represents a number in binary, then the $a_0$ qubit would be the LSB. The state $|a\rangle$ has no superposition, which means that all of its qubits are in the state $|0\rangle$ or $|1\rangle$. Now, let us define the state $|\phi_k(a)\rangle$, which describes the resulting state for each individual qubit after performing a Quantum Fourier Transformation on the state $|a\rangle$:

$$QFT|a\rangle = |\phi_{n-1}(a)\rangle|\phi_{n-2}(a)\rangle\cdots|\phi_0(a)\rangle$$

where

$$|\phi_k(a)\rangle = \frac{1}{\sqrt{2}}\Big(|0\rangle + e^{i\pi a_k.a_{k-1}\cdots a_0}|1\rangle\Big)$$

The notation above may seem a bit strange at first, so let's break it down. First off, since we are already familiar with the mathematics of a QFT, we should start by comparing these two lines with the QFT example at the top of this section. In essence, what this notation is doing is condensing all of the phases in the exponent on the $|1\rangle$ state down to a simplified math definition. The way in which phases get distributed amongst the qubits can escalate pretty rapidly as the number of qubits grows, so we need a shorthand way to express them. With that in mind, let's take a close look at the $e^{i\pi a_k.a_{k-1}\cdots a_0}$ term. It is by no means obvious, but the decimal in this exponential stands for a binary decimal, something most of us rarely encounter. As we know, each number in a base-10 decimal represents a different power of $10^{-m}$, summed together. For example:

$$0.1234 = \frac{1}{10} + \frac{2}{100} + \frac{3}{1000} + \frac{4}{10000}$$

A binary decimal works in the same manner, except with powers of $2^{-m}$:

$$0.1011 = \frac{1}{2} + \frac{0}{4} + \frac{1}{8} + \frac{1}{16}$$

This binary decimal is exactly what we need to describe the contributing phases for each qubit. The definition for $|\phi_k(a)\rangle$ uses the values of each qubit in a binary decimal to encode which qubits contribute phases. Qubits in the $|1\rangle$ state contribute a phase term like $\frac{i\pi}{2^p}$, where $p$ is the qubit's place in the decimal. Conversely, if a qubit is in the $|0\rangle$ state then it's contribution will be $e^0$ as intended. Let's see an example:

$$|a\rangle = |110\rangle = |1\rangle|1\rangle|0\rangle$$

$$|\phi_2(a)\rangle = \frac{1}{\sqrt{2}}\Big(|0\rangle + e^{i\pi 1.10}|1\rangle\Big) \qquad 1.10 = 1 + \frac{1}{2} + \frac{0}{4}$$

$$|\phi_1(a)\rangle = \frac{1}{\sqrt{2}}\Big(|0\rangle + e^{i\pi 1.0}|1\rangle\Big) \qquad 1.0 = 1 + \frac{0}{2}$$

$$|\phi_0(a)\rangle = \frac{1}{\sqrt{2}}\Big(|0\rangle + e^{i\pi 0.}|1\rangle\Big) \qquad 0. = 0$$

$$\begin{aligned} QFT|110\rangle &= |\phi_2(110)\rangle\otimes|\phi_1(110)\rangle\otimes|\phi_0(110)\rangle \\ &= \Big(|0\rangle + e^{i\pi 1.10}|1\rangle\Big)\Big(|0\rangle + e^{i\pi 1.0}|1\rangle\Big)\Big(|0\rangle + e^{i\pi 0.}|1\rangle\Big) \\ &= \Big(|0\rangle + e^{i\pi\frac{3}{2}}|1\rangle\Big)\Big(|0\rangle + e^{i\pi}|1\rangle\Big)\Big(|0\rangle + |1\rangle\Big) \end{aligned}$$

Let's confirm the math above with two examples in Qiskit. First we will implement the phases shown above by hand:

In [ ]:
```
q = QuantumRegister(3,name='q')
qc= QuantumCircuit(q,name='qc')
#------------------------------
qc.h( q[0] )
qc.h( q[1] )
qc.h( q[2] )

qc.u1( 3*m.pi/2, q[0] )
qc.u1( m.pi, q[1] )
qc.u1( 0, q[2] )

oq.Wavefunction(qc)
```

```
0.35355 |000>    -0.35355j |100>    -0.35355 |010>    0.35355j |110>    0.35355 |001>    -0.35355j |101>    -0.35355 |011>
0.35355j |111>
```

Next, to verify that the phases above have indeed performed a Quantum Fourier Transformation, we will use our custom $\mathrm{QFT}()$ function:

In [ ]:
```
q = QuantumRegister(3,name='q')
qc= QuantumCircuit(q,name='qc')
#------------------------------
qc.x( q[0] )
qc.x( q[1] )

oq.QFT(qc,q,3)

oq.Wavefunction(qc)
```

```
0.35355 |000>    -0.35355j |100>    -0.35355 |010>    0.35355j |110>    0.35355 |001>    -0.35355j |101>    -0.35355 |011>
0.35355j |111>
```

As we can see, the two examples are in agreement. Now that we have confirmed our notation yields us the correct results, we can move on to the real purpose for introducing it. Specifically, this new notation allows us to see the effect of a QFT on each individual qubit, rather than just the system as a whole. This in turn is exactly what we need to construct our quantum adder: the ability to manipulate individual qubits within two QFTs.

## Reverse Engineering The Quantum Adder

Recall back to lesson 5.2, the Bernstein-Vazirani Algorithm, and the way in which we were able to successfully pick out the desired state. The trick to that algorithm boiled down to creating a specific state within two $H^N$ transformations, such that the second transformation perfectly maps us to the desired state. For the B-V algorithm, creating the specific state inside the $H^N$ transformations was simply a matter of putting the correct negative phases on certain states, which we were able to achieve using control gates and an ancilla qubit in the $|-\rangle$ state. Here, our Quantum Adder Algorithm will follow the same principle, where the core of the algorithm relies on creating a specific state in anticipation for a QFT$^\dagger$ to transform us to the desired final state.

When studying a quantum algorithm, working backwards from the desired final state is often a good place to start. If we know that our algorithm plans to use a $U^N$ transformation as the final step, then our problem boils down to creating the state $|\Psi\rangle$ just before the transformation, such that $U^N|\Psi\rangle = |\mathrm{desired}\rangle$. Thus, if you know the desired final state $|\mathrm{desired}\rangle$, and the unitary transformation $U^N$, then you can calculate what $|\Psi\rangle$ must be:

$$|\Psi\rangle \quad = \quad U^{N\dagger}\,|\mathrm{desired}\rangle$$

For our quantum adder problem, the desired state will be the binary addition of two states $|a\rangle$ and $|b\rangle$, and the unitary transformation will be QFT$^\dagger$:

$$|\Psi\rangle \quad - \; QFT^\dagger \rightarrow \quad |a+b\rangle$$

And, now that we have our new notation for Quantum Fourier Transformations, we know exactly what the state $|\Psi\rangle$ needs to be for each qubit:

$$|\phi_k(a+b)\rangle \;\otimes\; |\phi_{k-1}(a+b)\rangle \;\otimes \cdots \otimes\; |\phi_0(a+b)\rangle \quad - \; QFT^\dagger \rightarrow \quad |a+b\rangle$$

Hopefully the problem of this quantum adder is becoming a bit clearer now. The final step of our algorithm will be to implement an inverse Quantum Fourier Transformation, which means that the state going *into* that transformation must be of the form shown above. Then, when we go to measure our system, we will find the state $|a+b\rangle$ with $100\%$ probability, which will be a successful addition of the numbers $a$ and $b$. Describing how the algorithm will work is important, but now we are tasked with the problem of how to create the states $|\phi_k(a+b)\rangle$ for each qubit.

## Rotations Between QFTs

Before tackling the full quantum adder, we shall first focus our attention to the case of just two qubits. Suppose the two numbers we want to add together are $1$ and $2$, letting $a=1$ and $b=2$. We begin by expressing these numbers as quantum states: $a=|01\rangle$ and $b=|10\rangle$. Using a classical adder, we know that the desired final answer should be $|11\rangle$. Thus, in order for our final QFT$^\dagger$ to successfully produce this state, we must have it act on the state $|\phi_1(11)\rangle \otimes |\phi_0(11)\rangle$. But to begin our algorithm, we must first create $|a\rangle$ and $|b\rangle$:

In [ ]:
```
qa = QuantumRegister(2,name='a')
qb = QuantumRegister(2,name='b')
qc = QuantumCircuit(qa,qb,name='qc')

qc.x( qa[1] )
qc.x( qb[0] )

oq.Wavefunction(qc,systems=[2,2])
```

```
1.0 |01>|10>
```

In the coding example above, the first thing you may notice is that our system already has four qubits, but according to a previous discussion the final QFT$^\dagger$ will only act on two. Thus, we have a choice to make on how we want to approach this problem, and which qubits we want to be the target of our final answer. For this lesson, we will pick the convention that our final answer will be stored in the qubits composing the state $|a\rangle$. We could just as easily store our answer on the qubits of $|b\rangle$, and I encourage you to construct such a circuit for yourself at the conclusion of this lesson if you're interested in some extra practice.

Now that we've picked where to apply our QFT$^\dagger$, our problem boils down to getting those two qubits into the state $|\phi_1(11)\rangle \otimes |\phi_0(11)\rangle$. Since $|a\rangle$'s qubits will be receiving a QFT$^\dagger$ at the end of the algorithm, they also need to receive the initial QFT at the beginning. Meanwhile, since the qubits of $|b\rangle$ play no role in either of these transformations, their influence on the algorithm will be to serve as control qubits for operations acting on $|a\rangle$ (which we haven't revealed what they are yet, but you might have a hunch based on the title of this section). Visualizing the algorithm as whole, so far we have:

$$|a\rangle|b\rangle \quad -QFT_a \rightarrow \quad \Big(|\phi_1(a)\rangle \otimes |\phi_0(a)\rangle\Big)|b\rangle \quad -??? \rightarrow \quad \Big(|\phi_1(a+b)\rangle \otimes |\phi_0(a+b)\rangle\Big)|b\rangle \quad -QFT_a^\dagger \rightarrow \quad |a+b\rangle|b\rangle$$

The mechanic by which the $|b\rangle$ qubits will influence the system is through control-rotation gates, $R_\phi$. They will serve as control qubits for these $R_\phi$ gates, acting on the qubits of $|a\rangle$ as targets. Recall from lesson 6 the way in which we achieved QFTs on our quantum systems, namely through a structured pattern of Hadamard and $R_\phi$ gates. These control rotation gates are what give rise to the $e^{i\pi a_k \cdot a_{k-1} \cdots a_0}$ phases discussed earlier, and they are what will allow us to tweak them as well. As shown in the visualization above, the final piece to this algorithm is picking the correct rotations such that we transformation the states $|\phi_k(a)\rangle$ to $|\phi_k(a+b)\rangle$. To understand the process, let's take a look at our 2-qubit example:

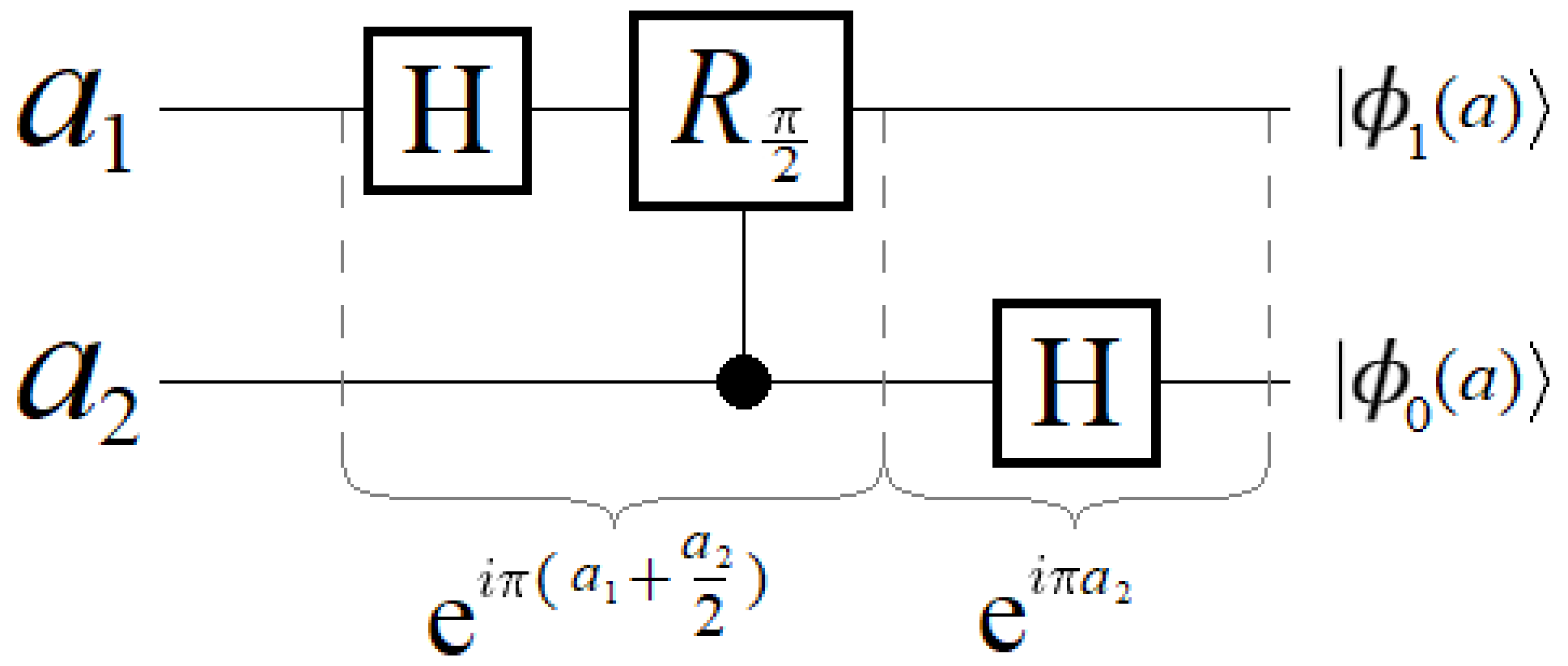

The diagram above shows the breakdown of a 2-qubit QFT, emphasizing which gates contribute to the resulting phase terms on each qubit. The Hadamard gates are providing the conditional $\pi$ phases on each qubit, while the $R_{\pi/2}$ gate is providing the conditional $\frac{\pi}{2}$ phase on the $a_1$ qubit. After implementing the QFT as shown above, the next step is how to incorporate the $|b\rangle$ qubits and their contributions to $|a\rangle$. Working backwards, we can write out the states $|\phi_1(a+b)\rangle$ and $|\phi_1(a+b)\rangle$ for some insight:

$$|\phi_1(a+b)\rangle \qquad\qquad |\phi_0(a+b)\rangle$$

$$\frac{1}{\sqrt{2}}\Big(|0\rangle + e^{i\pi\left( (a_1+b_1) + (a_2+b_2)/2 \right)}|1\rangle\Big) \qquad \frac{1}{\sqrt{2}}\Big(|0\rangle + e^{i\pi\left( a_2+b_2 \right)}|1\rangle\Big)$$

The states shown here are generated by directly plugging in the quantity $a+b$ into our definition of $|\phi_k(a)\rangle$ from earlier. As we can see, the phase contributions from the $b$ qubits perfectly match with those from $a$. We can decompose these states further still, by using the property of exponents:

$$e^{A+B} = e^A \cdot e^B$$

$$\frac{1}{\sqrt{2}}\Big(|0\rangle + e^{i\pi\left( a_1 + a_2/2 \right)} e^{i\pi\left( b_1 + b_2/2 \right)}|1\rangle\Big) \qquad \frac{1}{\sqrt{2}}\Big(|0\rangle + e^{i\pi a_2} e^{i\pi b_2}|1\rangle\Big)$$

By rearranging the phases inside the exponents, we have separated out the contributions from $a$ and $b$ respectively. The important thing to note here is that written in this way, we can now clearly see what the phase contributions from $b$ must be. Now that we've worked backwards from the final answer to arrive at this point, we can now tackle the problem from the opposite direction. We know that the states written above, with all of the phase terms from $a$ and $b$, are what we need to prepare before the final QFT$^\dagger$ And we also know what the state of our $|a\rangle$ qubits will be as a result of the first QFT:

$$QFT\,|a\rangle \quad = \quad \frac{1}{\sqrt{2}}\Big(|0\rangle + e^{i\pi\left( a_1 + a_2/2 \right)}|1\rangle\Big) \otimes \Big(|0\rangle + e^{i\pi a_2}|1\rangle\Big)$$

Comparing these states with the ones just above, it should be clear that the QFT is providing all of the phase contributions needed of $a$ for our desired state. If we remove these contributions from $a$, visually we can see what remaining phases are needed in order to achieve our $|\phi_k(a+b)\rangle$ states:

$$\frac{1}{\sqrt{2}}\Big(|0\rangle + e^{i\pi\left( b_1 + b_2/2 \right)}|1\rangle\Big) \qquad \frac{1}{\sqrt{2}}\Big(|0\rangle + e^{i\pi b_2}|1\rangle\Big)$$

At last, we've successfully boiled our problem down to the contributions necessary from $|b\rangle$. All we need to do is implement the phases shown here, onto the states $|a_1\rangle$ and $|a_2\rangle$. Looking at the phases in each state, and which qubits are contributing, we need the following gates: $R_\pi(b_1, a_1)$, $R_{\pi/2}(b_2, a_1)$, and $R_\pi(b_2, a_2)$, where the $b$ qubits are the control and the $a$ qubits are the target. If we add these additional gates after the first QFT and before the final QFT$^\dagger$, we should successfully produce the final state $|a+b\rangle$. Let's verify this with our code:

In [ ]:
```
qa = QuantumRegister(2,name='a')
qb = QuantumRegister(2,name='b')
qc = QuantumCircuit(qa,qb,name='qc')
#-------------------------------------
qc.x( qa[1] )
qc.x( qb[0] )
print('_____ States to Add Together _____')
oq.Wavefunction(qc,systems=[2,2])

oq.QFT(qc,qa,2)
#-----------------------------   phase contributions from |b>
qc.cu1( m.pi, qb[0], qa[0] )
qc.cu1( m.pi/2, qb[1], qa[0]  )
qc.cu1( m.pi, qb[1], qa[1] )
#-----------------------------
oq.QFT_dgr(qc,qa,2)

print('\n___ Sum Stored in |a> ___')
oq.Wavefunction(qc,systems=[2,2],show_systems=[True,False])
```

```
_____ States to Add Together _____
1.0 |01>|10>

___ Sum Stored in |a> ___
1.0 |11>
```

Success. The additional $R_\phi$ gates have allowed us to transform the states $|\,\phi_k\,(a)\rangle$ into $|\,\phi_k\,(a+b)\rangle$. Visually, the circuit for this 2-qubit quantum adder is as follows:

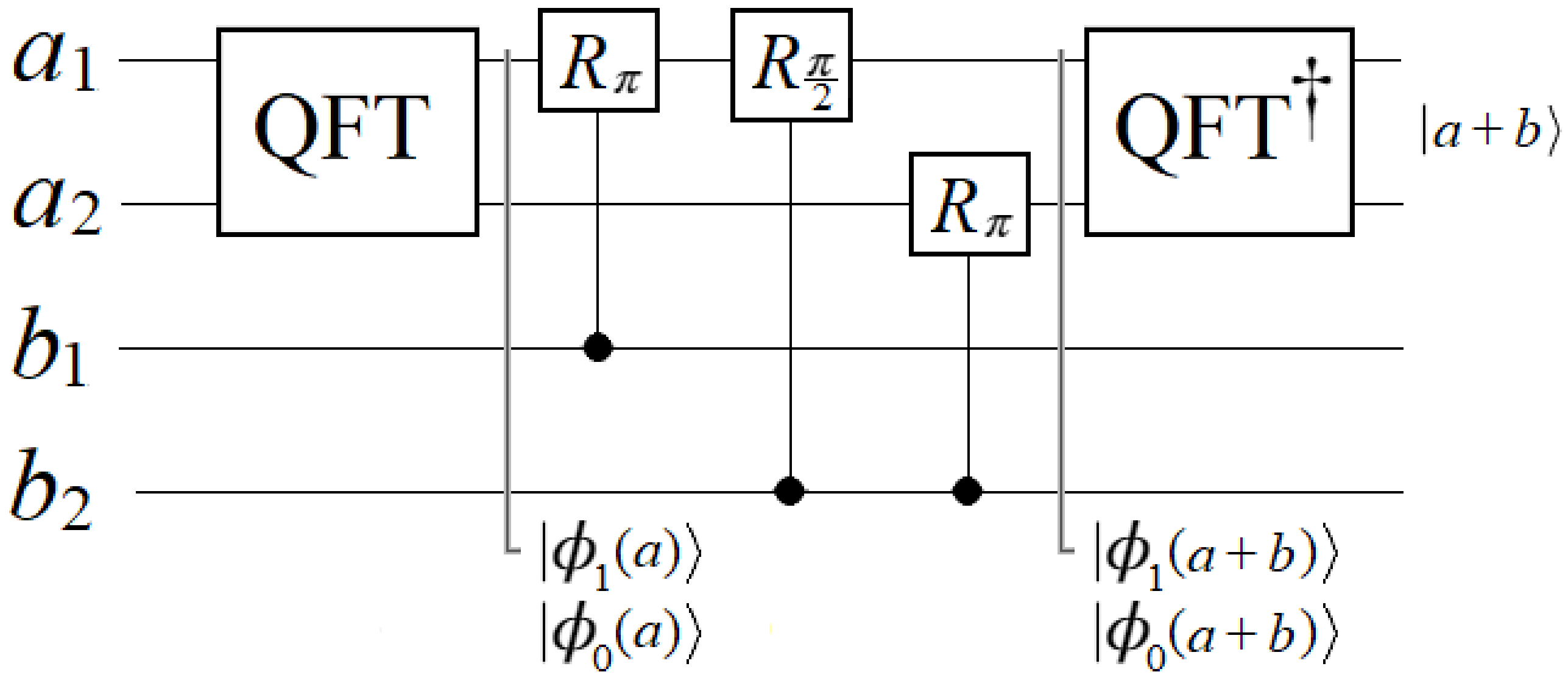

Having successfully completed the two qubit example, we will now want to extend our quantum adder to handle any sized numbers. Luckily, the pattern for implementing the additional rotation gates is straightforward, and very analogous to the way in which we construct the QFT itself. But first, let's revisit our code for the 2-qubit adder, and test a couple more possible combinations for $a$ and $b$:

In [ ]:
```
A_States = [[0,1],[1,0],[1,1]]
B_States = [[1,0],[1,1]]
#-------------------------------------
for a in np.arange(len(A_States)):
    A = A_States[a]
    for b in np.arange(len(B_States)):
        B = B_States[b]
        qa = QuantumRegister(2,name='a')
        qb = QuantumRegister(2,name='b')
        qc = QuantumCircuit(qa,qb,name='qc')
        #-----------------------------------
        if(A[0]==1):
            qc.x( qa[0] )
        if(A[1]==1):
            qc.x( qa[1] )
        if(B[0]==1):
            qc.x( qb[0] )
        if(B[1]==1):
            qc.x( qb[1] )
        oq.QFT(qc,qa,2)
        qc.cu1( m.pi, qb[0], qa[0] )
        qc.cu1( m.pi/2, qb[1], qa[0]  )
        qc.cu1( m.pi, qb[1], qa[1] )
        oq.QFT_dgr(qc,qa,2)

        print('\nA:',A,'  B:',B)
        oq.Wavefunction(qc,systems=[2,2],show_systems=[True,False])
```

```
A: [0, 1]   B: [1, 0]
1.0 |11>

A: [0, 1]   B: [1, 1]
1.0 |00>

A: [1, 0]   B: [1, 0]
1.0 |00>

A: [1, 0]   B: [1, 1]
1.0 |01>

A: [1, 1]   B: [1, 0]
1.0 |01>

A: [1, 1]   B: [1, 1]
1.0 |10>
```

Take a look at the six examples shown above, and you should notice something amiss. If you follow through each case and work out the expected binary addition of A + B, you'll find that five out of the six answers equal numbers greater than 3. However, we have only allocated two qubits to store our final answer, which means that we never properly gave our quantum system the resources needed to create the correct final states. But this doesn't mean that the answers above are wrong, in fact quite the opposite. If we take a closer look at the six examples, and put in the additional carry bit alongside each answer:

$$\begin{aligned}
|01\rangle + |10\rangle &\longrightarrow 0|10\rangle \\
|01\rangle + |11\rangle &\longrightarrow 1|00\rangle \\
|10\rangle + |10\rangle &\longrightarrow 1|00\rangle \\
|10\rangle + |11\rangle &\longrightarrow 1|01\rangle \\
|11\rangle + |10\rangle &\longrightarrow 1|01\rangle \\
|11\rangle + |11\rangle &\longrightarrow 1|10\rangle
\end{aligned}$$

As we can see, our quantum adder is correct in the two LSB qubits, even without the presence of a third qubit to store the carry. So then, all we need now is to extend our adder to handle one additional carry qubit.

## Complete Quantum Adder

In order to complete our quantum adder, and to handle the final carry qubit for cases that exceed the length of $|a\rangle$, we will work through our final example of this lesson to determine what additional gates are needed. Just as before, we will write out all of the phases needed of our final state, and slowly strip away ones already existing in our quantum adder up to this point. To store the additional carry qubit, we need to increase the size of our $|a\rangle$ state by one:

$$a = 2 = |010\rangle \qquad b = 3 = |11\rangle$$

In coming example, we will actually also represent $|b\rangle$ as $|011\rangle$, to match the length of $|a\rangle$. As we shall see however, it turns out that this will not be necessary. So then, we begin by writing out our desired $|\phi_k(a+b)\rangle$, as well as the decomposition of the phase contributions from $|a\rangle$ and $|b\rangle$:

$$|\phi_2(a+b)\rangle \qquad |\phi_1(a+b)\rangle \qquad |\phi_0(a+b)\rangle$$

$$|0\rangle + e^{i\pi\left((a_2+b_2) + (a_1+b_1)/2 + (a_0+b_0)/4\right)}|1\rangle \qquad |0\rangle + e^{i\pi\left((a_1+b_1) + (a_0+b_0)/2\right)}|1\rangle \qquad |0\rangle + e^{i\pi\left(a_0+b_0\right)}|1\rangle$$

Once again, we can remove the phases contributions from $|a\rangle$ due to the first QFT:

$$|0\rangle + e^{i\pi\left(b_2 + b_1/2 + b_0/4\right)}|1\rangle \qquad |0\rangle + e^{i\pi\left(b_1 + b_0/2\right)}|1\rangle \qquad |0\rangle + e^{i\pi b_0}|1\rangle$$

Shown above are all of the phases from $|b\rangle$ we need to add in order to construct our full quantum adder. If we compare these states with the ones derived earlier, we can see that there is no change to the states $|\phi_1(a+b)\rangle$ and $|\phi_0(a+b)\rangle$. Each one receives the exact same contributions from $|b\rangle$ as before, which means that we don't need to amend our quantum circuit for the qubits $|a_1\rangle$ and $|a_0\rangle$. All that's left then is the additional control-rotation gates needed for $|a_2\rangle$.

According to what we just derived then, the gates needed to complete our quantum adder are: $R_\pi(b_2, a_2)$, $R_{\pi/2}(b_1, a_2)$, and $R_{\pi/4}(b_0, a_2)$. But, remember that we only included the qubit $|b_2\rangle$ as a precaution, and set it to $|0\rangle$. And in general, if we choose to include this extra qubit just so $|b\rangle$ will be the same length as $|a\rangle$, and always set it to be $|0\rangle$, then there will never be an instance where the gate $R_\pi(b_n, a_n)$ ever has an effect on the system. Thus, we can drop the additional qubit on $|b\rangle$ and remove the $R_\pi(b_2, a_2)$ gate completely. This leaves us with just two additional gates to complete our quantum adder:

In [ ]:

```
qa = QuantumRegister(3,name='a')
qb = QuantumRegister(2,name='b')
qc = QuantumCircuit(qa,qb,name='qc')
#-----------------------------------
qc.x( qa[1] )
qc.x( qb[0] )
qc.x( qb[1] )
print('_____ States to Add Together _____')
oq.Wavefunction(qc,systems=[3,2])

oq.QFT(qc,qa,3)
#------------------------------  phase contributions from |b>
qc.cu1( m.pi/2, qb[0], qa[0]  )
qc.cu1( m.pi/4, qb[1], qa[0]  )
qc.cu1( m.pi, qb[0], qa[1]  )
qc.cu1( m.pi/2, qb[1], qa[1]  )
qc.cu1( m.pi, qb[1], qa[2] )
#------------------------------
oq.QFT_dgr(qc,qa,3)

print('\n___ Sum Stored in |a> ___')
oq.Wavefunction(qc,systems=[3,2],show_systems=[True,False])
```

```
_____ States to Add Together _____
1.0 |010>|11>

___ Sum Stored in |a> ___
1.0 |101>
```

Success! By including the additional qubit on $|a\rangle$, and the extra $R_\phi$ gates for the carry qubit, our quantum adder can now handle the complete addition of any two 2-digit binary numbers. And it is quite nice that the only additional gates necessary for the $a_n$ qubit follow the same pattern as all of the other $|\phi_k(a+b)\rangle$ states, minus one $R_\pi$ (which we showed was redundant because $|b_n\rangle$ is always $|0\rangle$). So long as we allocate one extra qubit to $|a\rangle$, our Quantum Adder Algorithm is complete!

For the final example below, all of the steps for the Quantum Adder studied in this lesson have been combined into a function called $\text{Quantum_Adder}$ from $\text{Our_Qiskit_Functions.py}$. Use the code below to test out the addition of any two numbers $a$ and $b$:

In [ ]:

```
A = [0,1,1,0]
B = [1,1,0,1]               #A and B need to be arrays of equal length  (don't include the extra 0 qubit for A)
print('States to Add Together:   ',A,' + ',B)

#=========================================

qa = QuantumRegister(len(A)+1,name='a')
qb = QuantumRegister(len(B),name='b')
qc = QuantumCircuit(qa,qb,name='qc')
#---------------------------------------
oq.Quantum_Adder(qc,qa,qb,A,B)

print('\n___ Sum Stored in |a> ___')
oq.Wavefunction(qc,systems=[len(A)+1,len(B)],show_systems=[True,False])
```

```
States to Add Together:    [0, 1, 1, 0]  +  [1, 1, 0, 1]

___ Sum Stored in |a> ___
1.0 |10011>
```

## Aside: The Quantum Subtractor

With our study of the quantum adder complete, this extra section will cover how to build a quantum subtractor using the same fundamentals, but with a slight tweak to the circuit. For the case of the adder, after the initial QFT on $|a\rangle$, we use control rotation gates to manipulate the state of each qubit in the $|a\rangle$ system such that they end up in the state $|\phi_k(a+b)\rangle$. And to do this, we essentially added phases to each $|a_n\rangle$ qubit by taking advantage of the fact that $e^A \cdot e^B = e^{A+B}$. If we now want to apply the same trick for subtraction, we can use the exact same control rotation gates to perform $e^A \cdot e^{-B} = e^{A-B}$ simply by changing the phases on our $R_\phi$ gates. For every rotation gate involving qubit $b$, we simply apply $R_{-\phi}$ instead of $R_\phi$:

In [ ]:

```
qa = QuantumRegister(3,name='a')
qb = QuantumRegister(3,name='b')
qc = QuantumCircuit(qa,qb,name='qc')
#------------------------------------
qc.x( qa[0] )
qc.x( qa[1] )
qc.x( qb[1] )
qc.x( qb[2] )
print('____  States to Subtract ____')
oq.Wavefunction(qc,systems=[3,3])

oq.QFT(qc,qa,3)
#------------------------------  phase contributions from |b>
qc.cu1( -m.pi, qb[0], qa[0]  )
qc.cu1( -m.pi/2, qb[1], qa[0]  )
qc.cu1( -m.pi/4, qb[2], qa[0]  )
qc.cu1( -m.pi, qb[1], qa[1]  )
qc.cu1( -m.pi/2, qb[2], qa[1]  )
qc.cu1( -m.pi, qb[2], qa[2] )
#------------------------------
oq.QFT_dgr(qc,qa,3)

print('\n___ Difference Stored in |a> ___')
oq.Wavefunction(qc,systems=[3,3],show_systems=[True,False])
```

```
____  States to Subtract ____
1.0 |110>|011>

___ Difference Stored in |a> ___
1.0 |011>
```

In the example above, we have successfully performed the operation $6 - 3$. First, we create the states $|a\rangle$ and $|b\rangle$ to represent the numbers $6$ and $3$ respectively $\left(|110\rangle \text{ and } |011\rangle\right)$. Then, by flipping the phase on all of the control rotations, we successfully produce the expected state $|011\rangle$. We won't be going into any further detail about how to construct a full Quantum Subtractor, as the emphasis here was just to show how a simple flipping of phases inside two QFTs can yield a completely different quantum process. And as we shall see in future lessons, the tricks one can perform inside QFT transformations are what lead to some impressive algorithms.

---

This concludes lesson 6.1! If your understanding of the Quantum Fourier Transformation felt a little lackluster after lesson 6, hopefully this lesson has helped reinforce some of the core concepts. In particular, understanding the role of the QFT and QFT$^{\dagger}$, written out in terms of the $|\phi_k(a)\rangle$ states, is an important milestone. It was what allowed us to dissect and understand the Quantum Adder in this lesson, and it will similarly serve as the foundation for future algorithms to come.

I hope you enjoyed this lesson, and I encourage you to take a look at my other .ipynb tutorials!

## Citations

**[1]** T. G. Draper, "Addition on a Quantum Computer", arXiv:0008033 (2000)

# Lesson 7 - Quantum Phase Estimation

In this tutorial, we will once again divert our attention away from studying full algorithms, and focus on something more akin to a subroutine. Just like how the Quantum Fourier Transformation (QFT) is used as a critical element to larger algorithms, the same will be true for the focus of our lesson here: Quantum Phase Estimation (QPE). Perhaps it's a bit unfair not to call QPE its own quantum algorithm (it's certainly derserving of the title), but the intention here is that its full potential is realized when used for more grand algorithms. At its core, the primary function of the QPE Algorithm is to find an approximate value to an eigenvalue phase of some unitary matrix $U$.

In order to make sure that all cells of code run properly throughout this lesson, please run the following cell of code below:

In [ ]:
```
from qiskit import ClassicalRegister, QuantumRegister, QuantumCircuit, Aer, execute, BasicAer
from qiskit.tools.visualization import plot_histogram
import Our_Qiskit_Functions as oq
import numpy as np
import math as m
import matplotlib.pyplot as plt
import random
```

## Solving for Eigenvalues

As the name of the algorithm suggests, the QPE subroutine is a technique for finding eigenvalue phases of a unitary matrix. First described in 1995 by Kitaev **[1]**, the goal is to compute the eigenvalue for a matrix $U$, with an eigenvector $|u\rangle$, where the eigenvalue can be written into the specific form $e^{2\pi i\theta}$:

$$U|u\rangle = e^{2\pi i\theta}|u\rangle \qquad 0 < \theta < 1$$

The final result from the algorithm will be to produce the value of $\theta$, either exactly or to some desired precision. Broadly speaking, the QPE algorithm is particularly useful for quantum computing because we work exclusively with unitary operators, for which we are always guaranteed that its eigenvalues satisfy our conditions. For example, below is a generalized single qubit unitary matrix, and an example value to demonstrate a pair of eigenvalues:

$$\alpha = \frac{2\pi}{3} \qquad U = \begin{bmatrix} \cos(\alpha) & \sin(\alpha) \\ -\sin(\alpha) & \cos(\alpha) \end{bmatrix}$$

$$\text{eigenvalue}: \frac{-1+\sqrt{3}i}{2} \qquad \text{eigenvector}: \frac{1}{\sqrt{2}}\begin{bmatrix} -i \\ 1 \end{bmatrix}$$

In [ ]:
```
alpha = 2*m.pi/3
U = np.array([ [ np.cos(alpha)  , np.sin(alpha) ],
               [ -np.sin(alpha) , np.cos(alpha) ] ])
#-------------------------------------------------
e = 0.5*(-1+m.sqrt(3)*1.0j)
v = 1.0/m.sqrt(2)*np.array( [-1.0j,1] )
#-------------------------------------------------
print('____ Unitary Matrix ____\n',U)
print( '\n       U |u> = ',np.dot(U,v))
print( 'e^{2\u03C0i\u03C6} |u> = ',e * v)
```

```
____ Unitary Matrix ____
 [[-0.5        0.8660254]
 [-0.8660254 -0.5      ]]

       U |u> =  [ 0.61237244+0.35355339j -0.35355339+0.61237244j]
e^{2πiϕ} |u> =  [ 0.61237244+0.35355339j -0.35355339+0.61237244j]
```

The code above simply demonstrates the relation between a unitary matrix and its eigenvectors / eigenvalues. We can rewrite the eigenvalue from this example into the form shown above as well:

$$\frac{-1+\sqrt{3}i}{2} = e^{2\pi i/3}$$

$$\therefore \ \theta = \frac{1}{3}$$

Thus, if we were to successfully run our QPE algorithm using the matrix above, we would find with high probability the value of $\theta = \frac{1}{3}$. Although the capabilities of this algorithm may seem a little singular at first glance, only being a tool for finding eigenvalues, the applications for QPE make it one of the most important quantum subroutines to date.

## The QPE Circuit

To begin our study of the QPE algorithm, let's first take a look at its circuit diagram, taking special note of a new element which we've yet to encounter before:

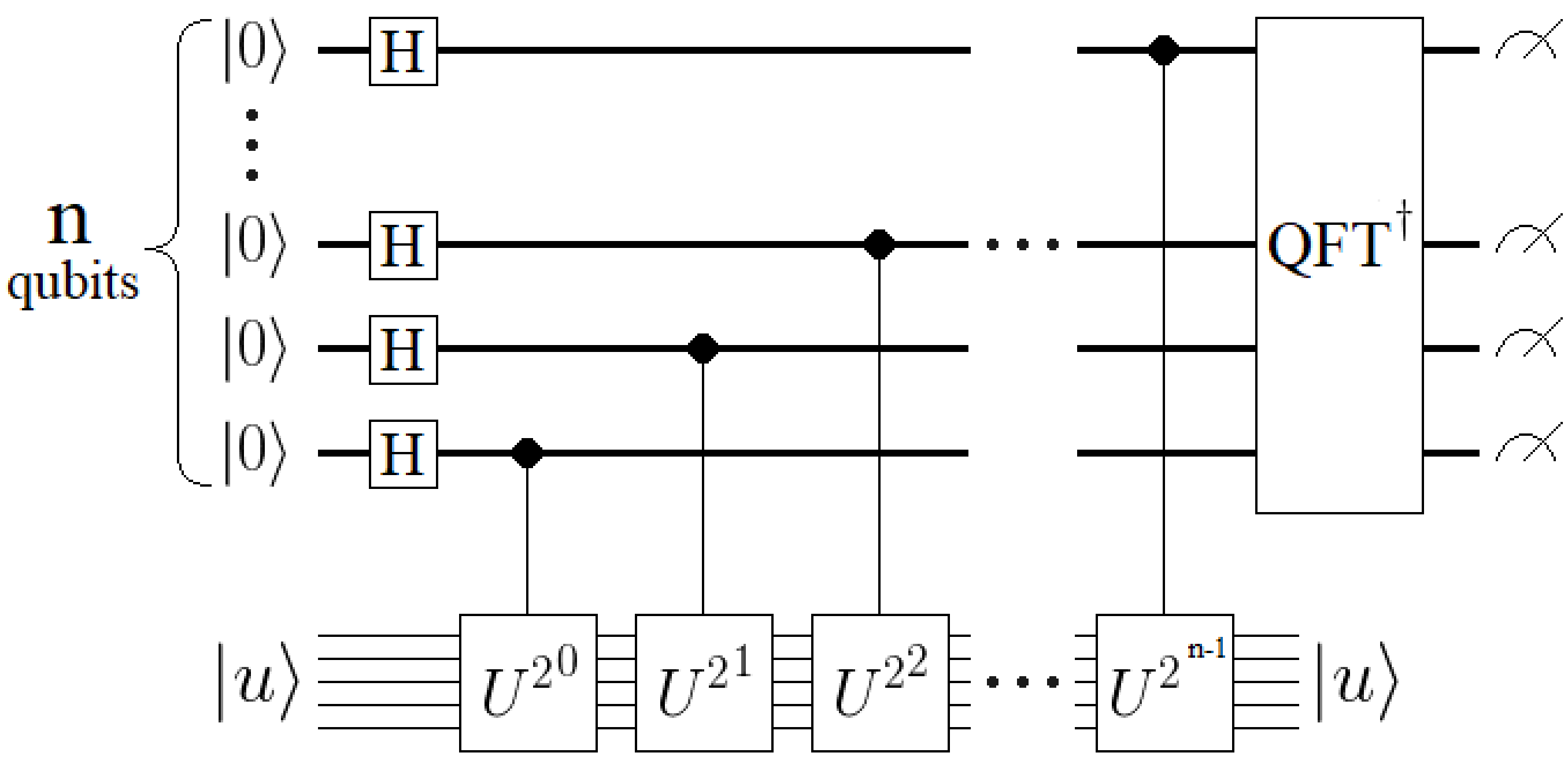

All of the elements in the circuit above should look familiar, with the exception of the control-$U$ operations in the middle. Like many algorithms, our circuit begins with $H$ gates on all of our qubits, creating an equal superposition of all states:

$$|\Psi\rangle_i \;=\; \frac{1}{\sqrt{n}} \sum_{i}^{2^n} |\, i \,\rangle$$

But even in this first step, there is already something fundamentally new about this algorithm, and that's the number $n$. In previous algorithms, the number of qubits was determined by certain constraints or conditions of our problem. For example, in Grover's Algorithm the number of qubits we used was determined by the total number of possible entries on which we were performing the search. Here, $n$ is actually a number of our choosing, representing the level of precision we would like out of the algorithm. In principle we can use as many or as few qubits as we want, and our QPE will fundamentally run all the same.

We will return to $n$ and its impact on the algorithm, but for now let's turn our attention to the second new feature: $|\, u \,\rangle$. At the bottom of our circuit diagram, we can see a second qubit register, one that needs to be prepared in the state of our eigenvector. Skipping forward, if we look at the state of this second register after all of the control-$U$ operations, our eigenvector $|\, u \,\rangle$ comes out unchanged. Conceptually, this agrees with our discussion of unitary matrices and eigenvectors from earlier: $U\,|\, u \,\rangle \;=\; e^{2\pi i\theta}\,|\, u \,\rangle$. All of the quantum operations that are being applied to the qubits in the second register are in principle applications of $U$, which means that the state of the qubits should remain unchanged. Consequently then, we must ask what is the result of all these operations?

If the effect of all the control-$U$ operations leaves the second register unchanged, then it would be fair to assume their effect must show up on the first register. This is indeed the case, and to see this, let's focus on a single application of a control-$U$:

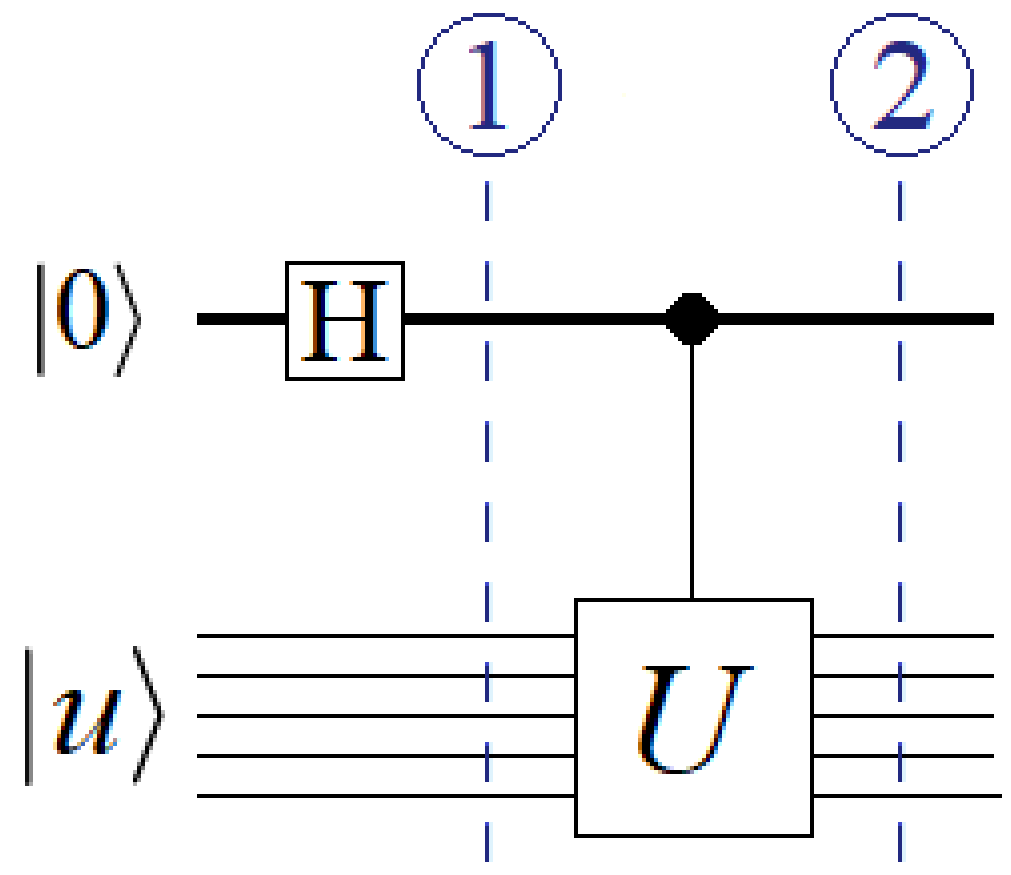

Following along with the diagram above, the state of our quantum system at $(1)$ is straightforward:

$$(1) \qquad \frac{1}{\sqrt{2}}\Big(\,|\,0\,\rangle \;+\; |\,1\,\rangle\,\Big) \;\otimes\; |\,u\,\rangle$$

Next let's take a look at what the control-$U$ operation is doing. Because our control qubit is sitting in the $|+\rangle$ superposition state, we know that the effect of the control-$U$ operation will only be applied to the $|\, 1 \,\rangle$ component:

$$\frac{1}{\sqrt{2}}\Big(\,|\,0\,\rangle|\,u\,\rangle \;+\; |\,1\,\rangle|\,u\,\rangle\,\Big)$$

$$(2) \qquad \frac{1}{\sqrt{2}}\Big(\,|\,0\,\rangle|\,u\,\rangle \;+\; e^{2\pi i\theta}|\,1\,\rangle|\,u\,\rangle\,\Big)$$

$$= \; \frac{1}{\sqrt{2}}\Big(\,|\,0\,\rangle \;+\; e^{2\pi i\theta}|\,1\,\rangle\,\Big) \;\otimes\; |\,u\,\rangle$$

As shown above, the $|\, 1 \,\rangle$ component of our first register qubit has picked up the $e^{2\pi i\theta}$ phase from $U$, leaving the $|\, 0 \,\rangle$ unchanged. Additionally, we can see in the last line that the effect of $U$ acting on $|\, u \,\rangle$ has left the eigenstate unaltered, just as we discussed earlier. The steps outlined above represent the individual action of each control-$U$ operation on the qubits, so now let's connect this result to the various unitary operations outlined in the complete circuit. In particular, all of the control-$U$ operations required of the full QPE circuit are essentially powers of $U^2$. Mathematically, the expected result from these operations is as follows:

$$U^n\,|\,u\,\rangle \;=\; UUU\ldots U\,|\,u\,\rangle \;=\; \Big(e^{2\pi i\theta}\Big)^n|\,u\,\rangle \;=\; e^{2\pi i n\theta}\,|\,u\,\rangle$$

Essentially, each application of $U$ on $|\, u \,\rangle$ results in a multiplication of $e^{2\pi i\theta}$ onto the $|\, 1 \,\rangle$ component of the respective control qubit. Thus, $n$ applications of $U$ results in $n$ copies of this phase term all being multiplied together, which can be combined nicely using exponential power laws. Putting all of the various control-$U$ operations outlined in the circuit together then, the state of our system just before the QFT$^\dagger$ will be:

$$|\Psi\rangle = \frac{1}{\sqrt{2^n}} \Big( |0\rangle + e^{2\pi i\,(2^{n-1})\,\theta}|1\rangle \Big) \Big( |0\rangle + e^{2\pi i\,(2^{n-2})\,\theta}|1\rangle \Big) \cdots \Big( |0\rangle + e^{2\pi i\,(2^{0})\,\theta}|1\rangle \Big) \otimes |u\rangle$$

If you follow along the circuit diagram, you should be able to correctly spot where each of the phase contributions are coming from in the state above. In total, we have all of the powers from $2^0$ through $2^n$ distributed amongst the phases of our $n$ qubits. And although we haven't seen anything exactly like this state before, the way in which these phases are so neatly distributed should remind you of some of the quantum states we studied in lessons 6 & 6.1. In the next section we are going to dissect the state above further and examine the effect of passing it through an inverse Quantum Fourier Transformation.

## Final Step: QFT$^\dagger$

Following from the state shown above, the final step of our QPE circuit requires a QFT$^\dagger$, which suggests that there is something special about the way in which all of the phase terms are arranged. For example, why does the circuit call for powers of $U^{2^n}$ rather than just $U^n$? To start, let's recall form lesson 6 the way in which a QFT$^\dagger$ applies phases on the various qubits in the system:

$$|\psi\rangle = |Q_1\rangle \otimes |Q_2\rangle \otimes |Q_3\rangle$$

where we will assume each of the states $|Q_i\rangle \in \Big\{ |0\rangle, |1\rangle \Big\}$

$$QFT^\dagger|\psi\rangle = \Big( |0\rangle + e^{-i\pi\left( Q_1 + \frac{Q_2}{2} + \frac{Q_3}{4} \right)}|1\rangle \Big) \otimes \Big( |0\rangle + e^{-i\pi\left( Q_2 + \frac{Q_3}{2} \right)}|1\rangle \Big) \otimes \Big( |0\rangle + e^{-i\pi Q_3}|1\rangle \Big)$$

In the previous two lessons we discussed how writing out the effect of the QFT$^\dagger$ as shown above was useful in illuminating where phase contributions were coming from. In particular, we can see exactly the way in which the phases on each qubit's $|1\rangle$ state component are determined by which qubits in $|\psi\rangle$ are in either $|0\rangle$ or $|1\rangle$. In a sense, we can think of these phase contributions as a "fingerprint" for the final quantum state, for example:

$$\Big( |0\rangle + A|1\rangle \Big) \otimes \Big( |0\rangle + B|1\rangle \Big) \otimes \Big( |0\rangle + C|1\rangle \Big)$$

$$= \Big( |000\rangle + A|100\rangle + AB|110\rangle + BC|011\rangle + \cdots \Big)$$

Notice in the example above how each state's phase is unique, and can be traced exactly to $|1\rangle$ states that it's composed of. In a similar way, the same can be said for the QFT$^\dagger$, and the way in which each $|\psi\rangle$ leads to a unique final combination of phases (not every individual state component from each $QFT^\dagger|\psi\rangle$ is unique, but rather the total combination of phases across all the state components). To see the resulting phase on a single state, we need only multiply out the $|0\rangle$ and $|1\rangle$ contributions that make up a particular computational basis state $|\psi_k\rangle$. For example:

$$|\psi_3\rangle = |0\rangle \otimes e^{-i\pi\left( Q_2 + \frac{Q_3}{2} \right)}|1\rangle \otimes e^{-i\pi Q_3}|1\rangle$$

$$= e^{-i\pi\left( Q_2 + \frac{Q_3}{2} + Q_3 \right)}|011\rangle$$

Note in the expression above the way in which $k = 3$ translates to the state $|011\rangle$, indicating that we are interpreting our quantum states as rightmost LSB binary numbers. Keeping this in mind, for a better understanding of how our QPE algorithm is going to work, we actually want to switch our QFT$^\dagger$ focus. Specifically, rather than focusing on where the phase contributions are coming from on an individual $|\psi_k\rangle$ component, we're interested in all the phases produced from a single QFT$^\dagger|\psi_j\rangle$:

$$QFT^\dagger|\psi_j\rangle = \frac{1}{\sqrt{2^n}} \sum_{k=0}^{2^n-1} e^{-\frac{2i\pi\,(j\cdot k)}{2^n}} |\psi_k\rangle$$

We can quickly confirm the expression above with the cell of code below:

In [ ]:

```
q = QuantumRegister(3,name='q' )
qc= QuantumCircuit(q,name='qc')
#------------------------------
qc.x( q[1] )
qc.x( q[2] )
print('____ Initial State ____')
oq.Wavefunction(qc)

qc.swap( q[0], q[2] )
oq.QFT_dgr( qc,q,3 )

print('\n____ After QFT\u2020 ____')
oq.Wavefunction(qc)
#=====================================
print('\n ____ QFT\u2020 Expression ____')
for k in np.arange( 8 ):
    phase = 1.0/(m.sqrt(8)) * np.exp( -1.0j*m.pi*( (3*k)/4 ) )
    print( 'state:  ',oq.Binary(int(k),8,'R'),'    phase:  ',round(phase.real,4)+round(phase.imag,4)*1.0j )
```

```
____ Initial State ____
1.0 |011>

____ After QFT† ____
0.35355 |000>    -0.35355 |100>    0.35355j |010>    -0.35355j |110>    -0.25-0.25j |001>    0.25+0.25j |101>    0.25-0.25j
|011>    -0.25+0.25j |111>

 ____ QFT† Expression ____
state:   [0, 0, 0]     phase:   (0.3536+0j)
state:   [0, 0, 1]     phase:   (-0.25-0.25j)
state:   [0, 1, 0]     phase:   0.3536j
state:   [0, 1, 1]     phase:   (0.25-0.25j)
state:   [1, 0, 0]     phase:   (-0.3536+0j)
state:   [1, 0, 1]     phase:   (0.25+0.25j)
state:   [1, 1, 0]     phase:   -0.3536j
state:   [1, 1, 1]     phase:   (-0.25+0.25j)
```

As we can see, the expression does indeed give us the correct phases produced from the QFT$^\dagger$. As an aside, please note the additional SWAP gate in our quantum circuit above, just before the QFT$^\dagger$. As we noted at the end of lesson 6, interpreting the quantum states from a Quantum Fourier Transformation as rightmost LSB binary numbers requires that we use SWAP gates at the end of a QFT. Consequently, when we want to perform a QFT$^\dagger$ using the same convention, these SWAP gates come *before* the QFT$^\dagger$ (remember that the QFT$^\dagger$ circuit is a mirrored version of the QFT circuit, with opposite phases). In the code above we put the SWAP gate in deliberately for demonstration purposes, but in future examples we will make use of our QFT_dgr's built in keyword "swap."

In a similar fashion to the expression shown above, we can describe the resulting state of our system after all of the control-$U$ operations in our QPE circuit in the same manner:

$$|\Psi\rangle_i \;=\; H^n|0\rangle^{\otimes n}$$

$$\text{After Control-}U\text{ Operations:} \qquad |\Psi\rangle \;=\; \frac{1}{\sqrt{2^n}}\sum_{j=0}^{2^n-1} e^{2\pi i\theta j}|\psi_j\rangle$$

As an example, let's take a closer look at a single $|\psi_j\rangle$ state, say $j = 6$ for $n = 3$. According to the formula above, the resulting phase on this state should be:

$$e^{2\pi i\theta(6)}|110\rangle$$

We can confirm that this is indeed the same phase that our QPE circuit will produce by carrying out the multiplication from earlier:

$$\frac{1}{\sqrt{2^3}}\Big(|0\rangle + e^{2\pi i\,(2^2)\,\theta}|1\rangle\Big)\otimes\Big(|0\rangle + e^{2\pi i\,(2^1)\,\theta}|1\rangle\Big)\otimes\Big(|0\rangle + e^{2\pi i\,(2^0)\,\theta}|1\rangle\Big)$$

focusing specifically on the phase on $|110\rangle$:

$$\begin{aligned} &= e^{2\pi i\,(2^2)\,\theta}|1\rangle \otimes e^{2\pi i\,(2^1)\,\theta}|1\rangle \otimes |0\rangle \\ &= e^{2\pi i\,(2^2+2^1)\,\theta}|110\rangle \\ &= e^{2\pi i\theta\,(6)}|110\rangle \end{aligned}$$

I encourage you to work through a few more $|\psi_j\rangle$ cases for yourself, proving that the formulation for the control-$U$ operations is indeed correct. Now, with the two equations above for expressing the control-$U$ and QFT$^\dagger$ operations in hand, we are ready to write down the final state of our system, the complete result of the QPE circuit. We will start by writing down the effect of the final QFT$^\dagger$ on one of these $|\psi_j\rangle$ states produced after the control-$U$ operations:

$$|\tilde{\psi}_j\rangle \;\equiv\; \frac{1}{\sqrt{2^n}}e^{2i\pi\theta j}|\psi_j\rangle \qquad\qquad j\in\left[0, 2^{n-1}\right]$$

$$QFT^\dagger|\tilde{\psi}_j\rangle \;=\; \frac{1}{2^n}\sum_{k=0}^{2^n-1} e^{2i\pi\theta j}e^{-\frac{2i\pi\,(j\cdot k)}{2^n}}|\psi_k\rangle$$

which we can rearrange into something more revealing about the final states:

$$=\; \frac{1}{2^n}\sum_{k=0}^{2^n-1} e^{-\frac{2i\pi j}{2^n}\left(k-2^n\theta\right)}|\psi_k\rangle$$

And lastly, we now add back in the eigenstate $|u\rangle$ and the summation over $j$, showing the full effect of the QFT$^\dagger$ on all of the $|\tilde{\psi}_j\rangle$ states:

$$|\Psi\rangle_{\text{final}} = \frac{1}{2^n}\sum_{j=0}^{2^n-1}\sum_{k=0}^{2^n-1} e^{-\frac{2i\pi j}{2^n}\left(k-2^n\theta\right)}|\psi_k\rangle\otimes|u\rangle$$

We have now finally reached the concluding state of our quantum system after the QPE circuit! What we've accomplished with the state above may not jump out at you straight away, but a measurement on this system is promised to reveal the phase $\theta$ with high probability. In the next section we are going to see exactly why this is, as well as the role of $n$ in increasing the precision of our result, but for now let's see this final state in action. The cell of code below is a classical version of the equation shown above, where $\theta = 0.52$ and $n = 3$:

In [ ]:

```
n = 3
theta = 0.52
print('Theta  =  ',theta,'   n  =  ',n,'\n---------------------------')
#===================
state  = []
bstate = []
bdec   = []
for i in np.arange(2**n):
    state.append(0)
    bstate.append(oq.Binary(int(i),2**n,'R'))
    bc = 0
    for i2 in np.arange(len(bstate[i])):
        bc = bc + ( 1.0/(2**(int(i2)+1)) )*int(bstate[i][i2])
    bdec.append(bc)
#-------------------------------------------------------------
for y in np.arange(2**n):
    for x in np.arange(2**n):
        state[int(y)] = state[int(y)] + 1.0/(2**n) * np.exp( (-2.0j*np.pi*x/(2**n))*(y-2**n*theta) )
#--------------------------------------------------------------
for j in np.arange(2**n):
    print('Probability: ',round( abs(state[j])**2,4 ),'        State: ',bstate[j],'        Binary Decimal: ',bdec[j])
```

```
Theta  =   0.52    n  =   3
---------------------------
Probability:  0.0036        State:  [0, 0, 0]        Binary Decimal:  0.0
Probability:  0.0041        State:  [0, 0, 1]        Binary Decimal:  0.125
Probability:  0.0064        State:  [0, 1, 0]        Binary Decimal:  0.25
Probability:  0.0187        State:  [0, 1, 1]        Binary Decimal:  0.375
Probability:  0.9198        State:  [1, 0, 0]        Binary Decimal:  0.5
Probability:  0.0346        State:  [1, 0, 1]        Binary Decimal:  0.625
Probability:  0.0083        State:  [1, 1, 0]        Binary Decimal:  0.75
Probability:  0.0045        State:  [1, 1, 1]        Binary Decimal:  0.875
```

## Approximating The Phase With $2^n\theta$

The coding example above represents the states and probabilities we will later expect from our quantum system (after we've covered how to physically implement our control-$U$ operators). If you haven't already, I encourage you to run the example above a few times, changing the values of $\theta$ and $n$ and seeing some of the various results. In particular, below are some recommended combinations to test, and the resulting probabilities you should find:

| $n$ | $\theta$ | Most Probable State | Probability |
|---|---|---|---|
| 3 | 0.50 | $\vert 100\rangle$ | 1.00 |
| 3 | 0.55 | $\vert 100\rangle$ | .578 |
| 4 | 0.55 | $\vert 1001\rangle$ | .876 |

Now, let's discuss the impact of the results found from these three combinations. In our first example, we are trying to find the phase $\theta = 0.5$, and our QPE algorithm is returning to us the state $|100\rangle$ with a $100\%$ probability. However, for the same number of qubits, $n = 3$, trying to find a slightly different phase of $\theta = 0.55$ results in a dramatic decrease in success, but the same most probable state. And lastly, by increasing the value of $n$ to $4$, we are able to significantly boost our success rate, but consequently get a different most probable state.

If we combine the results from all three of these examples, you may start to get a sense for how this algorithm is working. The first example suggests that for certain values of $\theta$ we can obtain perfect success rates, while the second and third are indicative that for other cases of $\theta$ we cannot. However, for these non-special values of $\theta$, we can boost our chances of measuring the most probable state by increasing $n$, but consequently must interpret the new state made up of more qubits. Keeping these observations in mind, let's now dig a little deeper into the answers our QPE is giving us, and more importantly $why$.

### Interpreting Our QPE Results

As crucial as it is to our understanding of QPE, up until this point we have made no comments as to what the final states of our system are meant to represent, and how we can tell a "correct" final state from a "wrong" one. The answer to this important question is that the final states of the QPE algorithm are designed to represent binary decimals, which we covered in lesson 6.1 to further our understanding of the QFT. As a quick reminder, a binary decimal is the equivalent to our standard base-$10$ decimal system, but for binary numbers. Whereas a typical decimal represents increasing powers of $\frac{1}{10}$, a binary decimal can be used to represent numbers between $0$ and $1$ in terms of powers of $\frac{1}{2}$. For example, take a look at the following numbers and their decimal representations:

| $N$ | base-10 | base-2 |
|---|---|---|
| $\frac{5}{8}$ | $0.625$ | $0.101$ |
| $\frac{7}{16}$ | $0.4375$ | $0.0111$ |
| $\frac{1}{5}$ | $0.2$ | $0.\overline{0011}$ |

Looking at the fraction $\frac{7}{16}$ in deeper detail:

$$\text{base-10}: \quad 0.625 = 6\times\frac{1}{10} + 2\times\frac{1}{100} + 5\times\frac{1}{1000}$$

$$\text{base-2}: \quad 0.0111 = 0\times\frac{1}{2} + 1\times\frac{1}{4} + 1\times\frac{1}{8} + 1\times\frac{1}{16}$$

For our QPE algorithm, after we make a measurement on the system in the computational basis and get our final state, we convert that state into a binary decimal and that's our approximation to $\theta$! With this new insight, I encourage you to return to the coding example above and see for yourself that indeed the states which were most probable correspond to the closest binary decimal representations to $\theta$.

## Accuracy of the Approximation

As our final topic before moving on to the physical implementation of the QPE algorithm into a quantum circuit (specifically the control-$U$ operators), we need to discuss the accuracy of QPE and the dependence on $n$. Originally, we said earlier that increasing $n$ will improve the accuracy of our algorithm, and the quick examples above seemed to suggest the same thing. Specifically, in attempting to discover $\theta = 0.55$, $n = 3$ gives us a probability of $0.578$ for $|100\rangle$ while $n = 4$ gave us a probability of $0.876$ for $|1001\rangle$. In interpreting these results as binary decimals, we get the values $0.5$ and $0.5625$ respectively. While this particular example demonstrated that increasing the number of qubits yielded a better final answer, it turns out that this is not always the case. In actuality, the real purpose in choosing an $n$ is to create a state closest to $\theta$, while simultaneously creating the largest separation possible from this closest state to all other states in the system. For example:

| $n$ | $\theta$ | Most Probable State | Probability | Binary Decimal |
|---|---|---|---|---|
| 3 | 0.52 | $\|100\rangle$ | .920 | 0.500 |
| 4 | 0.52 | $\|1000\rangle$ | .706 | 0.500 |
| 5 | 0.52 | $\|10001\rangle$ | .640 | 0.531 |
| 6 | 0.52 | $\|100001\rangle$ | .767 | 0.516 |

Take a look at the examples above, and the resulting most probable states as we increase $n$ from $3$ to $6$. In the first two examples, we find that bumping up our number of qubits to $4$ results in the same binary decimal answer, but with a significantly lower probability. Then as we increase to $n = 5$, we once again sacrifice some probability on the most probable state, but in turn our binary decimal interpretation is closer to the true value of $\theta$. And lastly, going from $n = 5$ to $6$ yields not only a better binary decimal approximation, but a higher probability as well! Suffice to say, the examples above demonstrate that more is not always better for QPE, as increasing $n$ isn't always guaranteed to yield a better final state. It's true that increasing $n$ allows for closer approximations to $\theta$, but we must always keep in mind the costs in doing so, both in probabilities and circuit size.

In determining how effective a particular value of $n$ will be in approximating a phase, we must return to the final result from our derivations earlier:

$$|\Psi\rangle_{\text{final}} = \frac{1}{2^n}\sum_{x=0}^{2^n-1}\sum_{y=0}^{2^n-1} e^{-\frac{2\pi i x}{2^n}\left(y-2^n\theta\right)}|\psi_y\rangle \otimes |u\rangle$$

Now let's suppose we are interested in the probability of measuring a particular state $|m\rangle$ in our final system, corresponding to some binary decimal approximation to $\theta$. We can calculate the probability of measuring this state $|m\rangle$ as:

$$\left|\langle m|\Psi\rangle_{\text{final}}\right|^2 = \left|\langle m|\frac{1}{2^n}\sum_x\sum_y e^{-\frac{2\pi i x}{2^n}\left(y-2^n\theta\right)}|\psi_y\rangle\right|^2$$

$$= \left|\langle m|\frac{1}{2^n}\sum_x e^{-\frac{2\pi i x}{2^n}\left(m-2^n\theta\right)}|m\rangle\right|^2$$

Notice how we've lost one of the summations in the expression above, and consequently set $|\psi_y\rangle = |m\rangle$. This is because $|m\rangle$ represents a state corresponding to some binary decimal value, which by definition means that it is a pure state in the computational basis. Thus, when we look at the $y$ summation over all the computational basis states, all of them except for $|\psi_y\rangle = |m\rangle$ will be orthogonal, canceling out to $0$. With only the summation over $x$ remaining then, our probability of measuring the state $|m\rangle$ becomes:

$$= \left|\frac{1}{2^n}\sum_x e^{-\frac{2\pi i x}{2^n}\left(m-2^n\theta\right)}\right|^2$$

We can see in the expression above that our probability of measuring the state $|m\rangle$ has boiled down to a summation of complex numbers squared, where the only remnant of $|m\rangle$ lies in the quantity $(m - 2^n\theta)$ (note that $m$ here is a base-$10$ number, corresponding to the binary number represented by $|m\rangle$). So then, what are we to make of this expression? Well, as a quick first example, let's see the special case where the binary decimal of $|m\rangle$ corresponds exactly to $\theta$, for which we then have $m = 2^n\theta$:

$$\left|\frac{1}{2^n}\sum_x^{2^n} e^{-2\pi i x(0)}\right|^2 = 1$$

Conceptually, the special case above is demonstrating that the QPE algorithm will give us the phase $\theta$ we're looking for with $100\%$ probability, *if* it can be perfectly represented as a binary decimal. Mathematically this results from the case where the base-$10$ value of $m$ perfectly cancels out with $2^n\theta$, resulting in every term in the summation going to $e^0$. This is why we chose to express the phase on each $|\psi_y\rangle$ state with the quantity $(y - 2^n\theta)$ separated out in our derivation earlier. The closer the number $y$ is to $2^n\theta$, the closer this quantity is to $0$, resulting in a higher probability of measurement. Conversely, the further $y$ is away from $2^n\theta$, the less probable that particular $|\psi_y\rangle$ state will be in the final quantum state.

To better understand the role of the quantity $(y - 2^n\theta)$, note that in our summation expression above this quantity is being divided by a factor of $2^n$. Thus, the bounds for their combined quantity are $\left[0, 1\right]$. We just saw the result for the lower limit of $0$, so now let's see what happens at the other extreme:

$$\Big| \frac{1}{2^n} \sum_x^{2^n} e^{-2\pi i x(1)} \Big|^2 \quad = \quad 0$$

As expected, for either of the extreme cases: $y = 0$ & $\theta = 1$ or $y = 2^n$ & $\theta = 0$, we should find these corresponding $|\psi_y\rangle$ states with exactly $0$ probabilities.

But now suppose we're interested in a state $|m\rangle$ that is not a perfect binary decimal representation of $\theta$, lying somewhere between the two extremes outlined above. If we are no longer dealing with a state $|m\rangle$ such that $2^n\theta = m$, then instead we can express our phase quantity as $2^n\theta = m + \phi$, where $\phi$ is essentially the difference between $\theta$ and our state $|m\rangle$. If we now substitute this back into our final state, let's see how the probability changes:

$$\big|\langle m|\Psi\rangle_{\text{final}}\big|^2 = \Big|\Big\langle m \Big| \frac{1}{2^n} \sum_x e^{-\frac{2\pi i x}{2^n}\left(m - 2^n\theta\right)} \Big| m \Big\rangle\Big|^2$$

$$= \Big| \frac{1}{2^n} \sum_x e^{-\frac{2\pi i x}{2^n}\left(-\phi\right)} \Big|^2$$

$$= \frac{1}{2^{2n}} \Big| \frac{-1 + e^{2\pi i\phi}}{-1 + e^{\frac{2\pi i\phi}{2^n}}} \Big|^2$$

Connecting the expression above to our discussion about the quantity $(y - 2^n\theta)$, the closer $\phi$ is to $0$, the more probable our state $|m\rangle$ should be. To verify this, let's plot the expression above as a function of $\phi$:

In [ ]:
```
x = []
y = []
n = 3
for k in np.arange( 1000 ):
    if( k != 500 ):
        phi = -5 + ( k/1000 ) * 10
        x.append( phi )
        y.append( 1/(2**(2*n)) * abs( (-1 + np.exp(2.0j*m.pi*phi) )/(-1 + np.exp(2.0j*m.pi*phi/(2**n))) )**2 )

plt.plot(x,y)
plt.axis([-5,5,0,1])
plt.show()
```

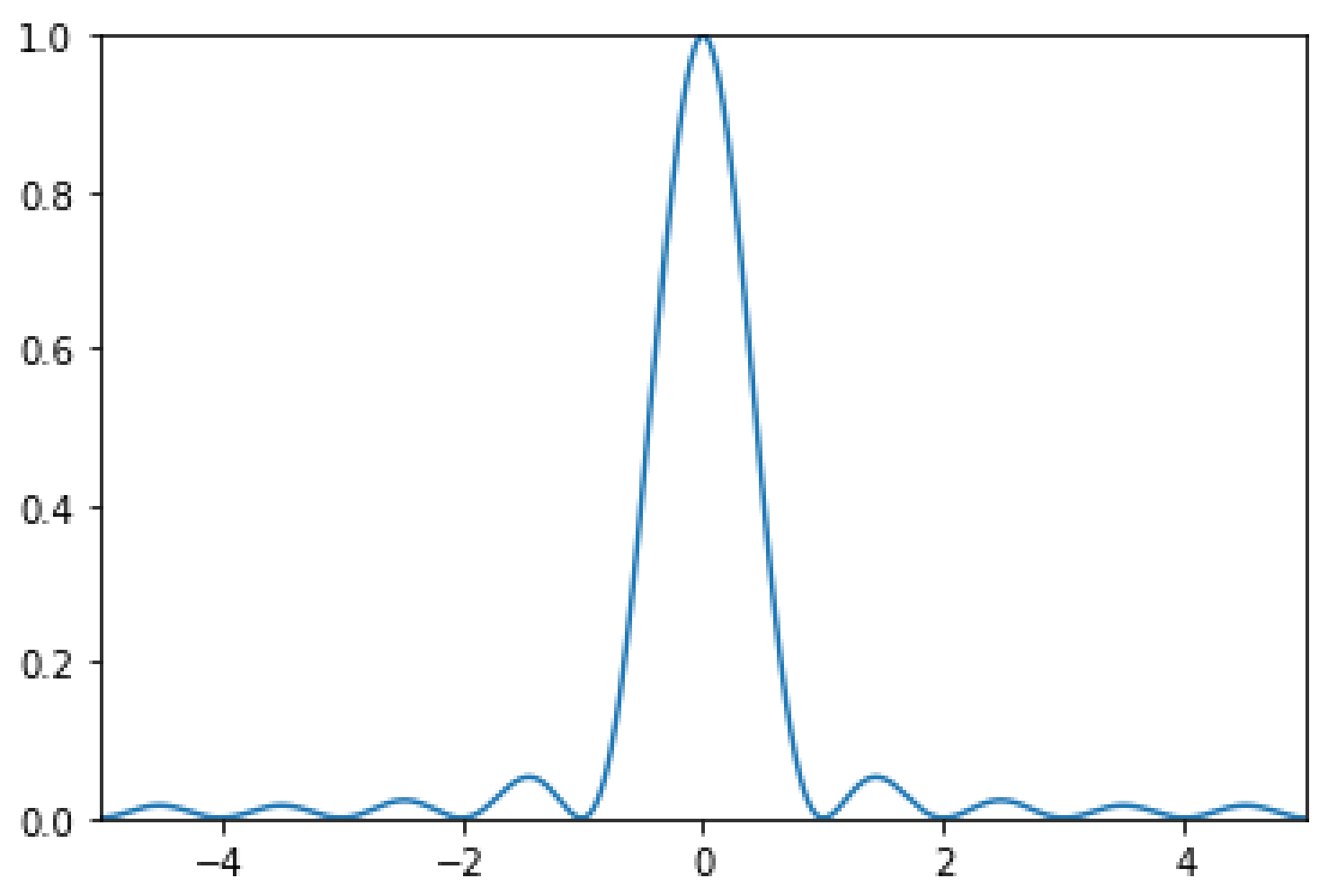

The plot above is revealing a few key features about $\phi$, which remember was an artificial parameter we created to represent the difference between $m$ and $2^n\theta$. But before jumping into our quantitative analysis, please note that the plot shown above is consistent for any value $n$, which right off the bat is telling us that all of the coming results are applicable to any sized QPE. Starting with the center of the plot then, $\phi = 0$, this is once again confirming our result from earlier for the case where $|m\rangle$ is a perfect binary decimal representation of $\theta$. Then as we start to move away from $\phi = 0$, we can see the probability of $|m\rangle$ start to rapidly fall off, reaching $0$ at $\phi = \pm 1$.

But now here is where things start to get interesting. What is the physical interpretation for all of the locations in the plot where we see the probability of $|m\rangle$ go to $0$, specifically when $\phi$ is equal to an integer value? To answer this, let's think about the case where $m = 2^n\theta$, noting once again that $m$ here is a base-$10$ integer. For this special case where the probability of $|m\rangle$ is exactly equal to $1$, this means that every other state in the system must have a probability of $0$. Or in other words, the states $|m-2\rangle$, $|m-1\rangle$, $|m+1\rangle$, $|m+2\rangle$... all have exactly $0$ probability. Returning to our plot above, this is exactly the situation described by $\phi$ equal to the various integer values, causing our state $|m\rangle$ to have a $0$% probability of being measured. If some other state $|m'\rangle$ is the perfect binary decimal representation of $\theta$, then the probability of measuring $|m\rangle$ must go to zero.

With all of the integer cases covered, what then can we say about all of the intermediate values of $\phi$ in the plot above? As we can see, even though the plot goes to $0$ periodically, in between these integer values we regain various peaks in probability of measuring $|m\rangle$. For these regions where the difference between $m$ and $2^n\theta$ is non-integer, we can conclude that there does $not$ exist a state $|m'\rangle$ in our system which is a perfect binary decimal representation of $\theta$. Thus, every state in the final system will have some non-zero probability of being measured. But what's more interesting is the way in which the peak probabilities in the plot seem to taper off with each successive integer increase. To connect this idea to some topics we've already discussed, let's revisit our classical code from earlier, this time for $\theta = 0.52$, $n = 4$:

In [ ]:

```
n = 4
theta = 0.52
print('Theta  =  ',theta,'   n  =  ',n,'\n---------------------------')
#===================
state  = []
bstate = []
bdec   = []
for i in np.arange(2**n):
    state.append(0)
    bstate.append(oq.Binary(int(i),2**n,'R'))
    bc = 0
    for i2 in np.arange(len(bstate[i])):
        bc = bc + ( 1.0/(2**(int(i2)+1)) )*int(bstate[i][i2])
    bdec.append(bc)
#--------------------------------------------------------------
for y in np.arange(2**n):
    for x in np.arange(2**n):
        state[int(y)] = state[int(y)] + 1.0/(2**n) * np.exp( (-2.0j*np.pi*x/(2**n))*(y-2**n*theta) )
#---------------------------------------------------------------
for j in np.arange(2**n):
    print('Probability: ',round( abs(state[j])**2,4 ),'        State: ',bstate[j],'        Binary Decimal: ',bdec[j])
```

```
Theta  =   0.52    n  =   4
---------------------------
Probability:  0.0028         State:  [0, 0, 0, 0]         Binary Decimal:  0.0
Probability:  0.0028         State:  [0, 0, 0, 1]         Binary Decimal:  0.0625
Probability:  0.0031         State:  [0, 0, 1, 0]         Binary Decimal:  0.125
Probability:  0.0037         State:  [0, 0, 1, 1]         Binary Decimal:  0.1875
Probability:  0.0049         State:  [0, 1, 0, 0]         Binary Decimal:  0.25
Probability:  0.0076         State:  [0, 1, 0, 1]         Binary Decimal:  0.3125
Probability:  0.0144         State:  [0, 1, 1, 0]         Binary Decimal:  0.375
Probability:  0.0424         State:  [0, 1, 1, 1]         Binary Decimal:  0.4375
Probability:  0.7063         State:  [1, 0, 0, 0]         Binary Decimal:  0.5
Probability:  0.1571         State:  [1, 0, 0, 1]         Binary Decimal:  0.5625
Probability:  0.0265         State:  [1, 0, 1, 0]         Binary Decimal:  0.625
Probability:  0.011         State:  [1, 0, 1, 1]         Binary Decimal:  0.6875
Probability:  0.0064         State:  [1, 1, 0, 0]         Binary Decimal:  0.75
Probability:  0.0044         State:  [1, 1, 0, 1]         Binary Decimal:  0.8125
Probability:  0.0035         State:  [1, 1, 1, 0]         Binary Decimal:  0.875
Probability:  0.003         State:  [1, 1, 1, 1]         Binary Decimal:  0.9375
```

Take a look at the probabilities of each state as we move away from $|1000\rangle$ in both directions, and the similarity to our $\phi$ plot above. With each successive state away from $|1000\rangle$, corresponding to $\phi$ either increasing or decreasing by $\pm 1$, we can see that our QPE algorithm is causing these states to be successively less and less probable. Additionally, take a look at the cases $|0111\rangle$ and $|1001\rangle$, which have the probabilities $0.042$ and $0.157$ respectively. Even though they are both exactly one state off from the most probable $|m\rangle$, the binary decimal representation for $|1001\rangle$ is closer to $\theta = 0.52$, which subsequently results in a higher probability of measurement. If we go back and ask what the corresponding $\phi$ values would be for both of these states, we find that the state $|1001\rangle$ has a $\phi$ value of 0.67, while $|0111\rangle$ has a $\phi$ of -1.33 (which means that our closest $|m\rangle$ has a $\phi = -0.33$). As you can see, the $\phi$ values for these states are all separated by exact integer numbers, corresponding to their respective base-$10$ $m$ values which show up in our probability formula from earlier.

The significance of the example above is the way in which the quantity $\phi$ translates into the decaying probabilities we find on our quantum states resulting from QPE. Remember that at the end of the QPE algorithm we are forced to make a measurement, which will return to us only a single $|m\rangle$. Thus, even though our circuit has created this wonderful final distribution of states, a single QPE run could yield an approximation to $\theta$ that is completely off. As an example, the figure below is a demonstration of the interplay between the $\theta$ we are trying to find, and the most probable state resulting from QPE. The plot below shows the total success probability on the states $|100\rangle$ and $|101\rangle$ as we vary $\theta$ from $0.5$ to $0.625$:

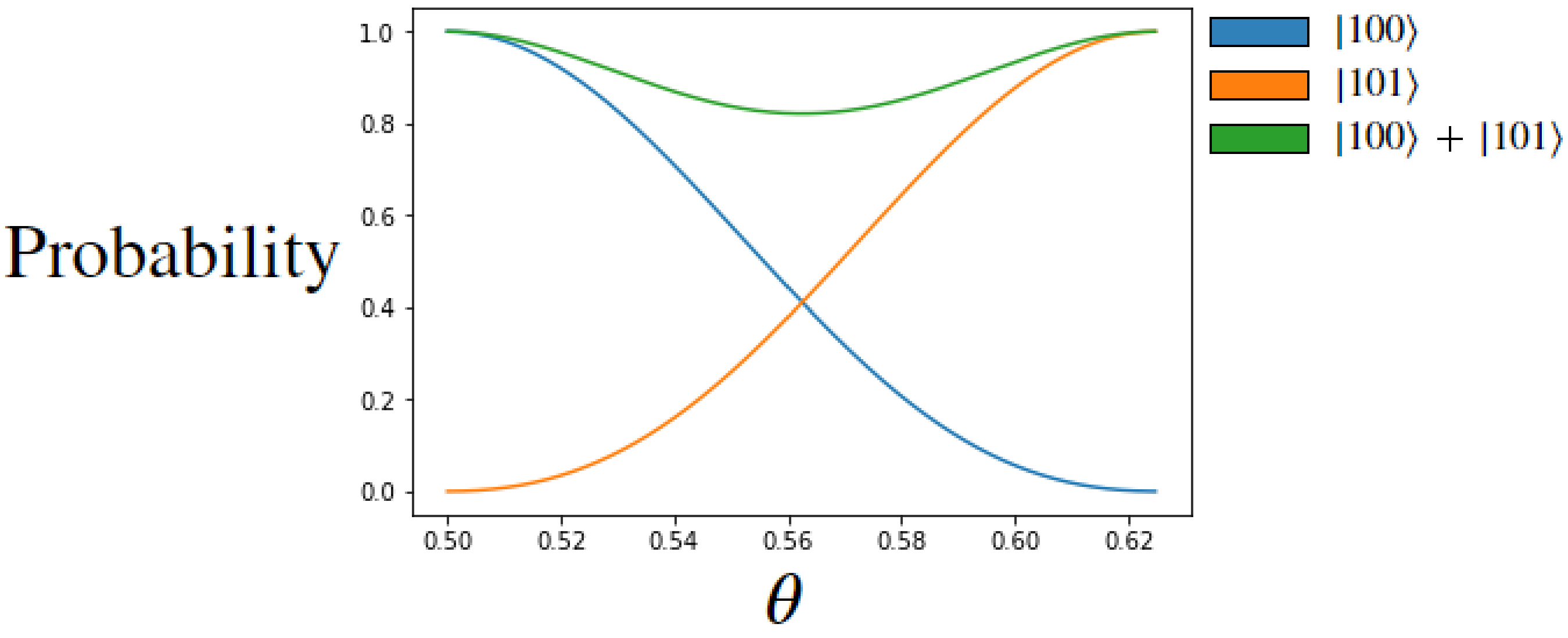

As you can see, when $\theta$ lines up exactly on either of the two states, the probability approaches $1$. For all values of $\theta$ in between however, the total probability between $|100\rangle$ and $|101\rangle$ no longer sums to $1$, meaning that it is being distributed amongst the other states in the system, reaching a minimum at $\theta = 0.5625$ (the exact halfway point between the two). Luckily, in worst case scenarios we still find roughly $80\%$ of the total probability in the system sitting on the closest two states.

### Determining $\theta$ Through Multiple Measurements

Having just seen how the accuracy of a QPE run can vary depending on the $\theta$ we are trying to find and the number of qubits we choose to use for our binary decimal representations, relying on a single measurement to give us our final answer can feel a bit unsettling. In particular, an unlucky measurement could give us a $\theta$ value that is completely wrong, essentially defeating the whole purpose of the algorithm. And even worse still, if we do run our QPE multiple times to properly determine the most probable $|m\rangle$, if the results clearly show that $\theta$ lies somewhere between some $|m\rangle$ and $|m'\rangle$, we have no means of getting a better approximation without increasing $n$ and starting all over.

Even though we're no strangers to the measurement dilemma in quantum algorithms, it would still be nice to have a more concrete methodology for determining $\theta$. And while in certain instances we may not have the luxury of running QPE multiple times (for example if we are using QPE as a subroutine inside of a larger algorithm), for now let's assume we do. As we shall see, even though certain values of $\theta$ are unfeasible for exact binary decimal representations, building up a statistic of measurements will allow us to reconstruct $\theta$ from the binary decimal states we *can* implement. To do this, we must first return to our formula for the probability of measuring $|m\rangle$:

$$\left|\langle m|\Psi\rangle_{\text{final}}\right|^2 = \frac{1}{2^{2n}}\left|\frac{-1+e^{2\pi i\phi}}{-1+e^{\frac{2\pi i\phi}{2^n}}}\right|^2$$

Let us now suppose we can obtain the left hand side of this equation through repeat measurements. Specifically, we run our QPE circuit enough times to determine the closest $|m\rangle$, and substitute its probability into the equation above. In doing so, we could then solve for $\phi$, which would ultimately allow us to solve for $\theta$:

$$\theta = \frac{m+\phi}{2^n}$$

The key here is that we need enough measurements to determine the closest $|m\rangle$ to $\theta$, that way we can be certain to solve for a value of $\phi$ that lies between $0$ and $1$. Specifically, since our $\left|\langle m|\Psi\rangle_{\text{final}}\right|^2$ function has the same value for multiple instances of $\phi$ (to see this, look back at our plot from earlier), we need to work within the region of $\phi$ where our function is one-to-one. Technically there is no where in our space where this is true, but if we know we are dealing with the most probable $|m\rangle$, then we can be certain $\phi \in \left[-0.5, 0.5\right]$. And if we look at the second most probable state $|m'\rangle$ in our system, we can determine whether or not our solved value for $\phi$ should be positive or negative.

To demonstrate this approximation technique, below is a simulated version of our $\theta = 0.52$, $n = 4$ example from earlier:

In [ ]:

```
Prob = [ 0.0028,0.0028,0.0031,0.0037,0.0049,0.0076,0.0144,0.0424,0.7063,0.1571,0.0265,0.011,0.0064,0.0044,0.0035,0.003 ]
Measured = np.zeros( len(Prob) )
trials = 10000
n = 4
#========================================          Simulate measurements on the final system
for t in np.arange(trials):
    M = random.random()
    p = 0
    for s in np.arange( len(Prob) ):
        if( p < M < (p + Prob[s]) ):
            Measured[s] = Measured[s] + 1
        p = p + Prob[s]
#----------------------------------------
MD = {}
for i in np.arange( len(Prob) ):
    state = oq.Binary(int(i),2**n,'R')
    state_str = ''
    for j in np.arange(len(state)):
        state_str = state_str+str(state[j])
    MD[state_str] = int( Measured[i] )
MP = oq.Most_Probable(MD,2)
for mp in np.arange(2):
    MP[0][mp] = MP[0][mp]/trials
#========================================          Compute phi using the most probable state
phi,theta = oq.QPE_phi(MP)
print('\nMost Probable State:   |'+str(MP[1][0])+'>      Probability:  ',round(MP[0][0],5))
print('\nCorresponding \u03A6:  ',phi,'\n\nApproximate \u03B8:  ',round(theta,5))
```

```
Most Probable State:   |1000>      Probability:   0.7042

Corresponding Φ:   0.3214

Approximate θ:   0.52009
```

Success! In the cell of code above we blindly measure our quantum system $10,000$ times to get an approximate probability for our closest $|m\rangle$. We then use this approximate probability to determine $\phi$ using the function $\textbf{QPE_phi}()$, and ultimately compute our approximate value to $\theta$. And as you can see, even though there isn't a quantum state in our system which is an exact binary decimal representation of $\theta$, our technique has determined $\theta$ with some impressive accuracy!

An important thing to point out about the technique above is that it is essentially independent of $n$, the qubit accuracy of our QPE. That is to say, increasing $n$ doesn't help us approximate $\theta$ any better, only increasing the number of measurements. Thus, we would ideally want to apply this approximation technique with the lowest $n$ we can in order to decrease our total gate count / run time. We will return to this approximation approach one more time in this lesson, after we can properly implement QPE circuits.

## Implementing the Unitary Operator

We have now finally reached the point in this lesson where we will actually be running QPE as a quantum circuit. Having now fully covered the QPE algorithm in mathematical detail, our last task is to construct it using quantum gates. If you're curious as to why we put off this section to the end, it's because unlike other quantum algorithms, there is no set way for implementing QPE. Aside from the general structure of the quantum circuit outlined earlier, how one chooses to implement the control-$U$ operators and eigenstate $|u\rangle$ is an open question. Thus, all of the mathematics for QPE hold true for any implementation, which is why we elected to cover them first, rather than diving straight into building a few example circuits.

To begin, we will first turn our attention towards the control-$U$ operators, for which we must determine their eigenvalue $\theta$. Before attempting to construct the full set of control-$U^{2^n}$ operators, we will start by focusing on implementing a single control-$U$ on one of its eigenstates $|u\rangle$. We certainly have no shortage of unitary operators available to us in our standard set of quantum gates from Qiskit, but implementing them as control gates and preparing $|u\rangle$ is often times more tricky. Luckily for us, there is at least one simple example that we can try:

$$R_\phi|11\rangle \;\;=\;\; e^{i\phi}|11\rangle$$

The control-phase gate is an ideal first example to test our algorithm, particularly because its eigenvalue is a free parameter. Thus, in order to make this gate match our eigenvalue template $2\pi i\theta$, all we need to do is factor the $2\pi$ into our $\phi$ (please note that this is *not* the same $\phi$ from our discussions earlier, but merely the standard notation for the control-phase gate):

$$\theta \;\;=\;\; \frac{\phi}{2\pi}$$

For the $R_\phi$ gate, the eigenstate we need to prepare is as simple as it gets: $|1\rangle$. And similarly, the various powers of our control-$U^2$ operators is achieved by repetitive uses of $R_\phi$. Let's first take a look at the state of the system after we apply all of these control operators:

In [ ]:
```
n = 3
q1 = QuantumRegister(n,name='q1')
q2 = QuantumRegister(1,name='q2')
qc = QuantumCircuit(q1,q2,name='qc')
theta = 0.52
phi = 2*m.pi*theta
#---------------------------------------------------
for i in np.arange(n):
    qc.h(q1[int(i)])
qc.x( q2[0] )

for j in np.arange(n):
    for k in np.arange(2**j):
        qc.cu1( phi, q1[int(n-1-j)], q2[0] )
print('\n___ After Control-U Operations ___')
oq.Wavefunction( qc, systems=[n,1] )
#---------------------------------------------------
Phases = [np.exp(4.0j*phi),np.exp(2.0j*phi),np.exp(1.0j*phi)]
print(' ')
for i in np.arange(8):
    state = oq.Binary(int(i),8,'R')
    phase = m.sqrt(1/8)
    for j in np.arange(3):
        if(state[j]==1):
            phase = phase*Phases[j]
    print('State: ',state,'   Phase: ',round(phase.real,5)+round(phase.imag,5)*1.0j)
```

```
___ After Control-U Operations ___
0.35355 |000>|1>    0.30982+0.17033j |100>|1>    0.34245+0.08793j |010>|1>    0.25773+0.24202j |110>|1>    -0.35077-0.04431
j |001>|1>    -0.28603-0.20781j |101>|1>    -0.32873-0.13015j |011>|1>    -0.22536-0.27242j |111>|1>

State:  [0, 0, 0]    Phase:  (0.35355+0j)
State:  [0, 0, 1]    Phase:  (-0.35077-0.04431j)
State:  [0, 1, 0]    Phase:  (0.34245+0.08793j)
State:  [0, 1, 1]    Phase:  (-0.32873-0.13015j)
State:  [1, 0, 0]    Phase:  (0.30982+0.17033j)
State:  [1, 0, 1]    Phase:  (-0.28603-0.20781j)
State:  [1, 1, 0]    Phase:  (0.25773+0.24202j)
State:  [1, 1, 1]    Phase:  (-0.22536-0.27242j)
```

The cell of code above achieves the desired phases from the control-$U^2$ operations on our quantum system, and then confirms each phase classically. Note how easily we were able to implement the control-$U^{2^n}$ operations, simply applying the $\mathrm{CU1}$ gate the correct number of times in succession.

With our control-$U$ operators successfully achieved, all that's left to do is apply the QFT$^\dagger$ and make a measurement on the system:

In [ ]:

```
n = 3
q1 = QuantumRegister(n,name='q1')
q2 = QuantumRegister(1,name='q2')
c  = ClassicalRegister(n,name='c')
qc = QuantumCircuit(q1,q2,c,name='qc')
theta = 0.52
#---------------------------------------------------
for i in np.arange(n):
    qc.h(q1[int(i)])
qc.x( q2[0] )

phi = 2*m.pi*theta
for j in np.arange(n):
    for k in np.arange(2**j):
        qc.cu1( phi, q1[int(n-1-j)], q2[0] )

print('\n___ After QFT_dgr ___')
oq.QFT_dgr( qc,q1,n,swap=True )
oq.Wavefunction( qc, systems=[n,1] )
#---------------------------------------------------
qc.measure(q1,c)
results = execute(qc, BasicAer.get_backend('qasm_simulator'), shots=10000).result()
plot_histogram(results.get_counts())
```

```
___ After QFT_dgr ___
0.02569-0.0546j |000>|1>    0.86777+0.40834j |100>|1>    0.07553-0.02719j |010>|1>    -0.03085-0.08568j |110>|1>    0.04708
-0.04284j |001>|1>    -0.12512-0.1375j |101>|1>    0.13673+0.00645j |011>|1>    0.00316-0.06698j |111>|1>
```

Out[19]:

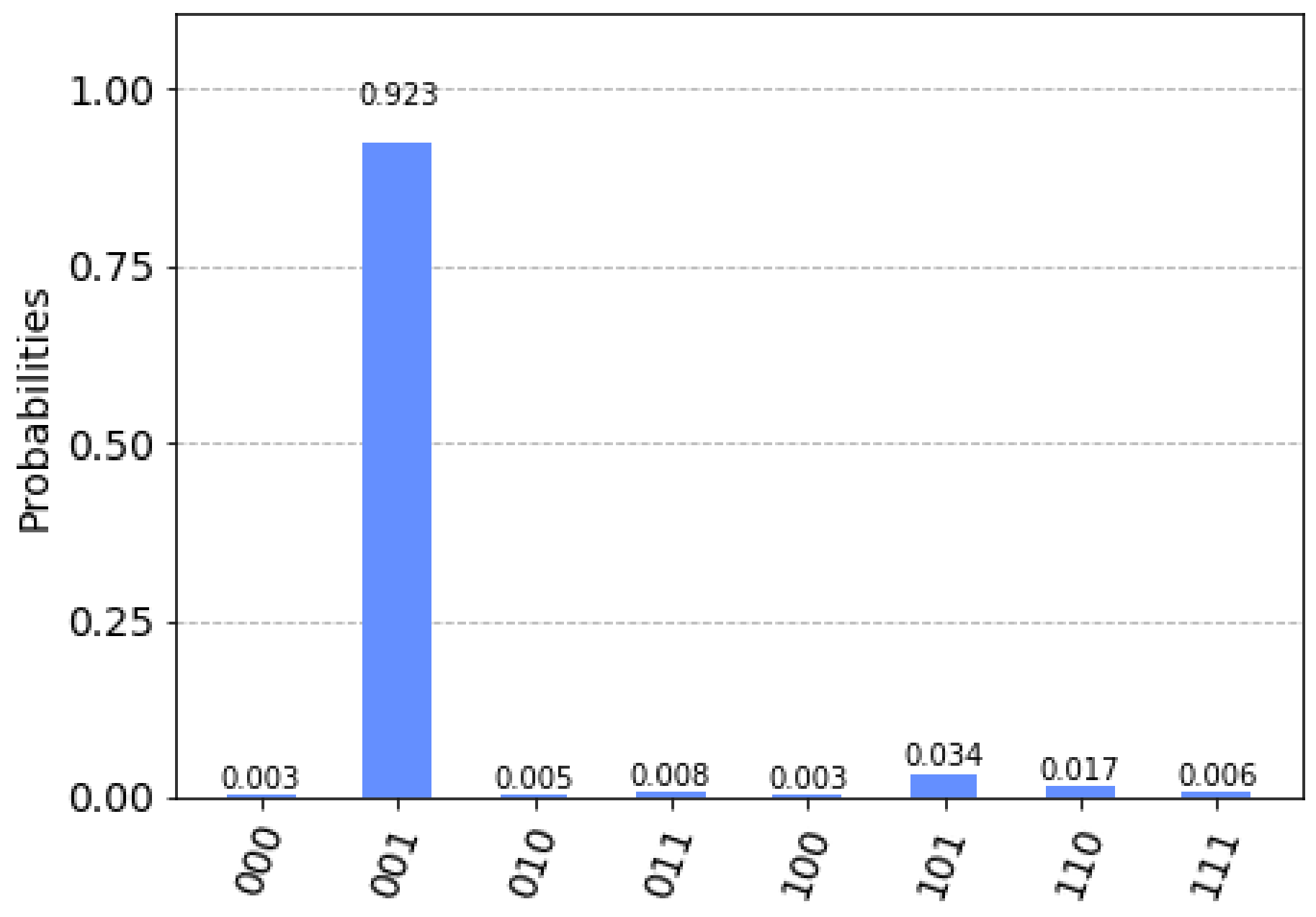

As we can see in the final amplitudes, the state $|100\rangle$ is overwhelmingly more probable than any other state in the system, and matches perfectly with our classical code from earlier. This is confirmed by the plot as well, although note the ordering on the qubits is backwards (just a small labeling discrepancy between our code and Qiskit). To recap, our code above demonstrates looking for the phase $\theta = 0.52$ with a qubit accuracy of $n = 3$. As we know, this means that we will not have a state in our system which is a perfect binary representation of $\theta$, so we expect that our most probable state $|100\rangle$, corresponding to $0.5$, will have a probability less than $1$.

The example above is the full QPE algorithm in action, although a bit contrived since we essentially put in the value of $\theta$ by hand into our $R_\phi$ gates. In our next example, we will once again be using $R_\phi$ gates, but this time we will look to implement a higher dimensional $|u\rangle$, as well as search for an unknown $\theta$. Specifically, we will let our code randomly pick out various $\phi$ phases, which will then be combined into our final unknown eigenphase:

$$\begin{aligned} R_{\phi_1} R_{\phi_2} |1\rangle |11\rangle &= e^{i\phi_1} e^{i\phi_2} |1\rangle |11\rangle \\ &= e^{i(\phi_1+\phi_2)} |1\rangle |11\rangle \\ &= e^{2i\pi\theta} |1\rangle |11\rangle \end{aligned}$$

As shown above, our QPE example will take advantage of the exponential nature of applying individual $R_\phi$ gates to create our higher order $|u\rangle$. But this time, let's see what running QPE looks like with an unknown phase:

In [ ]:

```
n1 = 3
n2 = 2
phases = []
for p in np.arange( n2 ):
    phases.append( round( 2*m.pi*random.random(),4 ) )
theta = round( sum(phases)/(2*m.pi),5)
if( theta > 1 ):
    theta = round( theta - m.floor(theta),5)
#================================
q  = QuantumRegister(n1,name='q')
qu = QuantumRegister(n2,name='qu')
c  = ClassicalRegister(n1,name='c')
qc = QuantumCircuit(q,qu,c,name='qc')
#---------------------------------
for i in np.arange(n1):
    qc.h( q[int(i)] )
for i2 in np.arange(n2):
    qc.x( qu[int(i2)] )
qc.barrier()
for j in np.arange(n1):
    for j2 in np.arange(2**j):
        for j3 in np.arange(n2):
            qc.cu1( phases[int(j3)], q[int(n1-1-j)], qu[int(j3)] )
    qc.barrier()

oq.QFT_dgr( qc,q,n1,swap=True )
#-------------------------------------------------
print('Phases:     ',phases)
print('\nCombined \u03B8:   ',theta)
qc.measure(q,c)
results = execute(qc, BasicAer.get_backend('qasm_simulator'), shots=10000).result()
plot_histogram(results.get_counts())
```

```
Phases:     [5.2892, 4.4193]

Combined θ:    0.54516
```

Out[52]:

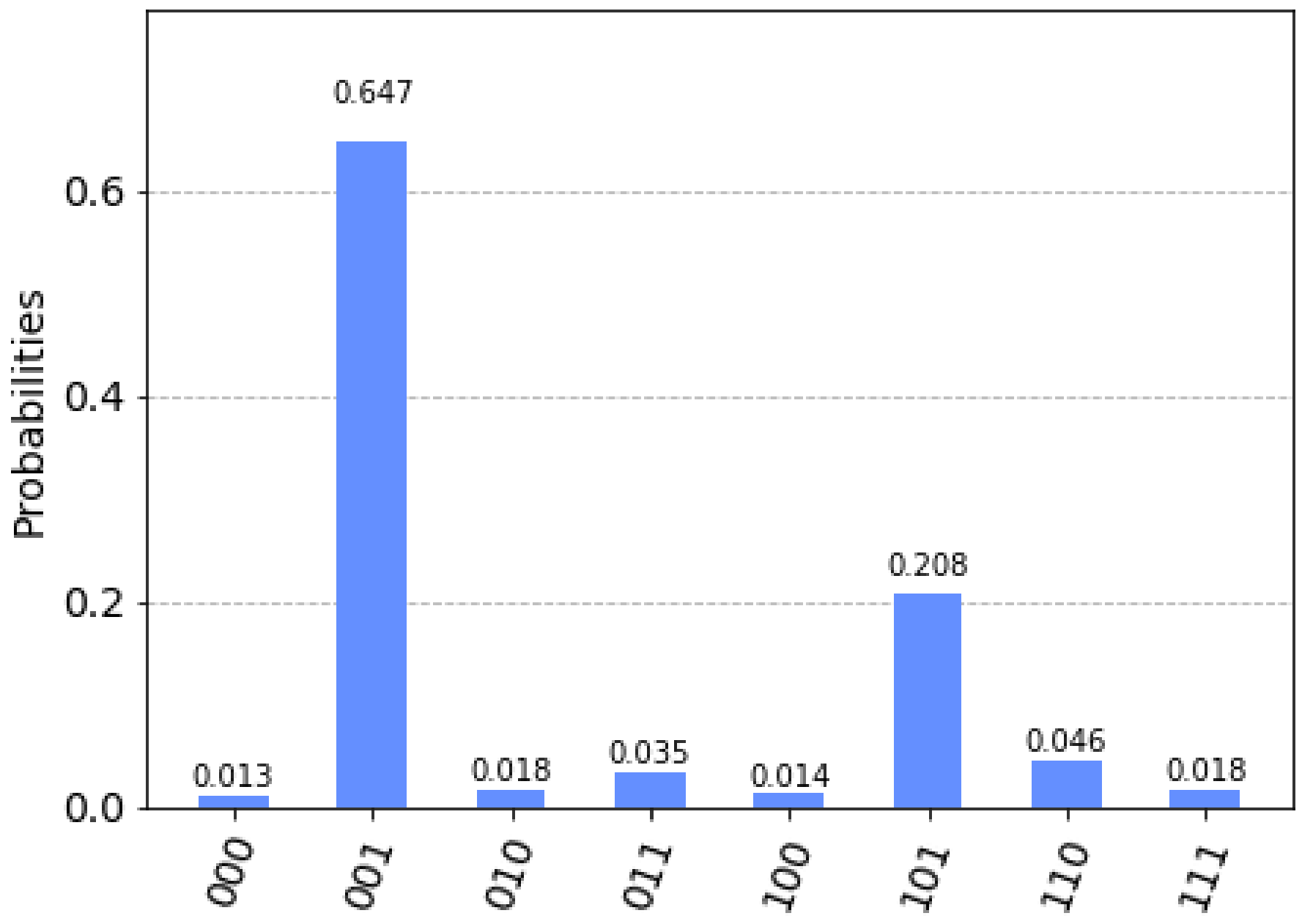

Take a look at the results of our code above, and confirm for yourself that our QPE circuit is indeed picking out the correct combined $\theta$. In particular, note the way in which the sum of our phases actually yields $\theta = 1.545$, but in our final answer our results show a successful QPE search for $0.545$. Mathematically this should make sense, as the two phases result in the same complex number:

$$e^{2i\pi\theta} \quad = \quad e^{2i\pi n\theta} \qquad\qquad n \in \left[\text{integers}\right]$$

Returning to our example then, we can see that for the combined unknown $\theta$ of $0.545$ our QPE has boosted the probability of the states $|\,100\,\rangle$ and $|\,101\,\rangle$. In particular, we see a larger probability on $|\,100\,\rangle$ over $|\,101\,\rangle$, which is in agreement with our $\theta$ being closer to $0.5$ than $0.625$. Conceptually this coding example doesn't offer anything new beyond the previous discussions, but the takeaway here is the way in which we are able to demonstrate a higher order QPE. If you haven't already, I encourage you to play around with the code above, changing the values of $n1$ & $n2$ and seeing the various final states that arise. For our two-dimensional example, it's worth while to see the full quantum circuit, in particular the way in which each control-$U^{2^n}$ operator is implemented:

```
In [ ]: print(qc)
```

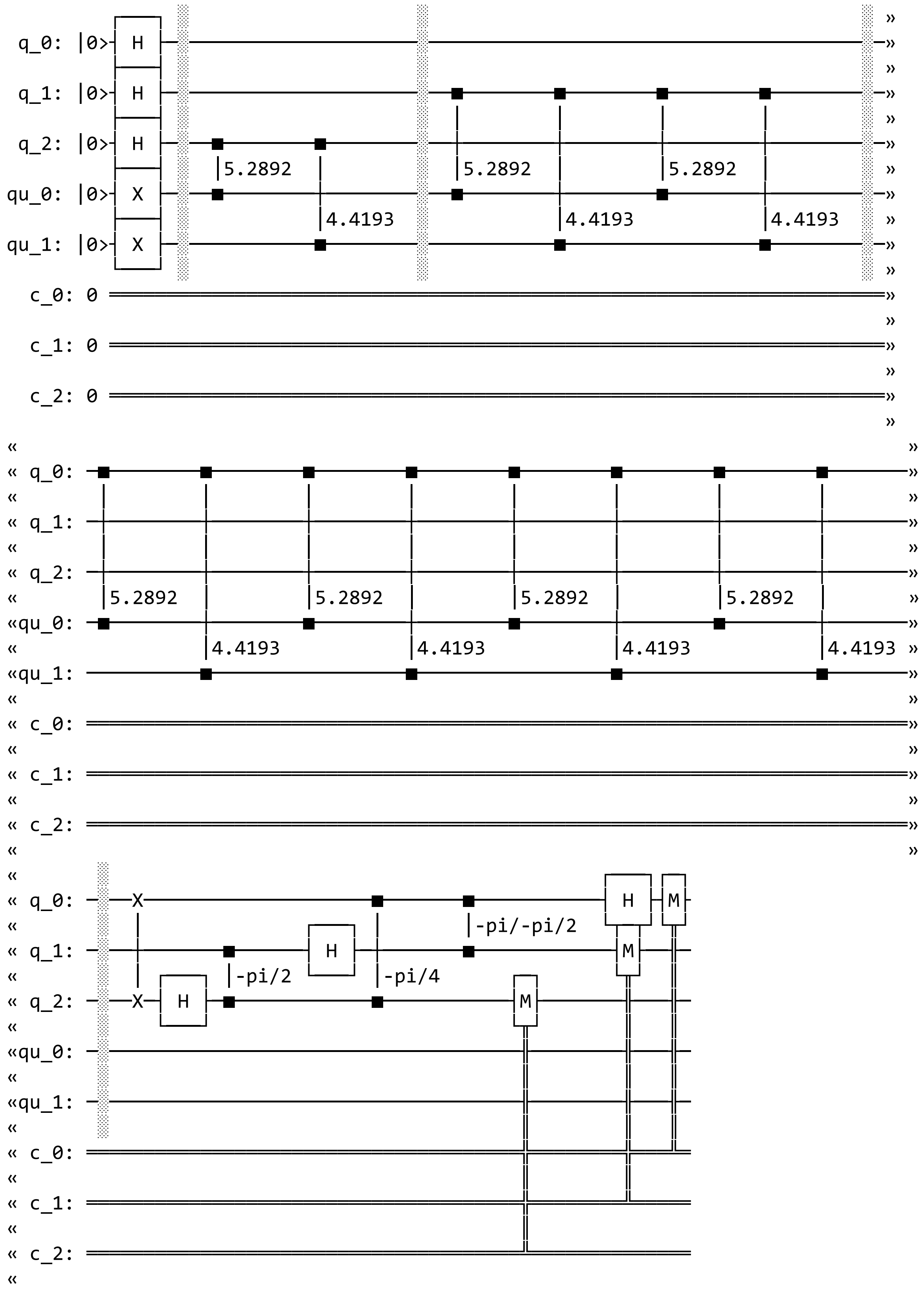

In the circuit diagram above, we've used the **barrier()** function to separate out each control-$U^{2^n}$ operator. As you can see, a single implementation of our control gate is achieved through $R_{\phi_1}$ and $R_{\phi_2}$, acting on their respective target qubits in $|u\rangle$. Then, each successive control-$U^{2^n}$ is simply a repetition of these two gates $2^n$ times.

With the code for implementing a random phase above now complete, our final example will look to implement the $\theta$ approximation technique from earlier. Specifically, we will take our measurement results and attempt to reconstruct $\theta$ using the $|\langle m|\Psi\rangle_{\text{final}}|^2$ formula:

In [ ]:

```
n1 = 5
n2 = 3
phases = []
trials = 10000
for p in np.arange( n2 ):
    phases.append( round( 2*m.pi*random.random(),4 ) )
theta = round( sum(phases)/(2*m.pi),5)
if( theta > 1 ):
    theta = round( theta - m.floor(theta),5)
#==================================================================    QPE Circuit
q  = QuantumRegister(n1,name='q')
qu = QuantumRegister(n2,name='qu')
c  = ClassicalRegister(n1,name='c')
qc = QuantumCircuit(q,qu,c,name='qc')
#-------------------------------------
for i in np.arange(n1):
    qc.h( q[int(i)] )
for i2 in np.arange(n2):
    qc.x( qu[int(i2)] )
for j in np.arange(n1):
    for j2 in np.arange(2**j):
        for j3 in np.arange(n2):
            qc.cu1( phases[int(j3)], q[int(n1-1-j)], qu[int(j3)] )
oq.QFT_dgr( qc,q,n1,swap=True )
#==================================================================
print('Phases:    ',phases,'     Combined \u03B8:   ',theta)
qc.measure(q,c)
M = oq.Measurement( qc,systems=[n1,n2],shots=trials,return_M=True,print_M=False )
MP = oq.Most_Probable(M,2)
for mp in np.arange(2):
    MP[0][mp] = MP[0][mp]/trials
phi,theta = oq.QPE_phi(MP)
print('\nMost Probable State:   |'+str(MP[1][0])+'>      Probability:  ',round(MP[0][0],5))
print('\nSecond Most Probable:  |'+str(MP[1][1])+'>      Probability:  ',round(MP[0][1],5))
print('\nCorresponding \u03A6:  ',phi,'\n\nApproximate \u03B8:  ',round(theta,5))
```

```
Phases:     [2.2115, 4.5131, 1.2023]      Combined θ:    0.26161

Most Probable State:   |01000>      Probability:   0.6221

Second Most Probable:  |01001>      Probability:   0.2121

Corresponding Φ:    0.371

Approximate θ:   0.26159
```

And there we have it, our complete QPE example using the function $\text{QPE_phi}()$ to approximate $\theta$ using repeat measurements! In the example shown above we've elected to use $5$ qubits for accuracy, and $3$ for our eigenstate $|\,u\,\rangle$. There's certainly a lot to unpack in this final example, so I encourage you to play around with it, dissecting it as needed to really get a strong understanding of QPE.

Like we mentioned at the top of this section, the ideal role for QPE would be to help find an eigenvalue of some unknown $U$. However, in order to do this, QPE requires that we:

$1)$ Have a means of physicalling implementing $U$ as a control operator in our quantum circuit

$2)$ Know the eigenstate $|\,u\,\rangle$ corresponding to the eigenvalue we're interested in

$3)$ Have a means for constructing $|\,u\,\rangle$ (arbitrary state preparation is no easy task in quantum computing)

These three reasons are why we put off creating the quantum code for QPE to the very end of this lesson. Our goal in this tutorial is to understand the fundamentals that make up QPE, which are circuit implementation agnostic. And now that we have these fundamentals, the real challenge lies in applying them to real world problems where QPE is viable.

---

This concludes lesson 7, and our first encounter with the Quantum Phase Estimation Algorithm. In this lesson we've seen all of the core elements that make up this powerful subroutine, as well as several ways to interpret the results it gives us. We will encounter this algorithm once again in the coming lesson, which will further reveal some of the finer details to QPE.

I hope you enjoyed this lesson, and I encourage you to take a look at my other .ipynb tutorials!

### Citations

**[1]** A. Y. Kitaev, "Quantum measurements and the Abelian Stabilizer Problem", arXiv:9511026 (1995)

# Lesson 7.1 - Quantum Counting

Having just studied Quantum Phase Estimation in lesson 7, we will now be looking at an algorithm that will help illustrate the power of the QPE. Developed by Gilles Brassard, Peter Hoyer, and Alain Tapp in 1998 **[1]**, the Quantum Counting Algorithm covered in this lesson will use the structure of QPE in combination with Grover's Algorithm to solve a problem that seeks to learn the total number of marked entries in a given list. As we shall see, this algorithm will shed new light on both the underlying mathematical nature of Grover's, as well as the power of QPE as a subroutine.

In order to make sure that all cells of code run properly throughout this lesson, please run the following cell of code below:

In [ ]:
```
from qiskit import ClassicalRegister, QuantumRegister, QuantumCircuit, Aer, execute
import Our_Qiskit_Functions as oq
import numpy as np
import math as m
import random
```

## Problem Overview

Let's begin by outlining the problem which we intend to solve using the Quantum Counting Algorithm: given a set $N$, we would like to know how many elements in $N$ belong to the subset of solutions $M$. To help us out, we are also given a function $f$ such that if $f(x) = 1$, $x$ is a part of $M$, otherwise $f(x) = 0$. For example:

$$N = \{2, 5, 8, 12, 13, 17, 21, 24, 30, 31, 32, 39\}$$

$$M = \{5, 12, 17, 21\}$$

Using our function $f(x)$, we test all of the elements in $N$ and count how many times we get $f(x) = 1$:

$$f(2) = 0 \qquad f(5) = 1 \qquad f(8) = 0 \quad \cdots$$

$$\text{solution:} \quad 4$$

The example above illustrates the exhaustive process of checking the $f$ function with every value in $N$, after which we conclude that the numbers $\left[5, 12, 17, 21\right]$ make up the solution to our problem. Below is an example of how a classical code would tackle the problem in this way:

In [ ]:
```
def f(x):
    if( (x==5) or (x==12) or (x==17) or (x==21) ):
        return 1
    else:
        return 0

#====================================

N = [2,5,8,12,13,17,21,24,30,31,32,39]
count = 0
M = []
#--------------------------
for i in np.arange(len(N)):
    if( f(N[i])==1 ):
        count = count + 1
        M.append( N[i] )
print('Solution:    ',count,'\n       M:    ',M)
```

```
Solution:    4
       M:    [5, 12, 17, 21]
```

The cell of code above arrives at the same result as our example, whereby the function $f(x)$ is hardcoded with all of the solutions. If we now consider the speed at which this classical approach arrives at the answer, we can see that the code must check every value in $N$ individually. In general, if $M$ isn't a set that has some underlying pattern or structure to it (for example, $M$ is all even numbers), then we are stuck searching through the entire set $N$. Thus, the classical approach to our counting problem scales like the order $O(N)$.

## An Extension to the Grover Problem

If we think back to some of the algorithms we have already covered in past lessons, this counting problem may feel a bit familiar. For example, consider what our problem would look like if we were told that the set $M$ only has one value in it. Then, our problem would be to try and find the single value in the set $N$ which satisfies our function $f(x) = 1$. Or in other words, our searching problem would simply be Grover's Algorithm! And as we already mentioned, Grover's Algorithm will have a critical role to play in the Quantum Counting Algorithm due to the similarity of the problems they solve.

As a first pass at solving our counting problem, a natural start would be to ask whether we could simply mimic Grover's Algorithm, but for multiple marked states. Specifically, the result of Grover's was a boost in probability to the desired state, so we should ask whether or not the same solution could work for the case of multiple marked states. If we are able to produce a final quantum state that boosts the probability of multiple marked states, could we extract any meaningful information through a measurement, or multiple measurements? Let's see what happens when we try marking three states in a $3$-qubit Grover circuit:

In [ ]:
```
marked = ['010','011','110']
Q = len(marked[0])
iters = 1
#========================================
q   = QuantumRegister(Q,name='q')
a1 = QuantumRegister(1,name='a1')
a2 = QuantumRegister(Q-2,name='a2')
qc  = QuantumCircuit(q,a1,a2,name='qc')
#-----------------------------------------
for j in np.arange(Q):
    qc.h( q[int(j)] )
qc.x( a1[0] )

for i in np.arange( iters ):
    for j in np.arange(len(marked)):
        M = list(marked[j])
        for k in np.arange(len(M)):
            if(M[k]=='1'):
                M[k] = 1
            else:
                M[k] = 0
        oq.Grover_Oracle(M, qc, q, a1, a2)
    oq.Grover_Diffusion(M, qc, q, a1, a2)

oq.Wavefunction(qc,systems=[Q,Q-2,1],show_systems=[True,False,False])
```

```
0.17678 |000>    0.17678 |100>    -0.53033 |010>    -0.53033 |110>    0.17678 |001>    0.17678 |101>    -0.53033 |011>
0.17678 |111>
```

In the example above, we have used the functions **Grover_Oracle** and **Grover_Diffusion** from Our_Qiskit_Functions (which we created in lesson 5.4) in order to handle all of the gate operations necessary to perform one Grover iteration of multiple marked states. In future coding examples we will combine these steps even further in a function called **Multi_Grover**, but for now it is worth while to see the full code. As a quick reminder to some of the structure for a single Grover iteration, note in the example our three quantum registers: $q$, $a_1$, and $a_2$. The first register $q$ holds the qubits which make up the final boosted quantum state, shown in the printed wavefunction. The two additional qubit registers, $a_1$ and $a_2$, represent the ancilla qubits needed to apply the Oracle and Diffusion operators. Specifically, $a_1$ holds the qubit which stays in the $|-\rangle$ state throughout the algorithm, while $a_2$ holds all of the additional ancilla qubits neccessary for the higher order CNOT operations (see lesson 5.4 for additional details on constructing $N$-dimensional Grover circuits).

Conceptually, there is no fundamental difference between the normal Grover's Algorithm, and the one shown above which boosts multiple marked states:

$1$) Use control gates to flip the phase on each marked state individually

$2$) Perform a reflection about the average (flip the phase on the state of all $0$'s).

As we can in the final wavefunction, the result of marking three states caues a boost in probability to each of them, equally in fact. However, if we now consider what a final measurement on this quantum state can tell us, it is still unclear whether or not these increases in amplitude are enough to be useful. But before jumping to make any measurements, we should check whether any further Grover iterations will yield a more desirable final state. After all, the optimal number of steps for a standard Grover's Algorithm is of the order $O(\sqrt{N})$. Let's see what happens to our quantum state after a second Grover iteration:

In [ ]:
```
marked = ['010','011','110']
Q = len(marked[0])
iters = 2
#========================================
q   = QuantumRegister(Q,name='q')
a1 = QuantumRegister(1,name='a1')
a2 = QuantumRegister(Q-2,name='a2')
qc  = QuantumCircuit(q,a1,a2,name='qc')
#-----------------------------------------
for j in np.arange(Q):
    qc.h( q[int(j)] )
qc.x( a1[0] )

qc,q,a1,a2 = oq.Multi_Grover(q,a1,a2,qc,marked,iters)

oq.Wavefunction(qc,systems=[Q,Q-2,1],show_systems=[True,False,False])
```

```
-0.44194 |000>    -0.44194 |100>    -0.08839 |010>    -0.08839 |110>    -0.44194 |001>    -0.44194 |101>    -0.08839 |011>
-0.44194 |111>
```

Taking a looked at our marked states in the code above, we can see that a second iteration has resulted in a dramatic decrease in probabilities. Therefore, for this particular example of three marked states, one Grover iteration is the optimal, giving rise to the following probability distribution:

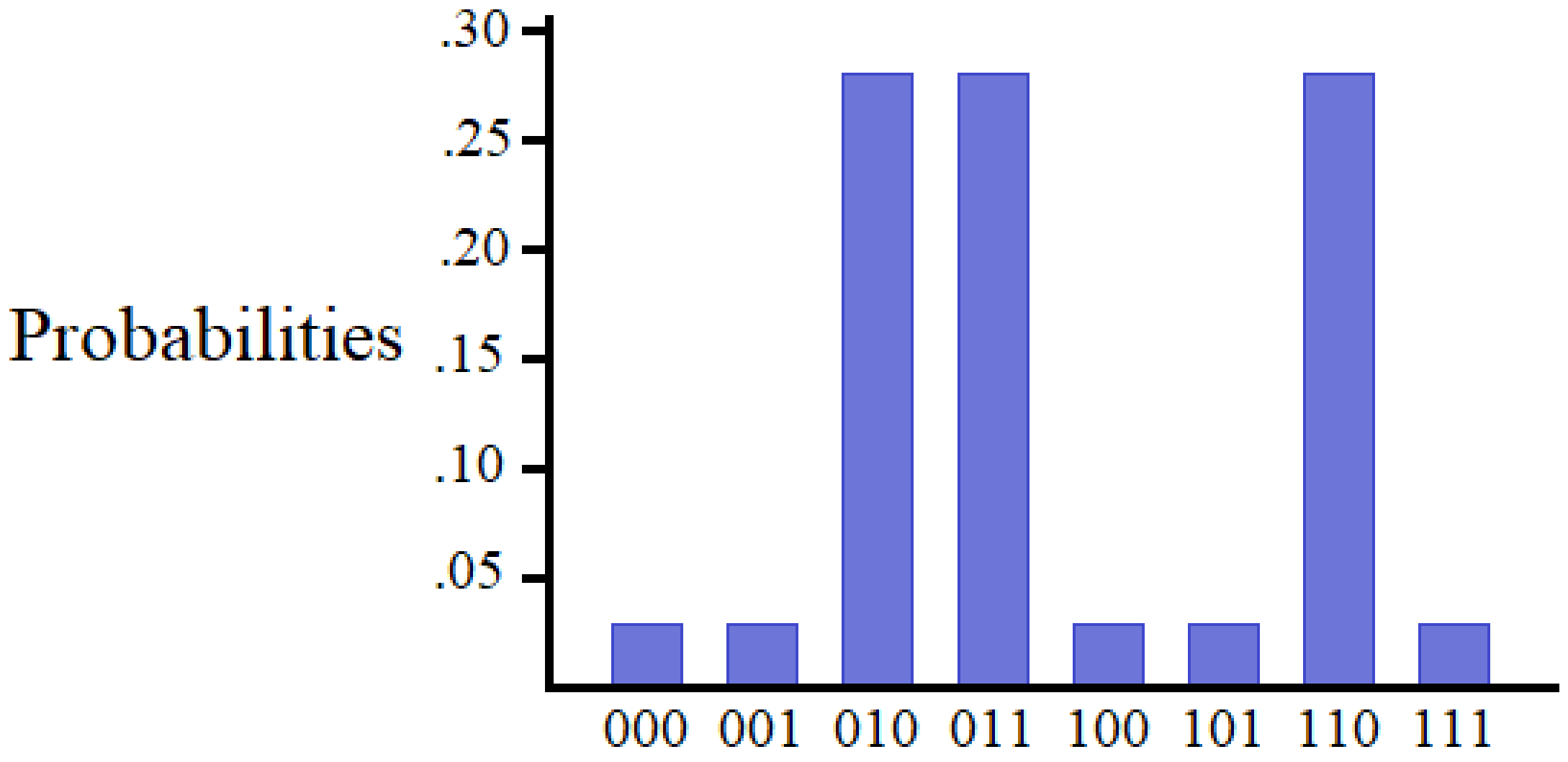

For our $M = 3$ example, our optimal state is one in which our marked states all possess a probability of roughly $28\%$, which means that a measurement on this system has a cumulative probability of just over $84\%$ of measuring one of the marked states. But now we must ask whether or not this probability distribution can provide us with any sort of speedup. For this particular case of three marked states in a space of eight, the answer is no. The bottleneck preventing a speedup comes from considering how many times we need to repeat the quantum circuit above in order to reveal all three desired states. Taking into account the number of steps needed for each preparation, along with the probabilities of failure / repeat measurements, there is simply no way our quantum computer can compete with the classical algorithm which arrives at the correct answer in at most $8$ checks.

But just like how the case of $N = 3$ isn't a speedup for the regular Grover's Algorithm, we should investigate cases of larger $N$ for potential speedups. To answer this, let's imagine a case where $N$ is sufficiently large, and consequently all of the states that make up $M$ contain nearly $100\%$ of the total probability in the system:

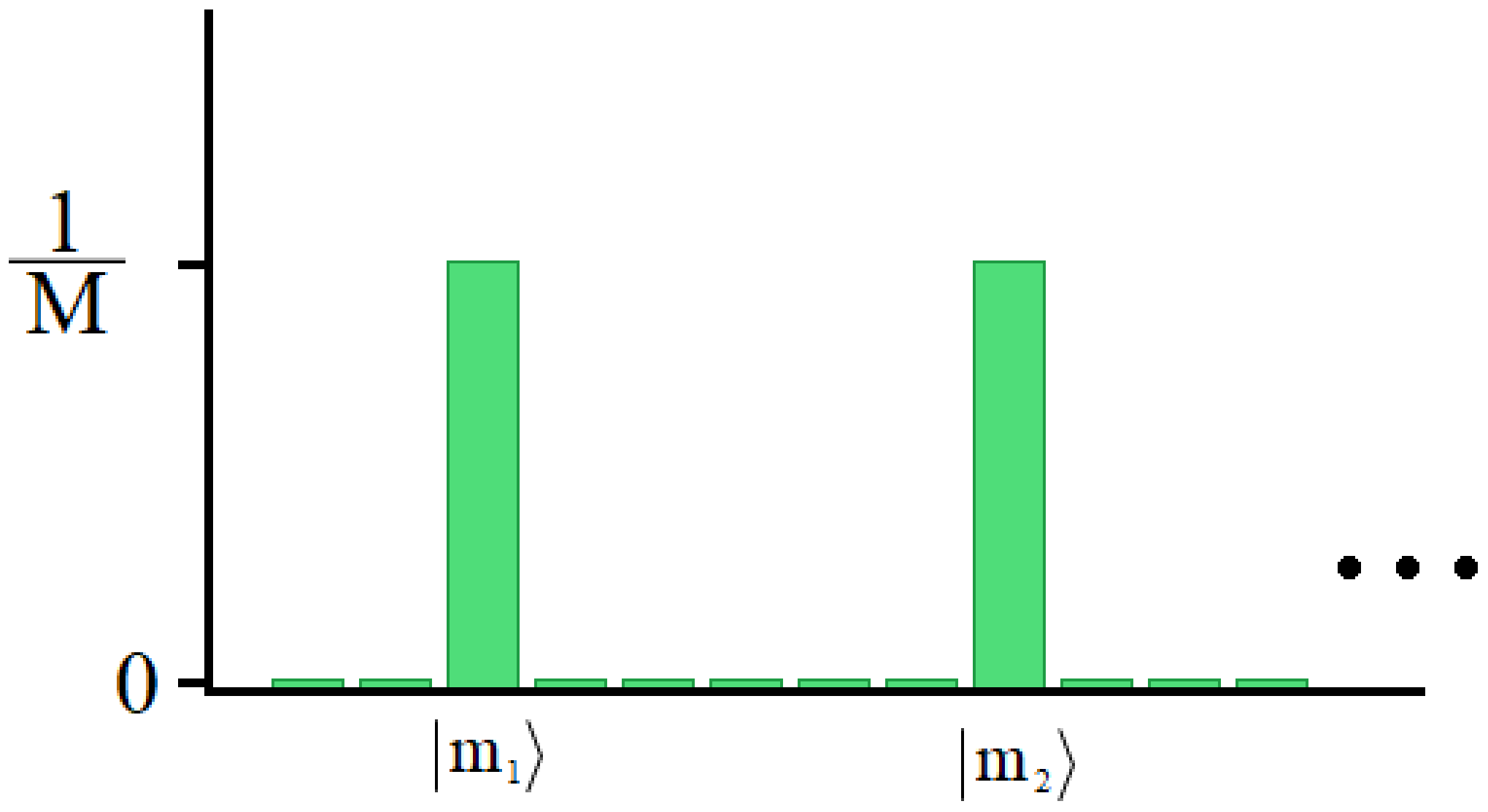

As a close example to the figure above, the cell of code below demonstrates the case for $5$ qubits (a space of $32$ possible states), with three marked states. After the optimal number of iterations, which turns out to be $2$, we reach a system where approximately $99.98\%$ of the total probability in the system is concentrated on the three marked states:

In [ ]:

```
marked = ['01010','01100','00101']
Q = len(marked[0])
iters = 2
#========================================
q   = QuantumRegister(Q,name='q')
a1  = QuantumRegister(1,name='a1')
a2  = QuantumRegister(Q-2,name='a2')
qc  = QuantumCircuit(q,a1,a2,name='qc')
#----------------------------------------
for j in np.arange(Q):
    qc.h( q[int(j)] )
qc.x( a1[0] )

qc,q,a1,a2 = oq.Multi_Grover(q,a1,a2,qc,marked,iters)

oq.Wavefunction(qc,systems=[Q,Q-2,1],show_systems=[True,False,False],column=True)
```

```
0.00276 |00000>     0.00276 |10000>
0.00276 |01000>     0.00276 |11000>
0.00276 |00100>     0.00276 |10100>
0.57729 |01100>     0.00276 |11100>
0.00276 |00010>     0.00276 |10010>
0.57729 |01010>     0.00276 |11010>
0.00276 |00110>     0.00276 |10110>
0.00276 |01110>     0.00276 |11110>
0.00276 |00001>     0.00276 |10001>
0.00276 |01001>     0.00276 |11001>
0.57729 |00101>     0.00276 |10101>
0.00276 |01101>     0.00276 |11101>
0.00276 |00011>     0.00276 |10011>
0.00276 |01011>     0.00276 |11011>
0.00276 |00111>     0.00276 |10111>
0.00276 |01111>     0.00276 |11111>
```

To see whether or not we can obtain a speedup on a system like the one shown above, we will consider two methodologies for determining the number of marked states $M$ In order to properly talk about speedups then, we need to know the optimal number of iterations to prepare our system in order to compare with the classical approach. We will derive this result later in this lesson, but for now we will simply give an approximate scaling:

$$\text{Grover iterations} \quad \approx \quad \frac{\pi}{4\sin^{-1}\left(\sqrt{\frac{M}{N}}\right)}$$

As expected, the number of iterations is dependent on both $M$ and $N$. And although it may look a bit different from our normal $O(\sqrt{N})$ scaling for Grover's if we substitute $M = 1$, as we shall see later on, this approximation is equally valid for the standard Grover's Algorithm as well.

### Grover's With Many Marked States

As the simplest approach to using our final quantum state as a means for determining $M$, let's examine the strategy of simply preparing and measuring the quantum system until we reveal all of the marked states. As you might suspect, there are several issues with this strategy, but for now let's see what an example would like, using our $N = 5$, $M = 3$ example:

$$\text{states:} \quad |m_1\rangle, |m_2\rangle, |m_3\rangle$$

$$1^{st} \ \text{ sample:} \ \ 99.9\% \text{ probability of measuring } \ |m_1\rangle$$

$$2^{nd} \ \text{ sample:} \ \ 66.6\% \text{ probability of measuring } \ |m_2\rangle$$

$$3^{rd} \ \text{ sample:} \ \ 33.3\% \text{ probability of measuring } \ |m_3\rangle$$

The probabilities shown above represent our chances of measuring *new* marked states after each subsequent successful measurement. Essentially, our first measurement is near guaranteed to reveal one of the three marked states, call it $|m_1\rangle$. But each subsequent measurement will have a $33.3\%$ chance of yielding $|m_1\rangle$ again, which results in wasted steps for our algorithm. The individual probabilities on each state never change, always remaining $33.3\%$ for each $|m_i\rangle$, but the odds of measuring a new marked state become probabilistically less and less, which in turn requires more measurements.

If we now extend the example above to some arbitrary $M$, we can imagine how the number of marked states directly slows down our algorithm. Not only will the final state take the longest to find, with a probability of $1/M$, but finding each previous marked state will also take a proportionally significant amount of repeat measurements as well. For example, consider the case for $M = 5$, once again assuming near $100\%$ probability accumulated on the marked states:

| Marked State: | 1 | 2 | 3 | 4 | 5 |
|---|---|---|---|---|---|
| Average Measurements to find: | 1 | $\frac{1}{0.8}$ | $\frac{1}{0.6}$ | $\frac{1}{0.4}$ | $\frac{1}{0.2}$ |
| Total Average: | | | 11.417 | | |

Just to confirm that this indeed the average number of times we would expect to prepare and measure our quantum system, below is a short code simulating our problem of measuring the five unique marked states:

In [ ]:

```
runs = 100000
Avgs = [0,0,0,0,0]
#======================================
for r in np.arange(runs):
    Measured = []
    Trials = []
    t = 0
    while( len(Measured) != 5 ):
        Measurement = round( 0.500000001+ random.random()*5 )
        t = t + 1
        new_state = True
        for i in np.arange( len(Measured) ):
            if( Measurement == Measured[i] ):
                new_state = False
        if(new_state == True):
            Measured.append(Measurement)
            Trials.append(t)
    for i in np.arange(5):
        Avgs[i] = Avgs[i] + Trials[i]
#======================================
for j in np.arange(5):
    Avgs[j] = round(Avgs[j]/runs,3)

print('Average Trials to Find Each State:    \n\n1) ',Avgs[0],'    2) ',Avgs[1],'   3)',Avgs[2],'   4) ',Avgs[3],'   5) ',Avgs[
```

```
Average Trials to Find Each State:

1)  1.0    2)  2.247   3) 3.913   4)  6.409   5)  11.402
```

In the example above, we simulate making measurements until we find all $5$ marked states, a total of $100,000$ times, giving us a sufficiently close approximation to the average number of measurements needed to find each $|m_i\rangle$. And as we can see, the numbers are in agreement with our theoretical predictions from earlier. One interesting thing to note, is that the number of average measurements here is only dependent on $M$. This is because we've assumed $N$ to be sufficiently large, resulting in nearly 100% of the total probability in the system being distributed amongst our marked states. Under this assumption, we can compute the average number of measurements needed to find $M$ marked states as:

$$\sum_{k=0}^{M-1} \frac{M}{M-k}$$

Below is a visual representation of the summation above, showing how the number of average measurements increases as a function of $M$:

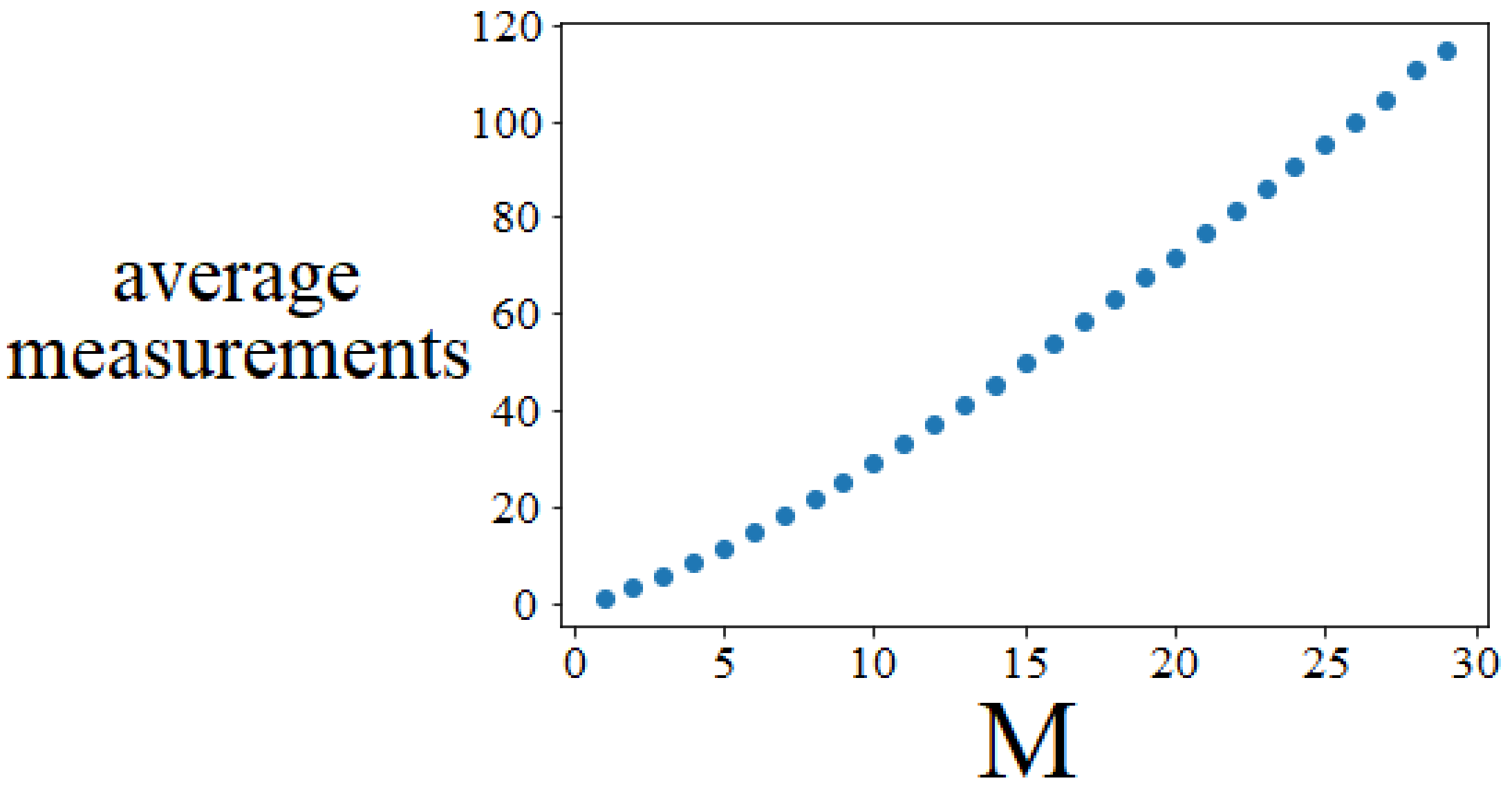

The plot above shows the steadily increasing rate at which we must keep sampling our quantum system in order to reveal the total number of marked states. Using $M = 20$ as an example, the average number of measurements will be $72$, which in turn means that the average speed of our quantum approach will be $72$ times the number of steps necessary to prepare our final quantum state. But as we showed earlier, the optimal number of Grover iterations also depends on $N$, which in turn means that in order to compare with the classical solution we need to compare the speed of both approaches as a function of $N$:

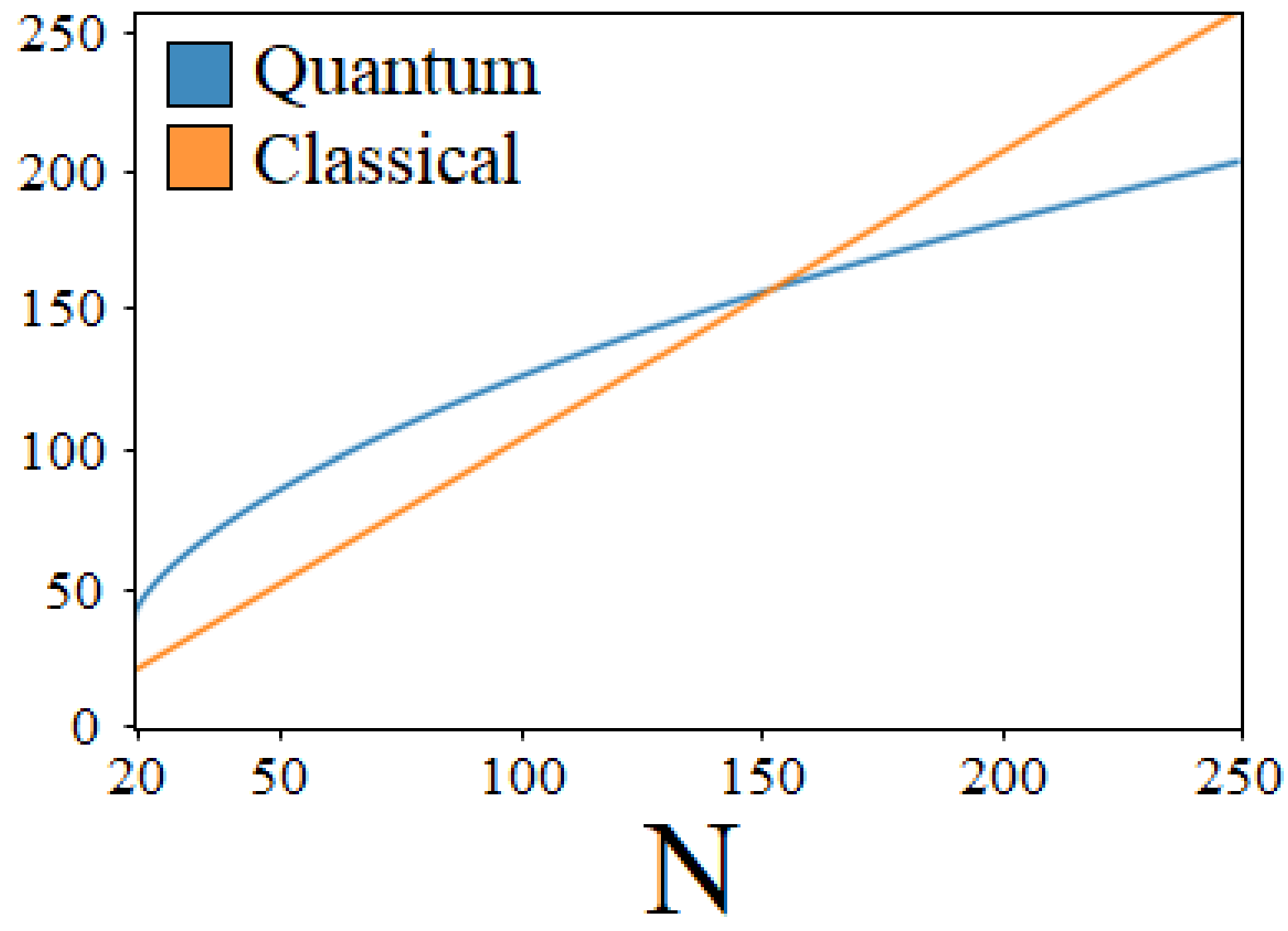

Plotted above is a comparison of classical vs quantum average solving speed for the case of $M = 20$. Shown in the figure as the location where the blue and orange plots overlap, we now have our answer as to the value of $N$ where we expect a quantum speedup: $154$. Thus, if our problem is to find $20$ marked elements out of a list of $154$ or more elements, our trial and error quantum approach should on average beat out the exhaustive classical technique.

Now, if the number $N = 154$ seemed surprisingly low to you, especially considering that our average number of measurements is expected to be $72$, your suspicions are well warranted. To be quite honest, the plot above is a bit like comparing apples and oranges. The main issue here is that we are comparing classical steps, checking an element in $N$ and seeing if it also exists in $M$, to quantum Grover iterations. But as we know, one Grover iteration can be a massively long and complex circuit depending on the size of $N$ and $M$. Conversely, our classical checking algorithm doesn't fundamentally change as we increase $N$ (and arguably $M$ as well if we are using an oracle function $f(x)$).

As a second major concern, just as with the standard Grover's Algorithm, a key piece of information needed for the quantum approach is the size of the problem: $N$ & $M$. But if we consider the more realistic scenario of our problem, we don't know $M$ (remember that the goal is to find out how many elements are shared between $N$ and $M$). In the standard Grover's Algorithm, this problem is averted because we have $M = 1$, thus only requiring information about the size of $N$ to determine the number of necessary Grover iterations. But if we now don't know the size of $M$, our multiple marked state Grover approach is essentially searching blind. If we relax this condition, and change the nature of the problem we are trying to solve:

"We know how many elements in $M$ are also in $N$, and the problem is to learn what these elements are."

then our quantum approach is viable. But this new problem is a bit contrived, and relies on the classical code not having access to $M$ either, only the oracle function $f(x)$ that tells us if an element exists in $M$. In light of these points brought up when comparing classical to quantum algorithms, the truth is that there is no concrete means for appropriately comparing the two. And while the topic is certainly rich with potential for discussion, it's outside our interests for this tutorial. Instead, we will continue on with the Quantum Counting Algorithm, and begin the main focus of this lesson.

## Quantum Counting Approach

Having just seen why using a multi-Grover approach can be problematic in solving for $M$, we will now turn our attention to the Quantum Counting Algorithm approach. At its core, this algorithm is essentially a combination of Quantum Phase Estimation and Grover's Algorithm. This combination may seem a bit odd at first, considering that in the previous lesson we discussed at length how the QPE circuit is designed for unitary operations, while Grover's Algorithm is a full blown circuit. We will return to this important idea later in the lesson, first electing to focus on the structure of the algorithm and the expected result first. The quantum circuit for this algorithm is shown below, and should look very familiar:

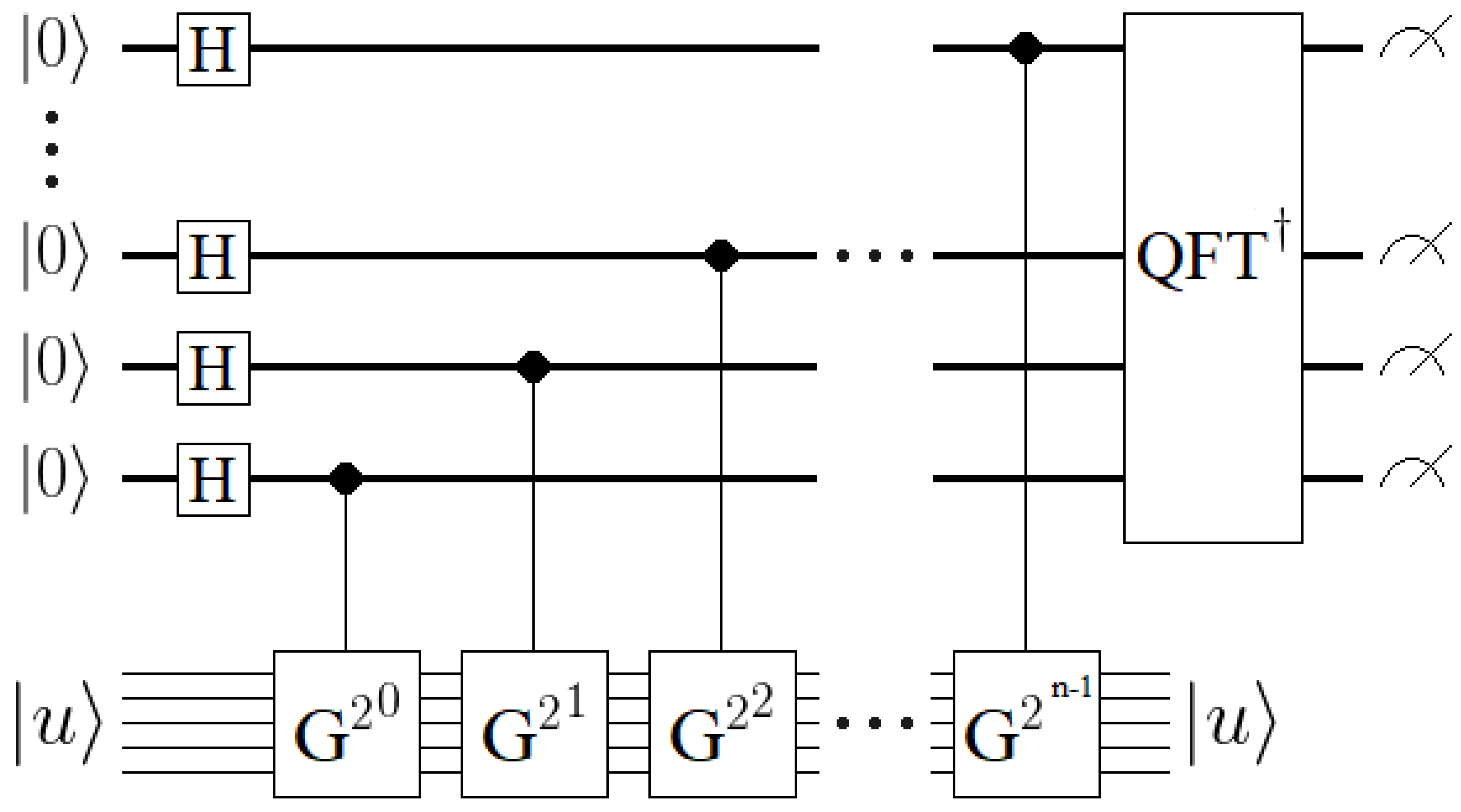

The circuit above is identical in every way to the QPE circuit studied in lesson 7, except that we have specified our unitary operator $U$ as the Grover operator $G$. For clarity, when we refer to the Grover operator here, what me mean is one full application of a single Grover iteration applied to the eigenstate $|u\rangle$. This operator $G$ is mathematically equivalent to all of the gate operations necessary to apply the Oracle and Diffusion operations making up one Grover iteration. Thus, for the purpose of understanding the QPE circuit shown above, the unitary operator representation of Grover's should be thought of as the combination of all of these gate operations into a single unitary matrix. Then, imagining we have this Grover operator $G$ available to us, we simply use it as we would any other $U$ in our QPE circuit.

As we know from our previous lessons up to this point, implementing arbitrary unitary matrices as operators is something not easily achievable (except when using Qiskit's simulator), and certainly something that we should consider out of the question if we want to run our algorithm on a real physical device. Thus, although it is convenient to talk about applying Grover's as a unitary operator $G$ in the QPE circuit, realizing such an operation is the true challenge. Thus, our first task will be to investigate the viability of implementing Grover's Algorithm as a control-$U$ operation.

## Constructing a Control-Grover Operation

For our Quantum Counting Algorithm, implementing Grover's is not quite as straightforward as simply applying some gate operation '$G$' to our qubits, it is an entire quantum circuit in itself. Mathematically, the ideal route would be to compile all of the individual unitary operations that make up the Grover circuit and express the full effect of the circuit as a single $2^N \times 2^N$ matrix operation. Then, with the mathematically equivalent matrix in hand, we would simply find the optimal number of gates to achieve this operation. While technically viable, we will instead be implementing our control-$G$ operators as adaptations of the standard Grover circuit, replacing every gate in the algorithm with a control version of the same gate. For example, cases where we require $X$ and $H$ gates for our Grover circuit will now become control-$X$ and control-$H$ gates:

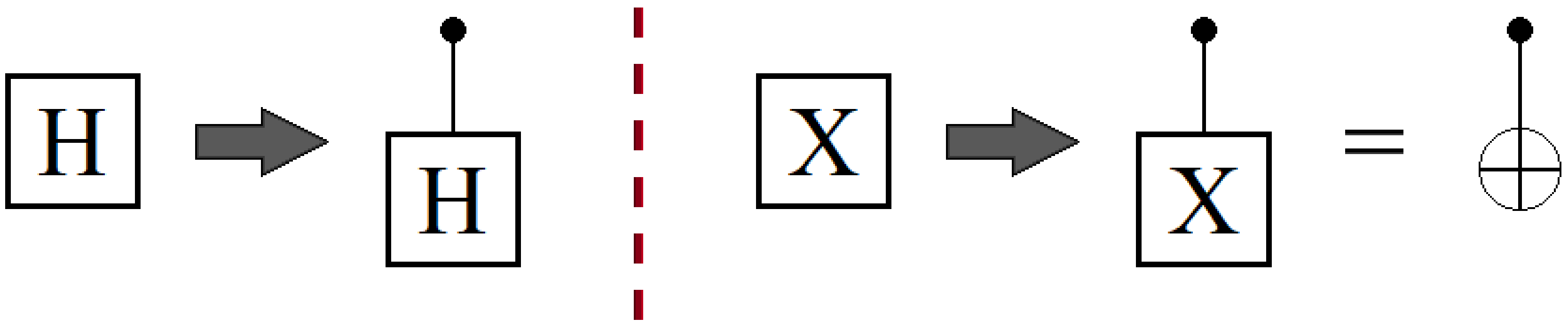

The control-$X$ gate should look familiar, as it is just the CNOT gate we've grown accustomed to. Conversely however, the control-Hadamard gate is something new to these lessons, and not necessarily as straightforward as one may think. If we ask what are the outputs from this new control gate, we would expect the following:

$$
\begin{aligned}
CH|00\rangle &= |00\rangle \\
CH|01\rangle &= |01\rangle \\
CH|10\rangle &= |1\rangle \otimes \frac{1}{\sqrt{2}}\big(|0\rangle + |1\rangle\big) \\
CH|11\rangle &= |1\rangle \otimes \frac{1}{\sqrt{2}}\big(|0\rangle - |1\rangle\big)
\end{aligned}
$$

which can be represented in matrix form as:

$$
\begin{bmatrix}
1 & 0 & 0 & 0 \\
0 & 1 & 0 & 0 \\
0 & 0 & \frac{1}{\sqrt{2}} & \frac{1}{\sqrt{2}} \\
0 & 0 & \frac{1}{\sqrt{2}} & -\frac{1}{\sqrt{2}}
\end{bmatrix}
$$

It can be shown with relative ease that the control-$H$ matrix above is indeed unitary, which means that it is a valid quantum operation. Essentially the control-$H$ gate works as you might expect, putting the target qubit in a $50 - 50$ superposition state when the control qubit is the $|1\rangle$ state, and nothing when the control qubit is $|0\rangle$. Let's see the four possible output states of the control-$H$, which comes as a standard gate in Qiskit:

In [ ]:
```
for i in np.arange(4):
    q = QuantumRegister(2,name='q')
    qc= QuantumCircuit(q,name='qc')

    if( (i == 1) or (i == 3) ):
        qc.x( q[1] )
    if( (i == 2) or (i == 3) ):
        qc.x( q[0] )

    print('\n___ Initial State ___')
    oq.Wavefunction(qc)

    qc.ch( q[0],q[1] )

    print('\n___ After Control-Hadamard ___')
    oq.Wavefunction(qc)
    if( i <= 2):
        print('\n------------------------------------')
```

```
___ Initial State ___
1.0 |00>

___ After Control-Hadamard ___
1.0 |00>

------------------------------------

___ Initial State ___
1.0 |01>

___ After Control-Hadamard ___
1.0 |01>

------------------------------------

___ Initial State ___
1.0 |10>

___ After Control-Hadamard ___
0.70711 |10>    0.70711 |11>

------------------------------------

___ Initial State ___
1.0 |11>

___ After Control-Hadamard ___
0.70711 |10>    -0.70711 |11>
```

The control-Hadamard gate turns out to be the only new operation needed for our construction of control-$G$. The remaining two gates that we need to upgrade with additional control qubits are the CNOT and CCNOT operations, both of which we have encountered several times before:

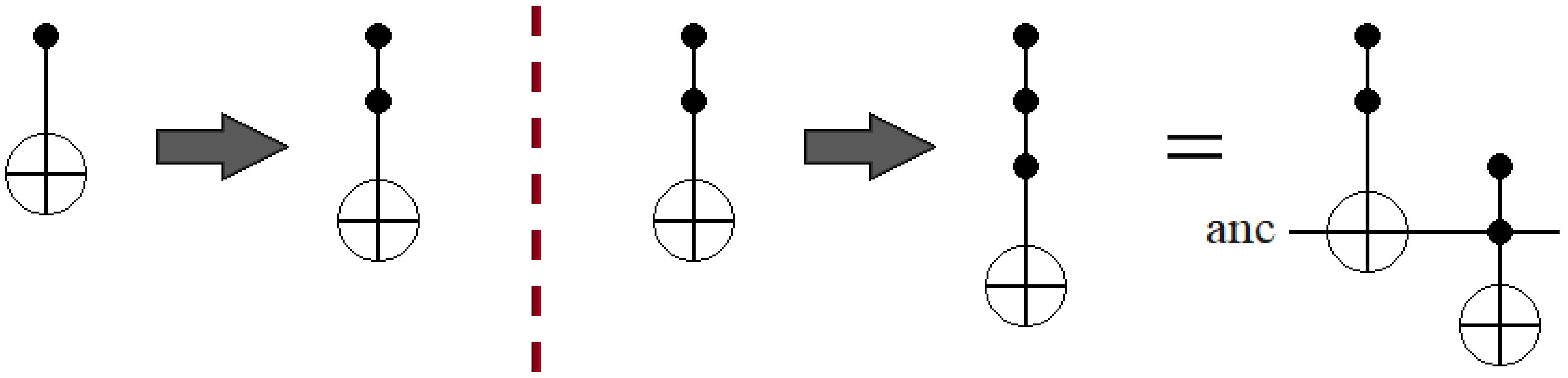

Easily enough, attaching an additional control qubit onto a CNOT gate simply transforms it into a CCNOT. And as for adding an additional control to CCNOT gates, we have already seen from past lessons that higher order control-not gates can be implemented using additional qubits, as shown above. After upgrading all of the $X$, $H$, and various orders of CNOT gates, overall the implementation of control-$G$ isn't really too bad (a bit costly perhaps in gates and qubits, but conceptually rather straightforward). As our first example, below are the circuit diagrams for a 2-qubit Grover circuit and its upgraded control-Grover, along with code examples showing the results of each circuit.

**Grover 2-qubit circuit:**

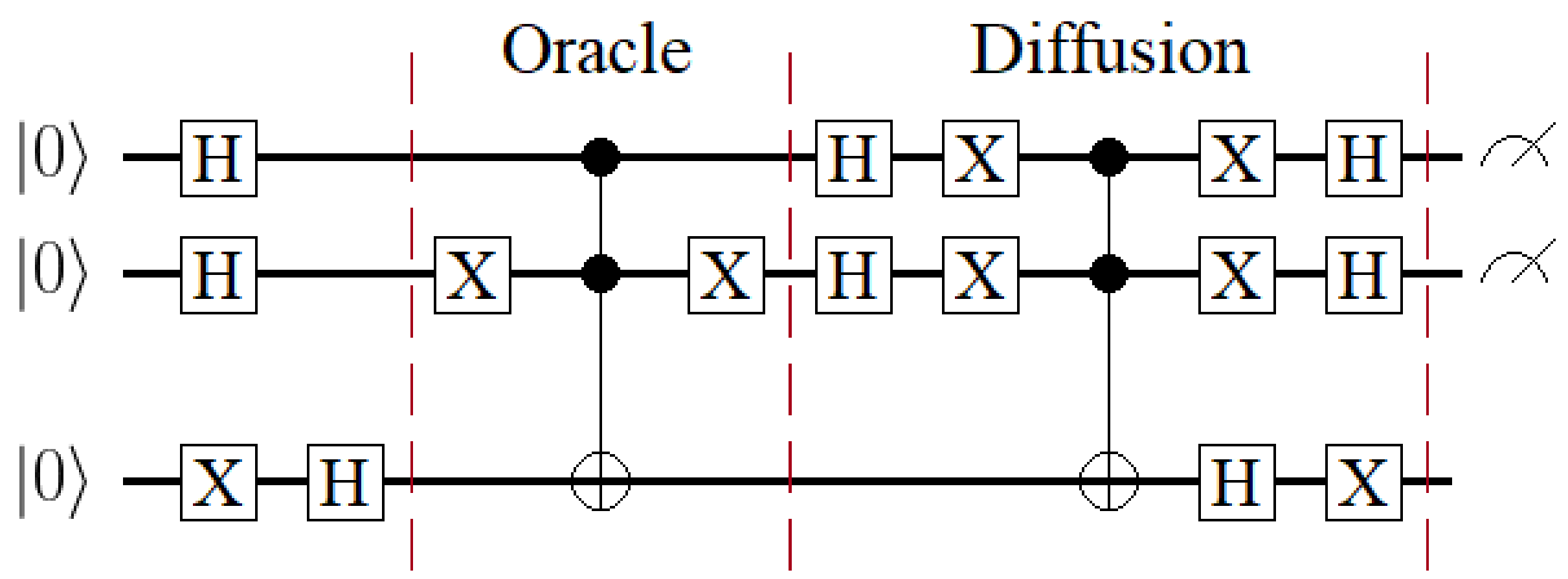

In [ ]:
```
q = QuantumRegister(2,name='q')
a = QuantumRegister(1,name='a')
qc= QuantumCircuit(q,a,name='qc')
#--------------------------------
qc.h( q[0] )
qc.h( q[1] )
print('__ Initial State __')
oq.Wavefunction(qc,systems=[2,1],show_systems=[True,False])
qc.x( a[0] )
qc.h( a[0] )
#-------------------------      Oracle
qc.x( q[1] )
qc.ccx( q[0],q[1],a[0] )
qc.x( q[1] )
#-------------------------      Diffusion
qc.h( q[0] )
qc.h( q[1] )
qc.x( q[0] )
qc.x( q[1] )
qc.ccx( q[0],q[1],a[0] )
qc.x( q[0] )
qc.x( q[1] )
qc.h( q[0] )
qc.h( q[1] )

qc.h( a[0] )
qc.x( a[0] )
print('\n__ After Grovers __')
oq.Wavefunction(qc,systems=[2,1],show_systems=[True,False])
```

```
__ Initial State __
0.5 |00>    0.5 |10>    0.5 |01>    0.5 |11>

__ After Grovers __
-1.0 |10>
```

**Control-Grover 2-qubit circuit:**

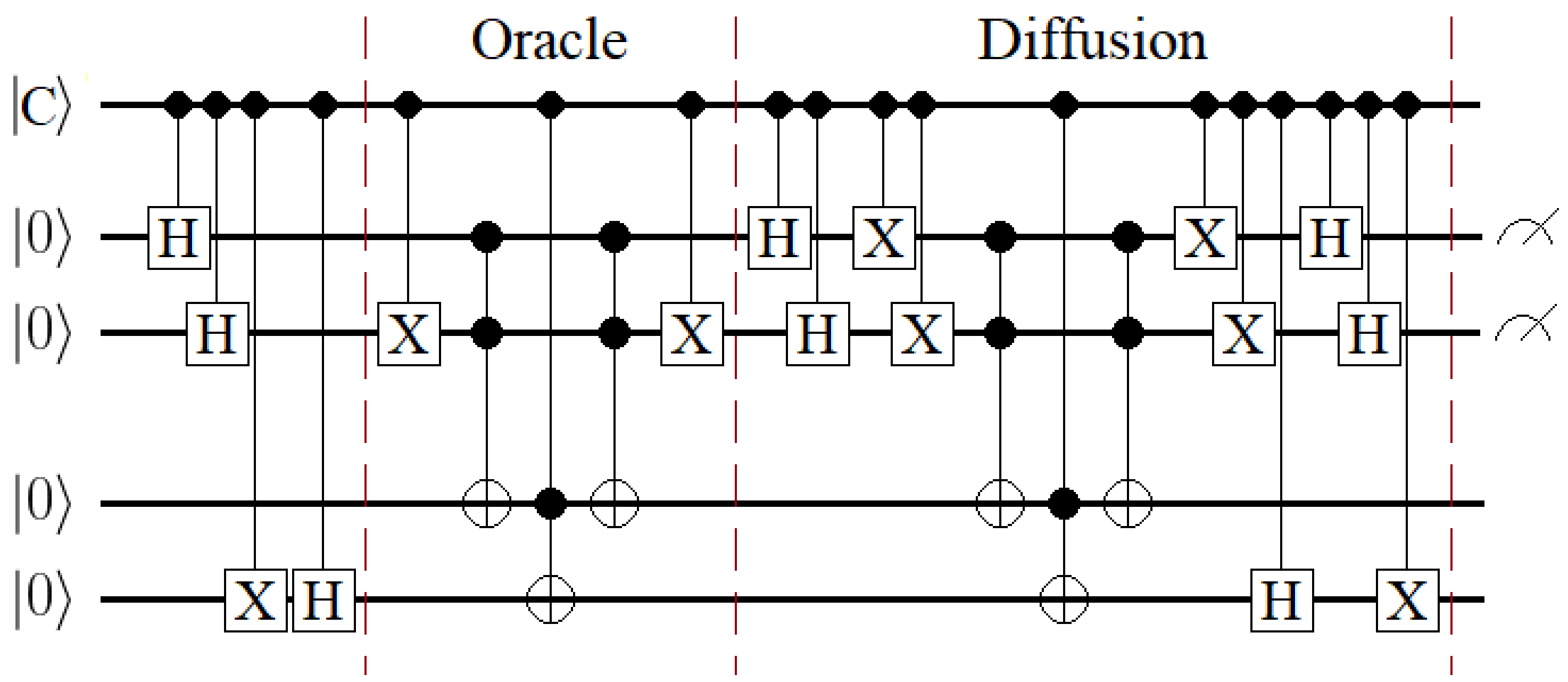

In [ ]:

```
c = QuantumRegister(1,name='c')
q = QuantumRegister(2,name='q')
a1 = QuantumRegister(1,name='a1')
a2 = QuantumRegister(1,name='a2')
qc= QuantumCircuit(c,q,a1,a2,name='qc')

qc.x( c[0] )

qc.ch( c[0], q[0] )
qc.ch( c[0], q[1] )
print('__ Initial State __')
oq.Wavefunction(qc,systems=[1,2,1,1],show_systems=[True,True,False,False])
qc.cx( c[0], a2[0] )
qc.ch( c[0], a2[0] )
#-------------------------     Oracle
qc.cx( c[0], q[1] )
qc.ccx( q[0], q[1], a1[0] )
qc.ccx( c[0], a1[0], a2[0]  )
qc.ccx( q[0], q[1], a1[0] )
qc.cx( c[0], q[1] )
#-------------------------     Diffusion
qc.ch( c[0], q[0] )
qc.ch( c[0], q[1] )
qc.cx( c[0], q[0] )
qc.cx( c[0], q[1] )
qc.ccx( q[0], q[1], a1[0] )
qc.ccx( c[0], a1[0], a2[0]  )
qc.ccx( q[0], q[1], a1[0] )
qc.cx( c[0], q[0] )
qc.cx( c[0], q[1] )
qc.ch( c[0], q[0] )
qc.ch( c[0], q[1] )

qc.ch( c[0], a2[0] )
qc.cx( c[0], a2[0] )

print('\n__ After c-Grovers __')
oq.Wavefunction(qc,systems=[1,2,1,1],show_systems=[True,True,False,False])
```

```
__ Initial State __
0.5 |1>|00>    0.5 |1>|10>    0.5 |1>|01>    0.5 |1>|11>

__ After c-Grovers __
-1.0 |1>|10>
```

As we can see in both of the coding examples above, each quantum circuit boosts the $|10\rangle$ state, successfully completing our search algorithm. While both circuits achieve the same result, in terms of circuit complexity it is clear that implementing the Grover Algorithm as a control-Grover comes at quite a costly price. While single qubit gates like $X$ and $H$ are relatively reliable on real quantum devices, the same cannot be said for 2-qubit operations such as CNOT. Our Qiskit simulator can handle these multi-qubit gates with no problem, but it is important to keep in mind that a circuit like the one shown above would be very noise-prone on a physical device.

## Grover's Operator: Rotational Representation

Now that we have a means for implementing a control version of our Grover circuit, albeit a bit bulky, the next step in our understanding of the Quantum Counting Algorithm requires us revisit our core understanding of Grover's. When we studied Grover's Algorithm in lesson 5.4, the primary emphasis was the algorithm's purpose as a means for searching. To that end, we studied in detail how a search algorithm can be encoded into the states of a quantum system by means of amplitude amplification. And while this is certainly still true, we are now going to view Grover's Algorithm from a different perspective, one in which the purpose of the algorithm is to perform rotations.

To provide some motivation for why we are interested in relearning Grover's in a new light, remember that our ultimate goal for this lesson is to implement Grover's as the $U$ in our QPE circuit. And as we studied in the previous lesson, the QPE circuit is designed to extract eigenvalues from unitary operators of the following form:

$$U|u\rangle = e^{2\pi i\theta}|u\rangle$$

One such operator that fits this mathematical formalism is a simple 2-dimensional rotation matrix:

$$U_\theta = \begin{bmatrix} \cos(\theta) & -\sin(\theta) \\ \sin(\theta) & \cos(\theta) \end{bmatrix}$$

The operator shown above rotates a unitary vector counter-clockwise by an angle of $\theta$, preserving the unitary length of the vector. Additionally, such a rotational operator has eigenvalues of the form $e^{\pm i\theta}$, which means that it is a viable candidate for phase extraction from our QPE circuit. Now then, to connect this operator with our discussion of Grover's, it turns out that we can represent any sized Grover operation as the $2 \times 2$ rotational matrix above by defining the following states:

$$|n\rangle = \frac{1}{\sqrt{N-M}}\sum_i |x_i\rangle$$

$$|m\rangle = \frac{1}{\sqrt{M}}\sum_i |y_i\rangle$$

In the states above, $N$ refers to the total number of entries in our search (always a space of $2^Q$ based on the number of qubits we are using) and $M$ corresponds to the number of marked states that get picked out by the oracle step (our multiple marked states Grover examples from earlier). As we know, a standard Grover search uses $M = 1$, but for the purpose of our Quantum Counting Algorithm we will typically be interested in cases where $M > 1$. And finally, in order to always reduce the mathematical space of our problem to that of the $2 \times 2$ rotational operator, we must assign all of the states in our system into one of the two sets $\{ |x\rangle \}$ or $\{ |y\rangle \}$, which correspond to the marked and non-marked states of our search:

$$|x_i\rangle \equiv |\text{non-marked}\rangle$$

$$|y_i\rangle \equiv |\text{marked}\rangle$$

Using the states $|n\rangle$ and $|m\rangle$ as defined above, we can now rewrite the starting state of our Grover search (the equal superposition state) in the following way:

$$\begin{aligned} |s\rangle &\equiv \frac{1}{\sqrt{N}} \sum_i^N |i\rangle \\ &= \sqrt{\frac{M}{N}}|m\rangle + \sqrt{\frac{N-M}{N}}|n\rangle \end{aligned}$$

Thus, regardless of the size or number of marked states in our problem, we can always rewrite the starting state to Grover's in terms of $|n\rangle$ and $|m\rangle$.

With our new rotational based context for describing Grover's, the question then becomes how to interpret the algorithm in this new light. Specifically, what does a Grover iteration look like in this new basis? Conceptually we know that the final product of our algorithm is a boost in amplitude to the marked states with each iteration, which we've now defined as $|m\rangle$. Simultaneously we expect our non-marked states to become less probable, defined as $|n\rangle$. Visually then, the result of applying the optimal number of Grover iterations should look something like:

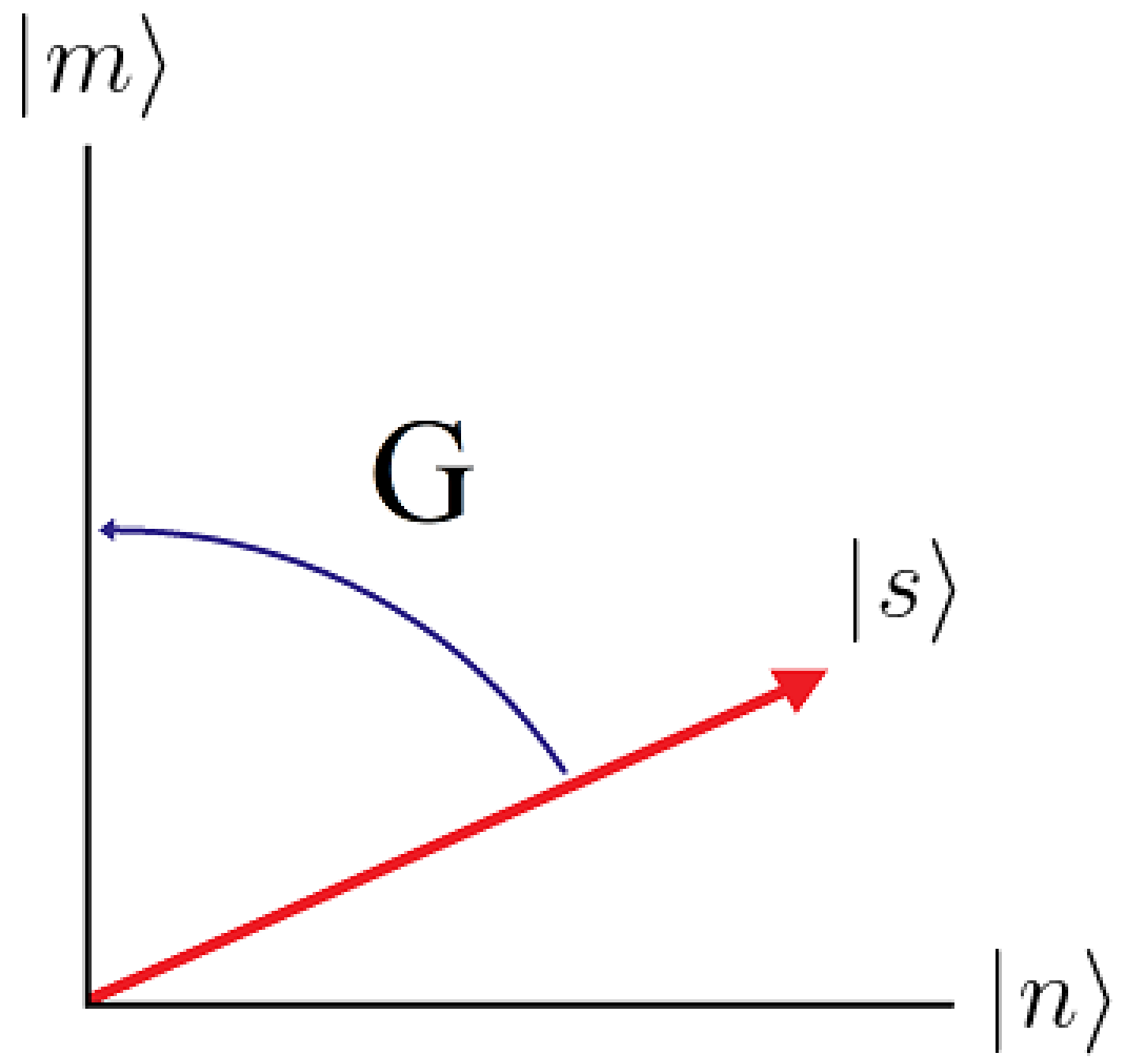

In the figure above, we can see that final result our Grover's Algorithm is a rotation from our starting state $|s\rangle$ to the $|m\rangle$ axis, which corresponds to the superposition of our marked states. Recalling our discussion for Grover's with multiple marked states, this rotational interpretation perfectly agrees with our result from earlier. Specifically, we saw that the effect of Grover's with a set of marked states results in a final distribution where all of the marked states equally share the amplitude amplification. Taking this interpretation one step further then, how close the final state aligns with the $|m\rangle$ axis mathematically translates to the total probability in the system accumulating on the marked states.

### Rotational Grover's Example: $M$ = 1

In order to help cement the connection between Grover's and the $2 \times 2$ rotational matrix, below are two examples which will hopefully prove that the two ideas are deeply connected. First, we will revisit a simple example of a standard Grover search of a single marked state on three qubits. Choosing the state $|101\rangle$ as our marked state, we have the following forms for $|n\rangle$, $|m\rangle$, and $|s\rangle$:

$$|n\rangle = \frac{1}{\sqrt{7}} \sum_{x \neq 101} |x\rangle \qquad |m\rangle = |101\rangle \qquad |s\rangle = \sqrt{\frac{7}{8}}|n\rangle + \sqrt{\frac{1}{8}}|m\rangle$$

Using this simple example, the goal will be to translate the states we find after each Grover iteration into the rotational description outlined above. Specifically, we will take the amplitudes we find after each Grover iteration and use them to derive a rotational angle $\theta$. We will later show that this angle can be theoretically predicted based on $N$ and $M$, matching the results of our example, but first let's start by observing the states produced after $1$ and $2$ Grover iterations:

In [ ]:

```
q = QuantumRegister(3,name='q')
a1= QuantumRegister(1,name='a1')
a2= QuantumRegister(1,name='a2')
qc= QuantumCircuit(q,a1,a2,name='qc')
marked = [1,0,1]

for i in np.arange(3):
    qc.h( q[int(i)] )
qc.x( a1[0] )

oq.Wavefunction(qc, systems=[3,1,1], show_systems=[True,False,False])
for i in np.arange(2):
    oq.Grover_Oracle(marked, qc, q, a1, a2)
    oq.Grover_Diffusion(marked, qc, q, a1, a2)
    print('\n____ Grover Iterations:  ',int(i+1),' ____')
    oq.Wavefunction(qc, systems=[3,1,1], show_systems=[True,False,False])
```

```
0.35355 |000>    0.35355 |100>    0.35355 |010>    0.35355 |110>    0.35355 |001>    0.35355 |101>    0.35355 |011>    0.35
355 |111>

____ Grover Iterations:   1  ____
-0.17678 |000>    -0.17678 |100>    -0.17678 |010>    -0.17678 |110>    -0.17678 |001>    -0.88388 |101>    -0.17678 |011>
-0.17678 |111>

____ Grover Iterations:   2  ____
-0.08839 |000>    -0.08839 |100>    -0.08839 |010>    -0.08839 |110>    -0.08839 |001>    0.97227 |101>    -0.08839 |011>
-0.08839 |111>
```

Tracking our marked state throughout the algorithm above, we find the following amplitudes on the $|\,101\,\rangle$ state:

$$|\,101\,\rangle: \quad 0.35355 \quad \longrightarrow \quad 0.88388 \quad \longrightarrow \quad .97227$$

Note that we've dropped the negative sign on the second amplitude above. If you recall from lesson 5.4, the proper implementation of Grover's (i.e. the reflection about the average step) always leaves the marked state with a positive after each Grover iteration. Here, we are basing our discussion of Grover's as a rotation on this formal mathematical description.

Following the formalism for $|\,n\,\rangle$ and $|\,m\,\rangle$ outlined above, we can write down our initial state $|\,s\,\rangle$, as well as its corresponding initial angle:

$$|\,s\,\rangle \quad = \quad \frac{1}{\sqrt{8}}|\,101\,\rangle \; + \; \sqrt{\frac{7}{8}}|\,n\,\rangle$$

$$\theta_i \quad = \quad \tan^{-1}\!\left(\frac{1}{\sqrt{7}}\right) \quad \approx \quad 20.7^\circ$$

Next let's track the amplitudes after the first Grover iteration, and once again derive the angle created by our resulting state in the $\big\{|\,n\,\rangle, |\,m\,\rangle\big\}$ basis:

$$\text{After 1 Grover Iteration:} \qquad \sqrt{\frac{25}{32}}|\,101\,\rangle \; + \; \frac{1}{\sqrt{32}}\sum_{x \neq 101}|\,x\,\rangle$$

$$= \quad \sqrt{\frac{25}{32}}|\,m\,\rangle \; + \; \sqrt{\frac{7}{32}}|\,n\,\rangle$$

$$\theta_1 \quad = \quad \tan^{-1}\!\left(\sqrt{\frac{25}{7}}\right) \quad \approx \quad 62.1^\circ$$

And lastly, the difference between $\theta_i$ and $\theta_1$ gives us the angle of rotation corresponding to our Grover iteration:

$$\text{Rotation Angle:} \quad \theta_1 \; - \; \theta_i \quad = \quad 41.4^\circ$$

It is important to note here that the angle $\theta \approx 41.4^\circ$, for which we are claiming is the rotational representation of one Grover iteration, was just derived from using purely geometric means. Later on we will reconfirm this rotational angle using only quantum states. Next however, our next goal is to verify that this angle is indeed intrinsically tied to a single Grover iteration (i.e. subsequent iterations won't result in different $\theta$'s). To do this, we will attempt to predict the resulting amplitude from a second Grover iteration using only $\theta$:

$$U_\theta^2 \begin{bmatrix} |\,n\,\rangle \\ |\,m\,\rangle \end{bmatrix} \quad = \quad \begin{bmatrix} \cos(\theta) & -\sin(\theta) \\ \sin(\theta) & \cos(\theta) \end{bmatrix}^2 \begin{bmatrix} \sqrt{\frac{7}{8}} \\ \frac{1}{\sqrt{8}} \end{bmatrix} \quad \approx \quad \begin{bmatrix} -.23385 \\ .97227 \end{bmatrix}$$

$$|\,101\,\rangle \;\; \text{state amplitude:} \quad 0.97227$$

Sure enough, compare the value shown here for the $|\,m\,\rangle$ component of our final vector to that of the amplitude residing on the $|\,101\,\rangle$ state above. As you can see, the two are equal, which means that the amplitude of our marked state is once again in agreement with the value predicted from our Grover rotation operator. This in turn confirms that the angle $\theta \approx 41.4^\circ$ is indeed the rotational angle associated with a single Grover iteration for this problem. With this angle in hand, we can predict the amplitude of our marked state after any number of iterations by using the rotational matrix. Additionally, this angle gives us a nice graphical picture of how the amplitude on our marked state is changing as a function of each iteration:

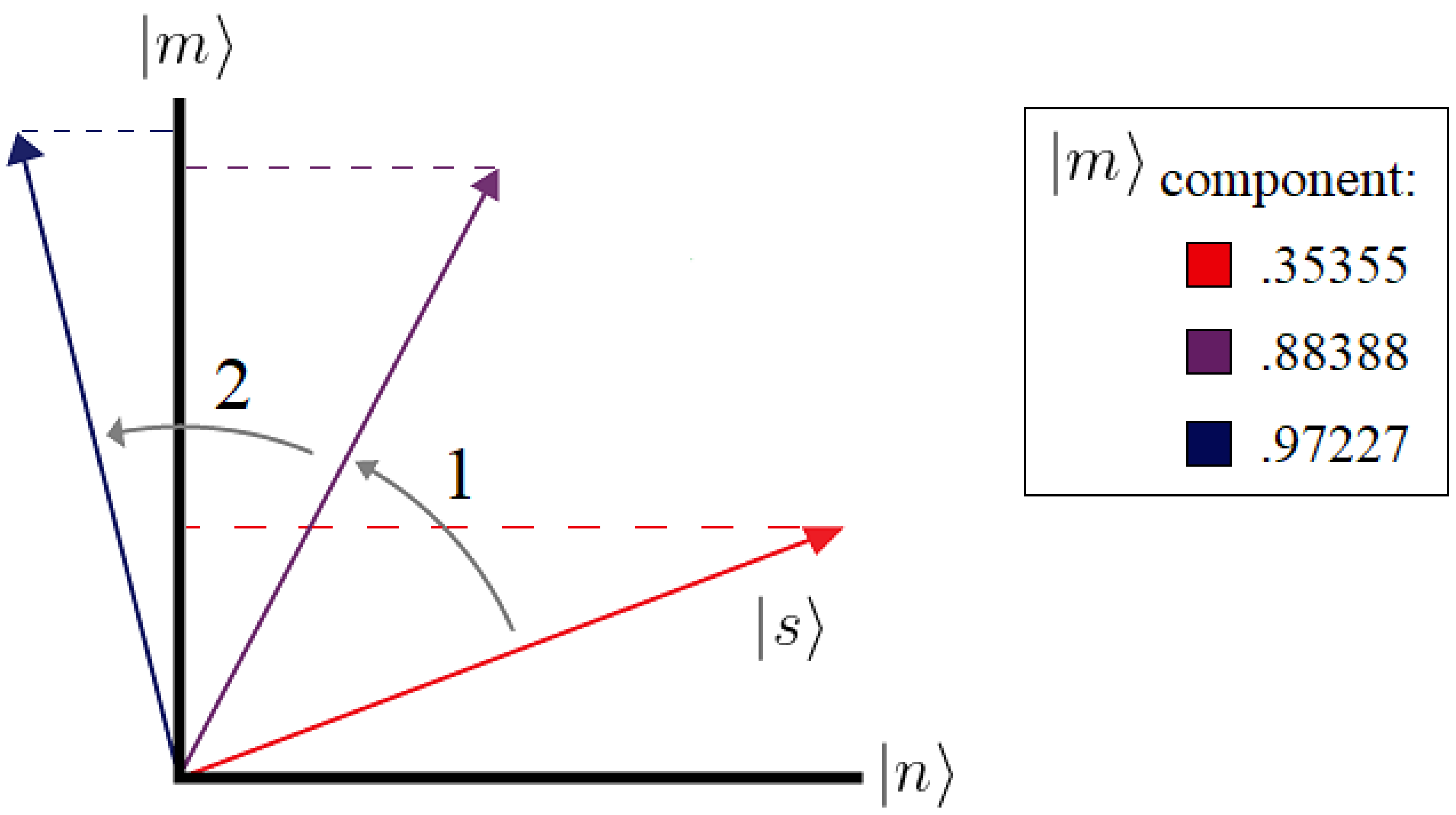

Hopefully viewing Grover's Algorithm in the rotational basis provides some additional insight into the nature of exactly what each Grover iteration is doing to our quantum system. As we shall see, this representation is the key to connecting Grover's to our Quantum Counting problem. To conclude this initial example then, there are two last points worth noting:

**(1)** Imagine the effect of the next application of our Grover operator in the diagram above, and the resulting $|m\rangle$ component of the vector. As we know, the Grover Algorithm has an optimal number of iterations for which we get the maximal probability of measuring our marked state. Beyond the optimal number, our chances of finding the marked state begin to diminish just as gradually as they built up. If we now picture the effect of one more rotation on the diagram above, it is clear that we will be rotating our vector *away* from $|m\rangle$, and consequently worsening our probability of measuring $|101\rangle$.

**(2)** In the example above we used the standard case for Grover's Algorithm in order to derive an easy result, namely $M = 1$. However, the steps we took to arrive at our result are just as equally valid for the case of multiple marked states. Regardless of how many marked states there are in the system, we can always map our problem to the $\{|n\rangle, |m\rangle\}$ basis where the same mathematics still hold. The only difference between the case of one or multiple marked states is the way in which the amplitudes are distributed. Remember that $|m\rangle$ is a superposition of all the marked states. Thus, whatever the final $|m\rangle$ component to our vector is, this amplitude is distributed evenly among all of the marked states.

## Rotational Grover's Example: $M$ = 3

To follow up on point **(2)** above, we will now work through an example using multiple marked states in our Grover's Algorithm. Just as before, we will look to rewrite the state of our system after each Grover iteration in the $\{|n\rangle, |m\rangle\}$ basis, and use the amplitudes we find to determine the angle of rotation for one Grover iteration. We will be working with the case $M = 3$, $N = 3$, for which we have the following initial superposition state:

$$|n\rangle = \frac{1}{\sqrt{5}}\big(|000\rangle + |001\rangle + |010\rangle + |100\rangle + |011\rangle\big) \qquad |m\rangle = \frac{1}{\sqrt{3}}\big(|101\rangle + |110\rangle + |111\rangle\big)$$

$$|s\rangle = \left(\sqrt{\frac{3}{8}}|m\rangle + \sqrt{\frac{5}{8}}|n\rangle\right)$$

For this example we've chosen the states $|101\rangle$, $|110\rangle$, and $|111\rangle$ to be our marked states. Note that even though we are starting from the same equal superposition state $|s\rangle$ as the $M = 1$ example (every state is initialized with an amplitude of $\frac{1}{\sqrt{8}}$), our initial state now has different amplitudes for $|n\rangle$ and $|m\rangle$. Consequently, the initial angle for our $M = 3$ case has changed:

$$\theta_i = \tan^{-1}\left(\sqrt{\frac{3}{5}}\right) \approx 37.8^\circ$$

Thus, without doing anything different to our initial quantum state (Hadamard gates on the three computational qubits), the mathematics of our rotational description have changed simply by the way in which we have organized our states into $|n\rangle$ and $|m\rangle$. Conceptually, this change reflects the fact that we are now three times more likely to find a marked state in $|s\rangle$, and thus our initial angle $\theta_i$ has increased, bringing the state closer to the $|m\rangle$ axis. Additionally, the optimal number of Grover iterations for this $M = 3$ case has changed, now only $1$, as compared to $2$ earlier. Let's now see the resulting amplitudes on our marked states using the cell of code below:

In [ ]:

```
marked = ['101','110','111']
Q = len(marked[0])
iters = 1
#==========================================
q   = QuantumRegister(Q,name='q')
a1 = QuantumRegister(1,name='a1')
a2 = QuantumRegister(Q-2,name='a2')
qc  = QuantumCircuit(q,a1,a2,name='qc')
#------------------------------------------
for j in np.arange(Q):
    qc.h( q[int(j)] )
qc.x( a1[0] )

qc,q,a1,a2 = oq.Multi_Grover(q,a1,a2,qc,marked,iters)
print('_____ 1 Grover Iteration     M = 3 _____ ')
oq.Wavefunction(qc,systems=[Q,Q-2,1],show_systems=[True,False,False])
```

```
_____ 1 Grover Iteration     M = 3 _____
0.17678 |000>    0.17678 |100>    0.17678 |010>    -0.53033 |110>    0.17678 |001>    -0.53033 |101>    0.17678 |011>    -
0.53033 |111>
```

Just as we did before, we will use the amplitudes of our final state after the Grover iteration to derive $\theta$ for this problem. Doing so will first require that we properly combine the amplitudes found on our marked and non-marked states, organizing them into the $\{ |n\rangle, |m\rangle \}$ basis:

$$
\begin{aligned}
G|s\rangle &\approx -0.17678 \left( |000\rangle + |001\rangle + |010\rangle + |100\rangle + |011\rangle \right) \quad + \quad 0.53033 \left( |101\rangle + |110\rangle + |111\rangle \right) \\
&= \left( -0.17678 \cdot \sqrt{5} \right) |n\rangle \quad + \quad \left( 0.53033 \cdot \sqrt{3} \right) |m\rangle \\
&= -0.3953 |n\rangle \quad + \quad 0.9186 |m\rangle
\end{aligned}
$$

Once again, take note of the amplitude sign changes shown above as compared to our code example, which are a consequence of the way in which we implement the "reflection about the average" step in our Grover's circuit. Now then, with our quantum state written out in the rotational basis, we can calculate our Grover iteration angle $\theta$ via trig:

$$
\theta_f = \tan^{-1}\left( -\frac{0.9186}{0.3953} \right) + 180^\circ \approx 113.284^\circ
$$

$$
\theta = \theta_f - \theta_i \approx 75.523
$$

The $+180^\circ$ above is to make sure our final angle properly lands in the II. quadrant when using the $\tan^{-1}$ function. As we've now come to expect, the presence of three marked states has drastically changed our rotational angle corresponding to one Grover iteration. And just for good measure, let's confirm that this angle does indeed reproduce our final quantum state by now tackling this same example from the reverse perspective:

$$
U_\theta \begin{bmatrix} |n\rangle \\ |m\rangle \end{bmatrix} = \begin{bmatrix} \cos(\theta) & -\sin(\theta) \\ \sin(\theta) & \cos(\theta) \end{bmatrix} \begin{bmatrix} \sqrt{\frac{5}{8}} \\ \sqrt{\frac{3}{8}} \end{bmatrix} \approx \begin{bmatrix} -.3953 \\ .9723 \end{bmatrix}
$$

The mathematics of our problem in the rotational basis are once again in agreement with our quantum code, reproducing the correct amplitudes on the $|n\rangle$ and $|n\rangle$ states. Visually, we can summarize our $M = 3$ example in the $\{ |n\rangle, |m\rangle \}$ basis as follows:

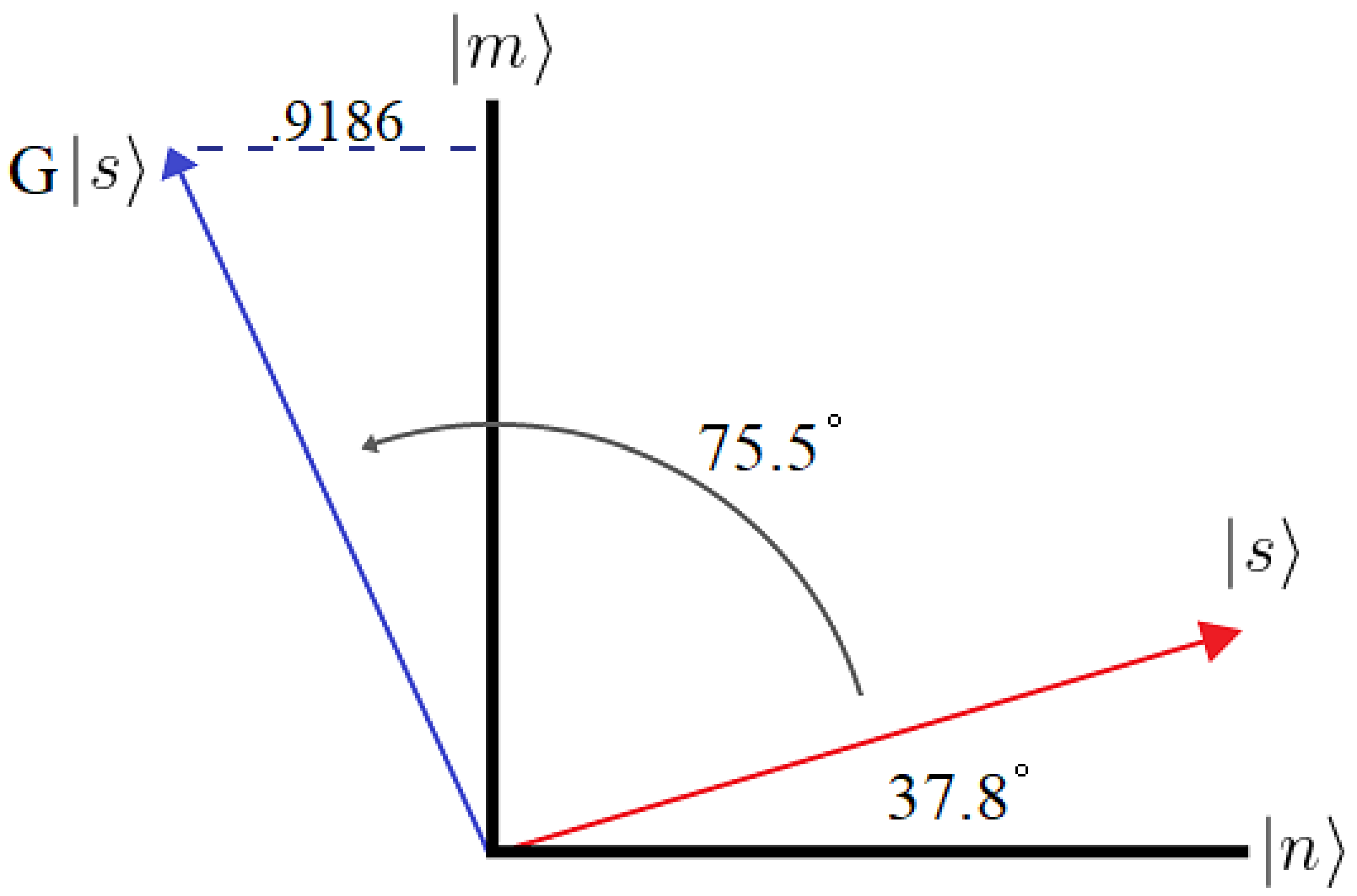

## Eigenstates & Eigenvalues of G

Having now seen two examples demonstrating that we can represent Grover iterations as rotations in the $\{ |n\rangle, |m\rangle \}$ basis, our next goal is to explore this new formalism further to see what meaningful insights it can provide us. To start, let's write the $2 \times 2$ rotational matrix from earlier, now defining it as $G_\theta$, for which we have shown to be equivalent to a single Grover iteration in the $\{ |n\rangle, |m\rangle \}$ basis:

$$G_\theta \equiv \begin{bmatrix} \cos(\theta) & -\sin(\theta) \\ \sin(\theta) & \cos(\theta) \end{bmatrix}$$

Our Grover operator $G_\theta$ has eigenvectors and eigenvalues of the following form:

$$\text{Eigenvectors:} \quad |\mathrm{E}_\pm\rangle = \begin{bmatrix} \pm i \\ 1 \end{bmatrix} \qquad \text{Eigenvalues:} \quad \lambda_\pm = e^{\pm i\theta}$$

Returning now to our original motivation for studying Grover's in this new light, namely implementing it into the QPE circuit as our control-$U$ operation, we are now equipped to discuss how $G_\theta$ will fit into our understanding of Quantum Phase Estimation. Ignoring for a moment that $G_\theta$ is actually a complex circuit composed of many gates, if we just think of it as a simple $2 \times 2$ operator, then implementing it into QPE would have very straightforward results. By preparing our eigenstate $|u\rangle$ as either $|\mathrm{E}_\pm\rangle$, the result of our QPE circuit would give us an approximation for $\theta$.

Now we take one step back and ask what are the challenges that prevent us from treating $G_\theta$ as a simple $2 \times 2$ operator and using it in QPE. Firstly, $G_\theta$ is only a $2 \times 2$ operator in the $\{|n\rangle, |m\rangle\}$ basis. But as we know, our quantum circuits live in the computational basis: $\{|0\rangle, |1\rangle\}$. More specifically, the issue here is that our true Grover iteration operator acts on $N$ computational qubits, whereas $G_\theta$ in theory only needs to act on a single qubit, where the two states of the qubit represent $|n\rangle$ and $|m\rangle$. Thus, our first issue to resolve will be the implementation of $|u\rangle$, and the way in which $|\mathrm{E}_\pm\rangle$ translate into the computational basis.

As it turns out, the eigenstate which we need to prepare for the Quantum Counting Algorithm is actually quite simple:

$$|u\rangle = |s\rangle = H^{\otimes n}|0\rangle$$

Thus, when we run our Quantum Counting circuit, all we need to do is prepare our eigenstate register in an equal superposition state. It is by no means obvious at first why this should be the case, but the reasoning is quite clever. With just a few steps, let's see how this superposition state $|s\rangle$ encodes our eigenstates of $G$. First let's write out the tools we need for our derivation, our eigenstates $|\mathrm{E}_\pm\rangle$ and $|s\rangle$ written out in the $\{|n\rangle, |m\rangle\}$ basis:

$$|\mathrm{E}_\pm\rangle = \begin{bmatrix} \pm i \\ 1 \end{bmatrix} = |m\rangle \pm i|n\rangle \qquad |s\rangle = \frac{1}{\sqrt{N}}\sum_i^N |i\rangle = \sqrt{\frac{M}{N}}|m\rangle + \sqrt{\frac{N-M}{N}}|n\rangle$$

Now we rewrite $|s\rangle$ in terms of the eigenstates to $G_\theta$:

$$\begin{aligned} |s\rangle &= \sigma_+|\mathrm{E}_+\rangle + \sigma_-|\mathrm{E}_-\rangle \\ &= \sigma_+\big(|m\rangle + i|n\rangle\big) + \sigma_-\big(|m\rangle - i|n\rangle\big) \\ &= \big(\sigma_+ + \sigma_-\big)|m\rangle + i\big(\sigma_+ - \sigma_-\big)|n\rangle \end{aligned}$$

If we now compare the two forms for $|s\rangle$ above, we get the following set of equations which will allow us to solve for $\sigma_\pm$:

$$\sigma_+ + \sigma_- = \sqrt{\frac{M}{N}} \qquad \sigma_+ - \sigma_- = -i\sqrt{\frac{N-M}{N}}$$

for which we get the following solutions:

$$\sigma_\pm = \frac{1}{2}\left(\sqrt{\frac{M}{N}} \mp i\sqrt{\frac{N-M}{N}}\right)$$

To recap, solving for the constants $\sigma_\pm$ allows us to rewrite our initial superposition state $|s\rangle$ in terms of the eigenstates of $G$. And as we can see, both of these constants are composed of the values $N$ and $M$, which means that we can use this formalism for any sized Grover's we like. And finally, when we apply a single Grover iteration to $|s\rangle$, we can track the corresponding effect in the eigenbasis of $G_\theta$ as follows:

$$\begin{aligned} G_\theta|s\rangle &= G_\theta\big(\sigma_+|\mathrm{E}_+\rangle + \sigma_-|\mathrm{E}_-\rangle\big) \\ &= e^{i\theta}\sigma_+|\mathrm{E}_+\rangle + e^{-i\theta}\sigma_-|\mathrm{E}_-\rangle \end{aligned}$$

Although the equal superposition state $|s\rangle$ is technically not an eigenstate of our $G_\theta$ operator, we have shown that it be rewritten as a superposition of the eigenstates of $G_\theta$. And while this result isn't anything new, as we can always translate states between different bases, the unique thing here will be the consequence of using $|s\rangle$ as our eigenstate $|u\rangle$ for the QPE circuit. In lesson 7, we made it a point that Quantum Phase Estimation is designed such that the effect of the control-$U$ operators leaves the eigenstate $|s\rangle$ unchanged, which in turn causes the effect from these operations to be absorbed into the control qubits. Here however, using $|s\rangle$ as our eigenstate register will not have the same effect:

$$G_\theta|s\rangle \neq e^{i\theta'}|s\rangle$$

The significance of the statement above is that our QPE circuit is still usable even if our eigenstate register $|u\rangle$ isn't an eigenstate of our control-$U$ operation. The consequence of choosing such a $|u\rangle$ then, as we shall see, is that our QPE algorithm will not return a single $\theta$ phase approximation, but rather a superposition of phases. How the final QFT$^\dagger$, and ultimately our measurements handle this superposition of phases will be a later topic in this lesson, but first we must discuss the rotational operator $G_\theta$ in a bit more detail.

## Deriving $\theta$

With the formalism outlined in the previous section, we are nearly ready to discuss the results obtained from applying $G_\theta$ to the QPE circuit. However, the final ingredient missing for our analysis is a formula for $\theta$, the angle of rotation corresponding to a single Grover iteration. In our previous examples we calculated $\theta$ by looking at amplitudes and using trig, but in general we would like an equation that allows us to calculate $\theta$ without having to implement any quantum circuits. To do this, we will start by defining $G_\theta$ in the following way:

$$U_m = I - 2|m\rangle\langle m| \qquad U_s = 2|s\rangle\langle s| - I$$

$$G_\theta \;=\; U_s U_m \;=\; \Big(2|s\rangle\langle s| - I\Big)\Big(I - 2|m\rangle\langle m|\Big)$$

Here we've written out $G_\theta$ as a decomposition into its Oracle and Diffusion components (see lesson 5.4 for a refresher on reflections within the Grover Algorithm), where $U_m$ achieves the phase flipping about the marked states $|m\rangle$, and $U_s$ is a reflection about the superposition state $|s\rangle$ (the proper reflection). Now, from the previous sections we know that the effect of $G_\theta|s\rangle$ is equivalent to a rotation by some angle $\theta$ in the $\Big\{|n\rangle, |m\rangle\Big\}$ basis. Thus, in order to derive the relation between $\theta$, $M$, and $N$ we are going to calculate the angle between $|s\rangle$ and $G_\theta|s\rangle$:

$$\begin{aligned} G_\theta|s\rangle \;&=\; \Big(2|s\rangle\langle s| - I\Big)\Big(I - 2|m\rangle\langle m|\Big)|s\rangle \\ &=\; 2|s\rangle - |s\rangle - 4|s\rangle\langle s|m\rangle\langle m|s\rangle + 2|m\rangle\langle m|s\rangle \end{aligned}$$

where from earlier we know that $\langle m|s\rangle \;=\; \sqrt{\frac{M}{N}}$:

$$=\; \Big(1 - \frac{4M}{N}\Big)|s\rangle + 2\sqrt{\frac{M}{N}}|m\rangle$$

In order to obtain our angle $\theta$, we need to rewrite the $|s\rangle$ state in terms of $|n\rangle$ and $|m\rangle$:

$$\begin{aligned} &=\; \Big(1 - \frac{4M}{N}\Big)\Big(\sqrt{\frac{M}{N}}|m\rangle + \sqrt{\frac{N-M}{N}}|n\rangle\Big) + 2\sqrt{\frac{M}{N}}|m\rangle \\ &=\; \Big(\frac{3N - 4M}{N}\sqrt{\frac{M}{N}}\Big)|m\rangle + \Big(\frac{N - 4M}{N}\sqrt{\frac{N-M}{N}}\Big)|n\rangle \end{aligned}$$

To recap, the state above is the result of $G_\theta|s\rangle$, written out in the $\Big\{|n\rangle, |m\rangle\Big\}$. Geometrically, we know that this state represents a rotation of $|s\rangle$ by some angle $\theta$. Thus, our final step is to throw these two states together with some trig, computing the angle between them:

$$\cos(\theta) \;=\; \frac{\vec{v_1}\cdot\vec{v_2}}{|\vec{v_1}||\vec{v_2}|}$$

where $\vec{v_1}$ and $\vec{v_2}$ are $|s\rangle$ and $G_\theta|s\rangle$. Conveniently for us, because we are dealing with quantum states, we are guaranteed that the magnitudes $\Big||s\rangle\Big|$ and $\Big|G_\theta|s\rangle\Big|$ are both equal to one. All that leaves then is the dot product between these two states, or equivalently $\langle s|G|s\rangle$:

$$\cos(\theta) \;=\; \Big(\frac{3N - 4M}{N}\sqrt{\frac{M}{N}}\Big)\Big(\sqrt{\frac{M}{N}}\Big)\langle m|m\rangle + \Big(\frac{N - 4M}{N}\sqrt{\frac{N-M}{N}}\Big)\Big(\sqrt{\frac{N-M}{N}}\Big)\langle n|n\rangle$$

Sparing you a little algebra, we arrive at our final result:

$$\cos(\theta) \;=\; \frac{N - 2M}{N}$$

And while this formula for $\theta$ is correct, we will actually opt to perform one final trig identity in order to derive a form for $\theta$ that is more commonly used:

$$\sin\Big(\frac{\theta}{2}\Big) \;=\; \pm\sqrt{\frac{1 - \cos(\theta)}{2}}$$

By plugging in our result for $\cos(\theta)$, it's only a couple of algebra steps to our final equation for $\theta$:

$$\theta \;=\; 2\sin^{-1}\Big(\sqrt{\frac{M}{N}}\Big)$$

The significance of the result above is that we now have a formula for $\theta$, representing a single Grover iteration for any sized problem of $N$ and $M$. With this equation, we can now easily calculate the optimal number of Grover iterations needed for any sized Grover Algorithm:

$$\theta_i = \tan^{-1}\Big(\sqrt{\frac{M}{N-M}}\Big) \qquad \theta_r = 2\sin^{-1}\Big(\sqrt{\frac{M}{N}}\Big)$$

$$\text{optimal steps:} \qquad \frac{\frac{\pi}{2} - \theta_i}{\theta_r}$$

Recalling our geometric illustrations from earlier, the optimal number of steps for Grover's corresponds to the number of rotations which will bring our quantum state as close as possible to the $|m\rangle$ axis. Mathematically, this is equivilant to dividing $90^\circ$ by our angle of rotation $\theta_r$, which we just calculated, taking into account the initial offset angle $\theta_i$ corresponding to $|s\rangle$. Earlier when discussing Grover's with multiple marked states, we stated the approximate optimal number of iterations without any justification. We can now arrive at this approximation from earlier using the result we just derived, letting $\theta_i \approx 0$ for the $N >> M$ limit.

To test this new result, the code below uses our derived expression for $\theta$ to determine the optimal number of Grover iterations, resulting in peak probabilities for the marked states:

In [ ]:

```
marked = ['10111','11000']
Q = len(marked[0])
N = 2**Q
M = len(marked)
iters = int(round( (m.pi/2 - np.arctan( m.sqrt( M/(N-M) ) ) ) / ( 2 * np.arcsin( m.sqrt(M/N) ) ) ) )
print('N: ',N,'     M: ',M,'     Optimal Iterations: ',iters,'\n\n')
#========================================
q   = QuantumRegister(Q,name='q')
a1 = QuantumRegister(1,name='a1')
a2 = QuantumRegister(Q-2,name='a2')
qc  = QuantumCircuit(q,a1,a2,name='qc')
#-----------------------------------------
for j in np.arange(Q):
    qc.h( q[int(j)] )
qc.x( a1[0] )

qc,q,a1,a2 = oq.Multi_Grover(q,a1,a2,qc,marked,iters)
oq.Wavefunction(qc,systems=[Q,Q-2,1],show_systems=[True,False,False],column=True)
```

```
N:  32      M:  2      Optimal Iterations:  3


0.03591 |00000>
0.03591 |10000>
0.03591 |01000>
-0.6933 |11000>
0.03591 |00100>
0.03591 |10100>
0.03591 |01100>
0.03591 |11100>
0.03591 |00010>
0.03591 |10010>
0.03591 |01010>
0.03591 |11010>
0.03591 |00110>
0.03591 |10110>
0.03591 |01110>
0.03591 |11110>
0.03591 |00001>
0.03591 |10001>
0.03591 |01001>
0.03591 |11001>
0.03591 |00101>
0.03591 |10101>
0.03591 |01101>
0.03591 |11101>
0.03591 |00011>
0.03591 |10011>
0.03591 |01011>
0.03591 |11011>
0.03591 |00111>
-0.6933 |10111>
0.03591 |01111>
0.03591 |11111>
```

Even though solving for the optimal number of Grover iterations is an important result, it's actually not why we are interested in $\theta$. For the purpose of solving our Quantum Counting problem, $\theta$ will serve as a stepping stone to arrive at the answer we really care about: $M$. Thus, if we are able to obtain the values of $N$ and $\theta$, we can rearrange our result from above to give us $M$:

$$M \;=\; N\sin^2\!\left(\frac{\theta}{2}\right)$$

The equation above represents our solution to the Quantum Counting Algorithm. Through the implementation of $G_\theta$ as our unitary operator in the QPE circuit, we will be able to obtain an approximate value for $\theta$. Then, using this value in combination with $N$, we will be able to determine $M$, the total number of marked states. The tricky part will be in extracting $\theta$, for which we will now spend the rest of this lesson discussing.

## Extracting $\theta$ from $|\text{E}_\pm\rangle$

We've already noted that $|\,s\,\rangle$ will ultimately be the state used in our eigenstate register, but that doesn't mean it's necessarily our first choice. That is to say, if implementing $|\,\text{E}_\pm\rangle$ as our eigenstate $|\,u\,\rangle$ was an easy task, there would be no reason to bother with using $|\,s\,\rangle$. Luckily for us however, working with Qiskit's simulator allows us to study quantum circuits that may be otherwise impossible on a real device. We can sidestep the issue of implementing these difficult eigenstates by using the $\textbf{initialize}()$ function, which will allow us to prepare $|\,\text{E}_\pm\rangle$ so long as we provide their correct normalized forms:

$$|\,\text{E}_\pm\rangle_{\text{norm}} \;=\; \frac{1}{\sqrt{2}}\Big(\,|\,m\rangle \,\pm\, i\,|\,n\,\rangle\Big)$$

The advantage to using the eigenstates of $G_\theta$ first will allow us to isolate the effects of passing each one through our QPE circuit, so that we may study their expected results individually. Then, in our final solution when we use $|\,s\,\rangle$ as our eigenstate $|\,u\,\rangle$, we will be able to better appreciate the consequence of passing a superposition of $|\,E_\pm\rangle$ through the eigenstate register. As our first example, below is the implementation of $|\,\text{E}_+\rangle$ for the case of $N\,=\,8,\ M\,=\,3$ from earlier:

$$|n\rangle = \frac{1}{\sqrt{5}}\Big(|000\rangle + |001\rangle + |010\rangle + |100\rangle + |011\rangle\Big) \qquad |m\rangle = \frac{1}{\sqrt{3}}\Big(|101\rangle + |110\rangle + |111\rangle\Big)$$

$$|\text{E}_+\rangle_{\text{norm}} = \frac{i}{\sqrt{10}}\Big(|000\rangle + |001\rangle + |010\rangle + |100\rangle + |011\rangle\Big) + \frac{1}{\sqrt{6}}\Big(|101\rangle + |110\rangle + |111\rangle\Big)$$

In [ ]:

```
E_plus = [ 1.0j/m.sqrt(10),1.0j/m.sqrt(10),1.0j/m.sqrt(10),1.0/m.sqrt(6),
           1.0j/m.sqrt(10),1.0/m.sqrt(6),1.0j/m.sqrt(10),1.0/m.sqrt(6) ]
#=========================================================================
q = QuantumRegister(3,name='q')
qc= QuantumCircuit(q,name='qc')
#------------------------------
qc.initialize( E_plus, q )
oq.Wavefunction( qc )
```

```
0.31623j |000>    0.31623j |100>    0.31623j |010>    0.40825 |110>    0.31623j |001>    0.40825 |101>    0.31623j |011>
0.40825 |111>
```

As you can see, the **initialize**() function has successfully implemented $|\text{E}_+\rangle$ as a quantum state, using the Qiskit simulator. With this tool, we will now demonstrate our first solution to the Quantum Counting Algorithm. Below is a code example which implements $G_\theta$ as the control-$U$ operator in the QPE circuit, acting on the eigenstate $|\text{E}_+\rangle$:

In [ ]:

```
Marked  = ['101','110','111']
Q = len(Marked[0])
E_plus = [ 1.0j/m.sqrt(10),1.0j/m.sqrt(10),1.0j/m.sqrt(10),1.0/m.sqrt(6),
           1.0j/m.sqrt(10),1.0/m.sqrt(6),1.0j/m.sqrt(10),1.0/m.sqrt(6) ]
#========================================================================
q = QuantumRegister(Q,name='q')
u = QuantumRegister(Q,name='u')
a1 = QuantumRegister(Q-1,name='a1')
a2 = QuantumRegister(1,name='a2')
c = ClassicalRegister(Q,name='c')
qc= QuantumCircuit(q,u,a1,a2,c,name='qc')
#-------------------------------------
qc.h( q[0] )
qc.h( q[1] )
qc.h( q[2] )
qc.initialize(E_plus,u)
qc.x( a2[0] )
qc.h( a2[0] )
#------------------------
for i in np.arange(Q):
    for j in np.arange(2**i):
        oq.C_Grover(qc,q[int(3-(i+1))],u,a1,a2,Marked,proper=True)
#------------------------
qc.h( a2[0] )
qc.x( a2[0] )
oq.QFT_dgr( qc,q,3,swap=True )
#-------------------------
qc.measure(q,c)
oq.Measurement( qc, shots=10000 )
```

```
186|100>    1623|001>    121|110>    119|101>    6994|010>    310|000>    162|111>    485|011>
```

The coding example above represents the following circuit diagram:

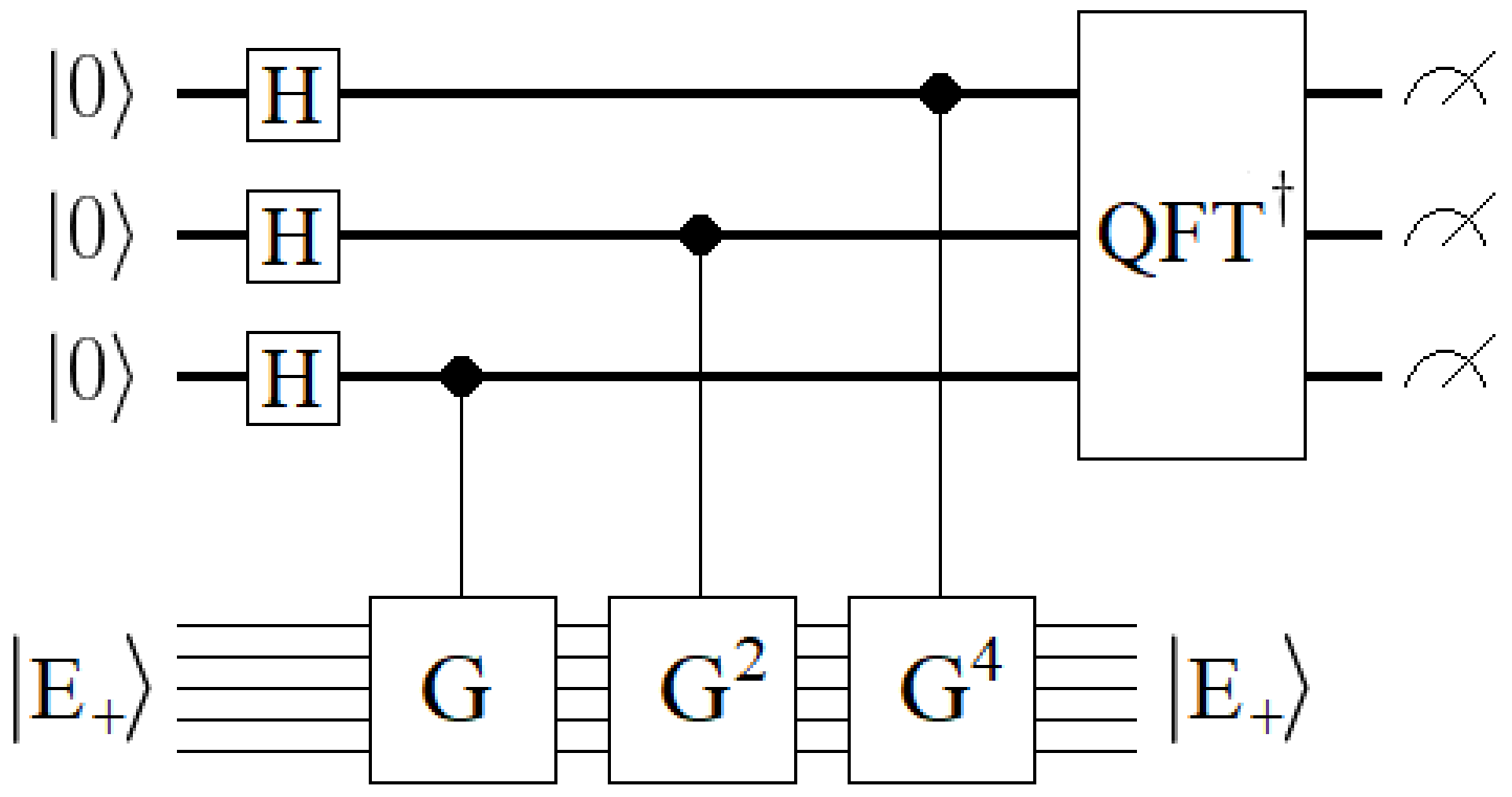

By using the initialize() function to set our eigenstate register to $|\text{E}_+\rangle$, the effect of our QPE circuit using $G_\theta$ has resulted in a significant boost to the state $|010\rangle$. Using our expression for $\theta$ from earlier, the theoretical output from this circuit should be an approximation to the phase:

$$\theta = 2\sin^{-1}\left(\sqrt{\frac{3}{8}}\right) \approx 1.318116$$

But because our eigenvalue of $G_\theta$ is of the form $e^{i\theta}$, and our QPE circuit is designed for eigenvalues of the form $e^{2\pi i\theta}$, we must first do a quick computation:

$$e^{i\theta} = e^{\frac{2\pi}{2\pi}i\theta} = e^{2\pi i\theta'}$$

$$\therefore \; \theta' \;=\; \frac{\theta}{2\pi} \;\approx\; 0.20978$$

Thus, when we take into account the extra factor of $2\pi$, the phase we expect to get out from our QPE circuit is approximately $0.21$. And because this value of $\theta$' does not fall on an exact multiple of $1/8$ (our precision register is three qubits), we expect to find the majority of the probability in the system on the two closest states: $|\,001\,\rangle$ and $|\,010\,\rangle$, representing the binary decimal values $0.125$ and $0.250$ respectively. And sure enough, when we look at the probability distribution generated by our coding example above, we find that the most probable state is $|\,010\,\rangle$, with $|\,001\,\rangle$ coming in second. And finally, using the state $|\,010\,\rangle$ as our approximation to $\theta$, we obtain the following answer for determining $M$:

$$M \;=\; 8\sin^2\Big(\frac{0.25}{2}\Big)$$

$$\approx \; 4$$

Obviously we know that $M = 4$ is incorrect, but such is the consequence of using only three qubits for our binary decimal precision. As it turns out, the fewest number of qubits needed in order to get a correct approximation of $M = 3$ for this particular example is $5$. Alternatively, we could have used the $\phi$ method discussed in lesson 7, using the probabilities generated through repeat measurements to produce an approximation to $\theta$ much closer to $0.21$, which would in turn give us the correct answer of $M = 3$.

Accuracy aside, which are issues inherit to working with QPE, the example above has indeed demonstrated that using $|\,\mathrm{E}_+\,\rangle$ as our eigenstate register will successfully result in solving our Quantum Counting problem. Extracting $\theta$ from $G_\theta$ allows us to solve for $M$, successfully determining the number of marked elements in $N$ without ever knowing what they are. The next question then, is what should we expect from $|\,\mathrm{E}_-\,\rangle$? Mathematically we know that the eigenvalue for this state is $e^{-i\theta}$, but it's not immediately obvious what the effect of -$\theta$ will have on our QPE circuit. In particular, Quantum Phase Estimation is designed such that all of the individual states in the final wavefunction represent binary decimal values from $0$ to $1$. In the previous lesson we saw cases where $\theta$ exceeded $1$, but never an example for $\theta < 0$. Thus, let's test $|\,\mathrm{E}_-\,\rangle$ as our eigenstate register and see if we can make sense of the final answer:

$$|\,\mathrm{E}_-\,\rangle_{\text{norm}} \;=\; \frac{-i}{\sqrt{10}}\Big(|000\rangle + |001\rangle + |010\rangle + |100\rangle + |011\rangle\Big) \;+\; \frac{1}{\sqrt{6}}\Big(|101\rangle + |110\rangle + |111\rangle\Big)$$

In [ ]:

```
Marked  = ['101','110','111']
Q = len(Marked[0])
E_minus = [ -1.0j/m.sqrt(10),-1.0j/m.sqrt(10),-1.0j/m.sqrt(10),1.0/m.sqrt(6),
           -1.0j/m.sqrt(10),1.0/m.sqrt(6),-1.0j/m.sqrt(10),1.0/m.sqrt(6) ]
#=====================================================================
q = QuantumRegister(Q,name='q')
u = QuantumRegister(Q,name='u')
a1 = QuantumRegister(Q-1,name='a1')
a2 = QuantumRegister(1,name='a2')
c = ClassicalRegister(Q,name='c')
qc= QuantumCircuit(q,u,a1,a2,c,name='qc')
#------------------------------------
qc.h( q[0] )
qc.h( q[1] )
qc.h( q[2] )
qc.initialize(E_minus,u)
qc.x( a2[0] )
qc.h( a2[0] )
#-------------------------
for i in np.arange(Q):
    for j in np.arange(2**i):
        oq.C_Grover(qc,q[int(3-(i+1))],u,a1,a2,Marked,proper=True)
#-------------------------
qc.h( a2[0] )
qc.x( a2[0] )
oq.QFT_dgr( qc,q,3,swap=True )
#--------------------------
qc.measure(q,c)
oq.Measurement( qc, shots=10000 )
```

```
181|100>    156|001>    7072|110>    455|101>    120|010>    315|000>    1565|111>    136|011>
```

Looking at the example for $|\,\mathrm{E}_-\,\rangle$ above, it is clear that our measurement distribution does not yield the same probabilities for the states $|\,010\,\rangle$ and $|\,001\,\rangle$ like before, which we already showed are the binary decimal representations closest to our angle $\theta$. Rather, the distribution for $|\,\mathrm{E}_-\,\rangle$ shows the states $|\,110\,\rangle$ and $|\,111\,\rangle$ possessing the greatestest probabilities. And upon closer inspection, we can see that their numerical values are quite similar to the $|\,\mathrm{E}_+\,\rangle$ case, where $|\,110\,\rangle$ is just as probable as $|\,010\,\rangle$ was, followed in suit by $|\,111\,\rangle$.

If we analyze what our probability distribution is telling us about the angle $\theta'$, the adjusted value for $\theta$ by $2\pi$, the results suggest that our phase lies somewhere between $0.750 < \theta' < 0.875$, closer to $0.750$, the binary decimal values for $|\,110\,\rangle$ and $|\,111\,\rangle$ respectively. And, if we take our analysis one step further and assume that because the distribution between $|\,110\,\rangle$ and $|\,111\,\rangle$ is nearly identical to that of $|\,010\,\rangle$ and $|\,001\,\rangle$ from before, we might be inclined to guess that the phase we are approximating is somewhere around $0.79$. Clearly there is a connection between these two cases, and to help illuminate exactly what it is, below is a coding example showing the distributions produced from passing $\theta$ versus -$\theta$ eigenphases through QPE:

In [ ]:

```
theta_vec = [0.79,-0.79]
for i in np.arange(len(theta_vec)):
    theta = theta_vec[i]
    n = 3
    #=====================================
    q1 = QuantumRegister(n,name='q1')
    q2 = QuantumRegister(1,name='q2')
    c  = ClassicalRegister(n,name='c')
    qc = QuantumCircuit(q1,q2,c,name='qc')
    #--------------------------------------
    for i in np.arange(n):
        qc.h(q1[int(i)])
    qc.x( q2[0] )
    phi = 2*m.pi*theta
    for j in np.arange(n):
        for k in np.arange(2**j):
            qc.cu1( phi, q1[int(n-1-j)], q2[0] )
    oq.QFT_dgr( qc,q1,n,swap=True )
    #--------------------------------------
    print('\n___ QPE for Theta = ',theta,' ___')
    qc.measure(q1,c)
    oq.Measurement( qc, shots=10000, systems=[n,2] )
```

```
___ QPE for Theta =  0.79  ___
161|100>    138|001>    7135|110>    453|101>    121|010>    273|000>    1585|111>    134|011>

___ QPE for Theta =  -0.79  ___
151|100>    1577|001>    111|110>    128|101>    7109|010>    306|000>    152|111>    466|011>
```

To summarize the results of the code above, when we run the QPE circuit looking for the positive $\theta$ value (using $R_\phi$ gates as our control-$U$ operators so we can manually control $\theta$), we find the most probable state in our system is $|110\rangle$, corresponding to the binary value $0.75$, followed by the state $|111\rangle$. This is in agreement with what we would expect when looking for a $\theta$ value of $0.79$. Then, when we run the same circuit looking for -$\theta$, we find that the most probable state in our system has switched to $|010\rangle$, followed by $|001\rangle$. And if we compare these probability distributions to those from our previous two examples, you will find (not coincidentally) that they are nearly identical to our cases for $|\text{E}_\pm\rangle$. You may already have a hunch as to what is going on here, but if not, I encourage you to run the cell of code above a few more times with various values for $\theta$. If you do, you will find results along the lines of:

| $\theta$ : | $0.64$ | $-0.64$ | $0.14$ | $-0.14$ |
|---|---|---|---|---|
| Most Probable State: | $\|101\rangle$ | $\|011\rangle$ | $\|001\rangle$ | $\|111\rangle$ |
| Binary Decimal: | $0.625$ | $0.375$ | $0.125$ | $0.875$ |

As the pattern above suggests, using the QPE Algorithm to search for an eigenphase of -$\theta$ results in a binary decimal approximation to $1-\theta$. Conceptually, this result can be explained if we visualize the way in which Quantum Phase Estimation handles eigenvalues outside the bounds of $\left[0,1\right]$:

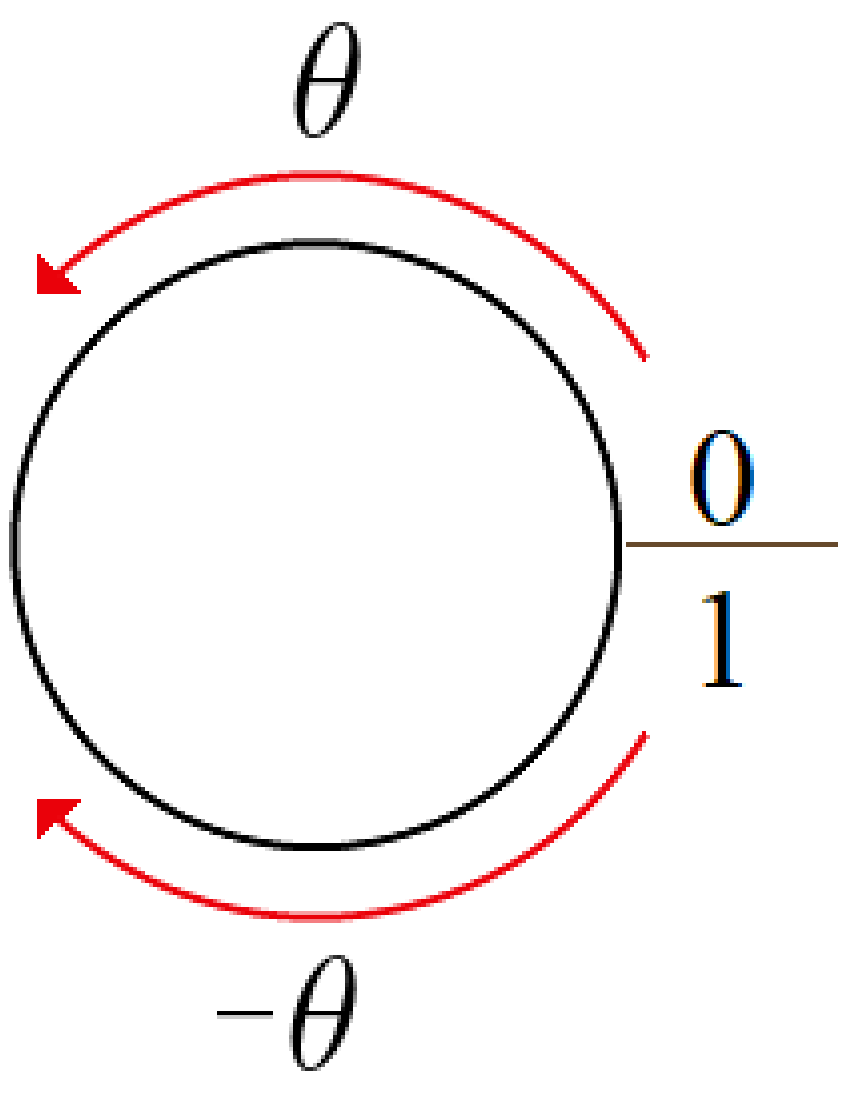

Mathematically, the illustration above is a reflection of the fact that QPE is designed around unitary operators of the form:

$$U|u\rangle \;=\; e^{2\pi i\theta}|u\rangle$$

Because $e^{2\pi i\theta}$ repeats in value for multiples of $2\pi$, this means that we can always choose to represent $\theta$ as a value somewhere in the range $\left[0,1\right]$. As we saw in lesson 7, values of $\theta$ that are greater than $1$ can be reduced by simply subtracting multiples of $1$ until landing somewhere between $0$ and $1$. Similarly, we have now just shown that the same principle applies to values of $\theta < 0$, whereby we can obtain the equivalent phase by adding the necessary amount of multiples of $1$. For example:

$$e^{2\pi i(1.8)} \;=\; e^{2\pi i(1+0.8)} \;=\; e^{2\pi i}e^{2\pi i(0.8)} \;=\; e^{2\pi i(0.8)}$$

$$e^{2\pi i(-0.7)} \;=\; e^{2\pi i(-1+0.3)} \;=\; e^{-2\pi i}e^{2\pi i(0.3)} \;=\; e^{2\pi i(0.3)}$$

Thus, to conclude our example of passing $|\text{E}_-\rangle$ through the QPE circuit from earlier, the result we were seeing was $1-\theta'$, roughly equal to $0.79$. This explains why we were seeing a boost in the state $|110\rangle$, and a mirroring in distribution to that of the positive $\theta'$ case. If we then wanted to use the result from $|\text{E}_-\rangle$ to solve our Quantum Counting problem, all we would need to do is subtract our binary decimal result from $1$.

$$1-\theta' \;\approx\; 0.79$$

$$\therefore\; \theta' \;\approx\; 0.21$$

$$\vdots$$

$$M = 3$$

So long as we know that the eigenstate we are passing through the QPE circuit has a negative $\theta$ eigenvalue, we can simply correct for this negative phase afterwards. However, if we *don't* have information about the nature of our eigenstate / eigenvalue, then post-processing the information becomes much more difficult. We shall see this issue arise in the next section when we complete the Quantum Counting Algorithm using the state $|s\rangle$, and the way in which we distinguish our measurement result as coming from $\theta$ or -$\theta$.

## Extracting $\theta$ Using $|s\rangle$

Having now see the individual cases of using $|E_\pm\rangle$ as our eigenstate register, we are now equipped to study their combined effect through the implementation of $|s\rangle$. To quickly recap, either eigenstate can be used as a viable solution for determining $M$, so long as we know how to interpret the binary decimal representation of our answer. Thus, to start our study of what will happen when we pass a superposition of these states through the eigenstate register, let's rewrite the decomposition of the equal superposition state in terms of the eigenbasis of $G_\theta$:

$$|s\rangle = \sigma_+|E_+\rangle + \sigma_-|E_-\rangle$$

$$\sigma_\pm = \frac{1}{2}\left(\sqrt{\frac{M}{N}} \mp i\sqrt{\frac{N-M}{N}}\right)$$

Individually, $G_\theta$ acting on either of the eigenstates $|E_\pm\rangle$ results in an application of the phase $e^{\pm i\theta}$. Thus, when applied to a superposition of these state, each one receives their respective eigenphase:

$$G^n|s\rangle = e^{in\theta}\sigma_+|E_+\rangle + e^{-in\theta}\sigma_-|E_-\rangle$$

As we've already alluded to in previous sections, the final result from using $|s\rangle$ as our eigenstate will be a superposition of phases picked out by our QPE circuit. As shown in the expression above, multiple applications of $G_\theta$ results in $e^{\pm i\theta}$ phase applications on the respective eigenstates. And while this may look similar to the way in which we normally think of applications of the control-$U$ operator acting on $|u\rangle$, it is important to note that each application of $G_\theta$ is changing the quantum state of $|u\rangle$. But, if we follow our eigenstate register in the $\{|E_+\rangle, |E_-\rangle\}$ basis, we can see that our superposition is essentially preserved throughout the entire circuit. To show this, let's define two quantum states, corresponding to our precision qubit register after the control-$G_\theta$ and QFT$^\dagger$ steps:

$$\text{Initial State}: \quad |s\rangle \otimes |E_\pm\rangle$$

$$\text{After control-}G_\theta: \quad |G_\pm\rangle \otimes |E_\pm\rangle$$

$$\text{After } QFT^\dagger: \quad |\omega_\pm\rangle \otimes |E_\pm\rangle$$

Using $|G_\pm\rangle$ and $|\omega_\pm\rangle$ to represent the state of our precision qubits for the cases corresponding to $|E_\pm\rangle$, we can track the state of our system when using $|s\rangle$ as it moves through the QPE circuit as follows:

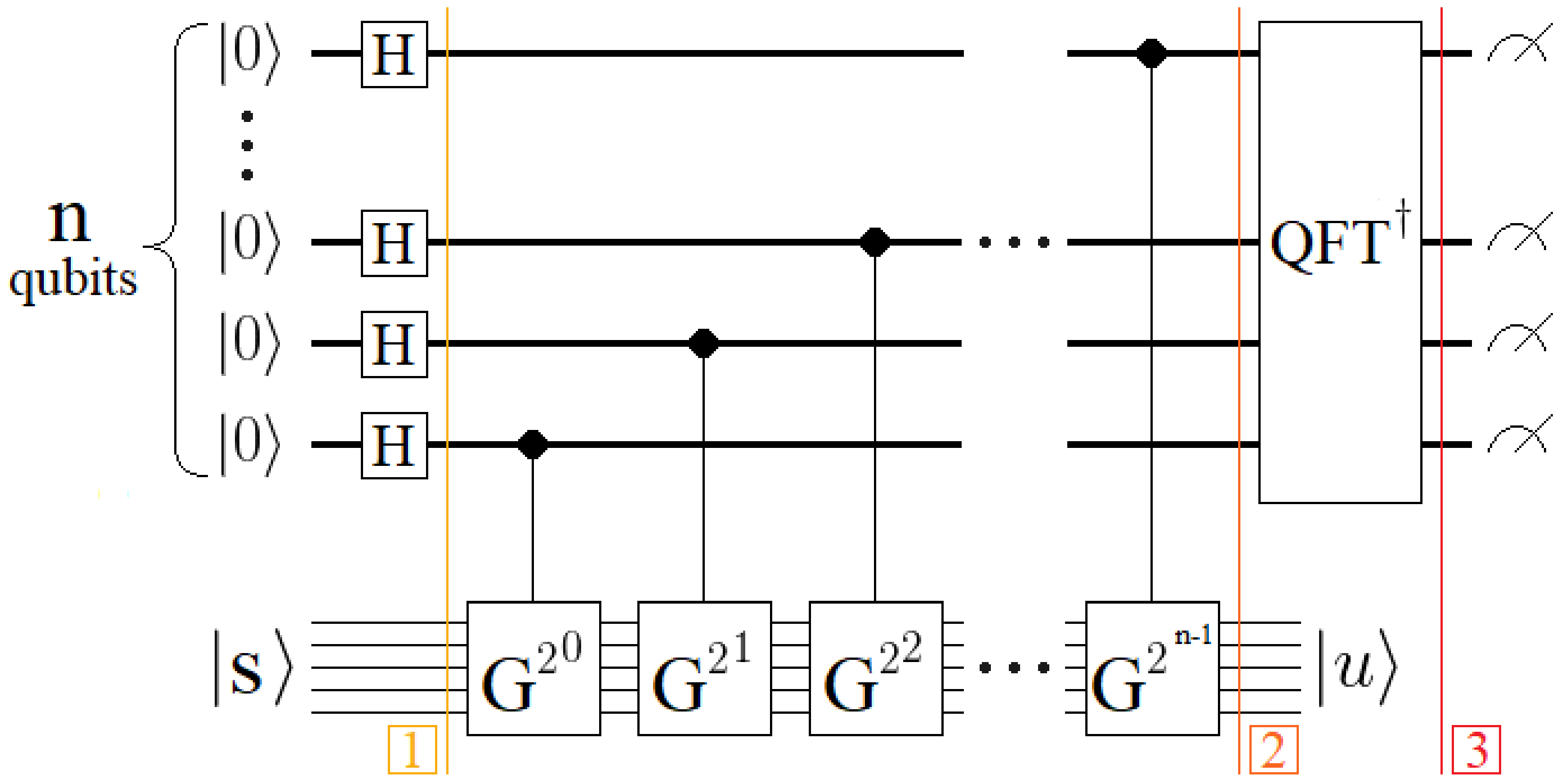

$$① \quad |s\rangle \otimes (\sigma_+|E_+\rangle + \sigma_-|E_-\rangle)$$

$$② \quad \sigma_+|G_+\rangle|E_+\rangle + \sigma_-|G_-\rangle|E_-\rangle$$

$$③ \quad \sigma_+|\omega_+\rangle|E_+\rangle + \sigma_-|\omega_-\rangle|E_-\rangle$$

As shown in the final state above (3), the superposition between $|E_\pm\rangle$ persists through each step of the algorithm, resulting in a state on our precision qubits that contains both $|\omega_+\rangle$ and $|\omega_-\rangle$. These $|\omega_\pm\rangle$ states represent the boosted probability wavefunctions which we've already seen, now combined together through superposition. This combined state contains the necessary information for extracting $\theta$ from either $|\omega_+\rangle$ or $|\omega_-\rangle$, but now we must be careful in how we distinguish between the two. One state contains a highly probable approximation to $\theta$, while the other -$\theta$. And because we are always confined to the $\{|0\rangle, |1\rangle\}$ basis, there's potential that our measurement result can come from either state.

As it turns out, if we perform our Quantum Counting Algorithm as shown above and are asked to make a measurement on the precision register, there is a simple way we could conclude whether the result of our measurement should be interpreted as coming from $|\omega_+\rangle$ or $|\omega_-\rangle$. Even though the final state is essentially a $50$-$50$ superposition between both answers, the way in which each eigenvalue manifests itself through the QPE circuit ends up being distinct. Or more specifically, $|\omega_\pm\rangle$ never produce a boost in probability to the same state. To show this, below are all of the values for $M$ from $0$ through $8$ for our $N = 3$ examples from earlier, and the corresponding $\theta'$ values from using either $|\text{E}_\pm\rangle$ as the eigenstate:

| M : | 0 | 1 | 2 | 3 | 4 | 5 | 6 | 7 | 8 |
|---|---|---|---|---|---|---|---|---|---|
| $\theta$ : | 0.0 | 0.7227 | 1.0472 | 1.3181 | 1.5708 | 1.8235 | 2.0944 | 2.4189 | 3.1416 |
| $\lvert\text{E}_+\rangle\ \ \theta'$ : | 0.0 | 0.1150 | 0.1667 | 0.2098 | 0.2500 | 0.2902 | 0.3333 | 0.3850 | 0.5000 |
| $\lvert\text{E}_-\rangle\ \ \theta'$ : | 1.0 | 0.8850 | 0.8333 | 0.7902 | 0.7500 | 0.7098 | 0.6667 | 0.6150 | 0.5000 |

As you can see, regardless of the value of $M$, the corresponding $\theta'$ that we will get out from our algorithm is always unique. Thus, depending on whether the binary decimal value of our final measurement lies above or below $0.5$ perfectly determines whether we interpret our eigenphase as coming from $|\text{E}_+\rangle$ or $|\text{E}_-\rangle$, fully solving our Quantum Counting Algorithm! To test this for yourself, below is a code example which randomly selects a value of $M$ and tries to correctly identify it based on the final measurement:

In [ ]:

```
Precision = 4
Grover_Size = 4
#---------------
N = 2**Grover_Size
M = int( 1 + (2**Grover_Size-2)*random.random() )
Marked = []
for m1 in np.arange(2**Grover_Size):
    Marked.append( oq.Binary(int(m1),2**Grover_Size,'R') )
for m2 in np.arange( (int(2**Grover_Size)-M) ):
    Marked.pop()
#=========================================
q = QuantumRegister(Precision,name='q')
u = QuantumRegister(Grover_Size,name='u')
a1 = QuantumRegister(Grover_Size-1,name='a1')
a2 = QuantumRegister(1,name='a2')
c = ClassicalRegister(Precision,name='c')
qc= QuantumCircuit(q,u,a1,a2,c,name='qc')
#-----------------------------------------
for p in np.arange(Precision):
    qc.h( q[int(p)] )
for g in np.arange(Grover_Size):
    qc.h( u[int(g)] )
qc.x( a2[0] )
qc.h( a2[0] )
#-------------------------
for i in np.arange(Precision):
    for j in np.arange(2**i):
        oq.C_Grover(qc,q[int(Precision-(i+1))],u,a1,a2,Marked,proper=True)
#-------------------------
qc.h( a2[0] )
qc.x( a2[0] )
oq.QFT_dgr( qc,q,Precision,swap=True )
qc.measure(q,c)
Meas = oq.Measurement( qc, shots=30000, return_M=True, print_M=True )
#-------------------------
print('\nCorrect M:  ',len(Marked),'   \u03B8:  ',round(2*np.arcsin(m.sqrt(len(Marked)/(2**Grover_Size)))/(2*m.pi),5))
C,S = oq.Most_Probable(Meas,1)
theta_QPE = oq.From_Binary(S[0],'R')/N
print('\nMost Probable State:  |'+str(S[0])+'>      ---->      \u03B8:  ',round(theta_QPE,4))
if( theta_QPE >= 0.5 ):
    theta_QPE = 1 - theta_QPE
    print('\nInterpretting Measurement as  1 - \u03B8:  ',round(theta_QPE,4),'     Corresponding M:  ',int(round(2**Grover_
else:
    print('\nInterpretting Measurement as  \u03B8:  ',round(theta_QPE,4),'     Corresponding M:  ',int(round(2**Grover_Size
```

```
53|1101>    50|1110>    304|0101>    309|1011>    93|0100>    34|0010>    560|1001>    52|0011>    30|1111>    105|1100>
13876|0110>    218|1000>    13703|1010>    31|0001>    554|0111>    28|0000>

Correct M:   14    θ:   0.38497

Most Probable State:  |0110>      ---->      θ:   0.375

Interpretting Measurement as  θ:   0.375      Corresponding M:   14
```

Take a look at the measurement results in the example above, showing that the states $|0110\rangle$ and $|1010\rangle$ have received the largest boosts in probability as a result of our full Quantum Counting circuit. As promised, these two states represent binary decimals which are above and below $0.5$, and both lead to the same approximation to $M$:

$$|0110\rangle \longrightarrow \theta' = 0.375 \longrightarrow M = 14$$

$$|1010\rangle \longrightarrow \theta' = 0.625 \longrightarrow 1 - \theta' = 0.375 \longrightarrow M = 14$$

To understand why the eigenstates $|\text{E}_\pm\rangle$ only result in $\theta'$ values above or below $0.5$, we need only return to our expression for $\theta$ from earlier:

$$\theta = 2\sin^{-1}\Big(\sqrt{\frac{M}{N}}\Big)$$

The $\theta$ shown above is the eigenvalue to $G_\theta$, corresponding to a single Grover iteration. Under normal circumstances the eigenvalue range of the $2 \times 2$ rotational matrix is from $0$ to $2\pi$, or $0$ to $1$ when converted into the eigenvalue form for QPE. However, the context of our counting problem poses different constraints, namely the number of possible marked elements in $N$:

$$0 \leq M \leq N$$

Conceptually the statement above should make sense, as we can only ever have as many marked elements in our list as we do total elements. Mathematically, these bounds are in agreement with the $\sin^{-1}()$ function in the expression above, which is argument bounded between $0$ and $1$. Substituting these limits into our $\theta$ expression, we arrive at our answer:

$$0 \leq \theta \leq \pi$$

which when converted into the form of our QPE algorithm:

$$0 \leq \theta' \leq 0.5$$

Thus, when we use either $|\,\mathrm{E}_\pm\rangle$ state as the eigenstate for our QPE circuit, $\theta$ will result in a binary decimal approximation between $0$ and $0.5$, while -$\theta$ will be between $0.5$ and $1$. This in turn means that the quantum states of our system are split right down the middle, half susceptible to a boost in probability from $|\,\mathrm{E}_+\rangle$, the other half from $|\,\mathrm{E}_-\rangle$, with no overlap between the two.

## Final Remarks

The example above represents the complete Quantum Counting Algorithm, so I encourage you to try out several different sized problems, both in the size of the search as well as the number of precision qubits. Just as with the examples in lesson 7, you will find that the approximate value for $M$ is not always correct, typically off by at most $1$ but sometimes more. We can always improve the accuracy of our results using the techniques discussed in lesson 7, such as increasing the number of precision qubits or determining $\phi$ through repeat measurements. But it is important to point out that because of the way in which our Quantum Counting Algorithm splits the correct final answer between two states, coming from $|\,\omega_\pm\rangle$ respectively, the accuracy of our QPE circuit is essentially cut in half as well. More specifically, our problem boils down to distinguishing between $M$ different possible $\theta'$ values using a quantum space of $2^Q$ states, where $Q$ is the number of qubits in the precision register. But because there are two possible $\theta'$ values for each $M$, we end up being forced to identify $M$ phases using only $2^{Q-1}$ states on either half of the $\theta' = 0.5$ divide. Thus, the true cost for using $|\,s\,\rangle$ as our eigenstate register is a $50\%$ reduction in accuracy, which we must compensate for by adding an additional precision qubit.

As a final note for this lesson, it is worth pointing out the significance of two key concepts that we have demonstrated about the QPE algorithm, namely that it is not necessarily limited to the restrictions we initially placed on it in lesson 7. Firsty, we have shown that the control-$U$ operators in QPE do not necessarily have to be operators, but rather can be full quantum circuits themselves. Essentially, the net effect of any number of gates in succession can be mathematically represented as a single operation $U$, for which our QPE circuit can then find its eigenphase. And secondly, we have also shown that we do not always have to use an eigenstate $|\,u\,\rangle$ of our control-$U$ operator in order to extract eigenphases. In fact, we can in principle use $any$ state as our $|\,u\,\rangle$, and mathematically translate it into the eigenbasis of $U$ to see the expected outcome. This is particularly powerful because it means that we are not barred from using control-$U$ operators that may have eigenstates which are otherwise too complicated or difficult to prepare. Overall, Quantum Counting has demonstrated that there is certainly more than meets the eye when it comes to Quantum Phase Estimation and the flexibility in problems it can solve.

---

This concludes lesson 7.1, and our discussion about the Quantum Counting Algorithm. Additionally, this marks our second encounter with the Quantum Phase Estimation Algorithm, hopefully providing a deeper understanding of some of its inner workings and potential. Throughout this lesson we saw the power of merging two major algorithms together: Grover's and QPE, and how the combination of two unique quantum algorithms can be combined to solve an entirely new problem!

I hope you enjoyed this lesson, and I encourage you to take a look at my other .ipynb tutorials!

## Citations

**[1]** G. Brassard, P. Hoyer, A. Tapp, "Quantum Counting", 25th Intl. Colloquium on Automata, Languages, and Programming **1443** (1998)

# Lesson 8 - Shor's Algorithm

In this lesson, we will cover the famed Shor's Factoring Algorithm, as well as its implications. At its core, we shall see that Shor's Algorithm is really a quantum algorithm for period finding, which cleverly allows for the factoring of large numbers. Due to its complexity, this algorithm will require us to explore numerous mathematical side topics in order to understand all of the moving parts that make up Shor's. If you have not already, I recommend reading lesson 6 before proceeding, as the Quantum Fourier Transformation will play an important role:

Lesson 6 - Quantum Fourier Transformation

---

In order to make sure that all cells of code run properly throughout this lesson, please run the following cell of code below:

In [ ]:
```
from qiskit import ClassicalRegister, QuantumRegister, QuantumCircuit, Aer, execute
import Our_Qiskit_Functions as oq
import numpy as np
import math as m
import random
import matplotlib
import matplotlib.pyplot as plt
S_simulator = Aer.backends(name='statevector_simulator')[0]
```

## Shor and RSA Encryption

When Peter Shor published his algorithm for factoring in 1995 **[1]**, it ignited the field of quantum computer science, and is still a large motivator for many of today's leading quantum computing efforts. And yet even after two decades, we still aren't anywhere close to being able to realize Shor's Algorithm in any practical sense. And if we're being honest, we're probably a couple decades away still. So how is it that one factoring algorithm changed the course of computer science so drastically? The answer: because the realization of no other quantum algorithm would be as globally impactful as Shor's. So much so, that many see the realization of Shor's Algorithm as a venture worth pursuing regardless of how many years it will take. And the reason for that, is security.

In 1978, Ron Rivest, Adi Shamir, and Leonard Adleman were the first to publicly describe an algorithm for how to encrypt and decrypt information in a novel manner **[2]**. Their algorithm describes a process for encryption which allows for one-way transmission of information with high security. This encryption scheme, now referred to as RSA after the founders, is the most widely used encryption method in modern technology. Thus, because this RSA cryptosystem is so abundantly used, it would be a natural alarming reaction if someone were to come along and claim they had a way to crack the encryption. Without trying to sound overly dramatic, that is essentially what Peter Shor did when he came up with his algorithm. And as we already mentioned, that's why it sparked tremendous interest into the field of quantum computing. Shor showed that through the use of a fully functioning quantum computer, one could crack RSA encryption exponentially faster than any classical means.

## Simple Example of RSA

Now that our brief history detour is over, let's look at the way in which RSA encryption works, and the core math element behind it. If you would like to jump straight to the beginning of Shor's Algorithm instead, feel free to skip to the section "Overview of Shor's Algorithm". Ultimately our goal is to show why Shor's Algorithm is a threat to the RSA scheme, and to do that we'll need to work through a simple example. The trick to encrypting messages in RSA is through modular exponentiation:

$$(m^e)^d \pmod{n}$$

The four quantities in the equation above: $m$, $e$, $d$, and $n$, all represent the components that go into the process of encrypting/ decrypting a message. Starting in order, $m$ is the message that we want to encrypt. Because we are working with numbers, all messages must be converted to some kind of numerical representation before being encrypted. As you shall see in the coming code example, representing a message with numbers can be as simple as 'a'= $0$, 'b'= $1$, ... etc. There are of course more complex ways to encode messages with numbers, but that is not our interest with RSA here.

The next two quantities, $e$ and $d$, represent the "keys" to a given RSA encryption. In combination with $n$, they are numbers which allow a person access to the encrypting / decrypting of messages for a chosen RSA. The number $e$ is known as the "public key", and is given to anyone who wishes to encrypt their message. The number $d$ is known as the "private key", and is only kept by those who are trusted to decrypt messages. People with $e$ have the ability to $encrypt$ and send messages, but only people with $d$ have the ability to $decrypt$ and read messages, hence the choice in variables $e$ and $d$.

The last quantity, $n$, is a number which can be thought of as the identity of a given RSA encryption. Suppose we have two parties who wish to simultaneously use RSA encryption, and may even accidentally choose the same $e$ or $d$ values. As long as the two encryptions have unique $n$'s, and both parties keep their $n$ values secret, then even if their messages were to somehow get crossed, neither one could eavesdrop on the other. And as we already mentioned, even if one party were to steal the $n$ from the other, they still couldn't decrypt any messages without the private key $d$.

To summarize, the security of RSA encryption is as strong as a party's ability to keep the quantities $e$, $d$, and $n$ secret ($d$ most of all). Those trusted with encrypting and sending messages are given the quantities $e$ and $n$, while those trusted with decrypting messages are given $d$ and $n$. The full mathematical process for encrypting and decrypting messages is a bit more cumbersome than simply picking values for $n$, $e$, and $d$, so let's now take a look at an example of the steps behind RSA encryption:

**Step 1 :** **Select two prime numbers $p$, $q$**

$$\begin{aligned} p &= 3 \\ q &= 11 \end{aligned}$$

**Step 2 :** **Compute $n$ and $\phi$**

$$n = p \cdot q = 33$$
$$\phi = (p-1)(q-1) = 20$$

❗ Important: $p$, $q$, and $\phi$ are all kept secret ❗

**Step 3 :** **Choose $e$** **(public key)**

• $e$ must be satisfy the conditions:

$$GCD(e, p-1) = GCD(e, q-1) = 1$$

$\therefore\ e = 3$ (multiple possible choices for $e$)

**Step 4 :** **Compute $d$** **(private key)**

• $d$ must be satisfy the condition:

$$e \cdot d = 1 \pmod{\phi}$$
$$\therefore\ d = 7$$

Once we have $n$, $e$, and $d$ subject to the constraints outlined above, we are ready to encrypt and decrypt messages. For example, let's suppose that the message we would like to send is $m = 6$ (remember all messages must be converted to numbers in the RSA protocol). Anyone with the public key $e$ can encrypt a message as follows:

$$\begin{aligned} m^* &= m^e \pmod{n} \\ &= 6^3 \pmod{33} \\ &= 18 \end{aligned}$$

Once the message has been encrypted, only those with access to the private key $d$ can decrypt the message:

$$\begin{aligned} m' &= m^{*d} \pmod{n} \\ &= 18^7 \pmod{33} \\ &= 6 \end{aligned}$$

In the example above, the message $m$ was successfully encrypted into $m^*$, and then decrypted back to $m'$. And as we can see, $m = m'$ indicates that our encryption scheme was successful. Now let's see the RSA scheme in action with some code, where we can choose an arbitrary message to send:

In [ ]:

```python
def Letter_Code(x):
    '''
    Input: integer  -->  Converts an integer between 0 and 26 into a letter of the alphabet (26 for space)
    Input: string   -->  Converts a lower case letter or space to an integer
    '''
    if( type(x) == type(1) ):
        code = ['a','b','c','d','e','f','g','h','i','j','k','l','m','n','o','p','q','r','s','t','u','v','w','x','y','z',' '
        if( x < len(code) ):
            return code[x]
        else:
            return '?'
    if( type(x) == type('s') ):
        code = {'a':0,'b':1,'c':2,'d':3,'e':4,'f':5,'g':6,'h':7,'i':8,'j':9,'k':10,'l':11,'m':12,'n':13,'o':14,'p':15,'q':1
        return code[x]

#====================================
p = 3
q = 11
n = p*q
e = 3
d = 7
message = 'hello qubits how are you'
#------------------------------------   Encrypt the message
M = list(message)
M_rsa = []
for i in np.arange(len(M)):
    M[i] = Letter_Code(M[i])
    M_rsa.append( M[i]**e%n )
#------------------------------------   Decrypt the message
encrypted_message = ''
decrypted_message = ''
for j in np.arange(len(M_rsa)):
    encrypted_message = encrypted_message+Letter_Code( M_rsa[j] )
    decrypted_message = decrypted_message+Letter_Code( (M_rsa[j]**d)%n )

print('  Encoded Messege:  ',M,'\n\nEncrypted Messege:  ',M_rsa,'\n\n                    ',encrypted_message,'\n\nDecrypted
```

```
  Encoded Messege:   [7, 4, 11, 11, 14, 26, 16, 20, 1, 8, 19, 18, 26, 7, 14, 22, 26, 0, 17, 4, 26, 24, 14, 20]

Encrypted Messege:   [13, 31, 11, 11, 5, 20, 4, 14, 1, 17, 28, 24, 20, 13, 5, 22, 20, 0, 29, 31, 20, 30, 5, 14]

                     n?llfueobr?yunfwua??u?fo

Decrypted Messege:   hello qubits how are you
```

The cell of code above is a simple demonstration of sending an encrypted message using RSA. We represent each letter of the alphabet with an integer, and pass those values through our encryption step: $m^e \pmod{n}$. Both the coded and encrypted versions of the message are shown above. Then, we pass the list of encrypted values to the decryption step: $m^{*d} \pmod{n}$, which is only possible by someone with the private key $d$. And lastly we print the message after decryption, confirming that it does indeed match the message we sent.

### Cracking RSA Encryption

Now suppose we want to crack a given RSA encryption, essentially eavesdropping on messages not intended for us. The easiest two pieces of information to get hold of are the values of $e$ and $n$, which we will assume were stolen off someone with only a "public" level of access. And we will also assume that stealing $d$ is not an option (hence why it is only given to those considered trusted), thus leaving us with the question of what can we do with $e$ and $n$ to crack the encryption?

If you review the math steps above, the critical element to cracking an RSA encryption boils down to factoring $n$. If we are able to factor $n$, giving us the two prime numbers $p$ and $q$, then determining $d$ becomes a simple task for a classical computer (provided we also have $e$). But factoring $n$ is no small task, as typical RSA encryption works with $n$'s that are of the order hundreds to thousands of digits long, which are essentially impossible to factor with classical computing in any realistic time frame. Thus, after our lengthy discussion of RSA, we have finally come to the critical fact where our Shor's algorithm comes into play:

$\star$ Shor's algorithm allows for the factoring of $n$ exponentially faster than any classical means, consequently cracking RSA encrypted messages. $\star$

Thus, the purpose for going into RSA encryption in such detail was to give a very concrete example of where a quantum factoring algorithm could be impactful (probably *the* most impactful quantum algorithm to date). Because security is such an important issue, the realization of Shor's Algorithm is one of the top priorities for the progression of quantum computers.

## Classical Factoring - Congruence of Squares

In order to appreciate the speedup that can be obtained through Shor's Algorithm, and the reason why RSA encryption is so secure, let's see how a classical computer solves factoring problems. We will briefly discuss a technique known as 'Congruence of Squares', also known as Fermat's factorization method, which is the basis for several integer factorization algorithms. The congruence of squares technique is based around finding two numbers $X$ and $Y$ that satisfy the following:

$$X^2 - Y^2 = N$$

where $N$ is the number we are trying to factor. If we can successfully find two squares $X^2$ and $Y^2$ that satisfy this condition, then we can factor the equation shown above into:

$$(X+Y)(X-Y) = N$$

which gives us two factors of $N$. For example, suppose we wanted to factor the number $72$:

$$121 - 49 = 72$$

$$11^2 - 7^2 = 72$$

$$(11-7)(11+7) = 72$$

$$\text{factors:} \quad 4, 18$$

By using the squares $11^2$ and $7^2$, we are able to find two factors of $N$, which in turn require further factorization if either of them aren't prime numbers. Now, we must point out that this factoring technique is entirely reliant on finding a combination of $X$ and $Y$ that works, which in practice is very slow. Without any intuition about what combination of $X$ and $Y$ will solve your problem, we are stuck searching through all combinations exhaustively. And if we compare this to the most basic form of factoring: exhaustively checking prime numbers for factors of $N$, we have essentially just swapped out one blind search for another.

### mod N

For the purpose of our study of Shor's Algorithm, we will now look at a second classical factoring technique based on the Congruence of Squares. As we shall see, this second technique will be analogous to the way in which our quantum factoring algorithm works. Similar to the difference of squares equation shown above, this second approach begins with the following equation:

$$X^2 \pmod{N} = Y^2$$

Rather than looking for two perfect squares, this new technique will only require us to search for one: $X^2$. However, in searching for the right value we must satisfy the condition that $X^2$ is equal to a second perfect square, mod $N$. Once we find such an $X^2$, then the real quantity of interest is the following equation:

$$X^2 - Y^2 \pmod{N}$$

Now, it is very important to point out that the two equations above are *not* equivalent! If you were to remove the (mod $N$) condition, then they would be, but the presence of the (mod $N$) means that we must obey modulo algebra. Because they not equal, but in fact two independent equations describing our conditions on $X^2$ and $Y^2$, we can perform some modulo arithmetic steps in order to derive a new equation:

$$\begin{aligned} X^2 \;-\; Y^2 \pmod N \;&=\; \left[X^2 \pmod N\right] + \left[-Y^2 \pmod N\right] \pmod N \\ &=\; \left[Y^2\right] + \left[N - Y^2\right] \pmod N \\ &=\; N \pmod N \\ &=\; 0 \end{aligned}$$

Since modulo arithmetic isn't something most people use frequently, I encourage you to review some available resources and work through the steps above for yourself to verify the answer we've arrived at. After following the steps shown above, we now have the equation that will allow us to find factors of $N$:

$$X^2 \;-\; Y^2 \pmod N \;=\; 0$$

And just like our first congruence of squares technique, our next step will be to factor the left hand side of the equation:

$$(X+Y)(X-Y) \pmod N \;=\; 0$$

Note that because the entire quantity $X^2 + Y^2$ is under (mod $N$), we can factor the left hand side of the equation as normal. In order to appreciate what this modulo equation is telling us, let's quickly remind ourselves about the meaning behind (mod $N$). When a quantity $Q$ is equal to $0$ (mod $N$), we can say that " $N$ divides Q", or in other words the quantity $Q/N$ is an integer. The only way for this to be true is if all of the factors that make up $N$ are also in $Q$:

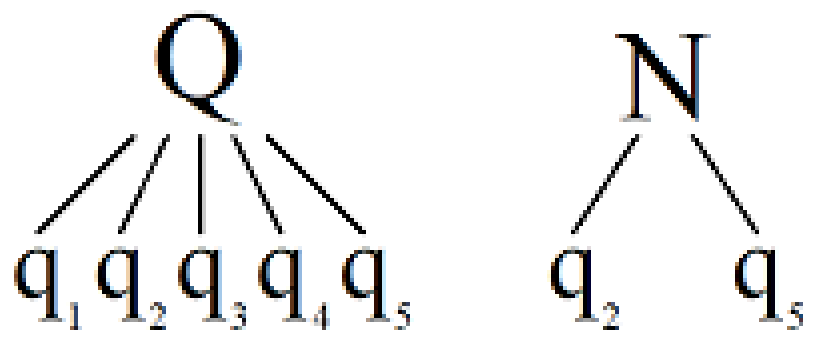

$$\frac{Q}{N} \;=\; \frac{q_1 \cdot q_2 \cdot q_3 \cdot q_4 \cdot q_5}{q_2 \cdot q_5} \;=\; q_1 \cdot q_3 \cdot q_4$$

So then, if we interpret what the previous two modulo equations are telling us:

- $N$ divides the quantity $X^2 - Y^2$
- The quantity $X^2 - Y^2$ has the factors: $(X+Y)$ and $(X-Y)$

$\therefore$ $N$ must have factors in common with $(X+Y)$ and $(X-Y)$

The points listed above form the core idea behind this factoring technique. To summarize, by finding a value $X^2$ that satisfies our perfect squares modulo condition, we are then guaranteed to have two numbers that share a common factor with $N$. After obtaining the two quantities $(X+Y)$ and $(X-Y)$, the $GCD$ (greatest common denominator) between these numbers and $N$ we will successfully solve our factoring problem.

To test this factoring technique, try putting in values for $N$ into the cell of code below:

In [ ]:

```
N = 1703
S = m.ceil(m.sqrt(N))
#====================
i = 0
found = False
while( (i<10000) and (found==False) ):
    Y2 = (S+i)**2 % N
    if( ( m.sqrt(Y2) == m.floor(m.sqrt(Y2)) ) and (Y2!=1) ):
        found = True
        X = int(S+i)
        X2 = int((S+i)**2)
        Y = m.sqrt(Y2)
        Y2 = int(Y2)
    else:
        i = i + 1
if( found==True ):
    print('N: ',N)
    print('\nX^2: ',X2,'    Y^2: ',Y2)
    print('\n(X+Y): ',int(X+Y),'    (X-Y): ',int(X-Y))
    print('\nfactors of N:  ',int(oq.GCD(N,X+Y)),'  ',int(oq.GCD(N,X-Y)) )
```

```
N:  1703

X^2:  9604     Y^2:  1089

(X+Y):  131

factors of N:   131
```

The example above successfully finds two factors of $N = 1703$, using the values $X = 98$ and $Y = 33$. It's important to note that neither of these values share a common factor with , but their sum and difference do.

Having now seen the success of this factoring technique, we must once again ask what is the bottleneck. As we already mentioned, the problem with a classical factoring algorithm is that it must exhaustively search for values. In the most basic factoring scheme, you start at $2$ and work your way up all of the prime numbers until you find a factor of $N$. In the modular Congruence of Squares method shown above, we are essentially stuck with the same issue of exhaustively searching for the quantity $X^2$. However, this blind search is typically faster than our previous technique thanks to the use of modulo $N$, and the fact that we only need to find one quantity.

To return to a point made at the beginning of this section, there is an important similarity between the classical factoring technique shown above and Shor's. The speedup we are able to obtain using modulo $N$ in the Congruence of Squares technique is largely due to the fact that we can determine factors of $N$ using $GCD$ computations. The reason for this is because computing $GCD$'s is very fast classically, which will be our next topic.

## Euclid's Algorithm

As we shall in the coming outline of Shor's Algorithm, computing GCDs is a critical ingredient necessary for the factorization of $N$. However, there is a seemingly paradoxical problem in saying that calculating GCDs is going to speed up our factorization problem:

"In order to find the greatest common denominator between two numbers, don't we need to know the factors of those numbers. And if so, won't our GCD computation boil right back down to another factorization problem?"

The statement above is a perfectly rational logic to follow if we assume that GCD computations require knowledge about the factors which make up the two numbers in question. However, it is for this exact reason why the algorithm we are about to discuss is so powerful, as it will allow us to compute GCDs without ever having to factor either number. First described by Euclid around 300 B.C. (and still going strong to this day), and consequently named after him, Euclid's Algorithm is based on the following observation between any two numbers and their GCD:

$$A > B$$

$$\text{GCD}(A, B) = \text{GCD}(A - B, B)$$

Stated simply, the GCD between two numbers $A$ and $B$ is the same as the GCD between their difference and the smaller of the two numbers. For example:

$$A = 78 \qquad B = 36 \qquad A - B = 42$$

$$\text{GCD}(78, 36) = 6$$

$$\text{GCD}(42, 36) = 6$$

As we can see, both $78$ and $42$ share the same GCD with $36$. And if we consider which of these two problems we would prefer to tackle, obviously we would always choose the GCD between the two smaller numbers. Thus, based on this observation by Euclid, his algorithm describes a process by which we can recursively keep taking the difference between numbers to reduce any GCD computation down to its minimum. To see this, suppose we were to take our example above even further:

$$42 - 36 = 6 \longrightarrow \text{GCD}(6, 36) = 6$$

$$36 - 6 = 30 \longrightarrow \text{GCD}(6, 30) = 6$$

$$30 - 6 = 24 \longrightarrow \text{GCD}(6, 24) = 6$$

$$\vdots$$

$$12 - 6 = 6 \longrightarrow \text{GCD}(6, 6) = 6$$

$$6 - 6 = 0$$

If we continuously follow Euclid's recursive method, we eventually arrive at the answer to our GCD problem when our subtractions lead us to 0. Note that in all of the examples above we included the GCD computation at each step just to show that Euclid's observation always holds, but in actuality we never compute any GCD's, only subtractions. Below is a simple code example which demonstrates Euclid's Algorithm in action:

In [ ]:
```
A_i = 462
B_i = 70
#---------------------
gcd = False
GCD = 0
#=====================
A = A_i
B = B_i
while( gcd == False ):
    r = A - B
    print('A:  ',A,'     B:  ',B,'      r:',r)
    if( r == 0 ):
        gcd = True
        GCD = B
    else:
        if( r > B ):
            A = r
        else:
            A = B
            B = r
print('------------------------------------------------------\nGreatest Common Denominator between',A_i,'and',B_i,': ',GCD)
```

```
A:   462      B:   70      r: 392
A:   392      B:   70      r: 322
A:   322      B:   70      r: 252
A:   252      B:   70      r: 182
A:   182      B:   70      r: 112
A:   112      B:   70      r: 42
A:   70      B:   42      r: 28
A:   42      B:   28      r: 14
A:   28      B:   14      r: 14
A:   14      B:   14      r: 0
------------------------------------------------------
Greatest Common Denominator between 462 and 70 :  14
```

Once again, the significance of the GCD algorithm shown above is that we *never* have to find any factors of the two numbers $A$ and $B$. Instead, our problem is 100$\%$ converted into arithmetic calculations, which are essentially what classical computers are designed to do faster than anything else! This is the power behind Euclid's Algorithm, and similarly Shor's as we shall see. In essence, if we can cleverly convert *any* computationally difficult problem into one which only requires GCD calculations, then we can expect enormous speedups.

Even though the code above arrives at the correct answer in relatively few steps, you've probably noticed in the two examples thus far that there is some redundancy in Euclid's Algorithm, namely subtracting by the same value numerous times. For small numbers like the calculations above, these extra steps aren't too costly, but once the numbers in question for the GCD computation become large enough, these redundant steps can become problematic. Thus, in the spirit of optimal algorithms, we can improve upon Euclid's algorithm further by making use of a second observation:

$$\text{GCD}(\,A, B) \;\; = \;\; \text{GCD}(\,A - q \cdot B, B\,)$$

where $q$ is an integer that satisfies:

$$A \;>\; q \cdot B$$

The two statements above are essentially guaranteeing the equivalency in GCDs when subtracting the same value numerous times consecutively in Euclid's algorithm. For example, when computing the GCD for $A = 462$ and $B = 70$ like in the cell of code above, the statement guarantees that the following two GCD computations are equal:

$$\text{GCD}(462, 70) \;\; = \;\; \text{GCD}(42, 70)$$

where the value $42$ is obtained by subtracting off $6$ multiples of $70$ from $462$:

$$q \;=\; 6 \quad\longrightarrow\quad 462 \;>\; 420 \quad \checkmark$$

In essence, we will arrive at this step following Euclid's Algorithm, subtracting $70$ six consecutive times in a row. Therefore, we can speed up the algorithm by noting how many times $70$ divides into $462$ cleanly and looking at the remainder. In this example, we would have:

$$462 \;\; = \;\; 70 \times 6 \; + \; 42$$

With this observation, our improved Euclid's Algorithm can skip the redundancy of repeat subtractions by using remainders. Whereas the computation for $\text{GCD}(462, 70)$ took $10$ steps previously, let's see its solution now:

$$462 \;\; = \;\; 70 \times 6 \; + \; 42$$

$$70 \;\; = \;\; 42 \times 1 \; + \; 28$$

$$42 \;\; = \;\; 28 \times 1 \; + \; 14$$

$$28 \;\; = \;\; 14 \times 2 \; + \; 0$$

$$\therefore \; \text{GCD}(462, 70\,) \;\; = \;\; 14$$

Using our improved Euclid's Algorithm, we terminate the recursive process when we find a remainder of $0$, which for this example occurs after only $4$ steps. The trick to this new method lies in finding remainders, which can easily be calculated using the modulo operator (%) in Python:

In [ ]:

```python
A = 123456
B = 789
print( A % B )
```

```
372
```

To conclude, we now have a powerful tool for computing greatest common denominators very quickly, which doesn't require any knowledge about factors of the two numbers used for the computation. In fact, if you look at the steps shown above for the improved Euclid's method for calculating $\text{GCD}(462, 70)$, we don't even need to know the multiples $q$ at each step! The only thing necessary for the recursive strategy to proceed are the remainders at each step, which we just showed are easily obtained using the modulo operator. There's a very good reason why Euclid's Algorithm is still implemented today, over 2300 years after it was first discovered, and hopefully our discussion has given sufficient insight into the computational power it can provide.

## Overview of Shor's Algorithm

Having now seen the context for why Shor's Algorithm is so significant, providing potential for cracking RSA encryption by converting the factoriziation of large numbers into something solvable through GCDs, we're now ready to switch gears and begin our discussion of Shor's. Because there are so many moving parts that make up Shor's, we will first look at all of the major steps of the algorithm, and then dissect components which require further explanation. Presented below is the general layout of the algorithm, followed by some short code examples.

## Step 1: Pick $N$ and $a$

The first step to Shor's Algorithm is picking $N$, the number we want to factor, and then a second number $a$, such that $a$ is within: $2 \leq a \leq N - 2$. $a$ is typically chosen at random, as there is no real strategy behind picking an $a$ that will ultimately lead to a successful factorization of $N$:

In [ ]:
```
N = 35
a = int( random.randint(2,N-2) )
print('N: ',N)
print('a: ',a)
```

```
N:  35
a:  23
```

## Step 2: Check GCD( $N$, $a$ )

Step two is to compute the greatest common denominator between $N$ and $a$. Doing so will yield one of two results:

**1)** The GCD($N$,$a$) $\neq 1$, in which case we have found a factor of $N$! This case is a rarity and a byproduct of pure luck. Recall from earlier that RSA encryption typically uses $N$'s on the order $100$ - $1000$+ digits, composed of two prime numbers $p$ and $q$. Thus, the probability of picking one of the two factors at random is astronomically small.

**2)** The GCD($N$,$a$) $= 1$, which is the expected result. The primary purpose for computing GCD($N$,$a$) here is to verify that $a$ is coprime with $N$, which is a necessary condition for proceeding to step 3.

In [ ]:
```
gcd = oq.Euclids_Alg(N,a)

if(gcd > 1):
    print('a has a common factor with N  --->  GCD:  ',gcd)
if(gcd == 1):
    print('a has no common factor with N  --->  GCD: ',gcd)
```

```
a has no common factor with N  --->  GCD:  1
```

## Step 3$^\star$: Find the period $r$

This is the most important step in the algorithm (marked with a $\star$ for emphasis). The reason this step is so important is because this is where our quantum computer comes in, exponentially outperforming any classical method. Mathematically, the goal is to find the period of the following modular function:

$$a^x \;=\; a^{x+r} \pmod{N}$$

The equation shown above is the heart of Shor's Algorithm, which we will discuss in great detail in the sections to come. Classically, this step is analogous to the bottleneck of the two classical factoring algorithms, whereby a classical computer is forced to exhaustively search for the value $r$. Using our quantum computer however, we will be able to find $r$ exponentially faster, which in turn will allow us to complete our factoring problem. For completeness, the cell of code below computes $r$ classacialy so that we can see the remaining two steps for Shor's Algorithm.

In [ ]:
```
r = oq.r_Finder(a,N)

print('period r: ',r)
print('\nVerify ',a,'^',r,' = 1 (mod ',N,'):  ',(a**r)%N)
```

```
period r:  12

Verify  23 ^ 12  = 1 (mod  35 ):   1
```

## Step 4: Check requirements of $r$

Once we have $r$, the algorithm still does not come with a $100\%$ guaranteed soluation rate. Step $4$ is to quickly check that the $r$ we found in step $3$ satisfies two conditions:

$$\text{GCD}\,(2, r) \;=\; 2 \qquad\qquad a^{r/2} \pmod{N} \neq N - 1$$

The condition on the left is simply saying that we require $r$ to be an even number. Meanwhile, the condition on the right is a bit more complicated, and requires further explanation from a later section. Mathematically, the condition is straightforward, but it's the reason $why$ we require this condition which is important. If we happen to get unlucky and find an $r$ that doesn't meet these conditions, then we are forced to start over at step $1$ and pick a new $a$. But if $r$ does meet both conditions, then we continue on to the final step.

In [ ]:
```
if( ((r)%2 == 0) and ( a**(int(r/2))%N != int(N-1) )):
    print('r is a good, proceed')
else:
    print('r is not good, start over')
    if((r)%2 != 0):
        print('r is odd: ',r)
    if(a**(int(r/2))%N == int(N-1)):
        print('a^(r/2) (mod N) = N-1')
```

```
r is a good, proceed
```

---

### Step 5: Compute GCD( $N, a^{r/2} \pm 1$ )

---

Once $a$ and $r$ have met all of the conditions for steps $1$ - $4$, computing the GCD between $N$ and $a^{\frac{r}{2}} \pm 1$ will yield factors of $N$:

In [ ]:
```
print('a^(r/2): ',int(a**(r/2.0)))

factor1 = oq.Euclids_Alg(int(a**(r/2.0)+1),N)
factor2 = oq.Euclids_Alg(int(a**(r/2.0)-1),N)

print('\nfactors of N: ',int(factor1),' ',int(factor2),'       ( N is',N,')')
```

```
a^(r/2):  148035889

factors of N:  5   7        ( N is 35 )
```

Following all of the steps outlined above, you should find that the final cell of code correctly returns the factors of $5$ and $7$ with some probability. It is important to note that the success of the algorithm is not strictly dependent on a single $a$, as there are many possible values that will complete the algorithm and lead to the correct answer. To see this, try running the cell of code below several times.

In [ ]:
```
N = 35
a = int( random.randint(2,N-2) )
print('a: ',a)
#--------------------------------#
if( oq.Euclids_Alg(N,a) == 1 ):
    r = oq.r_Finder(a,N)
    print( 'r: ',r )
    if( ((r)%2 == 0) and (a**(int(r/2))%N != int(N-1) )):
        print('\nfactors of N: ',int(oq.Euclids_Alg(int(a**(r/2.0)+1),N)),' ',int(oq.Euclids_Alg(int(a**(r/2.0)-1),N)),'  (
    else:
        if((r)%2 != 0):
            print('Condition Not Satisfied:    r is odd: ',r)
        if(a**(int(r/2))%N == int(N-1)):
            print('Condition Not Satisfied:    a^(r/2) (mod N) = N-1: ',a**(int(r/2))%N)
else:
    print('a  has a GCD with N:  ',oq.Euclids_Alg(N,a))
```

```
a:  13
r:  4

factors of N:  5   7   ( N is 35 )
```

I encourage you to test the cell of code several times before moving on. Not just to show cases that succeed, but more importantly cases that fail. For $N = 35$, the values for $a$ that will successfully lead to the factors $5$ and $7$ are:

$$\left[\, 2, 3, 4, 6, 8, 9, 12, 13, 17, 18, 22, 23, 26, 27, 29, 31, 32, 33 \,\right]$$

with following values also correctly lead to the factorization of $N$, simple by having GCD's not equal to 1:

$$\left[\, 5, 7, 10, 14, 15, 20, 21, 25, 28, 30 \,\right]$$

Thus, for the case of $N = 35$, the only values for $a$ which do not solve our factorization problem are $11, 16, 19$, and $24$. The values $11$ and $16$ lead to $r$'s which are odd, while $19$ and $24$ produce $r$'s which fail our second condition outlined above (note that $16$ is actually a special case that $does$ lead to the correct final answer, but only because it is a perfect square, which will make sense after our discussion about $r$ to come). Thus, when picking an $a$ at random to use for Shor's Algorithm, $N = 35$ has a nearly $88\,\%$ success rate.

Having now seen Shor's Algorithm in its entirety, we will next backtrack and revisit a couple topics in need of further explanation.

## Period Finding via Modulo Algebra

As we pointed out in the outline above, step $3$ is where our quantum computer comes in, making it the most important topic for this lesson. In demonstrating this step however, we didn't provide any motivation for $why$ we needed to compute the modulo period $r$, only that it was a necessary ingredient for the steps to come (which did in fact solve the factorization problem of $N = 35$). So then, the need for period finding stems from the following equation:

$$a^2 \ = \ 1 \ (\text{mod}\, N)$$

Starting from the equation above, we can rearrange the terms to create a difference of squares (obeying modulo arithmetic rules):

$$a^2 - 1 \ = \ 0 \ (\text{mod}\, N)$$

which can be once again factored:

$$(\,a + 1\,)(\,a - 1\,) \ = \ 0 \ (\text{mod}\, N)$$

The importance of the equation above is that it is telling us that $N$ divides $(\,a + 1\,)(\,a - 1\,)$ evenly, which we know means that all of the factors in $N$ also exist in $a^2 - 1$. Now, based on our previous experience with the classical congruence of squares technique, you may suspect that a GCD is coming in the next step, and you'd be correct. However, before computing the greatest common denominator, we must introduce one more condition on $a$:

$$a \pm 1 \ \neq \ 0 \ (\text{mod}\, N)$$

If we now look at the three equations above all together, what can we say about $N$ and $a$? The first two equations tell us the same story from earlier, $N$ divides the number $a^2 - 1$, which has the factors $(\,a + 1\,)$ and $(\,a - 1\,)$. But now, our new condition states that $N$ *does not* divide either $(\,a + 1\,)$ or $(\,a - 1\,)$. So then, the only way for these seemingly contradictory conditions to be satisfied is if $N$ has two components: one that divides $(\,a + 1\,)$ and one that divides $(\,a - 1\,)$. Or in other words, we can conclude that $N$ has at minimum two *unique* factors, one shared with each of the $(\,a \pm 1\,)$ terms. Thus, the consequence of introducing this condition means that the GCD of $N$ with $(\,a \pm 1\,)$ will yield these two unique factors:

$$\text{GCD}(\,a + 1, N\,) \ = \ f_1$$
$$\text{GCD}(\,a - 1, N\,) \ = \ f_2$$

The result here is more or less the same as before, whereby the use of GCD's leads us to $f_1$ and $f_2$, the solution to our factorization problem of $N$. The new aspect here however is that our conditions will lead to some stronger restrictions on the factors we find:

$$f_1 \ \neq \ f_2$$

$$f_1 \ \& \ f_2 \ \neq \ a \pm 1$$

The mathematical formalism outlined thus far will solve our factoring problem of $N$, under one condition: we find a correct value for $a$. Remember that $N$ is some number given to us, which means that all of the conditions laid out above must be enforced through the $a$ we pick. If we were to turn to our classical computers at this point, we know what they would suggest: perform a blind exhaustive search until we find an $a$ that meets all the criteria. However, we also know that when $N$ starts to get large, this technique bottlenecks quite disastrously. For sufficiently large factoring problems, the time required for a classical search to happen upon a correct value for $a$ becomes astronomically long. Thus, even using the best factoring techniques currently known, a classical computer is stuck behind the wall of exhaustively searching through a near-infinite list. So then, the question becomes what will be our new technique for finding an $a$ to solve our problem? Ironically, the answer is to just pick one at random!

As outlined in step $1$, our new technique begins by picking a random value for $a$, such that it is between $2$ and $N - 2$. After picking our random $a$, we do a quick GCD check to make sure that we didn't happen to pick a factor of $N$ by dumb luck (although the best case scenario *would* be to pick a factor I suppose). After confirming that $a$ is indeed coprime with $N$, our technique really begins with the following function:

$$a^n \ \ (\text{mod}\, N) \qquad\qquad n \in \{0, 1, 2, 3, \ldots\}$$

The equation above is a simple power function (mod $N$), but the key lies in the way these modulo power functions behave. Perhaps easier to show then describe, run the cell of code below a couple of times to see the patterns emerging:

In [ ]:

```
N = 35
coprimes = [4,6,8,9,11,13,16]
a = coprimes[int( random.randint(0,len(coprimes)-1) )]
print('N = ',N,'    a = ',a,'\n')
#-------------------------------
for i in np.arange(10):
    mod_N =  a**int(i) % N
    print('a^',i,' (mod N):  ',mod_N)
```

```
N =  35    a =  13

a^ 0  (mod N):   1
a^ 1  (mod N):   13
a^ 2  (mod N):   29
a^ 3  (mod N):   27
a^ 4  (mod N):   1
a^ 5  (mod N):   13
a^ 6  (mod N):   29
a^ 7  (mod N):   27
a^ 8  (mod N):   1
a^ 9  (mod N):   13
```

The pattern we are looking for is the way in which power functions (mod $N$) repeat themselves. Specifically, because $a^0$ is always $1$, the pattern is defined by the number of powers we must increase $a$ by until we get $1$ again. For example, below is the first six powers of $8$ (mod $35$):

$$\begin{aligned} 8^0 &= 1 = 1 \pmod{35} \\ 8^1 &= 8 = 8 \pmod{35} \\ 8^2 &= 64 = 29 \pmod{35} \\ 8^3 &= 512 = 22 \pmod{35} \\ 8^4 &= 4096 = 1 \pmod{35} \\ 8^5 &= 32768 = 8 \pmod{35} \end{aligned}$$

Between the cell of code above and this explicit example, hopefully it is clear that increasing the power of a number (mod $N$) will eventually lead to a repeating pattern. Moreover, we can define the period of this pattern as the number of powers we must increase $a$ by in order to get back to the same number. Mathematically, we can express this repeating function as:

$$f(a) = f(a+r)$$

where the quantity $r$ is the period. Now, remember that our primary goal is to find a value of $a$ such that we satisfy the modulo equations above, which will in turn lead us to the factorization of $N$. For now, we will assume that we have an efficient means for finding this period $r$ (which is the next topic in this lesson), such that we can continue with our discussion of its use. With $r$ in hand, assuming that we found an $r$ which is even, the final trick comes from rewriting our modulo equation of interest from earlier as follows:

$$\begin{aligned} a^0 &= 1 \pmod{N} \\ a^{0+r} &= 1 \pmod{N} \\ (a^{r/2})^2 &= 1 \pmod{N} \end{aligned}$$

Let us define $a^* \equiv a^{r/2}$:

$$(a^*)^2 = 1 \pmod{N}$$

Now compare the result we just derived with the equation at the beginning of this section. Following from all of the steps already outlined, if our technique can produce a value $a^*$, then it will lead us to the factorization of $N$ via GCDs. Additionally, we've now revealed the motivation for our first of two conditions on $r$, namely that we find a period which is even. As you can imagine, if we find an $r$ which is odd, then the quantity $a^{r/2}$ is no longer guaranteed to result in an integer, effectively nullifying our modulo math (with the special case exception that $a$ is a perfect square). But the most important thing to note in the final step shown above is that $a^*$ can come from numerous values of $a$. For example, suppose we had originally picked $a = 8$, and found the period $r = 4$:

$$\begin{aligned} 8^4 &= (8^2)^2 \\ \therefore\ a^* &= 64 \end{aligned}$$

Since we found an $r$ which is even, we can use $a^*$ to compute the GCDs of $(a^* + 1)$ and $(a^* - 1)$ with $N$:

$$\begin{aligned} \text{GCD}(65, 35) &= 5 \\ \text{GCD}(63, 35) &= 7 \end{aligned}$$

As promised, we have gotten two unique factors of $N$. And once again, the important concept to stress here is the fact that the $a$ we picked is nothing special. As demonstrated by the cell of code accompanying step $5$ earlier, over $50\%$ of the possible $a$ values we could have picked would have successfully lead us to an $a^*$ for factoring $N = 35$. One way of thinking about this new approach is that we've exchanged a search for $a$ in favor of a search for $r$. Before, we knew that there was some value $a$ that would solve our problem, but had no means other than a blind search for finding it. Here, once again we must search for an unknown value, $r$, but the advantage is that hopefully it is an easier quantity to find (but as we shall see, is still classically difficult).

Now then, before moving on to our next topic, which is where our quantum computer will come in, let's return to one subtle point we made earlier about a condition on $a^*$, namely:

$$a^{r/2} \pmod{N} \neq N - 1$$

Following from our derivation of $a^*$, we can now appreciate the reasoning behind this condition. If we find a period $r$ which is even, but doesn't satisfy the equation shown above, we get the following modulo equation:

$$a^* + 1 \pmod{N} = N$$

which means that the true modulo result is:

$$a^* + 1 \pmod{N} = 0$$

Thus, this requirement on $a^*$ protects the algorithm from leading to a dead end:

$$\text{GCD}(a^* + 1, N) = N$$

which means that the other unique factor, $f_2$, has to be:

$$\text{GCD}(a^* - 1, N) = 1$$

To now come full circle in our discussion, we must point out that even though finding the values $f_1 = N$ and $f_2 = 1$ don't solve our factorization problem, they technically *are* factors of $N$ *and* unique. That is to say, our technique of using $a^{r/2}$ still worked as promised, it just led us to an undesirable result, one that we already knew. And the reason this outcome can happen is because no where in our process did we ever actually enforce the condition: $a^* \pm 1 \neq 0 \pmod N$, outlined at the beginning of this section. As we discussed, this condition did not come from the modulo math, but was something *we* insisted be a property of $a$. Thus, the enforcement of this condition comes from our second criteria on $r$, complementing our first condition that $r$ be even, which together guarantee that Shor's Algorithm will lead to a nontrivial factorization of $N$.

The mathematics laid out in this section really are the backbone of Shor's Algorithm, and aren't exactly easy to fully appreciate at first since most of us don't work with modulo arithmetic too frequently. I encourage you to really take the time and process all of the different components which guarantee the success of the factorization of $N$ before moving on. For the remainder of this lesson we will be discussing the role of our quantum computer, namely for finding $r$, but fundamentally we've now covered all of the topics regarding the factorization algorithm as a whole.

## Quantum Period Finding

Now that we have seen how we can use modulo period finding to our advantage, we have arrived at the point in Shor's Algorithm where a quantum computer will provide us our speedup. But to properly motivate why we need a quantum computer, we must first ask how a classical computer would approach the period finding problem. As it turns out, when $N$ starts to get sufficiently large, the task of finding $r$ becomes increasingly more difficult, so much so in fact that we arrive at nearly the exact same bottleneck as before.

Previously we commented that a classical computer is stuck blindly searching for either the factors of $N$ or values for $a^2$, well the same turns out to be true for $r$. From the perspective of a classical algorithm, the only way to compute the period of these modulo power functions is to start with $a^0$, and continually increase the power until we once again get an output of $1 \pmod N$. However, for large $N$, the number of powers one must check before getting the correct output can be as high as the order $O(N)$, i.e. the same number of possible values we would normally check in the most basic exhaustive search. In fact, searching for $r$ could be *worse* than simply searching for $f_1$ and $f_2$, since at least an odd valued $N$ means that the factors must be smaller than $\frac{N}{2}$. And so, we have finally reached the end of the rope in terms of classical attempts at factoring, and now we turn to quantum.

### Quantum Modulo Operation

The first step to our quantum period finding subroutine is to initialize our quantum system by allocating our qubits into two seperate systems of equal length. To determine how many qubits are necessary for our circuit, we must first transform the number we are trying to factor into binary. Each of our systems must have enough qubits, which we will call $Q$, such that they can completely represent $N$ as a binary number. For example, if we would like to factor the number $15$, then we would initialize our system as follows:

$$|0000\rangle \otimes |0000\rangle$$

Written out in binary, the number $15$ is $1111$, which requires four qubits. Next, after allocating the necessary qubits, we create an equal superposition state on one of the two systems using Hadamard gates:

$$H^Q |0000\rangle \otimes |0000\rangle$$

Resulting from $Q$ Hadamard gates on the first set of qubits, we can think of this superposition state as a complete representation of all the binary numbers from $0$ to $2^Q$-1:

$$\frac{1}{2^{Q/2}} \left( |0000\rangle + |1000\rangle + |0100\rangle + |1100\rangle + \cdots + |1111\rangle \right) \otimes |0000\rangle$$

Now, the next operation we will look to perform is a bit tricky. In essence, our goal is to encode the modulo power function $a^n \pmod N$ into our quantum system. Specifically, this operation will take $n$ as the binary number from the first quantum system, and output the corresponding modulo power result onto the second quantum system. And since we have set up our first system in the superposition state, the resulting final state of the second system from this quantum modulo operation will contain all of the modulo power results from $0$ to $2^Q$-1. Visually, the final state of the combined quantum system will look like as follows, using $a = 8$ and $N = 15$ as our example:

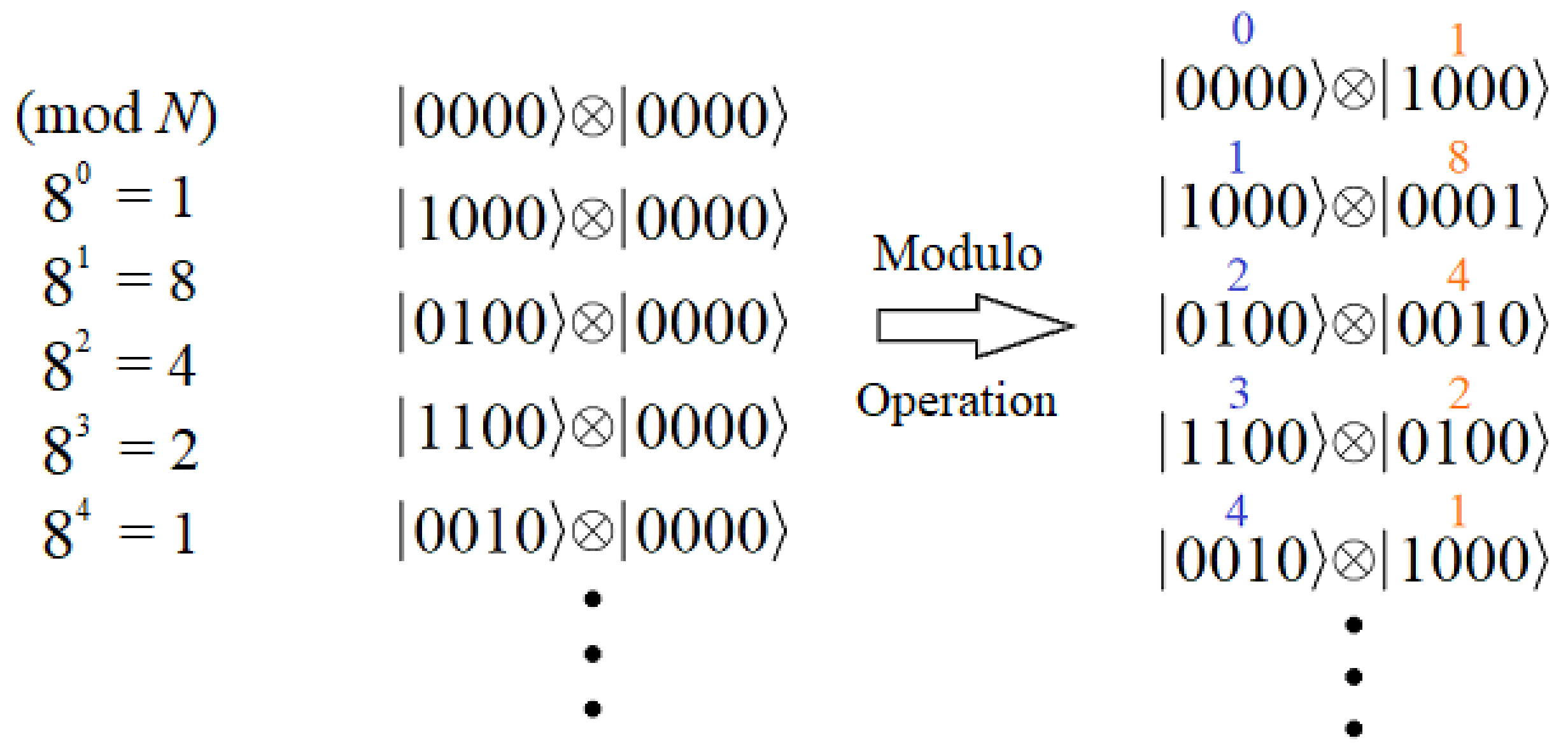

The implementation of the operation shown above is not a trivial one, and is still an active field of research. Mathematically however, hopefully it is clear that our goal is to use each state in the first system as an input, transforming the state of the second system to encode the modulo power function. Below is an example of the kind of final states we will aim to create:

In [ ]:

```
S1 = [round(random.random()),round(random.random()),round(random.random()),round(random.random())]
a = 8
N = 15
print('N: ',N,'    a: ',a)
#====================================
q1 = QuantumRegister(4,name='q1')
q2 = QuantumRegister(4,name='q2')
qc = QuantumCircuit(q1,q2,name='qc')
#------------------------------------
for i in np.arange(4):
    if(S1[i]==1):
        qc.x( q1[int(i)] )
print('\n_____ Initial State _____')
oq.Wavefunction(qc,systems=[4,4])
#------------------------------------
S1_num = S1[0] + 2*S1[1] + 4*S1[2] + 8*S1[3]
S2_num = a**(S1_num) % N
print('\nState 1:  ',S1_num,'     Desired State 2:  ',a,'^',S1_num,' ( mod ',N,') = ',S2_num)
#------------------------------------
for j in np.arange(4):
    if( S2_num >= 2**(3-j) ):
        qc.x( q2[int(3-j)] )
        S2_num = S2_num - 2**(3-j)
print('\n_____ After Modulo Operation_____')
oq.Wavefunction(qc,systems=[4,4])
```

```
N:  15     a:  8

_____ Initial State _____
1.0 |1110>|0000>

State 1:   7      Desired State 2:   8 ^ 7  ( mod  15 ) =  2

_____ After Modulo Operation_____
1.0 |1110>|0100>
```

The cell of code above creates a random state using four qubits, calculates what the modulo power function result should be, and then uses $X$ gates to create the desired state on the second system. For our purpose of learning Shor's Algorithm, handling the modulo operation in this way is sufficient. However, in order to obtain a real speedup, we would need a way of performing this task without the use of any classical computations. Such implementations do exist, but are unfortunately beyond the scope of this tutorial series here. That being said, it's not that we don't have the tools to understand such quantum circuits, which essentially combine Quantum Modulo Adders (very similar to our Quantum Adder Algorithm) with Quantum Phase Estimation, it's simply the fact that to properly discuss such circuits would take too long, basically warranting an entire lesson in itself. Moreover, our primary focus is to understand the theory behind Shor's Algorithm, which is the same regardless of how efficient / inefficient our circuit is at achieving the desirable quantum state.

If you're still curious about these more advanced quantum circuits for achieving the quantum modulo operation, I encourage you to check out some additional resources **[3]**. For our learning purposes here, we will be handling these modulo operations through the use of CNOT gates, analogous to the way in which we flip phases in Grover's Algorithm. Below is an example which demonstrates how we can create the desired states in our second quantum system, when our input system is in a superposition state:

In [ ]:

```
a = 8
N = 15
q1_state = [1,0,1,0]
q2_state = oq.Binary( a**(int( oq.From_Binary(q1_state,'L') ))%N, 2**4 ,'L' )
print('a  = ',a,'     N  = ',N)
print('\nInput State: ',q1_state,'      Desired Modulo State: ',q2_state)
#=====================================
q1 = QuantumRegister(4,name='q1')
q2 = QuantumRegister(4,name='q2')
an = QuantumRegister(3,name='a')
qc = QuantumCircuit(q1,q2,an,name='qc')
#--------------------------------------
qc.h(q1[0])
qc.h(q1[2])
qc.cx( q1[0], q1[1] )
qc.cx( q1[2], q1[1])
print('\n_____ Initial State _____')
oq.Wavefunction(qc,systems=[4,4,3],show_systems=[True,True,False])
qc.barrier()
#--------------------------------------   |1010> state
oq.X_Transformation(qc,q1,q1_state)
qc.ccx( q1[0], q1[1], an[0] )
qc.ccx( q1[2], an[0], an[1] )
qc.ccx( q1[3], an[1], an[2] )
for i in np.arange(len(q2_state)):
    if( q2_state[i]==1 ):
        qc.cx( an[2], q2[int(i)] )
qc.ccx( q1[3], an[1], an[2] )
qc.ccx( q1[2], an[0], an[1] )
qc.ccx( q1[0], q1[1], an[0] )
oq.X_Transformation(qc,q1,q1_state)

print('\n_____ After Modulo Operation _____')
oq.Wavefunction(qc,systems=[4,4,3],show_systems=[True,True,False])
print(qc)
```

```
a  =  8      N  =  15

Input State:  [1, 0, 1, 0]       Desired Modulo State:  [0, 0, 0, 1]

_____ Initial State _____
0.5 |0000>|0000>    0.5 |1100>|0000>    0.5 |1010>|0000>    0.5 |0110>|0000>

_____ After Modulo Operation _____
0.5 |0000>|0000>    0.5 |1100>|0000>    0.5 |0110>|0000>    0.5 |1010>|0001>
```

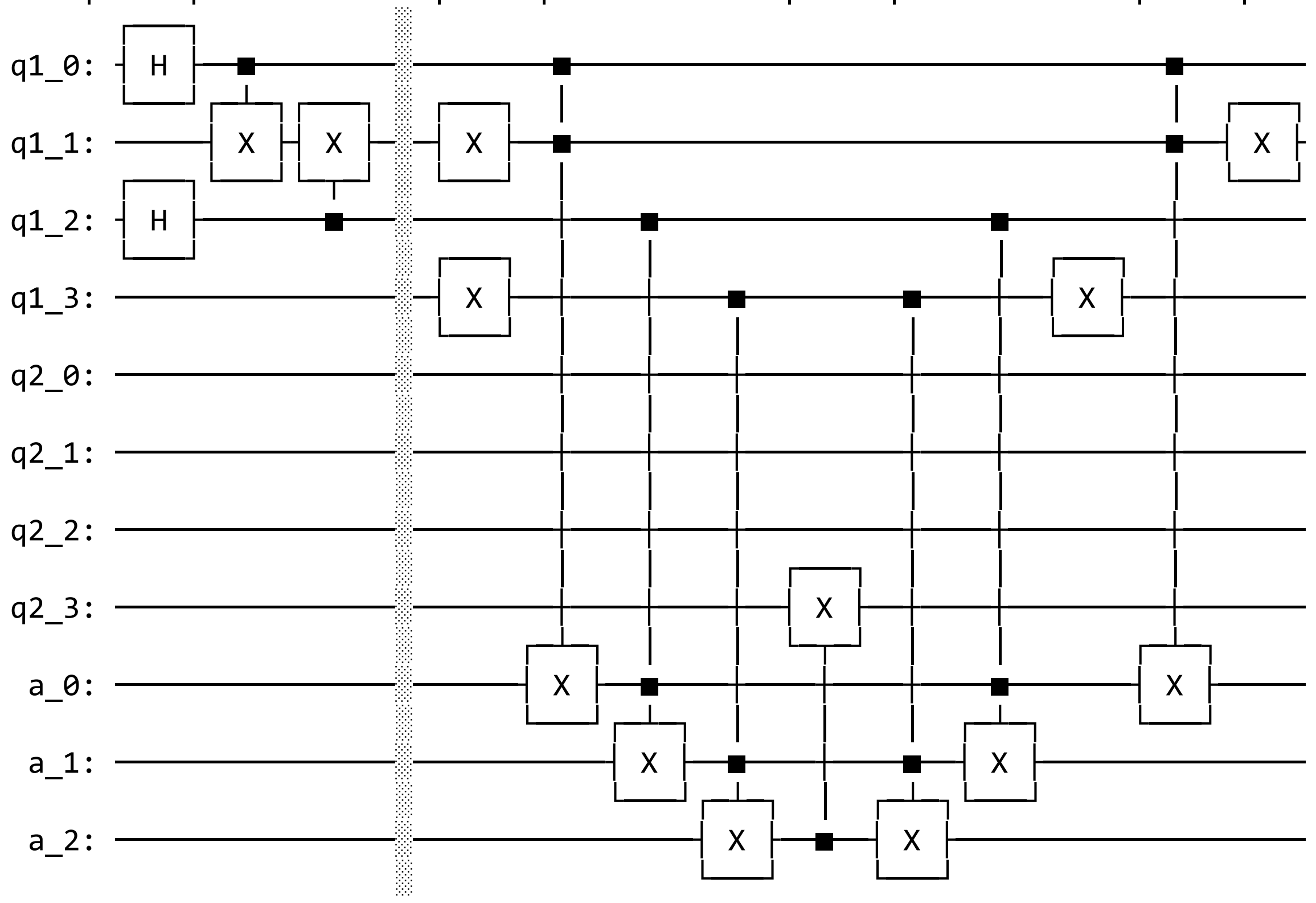

As shown above, the general approach to how we will be applying the quantum modulo operation is through the use of $X$ transformations and higher order CCX operations, the very same technique used in Grover's (see lesson $5.4$). The example above demonstrates how we will be able to create the desired modulo power result for each of the individual $2^Q$ basis states, despite being in a superposition. As shown in the results of the code, the circuit above creates the binary state representation of $8^5 \pmod{15}$ in system two, but only for the state $|1010\rangle$. In the coming examples, these steps will be handled by the **Mod_Op** function from $\text{Our_Qiskit_Functions}$, which will automatically assign all of the necessary gates for a complete modulo operation:

In [ ]:

```python
a = 8
N = 15
Q = 4
print('a:  ',a,'   N:  ',N)
#====================================
q1 = QuantumRegister(Q,name='q1')
q2 = QuantumRegister(Q,name='q2')
an = QuantumRegister(Q-1,name='a')
qc = QuantumCircuit(q1,q2,an,name='qc')
#-------------------------------------
for i in np.arange(Q):
    qc.h(q1[int(i)])
print('\n_____ Initial State _____')
oq.Wavefunction(qc,systems=[4,4,3],show_systems=[True,True,False])

oq.Mod_Op(Q,qc,q1,q2,an,a,N)

print('\n_____ After Modulo Operation _____')
oq.Wavefunction(qc,systems=[4,4,3],show_systems=[True,True,False])
```

```
a:   8    N:   15

_____ Initial State _____
0.25 |0000>|0000>    0.25 |1000>|0000>    0.25 |0100>|0000>    0.25 |1100>|0000>    0.25 |0010>|0000>    0.25 |1010>|0000>
0.25 |0110>|0000>    0.25 |1110>|0000>    0.25 |0001>|0000>    0.25 |1001>|0000>    0.25 |0101>|0000>    0.25 |1101>|0000>
0.25 |0011>|0000>    0.25 |1011>|0000>    0.25 |0111>|0000>    0.25 |1111>|0000>

_____ After Modulo Operation _____
0.25 |0000>|1000>    0.25 |0010>|1000>    0.25 |0001>|1000>    0.25 |0011>|1000>    0.25 |1100>|0100>    0.25 |1110>|0100>
0.25 |1101>|0100>    0.25 |1111>|0100>    0.25 |0100>|0010>    0.25 |0110>|0010>    0.25 |0101>|0010>    0.25 |0111>|0010>
0.25 |1000>|0001>    0.25 |1010>|0001>    0.25 |1001>|0001>    0.25 |1011>|0001>
```

The cell of code above is our complete quantum modulo operation, whereby we have successfully encoded the modulo power function of $a$ and $N$ into all $2^Q$ superposition states. I encourage you to work through some of the final states for yourself, verifying that the $\text{Mod_Op}$ function has indeed correctly created all of the appropriate $a^n \pmod{N}$ binary states.

## Partial Measurement

After successfully implementing the quantum modulo operation like shown above, the next step in our subroutine will be something new to these tutorials, namely a measurement. But unlike a typical measurement, here we will be performing a 'partial measurement', whereby the measurement only occurs on half of the qubits, specifically system two in this case (the qubits which received the effect of the modulo operation). By only measuring some of the qubits, but not all, we will in effect collapse our system only partially. That is to say, the act of the measurement will collapse the qubits in the second system into either the $|0\rangle$ or $|1\rangle$ state, but will still leave behind some superposition amongst the qubits in the first system. Let's see this step in action:

In [ ]:

```
a = 8
N = 15
Q = 4
print('a:  ',a,'   N:  ',N)
#====================================
q1 = QuantumRegister(Q,name='q1')
q2 = QuantumRegister(Q,name='q2')
an = QuantumRegister(Q-1,name='a')
c  = ClassicalRegister(Q,name='c')
qc = QuantumCircuit(q1,q2,an,c,name='qc')
#-------------------------------------
for i in np.arange(Q):
    qc.h(q1[int(i)])
print('\n_____ Initial State _____')
oq.Wavefunction(qc,systems=[Q,Q,Q-1],show_systems=[True,True,False])

oq.Mod_Op(Q,qc,q1,q2,an,a,N)
print('\n_____ After Modulo Operation _____')
oq.Wavefunction(qc,systems=[Q,Q,Q-1],show_systems=[True,True,False])

qc.measure(q2,c)
print('\n_____ After Partial Measurement _____')
oq.Wavefunction(qc,systems=[Q,Q,Q-1],show_systems=[True,True,False])
```

```
a:   8    N:   15

_____ Initial State _____
0.25 |0000>|0000>    0.25 |1000>|0000>    0.25 |0100>|0000>    0.25 |1100>|0000>    0.25 |0010>|0000>    0.25 |1010>|0000>
0.25 |0110>|0000>    0.25 |1110>|0000>    0.25 |0001>|0000>    0.25 |1001>|0000>    0.25 |0101>|0000>    0.25 |1101>|0000>
0.25 |0011>|0000>    0.25 |1011>|0000>    0.25 |0111>|0000>    0.25 |1111>|0000>

_____ After Modulo Operation _____
0.25 |0000>|1000>    0.25 |0010>|1000>    0.25 |0001>|1000>    0.25 |0011>|1000>    0.25 |1100>|0100>    0.25 |1110>|0100>
0.25 |1101>|0100>    0.25 |1111>|0100>    0.25 |0100>|0010>    0.25 |0110>|0010>    0.25 |0101>|0010>    0.25 |0111>|0010>
0.25 |1000>|0001>    0.25 |1010>|0001>    0.25 |1001>|0001>    0.25 |1011>|0001>

_____ After Partial Measurement _____
0.5 |1000>|0001>    0.5 |1010>|0001>    0.5 |1001>|0001>    0.5 |1011>|0001>
```

Running the cell of code above a few times, you should see the various possible outcomes for the second system. As promised, the partial measurement causes all of the qubits in the second system to collapse down to single values, but does not completely erase the superposition from the first system. More specifically, when one of the possible states in system two is measured, all of the corresponding states in system one which were attached to this state remain in the leftover superposition, but now with new amplitudes. Conceptually, this leftover superposition represents the remaining uncertainty in system one.

As an example, suppose states $|\,A\,\rangle$, $|\,B\,\rangle$, and $|\,C\,\rangle$ are all tensored to the same state $|\,Z\,\rangle$ in the initial grand superposition. If initially there are $100$ total states, then the probability of system one being in any of these three states is $1\%$ each. However, if a partial measurement were to collapse system two into the $|\,Z\,\rangle$ state, we must ask how this affects our uncertainty in system one. Since $|\,A\,\rangle$, $|\,B\,\rangle$, and $|\,C\,\rangle$ are the only possible candidates remaining for system one, probabilistically we know that a follow up measurement should pick out one of these states, each with 33.3% chance. Thus, the wavefunction corresponding to the state of our system after the partial measurement should reflect these probabilities, leaving $|\,A\,\rangle$, $|\,B\,\rangle$, and $|\,C\,\rangle$ all with amplitudes of $\frac{1}{\sqrt{3}}$.

The important thing to note in this made up example is the way in which the wavefunction of system one changes as a result of a measurement on system two. Initially the states $|\,A\,\rangle$, $|\,B\,\rangle$, and $|\,C\,\rangle$ all possess amplitudes of $\frac{1}{\sqrt{100}}$, but after the partial measurement their amplitudes jump to $\frac{1}{\sqrt{3}}$, and correspondingly all other amplitudes go to $0$. Similarly, take a look at the example code above, and note that the amplitudes for the leftover superposition states jump from $\frac{1}{4}$ initially, to $\frac{1}{2}$ after the measurement. This change in system one happens instantaneously once the measurement on system two occurs, and is yet another example of the weird nature of quantum mechanics (partial measurements and instantaneous wavefunction changing are essentially what Einstein could never reconcile with, calling it "spooky action at a distance"). For the purpose of Shor's Algorithm, this updating of the wavefunction after a partial measurement is exactly the ingredient needed for the next step.

Now, in order to properly understand the next and final operation in our algorithm, we need to notice something very particular happening with the quantum states after the partial measurement. As we shall see, the success of Shor's Algorithm isn't aided or hindered by any single partial measurement result on the second system. Take a moment and look carefully at all of the possible states in the first system that share the same tensored secondary state. What you should notice is a pattern something like this:

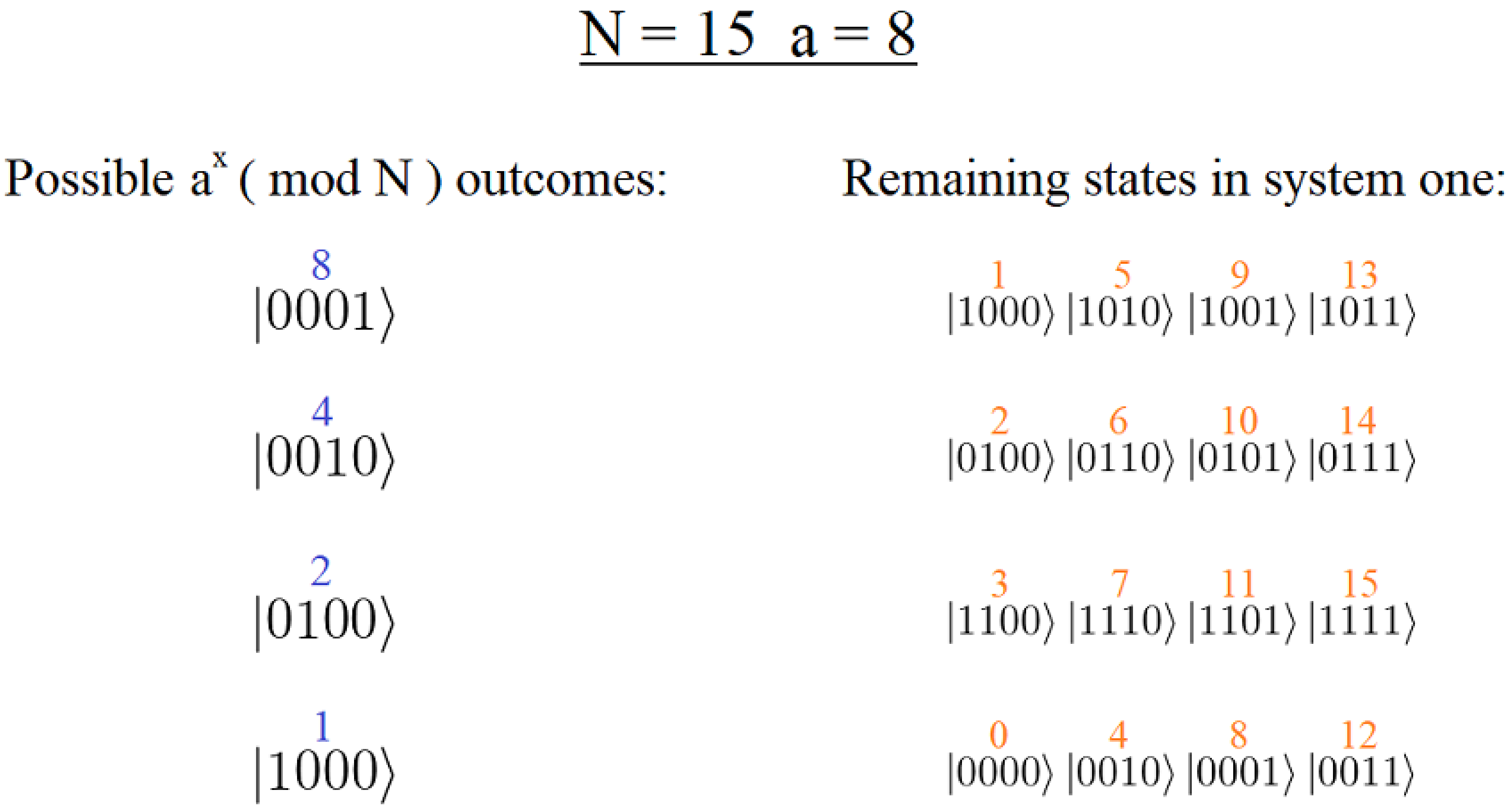

The important takeaway from the image above is the even spacing found in the states leftover from the partial measurement. No matter which state we find in our second system, the remaining states in system one will always be evenly spaced by our period $r$. Thus, resulting from the combination of the modulo operation along with a partial measurement, the period $r$ is systematically encoded into the final remaining superposition state.

Based on the result presented above, it may look like we've finally solved our problem of finding the period, but it is important to remember that these states are in a superposition. As such, a measurement on system one will only reveal a single state, which alone tells us nothing about $r$. Repeating the process numerous times will eventually reveal $r$, once a sufficient number of independent states have been found, but as we shall see, there is a better way.

### QFT$^\dagger$ on a Periodic Function

The role of the partial measurement in Shor's Algorithm is to set up a state on system one like shown above, where we have an equal superposition of evenly spaced states. Even though this superposition state has the period $r$ fundamentally woven into it, extracting $r$ out requires further work. The manner in which we are going to go about obtaining the period is quite clever, but will take some extra effort to fully digest. In particular, we must first revisit a property of the Inverse Discrete Fourier Transformation (DFT$^\dagger$), which in turn is the quantum mechanical equivalent to the QFT$^\dagger$ operation we will be using to extract $r$. The specific mathematical property we are interested in is the effect of a DFT$^\dagger$ on a discrete periodic function. Take a look at the structure of the example function below, which will serve as the template for an ideal quantum state:

$$f(x) = \begin{cases} 1 \;, & x \;=\; a_o + n \cdot r \\ 0 & \end{cases} \qquad\qquad n \in \left[0, 1, 2, \ldots, k-1\right]$$

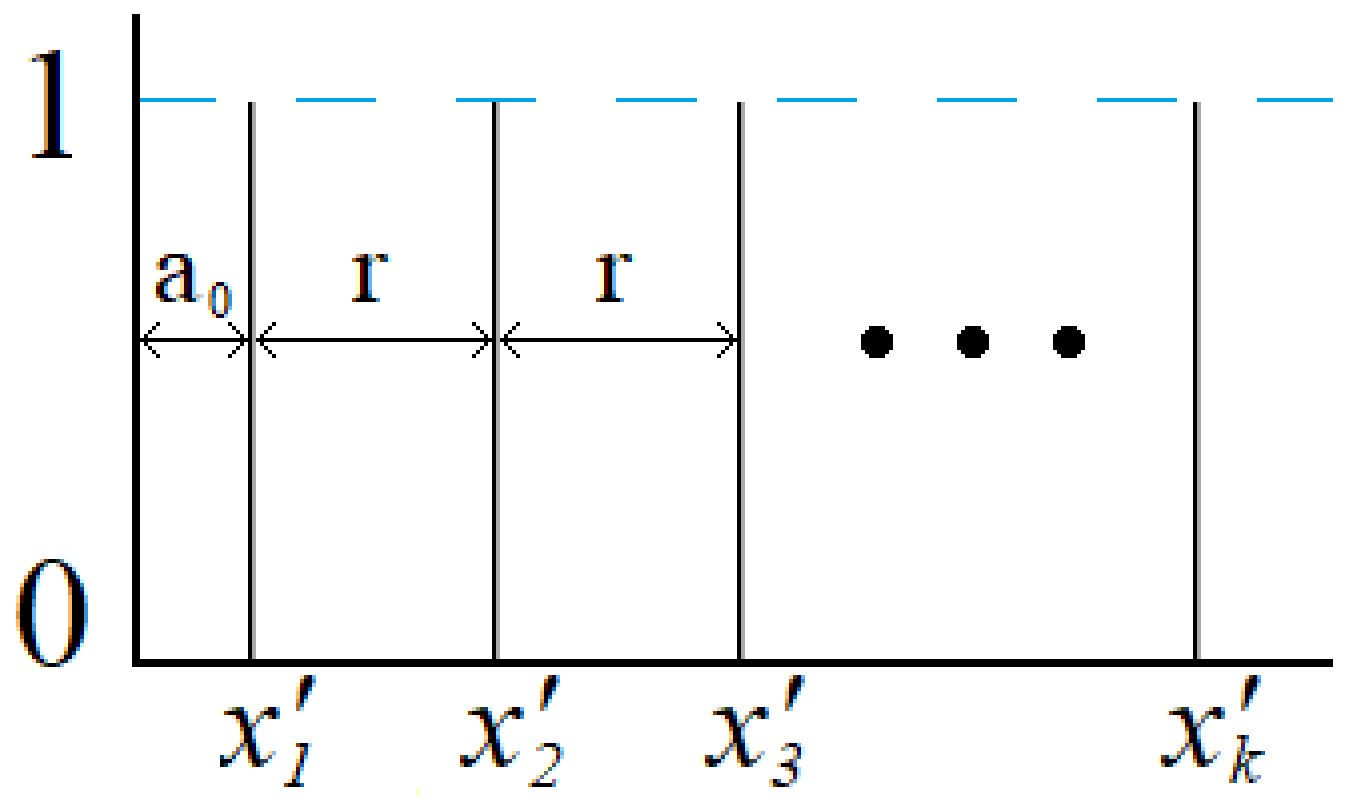

In the illustration shown above, the function $f(x)$ takes on the value $1$ only for select inputs $x_i'$. More specifically, these $x'$ values are all evenly spaced by a distance $r$, offset by some initial value $a_0$. If we return to the $N = 15$, $a = 8$ figure from the previous section, notice that this pattern is in perfect agreement with the quantum states produced from the partial measurement. Based on which of the four states is found by the partial measurement on system two, the remaining superposition state in system one contains all evenly spaced basis states, offset from $|0000\rangle$ by some value. The only difference between our quantum state and the $f(x)$ shown above is the fact that the amplitudes cannot equal $1$, but instead have normalized values. This slight difference however has no impact on the coming effect, only influencing the scale of the final numbers (keeping the resulting quantum state normalized after the QFT$^\dagger$). Thus, in studying the effect of a DFT$^\dagger$ on the function shown above, we can anticipate how the QFT$^\dagger$ will transform our leftover superposition states.

To begin our analysis, we must first note something missing from the $f(x)$ description above, namely its domain. The function provides the location of each $x'$, spanning a total of $k$ values, but doesn't say anything about how far beyond the final $x_k'$ the function continues (or if $f(x)$ extends to negative numbers for that matter). For the trick we are about to observe to work properly, a crucial ingredient is that our discrete function only exists over a finite domain:

$$x \in \left[0, L\right] \qquad\qquad L \equiv k \cdot r$$

One way to think about the requirement shown above is that we can break up our function $f(x)$ into 'cycles', where the length of each cycle is equal to $r$, and within each cycle there is exactly one $x'$ value. For example, if we have an $f(x)$ with a period of $r = 5$, $k = 3$, there are several ways in which we can define one cycle based on the value of $a_0$:

| $a_0$ | one cycle | $f(x)$ |
|---|---|---|

$$0 \qquad \begin{bmatrix} 1 & 0 & 0 & 0 & 0 \end{bmatrix} \qquad \begin{bmatrix} 1 & 0 & 0 & 0 & 0 & 1 & 0 & 0 & 0 & 0 & 1 & 0 & 0 & 0 & 0 \end{bmatrix}$$

$$1 \qquad \begin{bmatrix} 0 & 1 & 0 & 0 & 0 \end{bmatrix} \qquad \begin{bmatrix} 0 & 1 & 0 & 0 & 0 & 0 & 1 & 0 & 0 & 0 & 0 & 1 & 0 & 0 & 0 \end{bmatrix}$$

$$3 \qquad \begin{bmatrix} 0 & 0 & 0 & 1 & 0 \end{bmatrix} \qquad \begin{bmatrix} 0 & 0 & 0 & 1 & 0 & 0 & 0 & 0 & 1 & 0 & 0 & 0 & 0 & 1 & 0 \end{bmatrix}$$

As shown above, the value of $a_0$ defines where the location of $1$ is offset in each cycle, leading to distinct constructions of $f(x)$. In terms of our quantum period finding, $a_0$ is a problematic feature of our superposition state which makes determining $r$ through repeat sampling difficult. However, as we shall now see, regardless of the value of $a_0$, the effect of a DFT$^\dagger$ on any of the $f(x)$'s shown above will result in the same final function, $\tilde{f}(x)$. But before getting into any of the math details, let's first observe this effect with some classical code:

In [ ]:

```
k  = 3
r  = 5
a0 = int( (r)*random.random() )
L = k*r
C = np.zeros(r)
C[a0] = 1
print('k = ',k,'        r = ',r,'        a0 = ',a0,'\n\nOne Cycle:  ',C,'        L = ',L)
#------------------------------------------
f = []
for i in np.arange(k):
    for i2 in np.arange(r):
        f.append( int( C[i2] ) )
print('\nPeriodic Function:  ',f)

F = oq.DFT(f,inverse=True)
print('\nAfter DFT\u2020:  ',F)
#------------------------------------------
F2 = []
for j in np.arange( len(F) ):
    F2.append( round(abs(F[j]),3) )
print('\n       |F|:  ',F2 )
```

```
k =  3          r =  5          a0 =  4

One Cycle:   [0. 0. 0. 0. 1.]         L =  15

Periodic Function:   [0, 0, 0, 0, 1, 0, 0, 0, 0, 1, 0, 0, 0, 0, 1]

After DFT†:   [3.0, 0, 0, (0.9271-2.8532j), 0, 0, (-2.4271-1.7634j), 0, 0, (-2.4271+1.7634j), 0, 0, (0.9271+2.8532j), 0, 0]

        |F|:   [3.0, 0, 0, 3.0, 0, 0, 3.0, 0, 0, 3.0, 0, 0, 3.0, 0, 0]
```

The cell of code above uses the function **DFT** from Our_Qiskit_Functions to demonstrate the effect of a DFT$^\dagger$ on the discrete periodic function $f(x)$. I encourage you to run the cell several times, taking note of how the initial offset value $a_0$ affects the final $\tilde{f}(x)$. Although the final values may differ slightly, every construction of $f(x)$ leads to the same structure of $\tilde{f}(x)$, whereby all of the non-zero values are in the same locations, and always have a magnitude of $3.0$. Additionally, you should try changing the values of $k$ and $r$, and see if you can spot the underlying pattern going on here. Regardless of what values $a_0$, $k$, and $r$ take on, the cell of code above will produce a periodic function that fits our template $f(x)$, which in turn guarantees that the DFT$^\dagger$ operation will produce the desired final result.

With the cell of code above in hand, we can now be a bit more mathematically rigorous in describing what the DFT$^\dagger$ is accomplishing. For starters, let's quickly remind ourselves of the summation which determines the resulting values from a DFT$^\dagger$ (and similarly a QFT$^\dagger$):

$$\tilde{x}_n \;=\; \sum_{j=0}^{L-1} x_j \cdot e^{\frac{-2\pi i}{L} n \cdot j}$$

If now consider how the structure of $f(x)$ will affect this equation, all of the contributing $x_j$ terms in each of the summations shown above will be $0$ except for the $x'$ values, which are all equal to $1$. Thus, for the particular $f(x)$ we are interested in, the DFT$^\dagger$ summation shown above can be reduced to a summation over $k$ complex numbers of magnitude $1$. Using $k = 3$ as an example, every $\tilde{x}_n$ term can be computed with the following expression:

$$\begin{aligned} \tilde{x}_n \;&=\; e^{\frac{-2\pi i}{L} n \cdot a_0} \;+\; e^{\frac{-2\pi i}{L} n \cdot (a_0 + r)} \;+\; e^{\frac{-2\pi i}{L} n \cdot (a_0 + 2r)} \\ &=\; e^{\frac{-2\pi i}{L} n \cdot a_0} \cdot \left( e^0 \;+\; e^{\frac{-2\pi i}{L} n \cdot r} \;+\; e^{\frac{-2\pi i}{L} n \cdot 2r} \right) \end{aligned}$$

Notice how all of the contributing complex numbers only differ by increasing orders of $r$. Using the expression above as a template, we can write out the more general form as follows:

$$=\; e^{\frac{-2\pi i}{L} n \cdot a_0} \cdot \sum_{j=0}^{k-1} e^{\frac{-2\pi i}{L} n \cdot jr}$$

The significance of the expression shown above is the way in which the summation produces zero or non-zero $\tilde{x}_n$ terms. Beginning with $\tilde{x}_0$, we can see that the summation will reduce to $k$ terms of $e^0$ all added together, resulting in a total sum of $k$. Thus, the effect of DFT$^\dagger$ applied to $f(x)$ will always result in a $\tilde{f}(x)$ function beginning with the integer $k$, which will later be an important point. Moving on to $\tilde{x}_1$ then, the summation of exponentials will no longer all equal $e^0$, which means that we should expect contributions from complex numbers with various real and imaginary components, all with a magnitude of one. But now here's where things get interesting: because we have a finite discrete function with $L = k \cdot r$, the result of the summation above only ever has two outputs:

$$\sum_{j=0}^{k-1} e^{\frac{-2\pi i}{L} n \cdot jr} \;=\; \begin{cases} k \;, & j \;=\; 0 \,(\mathrm{mod}\, k) \\ 0 & \end{cases}$$

which we can confirm explicitly with the cell of code below:

In [ ]:

```
k = int( 2+4*random.random() )
r = int( 2+4*random.random() )
L = k*r
print('k: ',k,'      r: ',r,'      L: ',L,'\n------------------------------\n')
#=================================================
for q in np.arange(L):
    n = int(q)
    Q = 0
    for j in np.arange(k):
        Q = Q + np.exp( ((-2*m.pi*1.0j)/L) * n * j * r )
    print( 'n: ',n,'      \u03A3 =  ',round( Q.real,5 ),' + i',round(Q.imag,5) )
```

```
k:  4       r:  3       L:  12
------------------------------

n:   0       Σ =   4.0  + i 0.0
n:   1       Σ =   -0.0  + i -0.0
n:   2       Σ =   0.0  + i -0.0
n:   3       Σ =   0.0  + i -0.0
n:   4       Σ =   4.0  + i 0.0
n:   5       Σ =   -0.0  + i -0.0
n:   6       Σ =   0.0  + i -0.0
n:   7       Σ =   -0.0  + i -0.0
n:   8       Σ =   4.0  + i 0.0
n:   9       Σ =   -0.0  + i -0.0
n:   10       Σ =   0.0  + i -0.0
n:   11       Σ =   0.0  + i -0.0
```

The fact that the summation shown above always equals $0$ or $k$ is the driving force behind why we're interested in using a QFT$^\dagger$ on our quantum system. If you've never seen a Fourier Transformation on a periodic function before, the result can be quite surprising at first. In order to better appreciate why so many terms in the summation go to zero, the trick lies in the way in which the contributing complex numbers are equally spaced apart. Visually, these complex numbers can be represented as evenly spaced points along a circle of radius $1$ in the complex plane:

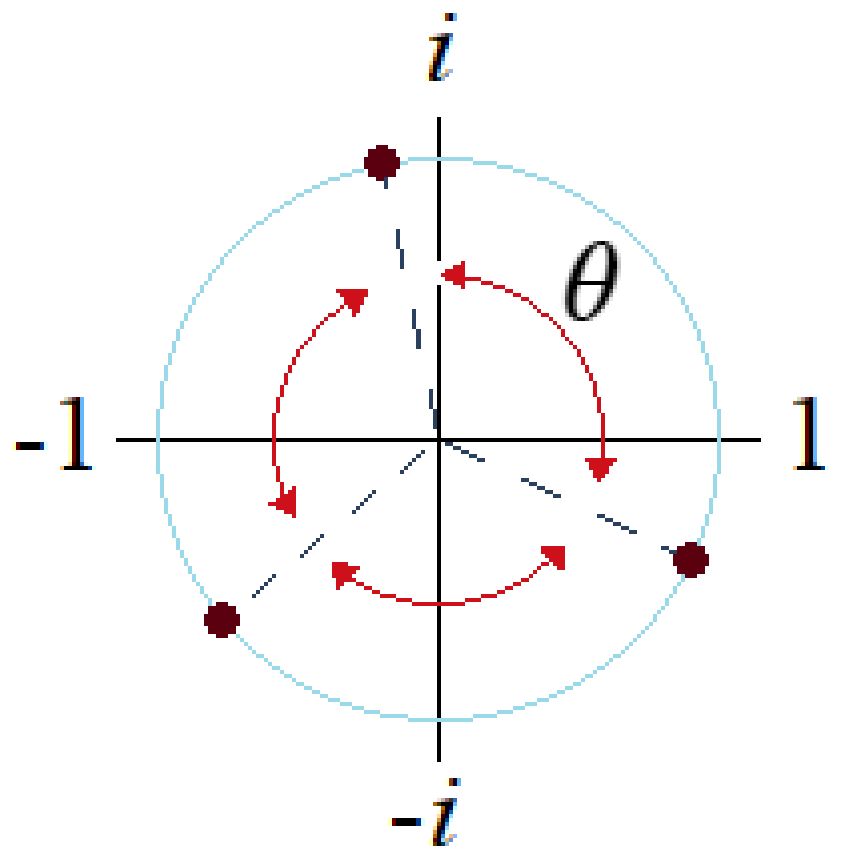

If you sum together $N$ complex numbers which are all separated by angles of $\frac{2\pi}{N}$, the net result is that the real and imaginary components will all cancel out to $0$. In terms of our DFT$^\dagger$ on a discrete periodic function, this is exactly what is happening for all instances of $\tilde{x}_n$ when $n \neq 0 \ (\mathrm{mod}\, k)$. For $\tilde{x}_0$, all $k$ of the contributing complex numbers are aligned at the point $1 + i0$, summing to a total of $k$. Then for each successive $\tilde{x}_n$ term, these $k$ complex numbers can be thought of as points moving around the complex unit circle with varying frequencies. The exact position and frequencies of these points all depend on the numbers $n$, $r$, $k$, and $a_0$, but their net result always produces the same effect. For values of $n = 0 \ (\mathrm{mod}\, k)$, all the points once again align to produce a non-zero sum with a magnitude of $k$. Conversely, for cases of $n \neq 0 \ (\mathrm{mod}\, k)$, the points are all evenly spaced apart by angles of $\frac{2\pi}{k}$, summing to $0$. Below is a visualization of this process for the case of $k = 5$, $r = 4$, and a random offset $a_0$:

In [ ]:

```
N_circle = 1000
x_c = []
y_c = []
#-----------------
for c in np.arange(N_circle+1):
    t = (2*m.pi*c)/N_circle
    x_c.append( np.cos(t) )
    y_c.append( np.sin(t) )
#================================================
k = 5
r = 4
L = k*r
a0= int(r*random.random())
print('k: ',k,'      r: ',r,'      L: ',L,'      a0: ',a0,'\n------------------------------------------\n')
for i in np.arange(L):
    p1 = np.exp( (-2*m.pi*1.0j/L) * (a0+0*r) * i)
    p2 = np.exp( (-2*m.pi*1.0j/L) * (a0+1*r) * i)
    p3 = np.exp( (-2*m.pi*1.0j/L) * (a0+2*r) * i)
    p4 = np.exp( (-2*m.pi*1.0j/L) * (a0+3*r) * i)
    p5 = np.exp( (-2*m.pi*1.0j/L) * (a0+4*r) * i)
    #=======================
    print('x_'+str(int(i))+' QFT term:   ',round((p1+p2+p3+p4+p5).real,4),'  +  i',round((p1+p2+p3+p4+p5).imag,4))
    fig = plt.figure(figsize=(4,4))
    plt.scatter( p1.real,p1.imag,s=40,color='blue' )
    plt.scatter( p2.real,p2.imag,s=40,color='orange' )
    plt.scatter( p3.real,p3.imag,s=40,color='red' )
    plt.scatter( p4.real,p4.imag,s=40,color='green' )
    plt.scatter( p5.real,p5.imag,s=40,color='purple' )
    plt.plot( x_c,y_c,linewidth=0.5 )
    plt.show()
```

```
x_4 QFT term:    -0.0   +  i 0.0
```

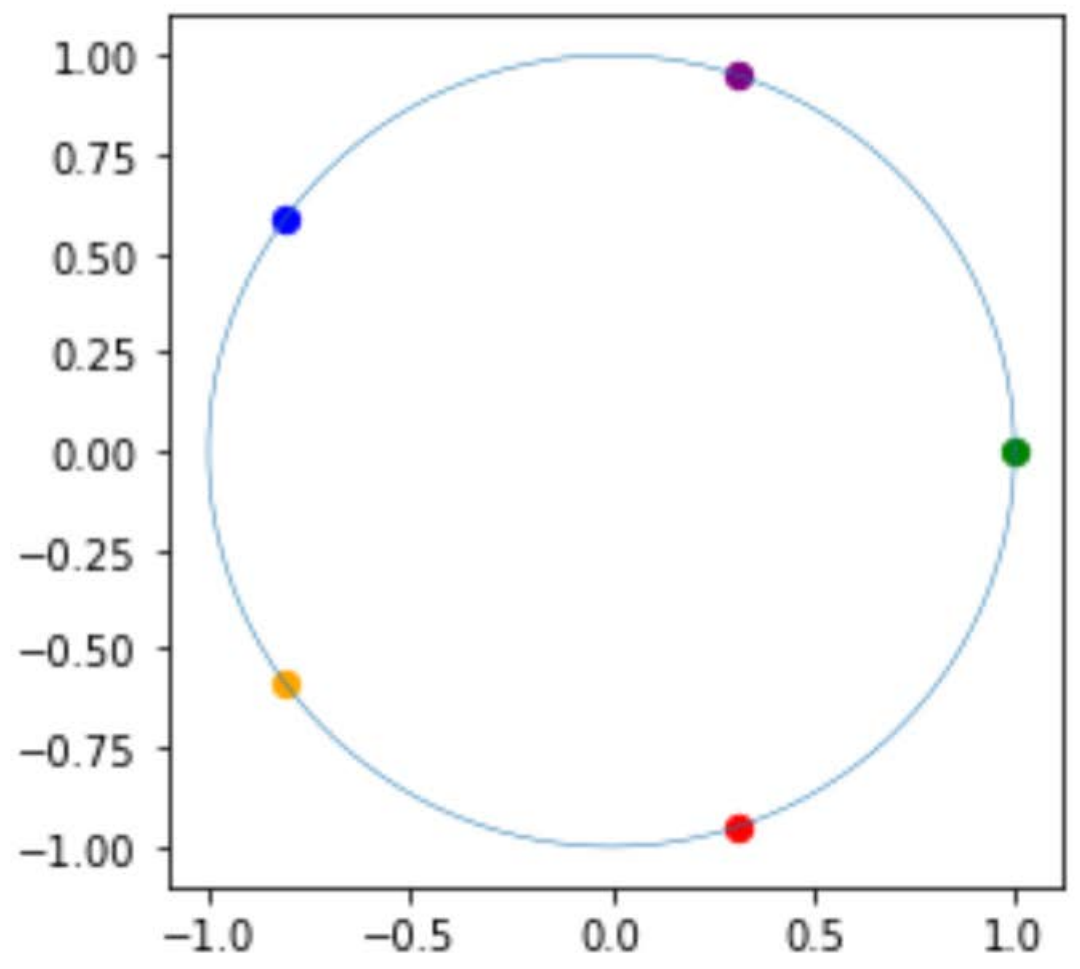

```
x_5 QFT term:    -0.0   +  i 5.0
```

As illustrated by the precession of the colored points, the cell of code above demonstrates how all of the summations resulting from the DFT$^\dagger$ produce either zero or non-term $\tilde{x}_n$ terms. The mathematical details which produce this effect are by no means obvious at first, so I encourage you to test out several examples to get a better feel for things. To summarize, the effect of a DFT$^\dagger$ on a periodic discrete function $f(x)$, which is mathematically equivalent to our quantum system up to normalization factors, can be seen below in the following figures:

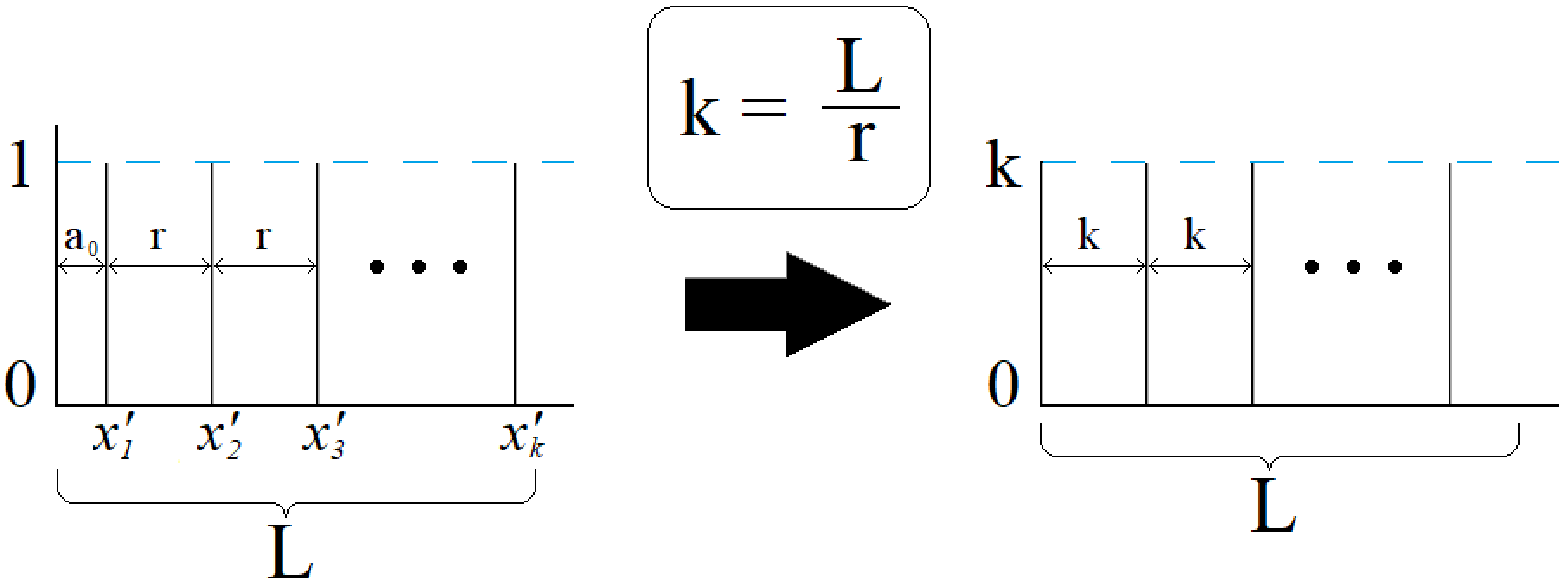

So long as the initial periodic function $f(x)$ matches the criteria illustrated above, the effect of the DFT$^\dagger$ will always transform the function according to the right plot. In terms of our quantum algorithm, so long as our partial measurement results in a superposition state on system one of the correct form, the resulting QFT$^\dagger$ will give us a predictable final state. And now that we know what to expect, let's return to our $N = 15$, $a = 8$ example from earlier, adding in the final QFT$^\dagger$. Based on our analysis thus far, since our quantum system has an underlying period of $r = 4$, with a total size of $L = 16$, we should find a final superposition state of four evenly spaced states, separated by a distance $k = 4$:

In [ ]:

```
a = 8
N = 15
Q = 4
print('a:  ',a,'   N:  ',N)
#===================================
q1 = QuantumRegister(Q,name='q1')
q2 = QuantumRegister(Q,name='q2')
an = QuantumRegister(Q-1,name='a')
c  = ClassicalRegister(Q,name='c')
qc = QuantumCircuit(q1,q2,an,c,name='qc')
#-------------------------------------
for i in np.arange(Q):
    qc.h(q1[int(i)])
print('\n_____ Initial State _____')
oq.Wavefunction(qc,systems=[Q,Q,Q-1],show_systems=[True,True,False])

oq.Mod_Op(Q,qc,q1,q2,an,a,N)
print('\n_____ After Modulo Operation _____')
oq.Wavefunction(qc,systems=[Q,Q,Q-1],show_systems=[True,True,False])

qc.measure(q2,c)
oq.QFT_dgr(qc,q1,Q)
print('\n_____ Partial Measurement + QFT\u2020_____')
oq.Wavefunction(qc,systems=[Q,Q,Q-1],show_systems=[True,True,False])
```

```
a:   8    N:   15

_____ Initial State _____
0.25 |0000>|0000>    0.25 |1000>|0000>    0.25 |0100>|0000>    0.25 |1100>|0000>    0.25 |0010>|0000>    0.25 |1010>|0000>
0.25 |0110>|0000>    0.25 |1110>|0000>    0.25 |0001>|0000>    0.25 |1001>|0000>    0.25 |0101>|0000>    0.25 |1101>|0000>
0.25 |0011>|0000>    0.25 |1011>|0000>    0.25 |0111>|0000>    0.25 |1111>|0000>

_____ After Modulo Operation _____
0.25 |0000>|1000>    0.25 |0010>|1000>    0.25 |0001>|1000>    0.25 |0011>|1000>    0.25 |1100>|0100>    0.25 |1110>|0100>
0.25 |1101>|0100>    0.25 |1111>|0100>    0.25 |0100>|0010>    0.25 |0110>|0010>    0.25 |0101>|0010>    0.25 |0111>|0010>
0.25 |1000>|0001>    0.25 |1010>|0001>    0.25 |1001>|0001>    0.25 |1011>|0001>

_____ Partial Measurement + QFT†_____
0.5 |0000>|0001>    -0.5 |1000>|0001>    -0.5j |0100>|0001>    0.5j |1100>|0001>
```

Take a look at the final superposition state produced on system one above, and note how the same four basis states show up every time, regardless of which state is found in the partial measurement. More specifically, rerunning the cell of code above to see the different partial measurement results reveals that the only difference is the phases on the four final basis states, which we can attribute to the displacement $a_0$ for each initial superposition. There is however, one slight difference between the result shown above and the states you may have expected based on our DFT$^\dagger$ derivations, namely the ordering of the qubits. Up to this point, Shor's Algorithm has interpreted the binary representation of each qubit state as leftmost LSB. Reading off the states in the final superposition shown above then, this would mean that our remaining four states are $|0\rangle$, $|1\rangle$, $|2\rangle$, and $|3\rangle$, which we know is wrong. Interpreted as rightmost LSB however, the final superposition state is composed of the states $|0\rangle$, $|4\rangle$, $|8\rangle$, and $|12\rangle$, yielding $k = 4$, which is the correct answer. Thus, in using QFT$^\dagger$ to transform our final superposition state, we can either use SWAP gates to rearrange our qubits into a leftmost LSB interpretation, or simply read off the final measurement on system one as rightmost LSB.

With the code demonstration above verifying we can translate our DFT$^\dagger$ results into quantum, we are now properly equipped to discuss $why$ the QFT$^\dagger$ transformation helps solve our period finding problem. The key lies in the fact that the QFT$^\dagger$ transforms the period of our superposition state from $r$ to $L/r$. Previously, if we were to make a measurement on system one after the partial measurement on system two, we would say that our state is some multiple of $r$, offset by an unknown $a_o$ (see the possible superposition states from the $N = 15$ $a = 8$ figure earlier). Thus, the result of our measurement contains $some$ information about the period, but not enough to be conclusive after just one measurement. Repeat measurements could in principle be used to eventually determine $r$, but the major issue is in state preparation. More specifically, because we have no control over the state we find from the partial measurement, we have no guarantee that repeating our quantum steps will lead to the same superposition state on system one. Thus, if we wanted to determine $r$ from sampling the superposition states without QFT$^\dagger$, we could be stuck preparing and measuring quantum systems over and over until we build up enough instances of the same partial measurement result on system two (keep in mind that each quantum modulo operation is very costly in terms of algorithm speed).

Alternatively, the question then becomes what new advantage does the QFT$^\dagger$ provide us? Based on our analysis from earlier, we know that the application of QFT$^\dagger$ onto system one will effectively result in a new superposition of states, one in which the separation between the basis states is no longer $r$, but now $L/r$. Additionally, because we know that the state $|00...0\rangle$ is guaranteed to be in the new superposition, this means that all other states must be multiples of $L/r$. Thus, the transformation has effectively removed the offset integer $a_0$, which means if we were to randomly measure a non-zero state (i.e. not the state of all $|0\rangle$'s), we can be certain that our state is some multiple $m$ of $L/r$. And with the help of a clever math trick yet to come and a bit of luck, it turns out that we can determine $r$ without any further measurements! If we convert the binary number representation of our measured state to base-$10$, call it $S$, then the result of our measurement tells us the following information:

$$S \;=\; m \cdot \frac{L}{r}$$

which can be rearranged as:

$$\frac{S}{L} \;=\; \frac{m}{r}$$

The expression shown above is the critical ingredient necessary for us to determine $r$. Working with our quantum system, we will always know the values of $S$ and $L$, which only leaves $m$ and $r$ as unknowns. But through the use of a technique known as Continued Fractions, we can approximate $m$ and $r$, which will in turn complete our period finding problem.

# Determining $r$ Through Continued Fractions

We have finally reached the final step of our quantum period finding subroutine, which is the classical post-processing of the information gathered from a measurement on system one. Since the value $S$ comes from a measurement after the QFT$^\dagger$, both parts of our quantum system have now been collapsed down to single states, which means that we have extracted all of the information we can. Thus, having completed all of the quantum steps, the only thing left is to finish the algorithm classically. And as we already pointed out, we are left with the following equation of two known, and two unknown integers:

$$\frac{S}{L} = \frac{m}{r}$$

The key idea behind our final step uses a technique known as Continued Fractions, which is a classical algorithm for approximating a fraction as two integer numbers:

$$N \approx \frac{p}{q}$$

$$N \neq \text{integer} \qquad p, q = \text{integers}$$

It is important to point out that this technique only works for non-integer numbers (if $N$ is an integer, then the obvious solution is $p = N$, $q = 1$). For our case, the number we want to approximate is $S/L$, which is mathematically guaranteed to be non-integer so long as $S \neq 0$. This is a subtle point, but it is worth noting that our quantum period finding technique essentially 'breaks down' if the state measured on system one is the state of all $|0\rangle$'s. Unfortunately, the zero state is guaranteed to always be in the final superposition, which means that we must rely on a bit of good luck not to measure it.

Returning now to our Continued Fractions technique, the methodology for approximating the integers $p$ and $q$ involves finding values $a_n$, which form the following repeating pattern:

$$a_0 + \cfrac{1}{a_1 + \cfrac{1}{a_2 + \cfrac{1}{\cdots}}}$$

A bit complicated looking at first, but the basis for the algorithm is actually quite straightforward. At each step we separate the number $N_n$ into its integer and decimal components, store the integer as $a_n$, and use the reciprocal of remaining decimal $b_n$ as the next number in the algorithm:

$$(1) \qquad N_o = a_0 + b_0 \qquad 0 \leq b_0 < 1$$

$$(2) \qquad N_1 = \frac{1}{b_0} = a_1 + b_1 \qquad 0 \leq b_1 < 1$$

$$(3) \qquad N_2 = \frac{1}{b_1} = a_2 + b_2 \qquad 0 \leq b_2 < 1$$

$$\vdots$$

The algorithm recursively continues as shown above, terminating when either a $b_n = 0$ is found, or a sufficient level of accuracy is reached. Below is an example of a full Continued Fractions process for which an exact approximation can be found:

$$N = 2.815$$

$$(1) \qquad N = 2 + 0.815 \qquad a_0 = 2 \quad b_0 = 0.815$$

$$(2) \qquad \frac{1}{b_0} = 1.2269938... \qquad a_1 = 1 \quad b_1 = 0.226...$$

$$(3) \qquad \frac{1}{b_1} = 4.4054054... \qquad a_2 = 4 \quad b_2 = 0.405...$$

$$(4) \qquad \frac{1}{b_2} = 2.4666666... \qquad a_3 = 2 \quad b_3 = 0.466...$$

$$(5) \qquad \frac{1}{b_3} = 2.1428571... \qquad a_4 = 2 \quad b_4 = 0.142...$$

$$(6) \qquad \frac{1}{b_4} = 7.0 \qquad a_5 = 7 \quad b_5 = 0$$

Putting all of the values of $a_n$ together, the final answer is as follows:

$$N = 2 + \cfrac{1}{1 + \cfrac{1}{4 + \cfrac{1}{2 + \cfrac{1}{2 + \cfrac{1}{7}}}}} = \frac{563}{200} = 2.815$$

Alternatively, the process shown above can be terminated early for an approximate value, which will be useful for us later on. For example, if we had stopped the algorithm after $a_3$, then our answer would be $31/11 = 2.8181...$ For this lesson, we will be implementing the Continued Fractions technique using the function **ConFrac** from Our_Qiskit_functions:

In [ ]:

```
N = 2.815
#=====================================
q,p,a = oq.ConFrac(N, return_a=True)
print('N  = ',N,'  = ',p,'/',q)
print('\na constants: ',a)
#-------------------------------------
accuracy = 4
q,p,a = oq.ConFrac(N, a_max=accuracy, return_a=True)
print('\n--------------------------------------\nN  = ',N,'  \u2248 ',p,'/',q)
print('\na constants: ',a)
```

```
N  =  2.815   =  563 / 200

a constants:  [2, 1, 4, 2, 2, 7]

--------------------------------------
N  =  2.815   ≈  31 / 11

a constants:  [2, 1, 4, 2]
```

Returning to our period finding problem, we now have the tool of Continued Fractions in our arsenal to complete our subroutine. We ended the previous section noting that our quantum system provided us with the information $S$ and $L$, which thanks to the QFT$^{\dagger}$ operation, guarantees the following equation holds:

$$\frac{S}{L} = \frac{m}{r}$$

Because the combination of $S/L$ is non-integer, we can now use Continued Fractions to approximate this value as the fraction $m/r$, which in turn means that the denominator of our answer is the solution to our period finding problem! Using the $\mathrm{ConFrac}$ function, let's now see our complete period finding subroutine in action:

In [ ]:

```
a = 8
N = 15
Q = 4
print('a:  ',a,'   N:  ',N)
#=====================================
q1 = QuantumRegister(Q,name='q1')
q2 = QuantumRegister(Q,name='q2')
an = QuantumRegister(Q-1,name='a')
c1 = ClassicalRegister(Q,name='c1')
c2 = ClassicalRegister(Q,name='c2')
qc = QuantumCircuit(q1,q2,an,c1,c2,name='qc')
#--------------------------------------
for i in np.arange(Q):
    qc.h(q1[int(i)])
oq.Mod_Op(Q,qc,q1,q2,an,a,N)
print('\n_____ After Modulo Operation _____')
oq.Wavefunction(qc,systems=[Q,Q,Q-1],show_systems=[True,True,False])

qc.measure(q2,c2)
oq.QFT_dgr(qc,q1,Q)
qc.measure(q1,c1)
M = oq.Measurement(qc,shots=1,print_M=False,return_M=True)
print('\n   Partial Measurement:   |'+list(M.keys())[0][5:9]+'>')
print('\nSystem One Measurement:   |'+list(M.keys())[0][0:4]+'>')
#--------------------------------------
S = int(oq.From_Binary(list(list(M.keys())[0][0:4]),'R'))
L = 2**Q
print('\nS = ',S,'        L = ',L)
if( S != 0 ):
    r,mult = oq.ConFrac(1.0*S/L)
    print('\nContinued Fractions Result:   m = ',mult,'   r = ',r)
```

```
a:   8    N:   15

_____ After Modulo Operation _____
0.25 |0000>|1000>    0.25 |0010>|1000>    0.25 |0001>|1000>    0.25 |0011>|1000>    0.25 |1100>|0100>    0.25 |1110>|0100>
0.25 |1101>|0100>    0.25 |1111>|0100>    0.25 |0100>|0010>    0.25 |0110>|0010>    0.25 |0101>|0010>    0.25 |0111>|0010>
0.25 |1000>|0001>    0.25 |1010>|0001>    0.25 |1001>|0001>    0.25 |1011>|0001>

   Partial Measurement:   |0100>

System One Measurement:   |1100>

S =  12          L =  16

Continued Fractions Result:   m =  3    r =  4
```

As promised, the Continued Fractions technique allows us to extract the value of $r$ with only a single measurement on system one. However, if you run the cell of code above several times, you will find that there are still instances where our code fails to produce the correct $r$. More specifically, based on the four possible measurement results for system one, our period finding subroutine produces the following answers:

| | $\lvert 0000 \rangle$ | $\lvert 1000 \rangle$ | $\lvert 0100 \rangle$ | $\lvert 1100 \rangle$ |
|---|---|---|---|---|
| $\frac{S}{L}$ : | $\frac{0}{16}$ | $\frac{4}{16}$ | $\frac{8}{16}$ | $\frac{12}{16}$ |
| $r$ : | 0 | 4 | 2 | 4 |

Thus, even though our final superposition state encodes the period $r$ through evenly spaced basis states of distance $L/r$, it appears that only $50\%$ of the possible final measurements will yield the correct $r$. However, when interpreting the values of $m$ and $r$ obtained from Continued Fractions, we must also consider the possibility that our final fraction may have been reduced past the intended values. For example, in the results shown above, a measurement of the state $\lvert 0100 \rangle$ leads to the Continued Fractions assessment of $8/16$, yielding $m = 1$ and $r = 2$. If we were to take this value of $r = 2$ at face value, we would find that it passes our requirements on $r$, but ultimately fails to lead to the full factorization of $N$:

$$\text{GCD}(2,2) = 2 \quad \checkmark \qquad\qquad 8^{2/2} \ (\text{mod}\ 15) \neq 14 \quad \checkmark$$

$$\text{GCD}(9,15) = 3$$
$$\text{GCD}(7,15) = 1$$

The fact that $r = 2$ leads to one of the factors of $N$ here can be thought of as a lucky coincidence, simply due to the fact that $3$ is a small factor which is easy to stumble upon accidentally. More generally, if one uses an incorrect value of $r$ for a given $a$, the result will very likely lead to a failed GCD attempt. As we can see above, here we have a case where our period finding subroutine has returned to us a value of $r$ which passes both requirements, and yet fails to factor $N$. Mathematically, our derivations from earlier proved that such a case shouldn't exist, which is a signal that one of two possibilities has occurred:

$1)$ Our Continued Fractions technique over reduced the quantity $\frac{m}{r}$

$2)$ Our quantum system measured a state $\lvert S \rangle$ which isn't a correct multiple of $\frac{L}{r}$

Based on the final superposition state for our particular problem of $N = 15,\ a = 8$, it should be clear that option $(2)$ isn't a possibility, unless we introduce measurement errors into the equation, which we are not. We will return to this second point in the next section, but for now let's focus on option $(1)$. If our subroutine returns to us values for $m$ and $r$ which lead to a failed GCD factorization of $N$, then our next move is to check whether any of the higher multiples of $m/r$ will lead to the correct period:

$$\text{check}: \quad \frac{l \cdot m}{l \cdot r} \qquad l \cdot r \ < \ N$$

So long as the quantity $l \cdot r$ is less than $N$, it is a viable candidate for the period of the modulo power function which we are trying to solve. Thus, if a measurement of the state $\lvert 0100 \rangle$ in our subroutine returned to us the quantity $m/r = 1/2$, then the combinations $2/4$, $3/6$, $4/8$, etc... are all valid possible solutions as well. After seeing that $r = 2$ did not solve our problem, the next value to check would then be $r = 4$:

$$\text{GCD}(2,4) = 2 \quad \checkmark \qquad\qquad 8^{4/2} \ (\text{mod}\ 15) \neq 14 \quad \checkmark$$

$$\text{GCD}(65,15) = 5$$
$$\text{GCD}(63,15) = 3$$

As anticipated, using the period $r = 4$ solves our factorization problem, which confirms that our Continued Fractions implementation did indeed over simplify the fraction $m/r$. To summarize then, knowing how to handle the possible interpretations of our Continued Fractions technique leads to a $75\%$ success rate of our quantum subroutine for this problem (any measurement other than $\lvert 0000 \rangle$). So long as we don't get an unlucky measurement on system one, our quantum period finding technique can successfully determine $r$ without any repeat measurements!

## Finding $r$ with $2^Q$ States

The example shown in the previous section demonstrates our first complete implementation of Shor's Algorithm. However, in taking the algorithm one step further, in which we are ready to tackle the factorization of any number $N$, there is one final feature we must discuss. In the example of $N = 15,\ a = 8$, the correct solution to the period finding subroutine turned out to be $r = 4$, which happens to be a factor of the quantity $2^Q$, where $Q$ is the number of qubits we used in our circuit. If we return to the beginning of our discussion regarding DFT$^{\dagger}$ on periodic functions, one subtle but very important point was the fact that $f(x)$ needed to have a precise structure. More specifically, here we are interested in the domain of $f(x)$, which needs to be over $\left[0, L\right]$, where $L$ is some integer multiple of the period $r$.

Now suppose that we remove the upper bound condition on $L$, but still keep the inherent periodic structure of $f(x)$ the same, producing a value of $1$ for $x'$ terms with a period of $r$, offset by some $a_0$, and $0$ for all other $x$ values. Even though we are still dealing with the same underlying periodic function, having a total length $L$ that is no longer an integer multiple of $r$ has dramatic consequences on the transformed function:

In [ ]:

```
k = int( 2+4*random.random() )
r = int( 2+4*random.random() )
L = k*r + 1
print('k: ',k,'      r: ',r,'      L: ',L,'\n------------------------------\n')
#==============================================
for q in np.arange(L):
    n = int(q)
    Q = 0
    for j in np.arange(k):
        Q = Q + np.exp( ((-2*m.pi*1.0j)/L) * n * j * r )
    print( 'n:  ',n,'      \u03A3 =  ',round( Q.real,5 ),' + i',round(Q.imag,5) )
```

```
k:  4       r:  3       L:  13
------------------------------

n:   0       Σ =   4.0   + i 0.0
n:   1       Σ =   -0.20501   + i -0.29701
n:   2       Σ =   0.166   + i -0.43772
n:   3       Σ =   0.78234   + i -0.19283
n:   4       Σ =   2.57406   + i 2.28042
n:   5       Σ =   0.24252   + i -1.99732
n:   6       Σ =   0.94009   + i -0.4934
n:   7       Σ =   0.94009   + i 0.4934
n:   8       Σ =   0.24252   + i 1.99732
n:   9       Σ =   2.57406   + i -2.28042
n:   10       Σ =   0.78234   + i 0.19283
n:   11       Σ =   0.166   + i 0.43772
n:   12       Σ =   -0.20501   + i 0.29701
```

Comparing the classical code shown above with the similar cell from earlier, we can see that extending the length of $L$ by one causes the intermediate $\tilde{x}$ terms between $L/r$ multiples to no longer sum to $0$. Mathematically, the condition that $L = k \cdot r$ is a critical ingredient for all of the exponentials to sum to zero in the DFT$^\dagger$ summation. Without it, the contributing non-zero terms from $f(x)$ no longer precess around the complex unit circle as evenly spaced points. By changing $L$, the frequencies of these complex numbers are effectively altered, no longer leading to perfectly synchronized moments of constructive interference.

Unfortunately, when it comes to constructing our quantum systems, $L$ is not something we can easily control. As we already know, the Hilbert space of our quantum system is determined by the number of qubits we use, $Q$, resulting in $L = 2^Q$. To emphasize our dilemma then, consider a realistic case in which one would look to implement Shor's Algorithm, whereby the goal is to factor some number $N$. Based on the $a$ we choose to use for our quantum modulo operation, we have no way of knowing what period $r$ we are fundamentally encoding into the quantum system. We know that for the algorithm to work $r$ must be even, but unless $r$ is exactly some number $2^n$, we are guaranteed that $r$ *won't* divide into $L = 2^Q$ evenly. And if we consider problems where $N$ is sufficiently large, the probability of picking an $a$ that will have a period of the form $2^n$ gets increasingly smaller as $N$ grows larger. Thus, it's fair to say that working with a quantum case where $L = k \cdot r$ is more of the exception than the rule.

In anticipation that our quantum system is more likely to encounter periods which don't divide into $2^Q$ evenly, what then should we expect in terms of diminishing algorithm accuracy? First off, because of the way in which our quantum modulo operation encodes the states of the system, we are guaranteed that $L = 2^Q$ will only ever be at most $r - 1$ states off from an optimal length. Secondly, states further away from $|00...0\rangle$, closer to $|11...1\rangle$ will feel the impact of an imperfect $L$ the most. To see this, take a look at the cell of code below, which compares the effect of a DFT$^\dagger$ operation on an $f(x)$ of length $L = k \cdot r$ versus $L = k \cdot r + 1$:

In [ ]:

```
k  = int( 4 + 2*random.random() )
r  = int( 3 + 5*random.random() )
a0 = int( (r-1)*random.random() )
print('k = ',k,'        r = ',r,'        a0 = ',a0)
#------------------------------------------
L = k*r
C = np.zeros(r)
C[a0] = 1
#------------------------------------------
f1 = []
f2 = []
for i in np.arange(k):
    for i2 in np.arange(r):
        f1.append( C[i2] )
        f2.append( C[i2] )
f2.append(0)

F1 = oq.DFT(f1,inverse=True)
F2 = oq.DFT(f2,inverse=True)

for q in np.arange( len(F1) ):
    F1[q] = round( abs(F1[q]/k)**2 ,4)
    F2[q] = round( abs(F2[q]/k)**2 ,4)
F2[-1] = round( abs(F2[-1]/k)**2 ,4)
#==========================================
x_bar = []
for j in np.arange(len(F1)):
    x_bar.append(int(j))
plt.bar(x_bar,F1)
x_bar.append(int(j+1))
plt.bar(x_bar,F2,width=0.5)
plt.legend(['Perfect L','L + 1'])
plt.axis([-1, len(F2), 0, 1.3])
plt.show()
```

```
k =  5          r =  7          a0 =  5
```

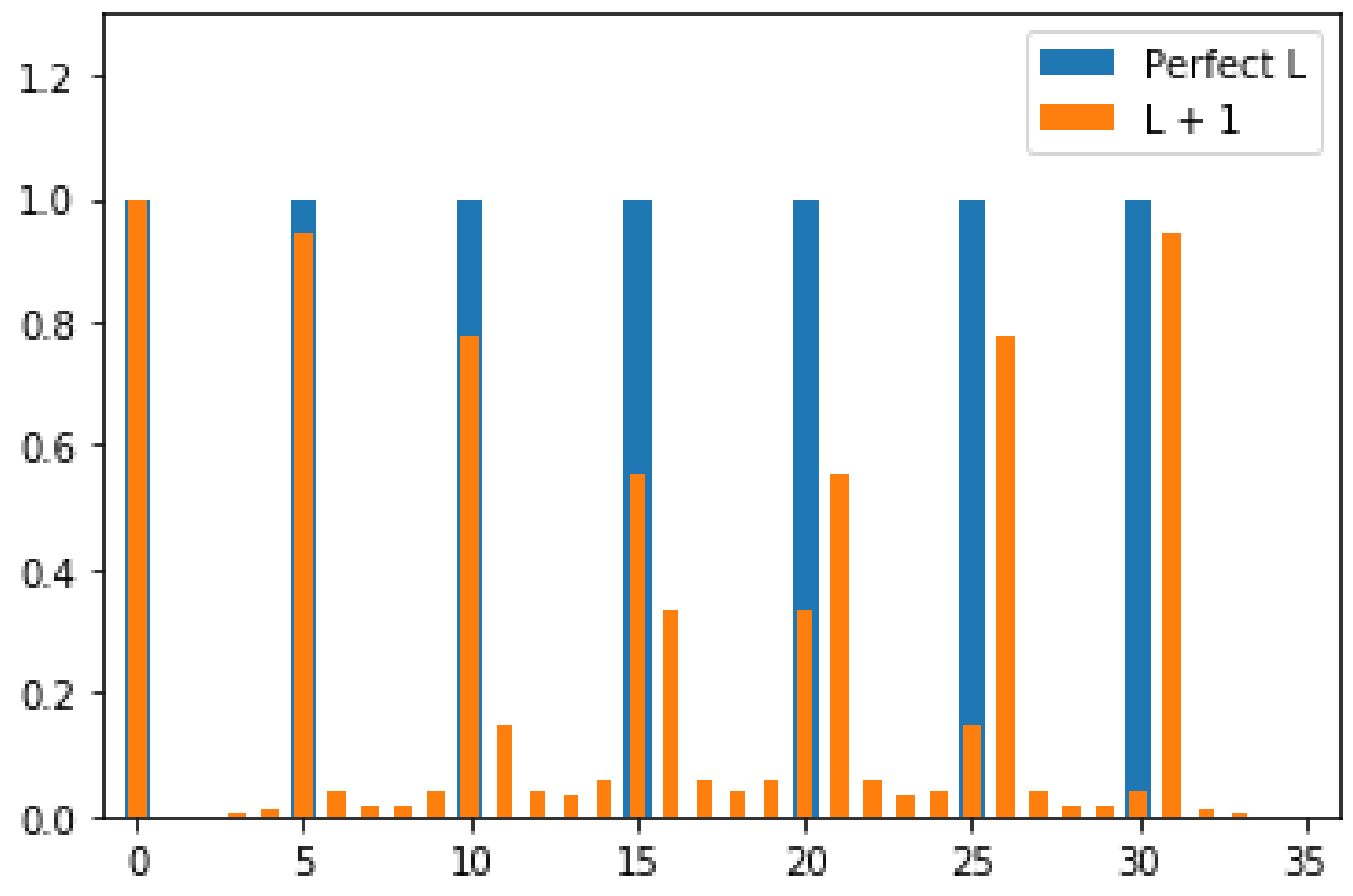

In the plot shown above, demonstrating the case of $r = 7$ for two functions of length $35$ and $36$, note how the probabilities of the two $\tilde{f}(x)$'s differ as the states in the system get closer to $S = L$. Although it may not look it, both DFT$^\dagger$ operations are actually correctly producing $\tilde{f}(x)$ functions that are periodic in $L/r$. Since for the imperfect case we have $L = 36$, the period $L/r$ is now a non-integer value: $5\,\frac{1}{7}$. Consequently, the resulting amplitudes for this new period cannot fall on exact integers, so their values are essentially distributed accordingly to the nearest available integer states (with some small overflow into further states). For example, consider the $4^{\text{th}}$ multiple of $L/R$ for the imperfect case, approximately $20.57$. Because this value is non-integer, the full amplitude of $1$, which would normally fall onto a single state like for the $L = 35$ case, is now distributed to the states $S = 20$ and $S = 21$, with a slight favor for $S = 21$. If we look at the plot above, we find exactly that: the states $S = 20$ and $S = 21$ are highly probable, with $S = 21$ being slightly higher.

Returning now to our quantum period finding problem, the example above is meant to illustrate the kinds of superposition states we can expect from encoding our modulo power function into a system of $L = 2^Q$ states. The effect of the QFT$^\dagger$ is *still* creating a final superposition that is periodic in $L/r$, it's just that $L/r$ is now no longer integer. Consequently, we must reevaluate how best to interpret the information obtained from the final measurement on system one, and how we can use it to extract $r$. In short, our Continued Fractions technique is still viable, but there are now some issues we must consider:

1) Our final measurement value, $S$, now has a non-zero probability of falling on a state that is more than one integer value away from $L/r$.

2) For ideal values of $S$ which *are* nearest integer values to $L/r$, a Continued Fractions attempt of $S/L$ may still fail to produce $r$.

To further differentiate the two problems outlined above, issue $(1)$ is a worst-case type scenario in which our final measurement yields a state which is nowhere near an $L/r$ value. Consequently, the Continued Fractions assessment of $S/L$ will almost certainly have no chance of yielding the correct $r$. Unfortunately, such instances are unavoidable due to the nature of superposition states, but are something we should come to expect when working with quantum algorithms in general. Unless our algorithm is designed such that $100\%$ of the probability in the system is distributed amongst desired states (and our quantum computer is perfectly noiseless), an unlucky measurement is always a possibility, effectively negating all of the work of the algorithm. The important thing is to always consider that a measurement result may be a dud, and to try and identify such instances as quickly as possible so that one can start over.

Moving on to issue $(2)$, which is actually the best-case scenario, even if our measurement falls on a state which is a nearest neighbor integer to a multiple of $L/r$, the Continued Fractions assessment of $S/L$ may still fail to yield $r$ directly. If we let the Continued Fractions technique continue to the full approximation, we will always simply find the integers $S$ and $L$, or their simplified fraction. And unless we get lucky and are searching for an $r$ which evenly divides our $L = 2^Q$, then chances are the quantity $S/L$ will *not* be an integer multiple of $L/r$. So then, in anticipation that the $S/L$ we find is some value slightly off from a true $L/r$ multiple, we can attack the problem from two different angles:

1) Try lower order approximations to the Continued Fractions assessment of $S/L$, and see if any of them yield $r$

2) Starting from the quantity $S/L$, try using the Continued Fractions technique on nearest integer values of $S$ and $L$, within a reasonable threshold

In essence, because of the way in which the final superposition boosts the probability of numerous states around the multiples of $L/r$, we need to introduce some flexibility in our classical analysis of the state $|\,S\,\rangle$. Strategy $(1)$ is to simply see if we get lucky enough such that a lower order approximation to $S/L$ will yield a multiple of $L/r$, while strategy $(2)$ is to thoroughly search in the general area around $S$. Depending on the number we are trying to factor and the value of $S$ we obtain, the combination of these two strategies is quite reliable. For example, in the $r = 7$, $L = 36$ code from above, 6 out of the $7$ most probable states, nearing $70\%$ of the total probability, will yield the correct $r$ with lower order approximations to $S/L$:

In [ ]:

```
r = 7
L = 36
print('r:  ',r,'      L: ',L,'\n==============================')
#=====================
S = 10
print('\nS  = ',S,'\n-----------------------------')
for i in np.arange(4):
    q,p = oq.ConFrac(S/L,a_max=int(i+2))
    print('order ',int(i+1),' Continued Fractions:  '+str(p)+'/'+str(q),' \u2248 ',round(p/q,4))
#---------------------
S = 21
print('\nS  = ',S)
for i in np.arange(4):
    q,p = oq.ConFrac(S/L,a_max=int(i+2))
    print('order ',int(i+1),' Continued Fractions:  '+str(p)+'/'+str(q),' \u2248 ',round(p/q,4))
```

```
r:   7       L:  36
==============================

S  =  10
-----------------------------
order  1  Continued Fractions:  1/3  ≈  0.3333
order  2  Continued Fractions:  1/4  ≈  0.25
order  3  Continued Fractions:  2/7  ≈  0.2857
order  4  Continued Fractions:  5/18  ≈  0.2778

S  =  21
order  1  Continued Fractions:  1/1  ≈  1.0
order  2  Continued Fractions:  1/2  ≈  0.5
order  3  Continued Fractions:  3/5  ≈  0.6
order  4  Continued Fractions:  7/12  ≈  0.5833
```

In the code example shown above, the Continued Fractions cases for $S = 10$ and $S = 21$ are shown to demonstrate instances where strategy $(1)$ does and does not work. For the case of a measurement yielding $S = 10$, the $3^{rd}$ order approximation to $10/36$ gives us the values $m = 2$ and $r = 7$, which in turn solves our period finding problem. Conversely, all of the lower order approximations to $21/36$ fail to produce a denominator of $r$, which we would have concluded by checking the unsuccessful values of $2$, $5$, and $12$ as well as their higher multiples. At this point, we would need to consider whether the state $|21\rangle$ is a dud measurement, or if we were simply unlucky in that strategy $(1)$ failed, but $S/L$ is still a close approximation to a multiple of $L/r$ (which for this case we know is true). Thus, our next move would be to employ strategy $(2)$ and consider nearby fractions to $21/36$ and see if any of them successfully yield $r$:

In [ ]:

```
S = 21
L = 36
order = 4
#==================
for i in np.arange(3):
    S_new = int( S - 1 + i)
    for j in np.arange(3):
        L_new = int( L - 1 + j)
        if( (S_new!=S) or (L_new!=L) ):
            print('\nS  = ',S_new,'    L  = ',L_new,'\n-----------------------------')
            for i in np.arange(4):
                q,p = oq.ConFrac(S_new/L_new,a_max=int(i+2))
                print('order ',int(i+1),' Continued Fractions:  '+str(p)+'/'+str(q),' \u2248 ',round(p/q,4))
```

```
S  =  20     L  =  35
-----------------------------
order  1  Continued Fractions:  1/1  ≈  1.0
order  2  Continued Fractions:  1/2  ≈  0.5
order  3  Continued Fractions:  4/7  ≈  0.5714
order  4  Continued Fractions:  4/7  ≈  0.5714

S  =  20     L  =  36
-----------------------------
order  1  Continued Fractions:  1/1  ≈  1.0
order  2  Continued Fractions:  1/2  ≈  0.5
order  3  Continued Fractions:  4/7  ≈  0.5714
order  4  Continued Fractions:  5/9  ≈  0.5556

S  =  20     L  =  37
-----------------------------
order  1  Continued Fractions:  1/1  ≈  1.0
order  2  Continued Fractions:  1/2  ≈  0.5
order  3  Continued Fractions:  6/11  ≈  0.5455
order  4  Continued Fractions:  7/13  ≈  0.5385

S  =  21     L  =  35
-----------------------------
order  1  Continued Fractions:  1/1  ≈  1.0
order  2  Continued Fractions:  1/2  ≈  0.5
order  3  Continued Fractions:  3/5  ≈  0.6
order  4  Continued Fractions:  3/5  ≈  0.6

S  =  21     L  =  37
-----------------------------
order  1  Continued Fractions:  1/1  ≈  1.0
order  2  Continued Fractions:  1/2  ≈  0.5
order  3  Continued Fractions:  4/7  ≈  0.5714
order  4  Continued Fractions:  17/30  ≈  0.5667

S  =  22     L  =  35
-----------------------------
order  1  Continued Fractions:  1/1  ≈  1.0
order  2  Continued Fractions:  1/2  ≈  0.5
order  3  Continued Fractions:  2/3  ≈  0.6667
order  4  Continued Fractions:  5/8  ≈  0.625
```

```
S  =  22     L  =  36
-----------------------------
order  1  Continued Fractions:  1/1  ≈  1.0
order  2  Continued Fractions:  1/2  ≈  0.5
order  3  Continued Fractions:  2/3  ≈  0.6667
order  4  Continued Fractions:  3/5  ≈  0.6

S  =  22     L  =  37
-----------------------------
order  1  Continued Fractions:  1/1  ≈  1.0
order  2  Continued Fractions:  1/2  ≈  0.5
order  3  Continued Fractions:  3/5  ≈  0.6
order  4  Continued Fractions:  22/37  ≈  0.5946
```

As we can see, three out of the eight combinations for $S$ and $L$ shown above correctly yield the period $r = 7$, once again completing our period finding subroutine. If however, none of these combinations had been successful, we would once again find ourselves with a decision of whether to continue searching around $S = 20$, or start the entire quantum process over for a new measurement. While there's no hard rule that can tell us when we've reached the limit of what a particular $S$ value can provide us, one optimistic viewpoint is that we can do both!

Like the expression "having your cake and eating it too", there's no reason why we can't simultaneously run a second quantum period finding subroutine while exploring all of the information we can extract from $S/L$. If our goal is to factor some number $N$ as quickly as possible, then the most efficient thing we could do is to continuously be running our quantum subroutine while simultaneously processing the information we get from measurements classically. At the very worst, a dud measurement sends us on a wild goose chase trying to approximate some $S/L$ which is nowhere near a multiple of $L/r$. But in the process, consider the fact that for every value of $r$ we check and fail, we are eliminating wrong answers and ultimately narrowing in on the correct period. And for context, it is very important to remember that checking GCD's and evaluating approximate Continued Fractions are both *very* fast classical processes compared to the bottleneck of our quantum subroutine: the modulo operation. Thus, in the time it would take to run a second full quantum subroutine, there's no reason *not* to try as many approximations to $S/L$ as we can!

## Full Shor's Example

Having now seen some techniques for evaluating the final measured state $|S\rangle$, we have reached the conclusion of our deep dive into Shor's Algorithm. While well over half of this lesson was spent just on just understanding the role of modulo period finding, it is important to remember the full context of the algorithm: factoring. Incorporating all of the topics we've covered thus far, below is a full code implementation of Shor's Algorithm for the case of $N = 55$. I encourage you to run the cell several times and see the various kinds of results one can get from the combinations of $a$ and $S$:

In [ ]:

```python
N = 55
Q = m.ceil( m.log(N,2) )
L = 2**Q
a = int( 2+ (N-3)*random.random() )
r = oq.r_Finder(a,N)
#=================================================
print('N  = ',N,'     Q  = ',Q,'     a  = ',a,'     Searching For:  r =',r)
if( oq.Euclids_Alg(a,N) > 1 ):
    print('\na happens to have a factor in common with N: ',oq.Euclids_Alg(a,N))
else:
    q1 = QuantumRegister(Q,name='q1')
    q2 = QuantumRegister(Q,name='q2')
    an = QuantumRegister(Q-1,name='a')
    c1 = ClassicalRegister(Q,name='c1')
    c2 = ClassicalRegister(Q,name='c2')
    qc = QuantumCircuit(q1,q2,an,c1,c2,name='qc')
    #----------------------------------------------
    for i in np.arange(Q):
        qc.h(q1[int(i)])
    oq.Mod_Op(Q,qc,q1,q2,an,a,N)
    qc.measure(q2,c2)
    oq.QFT_dgr(qc,q1,Q)
    qc.measure(q1,c1)
    M = oq.Measurement(qc,shots=1,print_M=False,return_M=True)
    S = int(oq.From_Binary(list(list(M.keys())[0][0:Q]),'R'))
    #----------------------------------------------
    print('\nSystem One Measurement:   |'+list(M.keys())[0][0:Q]+'>')
    print('\nS = ',S,'        L = ',L)
    if( S!= 0):
        r = oq.Evaluate_S(S,L,a,N)
        if( r!=0 ):
            print('\nFound the period r  = ',r)
            if( ((r)%2 == 0) and ( a**(int(r/2))%N != int(N-1) )):
                f1 = oq.Euclids_Alg(int(a**(int(r/2))+1),N)
                f2 = oq.Euclids_Alg(int(a**(int(r/2))-1),N)
                print('\nFactors of N:  ',int(f1),'  ',int(f2))
            else:
                if( (r)%2 != 0 ):
                    print('\nr does not meet criteria for factoring N:  r is not even')
                else:
                    print('\nr does not meet criteria for factoring N:  a^(r/2) (mod N) = N-1')
        else:
            print('\nCould not find the period using S,  start over')
    else:
        print('\nMeasured S = 0,  start over')
```

```
N  =  55      Q  =  6      a  =  46      Searching For:  r = 10

System One Measurement:   |000110>

S =  6         L =  64

Found the period r  =  10

Factors of N:   11    5
```

This concludes lesson 8, and our analysis of the famous Shor's Factoring Algorithm. Since it was first published 1995, Shor's Algorithm still remains one of the most important and influential quantum algorithms to date. As evidenced by the sheer number of mathematical tricks involved, Peter Shor's recognition of how to turn a factoring problem into a period finding problem makes this quantum algorithm an inspiration for the potential of quantum computing.

I hope you enjoyed this lesson, and I encourage you to take a look at my other .ipynb tutorials!

### Citations

**[1]** P. W. Shor, "Polynomial-Time Algorithms for Prime Factorization and Discrete Logarithms on a Quantum Computer", SIAM JSC. **26** (1997)

**[2]** R. Rivest, R. Shamir, L. Adleman, "A Method for Obtaining Digital Signatures and Public-Key Cryptosystems", Communications of the ACM. **21** (2): 120–126 (1978)

**[3]** S. Beauregard, "Circuit for Shor's algorithm using 2n+3 qubits", Quantum Information and Computation **3** (2): 175-185 (2003)

# Lesson 9 - Q-Means Clustering

In this lesson, we will be exploring a quantum approach to the classical learning algorithm $k$-Means Clustering. The aim of this algorithm is to take a large set of raw data and organize each data point as belonging to one of $k$ groupings, called clusters. The quantum analogue to this algorithm, $Q$-Means Clustering, will look to replace one the bottleneck of the classical algorithm, namely calculating distances between data points. At its core, this quantum distance calculating process will make use of a powerful subroutine: the SWAP Test, for which we will be delving deeply into.

Contributing author: Saahil Patel

In order to make sure that all cells of code run properly throughout this lesson, please run the following cell of code below:

In [ ]:

```
from qiskit import ClassicalRegister, QuantumRegister, QuantumCircuit, Aer, execute
import Our_Qiskit_Functions as oq
import numpy as np
import math as m
import scipy as sci
import random
import time
import matplotlib
import matplotlib.pyplot as plt
from itertools import permutations
S_simulator = Aer.backends(name='statevector_simulator')[0]
%matplotlib notebook
plt.rcParams['animation.html'] = 'jshtml'
```

## Analyzing Data Into Groups

In our modern era of smartphones, computers, etc., some of the most sought after algorithms are those which handle enormous amounts of data. Collecting data is easy, but making good use of it is tricky. So much so in fact that we've seen the rise of an entire field of science, "Data Scientists", who specialize solely in the advancement of techniques for handling data. And while classical computers have steadily been getting faster, their limits are starting to become clearer, and the need for new fundamental techniques are required if we want to tackle larger and more complex data analyzing problems.

When analyzing large sets of data, oftentimes we would like to categorically group data points of similar nature into what are known as "clusters", which in turn allows us to interpret, predict, and make use of the data in a much more compact manner. An example of this may include medical studies, whereby each individual person and their unique health conditions makes up a data point, and the goal is to categorically group people of similar dispositions to better prepare them for various potential health risks. Once enough data points have been properly sorted into clusters, allowing for a fast and efficient high-level interpretation of the data, newer points of data can then be quickly sorted in and analyzed. In the medical field, this can lead to life-saving decisions for patients who show health trends similar to previous cases.

In this tutorial, we will be focusing on a very well known data analyzing algorithm, $k$-means clustering, first published by Stuart Llyod in 1982 **[1]**, but developed over two decades prior. Mathematically, the core of this algorithm boils down to interpreting data points through calculating Euclidean distances:

$$A = \begin{bmatrix} A_1 \\ A_2 \\ \cdot \\ \cdot \\ A_n \end{bmatrix} \qquad B = \begin{bmatrix} B_1 \\ B_2 \\ \cdot \\ \cdot \\ B_n \end{bmatrix} \qquad D_{AB} = \Big( \sum_i^n (B_i - A_i)^2 \Big)^{\frac{1}{2}}$$

where the goal is to find a central point, known as a "centroid", which minimizes the distance from each point within the cluster:

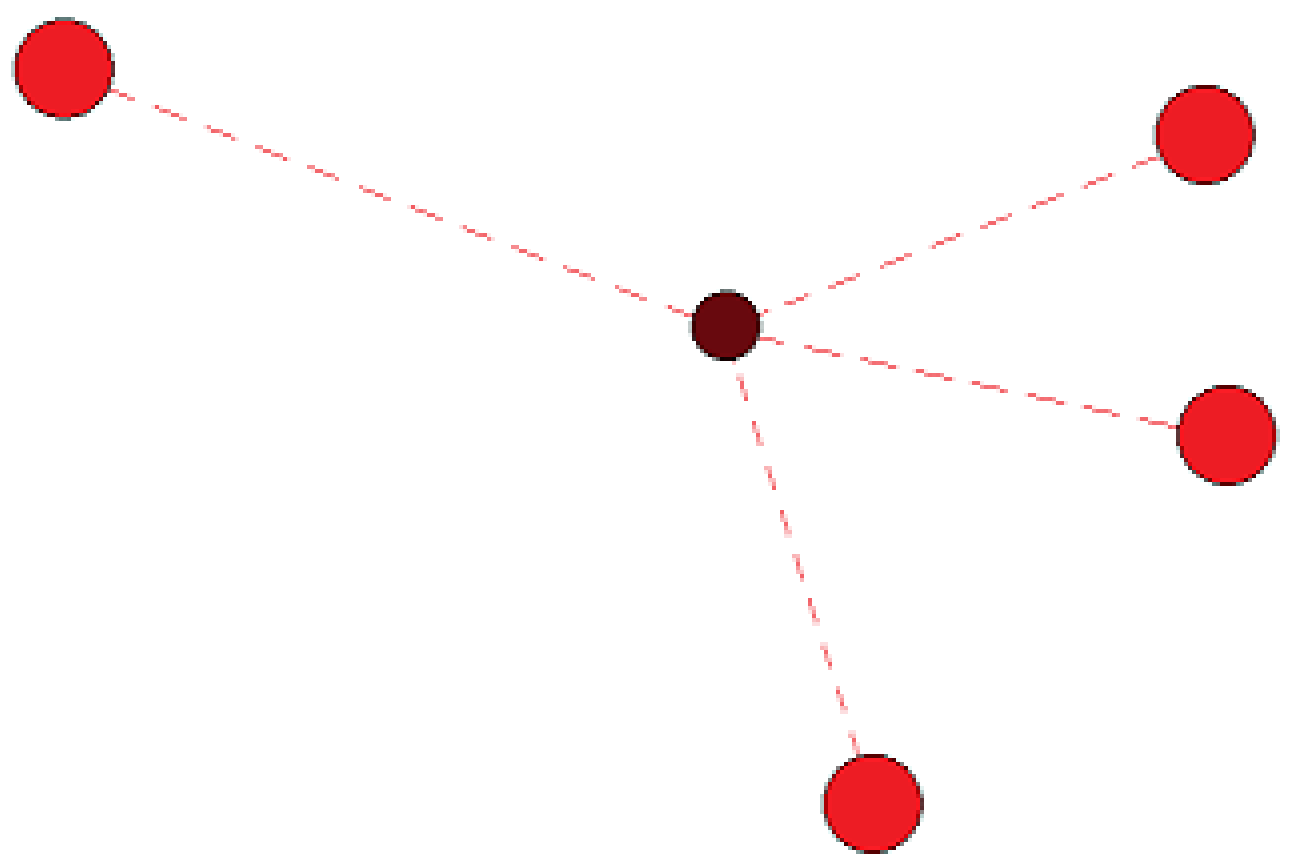

As illustrated above, the dark red point represents the centroid to our cluster. The dashed lines represent each data point's distance to the centroid, $D_{ij}$, whose sum we want to minimize. When calculating the centroid for a set of $N$ points, one approach is to use the mean $x$ and $y$ values computed from all of the data points making up the cluster:

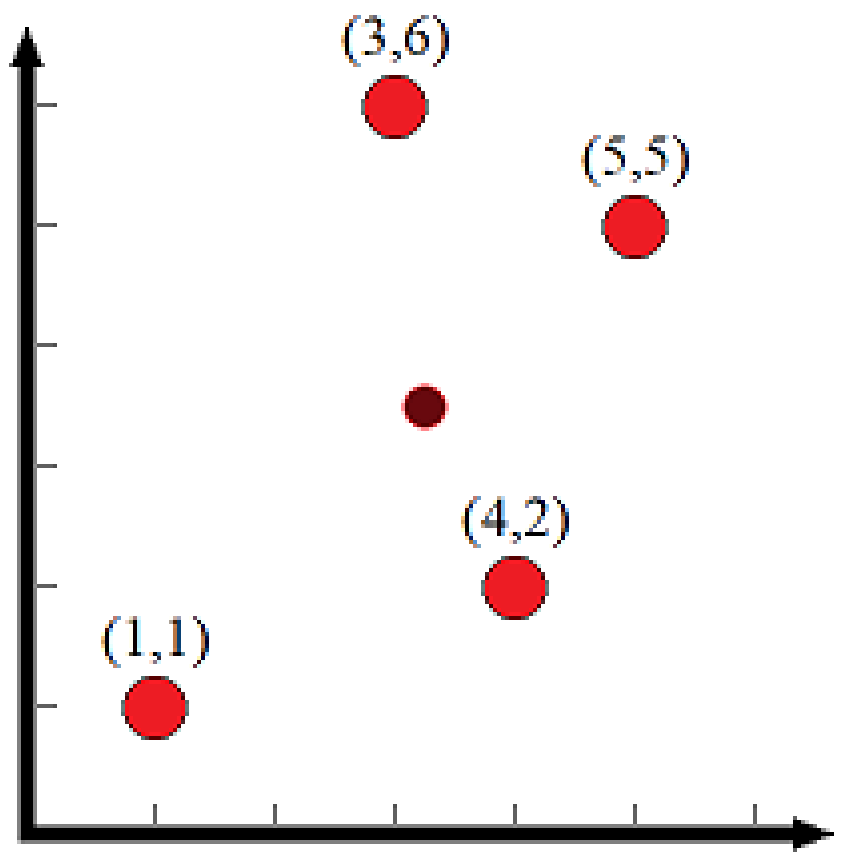

$$\bar{x} = \frac{1+3+4+5}{4} = 3.25$$

$$\bar{y} = \frac{1+2+5+6}{4} = 3.5$$

As shown in the example above, using the mean $x$ and $y$ values from all of the points in the cluster gives us a centroid that appears to be quite central. However, it turns out that computing a centroid in this manner does *not* give us the true center of our cluster (by which we mean minimizing the sum of distances $D_{ij}$). Instead, centroids computed using mean coordinate values minimize the total *squared* Euclidean distance:

$$\sum_i^N D_{ij}^2$$

To illustrate this difference, below is a quick code which exhaustively searches for the centroid points which minimize the sums of $D_{ij}$ versus $D_{ij}^2$, corresponding to the example above:

In [ ]:
```
Data = [ [1,1],[4,2],[5,5],[3,6] ]
D_min = [1000,0,0]
D2_min = [1000,0,0]
#------------------------------------------     Searches for centroids within   3 < x < 4   and   3 < y < 4
for j1 in np.arange(1000):
    X = 3.0 + j1/1000.0
    for j2 in np.arange(1000):
        Y = 3.0 + j2/1000.0
        D = 0
        D2 = 0
        for k in np.arange( len(Data) ):
            D = D + m.sqrt( (X-Data[k][0])**2 + (Y-Data[k][1])**2 )
            D2 = D2 + (X-Data[k][0])**2 + (Y-Data[k][1])**2
        if( D < D_min[0] ):
            D_min = [ D, X, Y ]
        if( D2 < D2_min[0] ):
            D2_min = [ D2, X, Y ]
#------------------------------------------
print('Minimum Distance:          ',round(D_min[0],2),'     coordinates:  (',D_min[1],D_min[2],')')
print('Minimum Distance Squared:  ',round(D2_min[0],2),'     coordinates:  (',D2_min[1],D2_min[2],')')
```

```
Minimum Distance:            9.78       coordinates:  ( 3.6 3.6 )
Minimum Distance Squared:    25.75      coordinates:  ( 3.25 3.5 )
```

As shown in the final answer to our code, the centroid computed using the mean coordinate values is indeed the location which minimizes the sum of $D_{ij}^2$, but not $D_{ij}$. Now presented with this difference, the natural first question to ask is: is this a problem? Does the success of our algorithm critically depend on finding centroids which minimize total Euclidean distance, or is distance squared sufficient? The short answer: not a problem at all. Working with centroids calculated from means is an entirely viable approach to data analysis (you probably suspected this answer, hence the name $k$-means clustering). In fact, there is an equally efficient algorithm known as $k$-medians clustering, which you guessed it, uses median coordinate values to compute cluster centroids. When using either approach, the important thing to note is the way in which we are able to efficiently represent a large group of data points with a single centroid, with relatively minimal computing power. By comparison, note how the coding example above went about computing the centroid that minimized $D_{ij}$, searching exhaustively within a given area for the optimal point. If our algorithm critically depended on this centroid, we would be in trouble, as the computing overhead would be too much for even realistically small data sets.

Moving right along with our $k$-means algorithm, representing an entire data set with a single centroid is nice, but is it helpful? Perhaps unsurprisingly, categorizing every data point into a single cluster isn't very useful, so our next step will be to group all of our data into $k$ different clusters. The question then becomes, which data points should belong to each cluster? Ideally we want to again minimize the distance (squared) of each data point to it's associated cluster centroid, but the challenge lies in determining which data points should contribute to which cluster. For example, take a look at the scatter plot of raw data below and think about how best to separate each point into one of two groups:

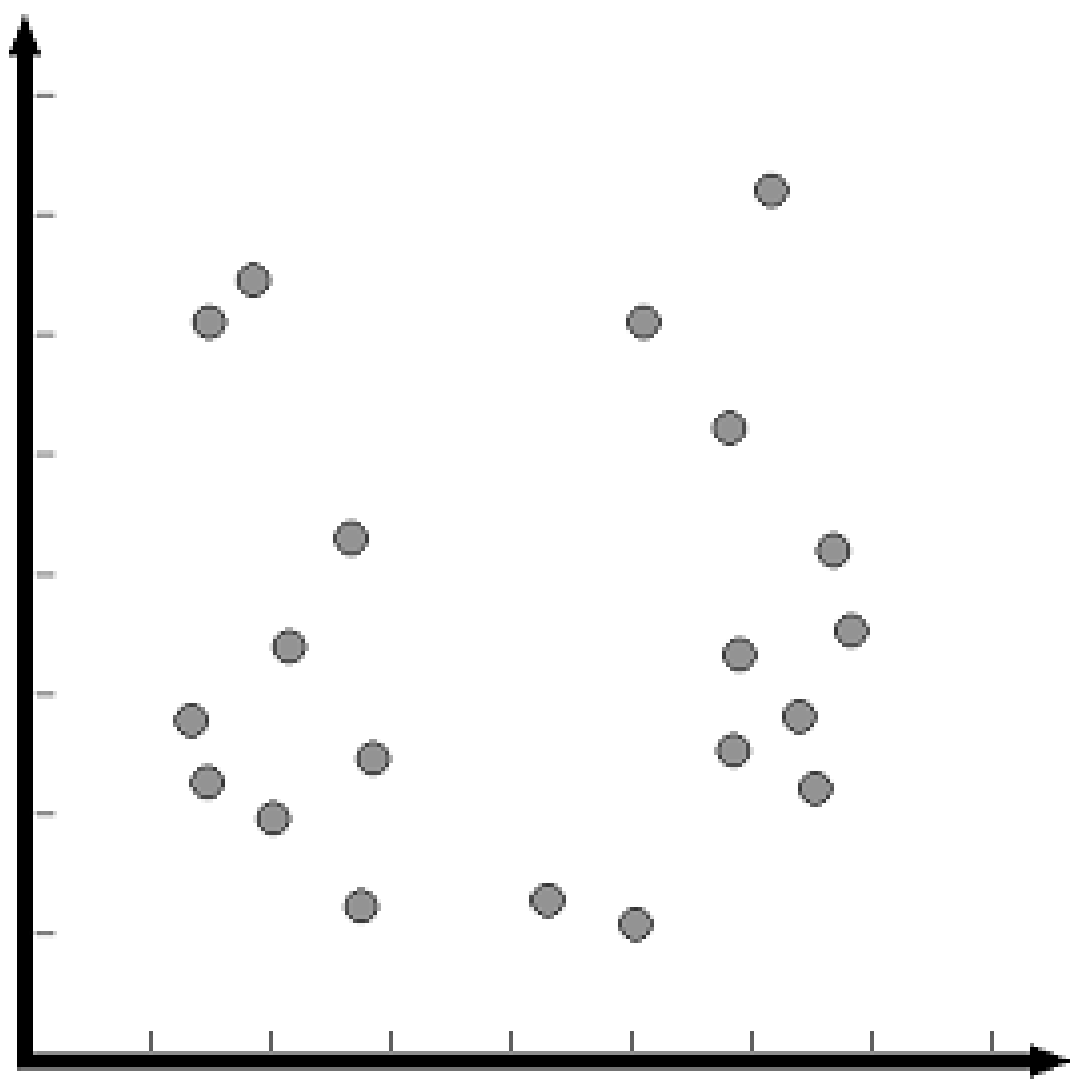

Visually, your eye may have a hunch as to the general shape of each potential cluster, but mathematically we need a way to be sure. Consider the two groupings below, each one representing a different interpretation of the same data:

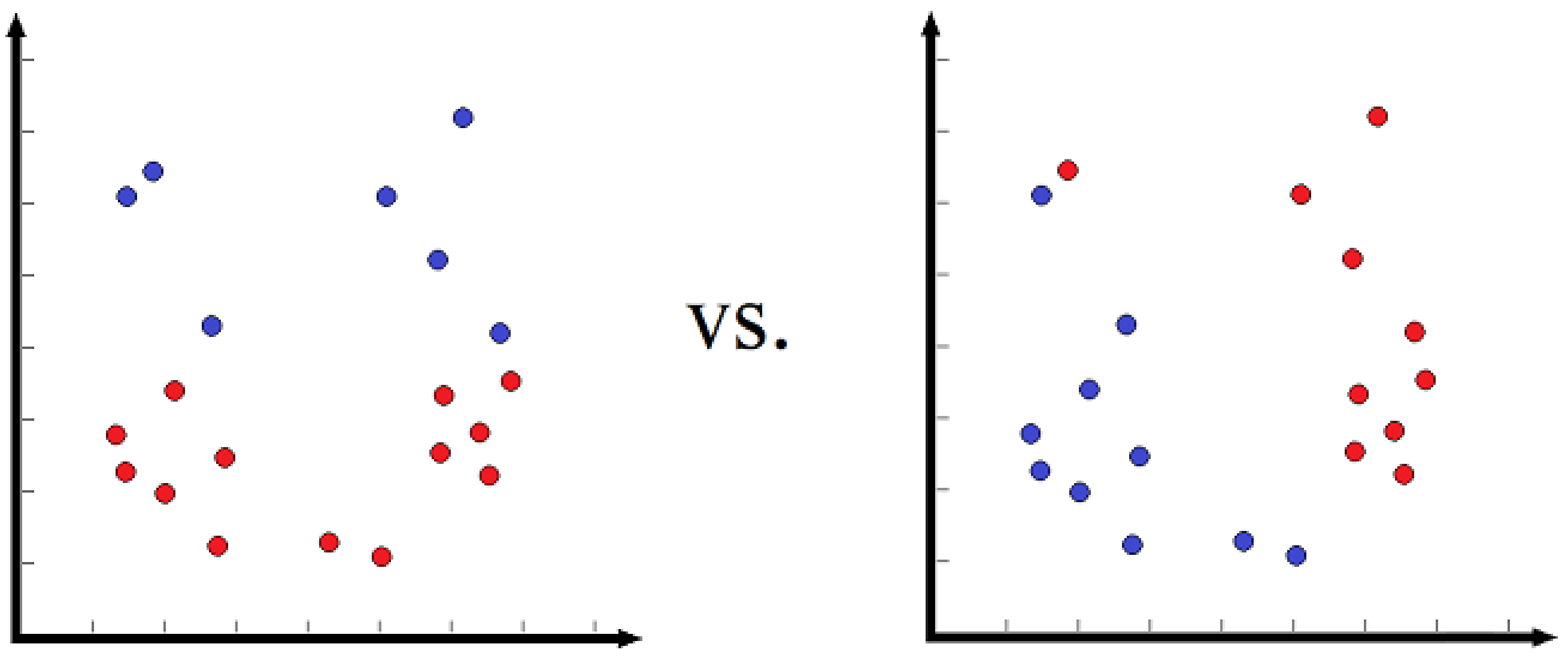

Sometimes 'eyeballing' the data can lead to good first intuitions about potential clusters, but in general there is no perfect methodology for solving this problem in one go. That is to say, there is no magical algorithm which can take in a complete set of data like the one shown above and perfectly sort it into $k$ clusters of similar data points. Thus, like many problems which can't be solved in a single step, our $k$-means algorithm will solve our cluster sorting problem iteratively. More specifically, each step of the algorithm will analyze all the data points and update each cluster and centroid accordingly.

## The $k$-Means Clustering Algorithm

As demonstrated in the previous section, we have the mathematical means for computing centroids with relative ease, but our problem lies in determining which data points to use for our $k$ clusters. More specifically, to begin our algorithm we need to select $k$ initial centroids, which in turn determine our initial clusters. As you might imagine though, where we choose to place our initial centroids can lead to drastically different clusters. Luckily, as we shall see in the coming demonstrations, the final answer provided by the $k$-means algorithm is only minimally influenced by our initial choices. Thus, for our implementation of the $k$-means algorithm in this lesson, we will be following a first step guess of centroids known as the Forgy method. Since we have no *a priori* information about the optimal structure of our clusters, we will simply pick $k$ points out of our data set at random and treat them as our first centroids. From there, the iterative flow of the algorithm works as follows:

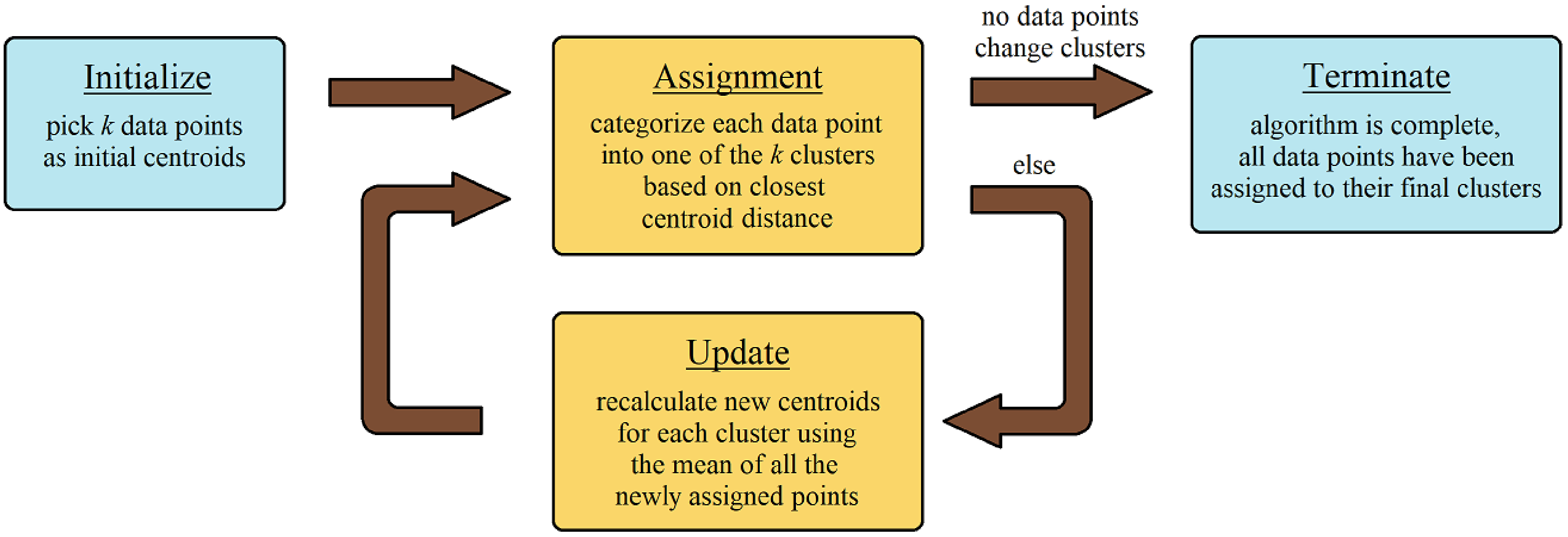

As shown above, the core of the algorithm is a back and forth flow between the "Assignment" and "Update" steps. After picking our initial $k$ centroids, by whichever means we choose, the algorithm then proceeds iteratively until finally being terminated after there are no changes in any of the data points. It's important to note that this process is not guaranteed to always terminate to the same final centroids / clusters. Although their influence is small, our final answer will have some remnants of the initial $k$ data points which were chosen at random. While this isn't problematic for the centroids, whose final values may differ slightly, ideally it would be nice if the algorithm were to always group the data points into the same final clusters, regardless of initial conditions. However, since this isn't a guarantee, one must always keep in mind that rerunning the algorithm again may lead to different classifications on some borderline points. That being said, overall the algorithm is quite stable, and in general will always lead to very similar final clusters, with only a small percentage of points that may differ from run to run.

Despite not having a guaranteed repeatable final solution, the $k$-means clustering algorithm, sometimes referred to as the "naive" $k$-means algorithm as we will be implementing it, is still a powerful and reliable tool from which more complex / precise algorithms can be constructed. For our tutorial here, the implementation of this classical unsupervised learning algorithm, as outlined above, is all we need in order to appreciate the quantum version to come. Remember, our true aim is to see how we can use our quantum computer to improve the bottleneck step of $k$-means clustering, which can then be implemented into simple / advanced clustering algorithms alike. But before jumping into the quantum side of things, let's conclude our discussion of the classical $k$-means algorithm with a visual demonstration. Below is a working example of the algorithm outlined above, designed to show how the centroids / clusters update with each step:

```
In [ ]:
fig = plt.figure()
ax = fig.add_subplot(1,1,1)
ax.axis([-0.2,8.2,-0.2,8.2])
fig.show()
colors  = ['red','limegreen','deepskyblue','gold']
colors2 = ['darkred','darkgreen','darkblue','darkorange']
#---------------------------------------------------------
N = 240
k = 4
#---------------------------------------------------------
Data = oq.k_Data(k,N)                                      # Comment out this line after one run to reuse the same data po
for d in np.arange(len( Data )):
    ax.scatter( Data[d][0], Data[d][1], color='black', s=10 )
fig.canvas.draw()
time.sleep(2)
#---------------------------------------------------------
Centroids = oq.Initial_Centroids( k, Data )
Clusters = []
Clusters,old_Clusters = oq.Update_Clusters( Data, Centroids, Clusters )
for c1 in np.arange(len(Clusters)):
    for c2 in np.arange( len( Clusters[c1] ) ):
        ax.scatter( Clusters[c1][c2][0],Clusters[c1][c2][1], color=colors[c1],s=10 )
    ax.scatter( Centroids[c1][0],Centroids[c1][1], color=colors2[c1], marker='x',s=50 )
    fig.canvas.draw()
    time.sleep(1)
time.sleep(2)
#---------------------------------------------------------
terminate = False
iters = 0
while( (terminate==False) and (iters<50)  ):
    Centroids,old_Centroids = oq.Update_Centroids(Centroids, Clusters)
    Clusters,old_Clusters = oq.Update_Clusters( Data, Centroids, Clusters )
    oq.Draw_Data( Clusters, Centroids, old_Centroids, fig, ax, colors, colors2 )
    terminate = oq.Check_Termination( Clusters, old_Clusters )
    iters = iters + 1

print( 'Clustering Complete:    ',iters,' Iterations' )
```

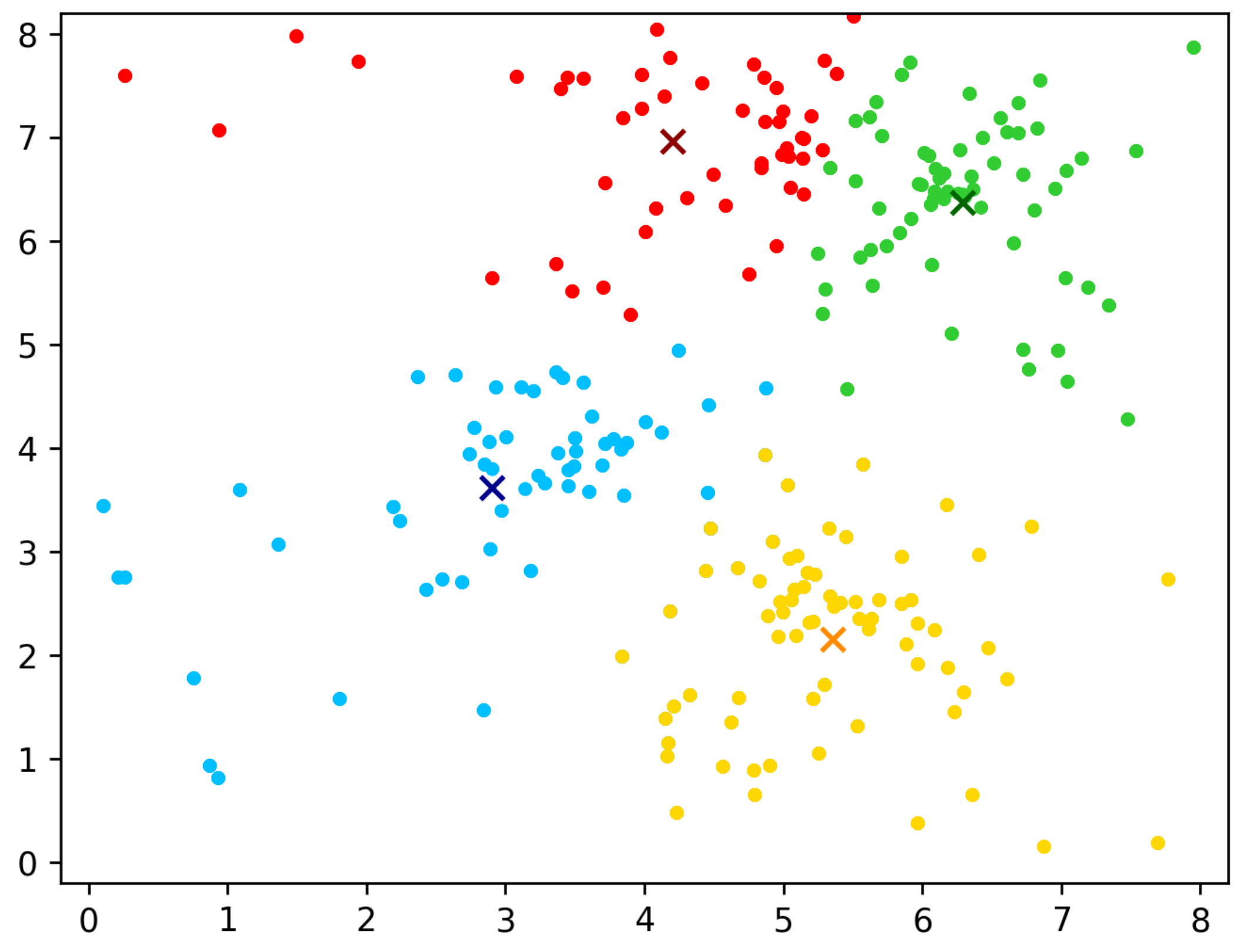

```
Clustering Complete:     6  Iterations
```

As demonstrated in the changing plots above, each iteration of the "Update" and "Assignment" steps in the algorithm result in a redistribution of the data points to one of the $k$ clusters, as well as the movement of any centroids associated with clusters which either lost or gained points. I encourage you to try several runs, with varying values for $N$ and $k$, and see for yourself the various ways in which the algorithm propagates. Additionally, if you're curious about how the algorithm performs using the same data points but with differing initial centroids, try commenting out the $\mathrm{k_Data(k,n)}$ line after running the code once. Below is an example of this, showing that indeed the algorithm will tend to converge towards the same final clusters regardless of initial conditions:

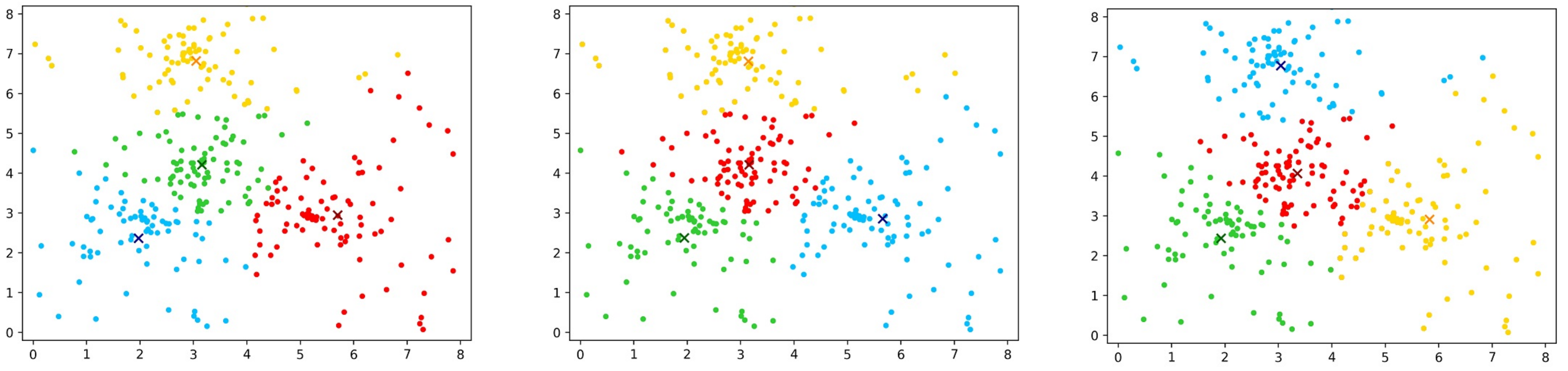

In comparing the three plots above, many of the borderline points between clusters fluctuate, but the overall location of the four centroids is consistent. If we think about how to interpret these fluctuations, it should make sense that borderline cases don't fully belong to one cluster or the other. In analyzing real-world data, the trickiest points to interpet are those which possess features of multiple distinct categories. For example, imagine we use our algorithm to determine TV show recommendations, and we want to categorize the new hit phenomenon "Baking Explosions". One iteration of the algorithm might recommend the show to users who like "Cooking", while a second run might recommend it to people who enjoy "Action". Although a somewhat silly example, the point here is that in real data analysis there will always be borderline cases which shouldn't fully belong to just one cluster.

## Classical to Quantum: $k$-Means $\longrightarrow$ $Q$-Means Clustering

Now that we have our baseline understanding of the classical $k$-means algorithm, it's time to investigate where a quantum computer can lend a hand. Returning to the work flow diagram from earlier, we can see that the majority of the runtime for this particular algorithm is contained within the "Assignment" and "Update" steps. And in particular, it is the "Assignment" step which is the most computationally costly, requiring the calculation of distance between every point and every centroid. By contrast, the "Update" step is considerably quicker, which we already pointed out as the main advantage to using means as our methodology for computing centroids.

Diving a bit deeper into the computational cost of the "Assignment" step, we can say that the bottleneck of our algorithm lies in the sorting of each point into clusters, requiring the computation of $k$ Euclidean distances per data point. Thus, if we are dealing with a data set of $N$ points, each "Assignment" step will cost us $O(kN)$ computations. Conversely, once all of the data points have been assigned to their respective clusters, the following "Update" step only requires $O(k)$ computations to generate the new centroids. This being the case, the natural first place to look for an algorithm improvement is in reducing the number of necessary distance calculations, or as we shall see, improve the speed at which we can calculate them!

## Quantum Subroutine: The SWAP Test

In order to properly discuss the role in which our quantum computer is going to provide us a speedup for the $k$-means clustering algorithm, we will table our discussion of calculating Euclidean distances for now, and instead turn our attention to two subroutines, the first of which being "The SWAP Test". Now, we've seen the SWAP gate in previous lessons, and as a quick reminder, below is its circuit, matrix, and mathematical description:

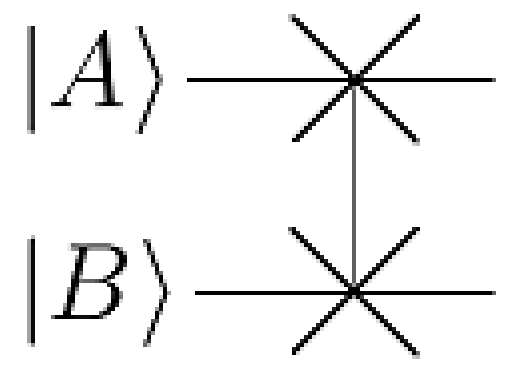

$$\text{SWAP} \;=\; \begin{bmatrix} 1 & 0 & 0 & 0 \\ 0 & 0 & 1 & 0 \\ 0 & 1 & 0 & 0 \\ 0 & 0 & 0 & 1 \end{bmatrix} \qquad\qquad \text{SWAP}\,|\,A\,\rangle \otimes |\,B\,\rangle \;=\; |\,B\,\rangle \otimes |\,A\,\rangle$$

The effect of the SWAP gate is an easy one to remember, as it "swaps" two qubits' quantum states. With that in mind, there is a special case where the effect of the SWAP gate is mute, namely when $|\,A\,\rangle \;=\; |\,B\,\rangle$. For the situation where two qubits possess the same quantum state, the action of a SWAP gate still switches them, but the overall effect on the joint quantum system is unchanged. Although this may seem like a trivial point, it turns out to be the basis for the SWAP Test subroutine for which we are interested in.

Let's suppose we are given two single qubit quantum states $|\,\Psi\,\rangle$ and $|\,\phi\,\rangle$, and we are interested in how similar they are. Similar how, you might ask. Well, we can define our metric of similarity based around two extremes: $|\,\Psi\,\rangle$ and $|\,\phi\,\rangle$ are exactly the same state, or they are perfectly orthogonal. Mathematically, this means that our measure of similarity ranges from $0$ to $1$:

$$0 \;\leq\; \langle\,\phi\,|\,\Psi\,\rangle \;\leq\; 1$$

As shown above, the inner product of $|\,\Psi\,\rangle$ and $|\,\phi\,\rangle$ is a perfect candidate for implementing our similarity measure. Unfortunately, the inner product of two states isn't a tool in our repertoire that we can easily put into practice. Because the result of an inner product is a scalar, this means that the only suitable candidate is a measurement (all gate operations are matrices which map vectors to vectors, and we need an operation that maps vectors to a scalar). If it were possible, the means by which we would go about implementing the inner product of $|\,\Psi\,\rangle$ and $|\,\phi\,\rangle$ would be to measure $|\,\phi\,\rangle$ in the $|\,\Psi\,\rangle$ basis, or vice versa. But as we already know, we are limited to measurements in the computational basis, which is where our problem lies.

To overcome the issue of being bound to the computational basis, and achieve our goal of performing a similarity measure between $|\,\Psi\,\rangle$ and $|\,\phi\,\rangle$, we will be performing the SWAP Test! In order to carry out this quantum subroutine, which will be the basis for several of the topics in this lesson yet to come, we only require two additional ingredients to accompany our quantum states $|\,\Psi\,\rangle$ and $|\,\phi\,\rangle$. First, we will need an additional ancilla qubit, which will ultimately store the result of our similarity measure. And secondly, we will need the control version of the SWAP gate: CSWAP, shown below:

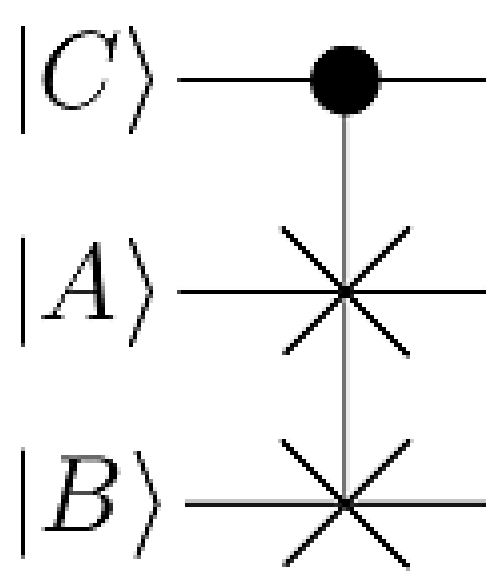

$$\text{CSWAP} \;=\; \begin{bmatrix} 1 & 0 & 0 & 0 & 0 & 0 & 0 & 0 \\ 0 & 1 & 0 & 0 & 0 & 0 & 0 & 0 \\ 0 & 0 & 1 & 0 & 0 & 0 & 0 & 0 \\ 0 & 0 & 0 & 1 & 0 & 0 & 0 & 0 \\ 0 & 0 & 0 & 0 & 1 & 0 & 0 & 0 \\ 0 & 0 & 0 & 0 & 0 & 0 & 1 & 0 \\ 0 & 0 & 0 & 0 & 0 & 1 & 0 & 0 \\ 0 & 0 & 0 & 0 & 0 & 0 & 0 & 1 \end{bmatrix} \qquad \text{CSWAP}\,|+\rangle \otimes |A\rangle \otimes |B\rangle \;=\; \frac{1}{\sqrt{2}}|0\rangle \otimes |A\rangle \otimes |B\rangle \;+\; \frac{1}{\sqrt{2}}|1\rangle \otimes |B\rangle \otimes |A\rangle$$

Just as the name implies, the CSWAP gate implements a SWAP gate between two qubits if and only if the control qubit is in the $|1\rangle$ state. This 3-qubit gate already comes standard with our qiskit library, and can be seen in the cell of code below:

In [ ]:
```
a = QuantumRegister(1,name='a')
q = QuantumRegister(2,name='q')
qc= QuantumCircuit(a,q)

qc.h( a[0] )
qc.x( q[1] )

print('____ Before CSWAP ____')
oq.Wavefunction(qc, systems=[1,2])

qc.cswap( a[0], q[0], q[1] )

print('\n____ After CSWAP ____')
oq.Wavefunction(qc, systems=[1,2])
```

```
____ Before CSWAP ____
0.70711 |0>|01>    0.70711 |1>|01>

____ After CSWAP ____
0.70711 |1>|10>    0.70711 |0>|01>
```

As promised, the code above illustrates how the effect of the CSWAP gate is only felt by the portion of our quantum state containing the control qubit in the $|1\rangle$ state, flipping the states of the two target qubits. With this CSWAP gate now in hand, we're ready to discuss the SWAP Test. As we shall see, the addition of the extra control qubit, in conjunction with CSWAP, is what is going to allow us to extract the scalar value needed for our similarity measure. Below is the complete quantum circuit for the SWAP Test:

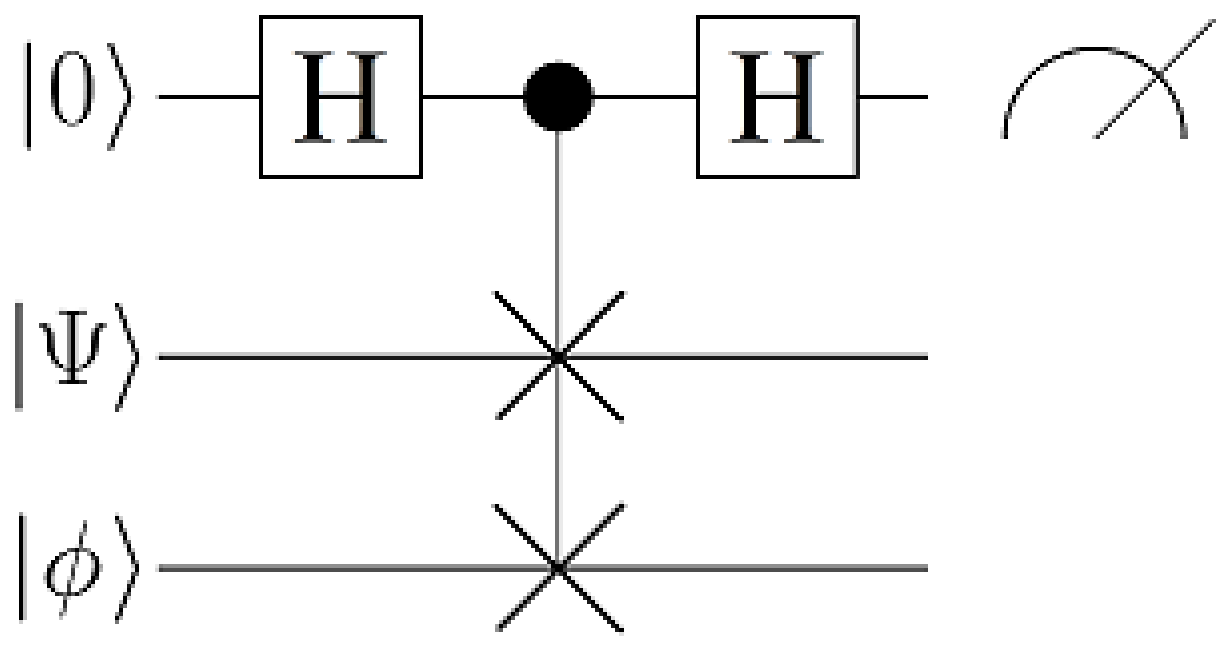

Simple and effective, the SWAP Test only requires three quantum gates (the CSWAP gate is actually composed of more gates however) and a measurement to determine the similarity between the states $|\Psi\rangle$ and $|\phi\rangle$. But in order to fully understand how this quantum circuit works, we will need to do a bit of math dissecting, shown in the steps below for a SWAP Test on the states $|A\rangle$ and $|B\rangle$:

$$|0\rangle \otimes |A\rangle \otimes |B\rangle \;\equiv\; |0AB\rangle$$

$$H \otimes I^2 \;\longrightarrow\; \frac{1}{\sqrt{2}}\Big( |0AB\rangle + |1AB\rangle \Big)$$

$$\text{CSWAP} \;\longrightarrow\; \frac{1}{\sqrt{2}}\Big( |0AB\rangle + |1BA\rangle \Big)$$

$$H \otimes I^2 \;\longrightarrow\; \frac{1}{2}\Big( |0AB\rangle + |1AB\rangle + |0BA\rangle - |1BA\rangle \Big)$$

$$= \frac{1}{2}\Big( |0\rangle \otimes \big[ |AB\rangle + |BA\rangle \big] \;+\; |1\rangle \otimes \big[ |AB\rangle - |BA\rangle \big] \Big)$$

Now we come to the final step, which is analyzing the result of our measurement on the control qubit. As shown in the last line of the derivation above, the final result of our SWAP Test circuit has put our quantum system into a state where the two components of the control qubit are tied to different quantities composed of $|A\rangle$ and $|B\rangle$. More specifically, we have:

$$|0\rangle \quad \longleftrightarrow \quad |A\rangle \otimes |B\rangle \;+\; |B\rangle \otimes |A\rangle \qquad\qquad |1\rangle \quad \longleftrightarrow \quad |A\rangle \otimes |B\rangle \;-\; |B\rangle \otimes |A\rangle$$

The key to interpreting this quantum state lies in the probability with which we can expect to measure our control ancilla qubit in the $|0\rangle$ state:

$$\begin{aligned}
\mathrm{P}\big(|0\rangle_C\big) &= \frac{1}{4}\Big|\,|AB\rangle + |BA\rangle\,\Big|^2 \\
&= \frac{1}{4}\Big[\,\langle AB|AB\rangle + \langle AB|BA\rangle + \langle BA|AB\rangle + \langle BA|BA\rangle\,\Big] \\
&= \frac{1}{2} + \frac{1}{2}\langle A|B\rangle\langle B|A\rangle \\
&= \frac{1}{2} + \frac{1}{2}\Big|\langle A|B\rangle\Big|^2
\end{aligned}$$

$\Big($ Note that $\langle AB|BA\rangle = \langle A|B\rangle \otimes \langle B|A\rangle = \langle B|A\rangle \otimes \langle A|B\rangle\Big)$

The full derivation above is by no means easy to digest at first, so I encourage you to spend some extra time fully grasping how the SWAP Test circuit leads to this final result. In summary, the last line shown above holds our final answer, giving us a clear relationship between the expected measurement rate of our control ancilla qubit and the inner product $\langle A|B\rangle$. Thus, through the measurement of the ancilla qubit, we can extrapolate the similarity between the states $|A\rangle$ and $|B\rangle$! The story doesn't quite end here however, but first let's see the SWAP Test in action with a coding example:

In [ ]:

```
q = QuantumRegister(3,name='q')
c = ClassicalRegister(1,name='c')
qc= QuantumCircuit(q,c,name='qc')

qc.h( q[1] )
qc.x( q[2] )
qc.barrier()
#------------------------------    The SWAP Test
qc.h( q[0] )
qc.cswap( q[0], q[1], q[2] )
qc.h( q[0] )
qc.measure( q[0], c[0] )
#-------------------------------
oq.Measurement( qc,shots=10000, )
print('  ')
print(qc)
```

```
2493|1>    7507|0>

               ░ ┌───┐   ┌───┐┌─┐
q_0: ──────────░─┤ H ├─■─┤ H ├┤M├
     ┌───┐     ░ └───┘ │ └───┘└╥┘
q_1: ┤ H ├─────░───────X───────╫─
     ├───┤     ░       │       ║
q_2: ┤ X ├─────░───────X───────╫─
     └───┘     ░               ║
c_0: ══════════════════════════╩═
```

The cell of code above performs a SWAP Test between the following two states:

$$|\Psi\rangle = \frac{1}{\sqrt{2}}\Big[\,|0\rangle + |1\rangle\,\Big] \qquad\qquad |\phi\rangle = |1\rangle$$

which results in the following inner product:

$$\langle\phi|\Psi\rangle = \frac{1}{\sqrt{2}}\langle 1|1\rangle$$

Thus, when we run our SWAP Test, we should find that the probability of measuring the $|0\rangle$ state is:

$$\begin{aligned}
\mathrm{P}\big(|0\rangle_C\big) &= \frac{1}{2} + \frac{1}{2}\Big|\langle\phi|\Psi\rangle\Big|^2 \\
&= \frac{1}{2} + \frac{1}{2}\Big|\frac{1}{\sqrt{2}}\Big|^2 \\
&= \frac{3}{4}
\end{aligned}$$

Comparing this result to our cell of code above, we do indeed find the $|0\rangle$ state with a $75\%$ probability! If you're interested in testing out more complex $|\Psi\rangle$ and $|\phi\rangle$ states, such as ones with complex phases on the different basis state components, I encourage you to play around with the code for yourself. In our next example, we're going to show how to extend the SWAP Test to states of more than one dimension. A simple enough task, all we need to do is add additional CSWAP gates to our cicuit, properly lining up which qubits to swap:

In [ ]:

```
a  = QuantumRegister( 1, name='a' )
q1 = QuantumRegister( 2, name='q1' )
q2 = QuantumRegister( 2, name='q2' )
c  = ClassicalRegister( 1, name='c' )
qc = QuantumCircuit( a, q1, q2, c )
#=========================================
qc.h( q1[0] )
qc.h( q1[1] )
qc.x( q2[1] )
oq.Wavefunction( qc, systems=[1,2,2], show_systems=[False,True,True] )
qc.barrier()
#---------------------------------          2-Qubit SWAP Test
qc.h( a[0] )
qc.cswap( a[0], q1[0], q2[0] )
qc.cswap( a[0], q1[1], q2[1] )
qc.h( a[0] )
qc.measure(a,c)
#---------------------------------
print('\n___ Measurement Probabilities on the Control Qubit ___')
oq.Measurement( qc, shots=8000 )
print('  ')
print(qc)
```

```
0.5  |00>|01>    0.5  |10>|01>    0.5  |01>|01>    0.5  |11>|01>

___ Measurement Probabilities on the Control Qubit ___
2977|1>    5023|0>

             ░ ┌───┐        ┌───┐┌─┐
  a_0: ──────░─┤ H ├─■──■───┤ H ├┤M├
       ┌───┐ ░ └───┘ │  │   └───┘└╥┘
 q1_0: ┤ H ├─░───────X──┼─────────╫─
       ├───┤ ░       │  │         ║ 
 q1_1: ┤ H ├─░───────┼──X─────────╫─
       └───┘ ░       │  │         ║ 
 q2_0: ──────░───────X──┼─────────╫─
       ┌───┐ ░          │         ║ 
 q2_1: ┤ X ├─░──────────X─────────╫─
       └───┘ ░                    ║ 
  c_0: ═══════════════════════════╩═
```

In the example above, the two states which we are applying the SWAP test to are:

$$|\Psi\rangle \;=\; \frac{1}{2}\Big[\,|00\rangle + |01\rangle + |10\rangle + |11\rangle\,\Big] \qquad\qquad |\phi\rangle \;=\; |01\rangle$$

which should result in the following inner product and corresponding SWAP Test probability:

$$\langle\phi|\Psi\rangle \;=\; \frac{1}{2}$$

$$\mathrm{P}\big(|0\rangle_C\big) \;=\; \frac{5}{8}$$

Returning now to the results of our code, sure enough we find roughly $5000$ out of the total $8000$ measurements to be in the $|0\rangle$ state, confirming that our SWAP Test has indeed worked. Taking a closer look at the code, you can see that we were able to implement our 2-dimensional CSWAP gate by using two individual CSWAPS stemming from the same control qubit (no need to bring in any additional ancilla qubits like we did for our higher order CCNOT implementations). Visually, we can implement a CSWAP gate between any two $N$-dimensional states by simply applying a series of $N$ CSWAP gates:

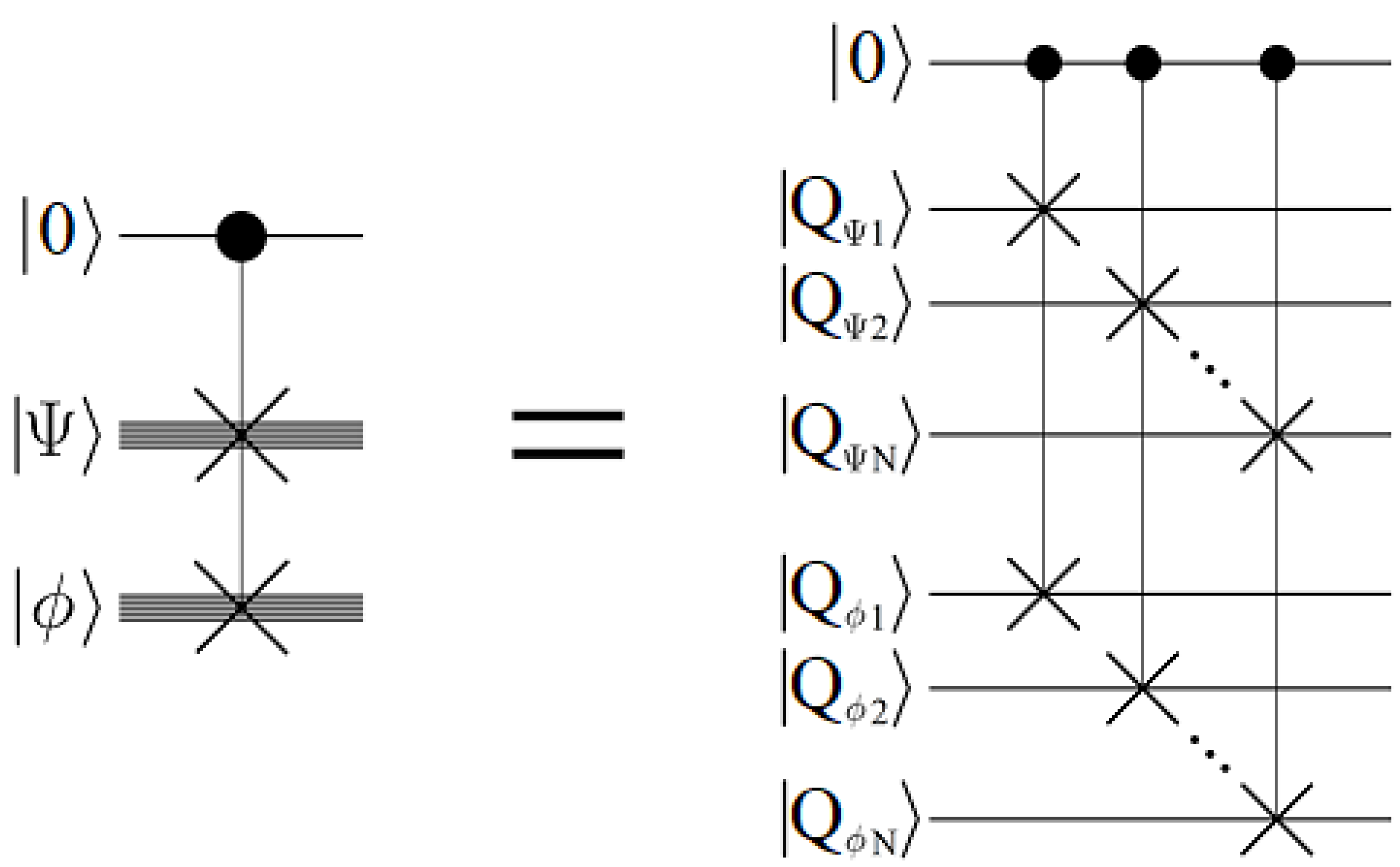

As the final point before moving on to our next topic in this lesson, it is important to note that the SWAP Test is a measurement based analysis, which means that we can only even achieve an approximate answer. To see this, consider how we would interpret the results of our cell of code above if we were to only perform $1$ shot. Regardless of whether we measured $|0\rangle$ or $|1\rangle$, we really couldn't say anything definitive about $\langle\phi|\Psi\rangle$. Thus, it's only through many runs of the SWAP Test can we build up a statistic about the probability of measuring $|0\rangle$. Although this may seem like a drawback to the SWAP Test, repeat measurements in order to gain an approximate answer is unfortunately one of the limitations of quantum computing we must accept. Luckily for us, the quantum circuit itself is so short that repeat trials are easy to do, so long as the state preparation for $|\Psi\rangle$ and $|\phi\rangle$ aren't terribly cumbersome.

## Quantum Subroutine: Distance Calculation

With the SWAP Test now in our toolbox for evaluating the inner product between two quantum states, we're ready to jump into our second quantum subroutine for this lesson: calculating distances. As we already mentioned earlier, computing Euclidean distances is the bottleneck of our $k$-means algorithm, so now we're going to see how we might improve $k$-means through the use of our quantum computer. In fact, we're going to take a look at *two* methods for calculating distances, each with their own pros and cons. Ultimately however, only one of them will be used in our final $Q$-means clustering algorithm, for reasons which will become clear after comparing the requirements necessary for each technique.

### DistCalc 1: Euclidean Distance

The first distance calculating subroutine (which we will nickname DistCalc) that we will be studying was first put forth by Llyod, Mohseni, and Rebentrost **[2]**, and demonstrates how one can use quantum states to compute the quantity $|A - B|^2$, the Euclidean distance squared between two vectors $\vec{A}$ and $\vec{B}$. And as we've already discussed, this is exactly the quantity that gets minimized when we compute centroids using the means. For the purpose of assigning data points to nearest clusters, once $|A - B|^2$ has been computed, taking the square root for the Euclidean distance between $\vec{A}$ and $\vec{B}$ is simple enough.

To begin this DistCalc subroutine, our first task is the encoding of the two vectors $\vec{A}$ and $\vec{B}$ as quantum states. Specifically, we want to represent the components of each vector as the different elements of our quantum state:

$$\vec{A} \quad \longrightarrow \quad |A\rangle \;=\; \frac{1}{|A|}\sum_i A_i\,|i\rangle$$

For example, suppose we wanted to encode the point $\vec{A} = \big[3,4\big]$ as a quantum state:

$$\big[3,4\big] \quad \longrightarrow \quad \frac{1}{5}\Big[\,3|0\rangle \;+\; 4|1\rangle\,\Big]$$

$$= \;\frac{3}{5}|0\rangle \;+\; \frac{4}{5}|1\rangle$$

As we can see by the way in which we've defined our encoding, we're always guaranteed to produce a normalized quantum state, which is good. The only drawback however is that points which are a constant multiple of each other get mapped to the same quantum state. For example:

$$\big[6,8\big] \quad \longrightarrow \quad \frac{1}{10}\Big[\,6|0\rangle \;+\; 8|1\rangle\,\Big]$$

$$= \;\frac{3}{5}|0\rangle \;+\; \frac{4}{5}|1\rangle$$

Even though we can't represent every data point as a unique quantum state, we shall see that this turns out to be a non-issue, one that can be corrected for with a single classical calculation. The bigger challenge lies in the quantum states themselves, as preparing arbitrary amplitudes is typically a tough ask of quantum computers. Luckily for us, Qiskit comes with the function **initialize()** already built into its simulator. So long as we provide Qiskit with a normalized vector, it will create the corresponding quantum state to the nearest accuracy it can:

In [ ]:

```
q = QuantumRegister(1)
qc= QuantumCircuit(q)

A = [ 3/5, 4/5 ]
qc.initialize( A, [q[0]] )
oq.Wavefunction( qc )
```

```
0.6 |0>    0.8 |1>
```

Interestingly enough, if you're curious about the quantum circuit that goes into producing a state created from initialize(), you can use the **decompose()** function to reveal the quantum gates. It may take a few layers of peeling back, but eventually you will find the elementary gate structure of the quantum circuit:

In [ ]:

```
print( qc.decompose().decompose().decompose() )
```

```
q10_0: |0>─|0>─┤ Ry(1.8546) ├┤ Rz(0) ├
```

Now that we can prepare arbitrary quantum states via initialize(), the next step in our subroutine will be to produce the following two quantum states:

$$|\Psi\rangle \;=\; \frac{1}{\sqrt{2}}\Big[\,|0\rangle\otimes|A\rangle \;+\; |1\rangle\otimes|B\rangle\,\Big]$$

$$|\phi\rangle \;=\; \frac{1}{\sqrt{Z}}\Big[\,|A||0\rangle \;-\; |B||1\rangle\,\Big]$$

where $Z \;=\; |A|^2 \;+\; |B|^2$

Now, let's discuss something interesting about the states $|\Psi\rangle$ and $|\phi\rangle$ above: they are not of the same dimension. Suppose we are trying to calculate the distance between two $N$-dimensional vectors $\vec{A}$ and $\vec{B}$. Encoding the states $|A\rangle$ and $|B\rangle$ will require $O(\log_2 N)$ qubits, while $|\phi\rangle$ is always a single qubit state. Thus in total, the preparation of these two states will require $O(\log_2 N)+1$ qubits to create. And the reason we've made a point to highlight their dimensional difference is because the next step in the subroutine will be to perform the SWAP Test between them, which will ultimately gives us our distance squared:

$$D^2 \;=\; 2Z\left|\langle\phi|\Psi\rangle\right|^2$$

We just jumped quite a few steps, but it's worthwhile to have seen the final answer first as we now backtrack through the math. Keeping in mind that so long as we are able to extract the quantity $\left|\langle\phi|\Psi\rangle\right|^2$ using our quantum computer, then the following derivation steps confirm the validity of our DistCalc subroutine:

$$\begin{aligned}\langle\phi|\Psi\rangle &= \frac{1}{\sqrt{2Z}}\Big[\,|A|\langle 0| \;-\; |B|\langle 1|\,\Big]\Big[\,|0\rangle\otimes|A\rangle \;+\; |1\rangle\otimes|B\rangle\,\Big]\\ &= \frac{1}{\sqrt{2Z}}\Big[\,|A||A\rangle \;-\; |B||B\rangle\,\Big]\\ &= \frac{1}{\sqrt{2Z}}\Big[\,|A|\cdot\frac{1}{|A|}\sum_i A_i|i\rangle \;-\; |B|\cdot\frac{1}{|B|}\sum_i B_i|i\rangle\,\Big]\\ &= \frac{1}{\sqrt{2Z}}\sum_i\Big(A_i \;-\; B_i\Big)|i\rangle\end{aligned}$$

Let's pause and take note of what the inner product $\langle\phi|\Psi\rangle$ above has produced. As shown in the final line, the inner product of these two dimensionally mismatched states has resulted in a new $N$-dimensional state, which mathematically represents the vector created from $\vec{A} - \vec{B}$. Intuitively, if this inner product represents the vector from $\vec{B}$ to $\vec{A}$, then it should make sense that the quantity $\left|\langle\phi|\Psi\rangle\right|^2$ will yield the distance between $\vec{A}$ and $\vec{B}$:

$$\begin{aligned}\left|\langle\phi|\Psi\rangle\right|^2 &= \frac{1}{2Z}\Big[\,\sum_i\Big(A_i^* \;-\; B_i^*\Big)\langle i|\,\Big]\Big[\,\sum_j\Big(A_j \;-\; B_j\Big)|j\rangle\,\Big]\\ &= \frac{1}{2Z}\sum_{i=j}\Big(A_i^* \;-\; B_i^*\Big)\Big(A_j \;-\; B_j\Big)\langle i|j\rangle\\ &= \frac{1}{2Z}\sum_i\Big|A_i \;-\; B_i\Big|^2\\ &= \frac{D^2}{2Z}\end{aligned}$$

And there you have it, the confirmation of our distance calculating subroutine. The scalar value obtained from $|\langle\phi|\Psi\rangle|^2$ represents the distance squared between $\vec{A}$ and $\vec{B}$, adjusted by the value of $2Z$ classically for the final answer. Simple and powerful, this subroutine can be best summarized as the encoding of $\vec{A}$ and $\vec{B}$ into two quantum states, whose inner product produces the vector $\vec{A} - \vec{B}$ as a new state, which in turn allows us to compute their distance via the SWAP Test. But now here comes the tricky part. Because $|\Psi\rangle$ and $|\phi\rangle$ are of different dimensions, what does it mean to perform a SWAP Test between them?

In our previous two implementations of the SWAP Test, for quantum states composed of either single or multiple qubits, the goal was to evaluate the inner product $\langle A|B\rangle$, which mathematically was a scalar quantity because $\vec{A}$ and $\vec{B}$ were of the same dimensionality. And as we've shown, we can use repeat measurements to approximate this quantity via the inner product squared resulting from our SWAP Test. But now if we consider the case where $|A\rangle$ and $|B\rangle$ of different dimensions, this in turn means that the inner product $\langle A|B\rangle$ is no longer a scalar, as demonstrated in the derivation above. Furthermore, since our SWAP Test is only able to tell us the value of $|\langle A|B\rangle|^2$, which is always a scalar, this means that the information pertaining to the leftover state from $\langle A|B\rangle$ is lost. If working with the state produced from the inner product of $|\Psi\rangle$ and $|\phi\rangle$ was necessary for our distance calculating subroutine, then unfortunately the SWAP Test would be of no use to us. Luckily however, the scalar quantity $|\langle A|B\rangle|^2$ is all we need!

In following the derivation steps above, which form the basis for why the quantity $|\langle\phi|\Psi\rangle|^2$ can in turn be used for computing Euclidean distances, there is one small detail we must revisit before we can construct our circuit. In computing the inner product $\langle\phi|\Psi\rangle$, notice that in the derivation we chose to carry out this operation between $|\phi\rangle$ and the *first* qubit within $|\Psi\rangle$, the one responsible for holding the superposition between the states $|A\rangle$ and $|B\rangle$. Mathematically, we could have chosen to enforce this inner product with *any* of the qubits that make up $|\Psi\rangle$, which means that the quantity $\langle\phi|\Psi\rangle$ can have up to $Q$ possible interpretations, where $Q$ is the number of qubits making up the state $|\Psi\rangle$. Consequently, this means that there are $Q$ different results we could obtain from our SWAP Test for evaluating $|\langle\phi|\Psi\rangle|^2$, but only one of them will correctly complete our Euclidean distance subroutine.

In order to correctly pick out the inner product $\langle\phi|\Psi\rangle$ we want, it turns out that all we need to do is enforce the SWAP Test between the two qubits whose inner product we are interested in. For our DistCalc subroutine here then, this means that we will be able to extract the $|\langle\phi|\Psi\rangle|^2$ quantity we are interested in with the use of a single CSWAP gate between the qubit containing the state $|\phi\rangle$ and the first qubit within $|\Psi\rangle$. It may come as a surprise that we can effectively evaluate an inner product between $2$ and $2^Q$ dimensional states the same way as between two single qubit states, so consider the derivation below which demonstrates how the choice in which qubits to swap determines the result of the SWAP Test:

$$|0\rangle\otimes|A\rangle\otimes|B_1\rangle|B_2\rangle \;\equiv\; |0AB_1B_2\rangle$$

$$\begin{aligned}H\otimes I^3 &\longrightarrow \frac{1}{\sqrt{2}}\Big(|0AB_1B_2\rangle \;+\; |1AB_1B_2\rangle\Big)\\ \text{CSWAP}(A,B_1) &\longrightarrow \frac{1}{\sqrt{2}}\Big(|0AB_1B_2\rangle \;+\; |1B_1AB_2\rangle\Big)\\ H\otimes I^3 &\longrightarrow \frac{1}{2}\Big(|0AB_1B_2\rangle \;+\; |1AB_1B_2\rangle \;+\; |0B_1AB_2\rangle \;-\; |1B_1AB_2\rangle\Big)\end{aligned}$$

$$= \frac{1}{2}\Big( |0\rangle \otimes \big[ |AB_1\rangle + |B_1A\rangle \big] \;+\; |1\rangle \otimes \big[ |AB_1\rangle - |B_1A\rangle \big] \Big) \otimes |B_2\rangle$$

Compare the steps of the SWAP Test shown above with those from the previous section (between two single qubit states $|A\rangle$ and $|B\rangle$). Note how we arrive at an identical result between the qubits $A$ and $B_1$, despite system $|B\rangle$ being composed of two qubits. If we now go ahead and compute the probability of measuring the $|0\rangle$ state on our control qubit, we will find that our measurement result correctly evaluates the desired inner product between qubits $A$ and $B_1$:

$$\begin{aligned} \mathrm{P}\big(|0\rangle_C\big) &= \frac{1}{4}\Big|\, |AB_1B_2\rangle + |B_1AB_2\rangle \,\Big|^2 \\ &= \frac{1}{4}\Big[ \big( \langle AB_1|AB_1\rangle + \langle AB_1|B_1A\rangle + \langle B_1A|AB_1\rangle + \langle B_1A|B_1A\rangle \big) \otimes \langle B_2|B_2\rangle \Big] \\ &= \frac{1}{2} + \frac{1}{2}\langle A|B_1\rangle\langle B_1|A\rangle \\ &= \frac{1}{2} + \frac{1}{2}\big|\langle A|B_1\rangle\big|^2 \end{aligned}$$

The important concept here is the way in which the state of qubit $B_2$ goes through the SWAP Test completely unaffected, causing the final measurement on the control qubit to exclusively extract the information of $\langle A|B_1\rangle$. From a quantum circuit perspective this should make sense, as the only operations ever applied to $|B_2\rangle$ are Identity gates in the derivation above. Based on this new result from the SWAP Test, hopefully it is clear that the derivation above could be applied to the inner product of $\langle A|B_2\rangle$ as well by simply changing the CSWAP gate. And more generally, so long as one properly lines up the correct qubits for the desired inner product via CSWAP gates, the SWAP Test can be used to compute the inner product between any two dimensionally mismatched quantum states. To demonstrate this, below is the complete implementation of our DistCalc subroutine for computing the Euclidean distance between two vectors $\vec{A}$ and $\vec{B}$:

In [ ]:

```
A = [1,0,-2,0]
B = [6,-4,0,0]
trials = 50000
#==============================
A_norm = 0
B_norm = 0
D = 0
for i in np.arange(len(A)):
    A_norm = A_norm + A[i]**2
    B_norm = B_norm + B[i]**2
    D = D + (A[i]-B[i])**2
D = m.sqrt(D)
A_norm = m.sqrt(A_norm)
B_norm = m.sqrt(B_norm)
Z = round( A_norm**2 + B_norm**2 )
#------------------------------
phi_vec = [A_norm/m.sqrt(Z),-B_norm/m.sqrt(Z)]
psi_vec = []
for i in np.arange(len(A)):
    psi_vec.append(  (A[i]/A_norm) /m.sqrt(2) )
    psi_vec.append(  (B[i]/B_norm) /m.sqrt(2) )

#==============================

a = QuantumRegister(1,name='a')
q = QuantumRegister(4,name='q')
c = ClassicalRegister(1,name='c')
qc= QuantumCircuit(a,q,c)
qc.initialize( phi_vec, q[0] )
qc.initialize( psi_vec, q[1:4] )
#------------------------------     The SWAP Test
qc.h( a[0] )
qc.cswap( a[0], q[0], q[1] )
qc.h( a[0] )
qc.measure(a,c)
#------------------------------
M = oq.Measurement(qc,shots=trials,return_M=True,print_M=False)
print('Euclidean Distance:   ',round(D,4))
print('\n DistCalc Distance:   ',round( m.sqrt((((M['0']/trials - 0.5)/0.5)*2*Z)),4) )
```

```
Euclidean Distance:    6.7082

 DistCalc Distance:    6.7094
```

The code example above showcases perhaps the most appealing aspect of this DistCalc subroutine: the ability to mathematically deliver on the *exact* distance between two vectors $\vec{A}$ and $\vec{B}$. And if we continually increase the number of repeat measurements for better and better approximations to $|\langle\phi|\Psi\rangle|^2$, our final answer will similarly become a more accurate approximation to the true Euclidean distance. Additionally, notice in the example that we actually computed the distance between two 3-dimensional vectors $\vec{A}$ and $\vec{B}$, demonstrating that this subroutine isn't limited to only working with vectors of length $2^N$.

In light of the successful code example above, showcasing that this subroutine can certainly be used as a replacement for distance calculating in our $k$-means algorithm, we must now discuss its major shortcoming and why we won't be implementing it going forward. As we already alluded to at the beginning of our discussion, the preparation of $|\phi\rangle$ and $|\Psi\rangle$ is no easy task, especially when we start to consider higher dimensional vectors. The strict requirements on structure for these states is precisely why this subroutine is able to yield exact distance measures, but it is also a huge barrier for implementing this technique on a real quantum computer. Thus, our next DistCalc subroutine will look to trade distance accuracy for ease of execution, representing a technique which is much more viable for current quantum computing technologies.

### DistCalc 2: A Quantum Distance Measure

If we return to our original motivation for studying the DistCalc subroutine above, it was to improve upon the "Assignment" step in our $k$-means clustering algorithm. Specifically, we were looking for a quantum subroutine which could quickly assess each data point and compute the closest centroid. Naturally then, we gravitated towards a quantum algorithm which promised to deliver us the same quantity we would normally have computed classically: Euclidean distance (or distance squared). But now suppose we could answer our question about nearest centroids $without$ necessarily computing data point distances. Sounds a bit counterintuitive, but in fact it's exactly how our next subroutine is going to determine "closeness" between data points.

In order for the $k$-means algorithm to successfully advance through each iteration, the only requirement is that each data point gets properly sorted into the nearest cluster. Whether we meet this criteria via computing Euclidean distances, distances squared, or by some other means is irrelevant, so long as at the end of the day each data point finds its way into the correct cluster and we are satisfied with the result. And as you may now suspect, our next subroutine will perform the "some other means" methodology for determining which cluster to assign each data point to. Essentially, we will once again be using SWAP Tests as our metric for closeness and data point assigning. But even though the quantity extracted from these SWAP Tests will have no direct connection to Euclidean distance, they will still serve the purpose of determining nearest centroids.

Just as before, in order to properly use our SWAP Test as a substitute metric for Euclidean distance, the critical element is the way in which we encode the data into quantum states. For example, we've already seen one way in which to encode cartesian data points as quantum states:

$$\vec{A} \quad \longrightarrow \quad |A\rangle \;=\; \frac{1}{|A|}\sum_i A_i\,|i\rangle$$

But now we must ask whether or not this encoding will serve our purpose for determining closeness between points. More specifically, what meaning can we extract from a SWAP Test performed between two states $|A\rangle$ and $|B\rangle$ encoded in this way? Previously, we needed to embed the states $|A\rangle$ and $|B\rangle$ into $|\phi\rangle$ and $|\Psi\rangle$ in order to obtain their distance. So let's now see what information we can extract from a SWAP Test between them directly:

In [ ]:

```
Shots = 10000
Points = [ [-1,1], [3,4], [7,7], [6,8] ]
Norm_Points = []
for p in np.arange( len(Points) ):
    Norm_Points.append( Points[p]/np.linalg.norm(Points[p]) )
#===================================================
for p2 in np.arange( len(Norm_Points)-1 ):
    q = QuantumRegister(3)
    c = ClassicalRegister(1)
    qc= QuantumCircuit(q,c)
    qc.initialize( Norm_Points[int(p2)], [q[1]] )
    qc.initialize( Norm_Points[-1], [q[2]] )
#---------------------------------------------------
    IP = oq.SWAP_Test( qc, q[0], q[1], q[2], c[0], Shots )
    print('\nComparing Points:  ',Points[p2],'   &   ',Points[-1],'\n',IP,'|0>     ',Shots-IP,'|1>')
```

```
Comparing Points:   [-1, 1]    &    [6, 8]
 5080 |0>      4920 |1>

Comparing Points:   [3, 4]    &    [6, 8]
 10000 |0>       0 |1>

Comparing Points:   [7, 7]    &    [6, 8]
 9904 |0>      96 |1>
```

The example above uses the SWAP Test to compare three different points to $\left[6, 8\right]$. Additionally, the code calls upon the function $\mathbf{SWAP_Test}$ from $\mathrm{Our_Qiskit_Functions}$, which handles all of the steps for a 1-qubit SWAP Test and returns the number of $|0\rangle$ counts. Starting from the top then, we can see that the point $\left[-1, 1\right]$ results in the highest number of $|1\rangle$ counts, signally that the quantum states representing these two points have the smallest overlap (practically completely orthogonal). At first glance this looks like a potentially positive result, reflecting the fact that $\left[-1, 1\right]$ is quite distant from $\left[6, 8\right]$. However, looking at the next two results confirms that our SWAP test isn't giving us a desirable metric.

If the goal of the example above was to determine which centroid is closest to our data point $\left[6, 8\right]$, then the results of the SWAP test would suggest that $\left[3, 4\right]$ is a perfect match, beating out $\left[7, 7\right]$. Obviously we know this is incorrect, thus signaling an issue with our quantum approach. Mathematically, it shouldn't be too hard to spot where the problem lies, as evidenced by the code below:

In [ ]:

```
Shots = 10000
Points = [ [2,3], [4,6], [8,12], [12,18] ]
Norm_Points = []
for p in np.arange( len(Points) ):
    Norm_Points.append( Points[p]/np.linalg.norm(Points[p]) )
#==================================================
for p2 in np.arange( len(Norm_Points)-1 ):
    q = QuantumRegister(3)
    c = ClassicalRegister(1)
    qc= QuantumCircuit(q,c)

    qc.initialize( Norm_Points[int(p2)], [q[1]] )
    qc.initialize( Norm_Points[-1], [q[2]] )
#--------------------------------------------------
    IP = oq.SWAP_Test( qc, q[0], q[1], q[2], c[0], Shots )
    print('\nComparing Points:  ',Points[p2],'   &   ',Points[-1],'\n',IP,'|0>     ',Shots-IP,'|1>')
```

```
Comparing Points:   [2, 3]    &    [12, 18]
 10000 |0>      0 |1>

Comparing Points:   [4, 6]    &    [12, 18]
 10000 |0>      0 |1>

Comparing Points:   [8, 12]    &    [12, 18]
 10000 |0>      0 |1>
```

At the beginning of our fist DistCalc discussion, we pointed out that points which were a constant multiple of each other resulted in the same quantum state. And while the complete subroutine could correct for this limitation using a quick classical computation to produce the desired final distance, unfortunately the SWAP Test alone cannot do the same. Never fear however, as the examples above are only meant to demonstrate the importance of choosing the correct mapping of classical data to quantum. In order for our SWAP Test to be an effective measure of distance, we will need to take care in the way we encode our classical space into quantum states.

### Mapping to the Bloch Sphere

In light of the examples we just saw, we're now ready to discuss the way in which we *will* be effectively computing centroid distances using the SWAP Test. Essentially, our goal will be to map our 2D cartesian space of data points onto the surface of a Bloch Sphere:

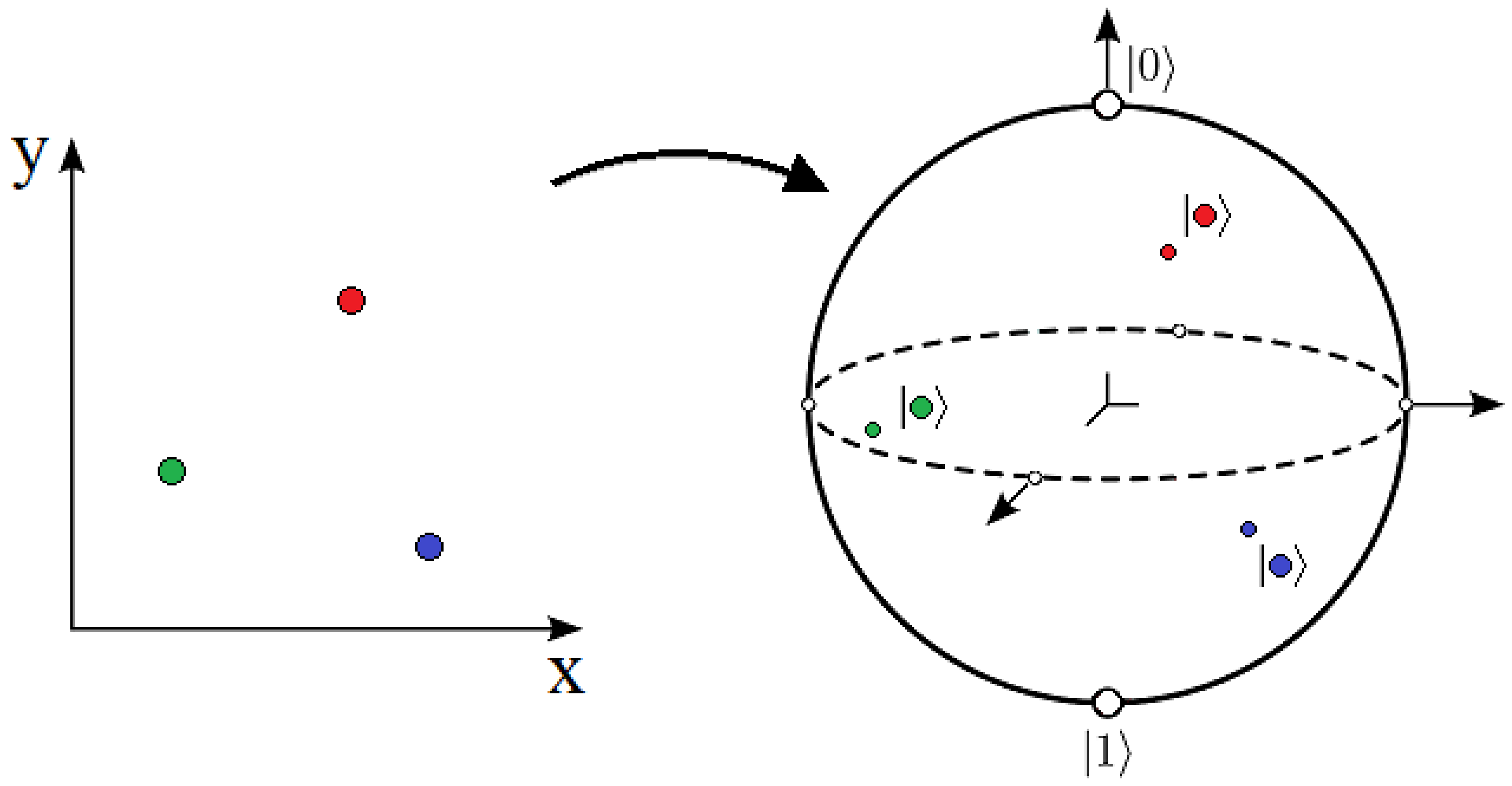

As represented in the illustration above, the conceptual goal is that we can use the surface of the Bloch Sphere as a space to mimic the Euclidean distances of our original 2D data space. Keeping in mind that the SWAP Test is our determining metric, we ideally then would like to choose a mapping such that points which are closer together have higher inner product overlaps, and subsequently distant points are more orthogonal. However, there are many possible ways we can map our data to meet these criteria, some more fitting to certain spaces of data than others. In the coming examples we're going to demonstrate two such choices, but it is important to keep in mind that the best way to encode one's data into quantum states is a case by case basis.

When we picture using the Bloch Sphere as our new data space, a natural change in metrics is to transform from the cartesian coordinate system $(x,y)$ to polar angles $(\theta,\phi)$. Once we have the polar coordinates for our data point, creating the corresponding quantum state is as simple as using the $U3$ gate:

$$U3(\theta,\phi,\lambda) = \begin{pmatrix} \cos\left(\frac{\theta}{2}\right) & -e^{-i\lambda}\sin\left(\frac{\theta}{2}\right) \\ e^{i\phi}\sin\left(\frac{\theta}{2}\right) & e^{i(\lambda+\phi)}\cos\left(\frac{\theta}{2}\right) \end{pmatrix}$$

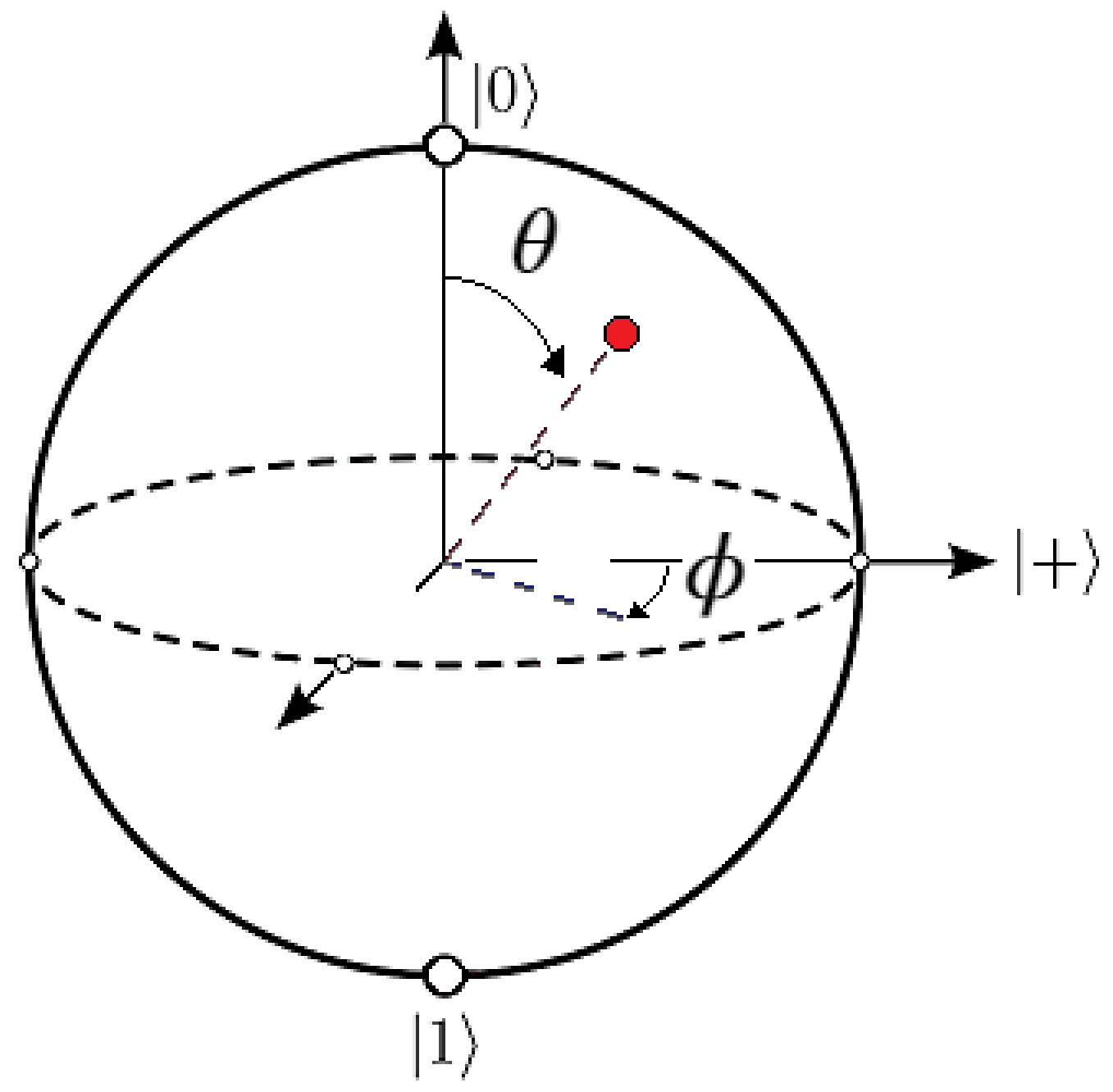

Compare the way in which we are able to prepare our states for the SWAP Test here via the $U3$ gate with $|\phi\rangle$ and $|\Psi\rangle$ from before. Whereas the previous DistCalc subroutine was reliant on elaborate state preparations, requiring complex circuit designs, here we are able to initialize our new subroutine with the use of a single gate! With the preparation of our quantum states easily taken care of, all that's left is to decide on our mapping: $(x,y) \longrightarrow (\theta,\phi)$ (in principle we could also include $\lambda$ in our encoding, but for now we will just work with $\theta$ and $\phi$). As our first example, let's take a look at a data space which spans from $-1$ to $1$ in both $x$ and $y$, and maps to the polar coordinates of the Bloch Sphere as follows:

$$\theta \;=\; \frac{(x+1)\cdot\pi}{2} \qquad\qquad \phi \;=\; \frac{(y+1)\cdot\pi}{2}$$

Using the mapping above, we are guaranteed that every point within our allowed data space is uniquely mapped to a single location on the Bloch Sphere. For reasons we've already seen, this $1$-to-$1$ mapping is an important quality that we will be enforcing from here on (no longer will we allow multiple data points to map to the same quantum state). Now, before we dissect the pros and cons of the mapping we've chosen, let's see it in action first. The two code examples below demonstrate how our transformation performs at determining which centroid a central data point is closest to:

In [ ]:

```
Point = [ 0.12, -0.15 ]
Centroids = [ [-0.38,0.61] ,  [-0.09,-0.34] ,  [0.52,0.29] ]
#-----------------------
fig = plt.figure(figsize=(4,4))
ax = fig.add_subplot(1,1,1)
ax.axis([-1,1,-1,1])
fig.show()
#-----------------------
plt.plot(Point[0], Point[1], 'ro')
markers = ['gx','bx','kx']
for c in np.arange( len(Centroids) ):
    plt.plot(Centroids[c][0], Centroids[c][1], markers[c])
```

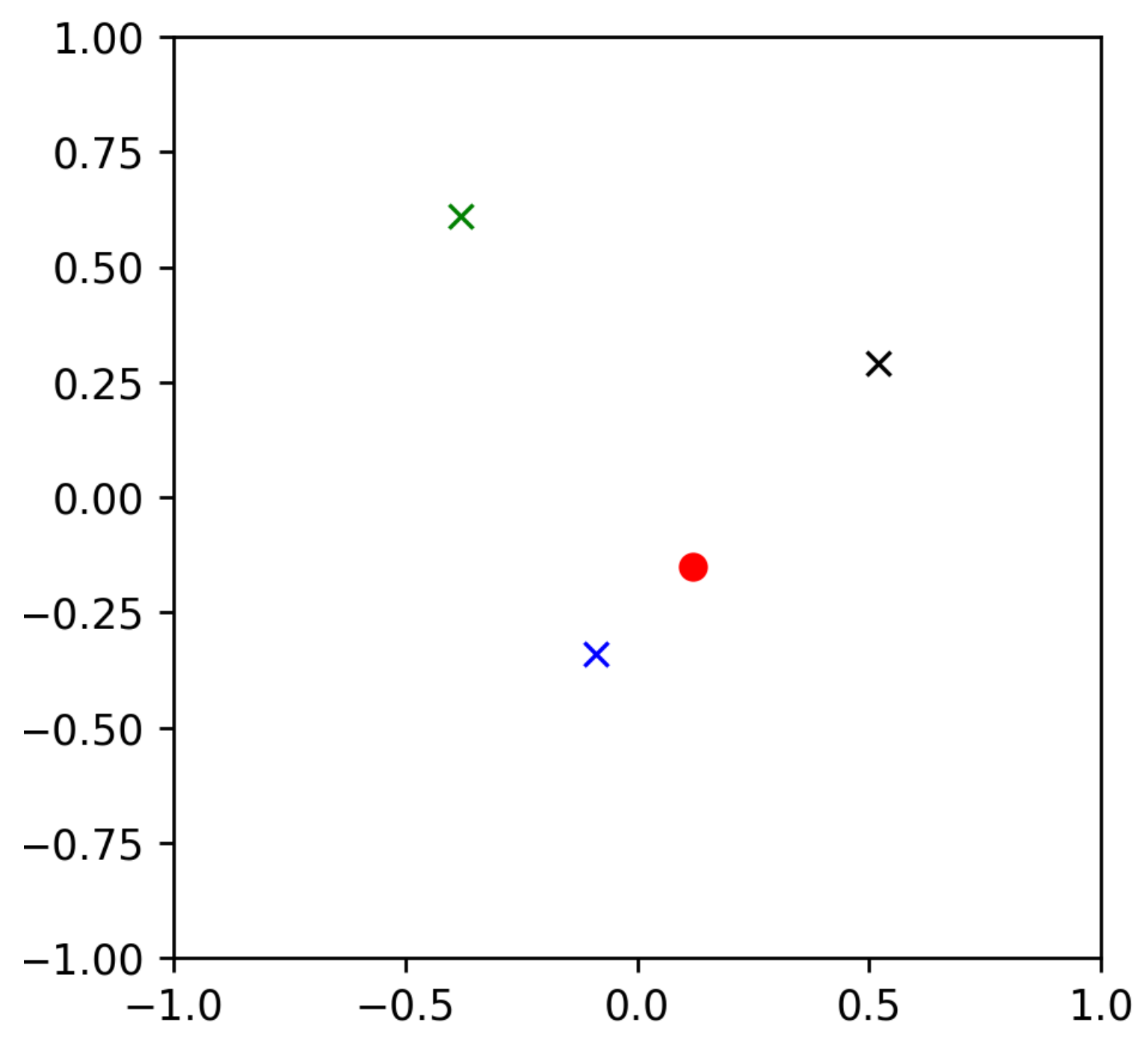

As we can see in the figure above, our central data point (red circle) is closest to the blue centroid. Thus, the ideal output from our quantum subroutine will be to classify this data point as belonging to the corresponding blue cluster. Let's see how we do:

In [ ]:
```
Shots = 10000
Point = [ 0.12, -0.15 ]
Centroids = [ [-0.38,0.61] ,  [-0.09,-0.34] ,  [0.52,0.29] ]
Bloch_Point = [ (Point[0]+1)*m.pi/2, (Point[1]+1)*m.pi/2 ]
Bloch_Cents = []
for c in np.arange( len(Centroids) ):
    Bloch_Cents.append( [ (Centroids[c][0]+1)*m.pi/2, (Centroids[c][1]+1)*m.pi/2 ] )
#====================================
colors = ['Green','Blue','Black']
for c2 in np.arange( len(Bloch_Cents) ):
    q = QuantumRegister(3)
    c = ClassicalRegister(1)
    qc= QuantumCircuit(q,c)
    qc.u3( Bloch_Point[0], Bloch_Point[1], 0, [q[1]] )
    qc.u3( Bloch_Cents[c2][0], Bloch_Cents[c2][1], 0, [q[2]] )
#------------------------------------------------
    IP = oq.SWAP_Test( qc, q[0], q[1], q[2], c[0], Shots )
    print('\n\nComparing Points:  ',Centroids[c2],'   &   ',Point,'     (',colors[c2],' Centroid )\n\n',IP,'|0>     ',Shots
```

```
Comparing Points:   [-0.38, 0.61]    &    [0.12, -0.15]      ( Green  Centroid )

 7980 |0>      2020 |1>


Comparing Points:   [-0.09, -0.34]    &    [0.12, -0.15]      ( Blue  Centroid )

 9749 |0>      251 |1>


Comparing Points:   [0.52, 0.29]    &    [0.12, -0.15]      ( Black  Centroid )

 9172 |0>      828 |1>
```

Just like we wanted, our SWAP Test results are telling us that the closest centroid to our data point is the blue cluster! Even better yet, we can see that the next closest centroid is black, and the differences in $|0\rangle$ counts between the three results is suggestively similar to differences in distances shown in the figure. Having now seen that our choice in mapping produces the desired results, we can now dive a bit deeper into why it works. For starters, let's take a look at the bounds of our data space, and the corresponding limits on our polar angles:

$$\big[-1, 1\big] \quad \longrightarrow \quad \big[0, \pi\big]$$

which when translated into the quantum states produced from the $U3$ gate gives us:

$$\theta \in \big[0, \pi\big] \quad \longrightarrow \quad \Big[\ |0\rangle\ ,\ e^{i\phi}\,|1\rangle\ \Big]$$

$$\phi \in \big[0, \pi\big] \quad \longrightarrow \quad \Big[\ \cos\Big(\frac{\theta}{2}\Big)|0\rangle\ +\ \sin\Big(\frac{\theta}{2}\Big)|1\rangle\ ,\ \cos\Big(\frac{\theta}{2}\Big)|0\rangle\ -\ \sin\Big(\frac{\theta}{2}\Big)|1\rangle\ \Big]$$

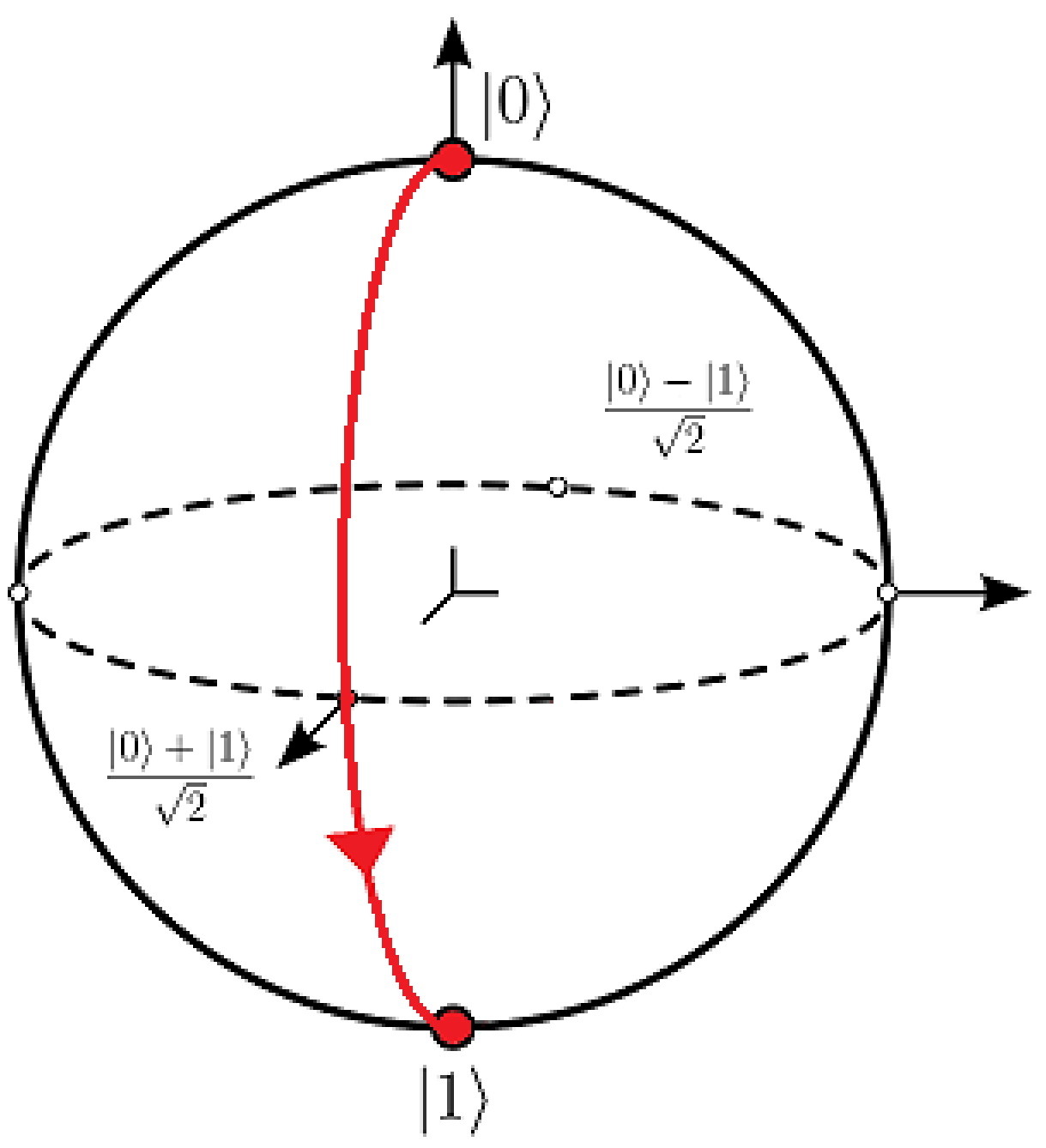

$$\theta \in \big[0, \pi\big]$$

$$\phi = 0$$

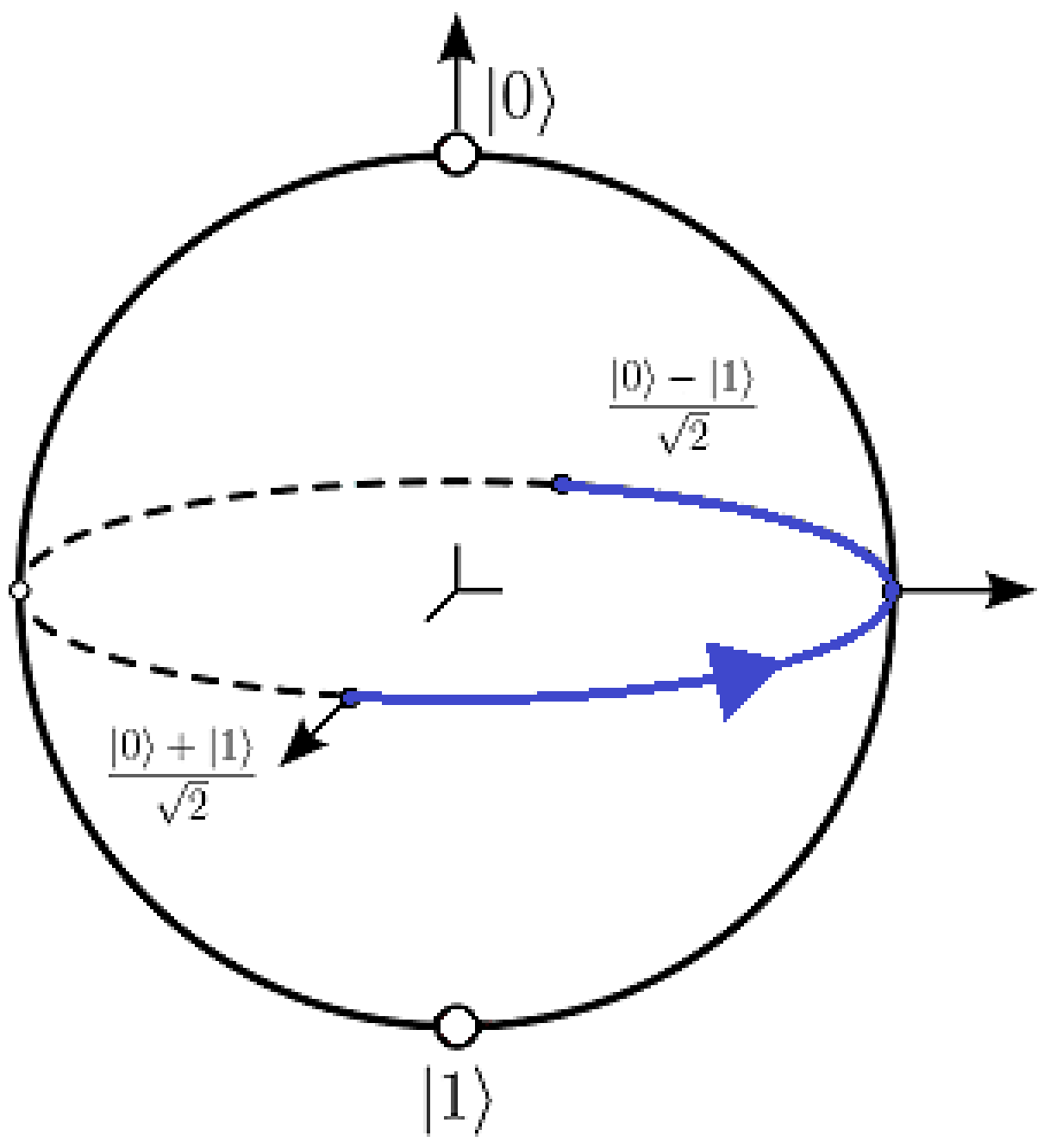

$$\phi \in \big[0, \pi\big]$$

$$\theta = \frac{\pi}{2}$$

As shown in the illustrations above, the bounds of our mapping result in 180˚ rotations along the surface of the Bloch Sphere. Consequently, these limits result in perfectly orthogonal states for $\theta$ (for any $\phi$), and for $\phi$ when $\theta = \frac{\pi}{2}$ (the $|+\rangle$ and $|-\rangle$ states along the equator). Translating these results back to our original data space, it means that points along the $x$ boundaries will be interpreted as maximally distant (regardless of their $y$ component), but only one instance of points along the $y$ boundaries will be maximally distant (for the case where $x = 0$).

Based on the results just discussed, it should be apparent that our mapping is asymmetric in its treatment of $x$ and $y$, which can be traced to the asymmetric way in which $\theta$ and $\phi$ influence the quantum state. Technically $\theta$ and $\phi$ both control the position of our state on the Bloch Sphere equally (the amount by which we rotate in each polar direction), but $\theta$ dictates the amplitude difference between the $|0\rangle$ and $|1\rangle$, while $\phi$ controls the phase difference. Ultimately, this difference in influence impacts the way in which our SWAP Test interprets the distance between points. To visualize this difference, the cell of code below displays the overlap (inner product) produced from the SWAP Test of a single point with every other point in the data space. As a first demonstration, let's see all of the the inner products produced from our example point from earlier. The cell of code below uses a function called **Heatmap** from Our_Qiskit_Functions to create a 2D display of the full inner product space:

In [ ]:

```
size = 100
Shots = 10000
Data_Space = [-1,1,-1,1]
Point = [0.12,-0.15]        # Example Point
t = (Point[0]+1)*m.pi/2
p = (Point[1]+1)*m.pi/2
OL_grid = np.zeros(shape=(size,size))
#=========================================================
for x in np.arange(size):
    t2 = (( (-1 + 2*(x/size)) + 1) * m.pi / 2)
    for y in np.arange(size):
        p2 = (( (-1 + 2*(y/size)) + 1) * m.pi / 2)
        #----------------------------------
        q = QuantumRegister( 3, name='q' )
        c = ClassicalRegister( 1, name='c' )
        qc= QuantumCircuit( q,c, name='qc' )
        qc.u3( t,   p, 0, q[1] )
        qc.u3( t2, p2, 0, q[2] )
        #----------------------------------
        IP = oq.SWAP_Test( qc, q[0], q[1], q[2], c[0], Shots )
        if( IP < 5000 ):
            IP = 5000
        OL_grid[int(size-y-1),int(x)] = m.sqrt((1.0*IP/Shots-0.5)*2)
#=========================================================
fig, ax = plt.subplots()
show_ticks = False
show_text  = False
oq.Heatmap(OL_grid, show_text, show_ticks, ax, "viridis", "Inner Product")
plt.plot((Point[0]+1)*size/2, size-(((Point[1]+1))*size/2), 'ro')
Centroids = [ [-0.38,0.61] ,  [-0.09,-0.34] ,  [0.52,0.29] ]
colors = ['green','blue','black']
for c in np.arange(len(Centroids)):
    plt.scatter((Centroids[c][0]+1)*size/2, size-((Centroids[c][1]+1)*size/2), color='white',   marker='s', s=50)
    plt.scatter((Centroids[c][0]+1)*size/2, size-((Centroids[c][1]+1)*size/2), color=colors[c], marker='x', s=50)
fig.tight_layout()
plt.show()
```

<IPython.core.display.Javascript object>

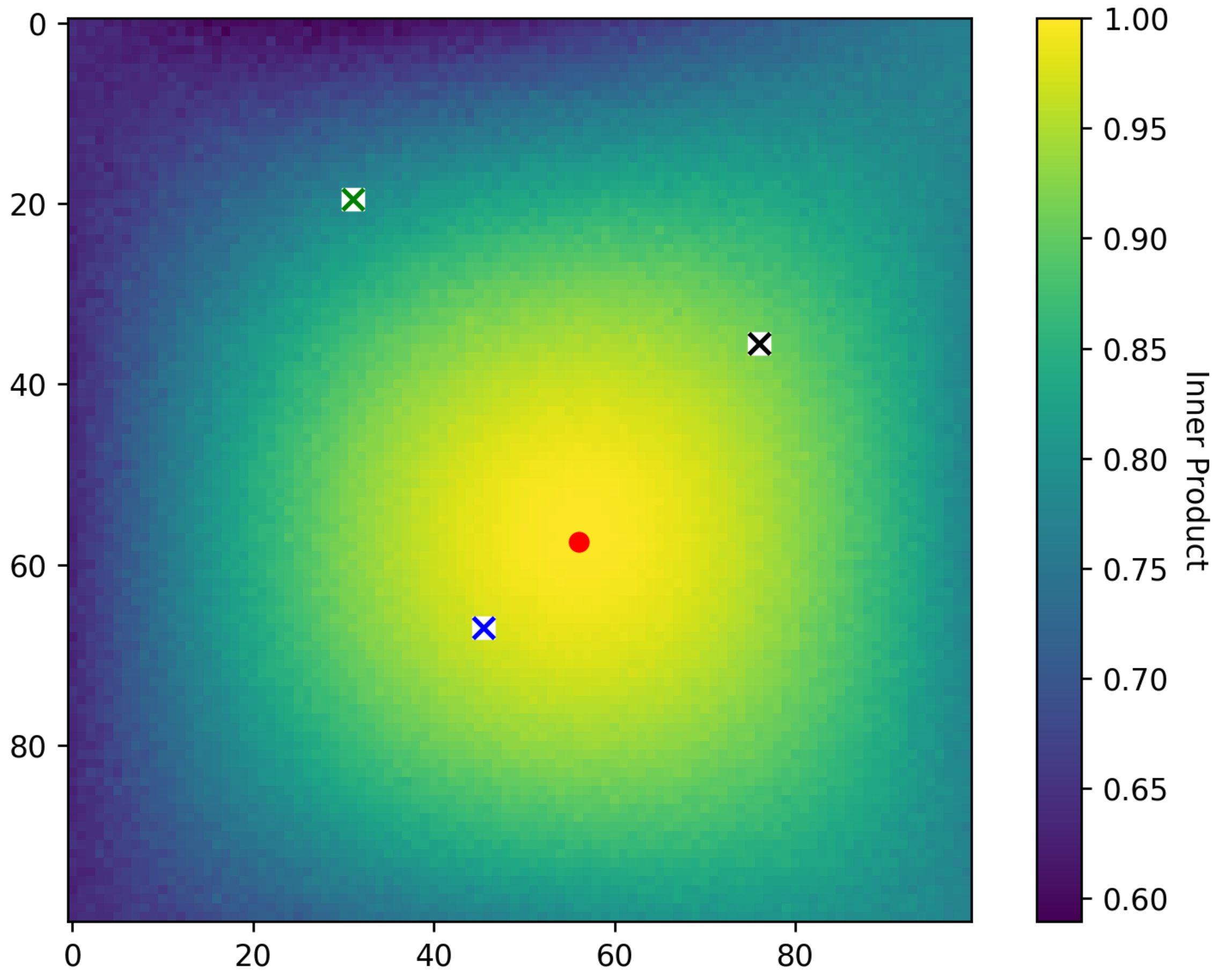

The plot above illustrates the inner product landscape associated with the point $[0.12, -0.15]$. Using the gradient scale on the right side as our reference for the percentage of $|0\rangle$ state measurements, we can once again see the order in which the SWAP Test will pick out centroid distances: blue, black, green. Looking at the overall shape of the inner product landscape above, the way in which the values appear to be radially decreasing around our point in question (red circle) is very promising for our quantum metric to replace Euclidean distance. As we noted earlier, so long as the SWAP Test can correctly distinguish closer and further data points, it doesn't matter whether or not our quantum metric has any direct relation to Euclidean distance. However, despite looking promising, a closer examination of the plot above will start to reveal a few problematic areas:

1) If we focus on the $x$ boundaries, we see the same inner product value all along both edges (a bit hard to tell at first glance, but both sides are monochromatic). This is telling us that our SWAP Test is interpreting all values along either boundary as equidistant.

2) Focusing now on the top left corner of the plot, it appears that the "furthest" point from our red dot is not $[-1, 1]$, but rather something closer to $[-0.8, 1]$. Consequently, points around this area of the data space will be incorrectly assessed for distance.

These two issues represent critical flaws with the transformation we've chosen to implement. Although they appear somewhat inconsequential for the example above, both problems become more prominent as the data point we're trying to assess approaches the $x$ and $y$ boundaries:

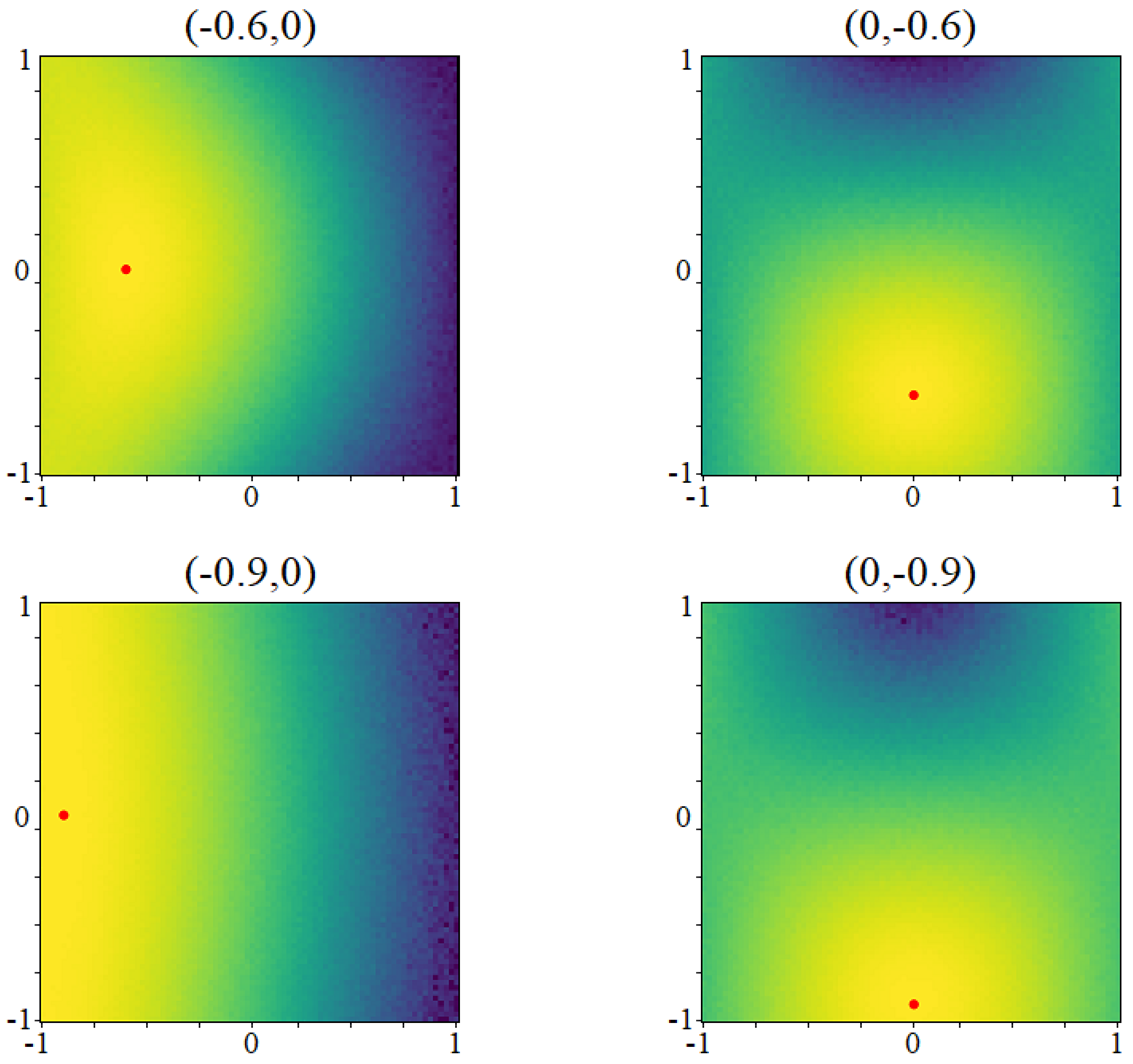

Starting with the left plots, it is clear that as our data point approaches the $x$ boundaries, the inner product landscape produced from the SWAP Test becomes heavily warped. At $[-0.6, 0]$ we can still see a bit of curvature, but by $[-0.9, 0]$ the inner product landscape has almost completely lost all $y$ dependence. Mathematically we can trace the source of this problem to the $\theta$ component of our quantum state:

$$(-0.9, 0) \quad \approx \quad 0.997\,|0\rangle \;+\; e^{i\phi}\,0.078\,|1\rangle$$

If we now consider the result of taking an inner product with the state above, it should be clear that the $|0\rangle$ component is dominant, which contains no $\phi$ dependence. Subsequently, for all data points with the same $x$ coordinate, the difference in inner product values produced from the full range of $y$ is practically negligible (at the extremes of $x = \pm 1$ we lose all $y$ dependence completely). Visually we can understand how much influence the $y$ component of our data point has by picturing rings of equal $\theta$ values on the Bloch Sphere:

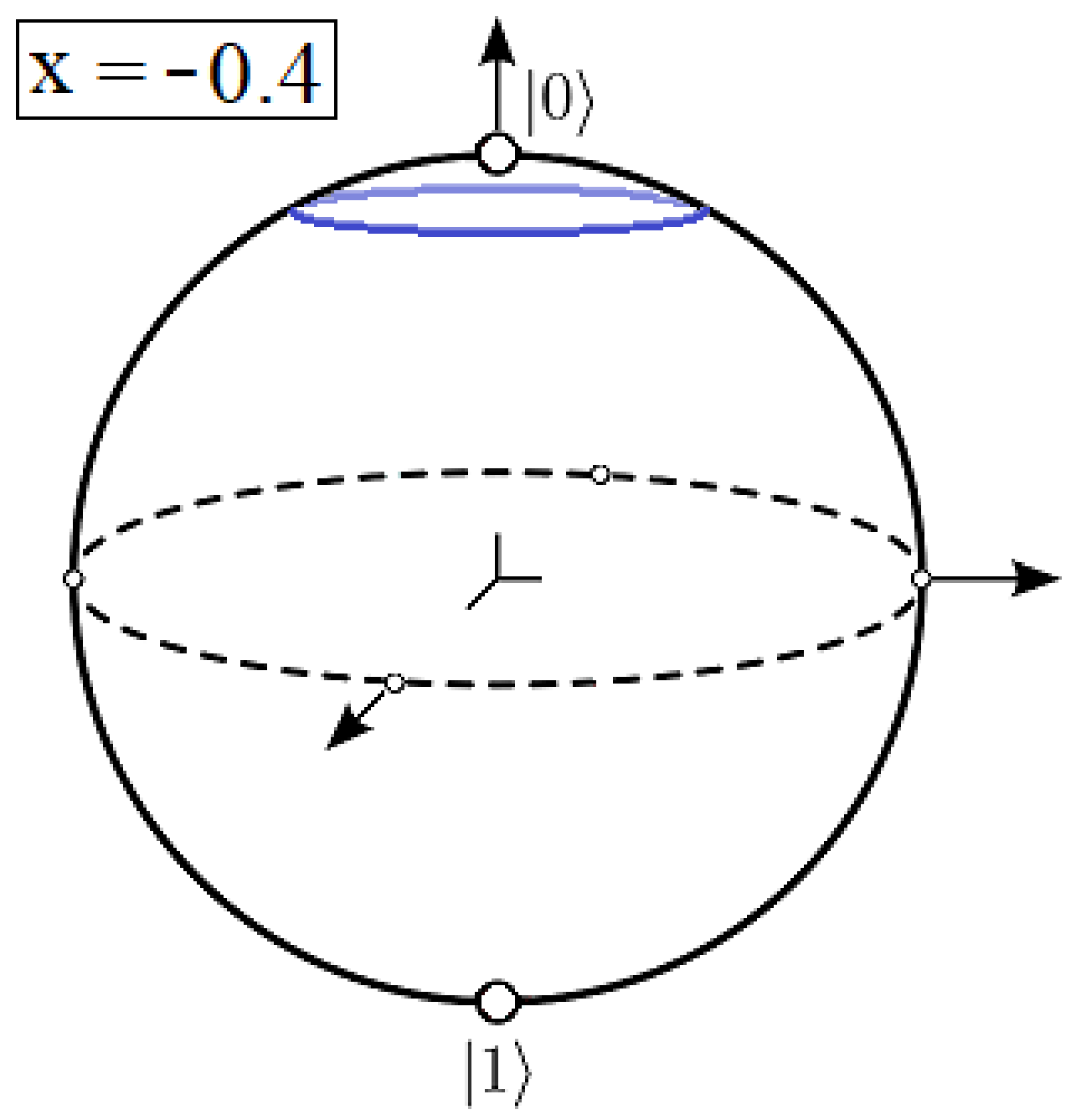

As $x$ approaches the boundaries, corresponding to $\theta$ nearing the $|0\rangle$ and $|1\rangle$ poles, the rings produced on the Bloch Sphere like shown above represent how the full range of $y$ values shrink (actually just half of the ring, because of our range from $0$ to $\pi$). As illustrated in the figure, even for a relatively centered value of $x = -0.4$, corresponding to $\theta \approx 20^\circ$, we can still see how small of a circle is carved out by the $y$ dependence. With this Bloch Sphere representation in mind then, the only location in our data space where $x$ and $y$ have equal influence is along the equator, or $\theta = \frac{\pi}{2}$. But, as we shall see next, this is precisely where our second problem is the most prominent.

Returning now to the plots for $[0, -0.6]$ and $[0, -0.9]$, the second issue resulting from our choice of mapping is that our inner product landscape is incorrectly determining the most distant point. For the reasons just discussed in problem 1, we can see why for example the inner product values at $[-1, 1]$ and $[-1, -1]$ are the same, but the issue here is that in both plots the interpreted "furthest" point from the red circle is $[0, 1]$. Once again, the problem can be traced to the fact that $y$ only influences the $|1\rangle$ component of our quantum state, specifically the phase. If we imagine any point along the surface of the Bloch Sphere and ask where the corresponding orthogonal state lies, the answer can be found by traversing directly through the center of the sphere:

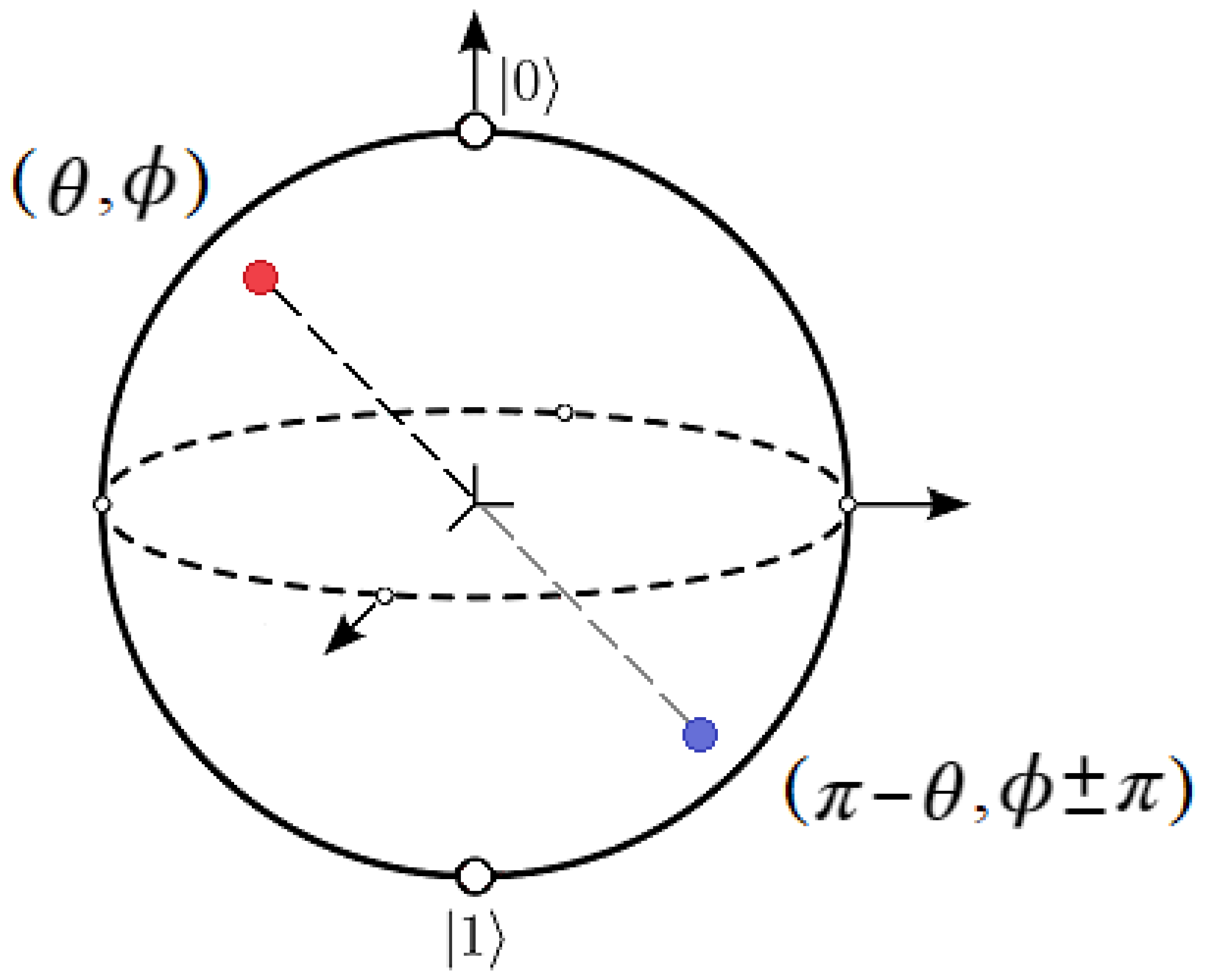

$$|P\rangle = \cos\left(\frac{\theta}{2}\right)|0\rangle + e^{i\phi}\sin\left(\frac{\theta}{2}\right)|1\rangle$$

$$|P_{\perp}\rangle = \cos\left(\frac{\pi-\theta}{2}\right)|0\rangle + e^{i(\phi\pm\pi)}\sin\left(\frac{\pi-\theta}{2}\right)|1\rangle$$

In [ ]:

```
Shots = 10000
Point = [2*random.random()-1,2*random.random()-1]
t = (Point[0]+1)*m.pi/2
p = (Point[1]+1)*m.pi/2
#=====================================================
q = QuantumRegister( 3, name='q' )
c = ClassicalRegister( 1, name='c' )
qc= QuantumCircuit( q,c, name='qc' )
qc.u3( t,   p, 0, q[1] )
qc.u3( m.pi-t, p+m.pi, 0, q[2] )
#--------------------------------
IP = oq.SWAP_Test( qc, q[0], q[1], q[2], c[0], Shots )
print('Inner Product Result Bewteen: [',round(t,3),' ,',round(p,3),']   &   [',round(m.pi-t,3),' ,',round(p+m.pi,3),']
print('\n',IP,' |0>')
```

```
Inner Product Result Bewteen: [ 1.28  , 1.386 ]   &   [ 1.861  , 4.527 ]     (θ,Φ)  &  (π-θ,π+Φ)

 4987  |0>
```

As shown above, the way to reach the maximally orthogonal point corresponding to any given data point is to subtract $\theta$ from $\pi$, and add $\pi$ to $\phi$. Consequently, the SWAP Test between these two orthogonal points will always yield a $|0\rangle$ state probability of 50%. Translated back into our Cartesian coordinate system, subtracting $\theta$ from $\pi$ is equivalent to a reflection of our $x$ coordinate about 0: $x \longrightarrow -x$. And in order to obtain the orthogonal $y$ component, adding or subtracting $\pi$ to $\phi$ is equivalent to $y \pm 2$. But since this would take us out of the bounds of our data space for any $y \neq -1$ or 1, the maximally orthogonal $y$ value will always just be 1 or -1 depending on whether $y$ is greater or less than 0 (or both if $y = 0$ exactly). All this however, just to point out that the maximally orthogonal point via our choice of transformation is *not* where we ideally want it to be, i.e. one of the four corners. As an example, notice how in the plot below our SWAP Test is telling us that the maximally distance point from $[0.3, -0.9]$ is $[-0.3, 1]$:

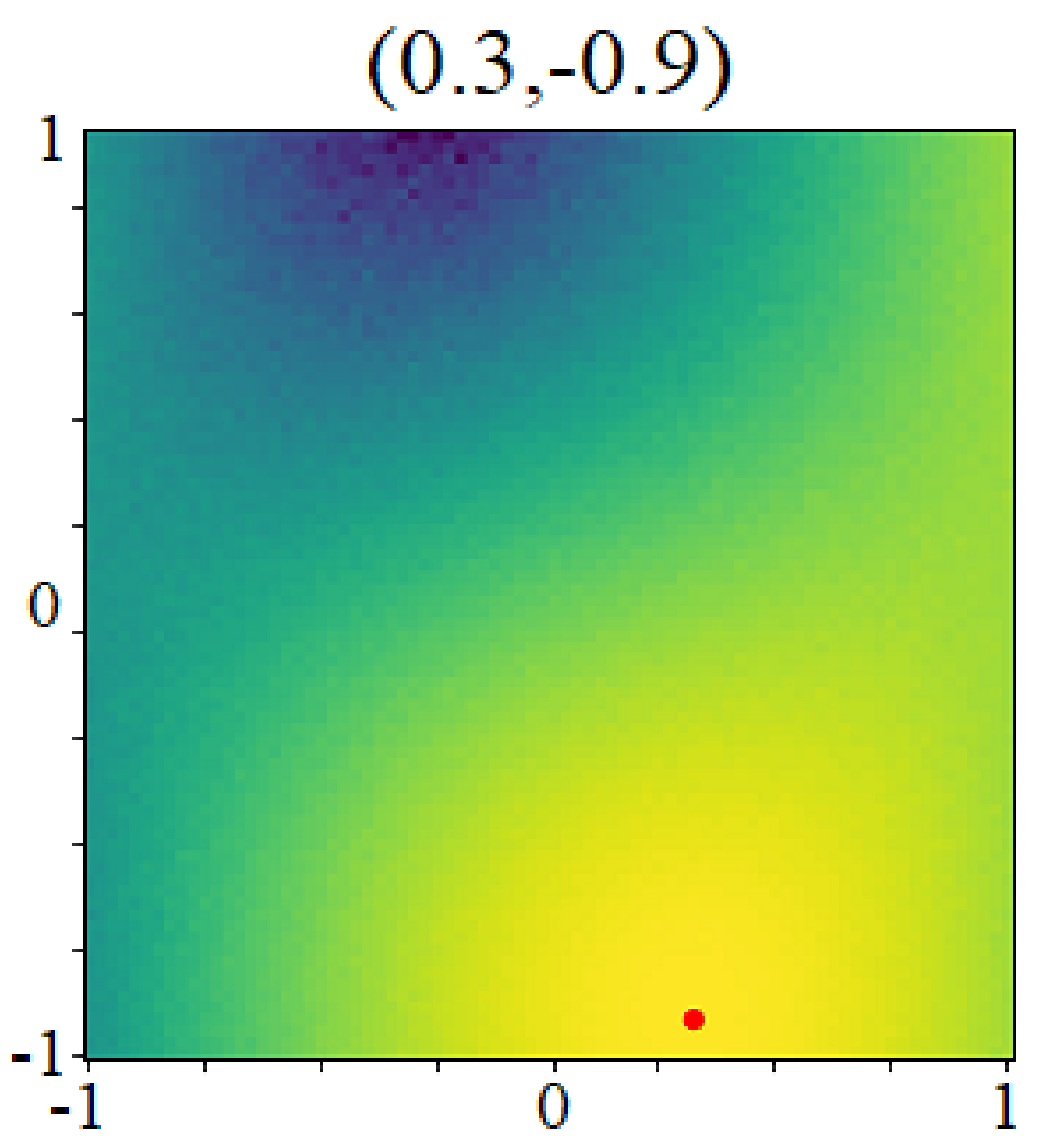

### Symmetric Mapping

Having now seen several examples of problems that can arise when mapping from Cartesian space to the Bloch Sphere, it's important to keep in mind that there is no "perfect" transformation that will exactly mimic Euclidean distances (unless we're able to create $|\,\phi\,\rangle$ and $|\,\Psi\,\rangle$ from earlier). We can design mappings which preserve certain qualities over others, but ultimately every transformation will have its shortcomings. As an example, we'll now turn our attention to a second transformation which is designed to be symmetric in $x$ and $y$, preserving the quantity that opposite corners of our data space be mapped to orthogonal states:

$$x_0 \equiv x_{\text{min}} \qquad\qquad y_0 \equiv y_{\text{min}}$$

$$x_1 \equiv x_{\text{max}} \qquad\qquad y_1 \equiv y_{\text{max}}$$

$$\theta = \frac{\pi}{2}\left[\frac{x - x_0}{x_1 - x_0} + \frac{y - y_0}{y_1 - y_0}\right]$$

$$\phi = \frac{\pi}{2}\left[1 + \frac{x - x_0}{x_1 - x_0} - \frac{y - y_0}{y_1 - y_0}\right]$$

Taking a look at the transformation above, the first thing you should notice is that both $x$ and $y$ equally factor into the values for $\theta$ and $\phi$, thus achieving our desired symmetry. Additionally, the transformation is adaptable to any sized data space, even ones that differ in $x$ and $y$ size. This is a nice feature, but we could have also done the same thing with our previous mapping. But most importantly, this transformation was designed specifically such that the four corners of our data space get mapped to the follow states:

$$\left[\, x_{\text{min}} \,,\, y_{\text{min}} \,\right] \longrightarrow |\,0\,\rangle$$

$$\left[\, x_{\text{max}} \,,\, y_{\text{max}} \,\right] \longrightarrow |\,1\,\rangle$$

$$\left[\, x_{\text{min}} \,,\, y_{\text{max}} \,\right] \longrightarrow |\,-\,\rangle$$

$$\left[\, x_{\text{max}} \,,\, y_{\text{min}} \,\right] \longrightarrow |\,+\,\rangle$$

The four quantities shown above represent the central design to this choice in mapping, and will hold regardless of the min and max values for $x$ and $y$. Just as before, we will breakdown the pros and cons of this transformation, but first let's see it in action using our example point from earlier:

In [ ]:

```
size = 100
#------------------------------------------------------
Data_Space = [-1,1,-1,1]
Point = [0.12,-0.15]       # Example Point
t,p = oq.Bloch_State( Point, Data_Space )
#------------------------------------------------------
OL_grid = np.zeros(shape=(size,size))
#=======================================================
for x in np.arange(size):
    Xp = Data_Space[0] + (x/size)*(Data_Space[1]-Data_Space[0])
    for y in np.arange(size):
        Yp = Data_Space[2] + (y/size)*(Data_Space[3]-Data_Space[2])
        t2,p2 = oq.Bloch_State( [Xp,Yp], Data_Space )
        #---------------------------
        q = QuantumRegister( 3, name='q' )
        c = ClassicalRegister( 1, name='c' )
        qc= QuantumCircuit( q,c, name='qc' )
        qc.u3( t,   p, 0, q[1] )
        qc.u3( t2, p2, 0, q[2] )
        #-------------
        IP = oq.SWAP_Test( qc, q[0], q[1], q[2], c[0], Shots )
        if( IP < 5000 ):
            IP = 5000
        OL_grid[int(size-y-1),int(x)] = m.sqrt((1.0*IP/Shots-0.5)*2)
#=======================================================
fig, ax = plt.subplots()
show_ticks = False
show_text  = False
oq.Heatmap(OL_grid, show_text, show_ticks, ax, "viridis", "Inner Product")
plt.plot((Point[0]-Data_Space[0])*size/(Data_Space[1]-Data_Space[0]), (Data_Space[3]-Point[1])*size/(Data_Space[3]-Data_Spa
Centroids = [ [-0.38,0.61] ,  [-0.09,-0.34] ,  [0.52,0.29] ]
colors = ['green','blue','black']
for c in np.arange(len(Centroids)):
    plt.scatter((Centroids[c][0]+1)*size/2, size-((Centroids[c][1]+1)*size/2), color='white',   marker='s', s=50)
    plt.scatter((Centroids[c][0]+1)*size/2, size-((Centroids[c][1]+1)*size/2), color=colors[c], marker='x', s=50)
fig.tight_layout()
plt.show()
```

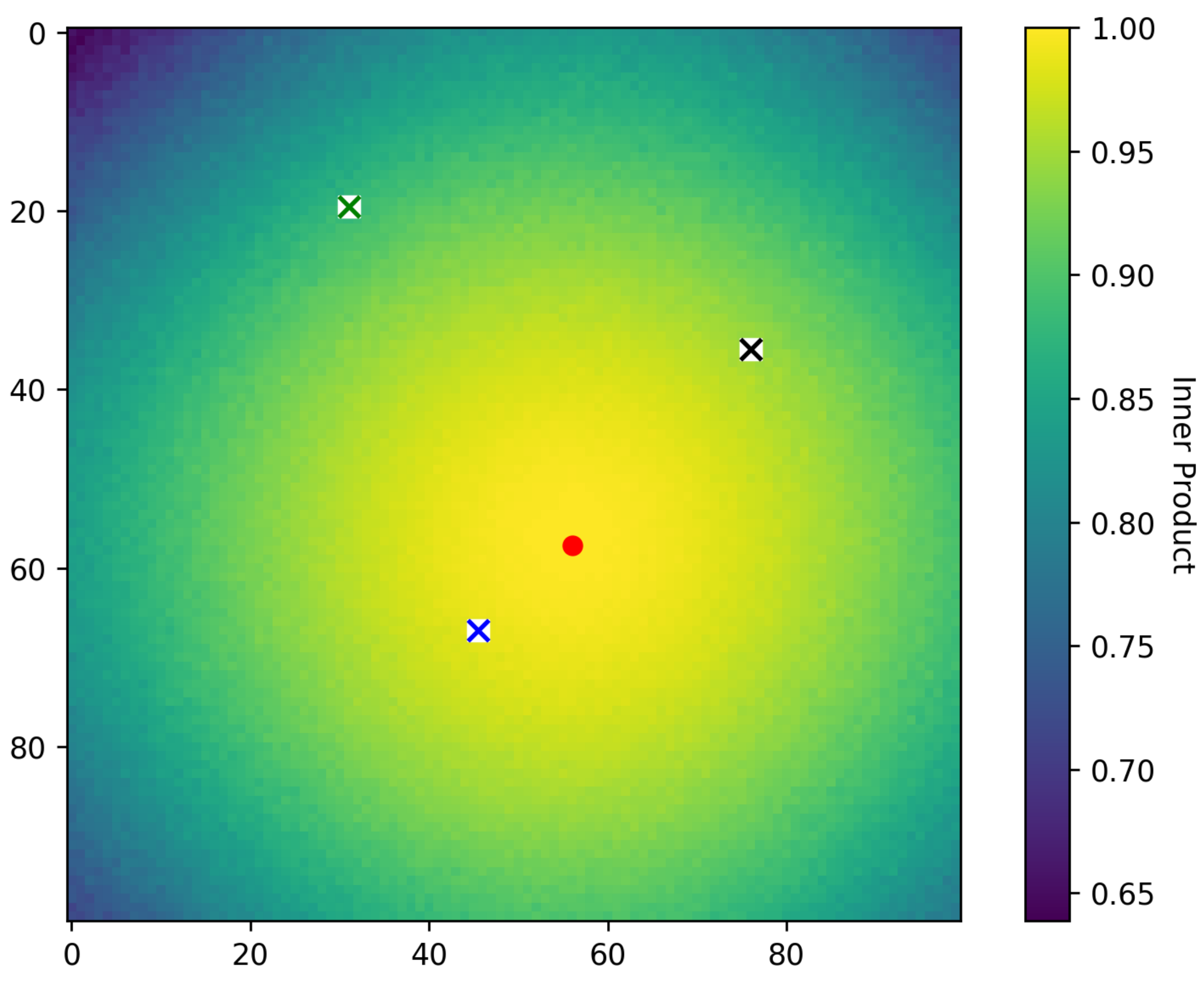

Just like the previous transformation, the inner product space looks promising for the three centroids, correctly assessing the order of nearest to furthest as: blue, black, green. So then, what can we say is different about this new mapping, and does it offer us any advantage over the old? For this particular example, both transformations perform perfectly fine at distinguishing the three clusters, so neither one is really advantageous over the other. That being said, let's investigate how this new transformation handles data points that were previously problematic, namely the boundaries:

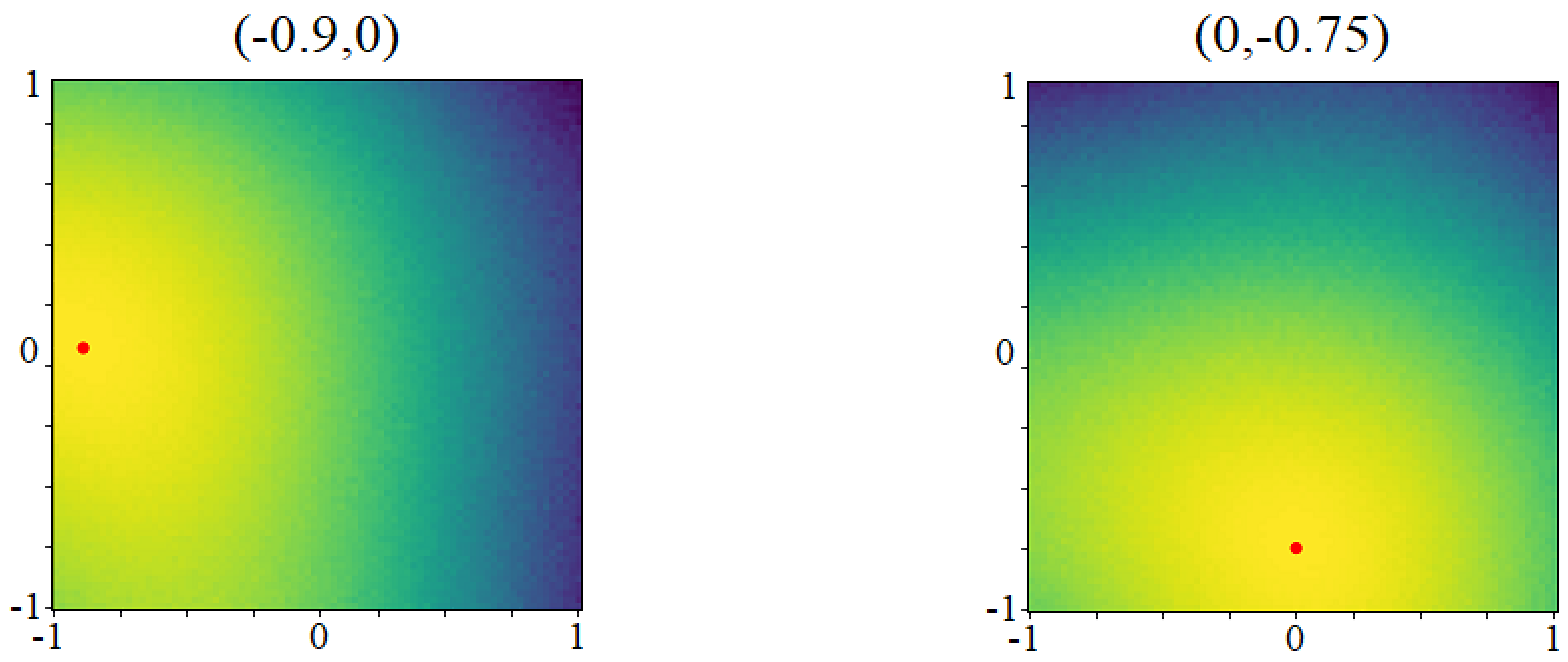

Taking a look at the two plots above, we will begin our analysis of this new transformation by noting that this mapping has indeed achieved the desired symmetry in $x$ and $y$, yielding nearly identically shaped plots with respect to the two axes. Compare the two plots shown here with those for $\left[-0.6, 0\right]$ and $\left[0, -0.6\right]$ from earlier, and you can see just how warped the previous inner product landscapes really were. Beyond the symmetry however, let's return to the two major problems we discussed with the previous mapping, and how our new transformation handles the same problematic points:

1) Looking at the plot for $\left[-0.9, 0\right]$ (both plots in fact), we can see that our inner product landscape no longer has the same value for all points along the $x$ boundaries. For example, if we focus on the boundary for $x = -1$, we can see a gradual decrease all along the border, emanating from $\left[0, 0\right]$ in both directions towards the corners $\left[-1, 1\right]$ and $\left[-1, -1\right]$.

2) Now turning our attention to the $\left[0, -0.75\right]$ plot, we can see that the $\left[-1, 1\right]$ and $\left[1, 1\right]$ corners correspond to the "furthest" points according to our inner product landscape (similarly for the $\left[-0.9, 0\right]$ plot as well). As a whole, both plots clearly show a decreasing radial trend centered from the data points, which is ideal for our quantum distance metric.

With both of the previous problems remedied, as well as achieving symmetry in $x$ and $y$, it is fair to say that this second choice in mapping is a better suited replacement for distance. And in the coming final example whereby we will implement the full $Q$-means clustering algorithm, we will indeed be using this transformation choice. But, despite the upgrade, we still must be aware of the potential cons of this mapping. For a square data space (the distance between $x_{\min}$ and $x_{\max}$ is the same as $y_{\min}$ and $y_{\max}$) like the one in the above examples, this mapping is ideal, but not so for an asymmetric space. By preserving that the four corners always map to the orthogonal states shown earlier, we can run into trouble if we are working with a data space that is noticeably longer in either $x$ or $y$. For example, suppose we have the four corners:

$$\left[0, 10\right] \qquad \left[2, 10\right]$$

$$\left[0, 0\right] \qquad \left[2, 0\right]$$

Using the symmetric mapping, our SWAP Test would tell us that the distances from $[0, 0]$ to $[2, 0]$ and $[0, 0]$ to $[0, 10]$ are the same. In essence, our quantum distance metric will be skewed by a factor of $5$ in the $y$ direction, which ultimately makes it unusable. One solution is to simply expand our data space to something square, like extending out to $x = 10$ for example. However, this comes at the cost of precision. If we place all of the same data points into a larger square space, this means that the differences in distances as determined by the SWAP Test will be smaller, in turn requiring more trials for better accuracy. Thus, as we already noted earlier, it is important to remember that there is no "perfect" transformation for a Euclidean distance substitute, and we must always be mindful of what kind of mapping will best suit a particular problem.

## Implementing $Q$-Means Clustering

With all of the tools for calculating distances via quantum states now in hand, it's time to construct our full $Q$-means clustering algorithm. As a reminder, the motivation for using quantum states and inner products is to improve the "Assignment" step in our classical $k$-means algorithm. Specifically, in the coding example to come, we will be replacing the **Update_Clusters** function with **Q_Update_Clusters**. Inside this new function will be our quantum system, using inner product values obtained from the SWAP Test via the symmetric mapping to determine the nearest centroid to each data point. Then after all of the data points have been sorted into their respective clusters, the remainder of the algorithm continues just as before. And without further ado, let's see our full $Q$-means clustering algorithm in action:

In [ ]:

```
fig = plt.figure()
ax = fig.add_subplot(1,1,1)
ax.axis([-0.2,8.2,-0.2,8.2])
fig.show()
colors  = ['red','limegreen','deepskyblue','gold']
colors2 = ['darkred','darkgreen','darkblue','darkorange']
#--------------------------------------------------------
n = 140
k = 4
shots = 500
Data_Space = [0,8,0,8]
#--------------------------------------------------------
Data = oq.k_Data(k,n)
for d in np.arange(len( Data )):
    ax.scatter( Data[d][0], Data[d][1], color='black', s=10 )
fig.canvas.draw()
time.sleep(2)
#--------------------------------------------------------
Centroids = oq.Initial_Centroids( k, Data )
Clusters = []
Clusters,old_Clusters = oq.Q_Update_Clusters( Data, Centroids, Clusters, Data_Space, shots )
for c1 in np.arange(len(Clusters)):
    for c2 in np.arange( len( Clusters[c1] ) ):
        ax.scatter( Clusters[c1][c2][0],Clusters[c1][c2][1], color=colors[c1],s=10 )
    ax.scatter( Centroids[c1][0],Centroids[c1][1], color=colors2[c1], marker='x',s=50 )
    fig.canvas.draw()
    time.sleep(1)
time.sleep(2)
#--------------------------------------------------------
terminate = False
iters = 0
while( (terminate==False) and (iters<50)  ):
    Centroids,old_Centroids = oq.Update_Centroids(Centroids, Clusters)
    Clusters,old_Clusters = oq.Q_Update_Clusters( Data, Centroids, Clusters, Data_Space, shots )
    oq.Draw_Data( Clusters, Centroids, old_Centroids, fig, ax, colors, colors2 )
    terminate = oq.Check_Termination( Clusters, old_Clusters )
    iters = iters + 1

print( 'Clustering Complete:    ',iters,' Iterations' )
```

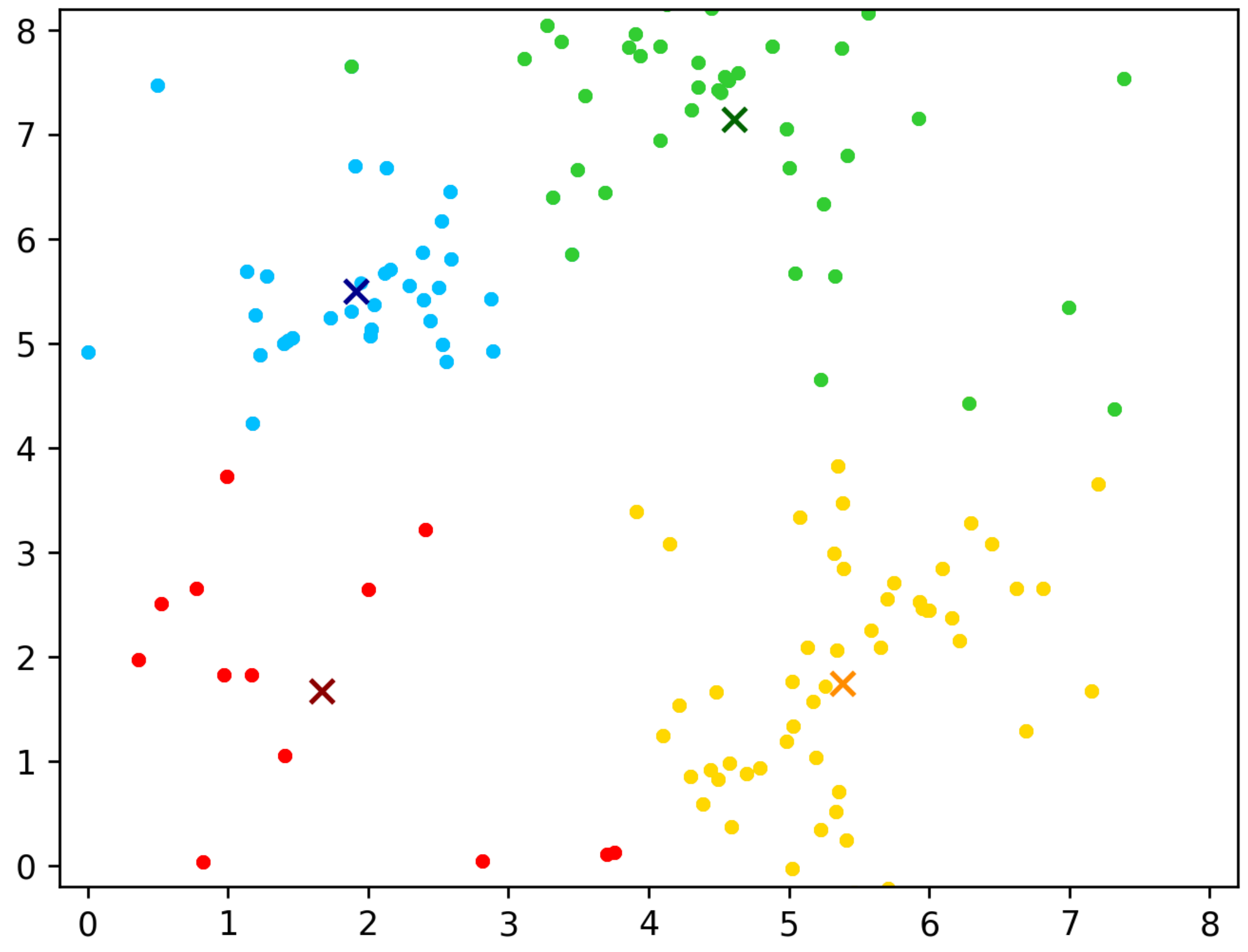

```
Clustering Complete:     12  Iterations
```

As demonstrated above, our $Q$-means algorithm is a success! By using the SWAP Test in combination with mapping each data point to the Bloch Sphere, we were able to correctly sort each data point into the appropriate cluster (within some minor exceptions for borderline points).

Aside from achieving the same end goal, you may notice that running the cell of code above takes quite a bit longer than our classical code from earlier. Although we made a point to highlight the potential speedup advantage gained from using $Q$-means over $k$-means, we can't ignore the fact that obtaining accurate inner product results requires the preparation and measurement of our quantum system numerous times. More accurate results require more measurements, which in turn means that we have a direct tradeoff in speed versus precision. Nevertheless, the demonstration of incorporating quantum into our unsupervised learning algorithm is a noteworthy step towards the advancement of quantum computing!

---

This concludes lesson $9$, and our discussion about the $Q$-means Clustering Algorithm. In this lesson we saw how to transform the data classifying algorithm $k$-means into the quantum upgraded version: $Q$-means. Additionally, we delve into the power of the SWAP Test, and just one of its many applications in quantum computing: calculating distances. At their core, our quantum distance calculating subroutines were heavily determined by the way in which we encoded our classical data into quantum states. The problem of how best to translate classical data into quantum is one of the biggest open questions in quantum computing, and hopefully this lesson has given you a taste of just some of the challenges / possibilities.

I hope you enjoyed this lesson, and I encourage you to take a look at my other .ipynb tutorials!

### Citations

**[1]** S. P. Lloyd, "Least Squares Quantization in PCM", IEEE Transactions on Information Theory **28** : 2 (1982)

**[2]** S. Lloyd, M. Mohseni, P. Rebentrost, "Quantum Algorithms for Supervised and Unsupervised Machine Learning",arXiv:1307.0411 (2013)

# Lesson 10 - QAOA

In this lesson, we will be taking a look at the first of two algorithms where we will be combining our quantum system with some classical optimization techniques. The Quantum Approximate Optimization Algorithm is a very promising quantum algorithm, one which has the potential for real speedups on near term quantum computers. Since its first proposal in 2014 **[1]**, the solving power of QAOA has been demonstrated on numerous problems and continues to be improved and refined.

In order to make sure that all cells of code run properly throughout this lesson, please run the following cell of code below:

In [1]:
```
from qiskit import ClassicalRegister, QuantumRegister, QuantumCircuit, Aer, execute
import Our_Qiskit_Functions as oq
import numpy as np
import math as m
import scipy as sci
import random
import matplotlib
import matplotlib.pyplot as plt
from itertools import permutations
S_simulator = Aer.backends(name='statevector_simulator')[0]
```

## Difficulty With Optimization Problems

In order to give the proper context for QAOA, we must first discuss the types of problems which it promises to help solve quicker. Optimization problems are very common in computer science, and their difficulty can range from driving instructions from point A to point B, to air traffic control keeping every plane at an airport on schedule. Newer and better classical optimization techniques are being developed all the time, but some problems are known to be just too computationally demanding for a classical computer.

Consider the following problem, all of your extended family has gathered for a party and you offer to take a photo of everyone. However, not everyone in your family is on good terms with each other, and some will be very upset if they have to stand next to each other. Simultaneously, some members of your family will be equally upset if they are not close enough to members they like. You realize that anyone whose conditions are met will give you a big smile for the photo, but if they are not, they will frown. After collecting everyone's requests, your task is to maximize the number of smiles for the photo, based on the following conditions:

**1)** Michelle, Betty, and Margaret all want to be at most 3 people away from each other

**2)** Cullen and Nate can't stand each other, and want to be at least 3 people away from one another

**3)** Clint is happy so long as he is not next to Will or Cullen

**4)** Nate wants to be next to Betty, but at least two people away from Michelle and Margaret

After a long $sigh$... you decide to try and make everyone happy by meeting their conditions. Alas, you quickly become overwhelmed by the number of possible combinations, so you turn to your computer for help. Rather than working through all 5040 possible arrangements (7!) yourself, you instead code up everyone's conditions, assign a happiness score for each condition met, and let your code find the best possible combination:

In [2]:

```
def Happiness( A ):
    '''
    Computes the total happiness score based on everyone's conditions
    '''
    happiness = 0
    for i in np.arange(len(A)):
        if( A[i] == 1):
            Mi = int(i)
        if( A[i] == 2):
            Be = int(i)
        if( A[i] == 3):
            Ma = int(i)
        if( A[i] == 4):
            Cu = int(i)
        if( A[i] == 5):
            Na = int(i)
        if( A[i] == 6):
            Cl = int(i)
        if( A[i] == 7):
            Wi = int(i)
    if( (abs(Mi - Be)<=3) and (abs(Mi - Ma)<=3) ):   # Michelle
        happiness += 1
    if( (abs(Be - Mi)<=3) and (abs(Be - Ma)<=3) ):   # Betty
        happiness += 1
    if( (abs(Ma - Mi)<=3) and (abs(Ma - Be)<=3) ):   # Margaret
        happiness += 1
    if( abs(Cu - Na)>=3 ):                           # Cullen
        happiness += 1
    if( (abs(Na - Cu)>=3) and (abs(Na - Be)==1) ):   # Nate
        happiness += 1
    if( (abs(Cl - Cu)>1) and (abs(Cl - Wi)>1) ):     # Clint
        happiness += 1
    happiness += 1                                   # Will
    return happiness

#==================================================================================
perm = permutations([1, 2, 3, 4, 5, 6, 7])
All_Happy = []
all_perm = 0
for i in list(perm):
    all_perm = all_perm + 1
    if( Happiness(list(i)) == 7 ):
        All_Happy.append( list(i) )
print('Total Combinations Where Everyone Is Happy:  ',len(All_Happy))
```

```
Total Combinations Where Everyone Is Happy:   72
```

Upon running your code above, you are pleasantly surprised to find a total of 72 possible arrangements for your family members. You pick one, line everyone up for the photo, and are gratefully met with 7 big smiles.

In the simple exercise of arranging family members, the size of the problem was small enough such that our classical code could work through all 5040 possible combinations with ease. However, each additional variable in the problem increases the space of all possible combinations by more and more:

$$8! \;=\; 40{,}320$$

$$9! \;=\; 362{,}880$$

$$10! \;=\; 3{,}628{,}800$$

As we can see, it doesn't take long before the classical brute force method is no longer viable, especially when we want to consider real-world problems with 100's or 1000's of variables. Luckily, computer scientists have developed a long list of techniques for efficiently solving optimization problems, and getting longer all the time. The hope then is that the assistance of quantum computers will unlock new techniques, allowing for solutions to previously unsolvable problems.

## Setting Up a QAOA Problem

Now, the example above is designed to illustrate the style of optimization problems which we are planning to solve using QAOA. Specifically, we are given some set of variables which make up a large space of possible combinations, and our problem is to find the combination which optimizes some criteria. We will refer to these criteria as our "Cost Function", $\mathbf{C}(\mathbf{z})$, where $\mathbf{z}$ is the set of all variables which we need to optimize:

$$\mathbf{z} \;\equiv\; \{z_1, z_2, \ldots, z_n\} \qquad\qquad z_j \;\in\; \{+1, -1\}$$

For example, consider the following cost function and its table of possible values:

$$\mathbf{C}(\mathbf{z}) \;=\; 3z_1 z_2 - z_2 z_3$$

| $z_1$ | $z_2$ | $z_3$ | $\mathbf{C}(\mathbf{z})$ |
|---|---|---|---|
| 1 | 1 | 1 | 2 |
| 1 | 1 | −1 | 4 |
| 1 | −1 | 1 | −2 |
| 1 | −1 | −1 | −4 |
| −1 | 1 | 1 | −4 |
| −1 | 1 | −1 | −2 |
| −1 | −1 | 1 | 4 |
| −1 | −1 | −1 | 2 |

Given a cost function like $\mathbf{C}(\mathbf{z})$ above, the first question we should ask is how do we translate it into a quantum system? Just like how we were able to encode all of the various criteria in the photograph example into if-statements about distances $\left( \middle| \text{person } A \text{ - person } B \middle| \right)$, we similarly need a way of encoding $\mathbf{C}(\mathbf{z})$ into quantum. The answer is that $\mathbf{C}(\mathbf{z})$ will become an operator $\mathbf{C}$, which when it acts on a quantum state will return a value corresponding to the cost function. More specifically, $\mathbf{C}$ is a diagonal matrix in the computational basis, where each element along the diagonal corresponds to a value of $\mathbf{C}(\mathbf{z})$ from a unique combination of $\mathbf{z}$:

$$\mathbf{C}\,|\,\mathbf{z}\rangle \;=\; \mathbf{C}\,|\,z_1 z_2 \ldots z_n\,\rangle \;=\; \mathbf{C}(\mathbf{z})\,|\,z_1 z_2 \ldots z_n\,\rangle$$

Using our table of values for $\mathbf{C}(\mathbf{z})$ from above, we could construct the matrix $\mathbf{C}$ by placing all of the $\mathbf{C}(\mathbf{z})$ values along the diagonal. However, let's derive $\mathbf{C}$ for this example another way. In case you were curious as to why we chose '$z$' to be the variable in our cost functions, it's because when we convert $\mathbf{C}(\mathbf{z})$ into an operator, we simply replace each $z_j$ with the operator $Z_j$. And, in agreement with our condition on each $z_j \in \{+1, -1\}$, the $Z$ operator similarly returns the values $\pm 1$ when acting on the states $|\,0\rangle$ and $|\,1\,\rangle$ respectively. By replacing each $z_j$ with $Z_j$, we can construct the matrix operator corresponding to $\mathbf{C}(\mathbf{Z})$:

$$\begin{aligned}
\mathbf{C}(\mathbf{Z}) \;&=\; 3\,Z_1 \otimes Z_2 \otimes I_3 \;-\; I_1 \otimes Z_2 \otimes Z_3 \\
&=\; 3 \begin{bmatrix} 1 & 0 \\ 0 & -1 \end{bmatrix} \otimes \begin{bmatrix} 1 & 0 \\ 0 & -1 \end{bmatrix} \otimes \begin{bmatrix} 1 & 0 \\ 0 & 1 \end{bmatrix} \;-\; \begin{bmatrix} 1 & 0 \\ 0 & 1 \end{bmatrix} \otimes \begin{bmatrix} 1 & 0 \\ 0 & -1 \end{bmatrix} \otimes \begin{bmatrix} 1 & 0 \\ 0 & -1 \end{bmatrix} \\
&=\; \begin{bmatrix} 2 & & & & & & & \\ & 4 & & & & & & \\ & & -2 & & & & & \\ & & & -4 & & & & \\ & & & & -4 & & & \\ & & & & & -2 & & \\ & & & & & & 4 & \\ & & & & & & & 2 \end{bmatrix}
\end{aligned}$$

Sure enough, the diagonal elements of the matrix above correspond exactly to the values from our table earlier (and in the same order as well). There is of course, one major problem with the matrix operator shown above, preventing us from physically implementing it on a real quantum system. And that problem is the fact that $\mathbf{C}(\mathbf{Z})$ is not a unitary operation. Parts of $\mathbf{C}(\mathbf{Z})$ are valid quantum operations we can perform: $Z_1 \otimes Z_2 \otimes I_3$ and $I_1 \otimes Z_2 \otimes Z_3$ , but their combination is not.

Given that we cannot physically implement $\mathbf{C}(\mathbf{Z})$, you may start to worry about the viability of using a quantum system to solve our optimization problem. Rightfully so, if we were to limit ourselves to only problems whereby the cost function perfectly corresponded with a unitary operator, the list of solvable problems would be quite small. Luckily however this is not the case. As we shall see in the coming section, QAOA cleverly encodes cost functions in a way such that the resulting quantum operator is always unitary!

## $U(\mathbf{C},\gamma)$ The Phase Operator

Although we were unable to use our cost function directly as an operator, the condition we put on $\mathbf{C}(\mathbf{Z})$ to be a diagonal matrix is actually one required of the operator we *will* be using. Specifically, we will be implementing $U(\mathbf{C},\gamma)$, which is defined as:

$$U(\mathbf{C}, \gamma) \;=\; e^{-i\gamma \mathbf{C}}$$

As a quick math reminder, the exponentiation of a matrix is as follows:

$$e^A \;=\; \sum_n^{\infty} \frac{A^n}{n!} \;=\; I \,+\, A \,+\, \frac{AA}{2!} \,+\, \frac{AAA}{3!} \,+\, \cdots$$

Example:

$$A = \begin{bmatrix} \alpha_1 & 0 \\ 0 & \alpha_2 \end{bmatrix} \qquad e^A \;=\; \begin{bmatrix} 1 & 0 \\ 0 & 1 \end{bmatrix} + \begin{bmatrix} \alpha_1 & 0 \\ 0 & \alpha_2 \end{bmatrix} + \frac{\begin{bmatrix} \alpha_1 & 0 \\ 0 & \alpha_2 \end{bmatrix}\begin{bmatrix} \alpha_1 & 0 \\ 0 & \alpha_2 \end{bmatrix}}{2!} + \cdots$$

$$= \begin{bmatrix} e^{\alpha_1} & 0 \\ 0 & e^{\alpha_2} \end{bmatrix}$$

The example above might take a moment to fully digest, especially if you've never seen matrix exponentiation before. To help complete the exercise, and because it is relevant to our next discussion:

$$\sum_{n}^{\infty} \frac{x^n}{n!} \; = \; e^x$$

As you can see, the form for matrix exponentiation is actually the same formula for regular numbers as well. Now, the reason this is relevant is because of the way in which diagonal matrices undergo matrix exponentiation. The key here that no matter what power it is raised to, a diagonal matrix will remain diagonal. And not only that, the diagonal elements do not mix, but rather simply get raised to the power of the operation:

$$\begin{bmatrix} \alpha & & & & \\ & \beta & & & \\ & & . & & \\ & & & . & \\ & & & & . \\ & & & & \end{bmatrix}^n = \begin{bmatrix} \alpha^n & & & & \\ & \beta^n & & & \\ & & . & & \\ & & & . & \\ & & & & . \\ & & & & \end{bmatrix}$$

With this in hand, I encourage you to work through the example above and verify for yourself that you indeed get the diagonal elements $e^{\alpha_1}$ and $e^{\alpha_2}$.

Returning now to $U(\mathbf{C},\gamma)$, let's discuss why this operator remedies the problem we encounter earlier with $\mathbf{C}(Z)$. So long as we meet the condition on our cost function $\mathbf{C}(Z)$ that its resulting matrix operator is diagonal, we are guaranteed that $U(\mathbf{C},\gamma)$ will also be diagonal, which is nice. But more importantly, because $U(\mathbf{C},\gamma)$ contains the imaginary element $i$ in the exponential, this $i$ carries through to all of the elements of $U(\mathbf{C},\gamma)$ as well:

$$B = \begin{bmatrix} \beta_1 & 0 \\ 0 & \beta_2 \end{bmatrix} \qquad\qquad e^{iB} \; = \; \begin{bmatrix} e^{i\beta_1} & 0 \\ 0 & e^{i\beta_2} \end{bmatrix}$$

This inclusion of the imaginary element $i$, in combination with $\mathbf{C}(Z)$ being diagonal, is what guarantees that our $U(\mathbf{C},\gamma)$ operator will always be unitary. In essence, each diagonal element of $\mathbf{C}(Z)$ becomes a corresponding phase term in $U(\mathbf{C},\gamma)$. To see this, let's return to our cost function example from earlier, and see what the corresponding $U(\mathbf{C},\gamma)$ matrix looks like:

$$\mathbf{C}(Z) \; = \; \begin{bmatrix} 2 & & & & & & & \\ & 4 & & & & & & \\ & & -2 & & & & & \\ & & & -4 & & & & \\ & & & & -4 & & & \\ & & & & & -2 & & \\ & & & & & & 4 & \\ & & & & & & & 2 \end{bmatrix} \qquad U(\mathbf{C},\gamma) \; = \; \begin{bmatrix} e^{i2\gamma} & & & & & & & \\ & e^{i4\gamma} & & & & & & \\ & & e^{-i2\gamma} & & & & & \\ & & & e^{-i4\gamma} & & & & \\ & & & & e^{-i4\gamma} & & & \\ & & & & & e^{-i2\gamma} & & \\ & & & & & & e^{i4\gamma} & \\ & & & & & & & e^{i2\gamma} \end{bmatrix}$$

If we now consider how this matrix would operate on a quantum state, we can see that every term will pick up some phase $e^{ix\gamma}$, which is certainly unitary!

By now you may be wondering what this extra $\gamma$ term is that comes with our $U(\mathbf{C},\gamma)$ operator. Why not just perform $e^{i\mathbf{C}}$ since it already contains all of our cost function's information? The answer is because implementing $e^{i\mathbf{C}}$ simply won't solve our problem. Jumping ahead a bit, there will be a second free parameter $\beta$, coming from a separate operation, which in combination with $\gamma$ will provide a platform for our QAOA to optimize. So far we've only discussed how to encode our cost function into a quantum operator, but we've actually said nothing about how QAOA finds the optimal solution! As we shall see, the goal of QAOA is to find a combination of $\gamma$ and $\beta$, which when applied to our quantum system produces a superposition state whereby a measurement will yield with high probability a state that solves our optimization problem.

As a final note before moving on to our second major operator, let's quickly show how one might implement $U(\mathbf{C},\gamma)$ as a quantum circuit. As shown above, the most general form for $U(\mathbf{C},\gamma)$ is one where each diagonal element is a unique phase term. For this general case, we can implement each phase term individually through the use of CNOT and $R_\phi$ gates, plus ancilla. For our coding example, let's implement the following matrix as a quantum circuit:

$$U(\mathbf{C},\gamma) \; = \; \begin{bmatrix} e^{i\frac{\pi}{4}\gamma} & & & \\ & e^{-i\frac{\pi}{2}\gamma} & & \\ & & e^{-i\frac{\pi}{4}\gamma} & \\ & & & e^{i\frac{\pi}{2}\gamma} \end{bmatrix}$$

which when applied to an equal superposition state should yield:

$$\begin{aligned} U(\mathbf{C},\gamma)|s\rangle &= \frac{1}{2}\left( e^{i\frac{\pi}{4}\gamma}|00\rangle + e^{-i\frac{\pi}{2}\gamma}|01\rangle + e^{-i\frac{\pi}{4}\gamma}|10\rangle + e^{i\frac{\pi}{2}\gamma}|11\rangle \right) \\ &= \frac{1}{2}\left( \frac{1+i}{\sqrt{2}}|00\rangle - i|01\rangle + \frac{1-i}{\sqrt{2}}|10\rangle + i|11\rangle \right) \end{aligned}$$

Here we have set $\gamma = 1$ for this example.

In [56]:

```
q = QuantumRegister(2, name = 'q')
a = QuantumRegister(1, name = 'a')
qc= QuantumCircuit(q,a, name = 'qc')
#==================================
qc.h( q[0] )
qc.h( q[1] )
qc.barrier()
print('__ Initial State __')
oq.Wavefunction(qc, systems=[2,1])
#-----------------------------------   Uc Operator
qc.x( q[0] )
qc.ccx( q[0], q[1], a[0] )
qc.cu1( -m.pi/2, a[0], q[0] )
qc.ccx( q[0], q[1], a[0] )
qc.x( q[0] )
#-----------------------------------
print('\n__ After Applying The |01> Phase Term __')
oq.Wavefunction(qc, systems=[2,1], show_systems=[True,False])
print('\n__ Circuit Diagram __\n')
print(qc)
```

```
__ Initial State __
0.5 |00>|0>    0.5 |10>|0>    0.5 |01>|0>    0.5 |11>|0>

__ After Applying The |01> Phase Term __
0.5 |00>    0.5 |10>    -0.5j |01>    0.5 |11>

__ Circuit Diagram __

        ┌───┐ ░ ┌───┐                     ┌───┐
q_0: |0>┤ H ├─░─┤ X ├──■────■────────■──┤ X ├
        ├───┤ ░ └───┘  │    │        │  └───┘
q_1: |0>┤ H ├─░────────■────┼────────■───────
        └───┘ ░      ┌─┴─┐  │-pi/2 ┌─┴─┐
a_0: |0>──────░──────┤ X ├──■──────┤ X ├─────
              ░      └───┘         └───┘
```

The circuit above illustrates how to implement the $e^{-i\frac{\pi}{2}\gamma}$ phase term from $U(\mathbf{C},\gamma)$ which falls on the $|01\rangle$ state, shown by the gates after the barrier. Essentially, we can use $X$ and CCNOT gates in combination with ancilla qubits to isolate any single state, apply a phase, and then undo all of the transformations. If you recall, this is the same technique we used in the Grover Algorithm to pick out our marked state. Here we can use this trick to mark individual states with arbitrary phase terms:

In [4]:

```
q = QuantumRegister(2, name = 'q')
a = QuantumRegister(1, name = 'a')
Uc_qc= QuantumCircuit(q,a, name = 'qc')

Uc_qc.h( q[0] )
Uc_qc.h( q[1] )
print('__ Initial State __')
oq.Wavefunction(Uc_qc, systems=[2,1])
#--------------------------------------   # |00> state
Uc_qc.x( q[0] )
Uc_qc.x( q[1] )
Uc_qc.ccx( q[0], q[1], a[0] )
Uc_qc.cu1( m.pi/4, a[0], q[0] )
Uc_qc.ccx( q[0], q[1], a[0] )
Uc_qc.x( q[1] )
Uc_qc.x( q[0] )
#--------------------------------------   # |01> state
Uc_qc.x( q[0] )
Uc_qc.ccx( q[0], q[1], a[0] )
Uc_qc.cu1( -m.pi/2, a[0], q[0] )
Uc_qc.ccx( q[0], q[1], a[0] )
Uc_qc.x( q[0] )
#--------------------------------------   # |10> state
Uc_qc.x( q[1] )
Uc_qc.ccx( q[0], q[1], a[0] )
Uc_qc.cu1( -m.pi/4, a[0], q[0] )
Uc_qc.ccx( q[0], q[1], a[0] )
Uc_qc.x( q[1] )
#--------------------------------------   # |11> state
Uc_qc.ccx( q[0], q[1], a[0] )
Uc_qc.cu1( m.pi/2, a[0], q[0] )
Uc_qc.ccx( q[0], q[1], a[0] )

print('\n__ After Applying U(C,gamma) __')
oq.Wavefunction(Uc_qc, systems=[2,1])
```

```
__ Initial State __
0.5 |00>|0>    0.5 |10>|0>    0.5 |01>|0>    0.5 |11>|0>

__ After Applying U(C,gamma) __
0.35355+0.35355j |00>|0>    0.35355-0.35355j |10>|0>    -0.5j |01>|0>    0.5j |11>|0>
```

As illustrated above, this technique is applicable for any sized system, so long as you have the necessary number of additional ancilla qubits. In general however, this technique is very costly and should typically be avoided if possible, but serves our learning purposes here just fine. In the coming examples, we will be working with $U(\mathbf{C},\gamma)$ operators that are much more gate efficient to implement.

## $U(\mathbf{B},\beta)$ The Mixing Operator

Having just seen how to encode our cost function $\mathbf{C}(\mathbf{z})$ into the unitary operator $U(\mathbf{C},\gamma)$, we will now turn our attention to the second operator of QAOA: $U(\mathbf{B},\beta)$. To motivate why we even need this second operation, consider the effect of $U(\mathbf{C},\gamma)$ applied to the even superposition state $|s\rangle$ (which is the first step in QAOA). Before applying $U(\mathbf{C},\gamma)$ we have an equal probability of measuring any state in the system, which can be thought of as the starting point for our optimization problem. Then we apply $U(\mathbf{C},\gamma)$, effectively distributing phases to each state in the system based on the cost function we encoded. But although we have achieved a quantum state representation of our cost function, a measurement on this system will still yield every state with equal probability. Thus, $U(\mathbf{C},\gamma)$ alone is not enough to solve our optimization problem, so we need to introduce a second operation, one that will effectively mix all of the amplitudes together so that we get constructive and destructive interference. Hence, $U(\mathbf{B},\beta)$, "The Mixing Operator", which has the following mathematical structure:

$$\mathbf{B} \;=\; \sum_{j}^{N} X_j \qquad\qquad U(\mathbf{B},\beta) \;=\; e^{i\beta\mathbf{B}}$$

Very similar in nature to our $U(\mathbf{C},\gamma)$ operator, $\mathbf{B}$ here is composed of $X$ gates instead of $Z$ gates. Consequently, when we apply matrix exponentiation to the operator $B$, we no longer get a diagonal matrix operator, which in turn means that we will get amplitude mixing. This mixing operator comes with another free parameter $\beta$, which we will discuss further in the next section, but essentially plays the same role as $\gamma$. Resulting from the fact that $B$ is a linear sum of $X$ gates, one applied to each qubit in the system, we can mathematically rewrite $U(\mathbf{B},\beta)$ as follows:

$$U(\mathbf{B},\beta) \;=\; \prod_{j}^{N} e^{i\beta X_j}$$

where $N$ is the number of qubits in the system, and each $e^{i\beta X_j}$ term has the following structure:

$$X \;=\; \begin{bmatrix} 0 & 1 \\ 1 & 0 \end{bmatrix} \qquad\qquad e^{i\beta X} \;=\; \begin{bmatrix} \cos(\beta) & i\sin(\beta) \\ i\sin(\beta) & \cos(\beta) \end{bmatrix} \;=\; R_x(\beta)$$

I encourage you to work through the matrix exponentiation math yourself, and you should arrive at the result that our $U(\mathbf{B},\beta)$ operator is just a product of $R_x(\beta)$ gates. As a reminder, $R_x(\beta)$ is a single qubit gate which rotates the state of the qubit about the $\hat{x}$ axis (on the Bloch Sphere) by an angle $\beta$. The fact that our $U(\mathbf{B},\beta)$ operator turns out to be implementable using single qubit gates on each qubit is very convenient! In particular, it means that the entire $U(\mathbf{B},\beta)$ operation can be performed with a circuit depth of one (all of the $R_x(\beta)$ gates performed in parallel).

## Expectation Value $F(\gamma,\beta)$

With both $U(\mathbf{C},\gamma)$ and $U(\mathbf{B},\beta)$ now in hand, as well as their physical implementation into quantum circuits, we have everything we need for QAOA! Below is a general overview of the quantum steps in a typical QAOA procedure, which we will go into further detail next:

**1)** Prepare the initial state of the system: $|\Psi\rangle_i \;=\; H^{\otimes N}|0\rangle \;=\; |s\rangle$

**2)** Choose your set of parameters $\gamma$ and $\beta$

**3)** Apply $U(\mathbf{B},\beta)\,U(\mathbf{C},\gamma)\;|s\rangle$

**4)** Measure in the computational basis

Now, the outline shown above is not the whole story of QAOA as we shall see. Ultimately, the goal of QAOA is to minimize (or maximize depending on your cost function) the expectation value of our cost function:

$$F(\gamma,\beta) \;=\; \langle \Psi_{\gamma\beta}|\,\mathbf{C}(Z)\,|\Psi_{\gamma\beta}\rangle$$

where

$$|\Psi_{\gamma\beta}\rangle \;=\; U(\mathbf{B},\beta)\,U(\mathbf{C},\gamma)\,|s\rangle$$

Working with expectation values in an algorithm is new to this tutorial series, and requires a bit of further explanation. First off, obtaining an expectation value using a quantum system requires repeat measurements, enough to approximate the wavefunction in question. Working with Qiskit, we have the luxury of being able to see our wavefunctions, but on a real quantum computer this is not the case. Thus, in order to determine the expectation value from some given state, one would need to prepare and measure the same state over and over for better accuracy. The cell of code below shows a quick example of this:

In [8]:

```
q = QuantumRegister(3,name='q')
c = ClassicalRegister(3,name='c')
qc = QuantumCircuit(q,c,name='qc')
#================================
qc.h( q[0] )
qc.h( q[1] )
qc.h( q[2] )
qc.u1( m.pi/10, q[0] )
qc.u1( m.pi/15, q[1] )
qc.u1( m.pi/20, q[2] )
qc.rx( m.pi/5, q[0] )
qc.ry( m.pi/6, q[1])
qc.rz( m.pi/7, q[2])
#----------------------
SV = execute( qc, S_simulator, shots=1 ).result().get_statevector()
qc.measure(q,c)
M = oq.Measurement( qc, shots=10000, return_M=True, print_M=False)
#-----------------------------------------------------------------
Energies = [ ['000',2],['100',-4],['010',-2],['110',4],['001',4],['101',-2],['011',-4],['111',2]]
EV1 = 0
EV2 = 0
for i in np.arange( len(Energies) ):
    EV1 = EV1 + M[ Energies[i][0] ]  * Energies[i][1]/10000
    EV2 = EV2 + np.real(SV[i]*np.conj(SV[i])) * Energies[i][1]
print('Energy Expectation Value From Measurements: ',round(EV1,4))
print('\nEnergy Expectation Value From Wavefunction: ',round(EV2,4))
```

```
Energy Expectation Value From Measurements:  -0.2576

Energy Expectation Value From Wavefunction:  -0.2665
```

The cell of code above creates a wavefunction with various amplitudes and phases, and then computes the expectation value of our quantum state two ways: one using the exact amplitudes of the wavefunction, and one using measurement results. If you rerun the cell several times, you will notice that the value obtained through measurements changes each time, but not so for the wavefunction method. As you may suspect, this is because measurement results change with each experimental run, but the quantum system we prepare is always the same, so its wavefunction never changes. We can then say that the expectation value obtained through the wavefunction is the $true$ value, while the other method gives us an approximation. Increasing the number of measurements will yield a more accurate approximation, but comes at the cost of increasing the length of the algorithm.

Returning back to QAOA, computing expectation values is a key ingredient for our algorithm. Whether we are trying to find the min or max value of $\mathbf{C}(\mathbf{z})$, QAOA achieves this goal by systematically trying to optimize $F(\gamma,\beta)$. To understand why this quantity leads to our solution, consider how the value of $F(\gamma,\beta)$ changes as we increase the amplitude of our solution state, call it $|\, z' \,\rangle$, corresponding to the maximal value $\mathbf{C}(\mathbf{z}')$. Starting from the equal superposition state $|\, s \,\rangle$, the expectation value obtained from this state will yield a value where all possible values of $\mathbf{C}(\mathbf{z})$ have been equally weighed together. This will include the contribution from $\mathbf{C}(\mathbf{z}')$, the smallest or largest possible value, but also every other value of $\mathbf{C}(\mathbf{z})$, giving us a final answer that is certainly not optimal. Taking the case where $\mathbf{C}(\mathbf{z}')$ is the maximal value of our cost function, the main point here is that if we prepare any state $|\,\Psi_{\gamma\beta}\rangle \neq |\, z' \,\rangle$, then our resulting expectation value $F(\gamma,\beta)$ will be smaller than its potential largest value.

In a real QAOA scenario, finding the maximal value of the classical cost function $\mathbf{C}(\mathbf{z})$ is the ultimate goal, so it is assumed that we don't know the value $\mathbf{C}(z')$ $a\ priori$. With each preparation and measurement of $|\,\Psi_{\gamma\beta}\rangle$, we can plug in the measured state to $\mathbf{C}(\mathbf{z})$ and see what values we get. Ultimately however, our problem boils down to searching through a variable space of $\gamma$ and $\beta$, for which we typically have no conclusive way of knowing whether any of the states we've checked are truly optimal, or close to optimal. This may sound a bit discouraging, but as we shall see in the next section, we can use various classical optimization techniques to give us a strong indication when we've reached an optimal $F(\gamma,\beta)$.

## QAOA Workflow

We've now come to the true bread and butter of the QAOA algorithm: finding the optimal $\gamma$ and $\beta$. Sometimes easier said than done, finding the values for $\gamma$ and $\beta$ that will optimize $F(\gamma,\beta)$, and ultimately lead to a measurement of $|\, z' \,\rangle$, will require the use of classical optimization techniques. The parameters $\gamma$ and $\beta$ both span the full range of values from $0$ to $2\pi$, so the solution to our problem of finding the optimal $\mathbf{C}(\mathbf{z}')$ boils down to a classical optimization search through the space of our free parameters. In this lesson we will cover one such classical optimization technique: gradient descent, which will serve as our example into the types of techniques one might employ. But first, let's take a look at the full workflow of QAOA:

1. Initialize the quantum system: $|\Psi\rangle_i = H^{\otimes N}|0\rangle$
2. Apply the phase and mixing operators: $|\Psi_{\gamma\beta}\rangle = U(B,\beta)U(C,\gamma)|\Psi\rangle_i$
3. Measure in the computational basis: $|\tilde{z}\rangle$
4. Check $\mathbf{C}(\tilde{z})$
5. Repeat steps 1 - 4 to compute: $F(\gamma,\beta) = \langle\Psi_{\gamma\beta}|\mathbf{C}(\mathbf{Z})|\Psi_{\gamma\beta}\rangle$
6. Use classical optimization technique to determine next $\beta$ and $\gamma$ for step 1
7. Conclude the algorithm based on classical optimization technique
   The best $\mathbf{C}(\tilde{z})$ obtained is the approximation to $\mathbf{C}(\mathbf{z}')$

In the diagram above, the different subsections of the algorithm are boxed together to illustrate repetition. Steps 1 - 4 are repeated until a sufficient accuracy for $F(\gamma,\beta)$ is obtained, while steps 1 - 6 are repeated until one reaches a convergence for the classical optimization technique of their choosing. For example, in the gradient descent technique which we are about to discuss next, the conclusion of the algorithm happens when a minima or maxima is reached. Importantly however, notice that step 7 is *not* "you have found the optimal value of the cost function $\mathbf{C}(\mathbf{z}')$", but rather an approximation to $\mathbf{C}(\mathbf{z}')$ based on the best state we were able to find through measurements. This is an important concept to keep in mind about QAOA: the algorithm does *not* guarantee that we will find $\mathbf{C}(\mathbf{z}')$, only that we will find the best approximation based on the classical optimization technique we choose.

Once again, it might seem a little discouraging that our QAOA has no guarantee on finding the optimal $\mathbf{C}(\mathbf{z}')$, but that doesn't mean it isn't an effective algorithm. In fact, quite the opposite! In many real-world scenarios where one might look to implement QAOA, finding an approximate solution may be of tremendous help, especially when we're talking about problems where blindly searching through the space of $\mathbf{C}(\mathbf{z})$ may be computationally infeasible. Ideally one may look to use QAOA to find an approximate solution, which could then be given to a classical computer to finish the problem and find the exact solution. Thus, the power of QAOA lies in its ability to assist in problems where classical techniques alone fail.

## Ising Model Cost Function

With the workflow and motivation for QAOA now laid out for us, it is time to code up our first full example. The cost function that we will be implementing is based on the Ising Energy Model, which describes the energy of a system composed of particles with two possible orientations for spin, and subject to an external magnetic field. In nature, systems such as these will always tend to their lowest energy state, so our goal will be to try and predict their ground state using QAOA. The cost function for this model is:

$$\mathbf{C}(\mathbf{Z}) = -\sum_{\langle ij \rangle} Z_i Z_j - \sum_i h_i Z_i$$

where each $Z_i$ can take on the values $\{+1, -1\}$, and the values for $h_i$ are the magnetic field strength and direction at each particle's location.

Looking at the cost function above, the notation under the first summation $\langle ij \rangle$ means that we only sum over nearest neighbors. Numerically, the quantity in this summation, $-Z_i Z_j$, takes on the value $+1$ if the values for $Z_i$ and $Z_j$ are opposite and $-1$ if they are the same. If we think about what this is physically describing, it means that the system will have a lower energy state if two neighboring particles have the same spin orientation. And since our goal is to minimize $\mathbf{C}(\mathbf{Z})$ in this case, at first glance it would seem that our optimal solution is to orient all of the particles in the system to be in the same direction. However, things are not quite that simple, as the second summation in our cost function can be thought of as the competing element: magnetic alignment. Even though arranging every particle's spin to align would optimize the first summation, depending on the direction and field strength at each particle's location, it may be energetically preferential to have some particles align according $h_i$. To illustrate this, consider the example below:

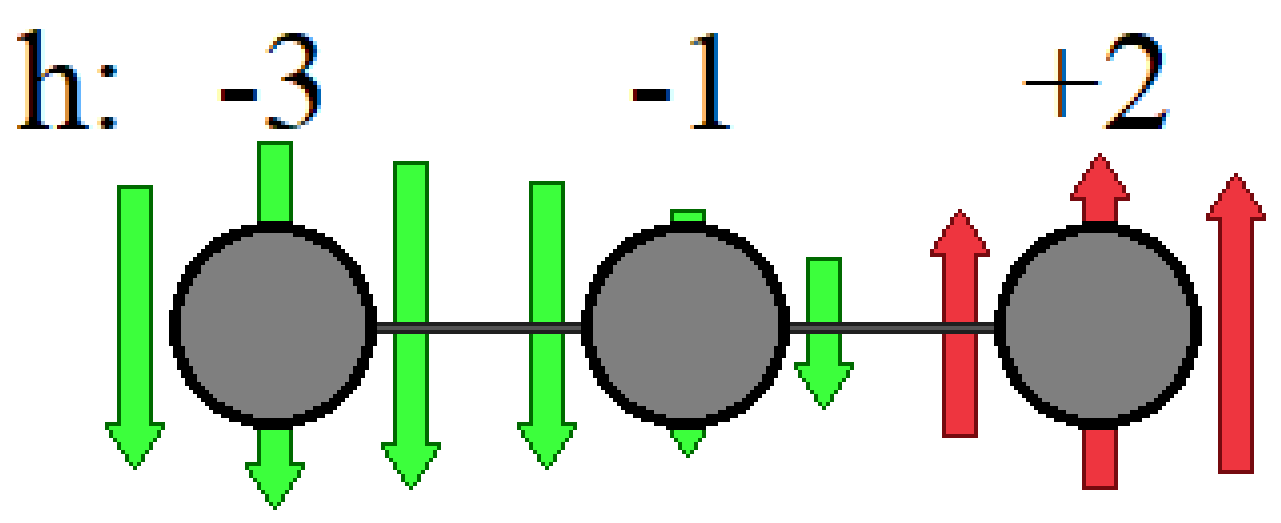

Here we have a system of three particles subject to a magnetic field which is different at each location. If we now try and arrange the spin of each particle to minimize the energy of the system, we must make a choice as to which particle(s) will be anti-aligned with the magnetic field. Below are two possible configurations and their corresponding energies:

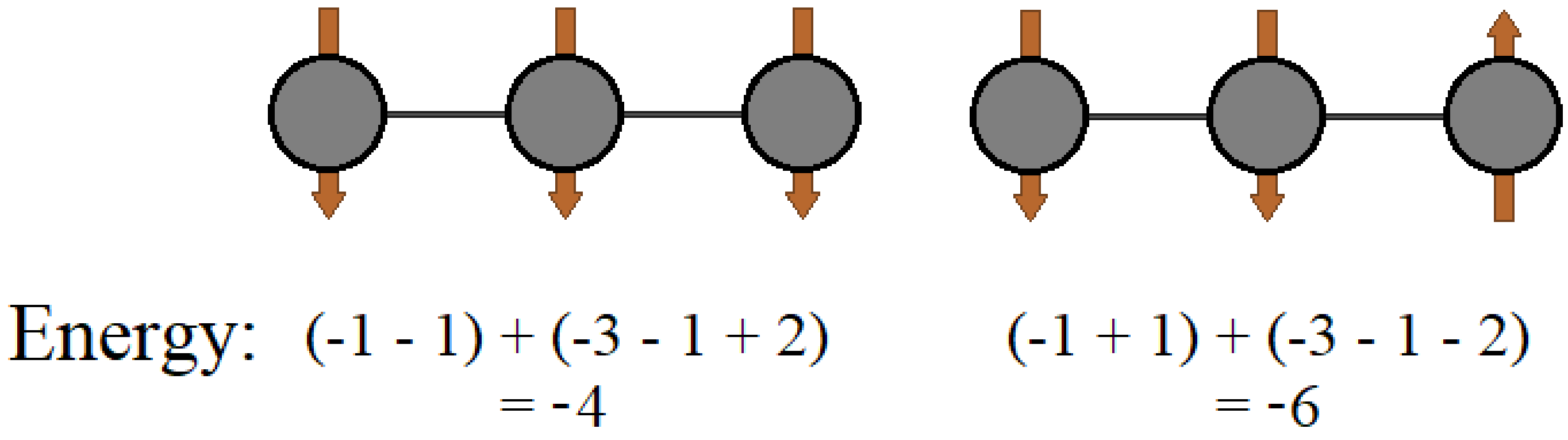

As you can see, the ground state energy of this system does not correspond to simply aligning every particle's spin in the same direction. The breakdown of the energy contributions beneath each orientation show how much of the total energy is coming from $\sum_{\langle ij \rangle} -Z_i Z_j$ versus $\sum_i -h_i Z_i$. For example, in the left orientation we can see that aligning all of the spins maximizes $\sum_{\langle ij \rangle} -Z_i Z_j$ , contributing a total of $-2$ to the energy. However, doing so comes at the cost of misaligning the rightmost particle with the magnetic field, causing a $+2$ energy contribution. Conversely, the right orientation sacrifices the $+1$ energy contribution from having the middle and right particle's opposite spins, instead gaining a $-2$ contribution from the right particle's alignment with the magnetic field, causing an overall lower energy state for the system.

For this particular example, the right orientation turns out to be the optimal combination of spins, making $-6$ the lowest possible energy for the system. In general, if we have an $N$ particle system, then there are $2^N$ possible orientations we must check in order to find the optimal combination of spins. As you might imagine, even this simple model can start to get quite computationally costly once $N$ gets large enough. A quantum computer on the other hand can effectively work with a space of $2^N$ orientations with only $N$ qubits.

Having now seen a simple example, let's put the Ising Energy Model to the test with QAOA. To do this, our first step is to encode our cost function into $U(\mathbf{C}, \gamma)$:

$$
\begin{aligned}
U(\mathbf{C},\gamma) \;=\; e^{-i\gamma C} \;&=\; e^{-i\gamma\left(-\sum_{\langle ij \rangle} Z_i Z_j \,-\, \sum_i h_i Z_i\right)} \\
&=\; e^{i\gamma \sum_{\langle ij \rangle} Z_i Z_j}\, e^{i\gamma \sum_i h_i Z_i} \\
&=\; \prod_{\langle ij \rangle} e^{i\gamma Z_i Z_j} \prod_i e^{i\gamma h_i Z_i}
\end{aligned}
$$

One nice thing about exponentiating our cost function is the way the exponents rule: $e^{A+B} = e^A \cdot e^B$ works in our favor here to separate the different components of the cost function. As shown above, the resulting $U(\mathbf{C},\gamma)$ operator can be implemented as two separate operations, which we will take advantage of when physically implementing $U(\mathbf{C},\gamma)$ in our quantum circuit. But first, let's finish our derivation of $U(\mathbf{C},\gamma)$ by seeing its components in matrix form:

$$
i\gamma Z_i Z_j \;=\; i\gamma \begin{bmatrix} 1 & 0 \\ 0 & -1 \end{bmatrix} \otimes \begin{bmatrix} 1 & 0 \\ 0 & -1 \end{bmatrix} \;=\; \begin{bmatrix} i\gamma & & & \\ & -i\gamma & & \\ & & -i\gamma & \\ & & & i\gamma \end{bmatrix} \qquad i\gamma h_i Z_i \;=\; \begin{bmatrix} i\gamma h & 0 \\ 0 & -i\gamma h \end{bmatrix}
$$

$$
e^{i\gamma Z_i Z_j} \;=\; \begin{bmatrix} e^{i\gamma} & & & \\ & e^{-i\gamma} & & \\ & & e^{-i\gamma} & \\ & & & e^{i\gamma} \end{bmatrix} \qquad e^{i\gamma h_i Z_i} \;=\; \begin{bmatrix} e^{i\gamma h} & 0 \\ 0 & e^{-i\gamma h} \end{bmatrix}
$$

The right matrix should be recognizable as the standard $R_z(\theta)$ gate:

$$
R_z(\theta) \;=\; \begin{bmatrix} e^{i\theta} & 0 \\ 0 & e^{-i\theta} \end{bmatrix}
$$

For implementing the left matrix, we've already seen one technique which allows us to manually apply each diagonal phase term individually. However, the structure of the $e^{i\gamma Z_i Z_j}$ matrix can be implemented in a much simpler way, which can be seen in the cell of code below:

In [57]:

```
q = QuantumRegister(2,name='q')
qc = QuantumCircuit(q,name='qc')
#==============================
qc.h( q[0] )
qc.h( q[1] )
qc.barrier()
print('___ Initial State ___')
oq.Wavefunction( qc )
#------------------------------
qc.cx( q[0], q[1] )
qc.u1( -m.pi/2, q[1] )
qc.cx( q[0], q[1] )
#------------------------------
print('\n___ After e^{ZZ} ___')
oq.Wavefunction( qc )
print( qc )
```

```
___ Initial State ___
0.5 |00>    0.5 |10>    0.5 |01>    0.5 |11>

___ After e^{ZZ} ___
0.5 |00>    -0.5j |10>    -0.5j |01>    0.5 |11>

q_0: |0>┤ H ├─░──■────────────────■──
             ░  │                │
q_1: |0>┤ H ├─░─┤ X ├┤ U1(-pi/2) ├┤ X ├
```

The example above implements an effective $e^{\frac{i\pi}{4}Z_iZ_j}$ gate, but the answer may not quite look like what we would expect based on our previous math derivations. Specifically, this circuit *should* have placed an $e^{\frac{i\pi}{4}}$ phase on the $|00\rangle$ and $|11\rangle$ states, and an $e^{-\frac{i\pi}{4}}$ phase on the $|01\rangle$ and $|10\rangle$ states. Instead, at first glance it looks like we got $e^{-\frac{i\pi}{2}}$ phases on the $|01\rangle$ and $|10\rangle$ states, and nothing on the $|00\rangle$ and $|11\rangle$ states. The trick here however, is that we have indeed achieved the desired state, up to a global phase:

$$
\begin{aligned}
|\Psi\rangle &= \frac{1}{2}(|00\rangle + e^{-\frac{i\pi}{2}}|01\rangle + e^{-\frac{i\pi}{2}}|10\rangle + |11\rangle) \\
&= \frac{1}{2}e^{-\frac{i\pi}{4}}(e^{\frac{i\pi}{4}}|00\rangle + e^{-\frac{i\pi}{4}}|01\rangle + e^{-\frac{i\pi}{4}}|10\rangle + e^{\frac{i\pi}{4}}|11\rangle)
\end{aligned}
$$

As you can see, by doing some rearranging of the phase terms, our circuit above has achieved the $e^{\frac{i\pi}{4}Z_iZ_j}$ operation, with a global phase of $e^{-\frac{i\pi}{4}}$. And as we've seen in past lessons, global phases are undetectable and have no overall impact when we ultimately go to make measurements. To recap then, we now have all of the tools necessary for implementing $U(\mathbf{C},\gamma)$ and $U(\mathbf{B},\beta)$:

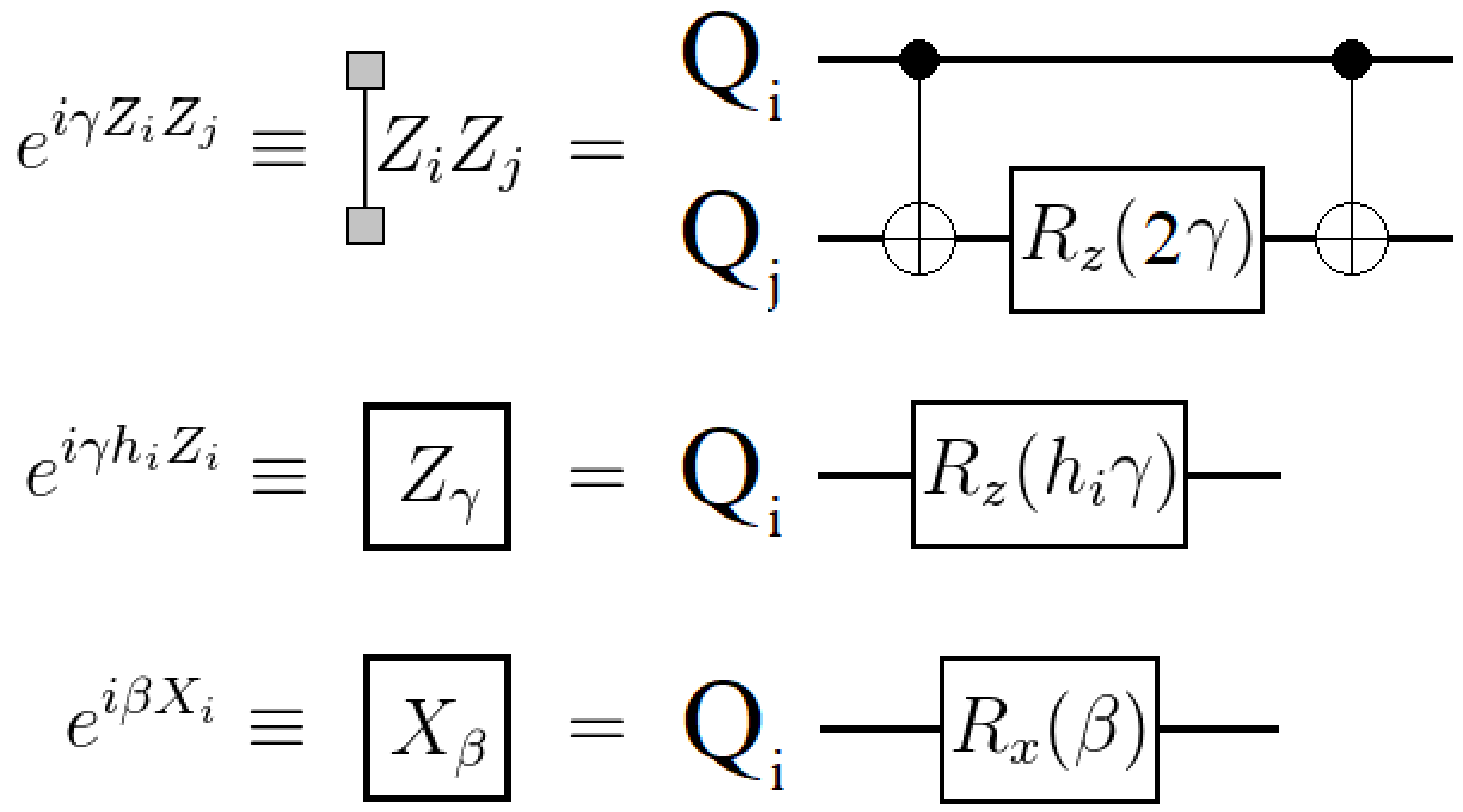

All that is left now is to pick a cost function and build our quantum circuit! As a first pass, we will not be implementing any classical optimization techniques for $\gamma$ and $\beta$, but rather simply look at the full space of these variables. Our code will span the full range of values for each parameter and build up a 2D map of the corresponding $F(\gamma,\beta)$ values. We'll start with a simple three particle system:

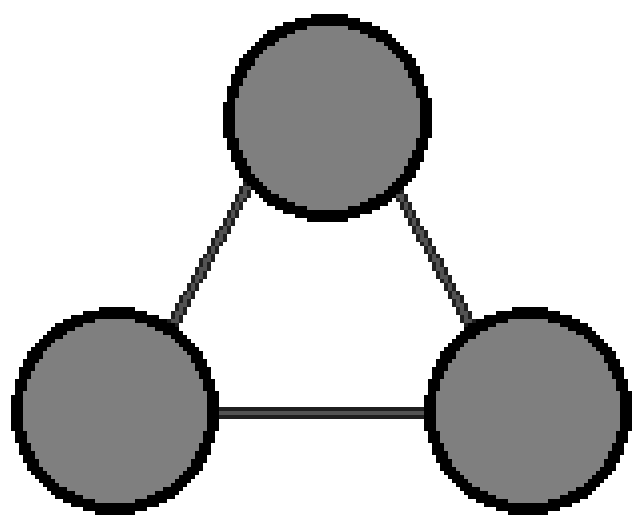

Just like our example from earlier, each particle will be subject to a different magnetic field strength and direction. We will leave these values as changeable parameters in our code so that we can try different combinations quickly. With our geometry picked out, let's take a look at the corresponding phase and mixing operators:

$$
\begin{aligned}
U(\mathbf{C},\gamma) &= \prod_{\langle ij\rangle} e^{i\gamma Z_iZ_j} \prod_i e^{i\gamma h_i Z_i} \\
&= (\, e^{i\gamma Z_1Z_2}\, e^{i\gamma Z_2Z_3}\, e^{i\gamma Z_1Z_3}\,)(\, e^{i\gamma h_1 Z_1}\, e^{i\gamma h_2 Z_2}\, e^{i\gamma h_3 Z_3}\,)
\end{aligned}
\qquad
\begin{aligned}
U(\mathbf{B},\beta) &= \prod_i e^{i\beta X_i} \\
&= e^{i\beta X_1}\, e^{i\beta X_2}\, e^{i\beta X_3}
\end{aligned}
$$

And lastly, the circuit implementation of $U(\mathbf{C},\gamma)$ and $U(\mathbf{B},\beta)$:

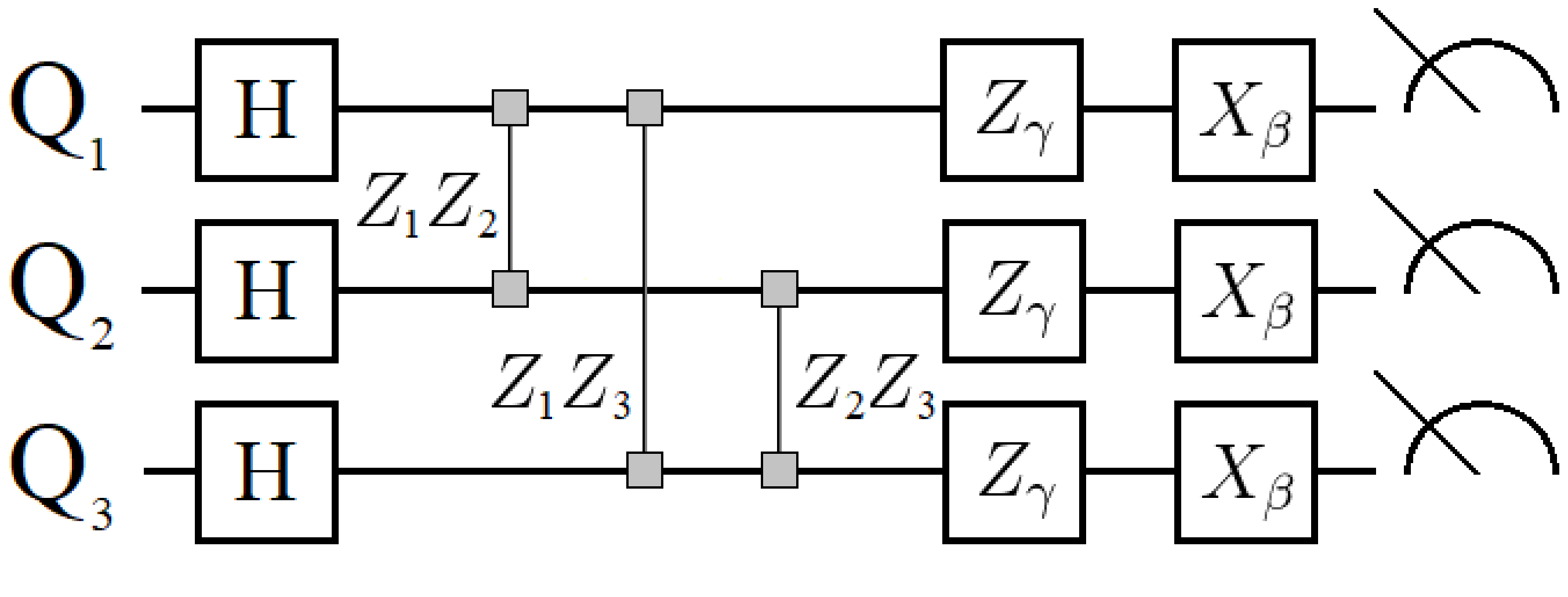

The circuit shown above represents our complete quantum circuit for this problem. We begin by initializing our qubits into the full equal superposition state using Hadamard gates, followed by $U(\mathbf{C},\gamma)$ and $U(\mathbf{B},\beta)$. The $U(\mathbf{C},\gamma)$ operation consists of three $Z_i Z_j$ and $R_z$ gates each, while the $U(\mathbf{B},\beta)$ operator is simply $R_x$ gates on each qubit. Note that in the circuit above we've elected to implement the $Z_i Z_j$ gates first, but switching their order with the $Z_\gamma$ gates works just as well, so long as we apply all of the $U(\mathbf{C},\gamma)$ gates before $U(\mathbf{B},\beta)$. Now then, let's build the circuit above and run it, picking a $\gamma$ and $\beta$ value at random:

In [5]:
```
gamma = 0.8
beta = 1.6
B = [-2.5,3.25,2.25]
#====================================
q = QuantumRegister(3,name='q')
c = ClassicalRegister(3,name='c')
qc = QuantumCircuit(q,c,name='qc')
#------------------------------------
for i in np.arange(3):
    qc.h( q[int(i)] )
#----------------------        # Z1Z2
qc.cx( q[0], q[1] )
qc.u1( 2*gamma, q[1] )
qc.cx( q[0], q[1] )
#----------------------        # Z1Z3
qc.cx( q[0], q[2] )
qc.u1( 2*gamma, q[2] )
qc.cx( q[0], q[2] )
#----------------------        # Z2Z3
qc.cx( q[1], q[2] )
qc.u1( 2*gamma, q[2] )
qc.cx( q[1], q[2] )
#----------------------        # Z_gamma gates
for j in np.arange(3):
    qc.u1( gamma*B[j], q )
#----------------------        # X_beta gates
for k in np.arange(3):
    qc.rx( beta, q )
#------------------------------------
qc.measure( q,c )
oq.Measurement( qc, shots = 1 )
```

```
1|111>
```

Running the cell of code above represents steps $1$ - $3$ in our workflow outline from earlier. As you can see, here we've elected to perform only a single measurement, which we would then use in step $4$ to check $\mathbf{C}(\mathbf{z})$. After obtaining the corresponding energy, the next step would be to repeat the process. However, since this is our first example, and the size of the problem is small enough, let's go ahead and simply compute all of the $\mathbf{C}(\mathbf{z})$ values so we know which state is the correct final answer:

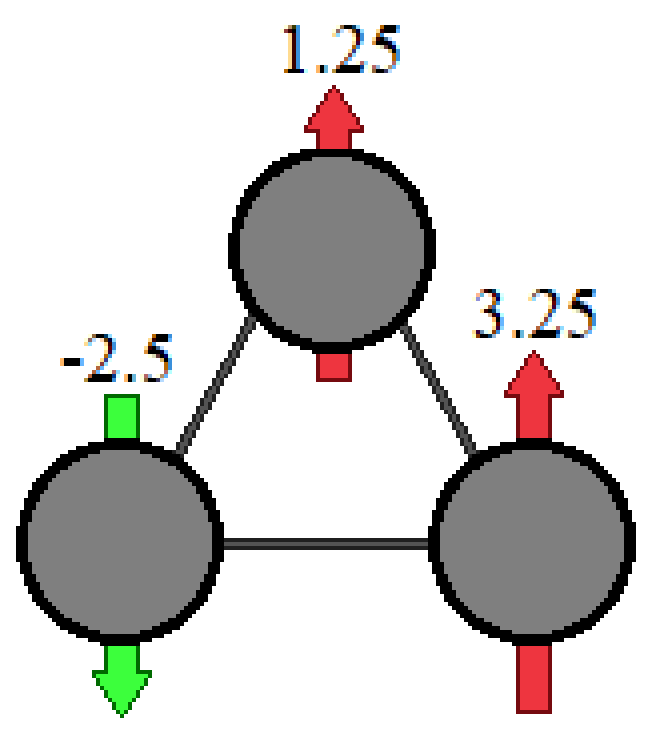

| $\mathbf{z}$ | $\mathbf{C}(\mathbf{z})$ |
|---|---|
| $\|000\rangle$ | -5.0 |
| $\|001\rangle$ | 1.5 |
| $\|010\rangle$ | 5.5 |
| $\|011\rangle$ | 8.0 |
| $\|100\rangle$ | -6.0 |
| $\|101\rangle$ | -3.5 |
| $\|110\rangle$ | 0.5 |
| $\|111\rangle$ | -1.0 |

In the calculations shown above, the state $|\,0\,\rangle$ corresponds to spin up, and similarly $|\,1\,\rangle$ for spin down. As we can see in the table of energies, the state $|\,100\rangle$ represents the minimal energy of our cost function for this example, with $|000\rangle$ as a close second. With these energies in mind then, the goal of QAOA is to find a combination of $\gamma$ and $\beta$ such that we end up with a state $|\,\Psi_{\gamma\beta}\rangle$ that has a high probability of measuring these states, preferably $|\,100\rangle$ being the highest probability. Since we picked our $\gamma$ and $\beta$ in the cell of code above essentially at random, let's now see the corresponding distribution from our $|\,\Psi_{\gamma\beta}\rangle$:

In [18]:
```
trials = 10000
M = oq.Measurement(qc, shots = trials, return_M=True)
```

```
931|010>    3547|111>    250|101>    248|011>    979|100>    211|110>    941|001>    2893|000>
```

Looking at the number of measurement counts in the cell above, it appears that our most probable state is $|\,111\rangle$, with $|000\rangle$ as a close second. Looking at the list of $\mathbf{C}(\mathbf{z})$ values above, it should be clear that this state isn't quite what we were looking for. Nevertheless, let's go ahead and use these measurement counts to compute $F(\gamma,\beta)$, the energy expectation value for our $|\,\Psi_{\gamma\beta}\rangle$ state:

In [19]:
```
K = list(M.keys())
Cz = { '000':-5.0, '001':1.5, '010':5.5, '011':8.0, '100':-6.0, '101':-3.5, '110':0.5, '111':-1.0 }
#---------------------------
F = 0
for k in np.arange( len(K) ):
    F = F + (M[K[k]]*Cz[K[k]])/trials
print('\u03B3 = ',gamma,'   \u03B2 = ',beta,'    Expectation Value: ',round(F,3))
```

```
γ =  0.8    β =  1.6     Expectation Value:  -1.614
```

Since we know that our optimal energy is $-6.0$, an expectation value around $-1.6$ tells us that we are quite a bit off from our optimal $\gamma$ and $\beta$. According to step 6 in our workflow then, our next task would be to use some classical optimization technique for choosing the next set of variables $\gamma$ and $\beta$ to test, repeating this process until we come to a satisfactory approximation. As a side note however, this example is clearly small enough such that we've already tested all of the possible combinations for $\mathbf{C}(\mathbf{z})$. That is to say, in the process of computing $F(\gamma,\beta)$ above, our quantum system has already measured all of the possible states producible from just three qubits, which in turn means that along the way we would have found our optimal $\mathbf{C}(\mathbf{z}')$, and concluded it is indeed optimal by process of elimination. In a more realistic QAOA however, the space of all possible states is typically so large that there's no guarantee that the optimal state $|\,z'\,\rangle$ will even show up in our initial measurement distribution for computing $F(\gamma,\beta)$. Thus, it may take numerous steps of improving $\gamma$ and $\beta$ before the probability of measuring $|\,z'\,\rangle$ becomes large enough to start becoming reliably measurable.

Improving $\gamma$ and $\beta$ through optimization techniques will be a topic later in this lesson, but for now we will take advantage of our example's small size to find the optimal parameters $\gamma$ and $\beta$ another way. As a learning demonstration, we are going to search the entire space of $\gamma$ and $\beta$ values, and find our optimal pair from there. Obviously this is a bit of an overkill, but the real merit to this exercise will be the visualization of our $[\gamma,\beta]$ space, which in turn will help motivate the 'Gradient Descent' technique to come. In order to clean up our code little bit, we will call upon the following functions from $\text{Our_Qiskit_Functions}$ to assist us in applying the necessary quantum gates / computations:

**Ising_Circuit** : Applies all the necessary gates for our quantum circuit using the Ising Energy model

**Ising_Energy** : Computes the cost function energy for each state of our geometry

**E_Expectation_Value** : Computes the energy expectation value of a state

**Heatmap** : Displays 2D heat map showing $F(\gamma,\beta)$ as a function of $\gamma$ and $\beta$

In [7]:
```
size = 100                                                  # Number of gamma and beta values to inspect
Vert = [ [0,-2.5] , [1,3.25] , [2,1.25] ]                   # Information about each qubit / magnetic field strength
Edge = [ [0,1],[0,2],[1,2] ]                                # Information about connections between particles
#--------------------------------------------------------
Energies,States = oq.Ising_Energy( Vert,Edge )
EV_grid = np.zeros(shape=(size,size))
EV_min = 10000
#=========================================================
for b in np.arange(size):
    beta = round(2*m.pi*(b/size),4)
    for g in np.arange(size):
        gamma = round(2*m.pi*(g/size),4)
        q = QuantumRegister(len(Vert))
        qc= QuantumCircuit(q)
        for hh in np.arange(len(Vert)):
            qc.h( q[int(hh)] )
        oq.Ising_Circuit( qc, q, Vert, Edge, beta, gamma  )
        EV = oq.E_Expectation_Value( qc, Energies )
        EV_grid[b,g] = EV
        if( EV < EV_min ):
            Params = [beta,gamma]
            EV_min = EV
print('Optimal Energy Expectation Value:  ',EV_min,'    \u03B3 = ',Params[1],'   \u03B2 = ',Params[0],'\n')
#=========================================================
fig, ax = plt.subplots()
show_text = False
show_ticks = False
oq.Heatmap(EV_grid, show_text, show_ticks, ax, "plasma", "Energy Expectation Value")
fig.tight_layout()
plt.show()
```

```
Optimal Energy Expectation Value:   -5.0     γ =  1.5708    β =  1.5708
```

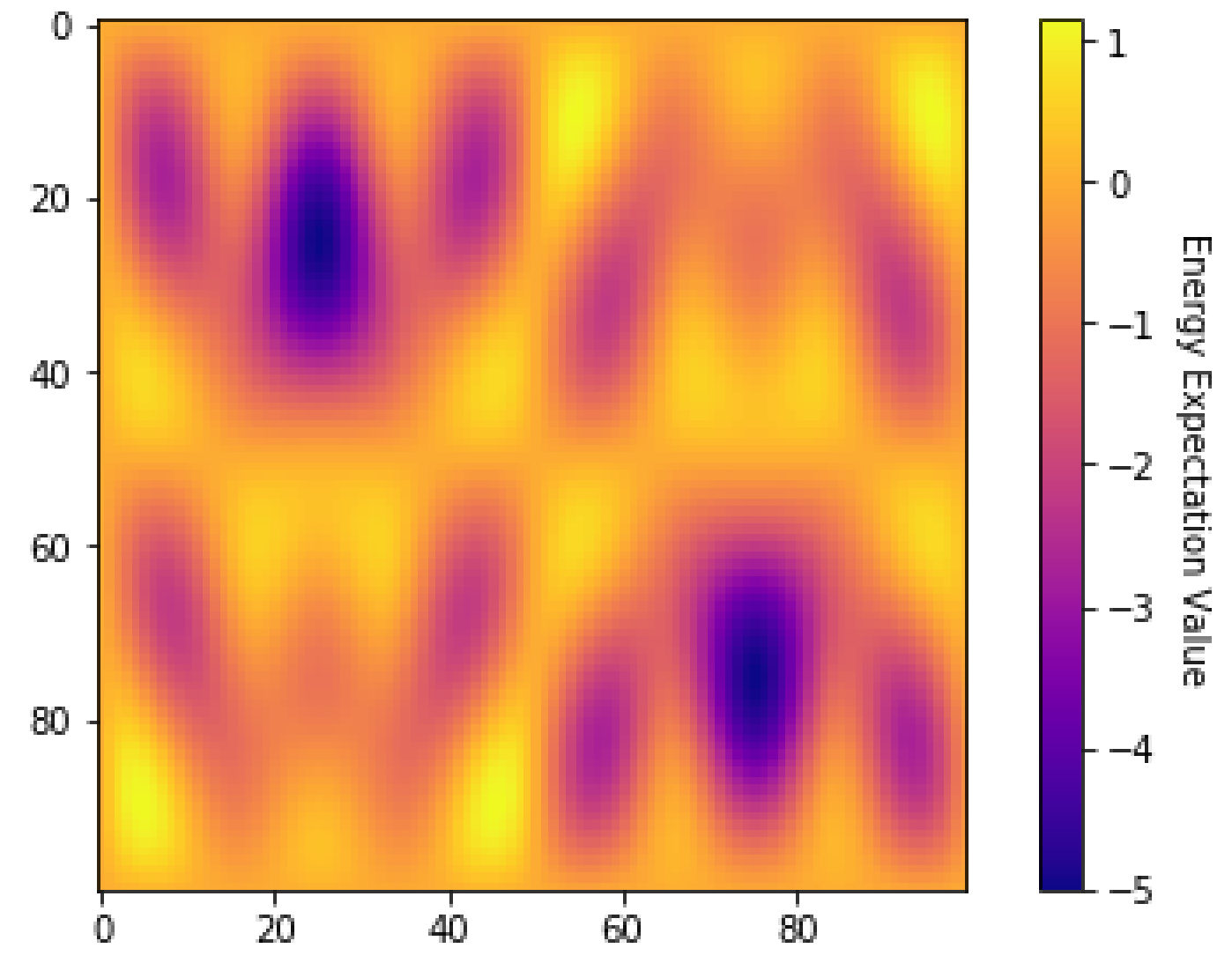

Shown above is a heatmap displaying all of the expectation values for the full range of $\gamma$ and $\beta$ values, spanning from $0$ to $2\pi$. As it turns out, our optimal $\gamma$ and $\beta$ values are $[\pi/2, \pi/2]$, which correspond to the center of the dark blue shape in the upper left corner. There are several new insights that this plot will provide us, but first let's take a look at what kind of state is produced from these optimal values:

In [8]:
```
beta = Params[0]                              # Params is required from the previous cell of code
gamma = Params[1]
Vert = [ [0,-2.5] , [1,3.25] , [2,1.25] ]
Edge = [ [0,1],[0,2],[1,2] ]
Energies,States = oq.Ising_Energy( Vert,Edge )
#=========================================================
q = QuantumRegister(len(Vert))
qc= QuantumCircuit(q)
for hh in np.arange(len(Vert)):
    qc.h( q[int(hh)] )
oq.Ising_Circuit( qc, q, Vert, Edge, beta, gamma  )
#==========================================================
print('Optimal Energy Expectation Value:  ',EV_min,'    \u03B3 = ',Params[1],'   \u03B2 = ',Params[0],'\n')
SV = execute( qc, S_simulator, shots=1 ).result().get_statevector()
oq.Top_States(States,Energies,SV,8)
```

```
Optimal Energy Expectation Value:   -5.0     γ =  1.5708    β =  1.5708

State  000      Probability:  100.0 %     Energy:  -5.0
State  110      Probability:  0.0 %     Energy:  0.5
State  011      Probability:  0.0 %     Energy:  8.0
State  101      Probability:  0.0 %     Energy:  -3.5
State  100      Probability:  0.0 %     Energy:  -6.0
State  001      Probability:  0.0 %     Energy:  1.5
State  010      Probability:  0.0 %     Energy:  5.5
State  111      Probability:  0.0 %     Energy:  -1.0
```

Perhaps not what we were expecting from our optimal $\gamma$ and $\beta$, but it certainly explains why our minimal $F(\gamma,\beta)$ turned out to be exactly $-5.0$. The quantum state produced from this particular combination of $\gamma$ and $\beta$ turns out to be simply $|000\rangle$. Despite being the optimal $F(\gamma,\beta)$, it is clear that this combination of $\gamma$ and $\beta$ will not solve our problem (the optimal state $|z'\rangle$ has an amplitude of zero). Thus, this example has highlighted an interesting, and seemingly counterintuitive problem: non-optimal combinations of $\gamma$ and $\beta$ will produce $|\Psi_{\gamma\beta}\rangle$ states which have non-zero probabilities for $|100\rangle$, but consequently their $F(\gamma,\beta)$ will be larger than the optimal $\gamma$ and $\beta$ pair.

## Expanding the $\gamma$, $\beta$ Space

The reason why our first QAOA attempt failed to find the correct optimal energy state can be summarized as follows: the state $|\Psi_{\gamma\beta}\rangle = |100\rangle$ does not exist within our parameter space $[\gamma, \beta]$. If it did, there would have been a combination of $\gamma$ and $\beta$ which produced an energy expectation value of $-6.0$ in our heatmap shown above. Similarly, if we take a look at the accompanying gradient scale to the right of the heatmap, we find that the limits are from $[1, -5]$. Three of the possible eight basis states correspond to energies greater than $1$, which means that these $|\Psi_{\gamma\beta}\rangle$ states do not exist within our parameter space either. Thus, in order to solve the underlying problem plaguing our QAOA, we must investigate $why$ our circuit is unable to reach these quantum states.

### Exploring a Complete Hilbert Space Through Mixing Operators

In analyzing our QAOA circuit, and the root of the problem whereby our $|\Psi_{\gamma\beta}\rangle$ states couldn't reach $|100\rangle$, the only two components of our algorithm which are candidates for improvement are $U(\mathbf{C},\gamma)$ and $U(\mathbf{B},\beta)$. However, when we break down the roles of two operators, $U(\mathbf{C},\gamma)$ is the operation responsible for encoding our cost function, and as we pointed out earlier, merely applies phases to each state in the system. Conversely, the mixing operator $U(\mathbf{B},\beta)$ was introduced specifically to cause constructive and destructive interference. Thus, we can conclude that the reason why our QAOA example above failed is because our mixing operator was insufficient, which we will now look to improve upon.

In order to fully appreciate why our mixing operator is in need of improvement, and how we might look to construct a more efficient one to solve our $3$-qubit QAOA example, we can start by looking at the case of only a single qubit. The question we need to answer is how to construct a quantum circuit which can reach $all$ possible quantum states that exist within a single qubit's Hilbert Space. Mathematically, this corresponds to all possible states of the form:

$$|\Psi\rangle \;=\; \alpha|0\rangle \;+\; \beta|1\rangle$$

for which we have a nice visual answer, the Bloch Sphere:

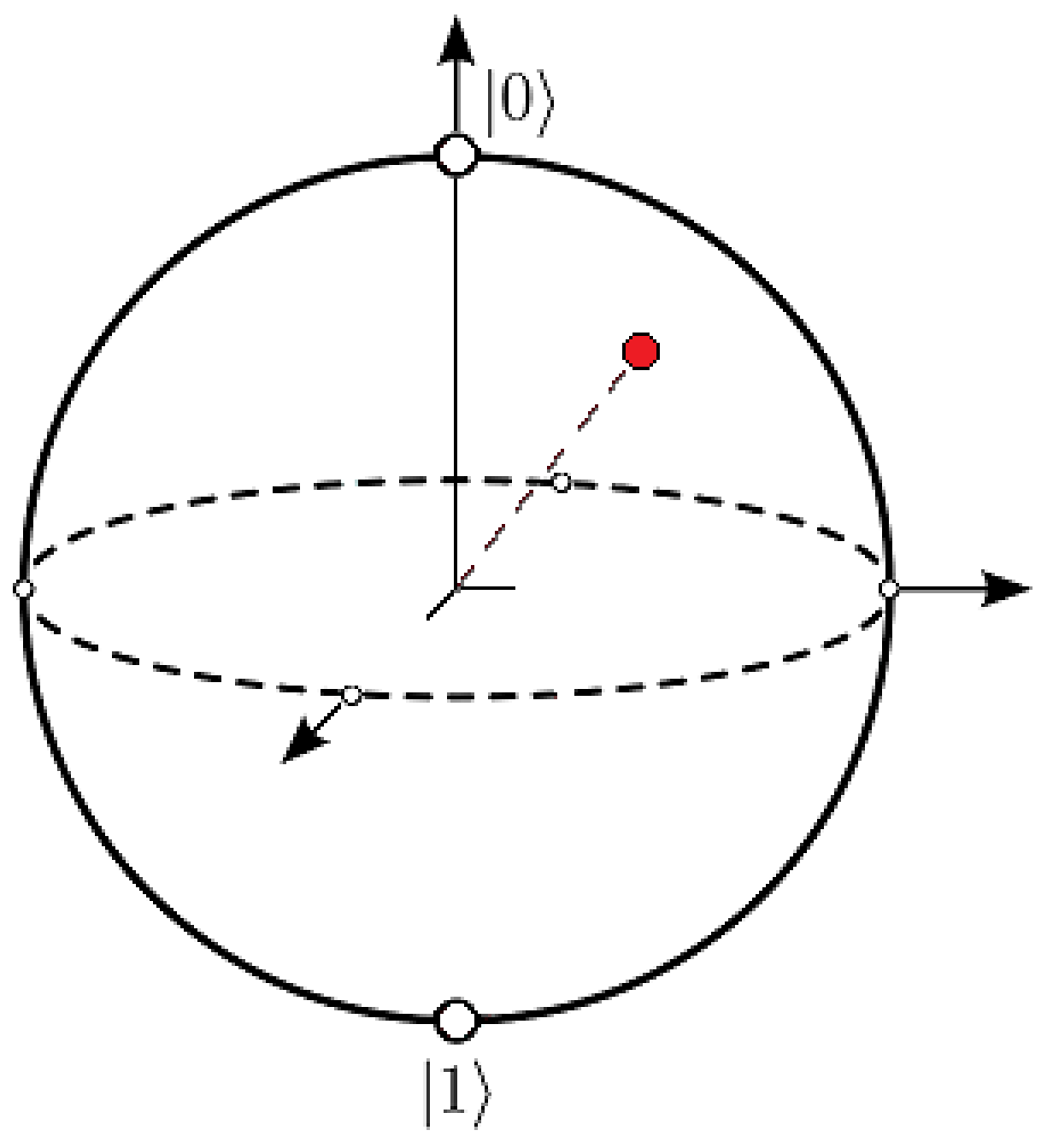

The surface of the Bloch Sphere represents the full Hilbert Space of a single qubit. Thus, in order to span the full $1$-qubit space, all we need are the gates $R_x(\theta)$ and $R_y(\theta)$ (or any two of the rotational operators), which together can rotate a qubit starting in the $|0\rangle$ state to any location along the surface (but not necessarily between any two arbitrary points). Take special note here that the full $1$-qubit space requires *two* different rotation gates, whereas our mixing operator up to this point only consists of one, $R_x(\beta)$. Thus, we have already found area of improvement for our mixing operator: adding the $R_y$ and $R_z$ operators.

While adding more rotation gates to our $U(\mathbf{B},\beta)$ operator will improve our ability to create a wider range of $|\Psi_{\gamma\beta}\rangle$ states, it's not the full story. To see the other critical missing ingredient, we must look to the $2$-qubit case and once again ask what gates are required to span the full Hilbert Space. Unlike the $1$-qubit case however, single qubit rotation gates alone are not enough, as demonstrated by the example below:

In [33]:

```
T = 8
Z = [m.sqrt(0.5),0,0,m.sqrt(0.5)]
Closest_IP = 0
#====================================
for i1 in np.arange(T+1):
    t1 =  i1*m.pi/T
    for i2 in np.arange(T+1):
        t2 =  i2*m.pi/T
        for i3 in np.arange(T+1):
            t3 =  i3*m.pi/T
            for i4 in np.arange(T+1):
                t4 =  i4*m.pi/T
                #---------------------
                q = QuantumRegister(2)
                qc= QuantumCircuit(q)
                qc.rx( t1, q[0] )
                qc.rx( t2, q[1] )
                qc.ry( t3, q[0] )
                qc.ry( t4, q[1] )
                SV = execute( qc, S_simulator, shots=1 ).result().get_statevector()
                IP = (SV[0]*Z[0]) + (SV[1]*Z[1]) + (SV[2]*Z[2]) + (SV[3]*Z[3])
                if( IP > Closest_IP ):
                    Closest_IP = IP

print( 'Largest Inner Product Overlap with the  |00> + |11>  state:  ',round(np.real(Closest_IP),4 ))
```

```
Largest Inner Product Overlap with the  |00> + |11>  state:   0.7071
```

The cell of code above represents the following quantum circuit:

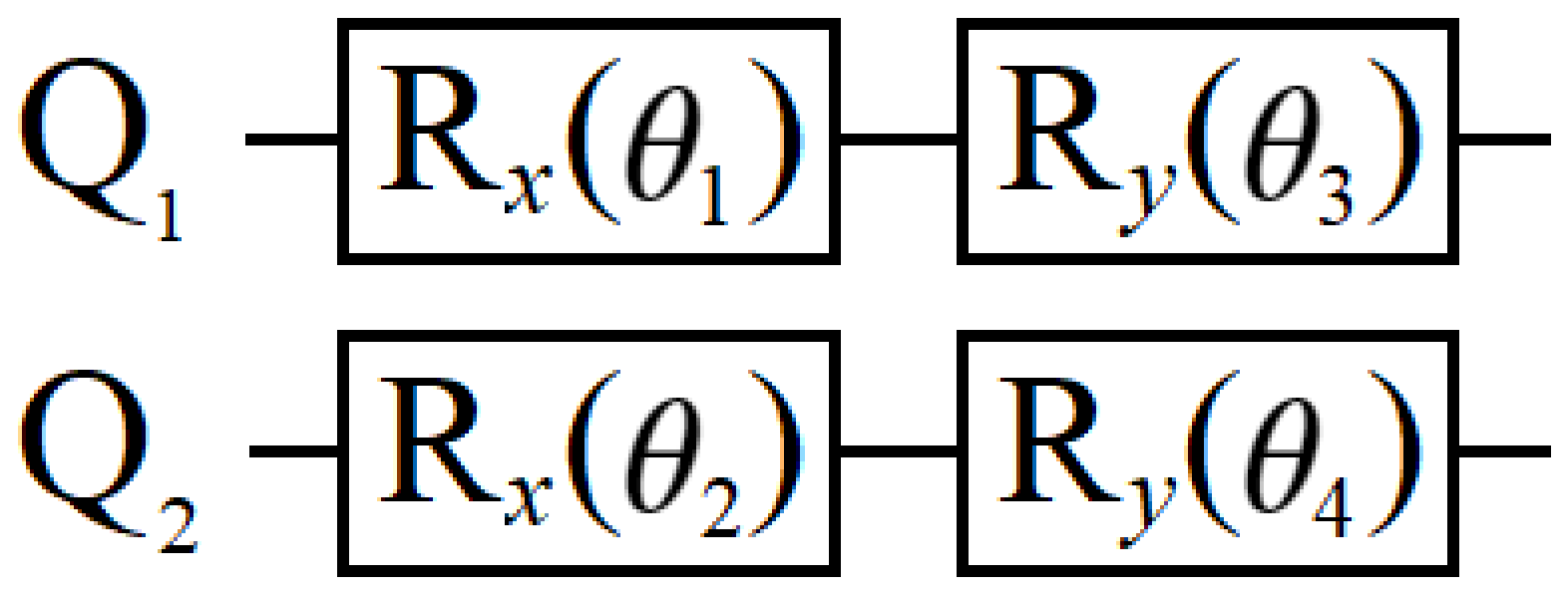

where the goal of the code is to find a combination of $\theta$'s which will produce the quantum state:

$$|\Psi\rangle \;=\; \frac{1}{\sqrt{}}\left(\,|00\rangle \;+\; |11\rangle\,\right)$$

In trying to find the state $|\Psi\rangle$ shown above, this example uses the two rotation gates $R_x$ and $R_y$ to fully explore the surface of each qubit's Bloch Sphere individually. Since we just showed that these two gates are sufficient for a single qubit's full Hilbert Space, applying them to the $2$-qubit case would be a natural first guess. However, as shown by the final result of the code, our quantum circuit is unable to create a state which produces and inner product of $1$ with $|\Psi\rangle$ (the only state that could do this is $|\Psi\rangle$ itself). Regardless of how many values of $\theta$ we check for the four rotation gates, the circuit above will never be able to produce $|\Psi\rangle$. Thus, spanning the complete Hilbert Space for each individual qubit is not enough, which means that our quantum circuit needs something extra in order to reach all states within the combined $2$-qubit Hilbert Space.

In short, the missing ingredient to our quantum circuit above is entanglement. While the rotation gates are sufficient in exploring all possible states for a single qubit, when applied to higher dimensional cases they lead to only separable states:

$$|\Psi_{\theta_1\theta_2}\rangle \;\equiv\; R_y(\theta_2)R_x(\theta_1)\,|0\rangle$$

$$|0\rangle \otimes |0\rangle \;\longrightarrow\; |\Psi_{\theta_1\theta_3}\rangle \otimes |\Psi_{\theta_2\theta_4}\rangle$$

By contrast, the state in the example above is non-separable, which explains why there is no amount of single qubit rotation gates which will be able to produce $|\Psi\rangle$. Thus, in order to explore the full $2$-qubit Hilbert Space, we must include at least one entangling operation, i.e. a control gate. Since any control operation will do, the usual choice is the CNOT gate, mainly due to its typically higher gate fidelity than other control operations. Let's now verify that adding a CNOT gate to our circuit will expand our reachable quantum state space:

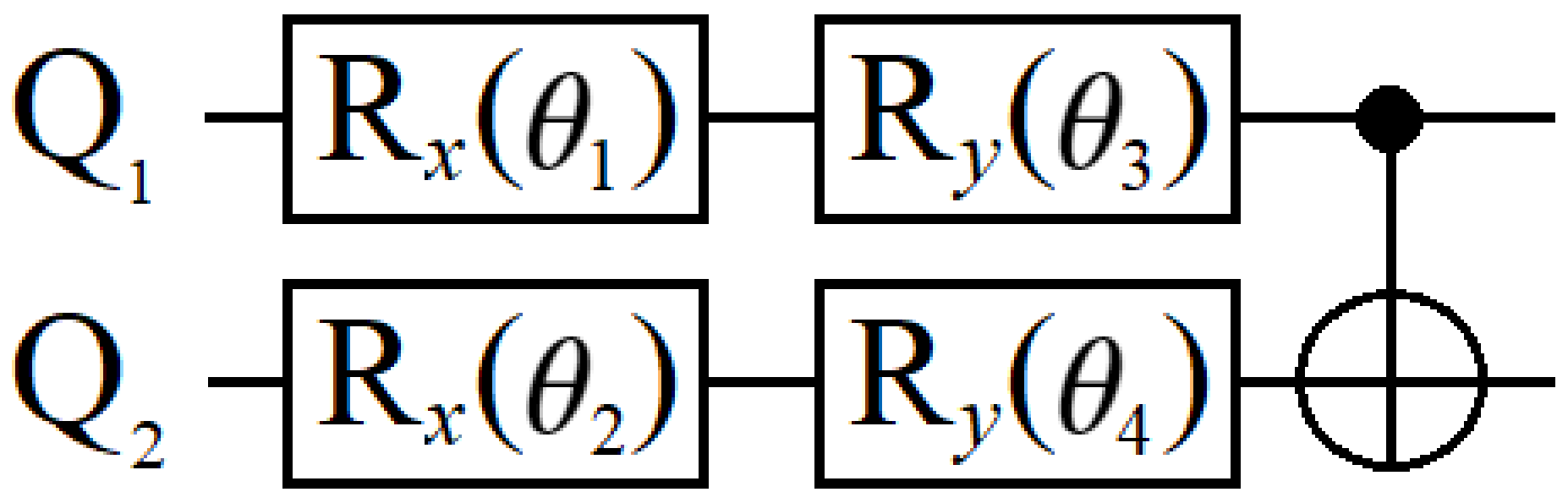

In [2]:

```
T = 8
Z = [m.sqrt(0.5),0,0,m.sqrt(0.5)]
Closest_IP = 0
#====================================
for i1 in np.arange(T+1):
    t1 =  i1*m.pi/T
    for i2 in np.arange(T+1):
        t2 =  i2*m.pi/T
        for i3 in np.arange(T+1):
            t3 =  i3*m.pi/T
            for i4 in np.arange(T+1):
                t4 =  i4*m.pi/T
                #---------------------
                q = QuantumRegister(2)
                qc= QuantumCircuit(q)
                qc.rx( t1, q[0] )
                qc.rx( t2, q[1] )
                qc.ry( t3, q[0] )
                qc.ry( t4, q[1] )
                qc.cx( q[0], q[1] )
                SV = execute( qc, S_simulator, shots=1 ).result().get_statevector()
                IP = (SV[0]*Z[0]) + (SV[1]*Z[1]) + (SV[2]*Z[2]) + (SV[3]*Z[3])
                if( IP > Closest_IP ):
                    Closest_IP = IP
                    Optimal_Thetas = [t1,t2,t3,t4]

print( 'Largest Inner Product Overlap with |\u03A8>:  ',round(np.real(Closest_IP),4 ))
```

```
Largest Inner Product Overlap with |Ψ>:   1.0
```

Success! By incorporating a single CNOT gate into our quantum circuit we are able to create previously unreachable quantum states, which in turn produce larger inner product overlaps with our desired $|\Psi\rangle$. Even better still, our quantum circuit was able to produce $|\Psi\rangle$ exactly, which for this example is achievable with the combination of $\theta$'s: $\left[0, 0, \frac{\pi}{2}, 0\right]$. For our $2$-qubit case here, a single CNOT gate is enough to create a fully entangled system, whereby no qubits are separable (except for certain combinations of $\theta$'s). Similarly, a system of $N$ qubits can reach a fully non-separable state with a minimum of $N-1$ CNOT gates. As a side note, here we are describing systems which are "fully" entangled (no qubits or groupings of qubits' states can be separated by a tensor product with the rest of the system), versus "maximally" entangled which has a very different meaning in quantum computing.

Returning now to our original motivation, exploring the full $2$-qubit Hilbert Space, it is still unclear whether the incorporation of a CNOT gate in combination with single qubit rotation gates is enough. Although our circuit example above was able to produce $|\Psi\rangle$, it turns out that there are still sections of the $2$-qubit Hilbert Space which it cannot reach. For example, trying putting the following state into the cell of code above, and you will once again find the largest inner product overlap to be $\frac{1}{\sqrt{2}}$, regardless of how many combinations of $\theta$ are searched:

$$|\Psi\rangle \;=\; \frac{1}{\sqrt{2}}\Big(\, -i|00\rangle \;+\; |11\rangle\,\Big)$$

Even though our quantum circuit is unable to produce the state shown above, the reason for its failure isn't because we are missing yet another fundamental ingredient, but because our circuit is simply too short. More specifically, single qubit rotation gates and CNOT *are* sufficient for exploring the full $2$-qubit Hilbert Space, all one has to do is assign enough of them in the correct order, with the necessary angles. However, for our purposes of QAOA, using a mixing operator $U(\mathbf{B},\beta)$ that is guaranteed to span a full $N$-dimensional Hilbert Space is an overkill. At the end of the day, our goal is to find the computational basis state $|\,z'\,\rangle$ that solves our optimization problem. Thus, we only really need a mixing operator which is 'good enough' to lead our search through the space of $\gamma$ and $\beta$ such that the state $|\,z'\,\rangle$ becomes increasingly probable.

### Higher Order QAOA Spaces: $p$

In light of the discussion above, hopefully it is clear that the true role of our mixing operator $U(\mathbf{B},\beta)$ is to reach as large of a Hilbert Space as necessary for solving our optimization problem. In achieving this goal however, the typical QAOA approach is to have a mixing operator which is only dependent on a single free parameter: $\beta$. But as we saw in the $2$-qubit example above, our quantum circuit of four free parameters wasn't even enough to explore the full Hilbert Space. The motivation for simplifying the mixing operator down to a single free parameter stems from the fact that ultimately QAOA requires a classical optimization technique for $\gamma$ and $\beta$. More free parameters may improve our ability to find a quantum state with a highly probable $|\,z'\,\rangle$ component, but it comes at the cost of slowing down the algorithm as a whole.

Despite our desire to keep things to a minimum, sometimes the complexity of a problem demands more than just the two parameters $\gamma$ and $\beta$. For cases such as these, rather than altering the mixing operator to incorporate more free parameters, the standard QAOA approach is to simply invoke more applications of $U(\mathbf{C},\gamma)$ and $U(\mathbf{B},\beta)$ in succession. This effectively increases the space of our search, which in turn will hopefully lead to a better final approximation. By applying $p$ rounds of our phase and mixing operators, we can create $|\Psi_{\gamma\beta}\rangle$ states which were previously unobtainable with only a single application of $U(\mathbf{C},\gamma)$ and $U(\mathbf{B},\beta)$:

$$|\Psi_{\gamma\beta}\rangle \;=\; U(\mathbf{B},\beta_p)\,U(\mathbf{C},\gamma_p)\ldots U(\mathbf{B},\beta_2)\,U(\mathbf{C},\gamma_2)\,U(\mathbf{B},\beta_1)\,U(\mathbf{C},\gamma_1)\,|\,s\rangle$$

The number of phase and mixing operators is denoted by the parameter $p$, which in turn gives us a $2p$-dimensional space of parameter values to search through. While increasing our search space opens up the opportunity for finding better $|\Psi_{\gamma\beta}\rangle$ states, as we already pointed out, it also comes at the cost of more states for our classical optimization techniques to handle. The tradeoff between choosing the smallest $p$ versus increased algorithm accuracy is something that needs to be considered on a case-to-case basis. Additionally, if a particular mixing operator isn't producing satisfactory results, we need to consider whether increasing $p$ will remedy the problem, or if a new $U(\mathbf{B},\beta)$ altogether is necessary. For example, if we return to our code for solving the Ising Energy problem from the end of the previous section, it turns out that increasing $p$ from 1 to 2 has no effect in producing a better final state (still reach a global minimum of $-5$). Thus, if we want to increase our $|\Psi_{\gamma\beta}\rangle$ space, we need to fundamentally change our mixing operator.

It is important to note that in many problems increasing $p$ *will* lead to better results, the unfortunate issue is that oftentimes there is no way of knowing beforehand. Classically, optimizing $2$ versus $2p$ variables can have dramatic runtime costs, which is why we must be cautious in choosing the order $p$. Using QAOA to solve optimization problems is a very case by case technique, which can be viewed as both a good and bad thing. On the one hand, it would be nice if we had mixing operators which were guaranteed to always solve certain optimization problems, but in reality the strength of QAOA lies in its flexibility. Because we essentially have limitless possibilities in the way in which we can construct mixing operators, QAOA can be adapted to solve a much larger breadth of problems than typical quantum algorithms.

To complete our discussion before moving on to the next topic, we will now look to upgrade our mixing operator for the Ising Energy example. Below is the quantum circuit for our new $U(\mathbf{B},\beta)$ operator, which once again only depends on a single parameter, but now incorporates CNOT gates as well as additional rotational operators:

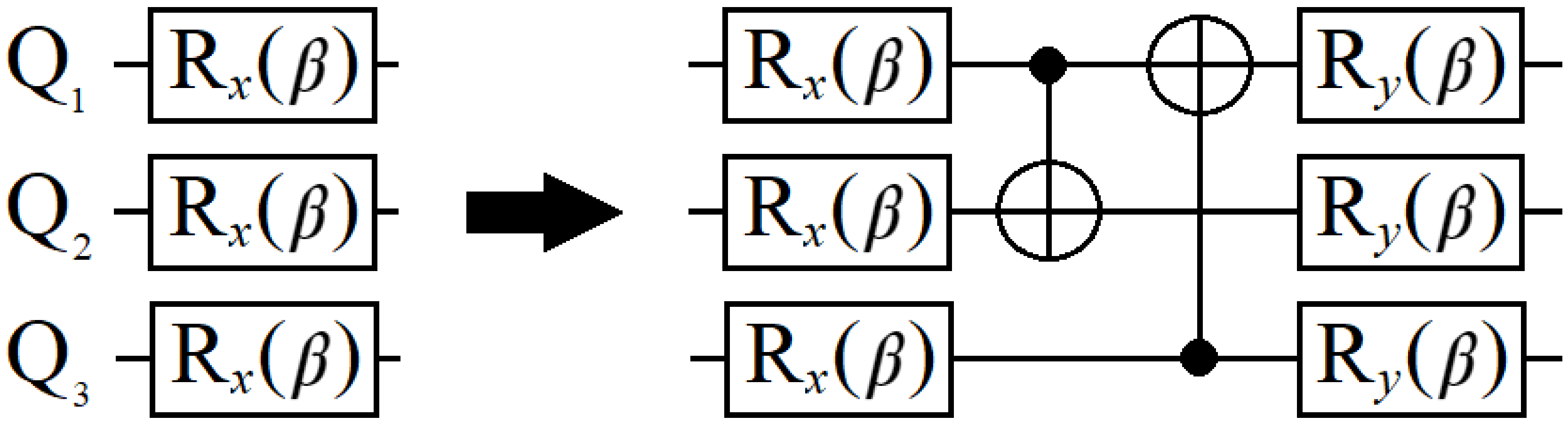

In [3]:

```
size = 100                                                # Number of gamma and beta values to inspect
Vert = [ [0,-2.5] , [1,3.25] , [2,1.25] ]                 # Information about each qubit / magnetic field strength
Edge = [ [0,1],[0,2],[1,2] ]                              # Information about connections between particles
#--------------------------------------------------------
Energies,States = oq.Ising_Energy( Vert,Edge )
EV_grid = np.zeros(shape=(size,size))
EV_min = 10000
#========================================================
for b in np.arange(size):
    beta = round(2*m.pi*(b/size),4)
    for g in np.arange(size):
        gamma = round(2*m.pi*(g/size),4)
        q = QuantumRegister(len(Vert))
        qc= QuantumCircuit(q)
        for hh in np.arange(len(Vert)):
            qc.h( q[int(hh)] )
        oq.Ising_Circuit( qc, q, Vert, Edge, beta, gamma, Mixing=2  )
        EV = oq.E_Expectation_Value( qc, Energies )
        EV_grid[b,g] = EV
        if( EV < EV_min ):
            Params = [beta,gamma]
            EV_min = EV
print('Optimal Energy Expectation Value:  ',EV_min,'    \u03B3 = ',Params[1],'   \u03B2 = ',Params[0],'\n')
#========================================================
fig, ax = plt.subplots()
show_text = False
show_ticks = False
oq.Heatmap(EV_grid, show_text, show_ticks, ax, "plasma", "Energy Expectation Value")
fig.tight_layout()
plt.show()
#========================================================

beta = Params[0]
gamma = Params[1]
q = QuantumRegister(len(Vert))
qc= QuantumCircuit(q)
for hh in np.arange(len(Vert)):
    qc.h( q[int(hh)] )
oq.Ising_Circuit( qc, q, Vert, Edge, beta, gamma, Mixing=2  )
SV = execute( qc, S_simulator, shots=1 ).result().get_statevector()
oq.Top_States(States,Energies,SV,8)
```

```
Optimal Energy Expectation Value:   -6.0
```

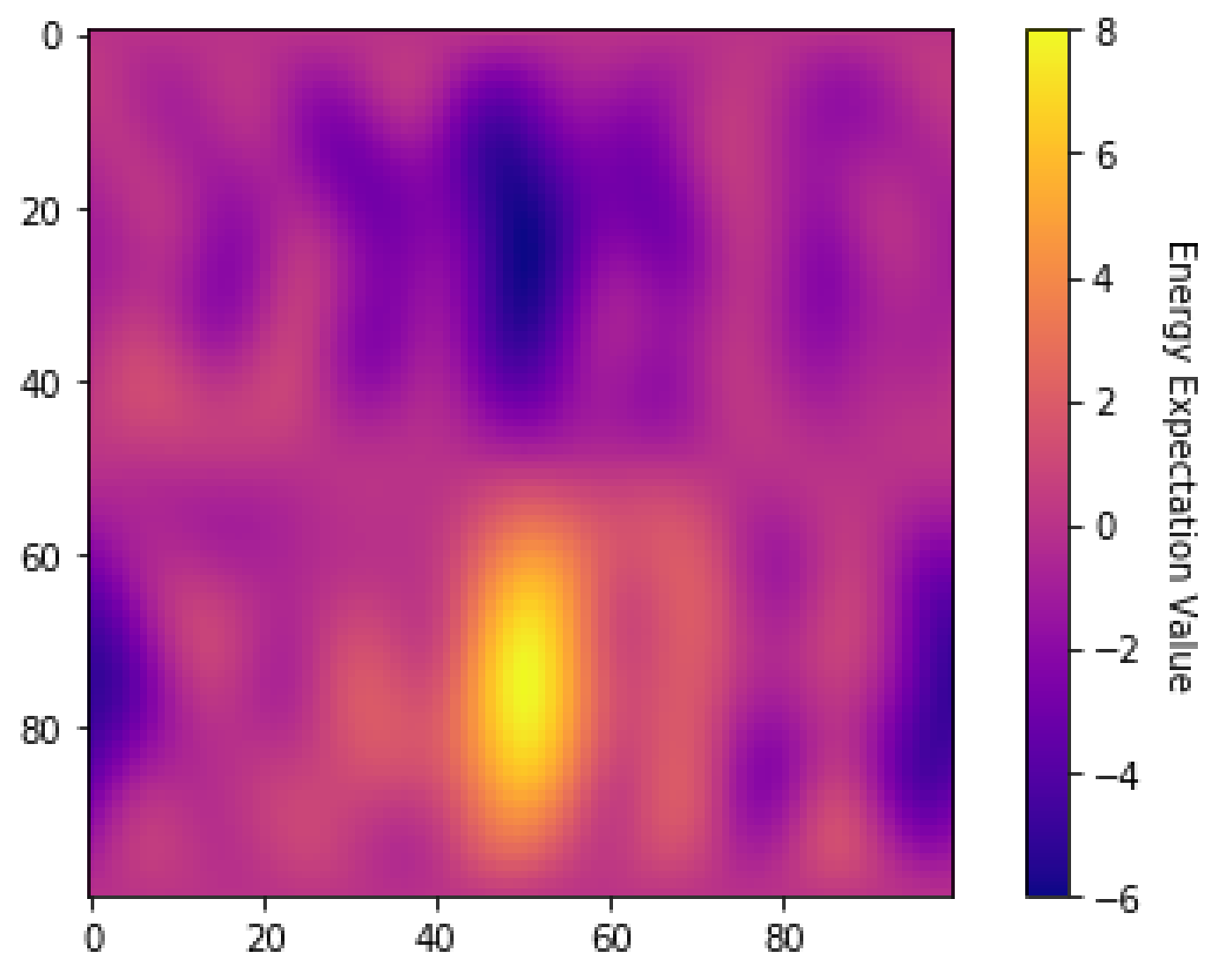

```
State  100      Probability:  100.0 %    Energy:  -6.0
State  110      Probability:  0.0 %    Energy:  0.5
State  000      Probability:  0.0 %    Energy:  -5.0
State  010      Probability:  0.0 %    Energy:  5.5
State  011      Probability:  0.0 %    Energy:  8.0
State  001      Probability:  0.0 %    Energy:  1.5
State  101      Probability:  0.0 %    Energy:  -3.5
State  111      Probability:  0.0 %    Energy:  -1.0
```

As we can see by this new heatmap, the addition of $R_y$ and CNOT gates to our $U(\mathbf{B},\beta)$ operator has resulted in a new $\left[\gamma, \beta\right]$ landscape, one which ranges froms $\left[8, -6\right]$. And if we recall all of the possible $\mathbf{C}(\mathbf{z})$ combinations, this range confirms that our parameter space can reach the $|\Psi_{\gamma\beta}\rangle$ states corresponding to the min/max of our Ising Energy problem. It is important to still keep in mind that our new circuit is unable to span the full $3$-qubit Hilbert Space. Increasing the order $p$ of our algorithm will in principle allow us to reach new states, but as you can see by the heatmap, $p = 1$ is already sufficient for solving our optimization problem. Thus, expanding the space of $|\Psi_{\gamma\beta}\rangle$ states further offers us no new advantage.

## Gradient Descent Optimization

Having now seen the extent to which QAOA is influenced by mixing operators and the order $p$, we have covered all of the quantum elements that make up this algorithm. In the coding examples shown above, we were able to illustrate the energy expectation value landscapes produced from $\gamma$ and $\beta$, revealing how these parameters influence the probability of measuring the desired optimal state $|z'\rangle$. However, in a real QAOA application we wouldn't want to exhaustively search through each possible $|\Psi_{\gamma\beta}\rangle$ state, as that would kill any chance of a speedup. Rather, one would instead look to implement a suitable classical optimization technique to aid in finding the optimal $\gamma$ and $\beta$ as quickly as possible.

To demonstrate the role a classical optimization technique plays in QAOA, we will be implementing the 'Gradient Descent' methodology for traversing the $\left[\gamma, \beta\right]$ landscape. First described by Augustin-Louis Cauchy in 1847, and later adapted by Haskell Curry in 1944 for solving optimization problems **[2]**, gradient descent finds the minimum (or maximum for gradient ascent) of some parameter space by using slopes to determine the next direction to search. Below is a one dimensional example:

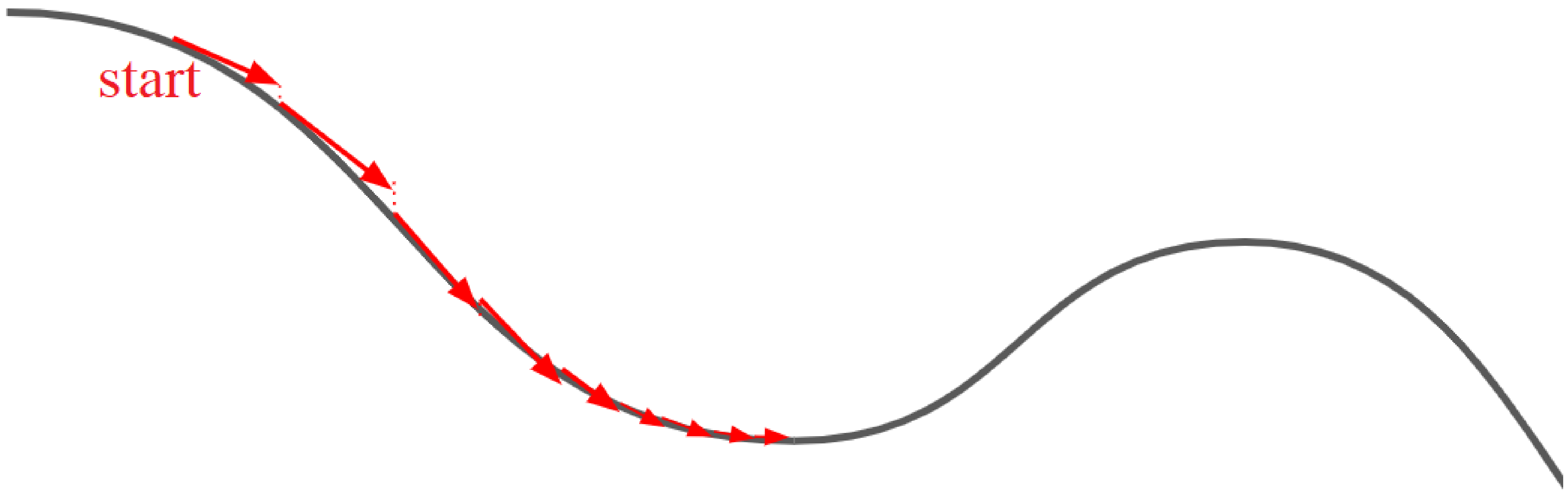

As illustrated above, the algorithm starts at the top of the left slope, which would correspond to the first $|\Psi_{\gamma\beta}\rangle$ state checked. The algorithm then proceeds by recursively determining the slope at each location and moving accordingly, with each iteration bringing the search closer to the local minimum. Note that the size of each arrow in the diagram represents the step size, which gets smaller as the algorithm approaches the minimum, terminating when a slope of zero is reached. For QAOA, we will be implementing this technique in two dimensions, which means each step must be calculated using the slope in two directions (and later four). For the $2$-dimensional version of the technique illustrated above, we will code the following formula for calculating the "slope" in the $\left[\gamma, \beta\right]$ space:

$$\vec{m}(\gamma, \beta) \quad = \quad \frac{F(\gamma+\epsilon, \beta) \;-\; F(\gamma-\epsilon, \beta)}{2\epsilon}\,\hat{\gamma} \;+\; \frac{F(\gamma, \beta+\epsilon) \;-\; F(\gamma, \beta-\epsilon)}{2\epsilon}\,\hat{\beta}$$

The formula above represents how we can approximate slope at a given location for $\gamma$ and $\beta$, where the parameter $\epsilon$ dictates how close of points to $F(\gamma,\beta)$ are used for determining $\vec{m}$. In practice, this method means that everytime our algorithm reaches a new point $F(\gamma,\beta)$, we need to also calculate four neighboring values of $F$ in order to progress to the next location. Similarly, if we want to scale this approach to higher orders of $p$, then each slope calculation requires the computation of $4p$ $F$ values. Not ideal in terms of speed, but this optimization strategy will serve its purpose for our learning goals.

As a final note before we code up our optimization scheme, take a look at the diagram above once more and notice how the function continues off to the right side, past the minimum where our gradient descent terminates. Using your eyes to guesstimate, you should be able to tell that the function continues on to a lower minimum than the one our algorithm produces. In terms of our search, this is meant to illustrate a problematic feature of the gradient descent algorithm: susceptible to getting stuck in local minima. This drawback can be quite problematic to work with, as it essentially means that we cannot $100\%$ trust the final answer of our optimization technique. More advanced versions of the gradient descent algorithm exist which aim to alleviate this problem in various ways, but for our implementation of QAOA we will simply accept the risk.

## Transverse Ising Model

As our first test for the gradient descent technique, we will be working with a slight variant of the Ising Energy model:

$$\mathbf{C}(\mathbf{Z}) \quad = \quad -\sum_{\langle ij \rangle} J_{ij} Z_i Z_j \;-\; \sum_i h_i Z_i$$

The new term in our cost function, $J_{ij}$, now places a strength value for each spin-spin interaction between particles. Whereas before the energy associated with the spin of each pair of particles was just $1$, now this value can vary across the geometry. Additionally, based on the sign of $J_{ij}$, neighboring pairs of particles may energetically prefer to have opposite or aligned spins. Similar to the magnetic field strength terms $h_i$, these new values add an extra degree of complication to the cost function. Mathematically, the consequence of this updated cost function leaves the $U(\mathbf{C},\gamma)$ operator essentially unchanged, only adding a scalar value to the exponentiation:

$$U(\mathbf{C},\gamma) \;\; = \;\; \prod_{\langle ij \rangle} e^{iJ_{ij}\gamma Z_i Z_j} \prod_i e^{i\gamma h_i Z_i}$$

All of the circuit implementations we previously developed hold for this case of the Ising Model as well, where the presence of these extra $J_{ij}$ terms simply change the values of the $R_z$ gates within the $Z_i Z_j$ construction. Before jumping straight into our optimization technique however, let's first see how this new cost function changes our $\left[\gamma,\beta\right]$ space. We will use the the same magnetic field strength values as before, now incorporating additional spin-spin coupling strengths, where a positive value for $J_{ij}$ means that parallel spin alignment is energetically preferred:

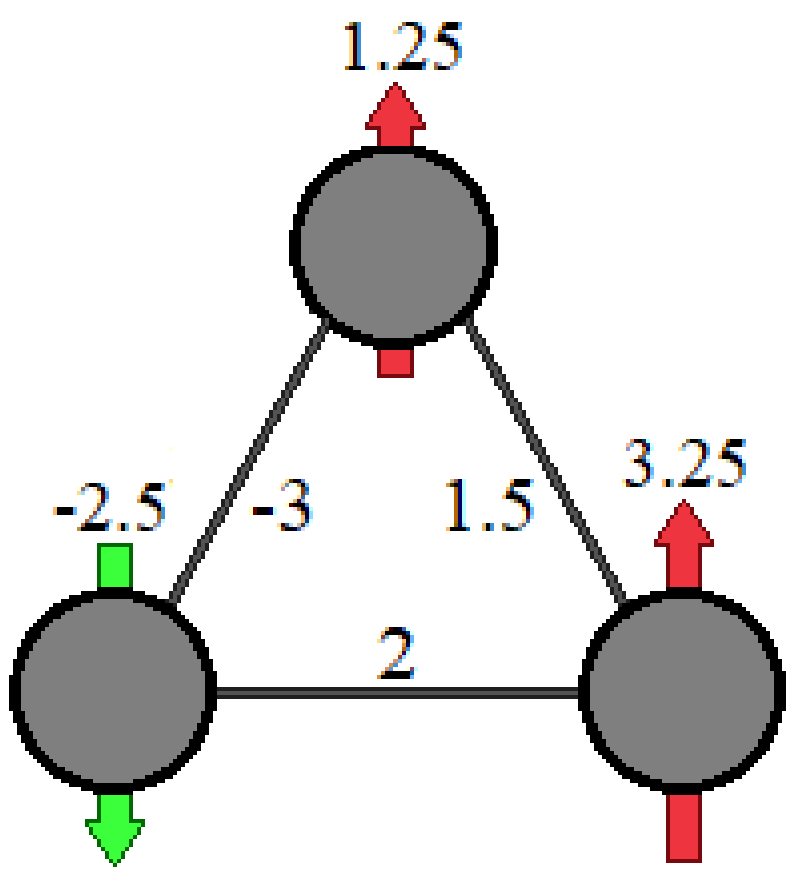

In [4]:
```
size = 100
#--------------------------------------------------------
Vert = [ [0,-2.5] , [1,3.25] , [2,1.25] ]
Edge = [ [0,1,2],[1,2,1.5],[2,0,-3] ]
#--------------------------------------------------------
Energies,States = oq.Ising_Energy( Vert,Edge,Transverse=True )
EV_grid = np.zeros(shape=(size,size))
EV_min = 10000
#========================================================
for b in np.arange(size):
    beta = round(2*m.pi*(b/size),4)
    for g in np.arange(size):
        gamma = round(2*m.pi*(g/size),4)
        q = QuantumRegister(len(Vert))
        qc= QuantumCircuit(q)
        for hh in np.arange(len(Vert)):
            qc.h( q[int(hh)] )
        oq.Ising_Circuit( qc, q, Vert, Edge, beta, gamma , Transverse=True, Mixing=2 )
        EV = oq.E_Expectation_Value( qc, Energies )
        EV_grid[b,g] = EV
        if( EV < EV_min ):
            Params = [beta,gamma]
            EV_min = EV
print('Optimal Energy Expectation Value:  ',EV_min,'    \u03B3 = ',Params[1],'   \u03B2 = ',Params[0],'\n')
#========================================================
fig, ax = plt.subplots()
show_text = False
show_ticks = False
oq.Heatmap(EV_grid, show_text, show_ticks, ax, "plasma", "Energy Expectation Value")
fig.tight_layout()
plt.show()
#========================================================

beta = Params[0]
gamma = Params[1]
q = QuantumRegister(len(Vert))
qc= QuantumCircuit(q)
for hh in np.arange(len(Vert)):
    qc.h( q[int(hh)] )
oq.Ising_Circuit( qc, q, Vert, Edge, beta, gamma, Transverse=True, Mixing=2  )
SV = execute( qc, S_simulator, shots=1 ).result().get_statevector()
oq.Top_States(States,Energies,SV,8)
```

```
Optimal Energy Expectation Value:   -6.0974     γ =  6.0319    β =  4.0841
```

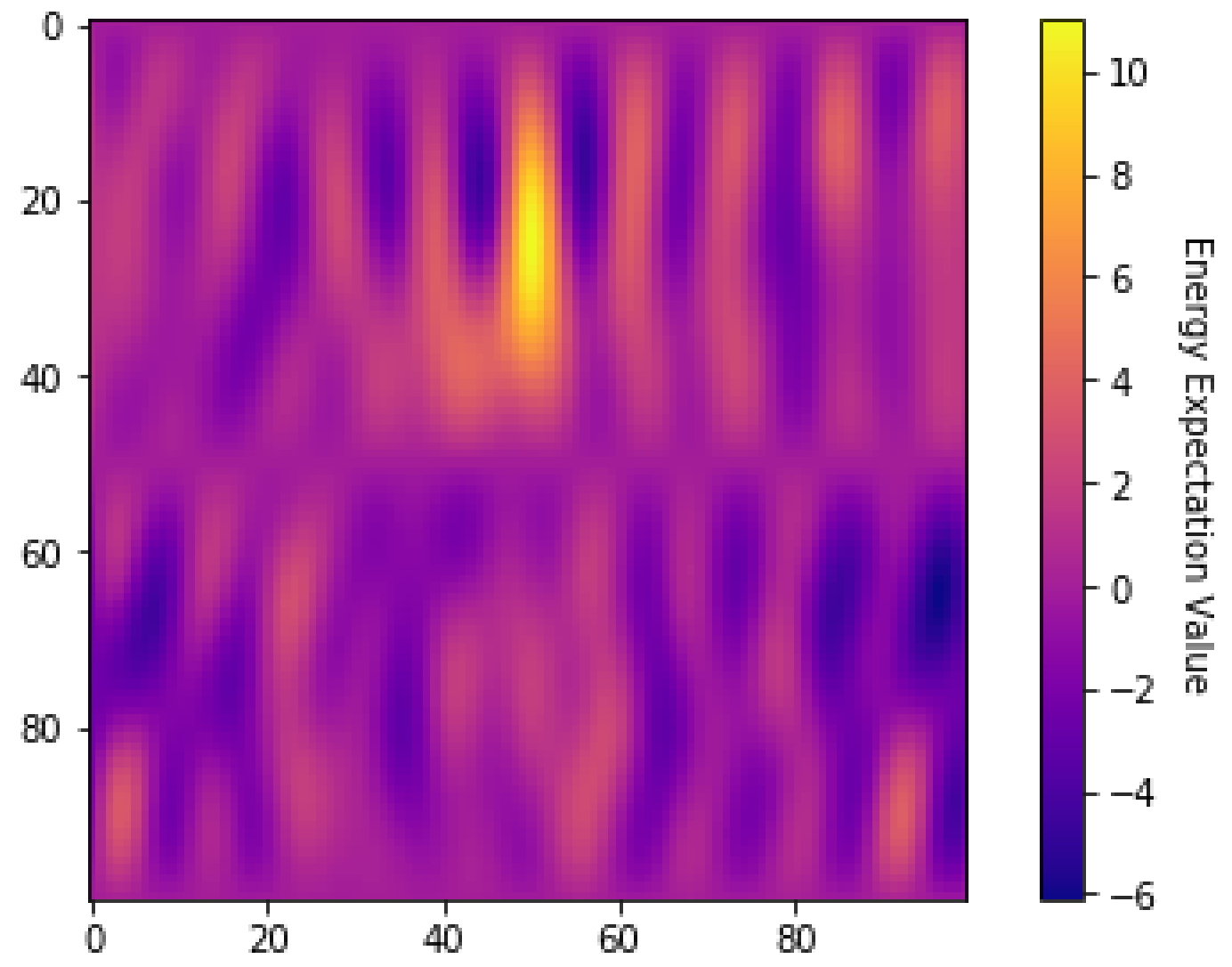

```
State  100      Probability:  58.68 %     Energy:  -9.5
State  000      Probability:  16.58 %     Energy:  -2.5
State  110      Probability:  9.68 %     Energy:  -4.0
State  101      Probability:  8.01 %     Energy:  2.0
State  001      Probability:  3.36 %     Energy:  -3.0
State  111      Probability:  1.57 %     Energy:  1.5
State  010      Probability:  1.55 %     Energy:  11.0
State  011      Probability:  0.57 %     Energy:  4.5
```

Having kept all of the magnetic field strengths the same for each particle, the plot above clearly demonstrates how the new spin-spin coupling strength terms have changed our $F(\gamma,\beta)$ landscape. For starters, the new optimal energy for our Transverse Ising model is now $-9$, but the lowest $F(\gamma,\beta)$ value produced above is only around $-6$. Note that we are using the same upgraded mixing operator as before, the one which previously was able to reach the state $|100\rangle$, but now cannot. Thus, due to the increased complexity of our new $U(\mathbf{C},\gamma)$, we either need a new mixing operator, or increase the order $p$ if we wish to reach more $3$-qubit states. But as we noted before, so long as our search leads us to a state where $|z'\rangle$ is highly probable ($58\%$ is indeed very probable), then our QAOA will be a success.

With the energy expectation value landscape shown above in mind, it is now time to turn our attention to our next task, implementing gradient descent into our QAOA. If we were to run our algorithm on a real quantum computer, our only means for computing slopes would be through measurements. Consequently, all of our calculations would be approximations, subject to the accuracy / noise of the measurements. For our learning purposes in this lesson however, we will be using wavefunctions for computing expectation values (just as we have been up to this point), which in turn will be used for computing slopes. Implementing gradient descent subject to measurements and noise comes with an entire set of new challenges, which we will put off for the next and final lesson in this series:  Lesson $11$ - VQE. For now, our primary goal is to see the theory behind incorporating an optimization technique into QAOA, which is better observed by using wavefunctions:

In [21]:

```
def Ising_Gradient_Descent(qc, q, Circ, V, E, beta, gamma, epsilon, En, step, **kwargs):
    '''
    Input:     qc (QuantumCircuit)    q (QuantumRegister)    Circ (Ising_Circuit function)    V (array)    E
               beta (float)    gamma (float)    epsilon    (float)    En (array)    step (float)
    Keyword Arguments:     Transverse (Bool) -  Changes to the Transve
                           Mixing (integer) - Denotes which mixing circuit to use for U(B,beta)
    Calculates and returns the next values for beta and gamma using gradient descent
    '''
    Trans = False
    if 'Transverse' in kwargs:
        if( kwargs['Transverse'] == True ):
            Trans = True
    Mixer = 1
    if 'Mixing' in kwargs:
        Mixer = int(kwargs['Mixing'])
    params = [ [beta+epsilon,gamma],[beta-epsilon,gamma],[beta,gamma+epsilon],[beta,gamma-epsilon] ]
    ev = []
    for i in np.arange( 4 ):
        q = QuantumRegister(len(V))
        qc= QuantumCircuit(q)
        for hh in np.arange(len(V)):
            qc.h( q[int(hh)] )
        Circ( qc, q, V, E, params[i][0], params[i][1], Transverse=Trans, Mixing=Mixer  )
        ev.append( E_Expectation_Value( qc, En ) )
    beta_next = beta - ( ev[0] - ev[1] )/( 2.0*epsilon ) * step
    gamma_next = gamma - ( ev[2] - ev[3] )/( 2.0*epsilon ) * step
    return beta_next, gamma_next
```

Shown above is the function we will be using to advance our gradient descent technique for the Transverse Ising Energy model. For this implementation of gradient descent, our slope computations at each step require two parameters of our choosing: $\epsilon$ and $\mathbf{step_size}$. The first parameter, $\epsilon$, dictates the distance used for approximating the slope at a given $[\gamma,\beta]$ location, as shown in the formula earlier. The second parameter $\text{step_size}$, in combination with the calculated slope, tells the algorithm how much distance to cover with each iteration. Changing this value can have dramatic effects on the progression of our gradient descent algorithm: too large of a $\text{step_size}$ value and the algorithm may overshoot a minima entirely, too small and the runtime of the algorithm will become too long for any chance of a speedup. I encourage you to play around with this parameter in the examples to come, and see how different values affect the termination of the algorithm.

As a first test for our gradient descent scheme, we will use the optimal values obtained from the cell of code above to help us pick our starting values for $\gamma$ and $\beta$. Obviously we don't want to start our search right on top of the global minimum, as that would defeat the purpose, so let's pick values slightly off. Additionally, our algorithm needs a criteria for termination, otherwise it will simply keep searching forever in whatever minima it finds its way into. For this, we will use the parameter $\delta$ to dictate the threshold for when we want our algorithm to stop searching:

In [7]:

```
epsilon = 0.001
step_size = 0.01
delta = 0.002
#--------------------------------------------------------
Vert = [ [0,-2.5] , [1,3.25] , [2,1.25] ]
Edge = [ [0,1,2],[1,2,1.5],[2,0,-3] ]
#--------------------------------------------------------
Energies,States = oq.Ising_Energy( Vert,Edge, Transverse=True )
EV = 100
EV_old = 1000
EV_min = 1000
#=========================================================
beta = 3.6
gamma = 6.4
s = 0
while( (abs( EV - EV_old ) > delta) and ( EV < EV_old ) ):
    q = QuantumRegister(len(Vert))
    qc= QuantumCircuit(q)
    for hh in np.arange(len(Vert)):
        qc.h( q[int(hh)] )
    if( s != 0 ):
        beta,gamma = oq.Ising_Gradient_Descent(qc,q,oq.Ising_Circuit,Vert,Edge,beta,gamma,epsilon,Energies,step_size,Transv
    oq.Ising_Circuit( qc, q, Vert, Edge, beta, gamma, Transverse=True, Mixing=2  )
    EV_old = EV
    EV = oq.E_Expectation_Value( qc, Energies )
    if( EV < EV_min ):
        Params = [beta,gamma]
        EV_min = EV
    s = int(s+1)
    print('F(\u03B3,\u03B2):  ',EV,'         \u03B3 = ',round(gamma,6),'        \u03B2 = ',round(beta,6),)
```

```
F(γ,β):   0.9857          γ =  6.4         β =  3.6
F(γ,β):   -0.6117          γ =  6.3005        β =  3.5935
F(γ,β):   -3.8804          γ =  6.0975        β =  3.609
F(γ,β):   -4.5557          γ =  6.061        β =  3.69
F(γ,β):   -5.073          γ =  6.0785        β =  3.7655
F(γ,β):   -5.4381          γ =  6.0545        β =  3.827
F(γ,β):   -5.6822          γ =  6.07         β =  3.8805
F(γ,β):   -5.8383          γ =  6.046        β =  3.9205
F(γ,β):   -5.936          γ =  6.064        β =  3.9555
F(γ,β):   -5.9964          γ =  6.04         β =  3.9795
F(γ,β):   -6.0323          γ =  6.06         β =  4.002
F(γ,β):   -6.0544          γ =  6.0355        β =  4.0155
F(γ,β):   -6.0688          γ =  6.057        β =  4.0305
F(γ,β):   -6.0767          γ =  6.033        β =  4.0375
F(γ,β):   -6.0836          γ =  6.0545        β =  4.0475
F(γ,β):   -6.0874          γ =  6.0325        β =  4.051
F(γ,β):   -6.0904          γ =  6.0525        β =  4.058
F(γ,β):   -6.0927          γ =  6.0325        β =  4.0595
F(γ,β):   -6.0947          γ =  6.0505        β =  4.065
```

As illustrated above, our gradient descent terminates after $19$ iterations, bringing us to the same $F(\gamma,\beta)$ global minimum we found earlier by scanning the entire space (which we acquired by checking $10000$ values). Thus, our gradient descent technique has indeed found the optimal pair of $\gamma$ and $\beta$ values in a significantly faster time. However, keep in mind that in this example we picked values which we already knew were close to the minimum we were looking for. If instead we had picked $\gamma$ and $\beta$ values completely at random, our results may be quite different:

In [15]:

```
epsilon = 0.001
step_size = 0.01
delta = 0.005
#--------------------------------------------------------
Vert = [ [0,-2.5] , [1,3.25] , [2,1.25] ]
Edge = [ [0,1,2],[1,2,1.5],[2,0,-3] ]
#--------------------------------------------------------
Energies,States = oq.Ising_Energy( Vert,Edge, Transverse=True )
EV = 100
EV_old = 1000
EV_min = 1000
#=========================================================
beta = 2*m.pi*random.random()
gamma = 2*m.pi*random.random()
s = 0
while( (abs( EV - EV_old ) > delta) and ( EV < EV_old ) ):
    q = QuantumRegister(len(Vert))
    qc= QuantumCircuit(q)
    for hh in np.arange(len(Vert)):
        qc.h( q[int(hh)] )
    if( s != 0 ):
        beta,gamma = oq.Ising_Gradient_Descent(qc,q,oq.Ising_Circuit,Vert,Edge,beta,gamma,epsilon,Energies,step_size,Transv
    oq.Ising_Circuit( qc, q, Vert, Edge, beta, gamma, Transverse=True, Mixing=2  )
    EV_old = EV
    EV = oq.E_Expectation_Value( qc, Energies )
    if( EV < EV_min ):
        Params = [beta,gamma]
        EV_min = EV
    s = int(s+1)
    print('F(\u03B3,\u03B2):  ',EV,'         \u03B3 = ',round(gamma,6),'        \u03B2 = ',round(beta,6),)
#=========================================================
print('\n------------------------------------------------------------\n')
q = QuantumRegister(len(Vert))
qc= QuantumCircuit(q)
for hh in np.arange(len(Vert)):
    qc.h( q[int(hh)] )
oq.Ising_Circuit( qc, q, Vert, Edge, beta, gamma, Transverse=True, Mixing=2  )
SV = execute( qc, S_simulator, shots=1 ).result().get_statevector()
oq.Top_States(States,Energies,SV,8)
```

```
F(γ,β):   1.6833          γ =  5.42361         β =  0.305828
F(γ,β):   0.0431          γ =  5.53111         β =  0.252828
F(γ,β):   -1.532          γ =  5.67661         β =  0.246828
F(γ,β):   -1.7773          γ =  5.72011         β =  0.290828
F(γ,β):   -1.9019          γ =  5.70411         β =  0.326328
F(γ,β):   -1.9808          γ =  5.70961         β =  0.355828
F(γ,β):   -2.032          γ =  5.70411         β =  0.379328
F(γ,β):   -2.0662          γ =  5.70661         β =  0.398828
F(γ,β):   -2.0884          γ =  5.70361         β =  0.414328
F(γ,β):   -2.1036          γ =  5.70511         β =  0.427328
F(γ,β):   -2.1137          γ =  5.70311         β =  0.437828
F(γ,β):   -2.1207          γ =  5.70411         β =  0.446828
F(γ,β):   -2.1253          γ =  5.70261         β =  0.453828

------------------------------------------------------------

State  001     Probability:  36.24 %     Energy:  -3.0
State  111     Probability:  24.87 %     Energy:  1.5
State  000     Probability:  16.7 %     Energy:  -2.5
State  110     Probability:  11.51 %     Energy:  -4.0
State  100     Probability:  6.94 %     Energy:  -9.5
State  101     Probability:  2.43 %     Energy:  2.0
State  011     Probability:  1.01 %     Energy:  4.5
State  010     Probability:  0.3 %     Energy:  11.0
```

In the example above, we can see that our gradient descent has gotten stuck in a local minima, yielding a $F(\gamma,\beta)$ value that is quite far off from what we know to be the global minimum. Consequently, we can see that the $|\Psi_{\gamma\beta}\rangle$ state corresponding to this particular local minima is dominated by its $|001\rangle$ component. If this had been our first attempt at solving the problem, and none of our measurements along the way yielded the true $|z'\rangle$, the final conclusion of our QAOA attempt would tell us that $|001\rangle$ corresponds to our problem's optimal configuration. Thus, using the solution provided from the gradient descent technique must always be taken with a grain of salt, unless we have additional information which can help guide the search. Alternatively, one could look to two different avenues for subsequent runs: 1) Increase the order $p$ and see if the search yields the same optimal $|z\rangle$ state 2) Keep the same order $p$, but implement the gradient descent search from several distant starting locations, with the hope that one of the searches will fall into the global minimum.

## Final Example: MaxCut

We have now seen a full QAOA run in action, hopefully giving you a flavor for the potential of this algorithm. As mentioned earlier, the most promising aspect about QAOA for near term quantum computers is its flexibility. The algorithm can tackle a wide array of problems through the encoding of cost functions into $U(\mathbf{C},\gamma)$, while simultaneously offering limitless possibilities for finding the optimal solution through $U(\mathbf{B},\beta)$ and the order $p$. Additionally, and perhaps most importantly, QAOA synchronizes with the best already known classical optimization techniques, which means that the algorithm only gets stronger as better classical optimizers are developed.

As our final example in this lesson, we will cover one additional graph optimization problem, which was the example demonstrated in the original QAOA paper: the MaxCut problem. Similar to our spin-spin interaction before, the MaxCut problem asks us to find the configuration of our system which maximizes the number of edges containing neighboring spin-up and spin-down particles. Below is an example graph, with several possible configurations and their accompanying cut values:

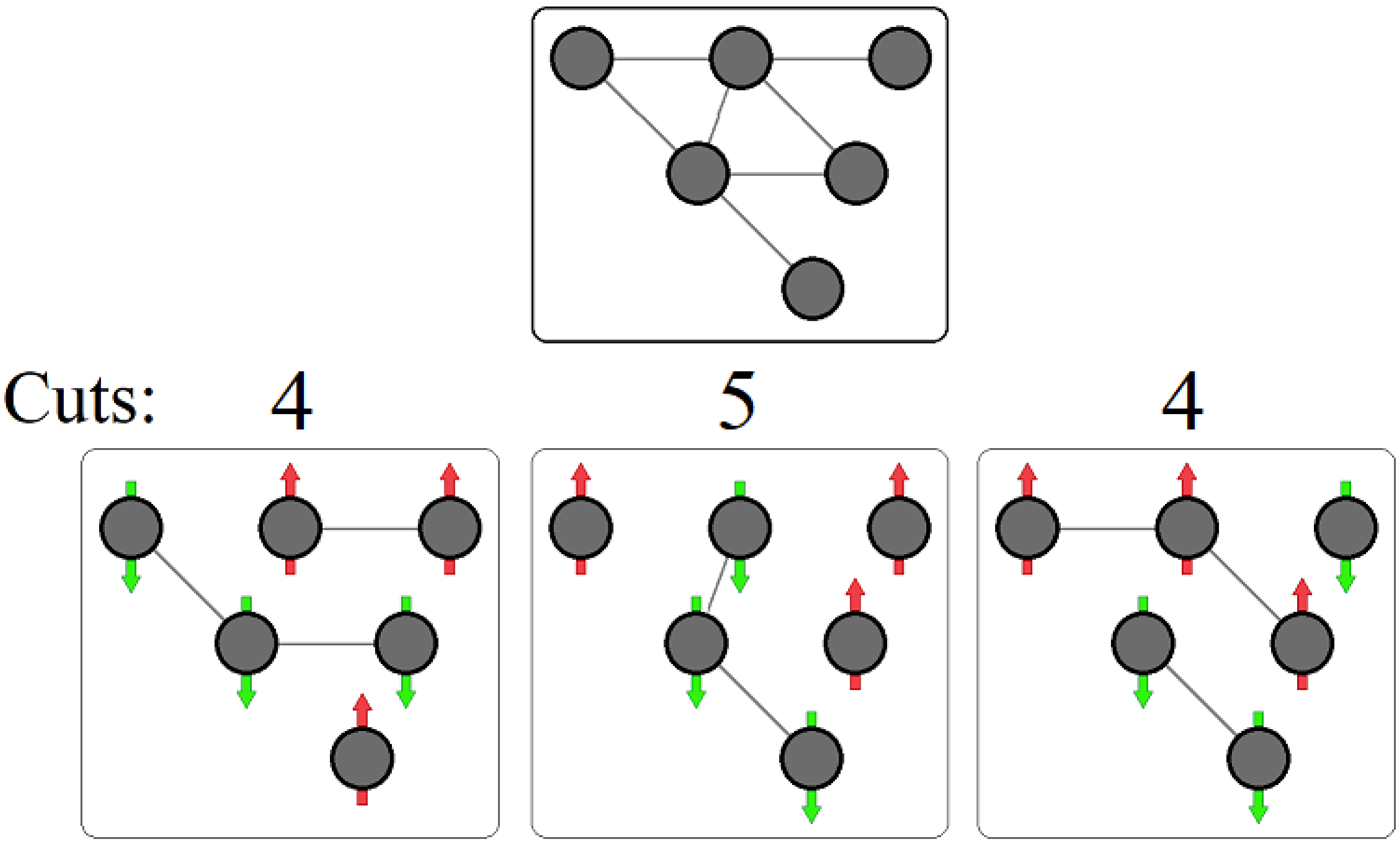

As shown in the figures, the energy value associated with each configuration is based on the number of 'cuts' made when every particle in the system is assigned a value of either spin up or down. Specifically, the number of cuts for a particular configuration is equivalent to the number of instances of adjacent particles (sharing a graph connection) with opposite spin. How we define energy for this problem can be done one of two ways: define a cut on the graph as $+1$ or $-1$. Based on this choice, our problem then becomes either a maximization or minimization problem, whereby we're looking for either the largest or smallest possible energy configuration. Since in our previous two examples we worked with minimization problems, here we are going to take the maximization approach to once again demonstrate the flexibility of QAOA. As such, our cost function for this problem then is defined as follows:

$$\mathbf{C}(\mathbf{Z}) \;=\; \frac{1}{2}\sum_{\langle ij\rangle} I \,-\, Z_i Z_j$$

In analyzing the cost function shown above, we once again find the familiar $Z_i Z_j$ term from before, which produces a negative energy contribution when particles $i$ and $j$ have the same spin. Specifically, when two particles have opposite spins, their contributing quantity in the cost function results in a $+1$: $\frac{1}{2}(1 - (-1)(+1))$. Conversely, neighbors with the same spin will result in a value of $-1$, which for this problem is the event which we're trying to minimize. All together, the phase operator for this new optimization problem is:

$$\begin{aligned} U(\mathbf{C},\gamma) &= e^{-i\gamma\left(\frac{1}{2}\sum_{\langle ij\rangle} I \,-\, Z_i Z_j\right)} \\ &= e^{-i\frac{\gamma}{2}\sum_{\langle ij\rangle} I}\;\; e^{i\frac{\gamma}{2}\sum_{\langle ij\rangle} Z_i Z_j} \\ &= \prod_{\langle ij\rangle} e^{-i\frac{\gamma}{2} I}\;\prod_{\langle ij\rangle} e^{i\frac{\gamma}{2} Z_i Z_j} \end{aligned}$$

Just like the Ising Energy models, we once again have a summation of $Z_i Z_j$ over all nearest neighbors, which we already know how to implement in our quantum circuit. As for the second term however, we have an exponentiation of $I_i \otimes I_j$, which results in the following matrix operator:

$$e^{-i\frac{\gamma}{2} I_i I_j} \;=\; \begin{bmatrix} e^{-i\frac{\gamma}{2}} & & & \\ & e^{-i\frac{\gamma}{2}} & & \\ & & e^{-i\frac{\gamma}{2}} & \\ & & & e^{-i\frac{\gamma}{2}} \end{bmatrix}$$

Now, this operation essentially places an $e^{-i\frac{\gamma}{2}}$ phase on every term corresponding to the $i^{th}$ and $j^{th}$ qubits. Mathematically, this is equivalent to applying a global phase when the states of $Q_i$ and $Q_j$ are separable, which we've already demonstrated has no measurable effect on a quantum system:

$$\begin{aligned} |\Psi\rangle &= |Q_1\rangle \otimes |Q_2 Q_3\rangle \\ &= |Q_1\rangle \otimes \left(\alpha_{00}|00\rangle + \alpha_{01}|01\rangle + \alpha_{10}|10\rangle + \alpha_{11}|11\rangle\right) \end{aligned}$$

$$\begin{aligned} I_1 \otimes e^{i\gamma I_2 I_3}|\Psi\rangle &= |Q_1\rangle \otimes \left(\alpha_{00}e^{i\gamma}|00\rangle + \alpha_{01}e^{i\gamma}|01\rangle + \alpha_{10}e^{i\gamma}|10\rangle + \alpha_{11}e^{i\gamma}|11\rangle\right) \\ &= e^{i\gamma}|Q_1\rangle \otimes \left(\alpha_{00}|00\rangle + \alpha_{01}|01\rangle + \alpha_{10}|10\rangle + \alpha_{11}|11\rangle\right) \\ &= e^{i\gamma}|\Psi\rangle \end{aligned}$$

Thus, if the $I_i I_j$ component of our phase operator has no meaningful impact on the quantum system, we can simply drop it from our quantum circuit. You may be wondering however, in light of our discussions earlier regarding full Hilbert Spaces, can we still ignore this Identity term if our system is in an entangled state? For our QAOA problem here, yes, but in general, no. If our system is no longer separable, then the phases can no longer be pulled out as a global phase, which in turn means that we have fundamentally changed our quantum state. However, the reason why we can still drop the phase gates corresponding to the $I_i I_j$ terms from our quantum circuit is because they have no impact on the final answer. That is to say, if we remove the Identity terms from the cost function, the configuration corresponding to the optimal energy will still be the same, even if the numeric value is different. With or without the Identity terms, the maximum energy state is going to correspond to the same configuration of spins, which means our QAOA is going to end up searching for the same $| z' \rangle$. It may feel a bit weird, but the important thing to remember here is that we aren't implementing $\mathbf{C}(\mathbf{Z})$, but rather the operator $U(\mathbf{C},\gamma)$. So long as our phase operator correctly encodes a cost function which will solve our optimization problem, dropping components of $\mathbf{C}(\mathbf{Z})$ which are inconsequential is perfectly okay.

When implementing this MaxCut problem into a quantum circuit, our $U(\mathbf{C},\gamma)$ operator is once again just the exponentiation of nearest neighbor $Z_i Z_j$ terms, which we already have the circuit construction for. Similarly, our mixing operator $U(\mathbf{B},\beta)$ can be either of the constructions laid out earlier (or any new construction of our choosing), which means we have everything we need to run QAOA. In fact, you may be wondering why we didn't start with this problem, considering it has the simplest cost function we've studied so far. The motivation for saving MaxCut for last is because it is simple enough such that we can practice incorporating everything we've covered in this lesson into a single working example: namely a higher order $p$ and the gradient descent algorithm (ascent in this case). Below is the example geometry we will look to solve, as well as its corresponding code implementation.

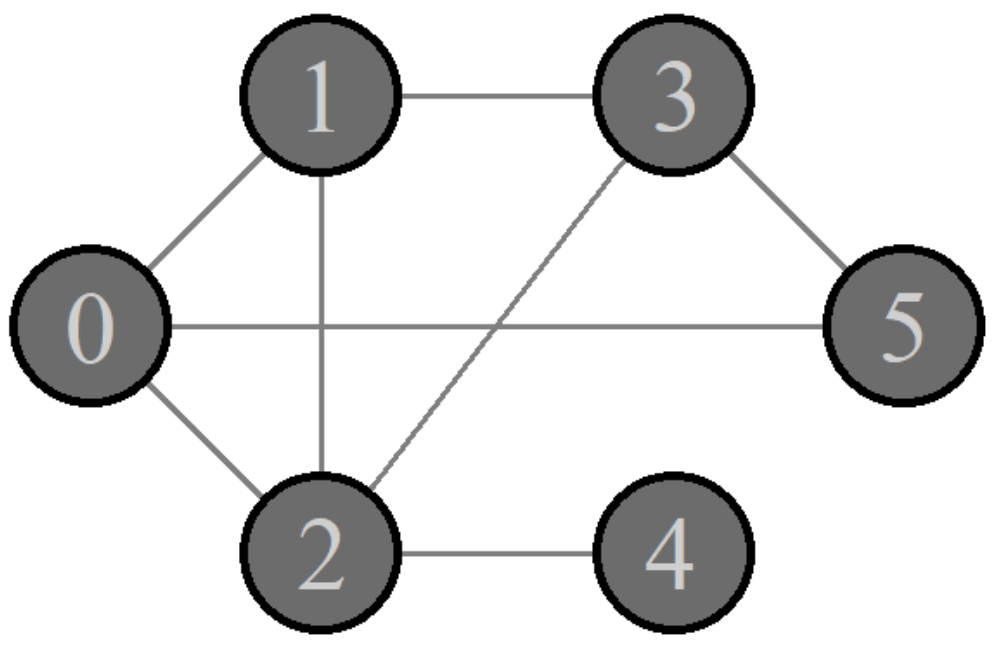

In [21]:

```
p = 2
epsilon = 0.001
step_size = 0.01
delta = 0.001
#--------------------------------------------------------
Vert = [ 0,1,2,3,4,5 ]
Edge = [ [0,1],[0,2],[0,5],[1,2],[1,3],[2,3],[2,4],[3,5] ]
#--------------------------------------------------------
Energies,States = oq.MaxCut_Energy( Vert,Edge )
EV = -100
EV_old = -1000
EV_max = -1
#========================================================
beta = []
gamma = []
for pp in np.arange(p):
    beta.append(2*m.pi*random.random())
    gamma.append(2*m.pi*random.random())
s = 0
while( abs( EV - EV_old ) > delta ):
    q = QuantumRegister(len(Vert))
    qc= QuantumCircuit(q)
    for hh in np.arange(len(Vert)):
        qc.h( q[int(hh)] )
    if( s != 0 ):
        beta,gamma = oq.p_Gradient_Ascent(qc,q,oq.MaxCut_Circuit,Vert,Edge,p,beta,gamma,epsilon,Energies,step_size)
    for i in np.arange(p):
        oq.MaxCut_Circuit( qc, q, Vert, Edge, beta[i], gamma[i] )
    #-------------------------------
    EV_old = EV
    EV = oq.E_Expectation_Value( qc, Energies )
    if( EV_old > EV ):
        EV_old = EV
    if( EV > EV_max ):
        Params = [beta,gamma]
        EV_max = EV
    s = int(s+1)
    #-------------------------------
    if( (m.floor( s/10 ) == s/10) or (s == 1) ):
        params_string = ''
        for ps in np.arange(p):
            params_string = params_string + '    \u03B3'+str(int(ps+1))+' = '+str(round(gamma[ps],6))+'    \u03B2'+str(int(
        params_string = params_string+'     steps: '+str(s)
        print('F(\u03B3,\u03B2):  ',EV,'    |',params_string)
print('\n _____ Terminated Gradient Ascent _____ \n')
params_string = ''
for ps in np.arange(p):
    params_string = params_string + '    \u03B3 +str(int(ps+1))+' = '+str(round(gamma[ps],6))+'    \u03B2'+str(int(ps+1))+'
params_string = params_string+'     steps:  +str(s)
print('F(\u03B3,\u03B2):  ',EV,'    |',params_string,'\n')
#========================================================

beta = Params[0]
gamma = Params[1]
p = len( Params[0] )
#-------------------------------
q = QuantumRegister(len(Vert))
qc= QuantumCircuit(q)
for hh in np.arange(len(Vert)):
    qc.h( q[int(hh)] )
for i in np.arange(p):
    oq.MaxCut_Circuit( qc, q, Vert, Edge, beta[i], gamma[i]  )
SV = execute( qc, S_simulator, shots=1 ).result().get_statevector()
oq.Top_States(States,Energies,SV,12)
```

```
F(γ,β):   3.8271    |     γ1 = 0.252951    β1 = 5.163677    γ2 = 4.936245    β2 = 5.134703     steps: 1
F(γ,β):   3.8815    |     γ1 = 0.299951    β1 = 5.129177    γ2 = 4.899745    β2 = 5.148203     steps: 10
F(γ,β):   3.9239    |     γ1 = 0.341451    β1 = 5.095177    γ2 = 4.885745    β2 = 5.181703     steps: 20
F(γ,β):   3.976    |     γ1 = 0.381451    β1 = 5.065177    γ2 = 4.889245    β2 = 5.232203     steps: 30
F(γ,β):   4.0494    |     γ1 = 0.427951    β1 = 5.038177    γ2 = 4.904245    β2 = 5.296703     steps: 40
F(γ,β):   4.141    |     γ1 = 0.478451    β1 = 5.019677    γ2 = 4.928745    β2 = 5.371703     steps: 50
F(γ,β):   4.2401    |     γ1 = 0.528951    β1 = 5.016177    γ2 = 4.955745    β2 = 5.452703     steps: 60
F(γ,β):   4.3392    |     γ1 = 0.576451    β1 = 5.030177    γ2 = 4.979245    β2 = 5.536203     steps: 70
F(γ,β):   4.4355    |     γ1 = 0.617451    β1 = 5.060677    γ2 = 4.994745    β2 = 5.618203     steps: 80
F(γ,β):   4.5326    |     γ1 = 0.649451    β1 = 5.103677    γ2 = 5.004245    β2 = 5.699703     steps: 90
F(γ,β):   4.6335    |     γ1 = 0.674451    β1 = 5.156177    γ2 = 5.007745    β2 = 5.781203     steps: 100
F(γ,β):   4.7378    |     γ1 = 0.690451    β1 = 5.214177    γ2 = 5.008245    β2 = 5.864203     steps: 110
F(γ,β):   4.8423    |     γ1 = 0.697451    β1 = 5.272677    γ2 = 5.006745    β2 = 5.948203     steps: 120
F(γ,β):   4.9439    |     γ1 = 0.697451    β1 = 5.332177    γ2 = 5.007245    β2 = 6.029203     steps: 130
F(γ,β):   5.0459    |     γ1 = 0.687451    β1 = 5.390177    γ2 = 5.009245    β2 = 6.111203     steps: 140
F(γ,β):   5.1509    |     γ1 = 0.668951    β1 = 5.447177    γ2 = 5.010745    β2 = 6.194203     steps: 150
F(γ,β):   5.2623    |     γ1 = 0.643451    β1 = 5.503177    γ2 = 5.012245    β2 = 6.280203     steps: 160
F(γ,β):   5.3782    |     γ1 = 0.613951    β1 = 5.558177    γ2 = 5.011745    β2 = 6.368203     steps: 170
F(γ,β):   5.4908    |     γ1 = 0.584451    β1 = 5.611177    γ2 = 5.005745    β2 = 6.455203     steps: 180
F(γ,β):   5.5871    |     γ1 = 0.561451    β1 = 5.660177    γ2 = 4.996245    β2 = 6.536203     steps: 190
F(γ,β):   5.6602    |     γ1 = 0.547951    β1 = 5.704177    γ2 = 4.985245    β2 = 6.607703     steps: 200
F(γ,β):   5.7097    |     γ1 = 0.543951    β1 = 5.741677    γ2 = 4.974745    β2 = 6.666703     steps: 210
F(γ,β):   5.7398    |     γ1 = 0.545451    β1 = 5.771177    γ2 = 4.964245    β2 = 6.711703     steps: 220
F(γ,β):   5.7576    |     γ1 = 0.548951    β1 = 5.793677    γ2 = 4.954245    β2 = 6.745703     steps: 230

 _____ Terminated Gradient Ascent _____

F(γ,β):   5.7638    |     γ1 = 0.551951    β1 = 5.803177    γ2 = 4.949745    β2 = 6.760203     steps: 235

State  011001      Probability:  15.0 %     Energy:  7.0
State  100110      Probability:  15.0 %     Energy:  7.0
State  101100      Probability:  8.44 %     Energy:  6.0
State  010011      Probability:  8.44 %     Energy:  6.0
State  001001      Probability:  4.92 %     Energy:  6.0
State  110110      Probability:  4.92 %     Energy:  6.0
State  100100      Probability:  3.39 %     Energy:  6.0
State  011011      Probability:  3.39 %     Energy:  6.0
State  110010      Probability:  2.47 %     Energy:  5.0
State  010110      Probability:  2.47 %     Energy:  5.0
State  101001      Probability:  2.47 %     Energy:  5.0
State  001101      Probability:  2.47 %     Energy:  5.0
```

The cell of code above represents starting our $p = 2$ gradient ascent from a random location in the parameter space. For completeness, the mixing operator implemented in this coding exercise corresponds to the first $U(\mathbf{B},\beta)$ we studied (only $R_x$ gates on all the qubits). Based on the energies printed at the end of the code, we can conclude that the true global maximum to our problem corresponds to $+7$ (you can verify for yourself that these are indeed the optimal configurations), which is quite a ways off from the $5.76$ we terminated at. Nevertheless, the probability distribution shows that our local maxima has the two optimal configurations as the most probable states, which means that we can consider this run of QAOA as a success.

Since this is our first time running QAOA with an order $p$ greater than $1$, it would be insightful to know what kind of solutions we could have expected from $p = 1$:

In [11]:

```
size = 100
#------------------------------------------------------
Vert = [ 0,1,2,3,4,5 ]
Edge = [ [0,1],[0,2],[0,5],[1,2],[1,3],[2,3],[2,4],[3,5] ]
#------------------------------------------------------
Energies,States = oq.MaxCut_Energy( Vert,Edge )
EV_grid = np.zeros(shape=(size,size))
EV_max = -1
#=======================================================
for b in np.arange(size):
    beta = round(2*m.pi*(b/size),4)
    for g in np.arange(size):
        gamma = round(2*m.pi*(g/size),4)
        q = QuantumRegister(len(Vert))
        qc= QuantumCircuit(q)
        for hh in np.arange(len(Vert)):
            qc.h( q[int(hh)] )
        oq.MaxCut_Circuit( qc, q, Vert, Edge, beta, gamma  )
        EV = oq.E_Expectation_Value( qc, Energies )
        EV_grid[b,g] = EV
        if( EV > EV_max ):
            Params = [beta,gamma]
            EV_max = EV
print('Energy Expectation Value:  ',EV_max,'    \u03B3 = ',Params[1],'   \u03B2 = ',Params[0],'\n')
#---------------------------------------
fig, ax = plt.subplots()
show_text = False
show_ticks = False
oq.Heatmap(EV_grid, show_text, show_ticks, ax, "plasma", "Energy Expectation Value")
fig.tight_layout()
plt.show()
#=======================================

beta = Params[0]
gamma = Params[1]
#---------------------------------------
q = QuantumRegister(len(Vert))
qc= QuantumCircuit(q)
for hh in np.arange(len(Vert)):
    qc.h( q[int(hh)] )
oq.MaxCut_Circuit( qc, q, Vert, Edge, beta, gamma  )
SV = execute( qc, S_simulator, shots=1 ).result().get_statevector()
oq.Top_States(States,Energies,SV,12)
```

```
Energy Expectation Value:   5.2934     γ =  5.6549    β =  0.6912
```

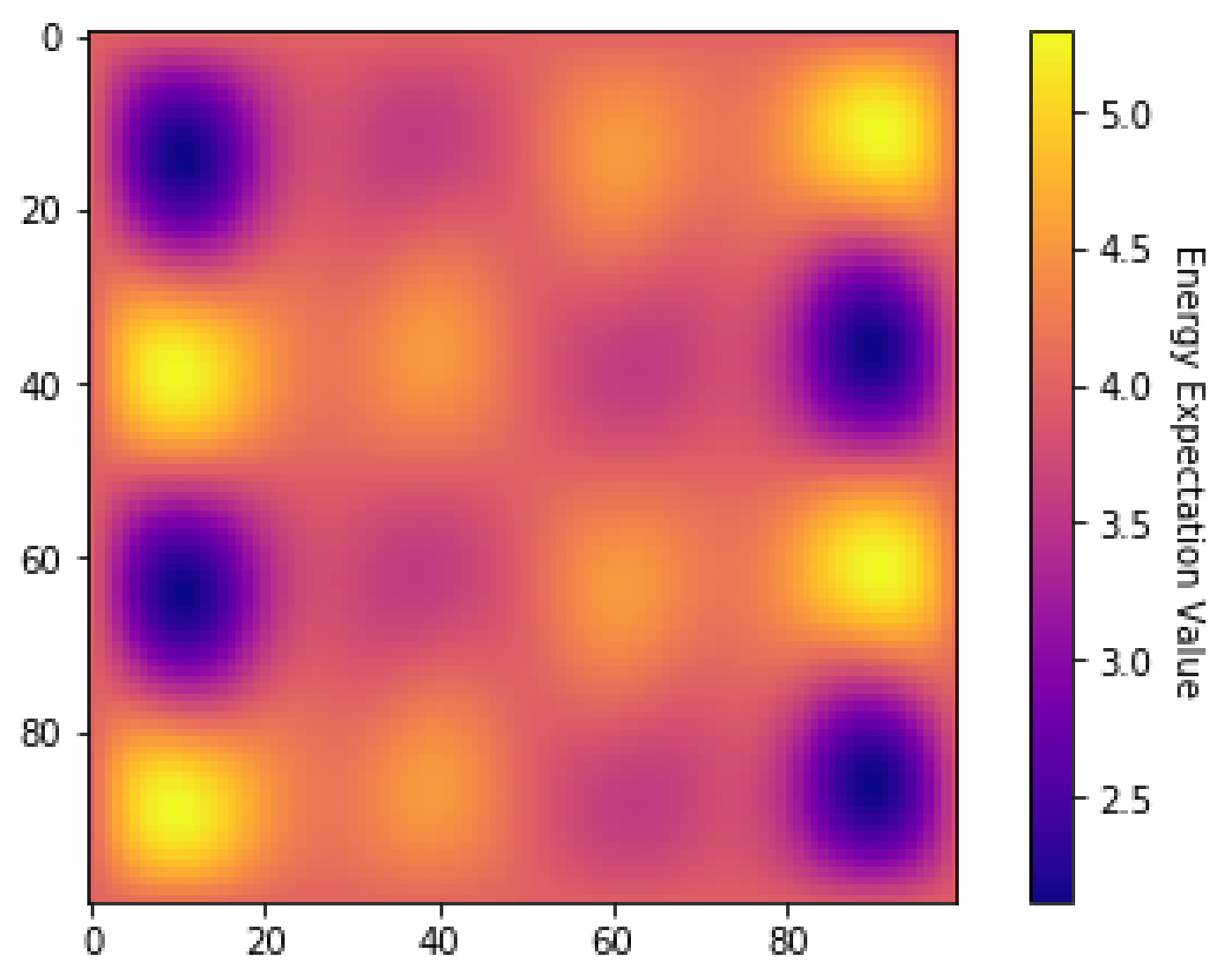

```
State  100110     Probability:  9.0 %     Energy:  7.0
State  011001     Probability:  9.0 %     Energy:  7.0
State  101100     Probability:  5.55 %     Energy:  6.0
State  010011     Probability:  5.55 %     Energy:  6.0
State  110110     Probability:  3.9 %     Energy:  6.0
State  001001     Probability:  3.9 %     Energy:  6.0
State  100100     Probability:  3.39 %     Energy:  6.0
State  011011     Probability:  3.39 %     Energy:  6.0
State  110010     Probability:  2.92 %     Energy:  5.0
State  010110     Probability:  2.92 %     Energy:  5.0
State  101001     Probability:  2.92 %     Energy:  5.0
State  001101     Probability:  2.92 %     Energy:  5.0
```

Comparing the results between the two runs, which both use the same mixing operator, the plot above shows that the optimal $F(\gamma,\beta)$ value for the $p = 1$ case is around $5.29$. By comparison, this value is *lower* than the one we obtained for the $p = 2$ case where we picked our starting parameters at random. Thus, this MaxCut problem has demonstrated the solving potential one can obtain through increasing the order of $p$, effectively unlocking new sections of the problem's Hilbert Space. I encourage you to try higher values of $p$ and other mixing operators, as well as various values for $\epsilon$ and $\mathrm{step_size}$ to see the different kinds of solutions QAOA can produce!

This concludes lesson $10$, and our deep dive into the Quantum Approximate Optimization Algorithm. We have now seen the first of two hybrid quantum algorithms which will be covered in this series, both of which draw their strength from encoding various optimization problems into quantum, which can then be solved with classical techniques. For our QAOA lesson here, the real workhorse that powers the algorithm is the way in which we are able to convert classical optimization problems into the $\left[\gamma, \beta\right]$ space through the phase and mixing operators. The speedup potential for this algorithm effectively comes from converting a search through classical graph combinations to one through $|\Psi_{\gamma\beta}\rangle$ states. For certain problems where the best classical technique is simply an exhaustive search, QAOA allows us to convert the problem to a space where better classical optimizers can be implemented.

I hope you enjoyed this lesson, and I encourage you to take a look at my other .ipynb tutorials!

## Citations

**[1]** E. Farhi, J. Goldstone, S. Gutmann, "A Quantum Approximate Optimization Algorithm", arXiv:1411.4028 (2014)

**[2]** H. B. Curry, "The Method of Steepest Descent for Non-linear Minimization Problems". Quart. Appl. Math. **2** : 3 (1944)

# Lesson 11 - Variational Quantum Eigensolver

In this final lesson, we will be taking a look at one of the most promising quantum algorithms in current research, the Variational Quantum Eigensolver Algorithm $\mathbf{[1]}$. As the name suggests, the goal of the algorithm is to find the eigenvalue(s) of some matrix, representing the Hamiltonian of some system. The strength of the VQE Algorithm, and the reason it has gained so much popularity so quickly, is the flexibility of the algorithm, allowing for a wide range of solvable problems. Similar to QAOA from the previous lesson, VQE translates classically challenging problems into a quantum space where advanced classical optimization techniques can then be used to solve the problem much faster.

In order to make sure that all cells of code run properly throughout this lesson, please run the following cell of code below:

In [2]:

```
from qiskit import ClassicalRegister, QuantumRegister, QuantumCircuit, Aer, execute
import Our_Qiskit_Functions as oq
import numpy as np
import math as m
import scipy as sci
import random
import matplotlib
import matplotlib.pyplot as plt
from itertools import permutations
S_simulator = Aer.backends(name='statevector_simulator')[0]
```

## Hamiltonians and Problem Complexity

Before diving into the Variational Quantum Eigensolver (VQE) Algorithm, we will start this lesson with a brief explanation about the kinds of problems that it promises to solve, and their importance. To begin, let's quickly review the role of a Hamiltonian in physics:

$$H(q,p)$$

$$\begin{aligned} q &\equiv \text{generalized position} \\ p &\equiv \text{generalized momentum} \end{aligned}$$

The Hamiltonian is an equation which describes the energy of a physical system in terms of position and momentum. For example, imagine we would like to model a system of many particles and compute the total energy of our system. In principle, if we knew the exact location and momentum of every single particle then we have everything needed to calculate the total energy of the system. With complete knowledge of position and momentum, we can compute all of the complexities associated with any particular physics model:

$$H \;=\; \sum_i p_i \dot{q}_i - L$$

where $L$ is the Lagrangian of the system:

$$L \;=\; T - V$$

$$\begin{aligned} T &\equiv \text{Kinetic Energy} \\ V &\equiv \text{Potential Energy} \end{aligned}$$

The important thing to note in the formulation above is that the Hamiltonian for a given system incorporates the associated Lagrangian, which in turn is composed of the kinetic and potential energy models for that system. In physics, kinetic energy is typically a much more straightforward concept to model (imagine all of your particles are simply tiny billiard balls: $T = \frac{1}{2}mv^2$), but potential energy is where things can quickly become complicated. For example, consider one of the simplest potential interactions between particles, one solely dependent on distances:

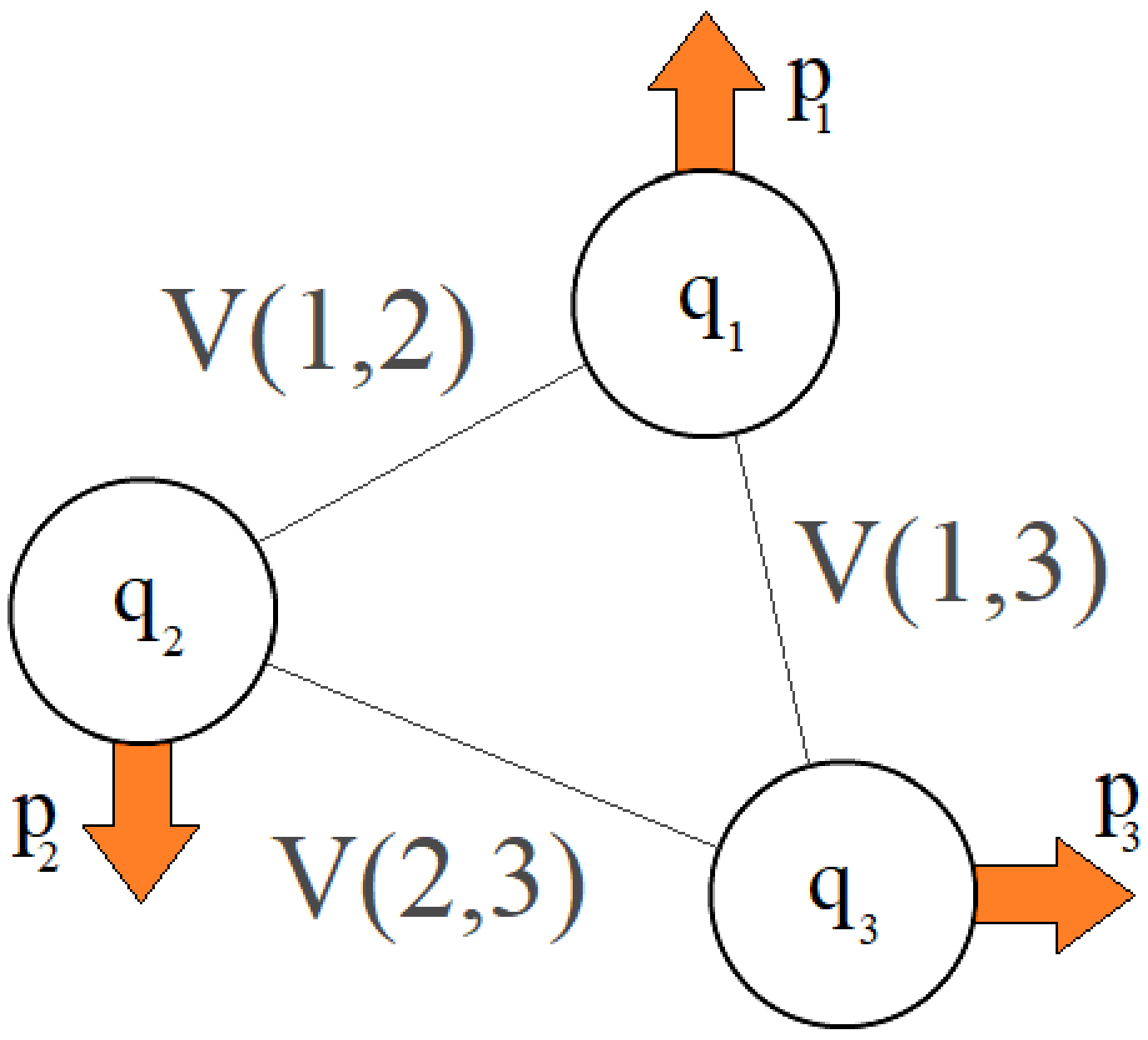

In the diagram above, we can see that all three particles in our system have their respective positions and momentum, $q_i, p_i$. Additionally, each particle in the system feels a force associated with the potential energy between every other particle in the system, $V(i, j)$. In this particular model, we can imagine that the potential energy between each particle pair is solely dependent on their positions, $V(q_i, q_j)$ (In more realistic models, this potential could also be dependent on things such as mass, charge, magnetic field strength, etc.). If we were to then compute the total potential energy of the system, we would need to sum together the potential energy of each particle-particle interaction:

$$V_{tot} = \sum_{i,j} V(q_i, q_j) \qquad (i \neq j)$$

Given the simple diagram shown above, the important thing to note is that each $q_i$ in the summation above is a continuous parameter (same for $p_i$), which technically means we can come up with an infinite number of combinations to solve. So then, given all possible arrangements, what sorts of quantities are of interest to us as physicists?

## Ground State Energy

When studying a problem like the diagram shown above, consisting of a system of particles subject to some potential energy dependent on all of their positions and momentums, one thing we could ask is: what is the combination of all $q_i$'s and $p_i$'s which yields the smallest energy? One motivation for looking for the smallest possible energy, often referred to as the ground state, is that in most cases there is in principle only one correct answer. By contrast, if we were to look for combinations of higher energy, there could in principle be an infinite number of solutions.

As another motivation for why ground state energies are of interest, it turns out that there are no shortage of challenging and worthwhile problems across the various sciences which seek the ground state of complex systems. These problems range from planetary motion to subatomic particles, and most every stop in between. For astrophysicists, determining ground state energies of distant solar systems can provide valuable insight into the potential for the earth-like planets. Meanwhile, for biochemists, calculating ground state energies of complex molecules can lead to breakthrough discoveries for new medicines.

Now that you're sufficiently convinced that the problem of solving for ground states is a worthwhile endeavor, let's take a look at a simple example which will help illustrate the difficulty classical computers face when solving such problems. Recall our example diagram from earlier, whereby we had three particles subject to a potential energy which is position dependent only. We will now investigate a similar problem in which the potential energy formula between any two particles is dependent on both position and mass:

$$V(m_1, m_2, d) = |m_1 - m_2| \cdot 2d^4 - |m_1 + m_2| \cdot 10d^2 + 25$$

For our example problem, we will use the following masses shown below, which give rise to the corresponding potential energy plots as a function of distance:

$$m_1 = 1 \qquad m_1 = 2 \qquad m_1 = 3$$

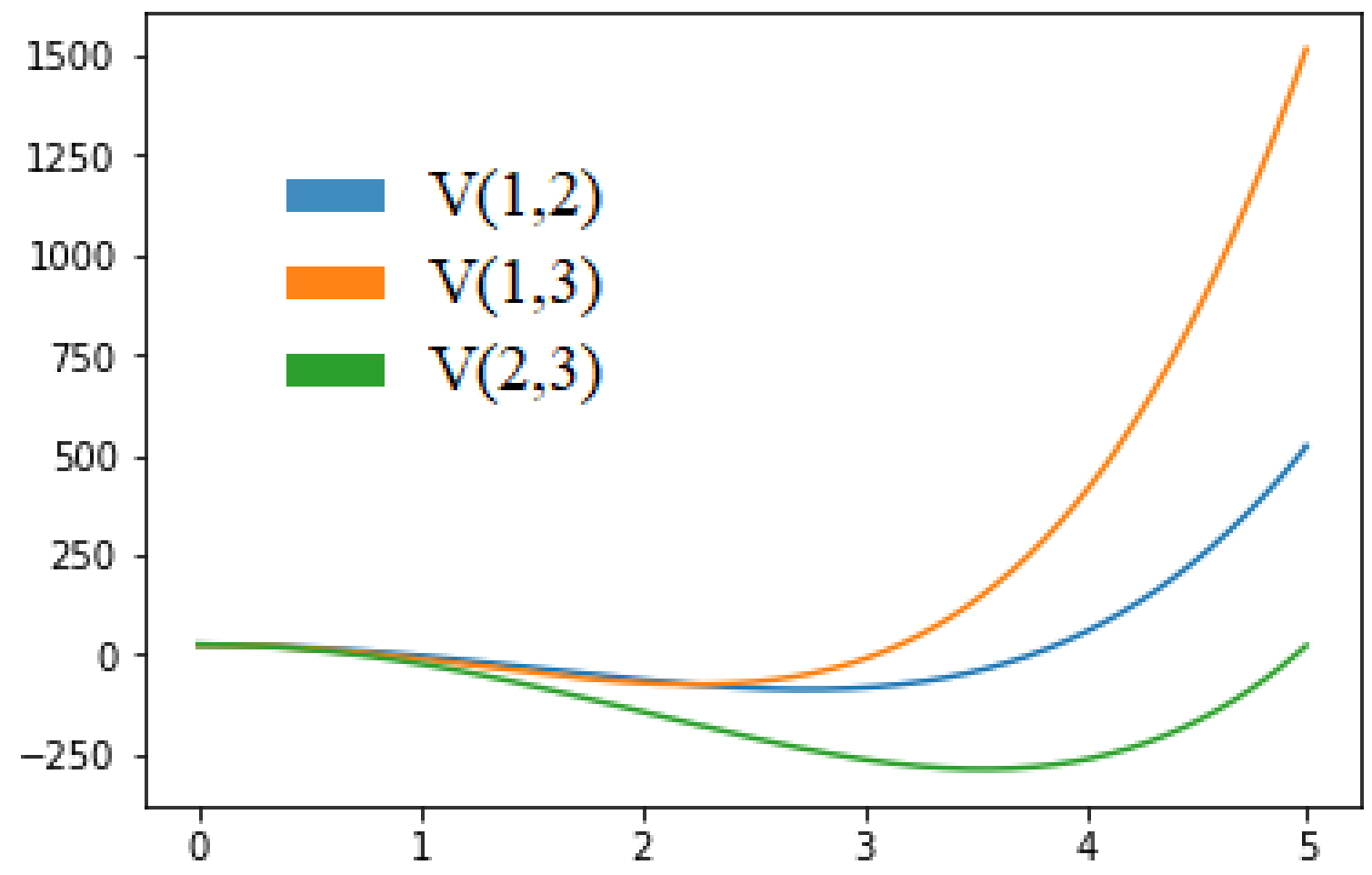

And now for the final piece to the problem: we will look to solve for the ground state energy of this system under the constraint that the total distance between all three particles is $D$. Defining the distance between any two particles as $d_{ij}$, we have the following condition for our system:

$$d_{12} + d_{13} + d_{23} = D$$

The constraint shown above represents the kinds of challenges that arise in real physical systems. Without the constraint, solving this problem becomes a trivial exercise of simply finding the minimum of each plot and summing them up. However, the presence of this constraint forces us to consider the interplay between setting the distances of the three particles. For example, the plot for $V(2, 3)$ has the potential for the largest negative energy contribution, thus suggesting that we should try and prioritize picking an optimal distance between particles $2$ and $3$. Conversely, the plot for $V(1, 3)$ illustrates that we must be cautious not to let $d_{13}$ become too large, otherwise the overwhelming positive energy contribution will wash away any advantages we gain from the other two.

Let's now see how to solve this problem classically through brute force means:

In [2]:
```
def E(m1,m2,q):
    return abs(m1-m2)*2*q**4 - abs(m1+m2)*10*q**2 + 25
#====================================================
D = 10
M = [1,2,3]
GS = [999,0,0,0]

N = 100
for i in np.arange(N):
    d12 = round(i/N,4)*D
    for j in np.arange(N):
        d13 = round((j/N)*(D-d12),4)
        d23 = round((D-d12-d13),4)
        Etotal = E(M[0],M[1],d12) + E(M[0],M[2],d13) + E(M[1],M[2],d23)
        if( Etotal < GS[0] ):
            GS =  [Etotal,d12,d13,d23]

print('Total Particle Distance:  ',D,'       Particle Masses      m1: ',M[0],'   m2: ',M[1],'   m3: ',M[2])
print('\nMinimal Total Energy: ',round(GS[0],4))
print('\nDistances Between Particles:       1-2: ',round(GS[1],3),'        1-3: ',round(GS[2],3),'        2-3: ',round(GS[3
print('\nEnergy Contributions:              1-2: ',round(E(M[0],M[1],GS[1]),2),'        1-3: ',round(E(M[0],M[2],GS[2]),2),
```

```
Total Particle Distance:   10        Particle Masses      m1:  1    m2:  2    m3:  3

Minimal Total Energy:  -381.5408

Distances Between Particles:       1-2:  3.3         1-3:  2.747         2-3:  3.953

Energy Contributions:              1-2:  -64.52         1-3:  -49.07         2-3:  -267.95
```

The cell of code above uses for-loops to iterate over the possible combinations of $[d_{12}, d_{13}, d_{23}]$, calculating the total energy from each combination and storing the smallest energy found. Then, after exhausting every possible energy configuration for the system (within the accuracy allowed by the for-loop), what we are left with is our ground state energy.

The classical approach shown above represents the most naive methodology for finding the ground state, simply looking at every possible combination in the system. As you might imagine, this technique is very limited in the sizes of problems it can tackle, mostly because a huge portion of the computational workload is spent testing combinations which are nowhere near the correct solution. As we've already seen in the previous lesson with gradient descent, more advanced techniques exist which can drastically improve optimization speeds by only searching in regions where a correct solution is more likely to exist. In this lesson we will once again be using such advanced optimization techniques, in combination with what our quantum system has to offer through VQE.

## The Variational Principle

The core idea behind the VQE Algorithm comes from the Variational Principle, for which the "V" in our algorithm gets its name. Simply put, the Variational Principle states that there is a single quantum state for which our system will yield the lowest possible energy expectation value: $|\Psi_{\text{min}}\rangle$

$$E_{\text{min}} \;=\; \langle \Psi_{\text{min}} | H | \Psi_{\text{min}} \rangle$$

If we prepare any other quantum state: $|\Psi(\vec{\theta})\rangle$, then we are guaranteed that the resulting energy expectation value will be higher than $E_{\text{min}}$:

$$E_{\text{min}} \;\leq\; E(\vec{\theta}) \;=\; \langle \Psi(\vec{\theta}) | H | \Psi(\vec{\theta}) \rangle$$

If this looks familiar to you, that's because it's essentially the same driving force behind QAOA from the previous lesson. In QAOA, we were guaranteed that by searching for the minimal or maximal expectation value in the $\left[\gamma, \beta\right]$ space, we would find the state which represented the solution to our particular problem (for example, the Ising Energy Model problem consisting of arranging particle spins). Similarly here, VQE is based around searching for the optimal state $|\Psi_{\text{min}}\rangle$ amongst all possible $|\Psi(\vec{\theta})\rangle$ states, commonly referred to as the ansatz states:

$$\text{solution:} \quad |\Psi(\vec{\theta})\rangle \;=\; |\Psi_{\text{min}}\rangle$$

Just like with QAOA, searching through the parameter space $\vec{\theta}$ comes with its own set of challenges. For starters, how do we know when we've found $|\Psi_{\text{min}}\rangle$? Supposing we are using VQE to solve a problem for which the ground state energy is unknown, how can we be certain that the best $|\Psi(\vec{\theta})\rangle$ we find is truly $|\Psi_{\text{min}}\rangle$. Unfortunately, our algorithm can't tell us that, which means that the smallest energy expectation value we find is simply the best approximation our algorithm can provide. However, one positive aspect of VQE as compared to QAOA is that every solution we find which yields a lower energy provides us with a new upper bound on the ground state energy for the given Hamiltonian. Thus, after running VQE, the best solution we find *could* be the ground state, but at the very least we can say that the true ground state energy of the system cannot be larger than our best $|\Psi(\vec{\theta})\rangle$. In turn, providing such information back to a classical computer could immensely help narrow down the search, ultimately leading to a completion of the problem which would have been otherwise impossible for a classical computer alone.

### Single Qubit Ansatz States

Having just covered how our VQE Algorithm will be driven by the Variational Principle, let's look at how to create ansatz states in more detail. For a comparison, recall from the QAOA lesson how the mixing operator $U(\mathbf{B},\beta)$, and order $p$ were largely responsible for determining the size of the parameter space in which our solution could lie. For QAOA, we saw that spanning the complete Hilbert Space for a $2^N$ dimensional problem wasn't necessary to solve the problem. Conversely, here our solution is dependent on finding $|\Psi_{\text{min}}\rangle$, which in principle could be anywhere within an $N$-qubit Hilbert space, so we must consider all possible $|\Psi(\vec{\theta})\rangle$ states.

As our first example, we will consider the case of a single qubit, and the quantum circuit necessary for spanning the complete $1$-qubit Hilbert Space. Having already covered this topic in the previous lesson, feel free to skip this section if you are already comfortable with the result previously derived for QAOA. Otherwise, if we want to consider all possible quantum states that a single qubit can occupy, we needn't look any further than the Bloch Sphere:

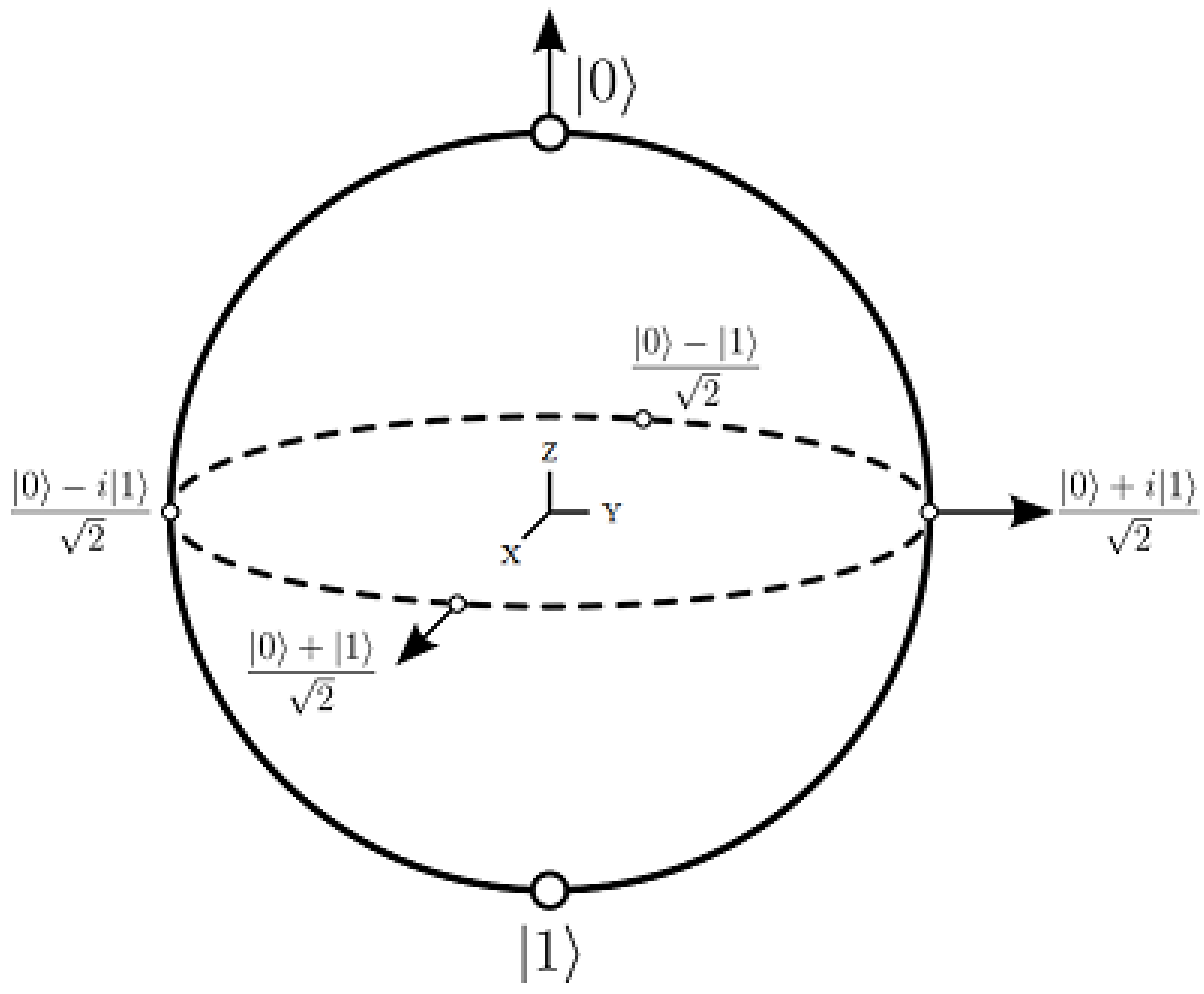

As shown above, the Bloch sphere is a great visual representation of the space of a single qubit. Specifically, the surface of the Bloch Sphere represents all possible $|\Psi(\vec{\theta})\rangle$ states we need for our ansatz. The question then becomes, given some arbitrary location on the surface of the Bloch Sphere, how do we create the corresponding quantum state? Or stated in another way, supposing we start our qubit in the $|0\rangle$ state, what gate operations will allow us to transform to all possible states $|\Psi(\vec{\theta})\rangle$.

The nice thing about the space of all possible single qubit states, and the reason why the Bloch Sphere representation is so great, is that the answer to our question comes from a simple geometric answer: any position along the surface of a unit sphere can be specified using two coordinates. More specifically, starting from the $|0\rangle$ state, we can transform to any possible state using two rotation operators:

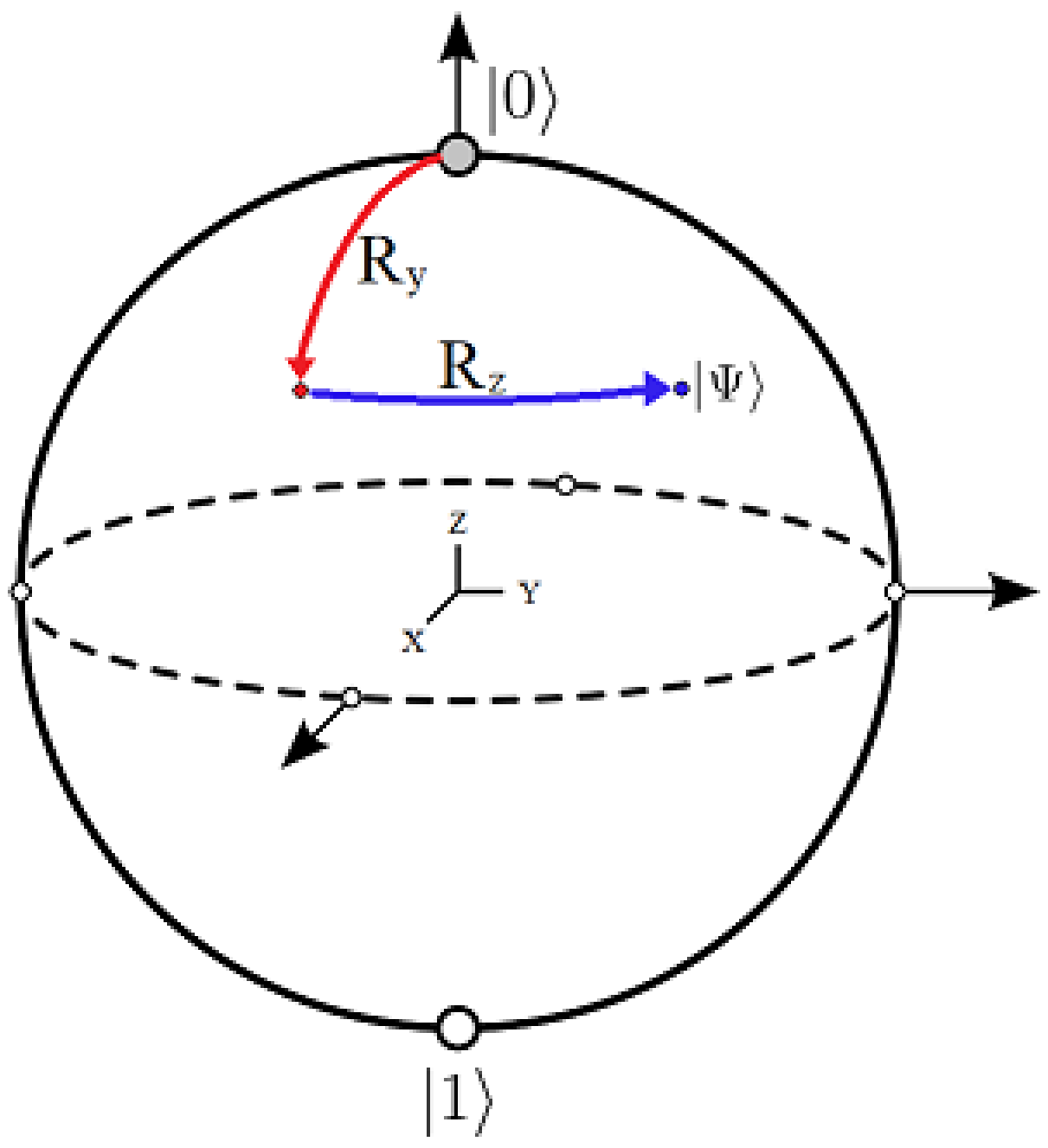

The figure above shows how one can use the $R_y$ and $R_z$ gates to transform our initial $|0\rangle$ state to anywhere along the surface of the Bloch Sphere (In the QAOA lesson we used $R_x$ and $R_y$, but any combination of the three will do). For completeness, below is an outline of the combination of these two operators in succession:

$$R_y(\theta) \;=\; \begin{bmatrix} \cos(\frac{\theta}{2}) & -\sin(\frac{\theta}{2}) \\ \sin(\frac{\theta}{2}) & \cos(\frac{\theta}{2}) \end{bmatrix} \qquad\qquad R_z(\phi) \;=\; \begin{bmatrix} e^{i\frac{\phi}{2}} & 0 \\ 0 & e^{-i\frac{\phi}{2}} \end{bmatrix}$$

$$\begin{aligned} U(\theta,\phi) \;&\equiv\; R_z(\phi) \otimes R_y(\theta) \\ &=\; \begin{bmatrix} e^{i\frac{\phi}{2}}\cos(\frac{\theta}{2}) & -e^{i\frac{\phi}{2}}\sin(\frac{\theta}{2}) \\ e^{-i\frac{\phi}{2}}\sin(\frac{\theta}{2}) & e^{-i\frac{\phi}{2}}\cos(\frac{\theta}{2}) \end{bmatrix} \end{aligned}$$

With this operation, we can create the full space of single-qubit ansatz states:

$$|\Psi(\vec{\theta})\rangle \;=\; U(\theta,\phi)|0\rangle \;=\; \cos(\theta/2)|0\rangle \;+\; e^{-i\phi}\sin(\theta/2)|1\rangle$$

$$\theta \in [0,\pi] \qquad\qquad \phi \in [0,2\pi]$$

Having now derived our ansatz state $|\Psi(\vec{\theta})\rangle$, let's discuss its components in a little more detail. As advertised, we can specify any location on the Bloch Sphere using only two parameters: a $\theta$ rotation from a $R_y$ gate followed by a $\phi$ rotation from $R_z$. Looking a little closer, we can see the role each of these parameters plays in our final state. Specifically, $\theta$ controls the relative amplitudes between the $|0\rangle$ and $|1\rangle$ components of $|\Psi(\vec{\theta})\rangle$, while the $\phi$ parameter controls the relative phase between them. Together, they span the complete space of possible states for a single qubit, which in turn is guaranteed to encompass the ground state $|\Psi_{min}\rangle$ for a given Hamiltonian.

Below is a simple coding example that creates an arbitrary ansatz state:

In [46]:
```
def Single_Qubit_Ansatz( qc, qubit, params ):
    qc.ry( params[0], qubit )
    qc.rz( params[1], qubit )

#================================================
q = QuantumRegister( 1, name='q' )
qc= QuantumCircuit( q, name='qc' )
theta = m.pi/3
phi = 3*m.pi/2

print('___ Initial State ___')
oq.Wavefunction( qc )

Single_Qubit_Ansatz( qc, q[0], [theta,phi] )

print('\n___ After U(\u03B8,\u03C6) ___')
oq.Wavefunction( qc )
print('  ')
print(qc)
```

```
___ Initial State ___
1.0 |0>

___ After U(θ,ϕ) ___
0.86603 |0>    -0.5j |1>

     ┌───────────┐┌────────────┐
q_0: ┤ RY(pi/3)  ├┤ RZ(3pi/2)  ├
     └───────────┘└────────────┘
```

As shown in the cell above, the function **Single_Qubit_Ansatz** handles the prepartion of our single qubit state given some parameter values for $\theta$ and $\phi$. Later in this lesson we will cover higher dimensional ansatz preparation, but for the coming discussions this single qubit function will be enough to implement our first VQE examples.

## Hamiltonian Decomposition and the Measurement Basis

With our ansatz state operator now in hand, the next topic for our VQE Algorithm is how to measure expectation values. As a reminder, the quantity of interest for which we are searching for is the energy expectation value, shown below:

$$E(\vec{\theta}) \;\;=\;\; \langle \Psi(\vec{\theta}) \,|\, H \,|\, \Psi(\vec{\theta}) \rangle$$

In the previous lesson, we briefly touched on the topic of using repeat measurements to determine expectation values, but ultimately chose to implement our QAOA examples via the exact wavefunctions. Simply put, if we are working with a real quantum computer, then the only means of computing the expectation value of a quantum state is to prepare and measure the state many times. Through this repetition we can build up the approximate probabilities of each individual state, and then multiply each state's weight with its associated energy contribution.

To begin our discussion, let's consider a single qubit system whose Hamiltonian is as follows:

$$H \;\;=\;\; X + Z$$

Not particularly exciting, but there is a lot to learn from this simple Hamiltonian. For starters, let's consider what our energy expectation value for this system will look like:

$$\begin{aligned} E(\vec{\theta}) \;\;&=\;\; \langle \Psi(\vec{\theta}) \,|\, X \,+\, Z \,|\, \Psi(\vec{\theta}) \rangle \\ &=\;\; \langle \Psi(\vec{\theta}) \,|\, X \,|\, \Psi(\vec{\theta}) \rangle \;+\; \langle \Psi(\vec{\theta}) \,|\, Z \,|\, \Psi(\vec{\theta}) \rangle \\ &=\;\; E_X(\vec{\theta}) \;+\; E_Z(\vec{\theta}) \end{aligned}$$

As shown above, when we have a Hamiltonian that can be expressed as a linear combination of terms, the total energy expectation value is equal to the sum of the individual expectation values. For our VQE Algorithm this is very important, as it means that we can study arbitrarily long Hamiltonians so long as we can decompose them linearly. However, there is a limitation to the kinds of Hamiltonians we can study with VQE, namely the operators which compose them. In our example $H$ above, the two elements making up this Hamiltonian are both Pauli Operators, and for good reason. As a quick refresher, below are the three Pauli Operators and their eigenstates:

$$X \;=\; \begin{bmatrix} 0 & 1 \\ 1 & 0 \end{bmatrix} \qquad\qquad Y \;=\; \begin{bmatrix} 0 & -i \\ i & 0 \end{bmatrix} \qquad\qquad Z \;=\; \begin{bmatrix} 1 & 0 \\ 0 & -1 \end{bmatrix}$$

$$\text{eigenstates}: \qquad\qquad \frac{|0\rangle \,\pm\, |1\rangle}{\sqrt{2}} \qquad\qquad \frac{|0\rangle \,\pm\, i\,|1\rangle}{\sqrt{2}} \qquad\qquad |0\rangle, |1\rangle$$

where the eigenvalues associated with each of the Pauli Operators is $\pm 1$. Plugging these eigenvalues and eigestates into our equation for $E(\vec{\theta})$, we can write out the full expression for the energy expectation value of our Hamiltonian:

$$E(\vec{\theta}) \;\;=\;\; (+1)\,|\langle + | \Psi(\vec{\theta}) \rangle|^2 \;\;+\;\; (-1)\,|\langle - | \Psi(\vec{\theta}) \rangle|^2 \;\;+\;\; (+1)\,|\langle 0 | \Psi(\vec{\theta}) \rangle|^2 \;\;+\;\; (-1)\,|\langle 1 | \Psi(\vec{\theta}) \rangle|^2$$

In the expression above we've written the shorthand form of the $X$ operator's eigenstates: $|\pm\rangle = \frac{|0\rangle \pm |1\rangle}{\sqrt{2}}$

Take a careful look at our energy expectation value above, and note the four unique terms and their physical interpretations. For example, the quantity $(+1)|\langle 0|\Psi(\vec{\theta})\rangle|^2$ represents the probability of measuring the $|0\rangle$ state, multiplied by its associated energy eigenvalue of $+1$. If asked to compute this quantity through repeat trials, we would take the number of $|\Psi(\vec{\theta})\rangle$ state measurements and divide by the total number of trials. This process gives us the value of $|\langle 0|\Psi(\vec{\theta})\rangle|^2$, for which we would them simply multiply by the eigenvalue $(+1)$ for our answer.

Because we can seperate our problem into the two quantities $E_X(\vec{\theta})$ and $E_Z(\vec{\theta})$, we can think of each of them as their own seperate problem, each requiring the preparation and measurement process just described. For the two quantities stemming from the $Z$ component of our Hamiltonian, determining the probabilities for $|0\rangle$ and $|1\rangle$ are straightforward, but how about the other two? If asked to compute the quantity $|\langle -|\Psi(\vec{\theta})\rangle|^2$ for example, how would we go about it on a real quantum computer (no looking at the wavefunction!)? To emphasize why this is a problem, remember that we are always bound to a measurement at the end of the day, but more specifically a measurement in the computational basis $\{|0\rangle, |1\rangle\}$.

To motivate why the quantity above in question is problematic for our computational basis bound quantum computer, consider the example below. Let's suppose we wanted to compute $|\langle +|\Psi(\vec{\theta})\rangle|^2$ through measurements on the state $|\Psi(\vec{\theta})\rangle$:

$$25,000 \text{ total measurements}$$

$$\text{Results:} \qquad |0\rangle: \ 9000 \qquad |1\rangle: \ 16000$$

Based on these measurement results, we can create the following approximate form for $|\Psi(\vec{\theta})\rangle$:

$$|\Psi(\vec{\theta})\rangle \approx \frac{3}{5}|0\rangle + e^{i\theta}\frac{4}{5}|1\rangle$$

The important thing to note in the equation above is the presence of $e^{i\theta}$, which represents the potential for a relative phase between the two basis states. Remember that when viewing the measurement results of a quantum state, phase terms in the wavefunction are essentially washed out, leaving only the magnitude of the amplitudes to determine probabilities. Thus, based on the measurement results in the example above, we can only approximate the amplitudes between $|0\rangle$ and $|1\rangle$ up to an unknown relative phase.

Returning now to our problem from earlier then, computing $|\langle +|\Psi(\vec{\theta})\rangle|^2$, let's see what happens when we try to use our approximate state to compute the expectation value:

$$|0\rangle = \frac{|+\rangle + |-\rangle}{\sqrt{2}} \qquad |1\rangle = \frac{|+\rangle - |-\rangle}{\sqrt{2}}$$

$$|\Psi(\vec{\theta})\rangle \approx \frac{3}{5}\frac{|+\rangle + |-\rangle}{\sqrt{2}} + e^{i\theta}\frac{4}{5}\frac{|+\rangle - |-\rangle}{\sqrt{2}}$$

$$= \frac{3+4e^{i\theta}}{5\sqrt{2}}|+\rangle + \frac{3-4e^{i\theta}}{5\sqrt{2}}|-\rangle$$

After rewriting our state in the $\{|+\rangle, |-\rangle\}$ basis, we can once again see the presence of this unknown relative phase term. Depending on the value of $\theta$, the amplitudes associated with the $|+\rangle$ and $|-\rangle$ components of our state can drastically change. So much so in fact, that if we consider the two extremes: $\theta = 0$ or $\pi$, the probability of measuring the $|+\rangle$ state can range anywhere from 2% to 98%! Thus, without proper knowledge of $\theta$, there is no way of determining the quantities $|\langle +|\Psi(\vec{\theta})\rangle|^2$ and $|\langle -|\Psi(\vec{\theta})\rangle|^2$ with measurements limited to the $\{|0\rangle, |1\rangle\}$ basis (at least not yet).

Hopefully the example above is enough to convince you of the problem at hand: determining expectation values which depend on eigenstates other than $|0\rangle$ and $|1\rangle$. Because we lose such a critical piece of information in phases when we measure, it prevents us from being able to reconstruct the true underlying wavefunction. However, fear not, as there is yet another way we can compute our quantity of interest, namely the overlap of the $|+\rangle$ state with our ansatz state. To do this, we will cleverly apply an additional gate just prior to the measurement, essentially transforming our measurement results from the computational basis to that of the eigenstate of our Hamiltonian. For the case of our Hamiltonian $H = X$, the required transformation is best visualized once again using the Bloch Sphere:

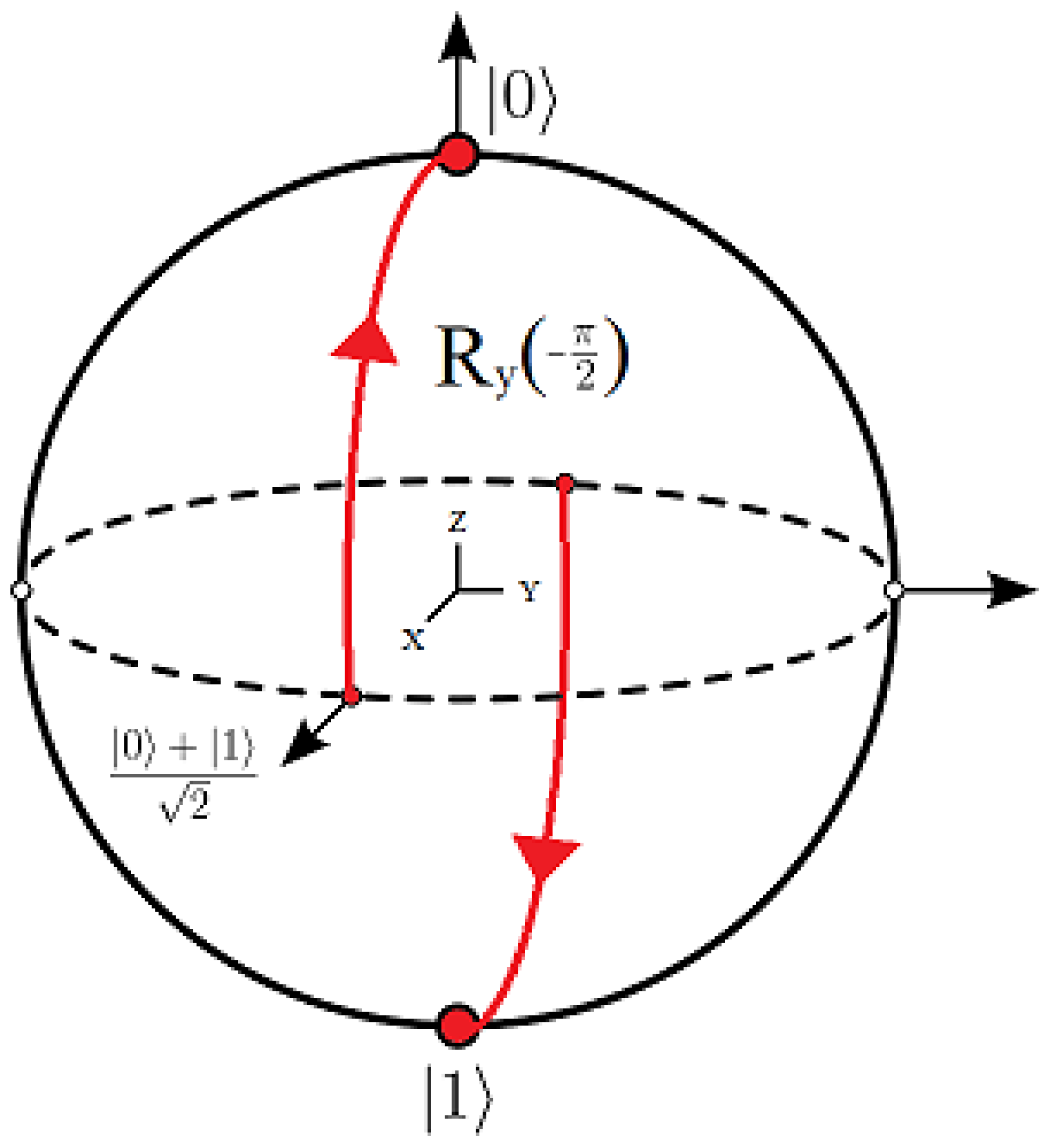

To compliment what the diagram above is illustrating, the $R_y(-\frac{\pi}{2})$ rotation gate transforms between the two sets of states: $\{|0\rangle, |1\rangle\}$ and $\{|+\rangle, |-\rangle\}$. Thus, by applying this gate just before our measurement, we will able to compute our energy expectation value in the $\{|+\rangle, |-\rangle\}$ basis. To show this is indeed true, consider the example below:

$$H = X$$

$$|\Psi\rangle \;=\; \frac{1}{2}(\,|+\rangle \;+\; \sqrt{3}|-\rangle\,)$$

Here we will compute the energy expectation value in the $\{\,|+\rangle, |-\rangle\,\}$ basis:

$$\begin{aligned} E_\Psi &= \langle\Psi|H|\Psi\rangle \\ &= \frac{1}{2}\langle\Psi|X(\,|+\rangle \;+\; \sqrt{3}|-\rangle\,) \\ &= \frac{1}{2}\langle\Psi|(\,|+\rangle \;-\; \sqrt{3}|-\rangle\,) \\ &= \frac{1}{4}\langle+|+\rangle \;-\; \frac{3}{4}\langle-|-\rangle \\ &= -\frac{1}{2} \end{aligned}$$

And now we show that we can arrive at the same value through a $R_y(-\frac{\pi}{2})$ rotation, followed by a measurement in the $\{\,|0\rangle, |1\rangle\,\}$ basis:

$$R_y(-\frac{\pi}{2})|\Psi\rangle \;\equiv\; |\Psi'\rangle \;=\; \frac{1}{2}(\,|0\rangle \;+\; \sqrt{3}|1\rangle\,)$$

$$\begin{aligned} E_{\Psi'} &= (+1)|\langle 0|\Psi'\rangle|^2 \;+\; (-1)|\langle 1|\Psi'\rangle|^2 \\ &= \frac{1}{4}\langle 0|0\rangle \;-\; \frac{3}{4}\langle 1|1\rangle \\ &= -\frac{1}{2} \end{aligned}$$

As promised, we arrive at the same energy expectation value. The important concept here is that using a $R_y(-\frac{\pi}{2})$ rotation just before a measurement essentially changes the way we interpret our measurement results. In doing so, the statistics gathered by measuring the $|0\rangle$ and $|1\rangle$ states now represent the eigenstates $|+\rangle$ and $|-\rangle$, and their corresponding eigenvalues. Below is a coding example showing the difference in results obtained by computing the energy expectation value in the two differing bases:

In [10]:

```
Shots = 10000
#==========================================
q = QuantumRegister(1,name='q')
c = ClassicalRegister(1,name='c')
qc= QuantumCircuit(q,c,name='qc')

qc.initialize( [m.sqrt(1/4),m.sqrt(3/4)], q[0] )
print('              ___ Ansatz State ___')
oq.Wavefunction( qc )
qc.measure(q,c)
M1 = oq.Measurement( qc, shots=Shots, print_M=False, return_M=True )
print( '\n{ |0> , |1> } Basis - Energy Expectation Value:   ',round( (M1['0']/Shots)+(-1.0*M1['1']/Shots) ,3) )

#==========================================

q2 = QuantumRegister(1,name='q2')
c2 = ClassicalRegister(1,name='c2')
qc2= QuantumCircuit(q2,c2,name='qc2')

qc.initialize( [m.sqrt(1/4),m.sqrt(3/4)], q[0] )
qc2.ry( -m.pi/2,  q2[0] )
qc2.measure(q2,c2)
M2 = oq.Measurement( qc2, shots=Shots, print_M=False, return_M=True )
print( '\n{ |+> , |-> } Basis - Energy Expectation Value:   ',round( (M2['0']/Shots)+(-1.0*M2['1']/Shots) ,3) )
```

```
           ___ Ansatz State ___
0.5 |0>    0.86603 |1>

{ |0> , |1> } Basis - Energy Expectation Value:    -0.514

{ |+> , |-> } Basis - Energy Expectation Value:    0.006
```

Additionally, this process of changing bases works for higher dimensional systems as well, so long as each qubit receives the proper rotation just prior to the final measurement. In fact, each qubit can receive a different rotation based on the Hamiltonian, for example:

$$H \;=\; X \otimes Y \qquad\qquad |\Psi\rangle \;=\; \alpha|-\rangle|+i\rangle \;+\; \beta|+\rangle|-i\rangle$$

$$R_y(-\frac{\pi}{2}) \otimes R_x(\frac{\pi}{2})|\Psi\rangle \;=\; \alpha|1\rangle|0\rangle \;+\; \beta|0\rangle|1\rangle$$

In the example above we've used the shorthand notation for the $Y$ operator's eigenstates: $|\pm i\rangle$. To emphasize what this example is demonstrating, so long as we have a Hamiltonian that is composed of Pauli operations on each qubit, we can compute the total energy expectation value by applying the necessary rotations on each qubit just prior to the measurement. These rotations correspond to each qubit's particular Pauli Operator in the Hamiltonian, and are given

below:

$$X \rightarrow R_y(-\frac{\pi}{2})$$

$$Y \rightarrow R_x(\frac{\pi}{2})$$

$$Z \rightarrow \text{no rotation}$$

Note that for the case of the $Z$ operator no rotation is necessary. This is because the eigenstate basis of the $Z$ operator is already $\{ |0\rangle, |1\rangle \}$.

In light of the examples provided in this section, demonstrating how to correct for a Hamiltonian composed solely of Pauli operators, it is important to note that the technique of changing bases right before a measurement is applicable to operators beyond just $X$, $Y$, and $Z$. In principle, if we wanted to compute the quantities $|\langle S_{\pm} | \Psi(\vec{\theta}) \rangle|^2$, corresponding to some operator $S$, then all we need is a second operation which transforms between the eigenstates of $S$ and the computational basis:

$$U_S^{\dagger} \begin{cases} |S_+\rangle \longrightarrow |0\rangle \\ |S_-\rangle \longrightarrow |1\rangle \end{cases}$$

$$|\Psi(\vec{\theta})\rangle = \alpha |S_+\rangle + \beta |S_-\rangle$$

$$|\langle S_+ | \Psi(\vec{\theta}) \rangle|^2 = \alpha^2 \qquad |\langle 0 | U_S^{\dagger} | \Psi(\vec{\theta}) \rangle|^2 = \alpha^2$$

So long as we know how to properly transform our quantum state to the computational basis just before the measurement, we can evaluate the inner product squared of any eigenbasis. However, working with $U_S^{\dagger}$ operators may be tricky, or sometimes even unimplementable. Luckily for us, there is a well known mathematical result that states that any hamiltonian $H$ can be decomposed into soley Pauli operators. We won't cover the derivation of this result here (as the math is a bit cumbersome, and not really necessary for our understanding of VQE), but I encourage you to check out additional resources if you're interested. In terms of our VQE Algorithm, it means that knowing how to handle the transformation and sampling of the three Pauli operators is enough to cover all hamiltonians. Thus, once we've finished our complete discussion of how the VQE Algorithm works, know that we can in principle take our understanding and apply it to any hamiltonian we want!

## The VQE Workflow

So far we've covered how to prepare our single qubit ansatz state, as well as compute energy expectation values for Hamiltonians composed of Pauli Operators, thus completing all the necessary tools needed for our first complete VQE run! Just like QAOA however, an important idea to understand about the VQE Algorithm is that it is fundamentally a hybrid algorithm. This means that in order to be successful, it requires synergy between the classical/quantum components. The figure below shows a rough outline of the roles each computer plays, as well as the information obtained through each and fed into the other for further processing.

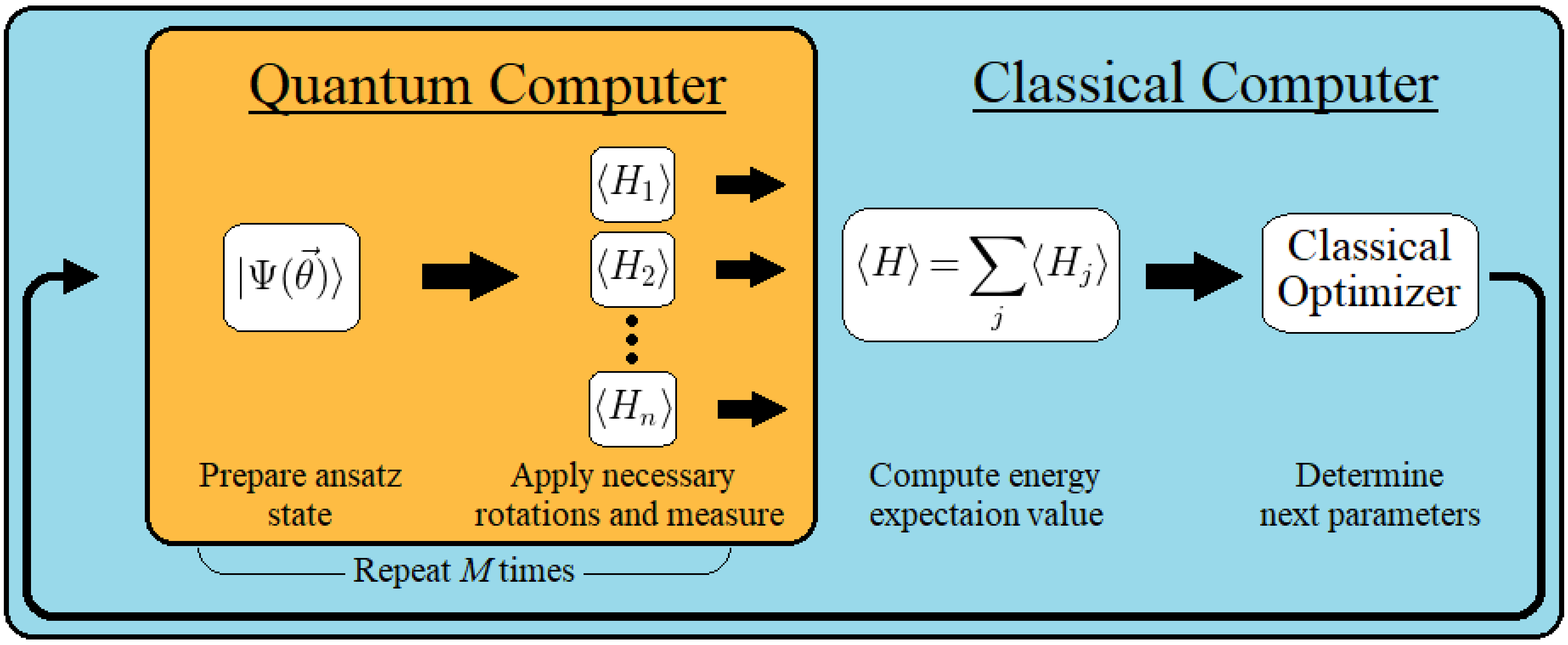

Shown above is the workflow for the VQE Algorithm. As illustrated by the boxes, the quantum component of our algorithm is best classified as a subroutine for the overarching classical component. Up to this point we've covered all of the necessary ingredients for the quantum computer, but have mentioned very little about the role of the classical optimization component. Essentially, the role of the classical computer is to determine which $|\Psi(\vec{\theta})\rangle$ states to feed the quantum computer. But in order to find our ground state solution as quickly as possible, this process for determining the best ansatz state needs to be optimal. Thus, just like in QAOA, buried within our hybrid quantum/classical eigen solving algorithm is a classical optimization technique. And this classical optimizer can really be thought of as the engine of the algorithm, largely responsible for how quickly the algorithm can converge on the correct solution. In principle, the best suited classical optimization technique will vary from problem to problem, which means we need to tap into the wealth of already known optimizers.

Returning now to the quantum subroutine component of our VQE Algorithm, it is worth pointing out how we've visually represented the step labeled "Apply necessary rotations and measure." As shown in the diagram, this consists of computing each of the individual components of our Hamiltonian: $\langle H_j \rangle$. Recall from our discussion earlier that so long as our Hamiltonian is linearly composed of Pauli Operator terms, then we can decompose our Hamiltonian and compute each term's energy expectation value individually. This is a powerful feature of our algorithm, allowing us to tackle arbitrarily large Hamiltonians, but it also comes at a computational cost. As we can see just below the quantum computer box, in order to compute our energy expectation values through measurements we need $M$ repeat trials. If we now consider a Hamiltonian composed of $n$ linear terms, then the number of quantum computations necessary for one set of parameters $\vec{\theta}$ becomes $M \times n$.

Unfortunately, due to the limiting nature of taking measurements on quantum systems, there is no way of getting around all of the tedious repeat trials. The number of Hamiltonian components $n$ is determined by our problem, and $M$ directly affects the accuracy of our energy expectation value. Thus, despite VQE's incredible flexibility in problem solving, we must always consider the true cost in speed when we factor in the use of our quantum subroutine. In principle however, we could consider that future quantum computers may be used in parallel, similar to current classical computing. Thus, if we had $n$ quantum computers running in parallel, the total time for our VQE quantum subroutine is once again only dependent on the desired accuracy of $M$ trials.

## Single Qubit VQE Example

The time has come to put all of our VQE pieces together and construct our first working example. We will start by defining the Hamiltonian for which we are seeking to find the ground state energy of:

$$H \;=\; 3X \;-\; 2Y \;+\; Z$$

As shown above, our Hamiltonian consists of all three Pauli Operators, as well as different weighing contributions from each. These weights don't actually affect the quantum component of our algorithm in any way, as we simply pass them along to our classical computer when we compute the total the expectation value. For this first example, we will not be implementing any classical optimizer, but rather will span the complete $\vec{\theta}$ parameter space:

In [12]:

```
t1 = 60
t2 = 60
Shots = 10000
Parameter_Space = np.zeros(shape=(t1,t2))
Ground_State = [100,0,0]
H = {'X':3,'Y':-2,'Z':1}
Hk = list( H.keys() )
#----------------------------------------------------
for i in np.arange( t1 ):
    theta = m.pi*(i/t1)
    for j in np.arange( t2 ):
        phi = 2*m.pi*(j/t2)
        Measures = []
        for k in np.arange(len(Hk)):
            q = QuantumRegister( 1, name='q' )
            c = ClassicalRegister( 1, name='c' )
            qc= QuantumCircuit( q, c, name='qc')
            oq.Single_Qubit_Ansatz( qc, q[0], [theta, phi] )
            if( Hk[k] == 'X' ):
                qc.ry( -m.pi/2, q[0])
            if( Hk[k] == 'Y' ):
                qc.rx(m.pi/2, q[0])
            qc.measure( q,c )
            M = {'0':0,'1':0}
            M.update( oq.Measurement( qc, shots=Shots, print_M=False, return_M=True ) )
            Measures.append( M )
        Parameter_Space[i,j] = H['X']*(Measures[0]['0'] - Measures[0]['1'])/Shots + H['Y']*(Measures[1]['0'] - Measures[1]|
        if( Parameter_Space[i,j] < Ground_State[0] ):
            Ground_State[0] = Parameter_Space[i,j]
            Ground_State[1] = theta
            Ground_State[2] = phi
#====================================================
print('Ground State Energy: ',round(Ground_State[0],5),'     \u03B8 = ',round(Ground_State[1],3),'     \u03C6 = ',round(Gro
fig, ax = plt.subplots()
show_text = False
show_ticks = False
oq.Heatmap(Parameter_Space, show_text, show_ticks, ax, "plasma", "Energy Expectation Value")
fig.tight_layout()
plt.show()
```

```
Ground State Energy:  -3.7892      θ =  1.78      ϕ =  2.618
```

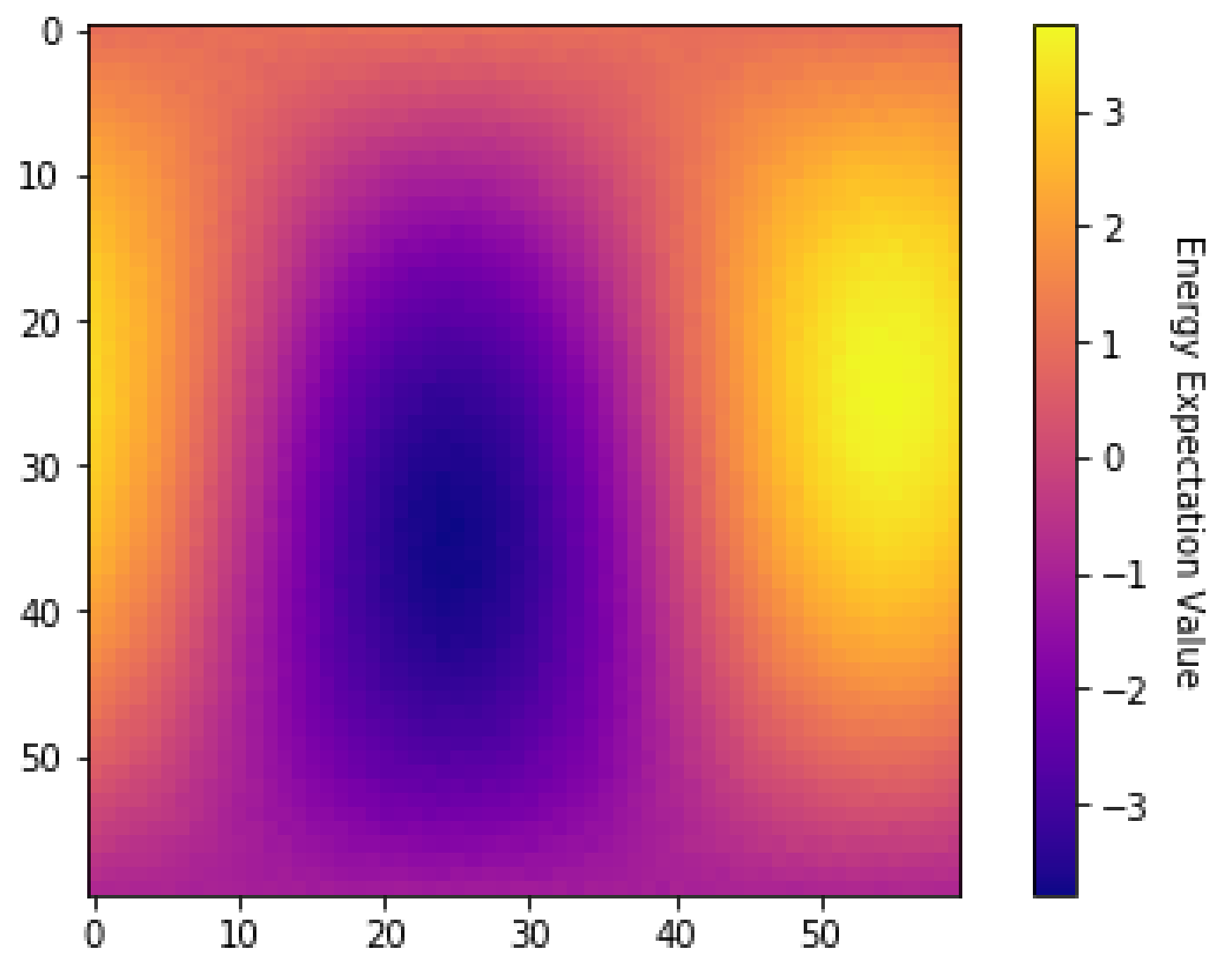

The energy landscape shown above represents all of the possible ansatz states we can create with the parameters $\theta$ and $\phi$, spanning the complete surface of the Bloch Sphere. Each combination of parameters creates a unique ansatz state, yielding the various energy expectation values shown above. Going forward, the next step in our algorithm is to use this parameter space as a guide for finding the ground state energy, corresponding to the center of the dark blue region. To do this, we will call upon the same technique we used in the previous lesson: Gradient Descent, to iteratively find our way into the global minima of this parameter space. For a review on the Gradient Descent optimization technique, I encourage you to take a look at lesson 10 - QAOA.

The cell of code below uses a function called **VQE_Gradient_Descent**, from Our_Qiskit_Functions, which handles all of the computations neccessary for computing the slope at each location, and providing the next set of $\theta$ and $\phi$ values for the ansatz state. The code starts off with a randomly chosen pair of $\theta$ and $\phi$ values, and the goal is to hopefully find our way to the oprtimal set of values shown above, corresponding to the ground state energy:

In [23]:

```
H = {'X':3,'Y':-2,'Z':1}
Hk = list( H.keys() )
Shots = 100000
Ground_State = [100,0,0]
epsilon = 0.001
step_size = 0.01
delta = 0.0001
M_bool = True
#-----------------------
EV = 100
EV_old = 1000
terminate = False
#=====================================================
theta = m.pi*random.random()
phi = 2*m.pi*random.random()
iters = 0
while( (abs( EV - EV_old ) > delta) and (terminate==False) ):
    EV_old = EV
    EV = 0
    for k in np.arange(len(Hk)):
        q = QuantumRegister( 1, name='q' )
        c = ClassicalRegister( 1, name='c' )
        qc= QuantumCircuit( q, c, name='qc')
        oq.Single_Qubit_Ansatz( qc, q[0], [theta, phi] )
        if( Hk[k] == 'X' ):
            qc.ry(-m.pi/2, q[0])
        if( Hk[k] == 'Y' ):
            qc.rx(m.pi/2, q[0])
        qc.measure( q,c )
        M = {'0':0,'1':0}
        M.update( oq.Measurement( qc, shots=Shots, print_M=False, return_M=True ) )
        EV = EV + H[Hk[k]]*(M['0']-M['1'])/Shots
    print('Iterations:  ',iters,'     EV:       ',round(EV,5),'     \u03B8 = ',round(theta,5),'     \u03C6 = ',round(phi,5)
    if( EV > EV_old ):
        terminate = True
    else:
        if( EV < Ground_State[0] ):
            Ground_State[0] = EV
            Ground_State[1] = theta
            Ground_State[2] = phi
        theta_old = theta
        phi_old = phi
        theta,phi = oq.VQE_Gradient_Descent(qc,q,H,oq.Single_Qubit_Ansatz,theta,phi,epsilon,step_size,measure=M_bool,shots=
        iters = iters + 1
if( (abs( EV - EV_old ) < delta) or (terminate==True) ):
    print('\n_____ Gradient Descent Complete _____\n')
    print('Iterations:  ',iters,'     EV:       ',round(Ground_State[0],5),'     \u03B8 = ',round(Ground_State[1],5),'
```

```
Iterations:   0      EV:        3.68106      θ =  1.21018      ϕ =  5.85229
Iterations:   1      EV:        3.6486      θ =  1.08018      ϕ =  5.60604
Iterations:   2      EV:        3.5419      θ =  0.99018      ϕ =  5.80479
Iterations:   3      EV:        3.59524      θ =  1.04893      ϕ =  5.82729

_____ Gradient Descent Complete _____

Iterations:   3      EV:        3.5419      θ =  0.99018      ϕ =  5.80479
```

Run the cell of code above several times, and you are likely to find that in most cases the algorithm never makes it past 5 iterations, and yields a final energy which is clearly quite far off from true ground state energy. So we must ask, why is this happening? In the QAOA lesson we were able to traverse much more complicated expectation value landscapes with the same technique, so why is our implementation here failing?

The root of problem demonstrated above turns out to have nothing to do with the Gradient Descent technique (which I can assure you we implemented correctly in our code), but rather with the information we are feeding the optimizer. More specifically, in the previous lesson we traversed expectation value landscapes which were computed using wavefunctions, while here we are doing everything strictly through measurements. Consequently, the values which we are providing the Gradient Descent optimizer are approximations subject to small fluxuations, which in turn are causing the technique to fail. To demonstrate this, the two cells of code below produce the same subsection of the energy landscape plot from earlier, but do so by calculating the expectation values two different ways:

## Method 1: Approximating $\langle H \rangle$ Using Measurements

In [34]:

```
t1 = 60
t2 = 60
Shots = 10000
Parameter_Space = np.zeros(shape=(t1,t2))
Ground_State = [100,0,0]
H = {'X':3,'Y':-2,'Z':1}
Hk = list( H.keys() )
#--------------------------------------------------
for i in np.arange( t1 ):
    theta = m.pi/2+ (m.pi/10)*(i/t1)
    for j in np.arange( t2 ):
        phi = m.pi+ (m.pi/10)*(j/t2)
        EV = 0
        for k in np.arange(len(Hk)):
            q = QuantumRegister( 1, name='q' )
            c = ClassicalRegister( 1, name='c' )
            qc= QuantumCircuit( q, c, name='qc')
            oq.Single_Qubit_Ansatz( qc, q[0], [theta, phi] )
            if( Hk[k] == 'X' ):
                qc.ry(-m.pi/2, q[0])
            if( Hk[k] == 'Y' ):
                qc.rx(m.pi/2, q[0])
            qc.measure( q,c )
            M = {'0':0,'1':0}
            M.update( oq.Measurement( qc, shots=Shots, print_M=False, return_M=True ) )
            EV = EV + H[Hk[k]]*(M['0']-M['1'])/Shots
        Parameter_Space[i,j] = EV
        if( Parameter_Space[i,j] < Ground_State[0] ):
            Ground_State[0] = Parameter_Space[i,j]
            Ground_State[1] = theta
            Ground_State[2] = phi
#==================================================
fig, ax = plt.subplots()
show_text = False
show_ticks = False
oq.Heatmap(Parameter_Space, show_text, show_ticks, ax, "plasma", "Energy Expectation Value")
fig.tight_layout()
plt.show()
```

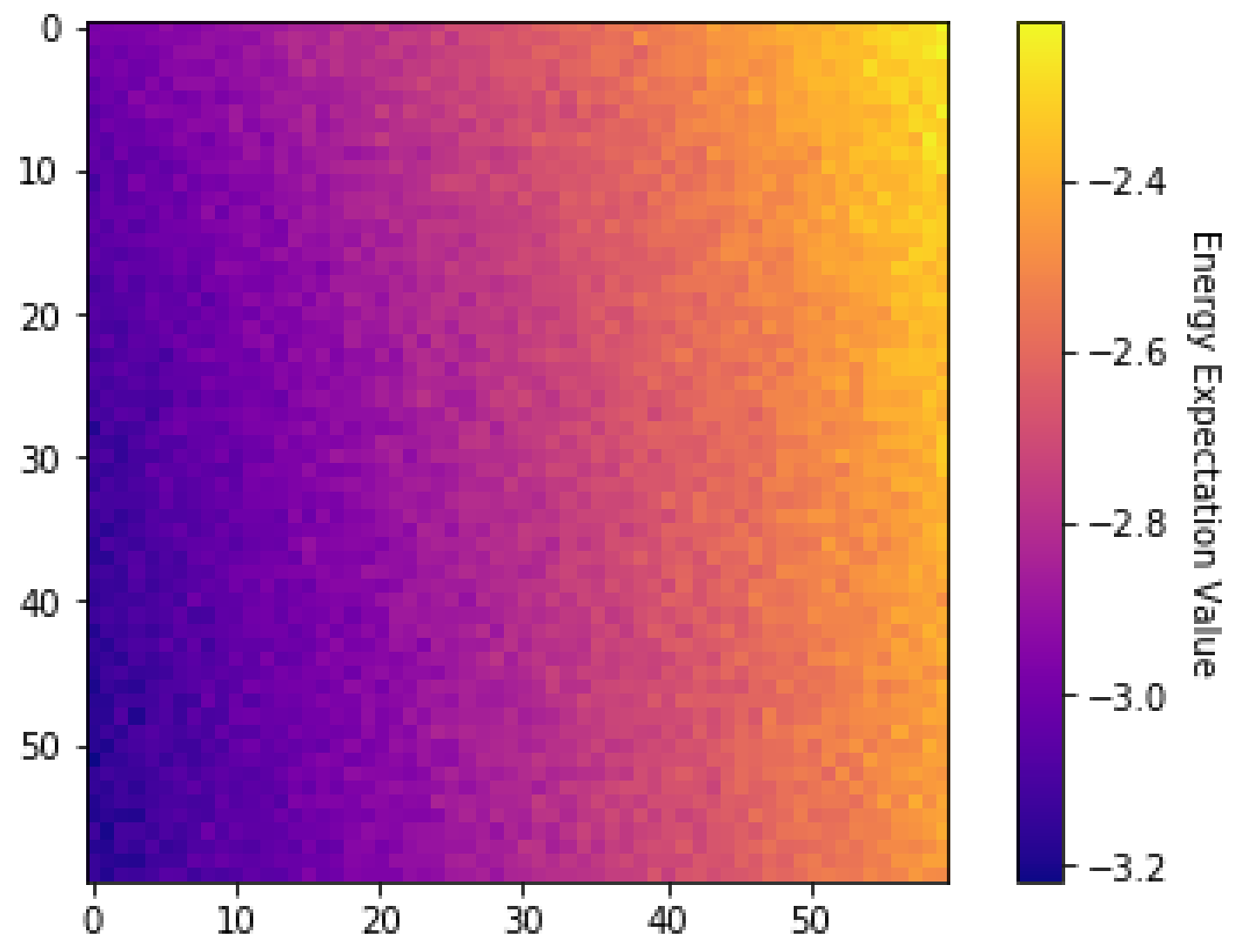

### Method 2: Approximating $\langle H \rangle$ Using The Wavefunction

In [35]:

```
t1 = 60
t2 = 60
Parameter_Space = np.zeros(shape=(t1,t2))
Ground_State = [100,0,0]
H = {'X':3,'Y':-2,'Z':1}
Hk = list( H.keys() )
#---------------------------------------------------
for i in np.arange( t1 ):
    theta = m.pi/2+ (m.pi/10)*(i/t1)
    for j in np.arange( t2 ):
        phi = m.pi+ (m.pi/10)*(j/t2)
        EV = 0
        for k in np.arange(len(Hk)):
            q = QuantumRegister( 1, name='q' )
            qc= QuantumCircuit( q, name='qc')
            oq.Single_Qubit_Ansatz( qc, q[0], [theta, phi] )
            sv0 = execute( qc, S_simulator, shots=1 ).result().get_statevector()
            if( Hk[k] == 'X' ):
                qc.x(q[0])
            if( Hk[k] == 'Y' ):
                qc.y(q[0])
            if( Hk[k] == 'Z' ):
                qc.z(q[0])
            sv = execute( qc, S_simulator, shots=1 ).result().get_statevector()
            ev = 0
            for k2 in np.arange(len(sv)):
                ev =  ev +  (np.conj(sv[k2])*sv0[k2]).real
            EV = EV + H[Hk[k]] * ev
        Parameter_Space[i,j] = EV
        if( Parameter_Space[i,j] < Ground_State[0] ):
            Ground_State[0] = Parameter_Space[i,j]
            Ground_State[1] = theta
            Ground_State[2] = phi
#===================================================
fig, ax = plt.subplots()
show_text = False
show_ticks = False
oq.Heatmap(Parameter_Space, show_text, show_ticks, ax, "plasma", "Energy Expectation Value")
fig.tight_layout()
plt.show()
```

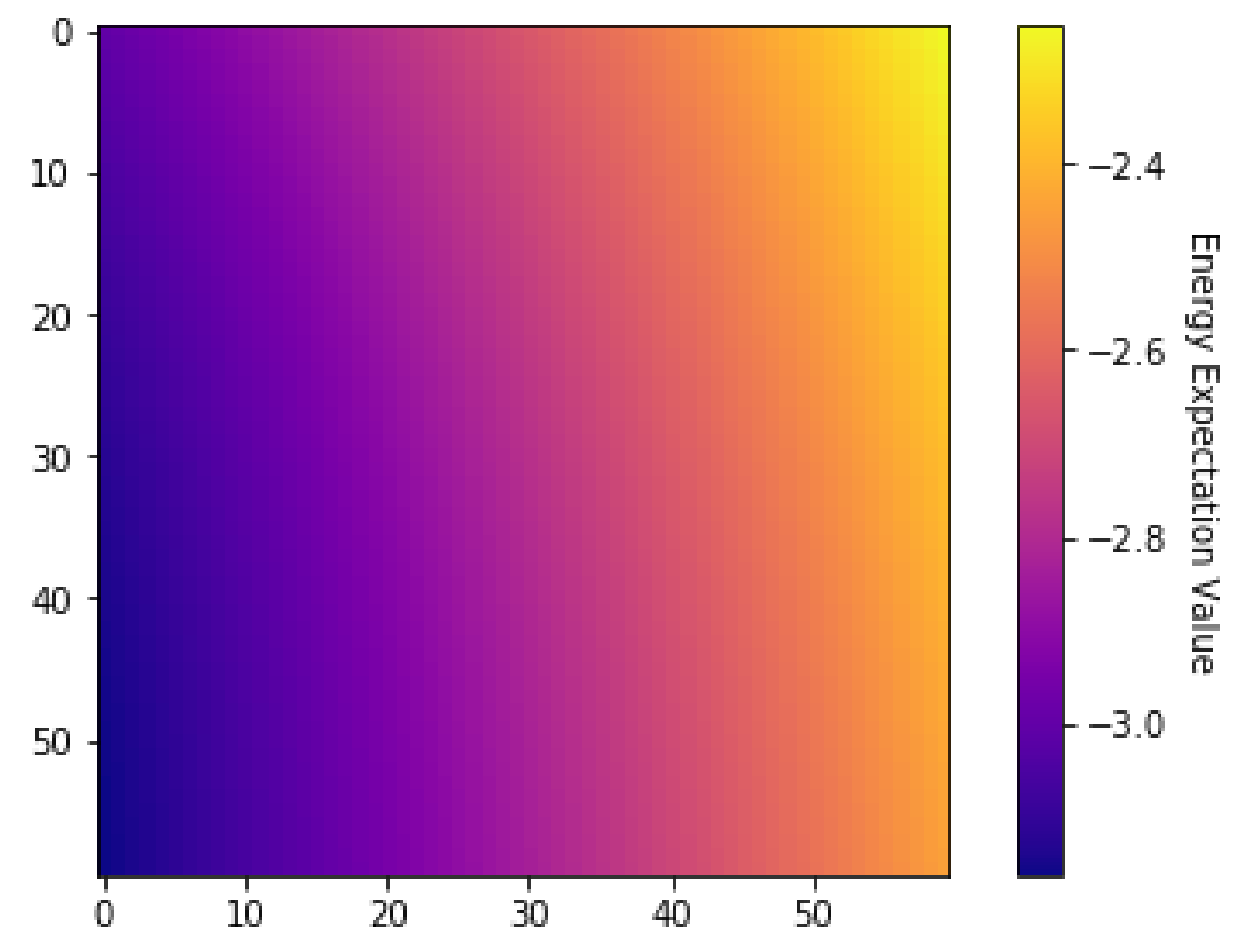

Now, you may notice that the two resulting plots above look very similar, but a side-by-side comparison should reveal the subtle difference between the two:

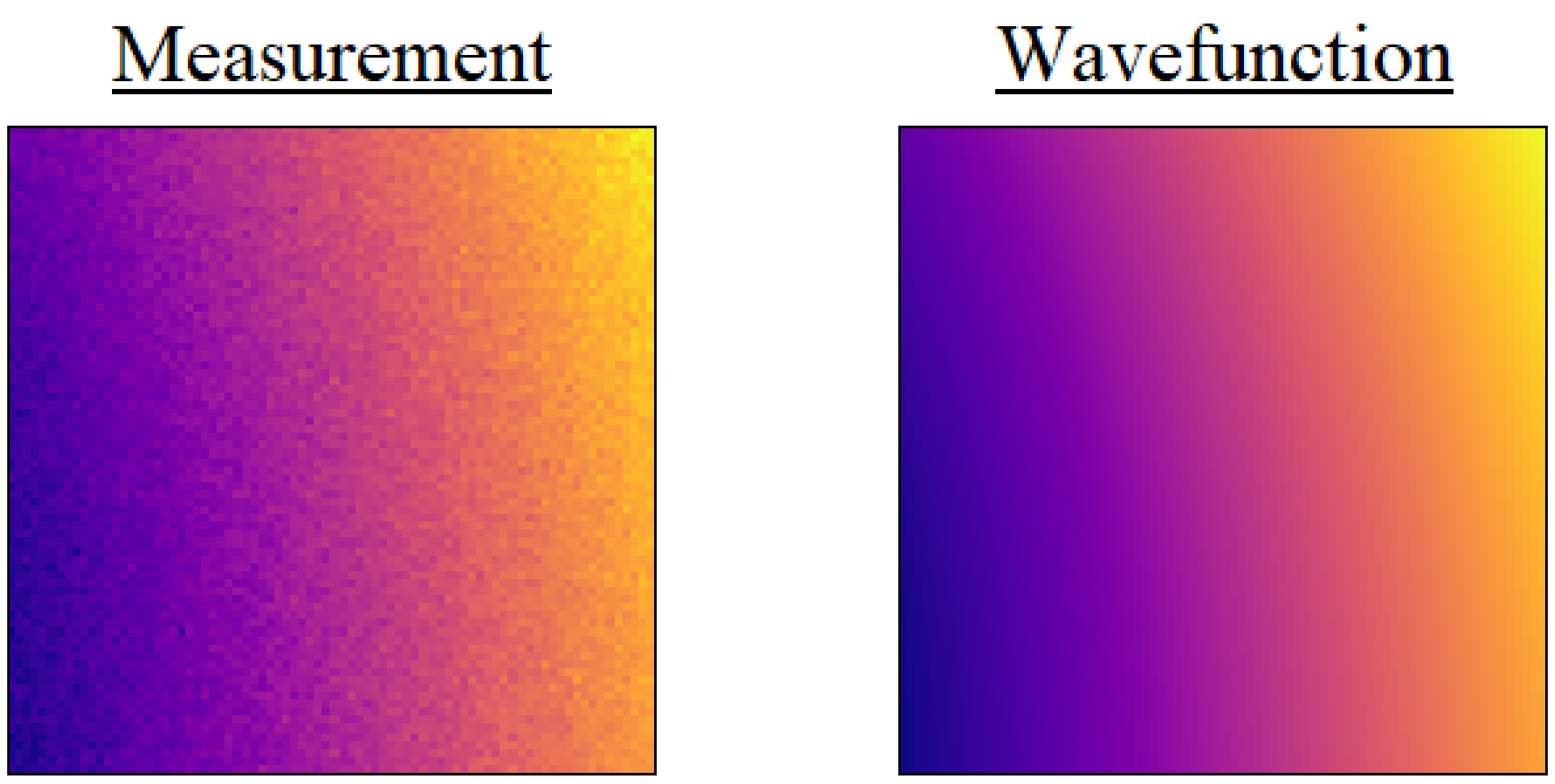

As we can see, both energy landscapes show the same general shape, but the one produced from using the quantum system's wavefunction is much smoother. Conversely, the plot produced from using simulated measurements shows slight deviations throughout, resulting in values at each location that are either slightly bigger or smaller than the true value (those obtained via the wavefunction method).

So to answer our question from earlier: why did our VQE optimizer fail to find the global minimum on what appeared to be a pretty straightforward energy landscape, we needn't look at further than the imperfections produced from the measurement technique. Inside our VQE Gradient Descent optimizer, we approximate the slope at any given point by using two nearby points in both directions, $\epsilon$ distance away, just as we did in the QAOA lesson. However, as shown by the "bumpy" energy landscape above, computing these slopes using measurements can lead to inaccurate approximations, sometimes even the complete wrong direction! For example, consider the diagrams shown below and their resulting approximations to the slope:

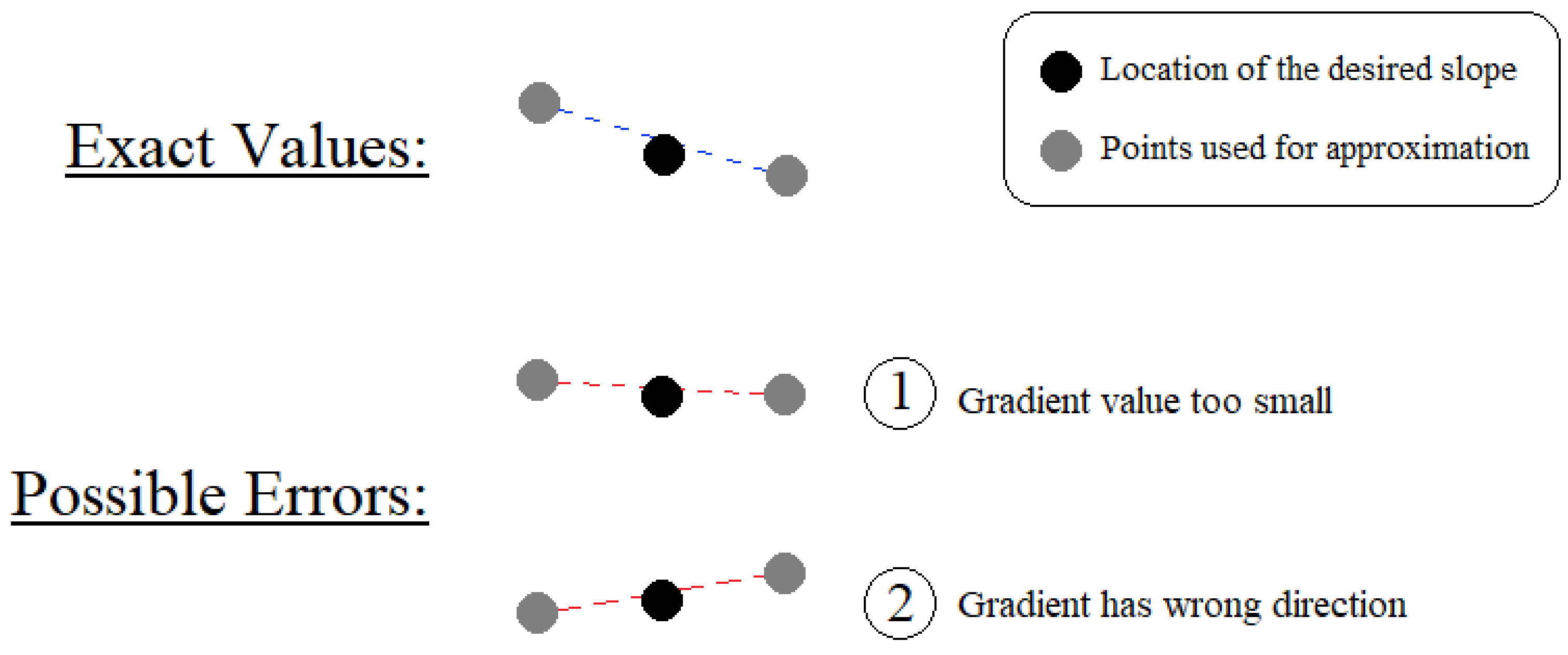

In both of the error configurations shown above, the slopes obtained using nearest neighbor points will result in early termination of our algorithm:

**1)** If the two neighboring points happen to be too close in value, the resulting gradient obtained will be smaller than our $\delta$ threshold for continuing. Consequently, the algorithm will interpret the small slope to mean that we've reached a minima, terminating prematurely.

**2)** If the values of the two neighboring points happen to result in a gradient that is in the opposite direction of the true slope, then the next step will be in the complete wrong direction! As a result, when we calculate the next energy expectation value, we've essentially climbed "uphill", resulting in a termination of our algorithm.

Now, in light of the errors outlined above, what are our options for remedying the situation? Unfortunately, the solution to our problem is one that can only be fixed by altering the way in which we compute gradients. Option one is to simply increase the number of measurements per step, thus giving us more accurate data points to work with, with less fluctuation. However, the results calculated above represent the case for $10,000$ measurements per data point, which is already way more than one would ideally use in a real VQE run on a problem of this size. Alternatively, we could change the way we compute our gradient together. For example, we could use four data points to compute the slope, extending our sampling space by one additional point in each direction, effectively reducing the potential for fluctuations to ruin the computation. However, this would require double the number of measurement samples per step in the optimization algorithm, once again negating any sort of speedup.

While Gradient Descent is certainly a powerful algorithm for finding minima in a parameter space, it is not the only technique in town. Remember that our goal is to pair the VQE Algorithm with the best suited classical optimizer to search through a given energy landscape. Therefore, the examples above serve to show that Gradient Descent probably shouldn't be our first choice. Alternatively, in the coming section we will be implementing a different optimization scheme: the Nelder-Mead method, which *will* be able to find minima despite fluctuations in the parameter space resulting from measurements. But before moving on to this second optimization technique, let's confirm that our Gradient Descent implementation would have worked if not for these fluctuations:

In [68]:

```
H = {'X':3,'Y':-2,'Z':1}
Hk = list( H.keys() )
Ground_State = [100,0,0]
epsilon = 0.05
step_size = 0.01
delta = 0.00005
M_bool = False
#------------------------
EV = 100
EV_old = 1000
terminate=False
EV_type = 'wavefunction'
#=====================================================
theta = m.pi*random.random()
phi = 2*m.pi*random.random()
iters = 0
while( (abs( EV - EV_old ) > delta) and (terminate==False) ):
    EV_old = EV
    EV = oq.VQE_EV([theta,phi],oq.Single_Qubit_Ansatz,H,EV_type)
    if( (iters/10.0)==m.ceil(iters/10.0) ):
        print('Iterations:  ',iters,'     EV:       ',round(EV,5),'     \u03B8 = ',round(theta,5),'     \u03C6 = ',round(ph
    if( EV > EV_old ):
        terminate = True
    else:
        if( EV < Ground_State[0] ):
            Ground_State[0] = EV
            Ground_State[1] = theta
            Ground_State[2] = phi
        theta_old = theta
        phi_old = phi
        theta,phi = oq.VQE_Gradient_Descent(qc,q,H,oq.Single_Qubit_Ansatz,theta,phi,epsilon,step_size,measure=M_bool)
        iters = iters + 1
if( (abs( EV - EV_old ) < delta) or (terminate==True) ):
    print('\n_____ Gradient Descent Complete _____\n')
    print('Iterations:  ',iters,'     EV:       ',round(Ground_State[0],5),'     \u03B8 = ',round(Ground_State[1],5),'
```

```
Iterations:   0      EV:        -0.21661      θ =  0.90901      ϕ =   1.27932
Iterations:   10      EV:        -0.55428      θ =  0.81959      ϕ =   1.47134
Iterations:   20      EV:        -0.84539      θ =  0.74813      ϕ =   1.68187
Iterations:   30      EV:        -1.13122      θ =  0.71605      ϕ =   1.90465
Iterations:   40      EV:        -1.49318      θ =  0.74184      ϕ =   2.13973
Iterations:   50      EV:        -1.96683      θ =  0.83517      ϕ =   2.39088
Iterations:   60      EV:        -2.4477      θ =  0.98911      ϕ =   2.65484
Iterations:   70      EV:        -2.72895      θ =  1.17464      ϕ =   2.91209

_____ Gradient Descent Complete _____

Iterations:   76      EV:        -2.75868      θ =  1.26673      ϕ =   3.02937
```

The example above represents what we could have originally expected from our Gradient Descent technique, but it critically relies on using wavefunctions to compute gradients (which is effectively cheating). Although the code was unable to find the true global minimum, instead getting stuck in a local minima, we can still see that the optimization technique *can* work. Thus, we can conclude that it is indeed the approximations from taking measurements which causes the Gradient Descent technique to ultimately be unusable.

## Nelder-Mead Optimization Method

---

The Nelder-Mead Optimization method, proposed by John Nelder and Roger Mead in 1965 **[2]**, is an optimization technique which is going to remedy the issue we previously encountered with Gradient Descent, namely the "bumpy" parameter space resulting from measurements. If we think about the main problem plaguing the Gradient Descent method, it's the fact that in order to get a good slope approximation, we want to use points as close as possible to the location in question. However, this is exactly where the issue of measurement fluxuations is the most problematic, leading to inaccurate gradient approximations. Thus, we either lose accuracy by using points which are further apart, or risk dealing with measurement fluctuations by using points which are too close (a lose-lose situation overall). Alternatively, the Nelder-Mead scheme isn't relient on using neighboring points, avoiding this problem entirely. Thus, even with small fluctuations in the parameter space due to measurements, the technique will still be able to efficiently converge towards a minima.

The basic premise for the Nelder-Mead technique, using a 2-parameter space as our example, is to use a triangle of points to iteratively step towards the minima of the space. At each step in the algorithm we compute the energy expectation value of the three vertices, and use these values to determine the next step. In essence, the goal of each step is to replace the worst valued vertex of our triangle with a new one closer to the minima. Below is rough outline of the technique, further explaining some of the possible steps one can take with each iteration:

**1)** Pick a starting location somewhere in the space and choose $N + 1$ relatively grouped points, where $N$ is the dimensionality of the parameter space. The geometric shape created from these points is known as our simplex, which is essentially a triangle extended to any dimensional space. For our $N = 2$ case, our simplex is exactly a triangle. Using the $N + 1$ points, we evaluate $F(x_n)$ at each vertex of the simplex and order them accordingly from highest to lowest: $F_h > F_{sh} > \cdots > F_l$. Here, the labels on each vertex correspond to 'highest': $F_h$, 'second highest': $F_{sh}$, and 'lowest': $F_l$. For our VQE Algorithm, these are the energy expectation values we compute through repeat measurements.

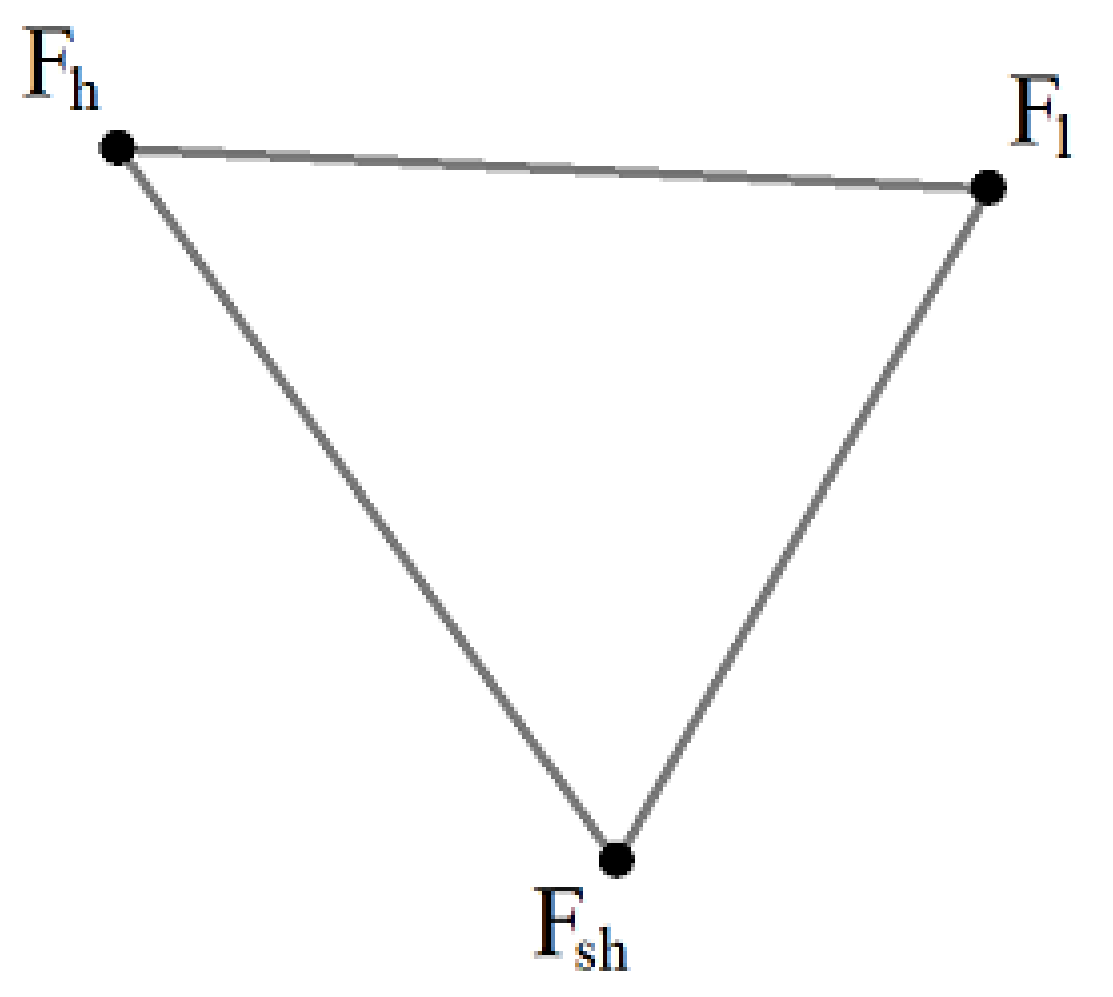

**2)** Next, we drop the highest value in our simplex, $F_h$, and calculate the centroid from the remaining vertices. For our two dimensional case, this means calculating the point directly between $F_l$ and $F_{sh}$. Once this point is obtained, we evaluate it and obtain $F_c$.

$$\text{Centroid } \vec{c} = \frac{\sum_i^n \vec{x_i}}{n}$$

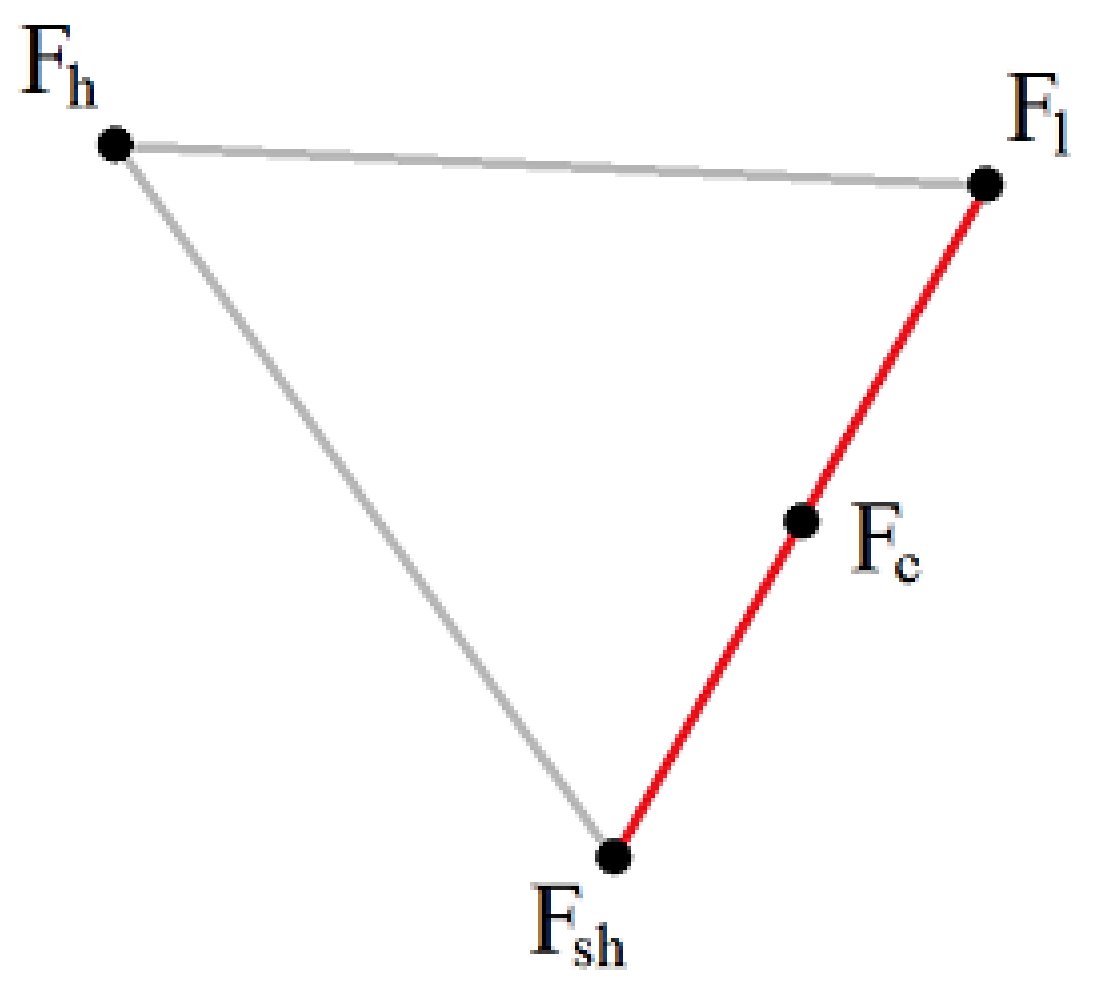

**3)** Using our centroid, we compute a new point $\vec{x_r}$ by performing a reflection of $\vec{x_h}$ about $\vec{c}$ by an amount $\alpha$ (standard value for $\alpha = 2$). The idea here is that if $\vec{x_h}$ is our worst point, and $\vec{c}$ is determined by using the remaining vertices in our simplex, then searching in the direction *away* from $\vec{x_h}$, towards $\vec{c}$, should lead us to a new point in our parameter space that is lower. After computing the reflect point $\vec{x_r}$, we then obtain $F_r$.

$$\text{Reflection Point } \vec{x_r} = \vec{c} + \alpha(\vec{c} - \vec{x_h})$$

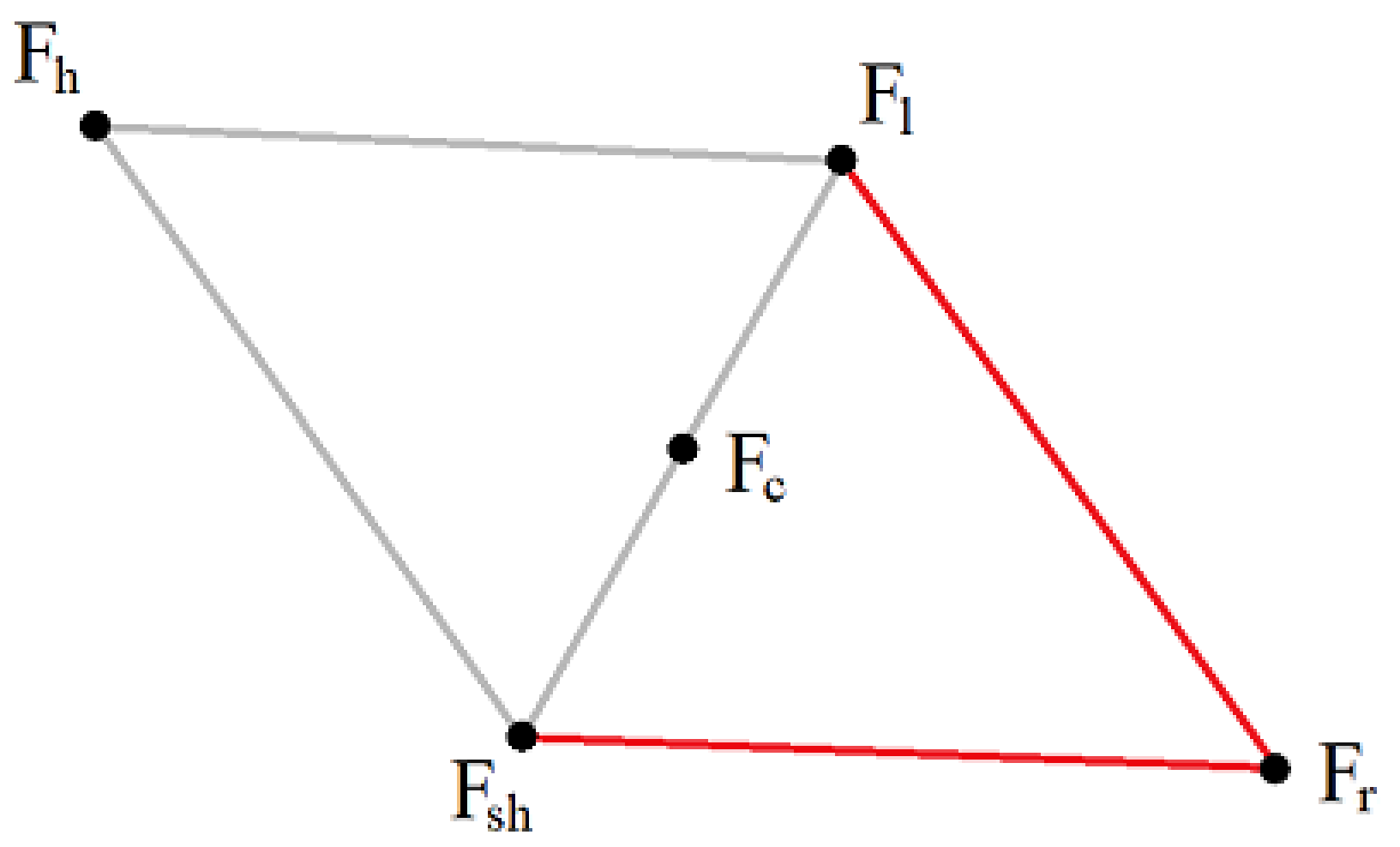

**4)** Based on the value we obtain from $F_r$, our algorithm diverges down a couple different possible routes. For example, if we find that our reflection point is the best so far, $F_r < F_l$, then we press on further in the same direction, obtaining what is called the expansion point $\vec{x_e}$. Conversely, if we find that $F_r$ turns out to be our worst point yet, we then look inwards for one of two possible next points to check, referred to as the contraction point $\vec{x_c}$. There are further possible avenues still, but hopefully these illustrations give you the general idea.

Contraction

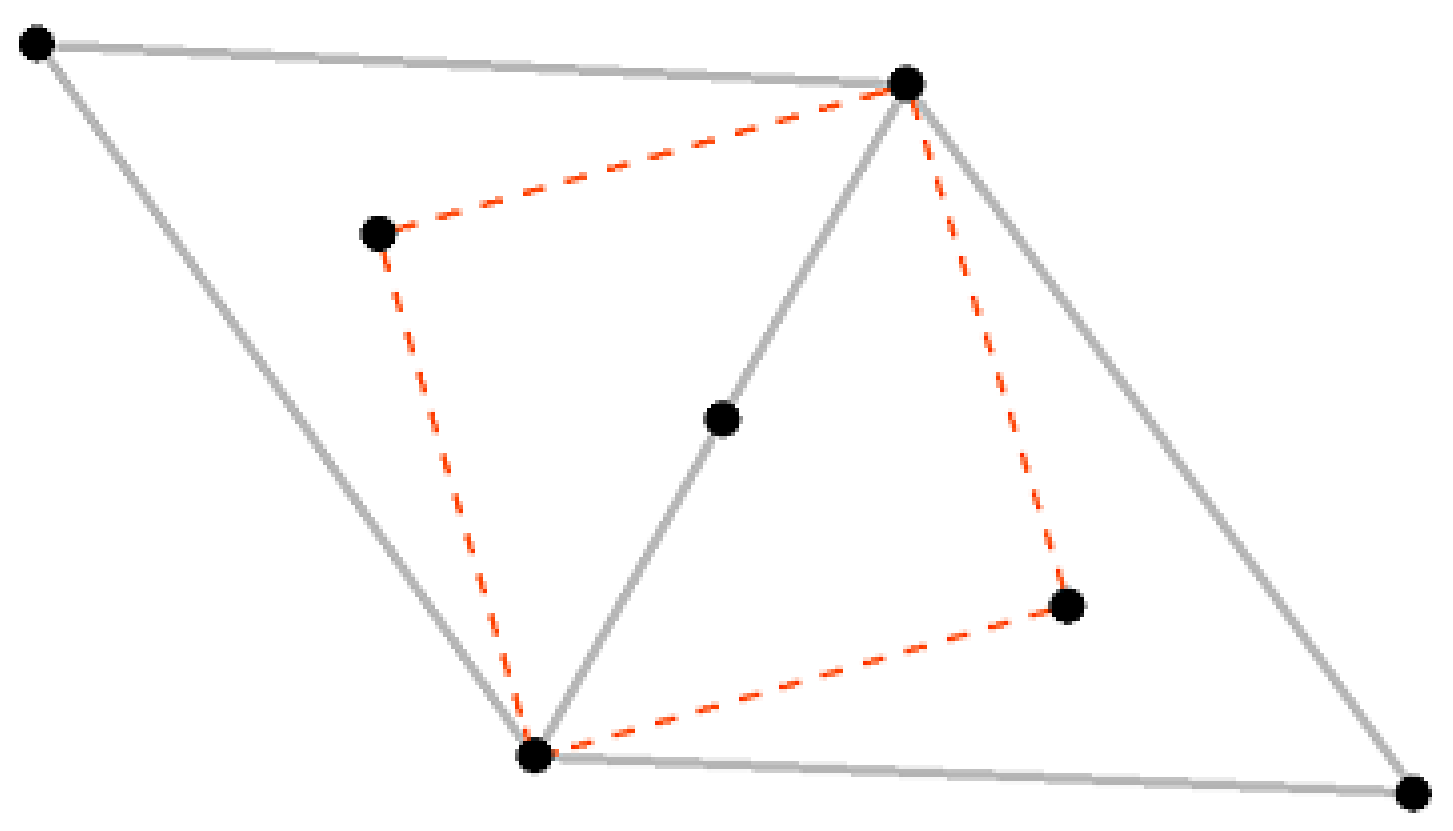

Expansion

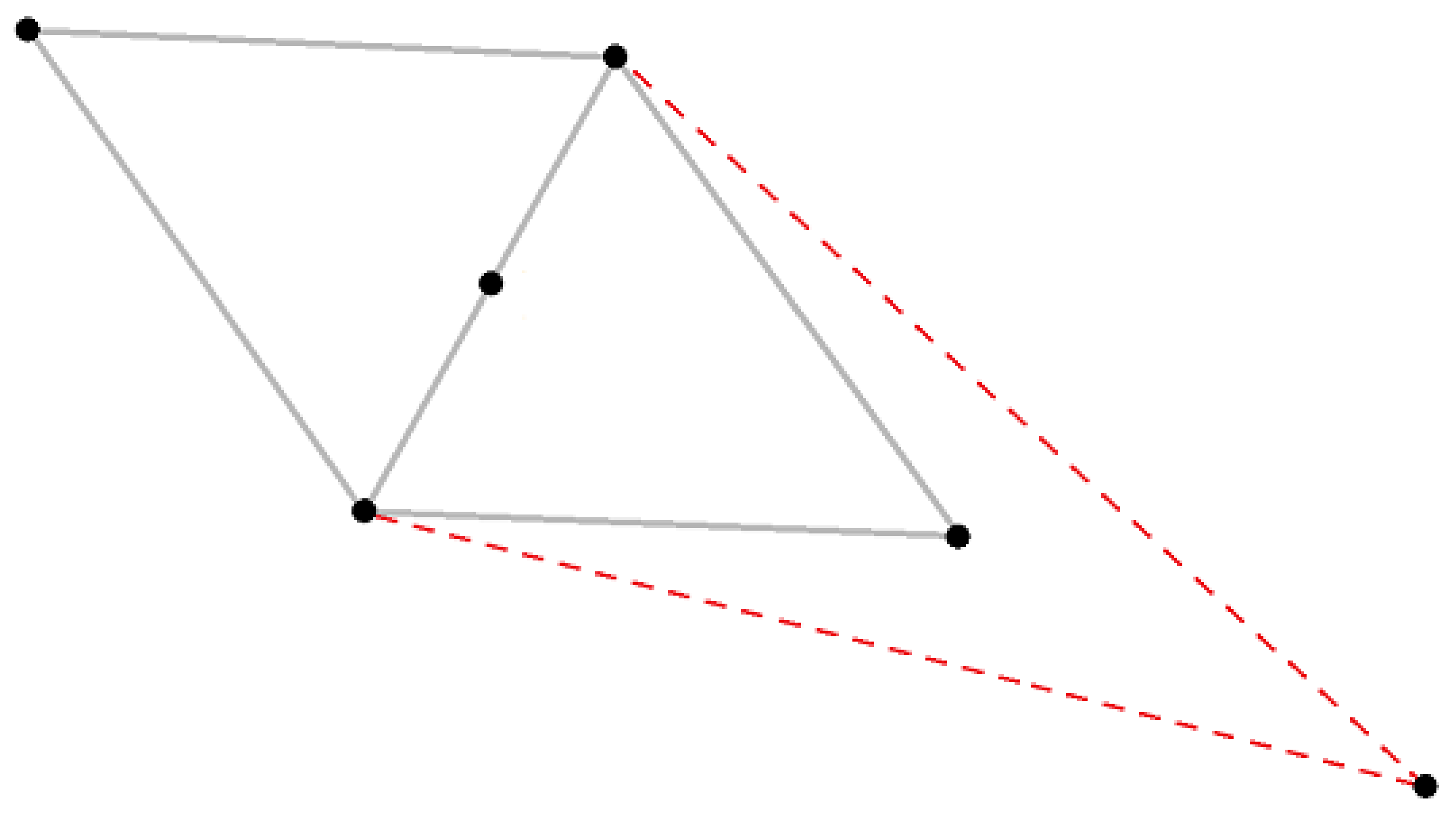

To summarize, the Nelder-Mead optimization technique converges on the minima of our parameter space by iteratively moving through simplexes. At each step we discard the vertex corresponding to the worst value, and replace it with a new vertex that brings us closer to the final minima. And, to return to the original motivation for introducing this optimization technique, hopefully the diagrams shown above illustrate that the separation between vertices is sufficiently large enough that we will avoid the issue we ran into with Gradient Descent. Even with slight fluctuations at each location resulting from taking measurements, our algorithm will still be able to converge on the minima of our parameter space.

In [6]:
```
H = {'X':3,'Y':-2,'Z':1}
EV_type = 'measure'
theta = random.random()*m.pi
phi = random.random()*2*m.pi
delta = 0.001
#------------------------------
Vertices = []
Values = []
radius = 0.35
R = random.random()*(2*m.pi/3)
for rr in np.arange(3):
    Vertices.append( [theta+radius*m.cos(R+(rr*2*m.pi/3)),phi+radius*m.sin(R+(rr*2*m.pi/3))] )
for v in np.arange(len(Vertices)):
    Values.append( oq.VQE_EV(Vertices[v],oq.Single_Qubit_Ansatz,H,EV_type) )
#------------------------------
terminate = False
iters = 0
terminate_count = 0
terminate_limit = 6
while( (terminate==False) and (iters < 100) ):
    iters = iters + 1
    low = oq.Calculate_MinMax( Values,'min' )
    oq.Nelder_Mead(H, oq.Single_Qubit_Ansatz, Vertices, Values, EV_type)
    new_low = oq.Calculate_MinMax( Values,'min' )
    if( abs( new_low[0] - low[0] ) < delta ):
        terminate_count = terminate_count + 1
    else:
        terminate_count = 0
    if( terminate_count >= terminate_limit ):
        terminate = True
        print('\n_____ Nelder-Mead Complete _____\n')
        print('Iteration: ',iters,'   Lowest EV:  ',round(low[0],6),'   \u03B8 = ',round(Vertices[low[1]][0],4),'     \u03
    if( ( (iters==1) or (m.ceil(iters/5))==m.floor(iters/5) ) and (terminate==False) ):
        print('Iteration: ',iters,'   Lowest EV:  ',round(low[0],6),'   \u03B8 = ',round(Vertices[low[1]][0],4),'     \u03
```

```
Iteration:  1    Lowest EV:   -3.529    θ =  1.6692      ϕ =  3.0195
Iteration:  5    Lowest EV:   -3.737    θ =  1.9316      ϕ =  2.6388
Iteration:  10    Lowest EV:   -3.7628    θ =  1.8499      ϕ =  2.5471
Iteration:  15    Lowest EV:   -3.7832    θ =  1.8108      ϕ =  2.5381
Iteration:  20    Lowest EV:   -3.7892    θ =  1.8145      ϕ =  2.5429

_____ Nelder-Mead Complete _____

Iteration:  23    Lowest EV:   -3.7892    θ =  1.8145      ϕ =  2.5429
```

As promised, our code implementation of the Nelder-Mead optimization technique above results in a converging search towards the global minimum of our parameter space. And most importantly, it works despite the fluctuations resulting from taking measurements. I encourage you to run the cell of code a few times to get a feel for the converging speed / success rate of this new technique. What you should find is that while the search isn't a guaranteed 100% success, it is still far more reliable than the Gradient Descent technique from earlier, even when using wavefunctions.

## Spanning Higher Dimensional Hilbert Spaces

At this point in the lesson, we have covered all of the key topics that make up the Variational Quantum Eigensolver Algorithm. However, all of our examples up until this point have been for single qubit systems, so as our final topic, we will now discuss how to apply the VQE Algorithm to problems requiring more qubits.

In order to appreciate the difficulty of scaling the VQE technique to larger problems, we must return back to our discussion from earlier about the Variational Principle and the role of ansatz states $|\Psi(\vec{\theta})\rangle$. Remember that the most important thing we require from our ansatz state is that it spans the complete Hilbert space, thereby giving our algorithm the necessary platform to converge on the ground state energy. For the case of a single qubit, we showed that reaching all possible states in the system is achievable through two parameters, corresponding to rotations on the Bloch Sphere. Extending this line of reasoning to two qubits then, a natural first guess is to suspect that we can similarly span the complete $2$-qubit Hilbert space using four parameters. However, this is unfortunately not the case. As we already showed in the previous QAOA lesson, no amount of single qubit rotation gates is enough to span the full $2$-qubit Hilbert Space. To illustrate this, consider the math exercise below:

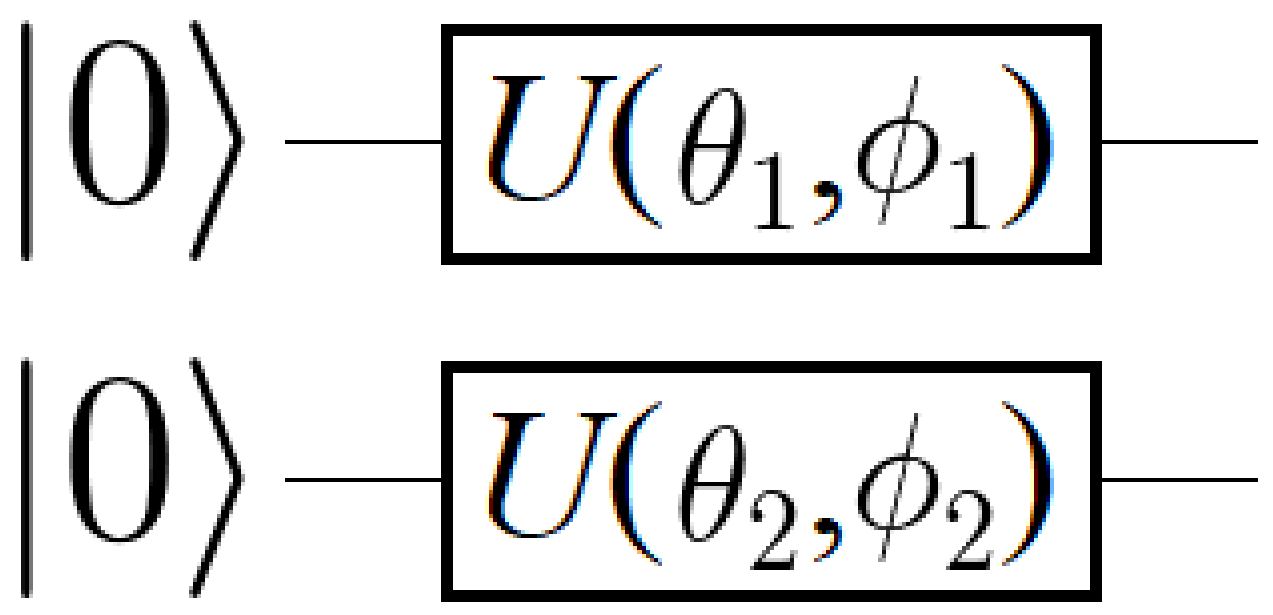

$$
\begin{aligned}
|\Psi(\vec{\theta})\rangle &= U(\theta_1,\phi_1)|0\rangle \otimes U(\theta_2,\phi_2)|0\rangle \\
&= \left[\cos\left(\frac{\theta_1}{2}\right)|0\rangle + e^{-i\phi_1}\sin\left(\frac{\theta_1}{2}\right)|1\rangle\right] \otimes \left[\cos\left(\frac{\theta_2}{2}\right)|0\rangle + e^{-i\phi_2}\sin\left(\frac{\theta_2}{2}\right)|1\rangle\right] \\
&= \cos\left(\frac{\theta_1}{2}\right)\cos\left(\frac{\theta_2}{2}\right)|00\rangle + e^{-i\phi_2}\cos\left(\frac{\theta_1}{2}\right)\sin\left(\frac{\theta_2}{2}\right)|01\rangle \\
&+ e^{-i\phi_1}\sin\left(\frac{\theta_1}{2}\right)\cos\left(\frac{\theta_2}{2}\right)|10\rangle + e^{-i(\phi_1+\phi_2)}\sin\left(\frac{\theta_1}{2}\right)\sin\left(\frac{\theta_2}{2}\right)|11\rangle
\end{aligned}
$$

At first glance, the state above may look general enough to span all possible 2-qubit states. However, because of the way in which we've created this state through only single qubit operations, this state can only cover all possible *non-entangled* quantum states. For example, suppose the ground state to a particular problem was:

$$
|\Psi_0\rangle = \frac{1}{\sqrt{2}}\Big(|00\rangle + |11\rangle\Big)
$$

If we tried to reach this state via our 4-parameters shown above, we run into the following issues:

$$
\text{condition 1:} \qquad \cos\left(\frac{\theta_1}{2}\right)\sin\left(\frac{\theta_2}{2}\right) = 0
$$

$$
\text{condition 2:} \qquad \sin\left(\frac{\theta_1}{2}\right)\cos\left(\frac{\theta_2}{2}\right) = 0
$$

In order to make sure that our ansatz state doesn't contain $|01\rangle$ or $|10\rangle$, we run into the contradictory conditions shown above. As an example, let's suppose we pick $\theta_1 = \pi$ and $\theta_2 = \pi$, which guarantees that both equations shown above will equal $0$. However, when we plug in these choices for $\theta_1$ and $\theta_2$, our resulting ansatz state becomes:

$$
e^{-i(\phi\ +\phi_2)}|11\rangle
$$

The conditions shown above are meant to illustrate that there is no combination of parameters which will yield the desired ansatz state. Physically, this is because the state $|\Psi_0\rangle$ contains entanglement, which means that we need to include at least one additional 2-qubit gate in our ansatz preparation circuit:

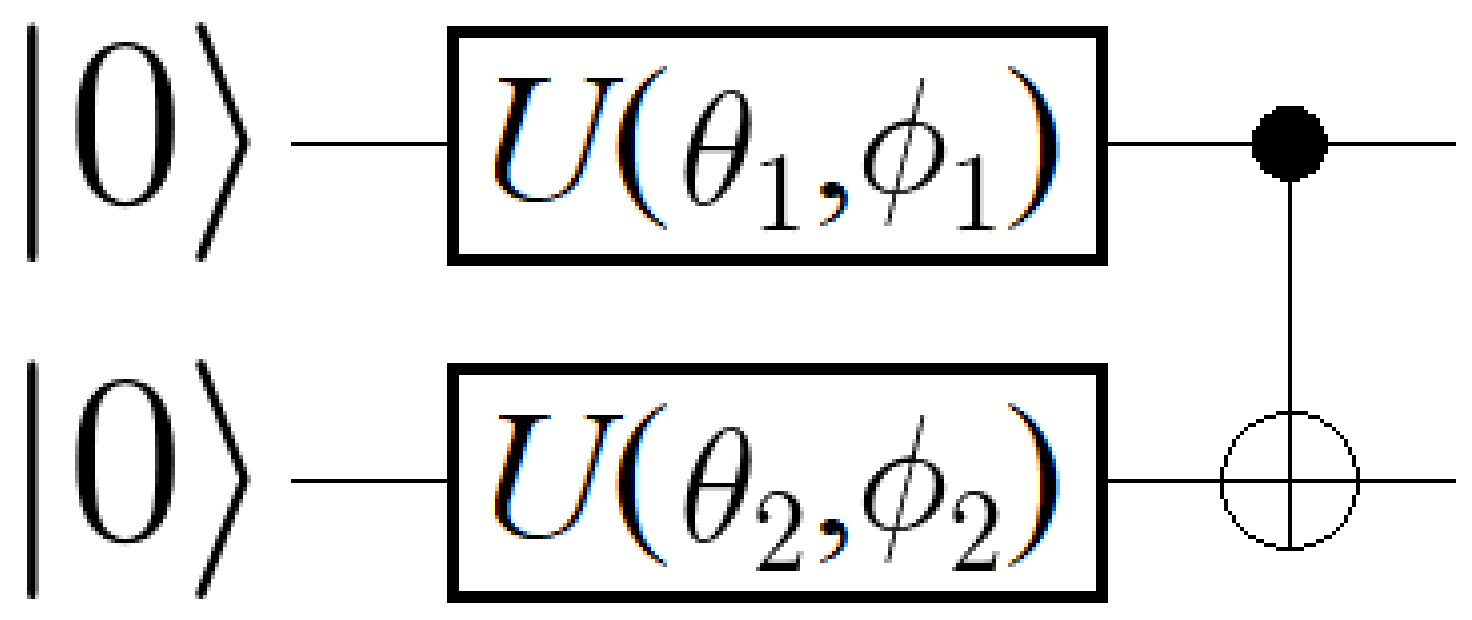

By adding a CNOT gate into our ansatz circuit, we are now able to reach states within the 2-qubit Hilbert space which were previously unavailable to us. For example, by selecting the parameters: $\theta_1 = \frac{\pi}{2}$, $\theta_2 = 0$, $\phi_1 = \frac{\pi}{2}$, we are now able to reach our target state above. However, the inclusion of this additional CNOT gate still isn't enough to reach our goal of spanning the complete $2$-qubit Hilbert space. To do this, we will need to expand our circuit further to include more parameters and CNOT gates:

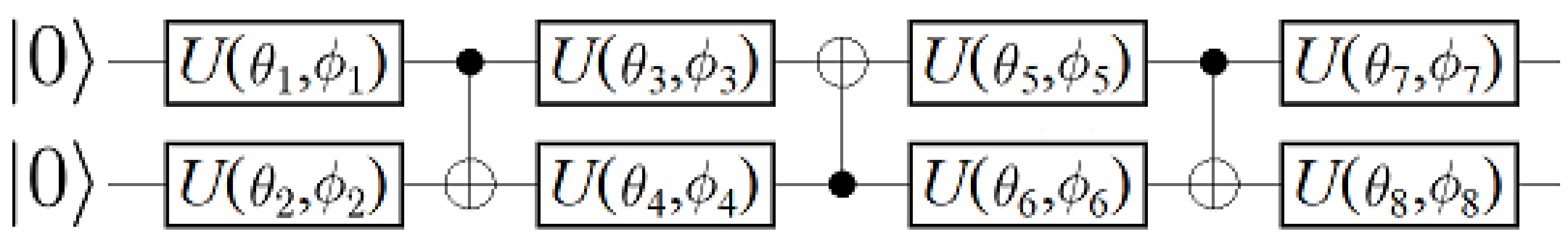

The circuit shown above represents a universal 2-qubit ansatz state, able to span the full $2$-qubit Hilbert space. Proving that the circuit above is truly universal is a bit beyond the scope of this lesson, but if you're interested I encourage you to check out **[3]** for further details. For the purpose of our VQE Algorithm, the circuit shown above will serve as our ansatz preparation circuit, leaving us with a total of $16$ free parameters. Jumping from $2$ to $16$ parameters by introducing a second qubit is quite a leap in our parameter space, and in truth it speaks volumes about the complexity of working with quantum states. It's subtle to appreciate at first, but these enormous Hilbert spaces serve as a demonstration of the potential computing power that quantum computers can achieve with minimal qubits.

## Final Example: 2-Qubit VQE

When implementing the $16$ variable quantum circuit shown above, spanning the complete $2$-qubit Hilbert space, we must be mindful that at the end of the day that VQE is ultimately a classical optimizer driven algorithm. At a certain point, we need to consider whether carrying extra free parameters is worth the additional computing costs, or if we should simply work with a smaller parameter space. In the coming final example, we will be working with an ansatz state that only contains the gates shown above up to the second CNOT gate (not including the second CNOT though), leaving us with a total of $8$ free parameters. Although reducing our parameter space by a factor of $2$ may seem like a lot, it actually saves our algorithm in terms of both speed and accuracy. It should make sense that optimizing $8$ versus $16$ parameters is much quicker, but also consider the difference in complexity between the two parameter spaces. While it's true the full ansatz circuit can reach a larger portion of the $2$-qubit Hilbert space, the tradeoff is that it creates more potential for our classical optimizer to get stuck in local minima. As a result, using the full $16$ parameter ansatz circuit can sometimes be *worse* at finding the true ground state energy to our problem, or a very close approximation.

Below is the code implementation of our Nelder-Mead based VQE algorithm for the following Hamiltonian:

$$
H = 3XY - 2ZZ
$$

which has a ground state energy of $-5$, corresponding to the state $\frac{1}{\sqrt{2}}\Big(i|00\rangle + |11\rangle\Big)$.

In [112]:
```
H = {'XY':3,'ZZ':-2}
EV_type = 'measure'
P = []
for p in np.arange(4):
    P.append( random.random()*m.pi )
    P.append( random.random()*2*m.pi )
delta = 0.001
#------------------------------
Vertices = []
Values = []
for v1 in np.arange(len(P)):
    V = []
    for v2 in np.arange(len(P)):
        R = round((0.4+random.random()*0.8)*(-1)**( round(random.random())),5)
        V.append( P[v2]+R )
    Vertices.append( V )
    Values.append( oq.VQE_EV(V,oq.Two_Qubit_Ansatz,H,EV_type) )
#------------------------------
terminate = False
iters = 0
terminate_count = 0
terminate_limit = 10
while( (terminate==False) and (iters < 100) ):
    iters = iters + 1
    low = oq.Calculate_MinMax( Values,'min' )
    oq.Nelder_Mead(H, oq.Two_Qubit_Ansatz, Vertices, Values, EV_type)
    new_low = oq.Calculate_MinMax( Values,'min' )
    if( abs( new_low[0] - low[0] ) < delta ):
        terminate_count = terminate_count + 1
    else:
        terminate_count = 0
    if( terminate_count >= terminate_limit ):
        terminate = True
        print('\n_____ Nelder-Mead Complete _____\n')
        print(' --------------------- \n Iteration: ',iters,'   Lowest EV:  ',round( low[0],6 ))
    if( ( (iters==1) or (m.ceil(iters/10))==m.floor(iters/10) ) and (terminate==False) ):
        print('Iteration: ',iters,'   Lowest EV:  ',round( low[0],6 ))
```

```
Iteration:  1    Lowest EV:   -0.7524
Iteration:  10    Lowest EV:   -1.3546
Iteration:  20    Lowest EV:   -2.6522
Iteration:  30    Lowest EV:   -4.2674
Iteration:  40    Lowest EV:   -4.69
Iteration:  50    Lowest EV:   -4.9458
Iteration:  60    Lowest EV:   -4.9806
Iteration:  70    Lowest EV:   -4.9962

_____ Nelder-Mead Complete _____

 --------------------- 
 Iteration:  80    Lowest EV:   -4.9996
```

As illustrated above, $8$ parameters is sufficient for coming very close to the true ground state energy. And if you're curious as to how the algorithm would have performed with the full $16$ parameter ansatz circuit, I encourage you to try for yourself by replacing the **Two_Qubit_Ansatz** function with the full circuit shown earlier. What you will likely find is that the Nelder-Mead search struggles to find the global minimum, oftentimes getting stuck in local minima or terminating early due to slow progression.

---

This concludes our final lesson, and our discussion of the Variational Quantum Eigensolver Algorithm. Now that we've seen both the QAOA and VQE algorithms in action, hopefully you have a deeper understanding as to why these algorithms have so quickly risen in popularity in recent years. In essence, the goal of both hybrid algorithms is the same, electing to use a quantum computer to generate a new parameter space versus the original classical space of possible solutions. But the most appealing aspect about both algorithms is their flexibility, particularly VQE and its ability to handle any Hamiltonian.

I hope you enjoyed this lesson, and I encourage you to take a look at my other .ipynb tutorials!

## Citations

**[1]** A. Peruzzo, J. McClean, P. Shadbolt, M-H. Yung, X-Q. Zhou, P. J. Love, A. Aspuru-Guzik, J. L. O'Brien, "A Variational Eigenvalue Solver on a Quantum Processor", Nature Communications **5** (2014)

**[2]** J. A. Nelder, R. Mead, "A Simplex Method for Function Minimization", The Computer Journal **7** (1965)

**[3]** V. V. Shende, I. L. Markov, S. S. Bullock, "Minimal Universal Two-qubit Controlled-NOT-based Circuits", APS Physical Review A **69**, 062321 (2004)

# Our_Qiskit_Functions.py

In [ ]:

```
from qiskit import ClassicalRegister, QuantumRegister, QuantumCircuit, Aer, execute
from qiskit.extensions.simulator import snapshot
from qiskit.tools.visualization import circuit_drawer
import numpy as np
import math as m
import scipy as sci
import random
import time
import matplotlib
import matplotlib.pyplot as plt
S_simulator = Aer.backends(name='statevector_simulator')[0]
M_simulator = Aer.backends(name='qasm_simulator')[0]



#==================================================
#-----------     Displaying Results     -----------
#==================================================

def Wavefunction(obj, **kwargs):
    '''
    Prints a tidier versrion of the array statevector corresponding to the wavefuntion of a QuantumCircuit object
    Keyword Arguments:     precision (integer) -  the decimal precision for amplitudes
                           column (Bool) -  prints each state in a vertical column
                           systems (array of integers) -  seperates the qubits into different states
                           show_systems (array of Bools) -  indictates which qubit systems to print
    '''
    if(type(obj) == QuantumCircuit  ):
        statevec = execute( obj, S_simulator, shots=1 ).result().get_statevector()
    if(type(obj) == np.ndarray):
        statevec = obj
    sys = False
    NL = False
    dec = 5
    if 'precision' in kwargs:
        dec = int( kwargs['precision'] )
    if 'column' in kwargs:
        NL = kwargs['column']
    if 'systems' in kwargs:
        systems = kwargs['systems']
        sys = True
        last_sys = int(len(systems)-1)
        show_systems = []
        for s_chk in np.arange(len(systems)):
            if( type(systems[s_chk])!=int ):
                raise Exception('systems must be an array of all integers')
        if 'show_systems' in kwargs:
            show_systems = kwargs['show_systems']
            if( len(systems)!=len(show_systems) ):
                raise Exception('systems and show_systems need to be arrays of equal length')
            for ls in np.arange(len(show_systems)):
                if((show_systems[ls]!=True)and(show_systems[ls]!=False)):
                    raise Exception('show_systems must be an array of Truth Values')
                if(show_systems[ls]==True):
                    last_sys = int(ls)
        else:
            for ss in np.arange(len(systems)):
                show_systems.append(True)
    wavefunction = ''
    qubits = int(m.log(len(statevec),2))
    for i in np.arange( int(len(statevec)) ):
        value = round(statevec[i].real, dec) + round(statevec[i].imag, dec) * 1j
        if( (value.real!=0) or (value.imag!=0) ):
            state = list(Binary(int(i),int(2**qubits),'L'))
            state_str = ''
            if( sys == True ):
                k = 0
                for s in np.arange(len(systems)):
                    if(show_systems[s]==True):
                        if(int(s)!=last_sys):
                            state.insert( int(k+systems[s]),'>|' )
                            k = int(k+systems[s]+1)
                        else:
                            k = int(k+systems[s])
                    else:
                        for s2 in np.arange(systems[s]):
                            del state[int(k)]
            for j in np.arange(len(state)):
                if(type(state[j])!=str):
                    state_str = state_str+str(int(state[j]))
                else:
                    state_str = state_str+state[j]
            if( (value.real!=0) and (value.imag!=0) ):
                if( value.imag > 0):
                    wavefunction = wavefunction +str(value.real)+'+'+str(value.imag)+'j |'+state_str+'>    '
                else:
                    wavefunction = wavefunction +str(value.real)+''+str(value.imag)+'j |'+state_str+'>    '
            if( (value.real!=0) and (value.imag==0) ):
                wavefunction = wavefunction +str(value.real)+'  |'+state_str+'>    '
```

```
            if( (value.real==0) and (value.imag!=0) ):
                wavefunction = wavefunction +str(value.imag)+'j |'+state_str+'>    '
            if(NL):
                wavefunction = wavefunction + '\n'
    print(wavefunction)


def Measurement(quantumcircuit, **kwargs):
    '''
    Executes a measurement(s) of a QuantumCircuit object for tidier printing
    Keyword Arguments:     shots (integer) -  number of trials to execute for the measurement(s)
                           return_M (Bool) -  indictaes whether to return the Dictionary object containng measurement resul
                           print_M (Bool) -  indictaes whether to print the measurement results
                           column (Bool) -  prints each state in a vertical column
    '''
    p_M = True
    S=1
    ret = False
    NL = False
    if 'shots' in kwargs:
        S = int(kwargs['shots'])
    if 'return_M' in kwargs:
        ret = kwargs['return_M']
    if 'print_M' in kwargs:
        p_M = kwargs['print_M']
    if 'column' in kwargs:
        NL = kwargs['column']
    M1 = execute(quantumcircuit, M_simulator, shots=S).result().get_counts(quantumcircuit)
    M2 = {}
    k1 = list(M1.keys())
    v1 = list(M1.values())
    for k in np.arange(len(k1)):
        key_list = list(k1[k])
        new_key = ''
        for j in np.arange(len(key_list)):
            new_key = new_key+key_list[len(key_list)-(j+1)]
        M2[new_key] = v1[k]
    if(p_M):
        k2 = list(M2.keys())
        v2 = list(M2.values())
        measurements = ''
        for i in np.arange( len(k2) ):
            m_str = str(v2[i])+'|'
            for j in np.arange(len(k2[i])):
                if( k2[i][j] == '0' ):
                    m_str = m_str+'0'
                if( k2[i][j] == '1' ):
                    m_str = m_str+'1'
                if( k2[i][j] == ' ' ):
                    m_str = m_str+'>|'
            m_str = m_str+'>    '
            if(NL):
                m_str = m_str + '\n'
            measurements = measurements + m_str
        print(measurements)
    if(ret):
        return M2


def Most_Probable(M,N):
    '''
    Input:     M (Dictionary)    N (integer)
    Returns the N most probable states accoding to the measurement counts stored in M
    '''
    count = []
    state = []
    if( len(M) < N ):
        N = len(M)
    for k in np.arange(N):
        count.append(0)
        state.append(0)
    for m in np.arange(len(M)):
        new = True
        for n in np.arange(N):
            if( (list(M.values())[int(m)] > count[int(n)]) and (new) ):
                for i in np.arange( int(N-(n+1)) ):
                    count[-int(1+i)] = count[-int(1+i+1)]
                    state[-int(1+i)] = state[-int(1+i+1)]
                count[int(n)] = list(M.values())[m]
                state[int(n)] = list(M.keys())[m]
                new = False
    return count,state

#================================================
#-----------     Math Operations     -----------
#================================================


def Binary(N, total, LSB):
```

```
    '''
    Input:     N (integer)   total (integer)   LSB (string)
    Returns the base-2 binary equivilant of N according to left or right least significant bit notation
    '''
    qubits = int(m.log(total,2))
    b_num = np.zeros(qubits)
    for i in np.arange(qubits):
        if( N/((2)**(qubits-i-1)) >= 1 ):
            if(LSB=='R'):
                b_num[i] = 1
            if(LSB=='L'):
                b_num[int(qubits-(i+1))] = 1
            N = N - 2**(qubits-i-1)
    B = []
    for j in np.arange(len(b_num)):
        B.append(int(b_num[j]))
    return B


def From_Binary(S, LSB):
    '''
    Input:     S (string or array)   LSB (string)
    Converts a base-2 binary number to base-10 according to left or right least significant bit notation
    '''
    num = 0
    for i in np.arange(len(S)):
        if(LSB=='R'):
            num = num + int(S[int(0-(i+1))]) * 2**(i)
        if(LSB=='L'):
            num = num + int(S[int(i)]) * 2**(i)
    return num


#==================================================
#-----------     Custom Operations     -----------
#==================================================


def X_Transformation(qc, qreg, state):
    '''
    Input:     qc (QuantumCircuit)   qreg (QuantumRegister)   state (array)
    Applies the neccessary X gates to transform 'state' to the state of all 1's
    '''
    for j in np.arange(len(state)):
        if( int(state[j])==0 ):
            qc.x( qreg[int(j)] )


def n_NOT(qc, control, target, anc):
    '''
    Input:     qc (QuantumCircuit)   control (QuantumRegister)   target (QuantumRegister[integer])   anc (QuantumRegister)
    Applies the neccessary CCX gates to perform a higher order control-X operation on the target qubit
    '''
    n = len(control)
    instructions = []
    active_ancilla = []
    q_unused = []
    q = 0
    a = 0
    while( (n > 0) or (len(q_unused)!=0) or (len(active_ancilla)!=0) ):
        if( n > 0 ):
            if( (n-2) >= 0 ):
                instructions.append( [control[q], control[q+1], anc[a]] )
                active_ancilla.append(a)
                a = a + 1
                q = q + 2
                n = n - 2
            if( (n-2) == -1 ):
                q_unused.append( q )
                n = n - 1
        elif( len(q_unused) != 0 ):
            if(len(active_ancilla)!=1):
                instructions.append( [control[q], anc[active_ancilla[0]], anc[a]] )
                del active_ancilla[0]
                del q_unused[0]
                active_ancilla.append(a)
                a = a + 1
            else:
                instructions.append( [control[q], anc[active_ancilla[0]], target] )
                del active_ancilla[0]
                del q_unused[0]
        elif( len(active_ancilla)!=0 ):
            if( len(active_ancilla) > 2 ):
                instructions.append( [anc[active_ancilla[0]], anc[active_ancilla[1]], anc[a]]  )
                active_ancilla.append(a)
                del active_ancilla[0]
                del active_ancilla[0]
                a =  a + 1
            elif( len(active_ancilla)==2):
```

```python
                instructions.append([anc[active_ancilla[0]], anc[active_ancilla[1]], target])
                del active_ancilla[0]
                del active_ancilla[0]
    for i in np.arange( len(instructions) ):
        qc.ccx( instructions[i][0], instructions[i][1], instructions[i][2] )
    del instructions[-1]
    for i in np.arange( len(instructions) ):
        qc.ccx( instructions[0-(i+1)][0], instructions[0-(i+1)][1], instructions[0-(i+1)][2] )


def n_Control_U(qc, control, anc, gates):
    '''
    Input:     qc (QuantumCircuit)   control (QuantumRegister)   anc (QuantumRegister)
               gates (array of the form [[string,QuantumRegister[i]],[],...])
    Performs the list of control gates on the respective target qubits as a higher order N-control operation
    '''
    if( len(gates)!=0 ):
        instructions = []
        active_ancilla = []
        q_unused = []
        n = len(control)
        q = 0
        a = 0
        while( (n > 0) or (len(q_unused)!=0) or (len(active_ancilla)!=0) ):
            if( n > 0 ):
                if( (n-2) >= 0 ):
                    instructions.append( [control[q], control[q+1], anc[a]] )
                    active_ancilla.append(a)
                    a = a + 1
                    q = q + 2
                    n = n - 2
                if( (n-2) == -1 ):
                    q_unused.append( q )
                    n = n - 1
            elif( len(q_unused) != 0 ):
                if(len(active_ancilla)>1):
                    instructions.append( [control[q], anc[active_ancilla[0]], anc[a]] )
                    del active_ancilla[0]
                    del q_unused[0]
                    active_ancilla.append(a)
                    a = a + 1
                else:
                    instructions.append( [control[q], anc[active_ancilla[0]], anc[a]] )
                    del active_ancilla[0]
                    del q_unused[0]
                    c_a = anc[a]
            elif( len(active_ancilla)!=0 ):
                if( len(active_ancilla) > 2 ):
                    instructions.append( [anc[active_ancilla[0]], anc[active_ancilla[1]], anc[a]]  )
                    active_ancilla.append(a)
                    del active_ancilla[0]
                    del active_ancilla[0]
                    a =  a + 1
                elif( len(active_ancilla)==2):
                    instructions.append([anc[active_ancilla[0]], anc[active_ancilla[1]], anc[a]] )
                    del active_ancilla[0]
                    del active_ancilla[0]
                    c_a = anc[a]
                elif( len(active_ancilla)==1):
                    c_a = anc[active_ancilla[0]]
                    del active_ancilla[0]
        for i in np.arange( len(instructions) ):
            qc.ccx( instructions[i][0], instructions[i][1], instructions[i][2] )
        for j in np.arange(len(gates)):
            control_vec = [ gates[j][0], c_a ]
            for k in np.arange( 1, len(gates[j])):
                control_vec.append( gates[j][k] )
            if( control_vec[0] == 'X' ):
                qc.cx( control_vec[1], control_vec[2] )
            if( control_vec[0] == 'Z' ):
                qc.cz( control_vec[1], control_vec[2] )
            if( control_vec[0] == 'PHASE' ):
                qc.cu1( control_vec[2], control_vec[1], control_vec[3] )
            if( control_vec[0] == 'SWAP' ):
                qc.cswap( control_vec[1], control_vec[2], control_vec[3] )
        for i in np.arange( len(instructions) ):
            qc.ccx( instructions[0-(i+1)][0], instructions[0-(i+1)][1], instructions[0-(i+1)][2] )


#===============================================
#-----------     Lesson 6 - QFT    ------------
#===============================================


def QFT(qc, q, qubits, **kwargs):
    '''
    Input:     qc (QuantumCircuit)   q (QuantumRegister)   qubits (integer)
    Keyword Arguments:     swap (Bool) -  Adds SWAP gates after all of the phase gates have been applied
    Assigns all the gate operations for a Quantum Fourier Transformation
```

```
    '''
    R_phis = [0]
    for i in np.arange(2,int(qubits+1)):
        R_phis.append( 2/(2**(i)) * m.pi )
    for j in np.arange(int(qubits)):
        qc.h( q[int(j)] )
        for k in np.arange(int(qubits-(j+1))):
            qc.cu1( R_phis[k+1], q[int(j+k+1)], q[int(j)] )
    if 'swap' in kwargs:
        if(kwargs['swap'] == True):
            for s in np.arange(m.floor(qubits/2.0)):
                qc.swap( q[int(s)],q[int(qubits-1-s)] )


def QFT_dgr(qc, q, qubits, **kwargs):
    '''
    Input:     qc (QuantumCircuit)   q (QuantumRegister)   qubits (integer)
    Keyword Arguments:     swap (Bool) -  Adds SWAP gates after all of the phase gates have been applied
    Assigns all the gate operations for a Quantum Fourier Transformation
    '''
    if 'swap' in kwargs:
        if(kwargs['swap'] == True):
            for s in np.arange(m.floor(qubits/2.0)):
                qc.swap( q[int(s)],q[int(qubits-1-s)] )
    R_phis = [0]
    for i in np.arange(2,int(qubits+1)):
        R_phis.append( -2/(2**(i)) * m.pi )
    for j in np.arange(int(qubits)):
        for k in np.arange(int(j)):
            qc.cu1(R_phis[int(j-k)], q[int(qubits-(k+1))], q[int(qubits-(j+1))] )
        qc.h( q[int(qubits-(j+1))] )


def DFT(x, **kwargs):
    '''
    Input:     x (array)
    Keyword Arguments:     inverse (Bool) - if True, performs a Inverse Discrete Fourier Transformation instead
    Computes a classical Discrete Fourier Transformation on the array of values x, returning a new array of transformed val
    '''
    p = -1.0
    if 'inverse' in kwargs:
        P = kwargs['inverse']
        if(P == True):
            p = 1.0
    L = len(x)
    X = []
    for i in np.arange(L):
        value = 0
        for j in np.arange(L):
            value = value + x[j]*np.exp(p*2*m.pi*1.0j * ( int(i*j)/(L*1.0) ) )
        X.append(value)
    for k in np.arange(len(X)):
        re = round(X[k].real,5)
        im = round(X[k].imag,5)
        if( (abs(im) == 0) and (abs(re) != 0) ):
            X[k] = re
        elif( (abs(re) == 0) and (abs(im) != 0)  ):
            X[k] = im*1.0j
        elif( (abs(re) == 0) and (abs(im) == 0)  ):
            X[k] = 0
        else:
            X[k] = re + im*1.0j
    return X


#==========================================================
#-----------     Lesson 6.1 - Quantum Adder    ------------
#==========================================================


def Quantum_Adder(qc, Qa, Qb, A, B):
    '''
    Input:  qc (QuantumCircuit)   Qa (QuantumRegister)   Qb (QuantumRegister)   A (array)   B (array)
    Appends all of the gate operations for a QFT based addition of two states A and B
    '''
    Q = len(B)
    for n in np.arange(Q):
        if( A[n] == 1 ):
            qc.x( Qa[int(n+1)] )
        if( B[n] == 1 ):
            qc.x( Qb[int(n)] )
    QFT(qc,Qa,Q+1)
    p = 1
    for j in np.arange( Q ):
        qc.cu1( m.pi/(2**p), Qb[int(j)], Qa[0] )
        p = p + 1
    for i in np.arange(1,Q+1):
        p = 0
        for jj in np.arange( i-1, Q ):
```

```
            qc.cu1( m.pi/(2**p), Qb[int(jj)], Qa[int(i)] )
            p = p + 1
    QFT_dgr(qc,Qa,Q+1)


#===============================================
#-----------     Lesson 7 - QPE    ------------
#===============================================


def QPE_phi(MP):
    '''
    Input:     array( [[float,float],[string,string]] )
    Takes in the two most probable states and their probabilities, returns phi and the approximate theta for QPE
    '''
    ms = [[],[]]
    for i in np.arange(2):
        for j in np.arange(len(MP[1][i])):
            ms[i].append(int(MP[1][i][j]))
    n = int(len(ms[0]))
    MS1 = From_Binary(ms[0],'R')
    MS2 = From_Binary(ms[1],'R')
    PHI = [99,0]
    for k in np.arange(1,5000):
        phi = k/5000
        prob = 1/(2**(2*n)) * abs( (-1 + np.exp(2.0j*m.pi*phi) )/(-1 + np.exp(2.0j*m.pi*phi/(2**n))) )**2
        if( abs( prob - MP[0][0] ) < abs( PHI[0] - MP[0][0]) ):
            PHI[0] = prob
            PHI[1] = phi
    if( (MS1 < MS2) and ( (MS1!=0) and (MS2!=(2**n-1)) ) ):
        theta = (MS1+PHI[1])/(2**n)
    elif( (MS1 > MS2) and (MS1!=0) ):
        theta = (MS1-PHI[1])/(2**n)
    else:
        theta = 1+(MS1-PHI[1])/(2**n)
    return PHI[1],theta

#==============================================================
#-----------     Lesson 7.1 - Quantum Counting    ------------
#==============================================================


def Grover_Oracle(mark, qc, q, an1, an2):
    '''
    Input:     mark (array)   qc (QuantumCircuit)   q (QuantumRegister)   an1 (QuantumRegister)   an2 (QuantumRegister)
    Appends the neccessary gates for a phase flip on the marked state
    '''
    qc.h( an1[0] )
    X_Transformation(qc, q, mark)
    if( len(mark) > 2 ):
        n_NOT( qc, q, an1[0], an2 )
    if( len(mark) == 2 ):
        qc.ccx( q[0], q[1], an1[0] )
    X_Transformation(qc, q, mark)
    qc.h( an1[0] )


def Grover_Diffusion(mark, qc, q, an1, an2):
    '''
    Input:     mark (array)   qc (QuantumCircuit)   q (QuantumRegister)   an1 (QuantumRegister)   an2 (QuantumRegister)
    Appends the neccessary gates for a Grover Diffusion operaton
    '''
    zeros_state = []
    for i in np.arange( len(mark) ):
        zeros_state.append( 0 )
        qc.h( q[int(i)] )
    Grover_Oracle(zeros_state, qc, q, an1, an2)
    for j in np.arange( len(mark) ):
        qc.h( q[int(j)] )


def Multi_Grover(q, a1, a2, qc, marked, iters):
    '''
    Input:     q (QuantumRegister)   a1 (QuantumRegister)   a2 (QuantumRegister)   qc (QuantumCircuit)
               marked (array)   iters (integer)
    Appends all of the gate operations for a multi-marked state Grover Search
    '''
    Q = int(len(marked))
    for i in np.arange( iters ):
        for j in np.arange(len(marked)):
            M = list(marked[j])
            for k in np.arange(len(M)):
                if(M[k]=='1'):
                    M[k] = 1
                else:
                    M[k] = 0
            Grover_Oracle(M, qc, q, a1, a2)
        Grover_Diffusion(M, qc, q, a1, a2)
    return qc, q, a1, a2
```

```
def C_Oracle(qc, c, q, a1, a2, state):
    '''
    Input:     qc (QuantumCircuit)   c (QuantumRegister[i])   q (QuantumRegister)
               a1 (QuantumRegister)   a2 (QuantumRegister)   state (array)
    Appends all of the gate operations for a control-Grover Oracle operation
    '''
    N = len(state)
    for i in np.arange(N):
        if( state[i]==0 ):
            qc.cx( c, q[int(i)] )
    #---------------------------------
    qc.ccx( q[0], q[1], a1[0] )
    for j1 in np.arange(N-2):
        qc.ccx( q[int(2+j1)], a1[int(j1)], a1[int(1+j1)] )
    qc.ccx( c, a1[N-2], a2[0] )
    for j2 in np.arange(N-2):
        qc.ccx( q[int(N-1-j2)], a1[int(N-3-j2)], a1[int(N-2-j2)] )
    qc.ccx( q[0], q[1], a1[0] )
    #---------------------------------
    for i2 in np.arange(N):
        if( state[i2]==0 ):
            qc.cx( c, q[int(i2)] )


def C_Diffusion(qc, c, q, a1, a2, Q, ref):
    '''
    Input:     qc (QuantumCircuit)   c (QuantumRegister[i])   q (QuantumRegister)   a1 (QuantumRegister)
               a2 (QuantumRegister)   Q (integer)    ref (Bool)
    Appends all of the gate operations for a control-Grover Diffusion operation
    '''
    N = 2**( Q )
    for j in np.arange(Q):
        qc.ch( c, q[int(j)] )
    if( ref ):
        for k in np.arange(1,N):
            C_Oracle(qc,c,q,a1,a2,Binary(int(k),N,'R'))
    else:
        C_Oracle(qc,c,q,a1,a2,Binary(0,N,'R'))
    for j2 in np.arange(Q):
        qc.ch( c, q[int(j2)] )


def C_Grover(qc, c, q, a1, a2, marked, **kwargs):
    '''
    Input:     qc (QuantumCircuit)   c (QuantumRegister[i])   q (QuantumRegister)
               a1 (QuantumRegister)   a2 (QuantumRegister)   marked (array)
    Keyword Arguments:     proper (Bool) -  Dicates how to perform the reflection about the average within the Diffusion Op
    Appends all of the gate operations for a control-Grover
    '''
    Reflection=False
    if 'proper' in kwargs:
        Reflection = kwargs['proper']
    M = []
    for m1 in np.arange( len(marked) ):
        M.append( list(marked[m1]) )
        for m2 in np.arange( len(M[m1]) ):
            M[m1][m2] = int( M[m1][m2] )
    L = len(M[0])
    N = 2**( L )
    for i in np.arange(len(M)):
        C_Oracle( qc,c,q,a1,a2,M[i] )
    C_Diffusion( qc,c,q,a1,a2,L,Reflection )


#================================================
#-----------     Lesson 8 - Shor's     -----------
#================================================

def GCD(a, b):
    '''
    Input:     a (integer)   b (integer)
    Computes the greatest common denominator between a and b using an inefficient exhaustive search
    '''
    gcd = 0
    if(a > b):
        num1 = a
        num2 = b
    elif(b > a):
        num1 = b
        num2 = a
    elif(a == b):
        gcd = a
    while( gcd == 0 ):
        i = 1
        while( num1 >= num2*i ):
            i = i + 1
        if( num1 == num2*(i-1) ):
```

```
            gcd = num2
        else:
            r = num1 - num2*(i-1)
            num1 = num2
            num2 = r
    return gcd


def Euclids_Alg(a, b):
    '''
    Input:     a (integer)   b (integer)
    Computes the greatest common denominator between a and b using Euclid's Algorithm
    '''
    if(a>b):
        num1 = a
        num2 = b
    if(b>a):
        num1 = b
        num2 = a
    if(b==a):
        gcd = a
        r_old = 0
    r_new = int( num1%num2 )
    r_old = int( num2 )
    while(r_new!=0):
        r_old = r_new
        r_new = int( num1%num2 )
        num1 = num2
        num2 = r_new
    gcd = r_old
    return gcd


def Modulo_f(Q, a, N):
    '''
    Input:     Q (integer)   a (integer)   N (integer)
    Produces an array of all the final modulo N results for the power function a^x (mod N)
    '''
    mods = [1]
    for i in np.arange(1,2**Q):
        if(i==1):
            mods.append(a**i%N)
            num = a**i%N
        if(i>1):
            mods.append((num*a)%N)
            num = (num*a)%N
    return mods


def Mod_Op(Q, qc, q1, q2, anc, a, N):
    '''
    Input:     Q (integer)   qc (QuantumCircuit)   q1 (QuantumRegister)   q2 (QuantumRegister)
               anc (QuantumRegister)   a (integer)   N (integer)
    Applies the Modulo Multiplication operator for Shor's algorithm
    '''
    mods = Modulo_f(Q,a,N)
    for j in np.arange( 2**Q ):
        q1_state = Binary( j, 2**Q, 'L' )
        q2_state = Binary( mods[j], 2**Q ,'L' )
        gates = []
        for k in np.arange(Q):
            if(q2_state[k]==1):
                gates.append(['X',q2[int(k)]])
        X_Transformation(qc,q1,q1_state)
        n_Control_U(qc, q1, anc, gates)
        X_Transformation(qc,q1,q1_state)


def ConFrac(N, **kwargs):
    '''
    Input:     N (float)
    Keyword Arguments:     a_max (integer) -  the maximum number of iterations to continue approximating
                           return_a (Bool) -  if True, returns the array a containing the continued fraction information
    Evaluates the non-integer number N as the quantity p/q, where p and q are integers
    '''
    imax = 20
    r_a = False
    if 'a_max' in kwargs:
        imax = kwargs['a_max']
    if 'return_a' in kwargs:
        r_a = kwargs['return_a']
    a = []
    a.append( m.floor(N) )
    b = N - a[0]
    i = 1
    while( (round(b,10) != 0) and (i < imax) ):
        n = 1.0/b
        a.append( m.floor(n) )
        b = n - a[-1]
```

```
        i = i + 1
    #------------------------------
    a_copy = []
    for ia in np.arange(len(a)):
        a_copy.append(a[ia])
    for j in np.arange( len(a)-1 ):
        if( j == 0 ):
            p = a[-1] * a[-2] + 1
            q = a[-1]
            del a[-1]
            del a[-1]
        else:
            p_new = a[-1] * p + q
            q_new = p
            p = p_new
            q = q_new
            del a[-1]
    if(r_a == True):
        return q,p,a_copy
    else:
        return q,p


def r_Finder(a, N):
    '''
    Input:     a (integer)   N (integer)
    Exhaustively computes the period r to the modulo power function a^x (mod N)
    '''
    value1 = a**1 % N
    r = 1
    value2 = 0
    while( (value1 != value2) or (r >1000) ):
        value2 = a**(int(1+r)) % N
        if( value1 != value2 ):
            r = r + 1
    return r


def Primality(N):
    '''
    Input:     N (integer)
    Returns True is N is a prime number, otherwise False
    '''
    is_prime = True
    if( (N==1) or (N==2) or (N==3) ):
        is_prime = True
    elif( (N%2==0) or (N%3==0) ):
        is_prime = False
    elif( is_prime==True ):
        p = 5
        while( (p**2 <= N) and (is_prime==True) ):
            if( (N%p==0) or (N%(p+2)==0) ):
                is_prime = False
            p = p + 6
    return is_prime


def Mod_r_Check(a, N, r):
    '''
    Input:     a (integer)   N (integer)   r (integer)
    Checks a value of r, returning True or False based on whether it correctly leads to a factor of N
    '''
    v1 = a**(int(2)) % N
    v2 = a**(int(2+r)) % N
    if( (v1 == v2) and (r<N) and (r!=0) ):
        return True
    else:
        return False


def Evaluate_S(S, L, a, N):
    '''
    Input:     S (integer)   L (integer)   a (integer)   N (integer)
    Attempts to use the measured state |S> to find the period r
    '''
    Pairs = [[S,L]]
    for s in np.arange(3):
        S_new = int( S - 1 + s)
        for l in np.arange(3):
            L_new = int( L - 1 + l)
            if( ((S_new!=S) or (L_new!=L)) and (S_new!=L_new) ):
                Pairs.append( [S_new,L_new] )
    #---------------------------                    Try 9 combinations of S and L, plus or minus 1 from S & L
    period = 0
    r_attempts = []
    found_r = False
    while( (found_r==False) and (len(Pairs)!=0) ):
        order = 1
        S_o = Pairs[0][0]
```

```
        L_o = Pairs[0][1]
        q_old = -1
        q = 999
        while( q_old != q ):
            q_old = int( q )
            q,p = ConFrac(S_o/L_o,a_max=(order+1))
            new_r = True
            for i in np.arange(len(r_attempts)):
                if( q == r_attempts[i] ):
                    new_r = False
            if(new_r):
                r_attempts.append( int(q) )
                r_bool = Mod_r_Check(a,N,q)
                if( r_bool ):
                    found_r = True
                    q_old = q
                    period = int(q)
            order = order + 1
        del Pairs[0]
    #---------------------------                    Try higher multiples of already attempted r values
    r_o = 0
    while( (found_r == False) and (r_o < len(r_attempts)) ):
        k = 2
        r2 = r_attempts[r_o]
        while( k*r2 < N ):
            r_try = int(k*r2)
            new_r = True
            for i2 in np.arange(len(r_attempts)):
                if( r_try == r_attempts[i2] ):
                    new_r = False
            if(new_r):
                r_attempts.append( int(r_try) )
                r_bool = Mod_r_Check(a,N,r_try)
                if( r_bool ):
                    found_r = True
                    k = N
                    period = int(r_try)
            k = k + 1
        r_o = r_o + 1
    #---------------------------                    If a period is found, try factors of r for smaller periods
    if( found_r == True ):
        Primes = []
        for i in np.arange(2,period):
            if( Primality(int(i)) ):
                Primes.append(int(i))
        if( len(Primes) > 0 ):
            try_smaller = True
            while( try_smaller==True ):
                found_smaller = False
                p2 = 0
                while( (found_smaller==False) and (p2 < len(Primes)) ):
                    #print('p2:  ',p2)
                    #print( 'period:  ',period,'   ',Primes[p2] )
                    try_smaller = False
                    if( period/Primes[p2] == m.floor( period/Primes[p2] ) ):
                        r_bool_2 = Mod_r_Check(a,N,int(period/Primes[p2]))
                        if( r_bool_2 ):
                            period = int(period/Primes[p2])
                            found_smaller = True
                            try_smaller = True
                    p2 = p2 + 1
    return period


#====================================================
#-----------     Lesson 9 - Q-Means     ------------
#====================================================


def k_Data(k,n):
    '''
    Input:     k (integer)    n (ineger)
    Creates a random set of data loosely centered around k locations
    '''
    Centers = []
    for i in np.arange(k):
        Centers.append( [1.5+np.random.rand()*5,1.5*random.random()*5] )
    count = round((0.7*n)/k)
    Data = []
    for j in np.arange(len(Centers)):
        for j2 in np.arange(count):
            r = random.random()*1.5
            Data.append( [ Centers[j][0]+r*np.cos(random.random()*2*m.pi) , Centers[j][1]+r*np.sin(random.random()*2*m.pi)
    diff = int( n - k*count )
    for j2 in np.arange(diff):
        Data.append( [random.random()*8,random.random()*8] )
    return Data
```

```python
def Initial_Centroids(k, D):
    '''
    Input:     k (integer)   D (array)
    Picks k data points at random from the list D
    '''
    D_copy = []
    for i in np.arange( len(D) ):
        D_copy.append( D[i] )
    Centroids = []
    for j in np.arange(k):
        p = random.randint(0,int(len(D_copy)-1))
        Centroids.append( [ D_copy[p][0] , D_copy[p][1] ] )
        D_copy.remove( D_copy[p] )
    return Centroids


def Update_Centroids(CT, CL):
    '''
    Input:     CT (array)   CL (array)
    Based on the data within each cluster, computes and returns new Centroids using mean coordinate values
    '''
    old_Centroids = []
    for c0 in np.arange(len(CT)):
        old_Centroids.append(CT[c0])
    Centroids = []
    for c1 in np.arange(len(CL)):
        mean_x = 0
        mean_y = 0
        for c2 in np.arange(len(CL[c1])):
            mean_x = mean_x + CL[c1][c2][0]/len(CL[c1])
            mean_y = mean_y + CL[c1][c2][1]/len(CL[c1])
        Centroids.append( [ mean_x,mean_y ] )
    return Centroids, old_Centroids


def Update_Clusters(D, CT, CL):
    '''
    Input:     D (array)   CT (array)   CL (array)
    Using all data points and Centroids, computes and returns the new array of Clusters
    '''
    old_Clusters = []
    for c0 in np.arange(len(CL)):
        old_Clusters.append(CL[c0])
    Clusters = []
    for c1 in np.arange( len(CT) ):
        Clusters.append( [] )
    for d in np.arange( len(D) ):
        closest = 'c'
        distance = 100000
        for c2 in np.arange( len(Clusters) ):
            Dist = m.sqrt( ( CT[c2][0] - D[d][0] )**2 + ( CT[c2][1] - D[d][1] )**2 )
            if( Dist < distance ):
                distance = Dist
                closest = int(c2)
        Clusters[closest].append( D[d] )
    return Clusters,old_Clusters


def Check_Termination(CL, oCL ):
    '''
    Input:     CL (array)   oCL (array)
    Returns True or False based on whether the Update_Clusters function has caused any data points to change clusters
    '''
    terminate = True
    for c1 in np.arange( len(oCL) ):
        for c2 in np.arange( len(oCL[c1]) ):
            P_found = False
            for c3 in np.arange( len(CL[c1]) ):
                if( CL[c1][c3] == oCL[c1][c2] ):
                    P_found = True
            if( P_found == False ):
                terminate = False
    return terminate


def Draw_Data(CL, CT, oCT, fig, ax, colors, colors2 ):
    '''
    Input:     CL (array)   CT (array)   oCT (array)   fig (matplotlib figure)
               ax (figure subplot)   colors (array of color strings)   colors2 (array of color strings)
    Using the arrays Clusters, Centroids, and old Centroids, draws and colors each data point according to its cluster
    '''
    for j1 in np.arange( len(CL) ):
        ax.scatter( oCT[j1][0],oCT[j1][1], color='white', marker='s',s=80 )
    for cc in np.arange(len(CL)):
        for ccc in np.arange( len( CL[cc] ) ):
            ax.scatter( CL[cc][ccc][0],CL[cc][ccc][1], color=colors[cc],s=10 )
    for j2 in np.arange( len(CL) ):
        ax.scatter( CT[j2][0],CT[j2][1], color=colors2[j2], marker='x',s=50 )
    fig.canvas.draw()
```

```python
    time.sleep(1)


def SWAP_Test( qc, control, q1, q2, classical, S ):
    '''
    Input:     qc (QuantumCircuit)   control (QuantumRegister[i])   q1 (QuantumRegister[i])   q2 (QuantumRegister[i])
               classical (ClassicalRegister[i])    S (integer)
    Appends the necessary gates for 2-Qubit SWAP Test and returns the number of |0> state counts
    '''
    qc.h( control )
    qc.cswap( control, q1, q2 )
    qc.h( control )
    qc.measure( control, classical )
    D = {'0':0}
    D.update( Measurement(qc,shots=S,return_M=True,print_M=False) )
    return D['0']


def Bloch_State( p,P ):
    '''
    Input:     p (array)   P(array)
    Returns the corresponding theta and phi values of the data point p, according to min / max paramters of P
    '''
    x_min = P[0]
    x_max = P[1]
    y_min = P[2]
    y_max = P[3]
    theta = np.pi/2*( (p[0]-x_min)/(1.0*x_max-x_min) + (p[1]-y_min)/(1.0*y_max-y_min) )
    phi   = np.pi/2*( (p[0]-x_min)/(1.0*x_max-x_min) - (p[1]-y_min)/(1.0*y_max-y_min) + 1 )
    return theta,phi


def Q_Update_Clusters(D, CT, CL, DS, shots):
    '''
    Input:     D (array)   CT (array)   CL (array)   DS (array)   shots (integer)
    Using all data points, Centroids, uses the SWAP Test to compute and return the new array of Clusters
    '''
    old_Clusters = []
    for c0 in np.arange(len(CL)):
        old_Clusters.append(CL[c0])
    Clusters = []
    for c1 in np.arange( len(CT) ):
        Clusters.append( [] )
    #-------------------------------------------------
    for d in np.arange( len(D) ):
        closest = 'c'
        distance = 0
        t,p = Bloch_State( D[d], DS )
        for c2 in np.arange( len(Clusters) ):
            t2,p2 = Bloch_State( CT[c2], DS )
            q = QuantumRegister( 3, name='q' )
            c = ClassicalRegister( 1, name='c' )
            qc= QuantumCircuit( q,c, name='qc' )
            qc.u3( t,   p, 0, q[1] )
            qc.u3( t2, p2, 0, q[2] )
            IP = SWAP_Test( qc, q[0], q[1], q[2], c[0], shots )
            if( IP > distance ):
                distance = IP
                closest = int(c2)
        Clusters[closest].append( D[d] )
    return Clusters,old_Clusters


def Heatmap(data, show_text, show_ticks, ax, cmap, cbarlabel, **kwargs):
    '''
    Input:     data (array)   show_text (Bool)   show_ticks (Bool)   ax (Matplotlib subplot)   cmap (string)   cbarlabel (s
    Takes in data and creates a 2D Heatmap
    '''
    valfmt="{x:.1f}"
    textcolors=["black", "white"]
    threshold=None
    cbar_kw={}
    #----------------------------
    if not ax:
        ax = plt.gca()
    im = ax.imshow(data, cmap=cmap, **kwargs)
    cbar = ax.figure.colorbar(im, ax=ax, **cbar_kw)
    cbar.ax.set_ylabel(cbarlabel, rotation=-90, va="bottom")
    ax.grid(which="minor", color="black", linestyle='-', linewidth=1)

    if( show_ticks == True ):
        ax.set_xticks(np.arange(data.shape[1]))
        ax.set_yticks(np.arange(data.shape[0]))
        ax.tick_params(which="minor", bottom=False, left=False)

    if threshold is not None:
        threshold = im.norm(threshold)
    else:
        threshold = im.norm(data.max())/2.
```

```
    kw = dict(horizontalalignment="center", verticalalignment="center")

    if isinstance(valfmt, str):
        valfmt = matplotlib.ticker.StrMethodFormatter(valfmt)
    if( show_text == True ):
        for i in range(data.shape[0]):
            for j in range(data.shape[1]):
                kw.update(color=textcolors[int(im.norm(data[i, j]) < threshold)])
                text = im.axes.text(j, i, valfmt(data[i, j], None), **kw)


#=================================================
#-----------     Lesson 10 - QAOA    ------------
#=================================================


def E_Expectation_Value( qc, Energies ):
    '''
    Input:     qc (QuantumCircuit)   Energies (array)
    Computes and returns the energy expectation value using the quantum system's wavefunction
    '''
    SV = execute( qc, S_simulator, shots=1 ).result().get_statevector()
    EV = 0
    for i in np.arange( len(SV) ):
        EV = EV + Energies[i] *abs( SV[i] * np.conj(SV[i]) )
    EV = round(EV,4)
    return EV


def Top_States(States, Energies, SV, top):
    '''
    Input:     States (array)   Energies (array)   SV (Qiskit statevector)   top (integer)
    Displays the top most probable states in the system, and their associated energy
    '''
    P = []
    S = []
    E = []
    for a in np.arange( top ):
        P.append(-1)
        S.append('no state')
        E.append('no energy')
    for i in np.arange(len(States)):
        new_top = False
        probs = abs(SV[i]*np.conj(SV[i]))*100
        state = States[i]
        energ = Energies[i]
        j = 0
        while( (new_top == False) and (j < top) ):
            if( probs > P[j] ):
                for k in np.arange( int( len(P) - (j+1) ) ):
                    P[int( -1-k )] = P[int( -1-(k+1) )]
                    S[int( -1-k )] = S[int( -1-(k+1) )]
                    E[int( -1-k )] = E[int( -1-(k+1) )]
                P[j] = probs
                S[j] = state
                E[j] = energ
                new_top = True
            j = int(j+1)
    for s in np.arange( top ):
        print('State ',S[s],'     Probability: ',round(P[s],2),'%','    Energy: ',round(E[s],2))


def Ising_Energy(V, E, **kwargs):
    '''
    Input:     V (array)   E (array)
    Keyword Arguments:     Transverse (Bool) -  Changes to the Transverse Ising Energy Model
    Calculates and returns the energy for each state according to either of the Ising Model Energy functions
    '''
    Trans = False
    if 'Transverse' in kwargs:
        if( kwargs['Transverse'] == True ):
            Trans = True
    Energies = []
    States = []
    for s in np.arange( 2**len(V) ):
        B = Binary(int(s),2**len(V),'L')
        B2 = []
        for i in np.arange(len(B)):
            if( B[i] == 0 ):
                B2.append(1)
            else:
                B2.append(-1)
        state = ''
        energy = 0
        for s2 in np.arange(len(B)):
            state = state+str(B[s2])
            energy = energy - V[s2][1]*B2[s2]
        States.append(state)
```

```
        for j in np.arange( len(E) ):
            if( Trans == False ):
                energy = energy - B2[int(E[j][0])] * B2[int(E[j][1])]
            else:
                energy = energy - B2[int(E[j][0])] * B2[int(E[j][1])] * E[j][2]
        Energies.append(energy)
    return Energies,States


def Ising_Circuit(qc, q, V, E, beta, gamma, **kwargs):
    '''
    Input:     qc (QuantumCircuit)   q (QuantumRegister)   V (array)   E (array)   beta (float)   gamma (float)
    Keyword Arguments:     Transverse (Bool) -  Changes to the Transverse Ising Energy Model
                           Mixing (integer) - Denotes which mixing circuit to use for U(B,beta)
    Constructs the quantum circuit for a given geometry, using the either of the Ising Model Energy functions
    '''
    Trans = False
    if 'Transverse' in kwargs:
        if( kwargs['Transverse'] == True ):
            Trans = True
    Mixer = 1
    if 'Mixing' in kwargs:
        Mixer = int( kwargs['Mixing'] )
    Uc_Ising(qc,q,gamma,V,E,Trans)
    if( Mixer == 2 ):
        Ub_Mixer2(qc,q,beta,V)
    else:
        Ub_Mixer1(qc,q,beta,V)


def Uc_Ising(qc, q, gamma, Vert, Edge, T):
    '''
    Input:     qc (QuantumCircuit)   q (QuantumRegister)   gamma (float)   Vert (array)   Edge (array)   T (Bool)
    Applies the neccessary gates for either of the Ising Energy Model U(C,gamma) operations
    '''
    for e in np.arange( len(Edge) ):                                           # ZZ
        if( T == False ):
            G = gamma
        else:
            G = gamma * Edge[e][2]
        qc.cx( q[int(Edge[e][0])], q[int(Edge[e][1])] )
        qc.rz( 2*G, q[int(Edge[e][1])] )
        qc.cx( q[int(Edge[e][0])], q[int(Edge[e][1])] )
    for v in np.arange( len(Vert) ):                                        # Z_gamma
        qc.rz( gamma, q[int(Vert[v][0])] )


def Ub_Mixer1(qc, q, beta, Vert):
    '''
    Input:     qc (QuantumCircuit)   q (QuantumRegister)   beta (float)   Vert (array)
    Applies the neccessary gates for a U(B,beta) operation using only Rx gates
    '''
    for v in np.arange( len(Vert) ):
        qc.rx( beta, q[int(v)] )


def Ub_Mixer2(qc, q, beta, Vert):
    '''
    Input:     qc (QuantumCircuit)   q (QuantumRegister)   beta (float)   Vert (array)
    Applies the neccessary gates for a U(B,beta) operation using Rx, Ry, and CNOT gates
    '''
    for v in np.arange( len(Vert) ):
        qc.rx( beta, q[int(Vert[v][0])] )
    qc.cx( q[0], q[1] )
    qc.cx( q[2], q[0] )
    for v2 in np.arange( len(Vert) ):
        qc.ry( beta, q[int(Vert[v2][0])] )


def Ising_Gradient_Descent(qc, q, Circ, V, E, beta, gamma, epsilon, En, step, **kwargs):
    '''
    Input:     qc (QuantumCircuit)   q (QuantumRegister)   Circ (Ising_Circuit function)   V (array)   E (array)
               beta (float)   gamma (float)   epsilon   (float)   En (array)   step (float)
    Keyword Arguments:     Transverse (Bool) -  Changes to the Transverse Ising Energy Model
                           Mixing (integer) - Denotes which mixing circuit to use for U(B,beta)
    Calculates and returns the next values for beta and gamma using gradient descent
    '''
    Trans = False
    if 'Transverse' in kwargs:
        if( kwargs['Transverse'] == True ):
            Trans = True
    Mixer = 1
    if 'Mixing' in kwargs:
        Mixer = int(kwargs['Mixing'])
    params = [ [beta+epsilon,gamma],[beta-epsilon,gamma],[beta,gamma+epsilon],[beta,gamma-epsilon] ]
    ev = []
    for i in np.arange( 4 ):
        q = QuantumRegister(len(V))
        qc= QuantumCircuit(q)
```

```
        for hh in np.arange(len(V)):
            qc.h( q[int(hh)] )
        Circ( qc, q, V, E, params[i][0], params[i][1], Transverse=Trans, Mixing=Mixer  )
        ev.append( E_Expectation_Value( qc, En ) )
    beta_next = beta - ( ev[0] - ev[1] )/( 2.0*epsilon ) * step
    gamma_next = gamma - ( ev[2] - ev[3] )/( 2.0*epsilon ) * step
    return beta_next, gamma_next


def MaxCut_Energy(V, E):
    '''
    Input:     V (array)   E (array)
    Calculates and returns the energy for each state according to the MaxCut Energy function
    '''
    Energies = []
    States = []
    for s in np.arange( 2**len(V) ):
        B = Binary(int(s),2**len(V),'L')
        B2 = []
        for i in np.arange(len(B)):
            if( B[i] == 0 ):
                B2.append(1)
            else:
                B2.append(-1)
        state = ''
        for s2 in np.arange(len(B)):
            state = state+str(B[s2])
        States.append(state)
        energy = 0
        for j in np.arange( len(E) ):
            energy = energy +  0.5* ( 1.0 - B2[int(E[j][0])]*B2[int(E[j][1])] )
        Energies.append(energy)
    return Energies,States


def MaxCut_Circuit(qc, q, V, E, beta, gamma):
    '''
    Input:     qc (QuantumCircuit)   q (QuantumRegister)   V (array)   E (array)   beta (float)   gamma (float)
    Constructs the quantum circuit for a given geometry, using the Maxcut Energy Model
    '''
    Uc_MaxCut( qc, q, gamma, E )
    Ub_Mixer1(qc,q,beta,V)


def Uc_MaxCut(qc, q, gamma, edge):
    '''
    Input:     qc (QuantumCircuit)   q (QuantumRegister)   gamma (flaot)   edge (array)
    Applies the neccessary gates for a U(C,gamma) operation for a MaxCut Energy Model
    '''
    for e in np.arange( len(edge) ):
        qc.cx( q[int(edge[e][0])], q[int(edge[e][1])] )
        qc.rz( gamma, q[int(edge[e][1])] )
        qc.cx( q[int(edge[e][0])], q[int(edge[e][1])] )


def p_Gradient_Ascent(qc, q, Circ, V, E, p, Beta, Gamma, epsilon, En, step):
    '''
    Input:     qc (QuantumCircuit)   q (QuantumRegister)   Circ (MaxCut_Circuit function)   V (array)   E (array)
               p (integer)   Beta (array)   Gamma (array)   epsilon (float)   En (array)   step (float)
    Computes the next values for beta and gamma using gradient descent, for a p-dimensional QAOA of the MaxCut Energy Model
    '''
    params = []
    for i in np.arange(2):
        for p1 in np.arange(p):
            if( i == 0 ):
                params.append( Beta[p1] )
            if( i == 1 ):
                params.append( Gamma[p1] )
    ep_params = []
    for p2 in np.arange( len( params ) ):
        for i2 in np.arange( 2 ):
            ep = []
            for p3 in np.arange( len(params) ):
                ep.append( params[p3] )
            ep[p2] = ep[p2] + (-1.0)**(i2+1)*epsilon
            ep_params.append( ep )
    ev = []
    for p4 in np.arange( len( ep_params ) ):
        run_params = ep_params[p4]
        q = QuantumRegister(len(V))
        qc= QuantumCircuit(q)
        for hh in np.arange(len(V)):
            qc.h( q[int(hh)] )
        for p5 in np.arange(p):
            Circ( qc, q, V, E, run_params[int(p5)], run_params[int(p5+p)] )
        ev.append( E_Expectation_Value( qc, En ) )
    Beta_next = []
    Gamma_next = []
    for k in np.arange( len( params ) ):
```

```
        if( k < len( params )/2 ):
            Beta_next.append( params[k] - (ev[int(2*k)] - ev[int(2*k+1)])/( 2.0*epsilon ) * step )
        else:
            Gamma_next.append( params[k] - (ev[int(2*k)] - ev[int(2*k+1)])/( 2.0*epsilon ) * step )
    return Beta_next, Gamma_next


#==============================================
#-----------     Lesson 11 - VQE   ------------
#==============================================


def Single_Qubit_Ansatz( qc, qubit, params ):
    '''
    Input:     qc (QuantumCircuit)   qubit (QuantumRegister[i])   params (array)
    Applies the neccessary rotation gates for a single qubit ansatz state
    '''
    qc.ry( params[0], qubit )
    qc.rz( params[1], qubit )


def VQE_Gradient_Descent(qc, q, H, Ansatz, theta, phi, epsilon, step, **kwargs):
    '''
    Input:     qc (QuantumCircuit)   q (QuantumRegister)   H (Dictionary)   Ansatz (Single_Qubit_Ansatz function)
               theta (float)   phi (float)   epsilon (float)   step (float)
    Keyword Arguments:     measure (Bool) -  Dictates whether to use measurements or the wavefunction
                           shots (integer) -  Dictates the number of measurements to use per computation
    Computes and returns the next values for beta and gamma using gradient descent, for a single qubit VQE
    '''
    EV_type = 'measure'
    if 'measure' in kwargs:
        M_bool = kwargs['measure']
        if( M_bool == True ):
            EV_type = 'measure'
        else:
            EV_type = 'wavefunction'
    Shots = 1000
    if 'shots' in kwargs:
        Shots = kwargs['shots']
    params = [theta,phi]
    ep_params = [[theta+epsilon,phi],[theta-epsilon,phi],[theta,phi+epsilon],[theta,phi-epsilon]]
    Hk = list( H.keys() )
    EV = []
    for p4 in np.arange( len( ep_params ) ):
        H_EV = 0
        qc_params = ep_params[p4]
        for h in np.arange( len(Hk) ):
            qc_params = ep_params[p4]
            q = QuantumRegister(1)
            c = ClassicalRegister(1)
            qc= QuantumCircuit(q,c)
            Ansatz( qc, q[0], [qc_params[0], qc_params[1]]  )
            if( Hk[h] == 'X' ):
                qc.ry(-m.pi/2,q[0])
            if( Hk[h] == 'Y' ):
                qc.rx(-m.pi/2,q[0])
            if( EV_type == 'wavefunction' ):
                sv = execute( qc, S_simulator, shots=1 ).result().get_statevector()
                H_EV =  H_EV + H[Hk[h]]*( (np.conj(sv[0])*sv[0]).real - (np.conj(sv[1])*sv[1]).real )
            elif( EV_type == 'measure' ):
                qc.measure( q,c )
                M = {'0':0,'1':0}
                M.update( Measurement( qc, shots=Shots, print_M=False, return_M=True ) )
                H_EV = H_EV + H[Hk[h]]*(M['0']-M['1'])/Shots
        EV.append(H_EV)
    theta_slope = ( EV[0]-EV[1] )/(2.0*epsilon)
    phi_slope =   ( EV[2]-EV[3] )/(2.0*epsilon)
    next_theta = theta - theta_slope*step
    next_phi = phi - phi_slope*step
    return next_theta,next_phi


def Two_Qubit_Ansatz(qc, q, params):
    '''
    Input:     qc (QuantumCircuit)   q (QuantumRegister)   params (array)
    Applies the neccessary rotation and CNOT gates for a two qubit ansatz state
    '''
    Single_Qubit_Ansatz( qc, q[0], [params[0], params[1]] )
    Single_Qubit_Ansatz( qc, q[1], [params[2], params[3]] )
    qc.cx( q[0], q[1] )
    Single_Qubit_Ansatz( qc, q[0], [params[4], params[5]] )
    Single_Qubit_Ansatz( qc, q[1], [params[6], params[7]] )


def Calculate_MinMax(V, C_type):
    '''
    Input:     V (vert)   C_type (string)
    Returns the smallest or biggest value / index for the smallest value in a list
    '''
```

```
    if( C_type == 'min' ):
        lowest = [V[0],0]
        for i in np.arange(1,len(V)):
            if( V[i] < lowest[0] ):
                lowest[0] = V[i]
                lowest[1] = int(i)
        return lowest
    if( C_type == 'max' ):
        highest = [V[0],0]
        for i in np.arange(1,len(V)):
            if( V[i] > highest[0] ):
                highest[0] = V[i]
                highest[1] = int(i)
        return highest


def Compute_Centroid(V):
    '''
    Input:     V (array)
    Computes and returns the centroid from a given list of values
    '''
    points = len( V )
    dim = len( V[0] )
    Cent = []
    for d in np.arange( dim ):
        avg = 0
        for a in np.arange( points ):
            avg = avg + V[a][d]/points
        Cent.append( avg )
    return Cent


def Reflection_Point(P1, P2, alpha):
    '''
    Input:     P1 (array)   P2 (array)   alpha (float)
    Computes a reflection point from P1 around point P2 by an amount alpha
    '''
    P = []
    for p in np.arange( len(P1) ):
        D = P2[p] - P1[p]
        P.append( P1[p]+alpha*D )
    return P


def VQE_EV(params, Ansatz, H, EV_type, **kwargs):
    '''
    Input:     params (array)   Ansatz( Single or Two Qubit Ansatz function)   H (Dictionary)   EV_type (string)
    Keyword Arguments:     shots (integer) -  Dictates the number of measurements to use per computation
    Computes and returns the expectation value for a given Hamiltonian and set of theta / phi values
    '''
    Shots = 10000
    if 'shots' in kwargs:
        Shots = int( kwargs['shots'] )
    Hk = list( H.keys() )
    H_EV = 0
    for k in np.arange( len(Hk) ):
        L = list( Hk[k] )
        q = QuantumRegister(len(L))
        c = ClassicalRegister(len(L))
        qc= QuantumCircuit(q,c)
        Ansatz( qc, q, params )
        sv0 = execute( qc, S_simulator, shots=1 ).result().get_statevector()
        if( EV_type == 'wavefunction' ):
            for l in np.arange( len(L) ):
                if( L[l] == 'X' ):
                    qc.x( q[int(l)] )
                if( L[l] == 'Y' ):
                    qc.y( q[int(l)] )
                if( L[l] == 'Z' ):
                    qc.z( q[int(l)] )
            sv = execute( qc, S_simulator, shots=1 ).result().get_statevector()
            H_ev = 0
            for l2 in np.arange(len(sv)):
                H_ev =  H_ev +  (np.conj(sv[l2])*sv0[l2]).real
            H_EV = H_EV + H[Hk[k]] * H_ev
        elif( EV_type == 'measure' ):
            for l in np.arange( len(L) ):
                if( L[l] == 'X' ):
                    qc.ry(-m.pi/2,q[int(l)])
                if( L[l] == 'Y' ):
                    qc.rx( m.pi/2,q[int(l)])
            qc.measure( q,c )
            M = Measurement( qc, shots=Shots, print_M=False, return_M=True )
            Mk = list( M.keys() )
            H_ev = 0
            for m1 in np.arange(len(Mk)):
                MS = list( Mk[m1] )
                e = 1
                for m2 in np.arange(len(MS)):
```

```
                if( MS[m2] == '1' ):
                    e = e*(-1)
            H_ev = H_ev + e * M[Mk[m1]]
        H_EV = H_EV + H[Hk[k]]*H_ev/Shots
    return H_EV


def Nelder_Mead(H, Ansatz, Vert, Val, EV_type):
    '''
    Input:     H (Dictionary)   Ansatz( Single or Two Qubit Ansatz function)   Vert (array)   Val (array)   EV_type (string
    Computes and appends values for the next step in the Nelder_Mead Optimization Algorithm
    '''
    alpha = 2.0
    gamma = 2.0
    rho   = 0.5
    sigma = 0.5
    add_reflect = False
    add_expand = False
    add_contract = False
    shrink = False
    add_bool = False
#-----------------------------------------
    hi = Calculate_MinMax( Val,'max' )
    Vert2 = []
    Val2 = []
    for i in np.arange(len(Val)):
        if( int(i) != hi[1] ):
            Vert2.append( Vert[i] )
            Val2.append( Val[i] )
    Center_P  = Compute_Centroid( Vert2 )
    Reflect_P = Reflection_Point(Vert[hi[1]],Center_P,alpha)
    Reflect_V = VQE_EV(Reflect_P,Ansatz,H,EV_type)
#--------------------------------------------------          # Determine if: Reflect / Expand / Contract / Shrink
    hi2 = Calculate_MinMax( Val2,'max' )
    lo2 = Calculate_MinMax( Val2,'min' )
    if( hi2[0] > Reflect_V >= lo2[0] ):
        add_reflect = True
    elif( Reflect_V < lo2[0] ):
        Expand_P = Reflection_Point(Center_P,Reflect_P,gamma)
        Expand_V = VQE_EV(Expand_P,Ansatz,H,EV_type)
        if( Expand_V < Reflect_V  ):
            add_expand = True
        else:
            add_reflect = True
    elif( Reflect_V > hi2[0] ):
        if( Reflect_V < hi[0] ):
            Contract_P = Reflection_Point(Center_P,Reflect_P,rho)
            Contract_V = VQE_EV(Contract_P,Ansatz,H,EV_type)
            if( Contract_V < Reflect_V ):
                add_contract = True
            else:
                shrink = True
        else:
            Contract_P = Reflection_Point(Center_P,Vert[hi[1]],rho)
            Contract_V = VQE_EV(Contract_P,Ansatz,H,EV_type)
            if( Contract_V < Val[hi[1]] ):
                add_contract = True
            else:
                shrink = True
#--------------------------------------------------     # Apply: Reflect / Expand / Contract / Shrink
    if( add_reflect == True ):
        new_P = Reflect_P
        new_V = Reflect_V
        add_bool = True
    elif( add_expand == True ):
        new_P = Expand_P
        new_V = Expand_V
        add_bool = True
    elif( add_contract == True ):
        new_P = Contract_P
        new_V = Contract_V
        add_bool = True
    if( add_bool ):
        del Vert[hi[1]]
        del Val[hi[1]]
        Vert.append( new_P )
        Val.append( new_V )
    if( shrink ):
        Vert3 = []
        Val3 = []
        lo = Calculate_MinMax( Val,'min' )
        Vert3.append( Vert[lo[1]] )
        Val3.append( Val[lo[1]] )
        for j in np.arange( len(Val) ):
            if( int(j) != lo[1] ):
                Shrink_P = Reflection_Point(Vert[lo[1]],Vert[j],sigma)
                Vert3.append( Shrink_P )
                Val3.append( VQE_EV(Shrink_P,Ansatz,H,EV_type) )
        for j2 in np.arange( len(Val) ):
```

```
        del Vert[0]
        del Val[0]
        Vert.append( Vert3[j2] )
        Val.append( Val3[j2] )
```